\documentclass[amsfonts,amsmath,amssymb,final,3p,times,nofootinbib]{elsarticle}

\usepackage{marginnote}
\usepackage{xcolor}
\usepackage{endnotes}
\definecolor{mycolor}{RGB}{139,0,255}
\definecolor{mycolor2}{RGB}{3,168,158}

\definecolor{red}{RGB}{0,0,0}
\newcommand{\mar}[3][0cm]{}
\newcommand{\com}[2][0cm]{}
 \newcommand{\marendnote}[3][0cm]{}

\usepackage{longtable,comment,braket}
\usepackage{graphicx}
\usepackage{amssymb}
\usepackage{amsfonts,amsmath,mathrsfs}
\usepackage{booktabs}
\usepackage{bm}
\usepackage{color}
\usepackage{multirow}
\usepackage{arydshln}
\usepackage{rotating}
\usepackage{threeparttable}
\usepackage{slashed}
\usepackage{epsfig}
\usepackage{mathtools}
\usepackage[resetlabels]{multibib}
\usepackage[colorlinks, citecolor=blue,anchorcolor=red,menucolor=red, linkcolor=red,filecolor=red,runcolor=red,urlcolor=blue,frenchlinks=red]{hyperref}
\usepackage{orcidlink}


\def\Tr{\textnormal{Tr}}
\def\SU{\textnormal{SU}}

\def\DsI{D_{s0}^*(2317)}
\def\DsII{D_{s1}(2460)}
\def\nopi{$\slashed{\pi}$EFT }
\def\xthn{D^0\bar{D}^{*0}/D^{*0}\bar{D}^0 }
\def\xthc{D^+D^{*-}/D^{*+}D^- }
\def\tth{D^{*+}D^0 }

\def\DastDbar{D\bar{D}^{*}/D^{*}\bar{D}}
\def\etal{\textit{et al }}
\def\lnb{\overleftarrow{\bm{\nabla}}}
\def\rnb{\overrightarrow{\bm{\nabla}}}
\def\lrnb{\overleftrightarrow{\bm{\nabla}}}
\def\cvI{convention-I }
\def\cvII{convention-II }

\allowdisplaybreaks[4]
\journal{Physics Reports}

\begin{document}


\begin{frontmatter}

\title{Chiral perturbation theory for heavy hadrons and chiral effective field theory for heavy hadronic molecules}

\author[RUB]{Lu Meng\,\orcidlink{0000-0001-9791-7138}\footnotemark[1]}
\ead{lu.meng@rub.de}

\author[HBU,KLHCAQFT,RCCPHP]{Bo Wang\,\orcidlink{0000-0003-0985-2958}\footnotemark[1]}
\ead{ wangbo@hbu.edu.cn}

\author[JAEA,KEK]{Guang-Juan Wang\,\orcidlink{0000-0001-9265-6741}\footnotemark[1]}
\ead{wgj@post.kek.jp}

\author[PKU]{Shi-Lin Zhu\,\orcidlink{0000-0002-4055-6906}\corref{cor1}}
\cortext[cor1]{Corresponding author.}
\ead{zhusl@pku.edu.cn}

\address[RUB]{Institute of theoretical physics II, Department of physics and astronomy, Ruhr-University Bochum, 44780 Bochum, Germany}
\address[HBU]{School of Physical Science and Technology, Hebei University, Baoding 071002, China}
\address[KLHCAQFT]{Key Laboratory of High-precision Computation and Application of Quantum Field Theory of Hebei Province, Baoding 071002, China}
\address[RCCPHP]{Research Center for Computational Physics of Hebei Province, Baoding, 071002, China}
\address[JAEA]{Advanced Science Research Center, Japan Atomic Energy Agency, Tokai, Ibaraki 319-1195, Japan}
\address[KEK]{ KEK Theory Center, Institute of Particle and Nuclear Studies (IPNS), High Energy Accelerator Research Organization (KEK), 1-1 Oho, Tsukuba, Ibaraki, 305-0801, Japan}
\address[PKU]{School of Physics and Center of High Energy Physics, Peking University, Beijing 100871, China}
\footnotetext[1]{These authors equally contribute to this work.}

\begin{abstract}
 
Chiral symmetry and its spontaneous breaking play an important role both in the light hadron and heavy hadron systems. The chiral perturbation theory ($\chi$PT) is the low energy effective field theory of the Quantum Chromodynamics. In this work, we shall review the investigations on the chiral corrections to the properties of the heavy mesons and baryons within the framework of $\chi$PT. We will also review the scatterings of the light pseudoscalar mesons and heavy hadrons, through which many new resonances such as the $D_{s0}^\ast(2317)$ could be understood.

Moreover, many new hadron states were observed experimentally in the past decades. A large group of these states is near-threshold resonances, such as the charged charmoniumlike $Z_c$ and $Z_{cs}$ states,  bottomoniumlike $Z_b$ states, hidden-charm pentaquark  $P_c$ and $P_{cs}$ states and the doubly charmed $T_{cc}$ state,  etc. They are very good candidates of the loosely bound molecular states composed of a pair of charmed (bottom) hadrons, which are very similar to the loosely bound deuteron. The modern nuclear force was built upon the chiral effective field theory ($\chi$EFT), which is the extension of the $\chi$PT to the systems with two matter fields. The long-range and medium-long-range interactions between two nucleons arise from the single- and double-pion exchange respectively, which are well constrained by the chiral symmetry and its spontaneous breaking. The short-distance interactions can be described by the low energy constants. Such a framework works very well for the nucleon-nucleon scattering and nuclei. In this work, we will perform an extensive review of  the progress on the heavy hadronic molecular states within the framework of $\chi$EFT. We shall emphasize that the same chiral dynamics not only govern the nuclei and forms the deuteron, but also dictates the shallow bound states or resonances composed of two heavy hadrons.

\end{abstract}


\end{frontmatter}


\tableofcontents 
\section{Brief overview of the heavy hadron spectroscopy}\label{sec:introduction}

 Quantum chromodynamics (QCD) is the theory of the strong interaction. Its basic constituents are the quarks and gluons. They carry the color degrees of freedom of the non-Abelian SU(3) color gauge group and interact with each other through the exchange of the gluons. QCD has three salient features: color confinement, asymptotic freedom, chiral symmetry and its spontaneous breaking. No isolated quarks and gluons have been observed. Instead, the quarks and gluons are confined within the colorless mesons and baryons. The interaction strength between the quarks and gluons increases as the energy scale of the interaction decreases. The perturbation in terms of the coupling constant is invalid in the low energy region.  {The QCD Lagrangian exhibits the exact $\text{SU(3)}_L\otimes \text{SU(3)}_R$ chiral symmetry when the current quark masses of the up, down and strange quarks vanish, which is spontaneously broken to the SU(3)$_V$. With the small current quark masses, QCD also exhibits the approximate SU(3) flavor symmetry. When the masses of the charm, bottom and top quarks become very large, QCD exhibits the approximate heavy quark spin and flavor symmetries. These symmetries play a crucial role in our understanding of the heavy hadrons and their strong interactions.}

The hadrons encode the underlying information of the QCD dynamics in the nonperturbative region. The hadron spectroscopy tightly connects the experimental measurements with various theoretical frameworks, such as the {\it ab-initio} lattice QCD simulations, the quark models, QCD sum rule, etc.
In the traditional quark model, the mesons and baryons are composed of the quark-antiquark pair ($q\bar q$) and three quarks $(qqq)$, respectively.
Most of the ground-state and excited hadrons can be successfully described by such a simple framework. 

In the past decades, the hadron spectroscopy has been revived by the observation of many excited heavy hadrons~\cite{ParticleDataGroup:2022pth}, the exotic $X,Y,Z$ states~\cite{ParticleDataGroup:2022pth}, and the hidden-charm pentaquark states~\cite{LHCb:2015yax,LHCb:2019kea} as well as the doubly charmed teraquark state $T_{cc}$~\cite{LHCb:2021vvq,LHCb:2021auc}, etc. Readers may find more experimental and theoretical details in the reviews~\cite{Chen:2016qju,Chen:2016spr,Lebed:2016hpi,Esposito:2016noz,Hosaka:2016pey,Guo:2017jvc,Ali:2017jda,Liu:2019zoy,Brambilla:2019esw,Lucha:2021mwx,Chen:2021ftn,Chen:2022asf}. The traditional quark model failed badly for most of these states. In general, QCD allows the existence of the more complicated forms of structures for the hadrons, such as the hybrid meson, the glueball, the tetraquark state, the pentaquark state, and hadronic molecule, etc. In fact, the multiquark states $(qq\bar q\bar q)$  and $(qqqq\bar q)$ were proposed together with the conventional mesons and baryons by Gell-Mann~\cite{Gell-Mann:1964ewy} and Zweig~\cite{Zweig:1964jf}. These newly observed structures might be the good candidates of the exotic states and contain the missing piece of knowledge about the nonperturbative strong interactions. 

Most of the new hadron states are in the heavy sector. Thus, the heavy hadron spectroscopy is of great interest. The heavy hadrons are composed of the light quarks $q=u,d,s$ ($m_q\ll\Lambda_\chi$), and the heavy quarks $Q=c,b$ ($m_Q\gg\Lambda_\text{QCD}$). {Note that the top quark is not considered because it decays into the bottom quark and $W$ boson rapidly and its life time is much shorter than the typical hadronization time scale.} Therefore, the heavy hadrons synchronously possess the chiral symmetry for the light quarks as well as the heavy quark symmetry for the  heavy quarks. The heavy hadrons are ideal objects to study the dynamics between the light and heavy quarks, and explore the chiral symmetry and heavy quark symmetry simultaneously. In the following, we will give a brief review about the heavy hadron spectroscopy, especially the states which will be analyzed carefully in the following sections. 

\subsection{Singly heavy mesons} \label{sec1.1}

\begin{figure}[h]
\begin{center}
\includegraphics[width=0.98\textwidth]{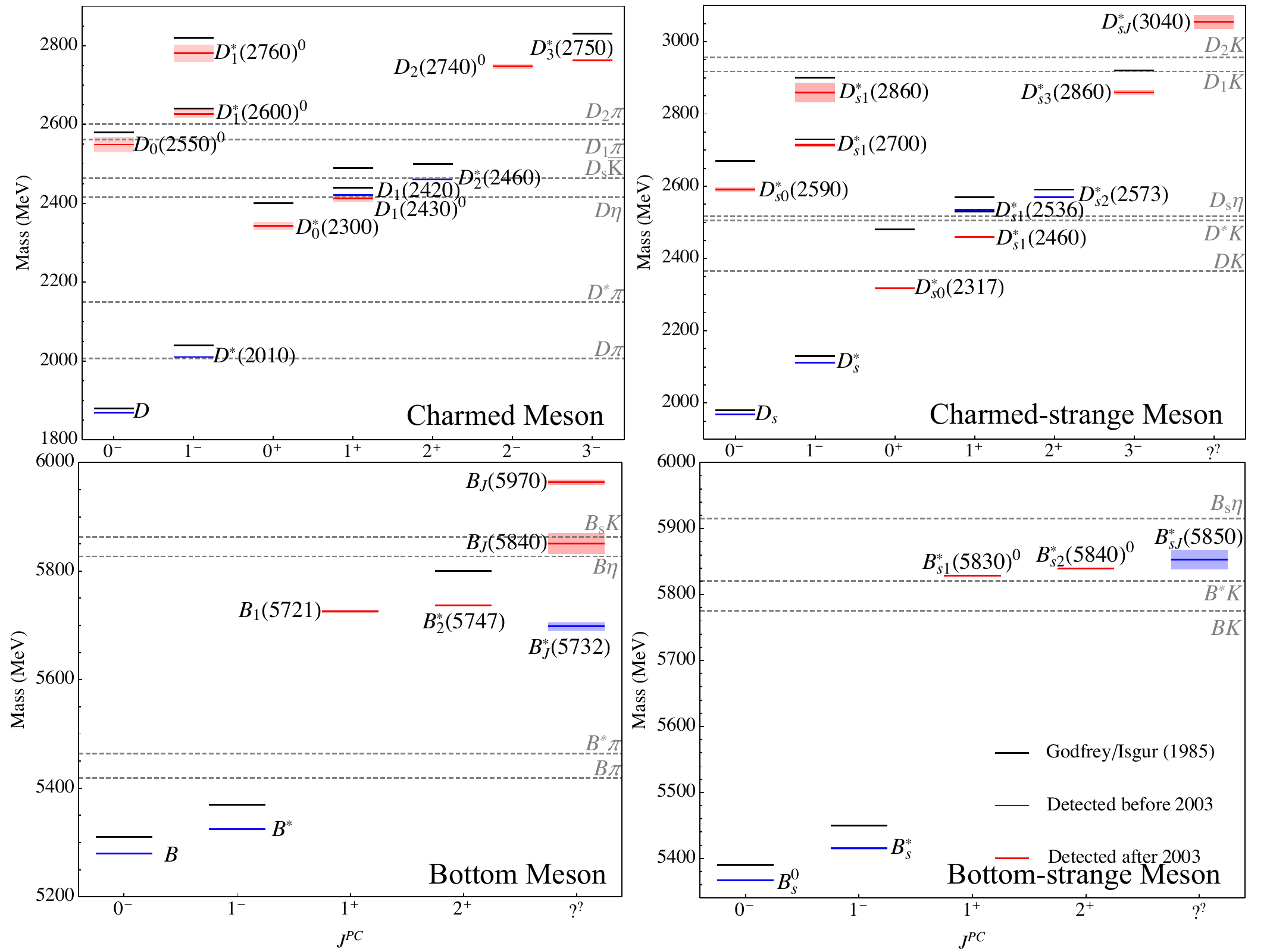}
\end{center}
\caption{The charmed and bottom mesons from the Review of Particle Physics~\cite{ParticleDataGroup:2022pth}. The colored band denotes the error of the mass. The dashed horizontal line stands for the threshold. The $x$- and $y$-axis represent the $J^P$ quantum numbers and masses (in units of MeV), respectively.}
\label{fig:hms}
\end{figure}

{A singly heavy meson is composed of a heavy quark $Q$  and a light antiquark $\bar{q}$ from the perspective of the quark model. From the point of view of heavy quark effective field theory (HQEFT), an open flavor meson contains a heavy quark and a light quark cloud with the quantum numbers of a light antiquark.} In Fig.~\ref{fig:hms}, we present the masses of the heavy mesons collected from the Review of Particle Physics (RPP)~\cite{ParticleDataGroup:2022pth}. For comparison, we also list the predictions of the Godfrey-Isgur quark model (GI model)~\cite{Godfrey:1985xj}. The hyperfine splittings between the ground $1^-$ and $0^-$ heavy mesons are measured precisely~\cite{ParticleDataGroup:2022pth}
\begin{eqnarray}
&&m_{D^{*}(2007)^{0}}-m_{D^{0}}=(142.014\pm0.030)~\text{MeV},\\ \label{eq:hmms1}
&&m_{D^{*}(2010)^{+}}-m_{D^{+}}=(140.603\pm0.015)~\text{MeV},\\ \label{eq:hmms2}
&&m_{D_{s}^{*\pm}}-m_{D_{s}^{\pm}}=(143.8\pm0.4)~\text{MeV},\\ \label{eq:hmms3}
&&m_{B^{*}}- m_{B} =(45.21 \pm 0.21) ~\text{MeV},\\  \label{eq:hmms4}
&&m_{B_{s}^{*}}-m_{B_{s}} =( 48.6_{-1.5}^{+1.8}) ~\text{MeV}, \label{eq:hmms5}
\end{eqnarray}
where the ratio between the hyperfine splitting for the charmed and the bottom mesons is around $m_c/m_b$. The strange quark mass is around 90 MeV, which induces the $\SU(3)$ flavor symmetry breaking mass 
\begin{eqnarray}
&&m_{D_{s}^{\pm}}-m_{D^{\pm}}= (98.69 \pm 0.05) \mathrm{MeV},\\ \label{eq:hmfms1}
&&m_{B_{s}^{0}}-m_{B^0} =(87.38 \pm 0.16)~\text{MeV}. \label{eq:hmfms2}
\end{eqnarray}

The $P$-wave excited states are of special interest. In the quark model, there are four possible spin-orbital configurations  $^{(2S+1)}L_J^P=$ $^1 P_1^+$, $^3 P_0^+$,$^3 P_1^+$, $^3 P_2^+$ for the $P$-wave heavy  mesons with $J^P=1^+,~0^+,~1^+,~2^+$. The $ ^1 P_1^+$ and $^3 P_1^+$ components may mix with each other to form the $J^P=1^+ $ states through the spin-orbital potential. 
Since the heavy quark is much heavier than the light quarks, the heavy quark symmetry (HQS) is a good symmetry, especially for the bottom hadrons. In the heavy quark limit $m_{Q}\rightarrow \infty$, the light spin $j_\ell$ decouples with the heavy quark spin $s_Q$, where $\bm{j}_\ell=\bm{s}_q+ \bm{L}$. The  quantum numbers $j_\ell^{P}$, $s_Q$ and $J$ are  conserved separately. Therefore, one can categorize the heavy mesons with the light $j_\ell$ and heavy degrees of freedom (d.o.f) $s_Q$. The two categorization methods based on different bases can always be related to each other through the Clebsch–Gordan (CG) coefficients.  For the $P$-wave heavy mesons,
\begin{eqnarray}
P\text{-wave}:~~j_\ell^{P}&=&\frac{1}{2}^{+}, ~~J^{P}=0^{+}~\text{or} ~1^{+},\\
~~~~~~~~~~~~~j_\ell^{P}&=&\frac{3}{2}^{+},~~ J^{P}=1^{+}~\text{or}~2^{+}.\label{eq:jlp32}
\end{eqnarray}
 In the heavy quark limit, the mesons with the same $j^P_\ell$ are degenerate and can be treated as the heavy spin doublet (see Sec.~\ref{sec:1.5:combChandHQ}). {The decay behavior of the two doublets could be distinct. For instance, given large enough phase space, the $j_\ell^P=\frac{1}{2}^+$ and  $j_\ell^P=\frac{3}{2}^+$ $c\bar q$ doublets decay into the $D^{(*)} \pi$ modes via  $S$-wave and $D$-wave in the heavy quark limit, respectively. Hence, the $j_\ell^P=\frac{1}{2}^+$ doublet is generally expected to be broad while the $j_\ell^P=\frac{3}{2}^+$ doublet is 
narrow.}

There are four non-strange $P$-wave charm mesons, the $D_{0}^{*}(2300)$ [{{known as $D_{0}^{*}(2400)$ previously}}], $D_{1}(2420)$, $D_{1}(2430)^{0}$ and $D_{2}^{*}(2460)$~\cite{ParticleDataGroup:2022pth}. 
The $D_{1}(2420)$ is narrow with a width around $(31.3 \pm 1.9)$ MeV, while the $D_{1}(2430)$ is very broad with a width around $(314 \pm 29)$ MeV. The $D^*_{0}(2300)$ is also very broad and its mass from different experimental collaborations varies from $2300$ MeV to $2400$ MeV~\cite{ParticleDataGroup:2022pth}.

In the charm-strange sector, the four $P$-wave states are $D_{s0}^*(2317)$, $D_{s 1}(2460)$, $D_{s 1}(2536)$, $D_{s 2}^{*}(2573)$. All the four states are very narrow. The masses and narrow widths of the $D_{s 1}(2536)$ and $D_{s 2}^{*}(2573)$ are consistent with the theoretical predictions for the $(1^+, 2^+)$ doublet~\cite{Godfrey:1986wj}.
But the low mass and extremely narrow width of 
the $D_{s0}^*(2317)$ and $D_{s 1}(2460)$ states are very puzzling.
 
In 2003, the $D_{s0}^*(2317)$ was first observed in the $ D_{s}^{+} \pi^{0}$ channel by BaBar Collaboration~\cite{BaBar:2003oey}. Later, its axial-vector partner state $D_{s 1}(2460)$ was observed by the CLEO Collaboration~\cite{CLEO:2003ggt}. They were confirmed by Belle and BaBar Collaborations~\cite{Belle:2003kup,BaBar:2006eep,BaBar:2003cdx}. Their masses are $(2317.8 \pm 0.5)$ MeV and $(2459.5 \pm 0.6)$ MeV, respectively~\cite{ParticleDataGroup:2022pth}. In particular, the $D_{s0}^*(2317)$ and $D_{s1}(2460)$ lie below the $DK$ and $D^*K$ thresholds, respectively. The only allowed strong decay channels are the $D^{(*)}_s\pi$ which break the isospin symmetry and their decay widths are therefore extremely narrow. Since the discoveries, these two states inspired strong interests in the charm-strange mesons due to the following puzzles: (i)
the masses of the $D_{s0}^*(2317)$ and $D_{s1}(2460)$ states are significantly  lighter than the quark model predictions; (ii)
the mass splitting $m_{D_{s 1}(2460)}-m_{D_{s0}^{*}(2317)}=141.7~\text{MeV}$ is equal to $ m_{D^*}-m_D$ within $2$ MeV [(]the $P$-wave $(0^+, 1^+)$ states and the ground $D_s$ states belong to different heavy quark spin doublets and their hyperfine splittings are unrelated by any symmetry]; (iii)
the mass hierarchy  $m_{D_{s 0}^{*}(2317)}-m_{D_{0}^{*}(2300)}=-25.2$ MeV and $M_{D_{s 1}(2460)}-M_{D_{1}(2430)}=47.5$ MeV are unnatural. 

In order to unveil the mysteries of the $D_{s0}^*(2317)$ and $D_{s1}(2460)$, there exist a lot of investigations in literature, including the quenched and unquenched $c\bar{s}$ quark model~\cite{Dai:2003yg, Hwang:2004cd,Simonov:2004ar,Cheng:2014bca,Song:2015nia,Dai:2006uz,Cheng:2017oqh,Luo:2021dvj,Zhou:2021uug,Tan:2021bvl}, the hadronic molecules ~\cite{Kolomeitsev:2003ac, Szczepaniak:2003vy,Hofmann:2003je,vanBeveren:2003kd,Barnes:2003dj,Gamermann:2006nm, Guo:2006rp, Guo:2006fu, Flynn:2007ki, Faessler:2007gv,Guo:2009ct,Xie:2010zza,Cleven:2010aw,Wu:2011yb,Guo:2015dha,Albaladejo:2016hae,Du:2017ttu,Guo:2018tjx,Albaladejo:2018mhb,Wu:2019vsy,Kong:2021ohg,Gregory:2021rgy}, the compact tetraquark states ~\cite{Cheng:2003kg,Chen:2004dy,Dmitrasinovic:2005gc,Kim:2005gt,Zhang:2018mnm}, and the mixing of the $c\bar{s}$ state and tetraquark state ~\cite{Terasaki:2003qa,Browder:2003fk,Maiani:2004vq,Dai:2006uz}. The readers may consult the reviews~\cite{Chen:2016spr,Dong:2017gaw,Guo:2017jvc,Yao:2020bxx,Chen:2022asf} for more details. We will review the $P$-wave states with the chiral unitary approaches in Sec.~\ref{sec:sec4}. 

\subsection{Heavy baryons}\label{sec1.2}

{In the language of the quark model}, the heavy baryons are composed of three quarks, which give rises to a very intricate spectroscopy. According to the number of the heavy quarks, the heavy baryons can be classified into the singly heavy baryon $Qqq$ ($B_Q$), doubly heavy baryon $QQq$ ($B_{QQ}$) and triply heavy baryon $QQQ$ $(B_{QQQ})$, respectively. {In the view of HQEFT, these three types of heavy baryons consist of the singly, doubly and triply heavy quarks accompanied by the light quark cloud.}

\begin{figure*}[h]
	\centering
	\includegraphics[width = 1.0\textwidth]{{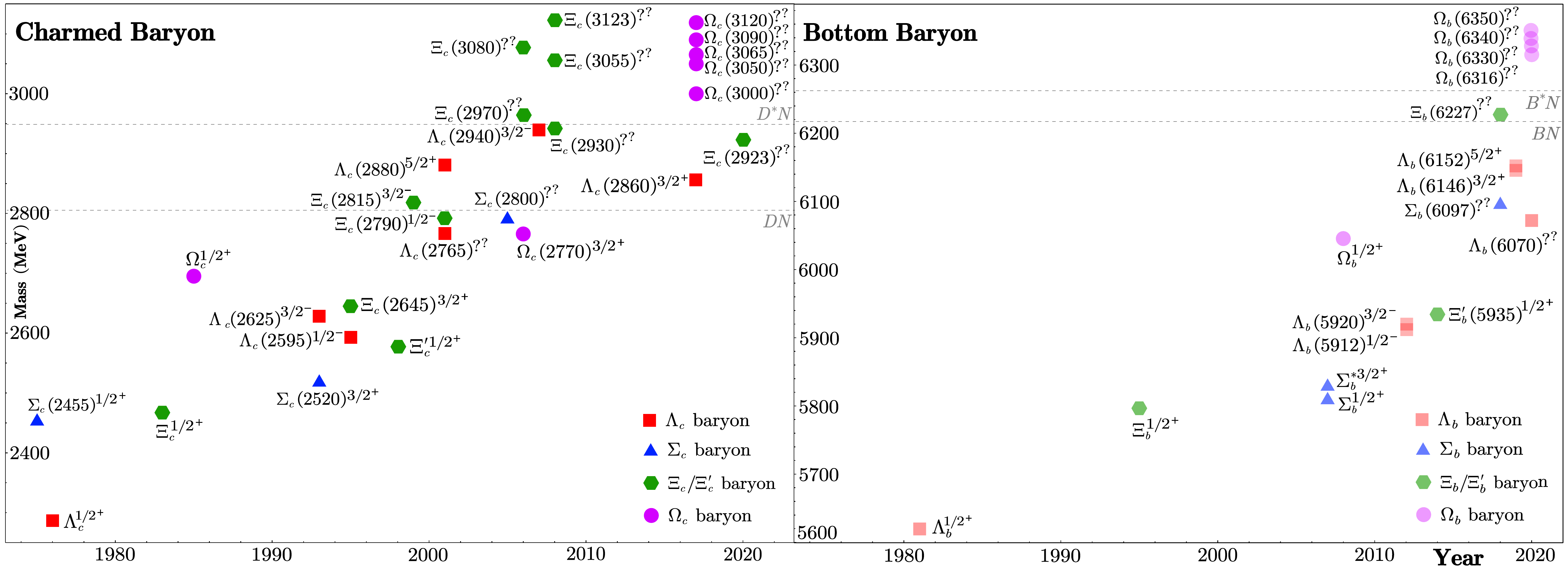}}
	 \caption{The charmed and bottom baryons from the Review of Particle Physics~\cite{ParticleDataGroup:2022pth}. The $x$- and $y$-axis denote the year of  the observation and mass (in units of MeV), respectively. Here, the mass errors are not presented.}\label{fig:hsbe}
\end{figure*}

For the singly heavy baryons, the two light quarks are in the antisymmetric $\bar {\bm3}_f$ or the symmetric $\bm{6}_f$ flavor representations in the SU(3) flavor symmetry. Experimentally, $30$ singly charmed baryons and $22$ bottom baryons have been established now~\cite{ParticleDataGroup:2022pth}, which are shown in Fig.~\ref{fig:hsbe}. In the review, we focus on the ground-state singly/doubly heavy baryons (see Sec.~\ref{sec.3}) and the possible hadronic molecules composed of the singly/doubly heavy baryons and other hadrons (see Sec.~\ref{sec:Pc} and Sec.~\ref{sec:othersystems}).

In the ground-state $S$-wave singly heavy baryons, the two light quarks form the diquark with the total spin $0$ and $1$ as shown in Table~\ref{tab:hsbd}. The diquark then combines with the heavy quark to form the spin-$\frac{1}{2}$ anti-triplet heavy baryons which include the ($\Lambda_{c}^{+}$, $\Xi_{c}^{0}$, $\Xi_{c}^{+}$ )/($\Lambda_{b}^{0}$, $\Xi_{b}^-$, $\Xi_{b}^{0}$), and the sextet spin-$\frac{1}{2}$ or $\frac{3}{2}$ heavy baryons which are denoted as  the ($\Sigma_{c}^{++}$, $ \Sigma_{c}^{+}$, $\Sigma_{c}^{0}$, $\Xi_{c}^{\prime+} $, $ \Xi_{c}^{\prime 0}$, $ \Omega_{c}^{0}$)/ ($\Sigma_{b}^{+}$,  $\Sigma_{b}^{0}$,  $\Sigma_{b}^{-}$, $\Xi_{b}^{\prime 0}$, $\Xi_{b}^{\prime-}$, $\Omega_{b}^{-}$) in the charmed/bottom sectors. 

 In the $\Sigma_Q$ sector, the {$(\Sigma_{c}(2455), \Sigma_{c}(2520))$} and $(\Sigma_{b}, \Sigma_{b}^{*})$, form the $S$-wave doublet $(\frac{1}{2}^{+},\frac{3}{2}^{+})$ as listed in Table~\ref{tab:hsbd}. In the $\Xi_Q$ sector, since the three quarks in the $\Xi_Q$ heavy baryons are different, the  possible configurations are quite rich. The $\Xi_Q$  can be in the antisymmetric $\bar{ \bm{3}}_f$ or symmetric $\bm{6}_f$ flavor representations as shown in Table~\ref{tab:hsbd}, which are analogous to the $\Lambda_Q$ and $\Sigma_Q$, respectively. 
The two lowest  $\Xi_Q$ baryons are in the  $\bar{ \bm{3}}_f$  representation and can only decay weakly.  The $\Xi_{c}^{\prime}$ and $\Xi_{c}(2645)$, as well as $\Xi_{b}^{\prime}(5935)$  and $\Xi_{b}(5955)^{-}$/$\Xi_{b}(5945)^{0}$ are in the symmetric $\bm{6}_f$ representation and form the $S$-wave  $\Xi'_c\left(\frac{1}{2}^{+}, \frac{3}{2}^{+}\right)$ doublet. Constrained by the phase space, the $\Xi_{c}^{\prime}$ has no strong decay modes but can decay radiatively. In the $\Omega_c$ sector, the lowest $\Omega_c$ and $\Omega_{c}(2770)^{0}$ form the $S$-wave doublet $\left(\frac{1}{2}^{+}, \frac{3}{2}^{+}\right)$  as shown in Table~\ref{tab:hsbd}. The former one can only decay weakly, and the latter one mainly decays into the $\Omega_{c}^{0} \gamma$ since the mass difference between the  $\Omega_{c}(2770)^{0}$ and $\Omega_{c}^{0}$ is too small to permit the strong decays. In the bottom sector, only the lowest $\Omega_b^-$ with $J^P=\frac{1}{2}^{+}$ state was observed~\cite{D0:2008sbw} while its partner state with $J^P=\frac{3}{2}^{+} $ is still absent. These ground-state singly heavy baryons are understood very well as the conventional $Qqq$ baryons in the quark model.

 { In quark model, a heavy baryon, composed of three quarks  $Qqq$, has}  two orbital excitation modes, the $\rho$-mode with the orbital excitation (${\bm L}_\rho$) between the two light quarks, and the $\lambda$-mode one with the orbital excitation ${\bm L}_\lambda$ between the heavy quark and the light diquark. They combine to form the total orbital angular momentum 
$\bm L={\bm L}_\rho+{\bm L}_\lambda$, which then couples with the spin of the two light quarks ($\bm s_{qq} $) and the heavy quarks ($\bm s_Q$) to form the total spin $\bm J$ of the heavy baryon. Due to the different excitation modes, there are multiple candidates for the singly baryons with specific $I(J^P)$ quantum numbers as shown in Table~\ref{tab:hsbd}. { In the HQEFT, only the quantum numbers of the light quark cloud $j^P_{\ell}$ matter. }

Up to now, the knowledge about the excited singly heavy baryons is still quite poor. Among them, the $\Lambda_{c}(2940)^{+}$ and $\Sigma_{c}(2800)$ attracted much attention since they are located  very close to the $ND^*$ ($2946$ MeV) and $ND$ ($2805$ MeV) thresholds, respectively, which leads to the molecular interpretations as discussed in Sec.~\ref{sec:5.8.1}. 

The $\Lambda_{c}(2940)^{+}$ was reported by the BaBar Collaboration in the invariant mass spectrum of the $D^{0}p$ channel~\cite{BaBar:2006itc}. No signal was seen in the $D^+p$ final state. Therefore the isospin of the $\Lambda_c(2940)^+$ equals $0$. Later, the Belle Collaboration confirmed the $\Lambda_{c}(2940)^{+}$ in the $\Lambda_{c}^{+} \pi^{+} \pi^{-}$ final state~\cite{Belle:2006xni}. In 2017, the LHCb Collaboration analyzed the amplitude  of the decay $\Lambda_{b}^{0} \rightarrow {D}^{0} {p} \pi^{-}$ and found the most likely spin-parity assignment of the $\Lambda_{c}(2940)^{+}$ is $\frac{3}{2}^-$, but the assignments with spin ${1\over 2}$ to ${7\over 2}$ cannot be excluded~\cite{LHCb:2017jym}. {Recently, the $\Lambda_c(2910)$ was reported by the Belle Collaboration in the decay process $\bar{B}^0\to\Sigma_c(2455)\pi \bar{p}$~\cite{Belle:2022hnm}. Its mass and width are measured to be $2913.8\pm5.6\pm3.8$ MeV and $51.8\pm20.0\pm18.8$ MeV, respectively}.

In 2005, the Belle Collaboration observed the $\Sigma_{c}(2800)$ in the $ \Lambda_{c} \pi $ channel~\cite{Belle:2004zjl}, which was confirmed by BaBar Collaboration in the  $\Lambda_{c}^{+} \bar{p}$ channel~\cite{BaBar:2008get}, but with the measured mass about $50$ MeV larger. Up to now, the spin-parity quantum numbers of the $\Sigma_{c}(2800)$ have not been determined yet. 

The interpretations of the $\Lambda_{c}(2940)$ and $\Sigma_{c}(2800)$ are still controversial. 
The mass of the $\Lambda_{c}(2940)^{+}$ is consistent with the first radial or $P$-wave excitation of the $\Lambda_c$ considering the uncertainty of the quark model around tens of MeV~\cite{Valcarce:2008dr, Ebert:2007nw,Roberts:2007ni,Lu:2018utx}. 
The $\Sigma_c(2800)$ may be the $P$-wave excitation~\cite{Ebert:2011kk,Cheng:2006dk,Chen:2016iyi,Garcilazo:2007eh,Wang:2017kfr}.
However, the $ \Lambda_{c}(2940)^{+}$ is only $6$ MeV lower than the $D^{*0}p$ threshold, which inspired various $D^\ast N$ molecular interpretations~\cite{He:2006is, He:2010zq,Ortega:2012cx,Dong:2009tg,Dong:2010xv,Zhang:2012jk,Zhao:2016zhf,Zhang:2019vqe, Wang:2020dhf}. Similar to the $ \Lambda_{c}(2940)^{+}$, the $\Sigma_c(2800)$ is located just below the $DN$ threshold and was proposed as a candidate of the $DN$ molecule~\cite{Zhao:2016zhf,Zhang:2019vqe,Wang:2020dhf,Dong:2010gu,Jimenez-Tejero:2009cyn}. 
More discussions are referred to reviews~\cite{Klempt:2009pi,Crede:2013kia,Cheng:2015iom,Chen:2016spr,Kato:2018ijx,Cheng:2018rkz,Chen:2022asf}. In the past several years, many excited heavy baryons including some $P$-wave states have been observed by LHCb Collaboration. An extensive review of these states can be found in Ref. \cite{Chen:2022asf}. 

In literature, the singly heavy baryon spectroscopy have been studied in different frameworks, such as the various quark models~\cite{Copley:1979wj,Ebert:2007nw,Yoshida:2015tia,Capstick:1986bm,Ebert:2011kk,Ebert:2007nw,Ebert:2005xj,Shah:2016nxi,Lu:2016ctt,Gandhi:2019xfw,Faustov:2020gun,Roberts:2007ni,Copley:1979wj,Garcilazo:2007eh,Migura:2006ep,Yang:2017qan}, QCD sum rule~\cite{Bagan:1991sc, Liu:2007fg,Chen:2015kpa,Chen:2016phw,Wang:2007sqa,Jin:2001jx,Aliev:2018lcs}, bag models~\cite{Hasenfratz:1980ka,Izatt:1981pt,Aliev:2010ev,Hwang:1986ee}, nonperturbative string approach~\cite{Narodetskii:2001bq}, effective field theory incorporating different symmetries~\cite{Jenkins:1996de,Heo:2018qnk,Ito:1996mr,Kim:2020imk,Harada:2019udr}, Skyrme model~\cite{Oh:1995ey}, the Faddeev method~\cite{Valcarce:2008dr}, the relativistic flux tube model~\cite{Chen:2014nyo}, lattice QCD~\cite{UKQCD:1996ssj,AliKhan:1999yb,Flynn:2003vz,Liu:2009jc,Detmold:2011bp,Lin:2011ti,Alexandrou:2012xk,Briceno:2012wt,Detmold:2012ge,Brown:2014ena,Olamaei:2021eyo,Alexandrou:2017xwd,Durr:2012dw,Brown:2014ena,PACS-CS:2013vie,Chen:2017kxr,Briceno:2012wt,Alexandrou:2014sha,Perez-Rubio:2015zqb,Bahtiyar:2020uuj,Mathur:2018rwu,Padmanath:2013zfa,Padmanath:2015jea,Padmanath:2013bla,Padmanath:2014lvr} and so on. Besides the mass spectrum, the other properties, such as the strong and radiative decays, magnetic moments have also been widely studied~\cite{Cheng:2006dk,Cheng:2015naa, Chen:2017aqm,Gong:2021jkb,Lu:2019rtg, Guo:2019ytq,Yao:2018jmc,Lu:2018utx,Yang:2019cvw,Chen:2017sci,Cheng:2015naa,Wang:2021bmz,Bijker:2020tns,Can:2021ehb,Shi:2021kmm,Bahtiyar:2021voz}. More detailed discussions about the singly heavy baryons can be found in Refs.~\cite{Chen:2016spr,Crede:2013kia,Cheng:2021qpd,Cheng:2015iom,Chen:2022asf}. 

\begin{table*}[htbp]
 \renewcommand\arraystretch{1.5}
  \caption{The possible configurations of the two light quarks in the singly heavy baryon {in the quark model.} The first five columns are the flavor-color-spatial-spin configurations of the two light quarks, which are antisymmetric under the light quark interchange. The {$L_{\rho}$ and $L_{\lambda}$ are the orbital angular momentum between the two light quarks and that between the light quark cluster and the heavy quark, respectively.} The $j^P_\ell$ and $s_Q$ denote the light and heavy d.o.f, respectively. The scripts ``S" and ``A"  represent the exchange symmetry and antisymmetry for the identical particles, respectively. }\label{tab:hsbd}
 \centering
 \setlength{\tabcolsep}{1.2mm}
 \begin{tabular}{cccccccccc}
\toprule[0.5pt]
\multicolumn{10}{c}{$S$-wave} \tabularnewline
\midrule[0.5pt]
Flavor & Color & $L_{\rho}$ &  $s_{qq}$ & $J_{qq}^{P}$ & $L_{\lambda}$ & $j_{\ell}^P$ & $s_{Q}$ & State & $J^{P}$\tabularnewline
$\bar{\bm{3}}_{f}$ (A) & $\bar{\bm{3}}_{c}$ (A) & 0(S) & $0$ (A) & $0^{+}$ & $0$ & $0^{+}$ & $\frac{1}{2}$ & $\Lambda_{Q}$, $\Xi_{Q}$ & $\frac{1}{2}^{+}$\tabularnewline
$\bm{6}_f$ (S) & $\bar{\bm{3}}_{c}$ (A) & 0 (S) & $1$ (S) & $1^{+}$ & $0$ & $1^{+}$ & $\frac{1}{2}$ & $\Sigma_{Q}$, $\Xi'_{Q}$, $\Omega_{Q}$ & $(\frac{1}{2}^{+},\frac{3}{2}^{+})$\tabularnewline
\midrule[0.5pt]
\multicolumn{10}{c}{$\lambda$-mode $P$-wave} \tabularnewline
\midrule[0.5pt]
Flavor & Color &  $L_{\rho}$ & $s_{qq}$ & $J_{qq}^{P}$ & $L_{\lambda}$ & $j_{\ell}^P$ & $s_{Q}$ & State & $J^{P}$\tabularnewline
$\bar{\bm{3}}_{f}$ (A) & $\bar{\bm{3}}_{c}$ (A) & 0 (S) & $0$ (A) & $0^{+}$ & $1$ & $1^{-}$ & $\frac{1}{2}$ & $\Lambda_{QS}$, $\Xi_{QS}$ & $(\frac{1}{2}^{-},\frac{3}{2}^{-})$\tabularnewline
$\bm{6}_f$ (S) & $\bar{\bm{3}}_{c}$ (A) & 0 (S) & $1$ (S) & $1^{+}$ & $1$ & $0^{-}/1^{-}/2^{-}$ & $\frac{1}{2}$ & $\Sigma_{QA}$, $\Xi'_{QA}$, $\Omega_{QA}$ & $\frac{1}{2}^{-}/(\frac{1}{2}^{-},\frac{3}{2}^{-})/ (\frac{3}{2}^{-},\frac{5}{2}^{-})$\tabularnewline
\midrule[0.5pt]
\multicolumn{10}{c}{$\rho$-mode $P$-wave} \tabularnewline
\midrule[0.5pt]
Flavor & Color & $L_{\rho}$ & $s_{qq}$ & $J_{qq}^{P}$ & $L_{\lambda}$ & $j_{\ell}^P$ & $s_{Q}$ & State & $J^{P}$\tabularnewline
\multirow{2}{*}{$\bar{\bm{3}}_{f}$ (A)} & \multirow{2}{*}{$\bar{\bm{3}}_{c}$ (A)} & \multirow{2}{*}{1 (A)} & \multirow{2}{*}{$1$ (S)} & $0^{-}$ & $0$ & $0^{-}$ & $\frac{1}{2}$ & $\Lambda_{QP}$, $\Xi_{QP}$ & $\frac{1}{2}^{-}$\tabularnewline
 &  &  &  & $1^{-}$ & $0$ & $1^{-}$ & $\frac{1}{2}$ & $\Lambda_{QV}$, $\Xi_{QV}$ & $(\frac{1}{2}^{-},\frac{3}{2}^{-})$\tabularnewline
$\bm{6}_f$ (S) & $\bar{\bm{3}}_{c}$ (A) & 1 (A) & $0$ (A) & $1^{-}$ & $0$ & $1^{-}$ & $\frac{1}{2}$ & $\Sigma_{QV}$, $\Xi'_{QV}$, $\Omega_{QV}$ & $(\frac{1}{2}^{-},\frac{3}{2}^{-})$\tabularnewline
\midrule[0.5pt]
\multicolumn{10}{c}{2-orbital excitations }  \tabularnewline
\midrule[0.5pt]
 Flavor & Color & $L_{\rho}$ & $s_{qq}$ & $J_{qq}^{P}$ & $L_{\lambda}$ & $j_{\ell}^P$ & $s_{Q}$ & State & $J^{P}$\tabularnewline
$\bar{\bm{3}}_{f}$ (A) & $\bar{\bm{3}}_{c}$ (A) & 2 (S) & $0$ (A) & $2^{+}$ & $0$ & $2^{+}$ & $\frac{1}{2}$ & $\Lambda_{Q2}$, $\Xi_{Q2}$ & $(\frac{3}{2}^{+},\frac{5}{2}^{+})$\tabularnewline
$\bm{6}_f$ (S) & $\bar{\bm{3}}_{c}$ (A) & 2 (S) & $1$ (S) & $1^{+}$/$2^{+}$/$3^{+}$ & $0$ & $1^{+}$/$2^{+}$/$3^{+}$ & $\frac{1}{2}$ & $\Sigma_{Q1/2/3}$, $\Xi'_{Q1/2/3}$, $\Omega_{Q1/2/3}$ & $\left(\frac{1}{2}^{+},\frac{3}{2}^{+}\right)/\left(\frac{3}{2}^{+},\frac{5}{2}^{+}\right)/\left(\frac{5}{2}^{+},\frac{7}{2}^{+}\right)$\tabularnewline
\midrule[0.5pt]  
Flavor & Color &$L_{\rho}$  & $s_{qq}$ & $J_{qq}^{P}$ & $L_{\lambda}$ & $j_{\ell}^P$ & $s_{Q}$ & State & $J^{P}$\tabularnewline
$\bar{\bm{3}}_{f}$ (A) & $\bar{\bm{3}}_{c}$ (A) & 0 (S) & $0$ (A) & $0^{+}$ & $2$ & $2^{+}$ & $\frac{1}{2}$ & $\tilde{\Lambda}_{Q2}$, $\tilde{\Xi}_{Q2}$ & $(\frac{3}{2}^{+},\frac{5}{2}^{+})$\tabularnewline
$\bm{6}_f$ (S) & $\bar{\bm{3}}_{c}$ (A) & 0 (S) & $1$ (S) & $1^{+}$ & $2$ & $1^{+}$/$2^{+}$/$3^{+}$ & $\frac{1}{2}$ & $\tilde{\Sigma}_{Q1/2/3}$, $\tilde{\Xi}'_{Q1/2/3}$, $\tilde{\Omega}_{Q1/2/3}$ & $\left(\frac{1}{2}^{+},\frac{3}{2}^{+}\right)/\left(\frac{3}{2}^{+},\frac{5}{2}^{+}\right)/\left(\frac{5}{2}^{+},\frac{7}{2}^{+}\right)$\tabularnewline
\midrule[0.5pt]  
Flavor & Color &
$L_{\rho}$ & $s_{qq}$ & $J_{qq}^{P}$ & $L_{\lambda}$ & $j_{\ell}^P$ & $s_{Q}$ & State & $J^{P}$\tabularnewline
\multirow{3}{*}{$\bar{\bm{3}}_{f}$ (A)} & \multirow{3}{*}{$\bar{\bm{3}}_{c}$ (A)} & \multirow{3}{*}{1 (A)} & \multirow{3}{*}{$1$ (S)} & $0^{-}$ & $1$ & $1^{+}$ & $\frac{1}{2}$ & $\tilde{\tilde{\Lambda}}_{Q1}^{0}$, $\tilde{\tilde{\Xi}}_{Q1}^{0}$ & $\left(\frac{1}{2}^{+},\frac{3}{2}^{+}\right)$\tabularnewline
 &  &  &  & $1^{-}$ & $1$ & $0^{+}$/$1^{+}$/$2^{+}$ & $\frac{1}{2}$ & $\tilde{\tilde{\Lambda}}_{Q0/1/2}^{1}$, $\tilde{\tilde{\Xi}}_{Q0/1/2}^{1}$ & $\frac{1}{2}^{+}/\left(\frac{1}{2}^{+},\frac{3}{2}^{+}\right)/\left(\frac{3}{2}^{+},\frac{5}{2}^{+}\right)$\tabularnewline
 &  &  &  & $2^{-}$ & $1$ & $1^{+}$/$2^{+}$/$3^{+}$ & $\frac{1}{2}$ & $\tilde{\tilde{\Lambda}}_{Q1/2/3}^{2}$, $\tilde{\tilde{\Xi}}_{Q1/2/3}^{2}$ & $\left(\frac{1}{2}^{+},\frac{3}{2}^{+}\right)/\left(\frac{3}{2}^{+},\frac{5}{2}^{+}\right)/\left(\frac{5}{2}^{+},\frac{7}{2}^{+}\right)$\tabularnewline
\midrule[0.5pt] 
$\bm{6}_f$ (S) & $\bar{\bm{3}}_{c}$ (A) & 1 (A) & $0$ (A) & $1^{-}$ & $1$ & $0^{+}$/$1^{+}$/$2^{+}$ & $\frac{1}{2}$ & $\tilde{\tilde{\Sigma}}_{Q0/1/2}$, $\tilde{\tilde{\Xi}}'_{Q0/1/2}$,
$\tilde{\tilde{\Omega}}_{Q0/1/2}$ & $\frac{1}{2}^{+}/\left(\frac{1}{2}^{+},\frac{3}{2}^{+}\right)/\left(\frac{3}{2}^{+},\frac{5}{2}^{+}\right)$\tabularnewline
\bottomrule[0.5pt]
\end{tabular}
\end{table*}

For the doubly heavy baryons, the SELEX Collaboration reported the $\Xi_{c c}(3520)^{+}$ in the $\Lambda_{c}^{+} K^{-} \pi^{+}$~\cite{SELEX:2002wqn} as well as the $p D^{+} K^{-}$ channels~\cite{SELEX:2004lln}. They also reported the $\Xi_{c c}^{++}(3460)$ and $\Xi_{c c}^{++}(3780)$ in the $\Lambda_{c}^{+} K^{-} \pi^{+} \pi^{+}$ channel~\cite{SELEX:2002xhn}. However, none of them has been confirmed by the other experimental collaborations~\cite{Ratti:2003ez,BaBar:2006bab,Belle:2006edu,LHCb:2013hvt}.

In 2017, the LHCb Collaboration reported the   $\Xi_{c c}^{++} $ in the $\Lambda_{c}^{+} K^{-} \pi^{+} \pi^{+}$  decay mode~\cite{LHCb:2017iph} and one year later in the  $\Xi_{c}^{+} \pi^{+}$ decay mode~\cite{LHCb:2018pcs}. Its mass is~\cite{LHCb:2019epo}
\begin{eqnarray}
m_{\Xi_{c c}^{++}}=3621.55 \pm 0.23 \text{ (stat) } \pm 0.30 \text{ (syst) } \mathrm{MeV},
\end{eqnarray}
and its lifetime~\cite{LHCb:2018zpl} is  
\begin{eqnarray}
\tau=(2.56 \pm 0.27) \times 10^{-13} \mathrm{~s}.
\end{eqnarray}

There have been extensive theoretical efforts to study the mass spectrum, the strong decay patterns, the electromagnetic properties, the life time and other properties of the doubly heavy baryon in different frameworks, such as the  bag models~\cite{Fleck:1989mb,Jaffe:1975us,Ponce:1978gk,He:2004px}, various quark models~\cite{Ebert:1996ec,Gerasyuta:1999pc,Itoh:2000um,Wang:2021rjk,Albertus:2006ya,Migura:2006ep,Roberts:2007ni}, Bethe-Salpeter equation~\cite{Weng:2010rb}, Born-Oppenheimer EFT~\cite{Castella:2021yru,Soto:2020xpm}, the Regge phenomenology~\cite{Wei:2015gsa,Oudichhya:2021yln}, QCD sum rule~\cite{Zhang:2008rt,Zhang:2009re,Wang:2010it,Wang:2010vn,Wang:2010hs,Wang:2021pua,Chen:2017sci},
heavy diquark effective theory~\cite{ Shi:2020qde}, and lattice QCD~\cite{Flynn:2003vz,Perez-Rubio:2015zqb,Bahtiyar:2020uuj}. 
 
In Sec.~\ref{sec.3}, we will review the investigations of the doubly heavy baryons using the chiral perturbative theory. Under the heavy diquark-antiquark symmetry (HDAS)~\cite{Savage:1990di}, the interactions between the doubly heavy baryons with the other hadrons can be related to the interactions of the heavy mesons. The doubly heavy  baryons may interact with other hadrons to form the exotic molecules similar to the heavy mesons. We will review these theoretical studies about the $B_{QQ}\varphi$ system in Sec.~\ref{sec:uchptothers}, and the $B_{QQ}D^{(*)}$, $B_{QQ}B_{Q}$ as well as $B_{QQ}B_{QQ}$ systems in Sec.~\ref{sec:othersystems}, respectively. 

\subsection{Heavy hadronic molecule candidates} \label{sec1.3}
\subsubsection{$X(3872)$ and $T_{cc}^+$} \label{sec1.3.1}

We focus on the experimental progresses after 2015. For the previous experimental results and huge theoretical efforts, we refer to Refs.~\cite{Chen:2016qju,Esposito:2016noz,Lebed:2016hpi,Hosaka:2016pey,Guo:2017jvc,Olsen:2017bmm,Liu:2019zoy,Brambilla:2019esw} for reviews.

{The $\chi_{c1}(3872)$, also known as $X(3872)$, was first observed by Belle Collaboration in 2003~\cite{Belle:2003nnu}. With the tremendous efforts of many experimental collaborations, its quantum numbers were measured to be $J^{PC}=1^{++}$ ~\cite{ParticleDataGroup:2022pth,LHCb:2013kgk}.} The $X(3872)$ is an excellent candidate of the exotic states. Its mass coincides exactly with the $\bar{D}^{*0} D^0/\bar{D}^0
D^{*0}$ threshold as $m_{D^0}+m_{D^{*0}}-m_{X(3872)}=(0.00\pm 0.18)$
MeV with a very narrow width $\Gamma=(1.19\pm 0.21)$ MeV~\cite{ParticleDataGroup:2022pth}. Whether the $X(3872)$ lies below or above the $\bar{D}^{*0} D^0/\bar{D}^0
D^{*0}$ threshold is still a pending question. {In fact, the above decay width was extracted from the Breit-Wigner fits~\cite{LHCb:2020xds,LHCb:2020fvo}.  Using a Fl\`atte-inspired model, the full width at the half maximum (FWHM) of the lineshape was determined to be $\mathrm{FWHM}  = 0.22^{\,+\,0.07\,+\,0.11}_{\,-\,0.06\,-\,0.13} \mathrm{MeV}$~\cite{LHCb:2020xds}.} As the $\chi_{c1}(2P)$ state, there is large discrepancy between the experimental mass of $X(3872)$ and the quark model predictions (e.g.~\cite{Godfrey:1985xj}). Another important feature of the $X(3872)$ is the large isospin
violating decay
patterns~\cite{Belle:2005lfc,BaBar:2010wfc,BESIII:2019qvy},
\begin{eqnarray}
&&\frac{{\cal B}[X\to J/\psi\pi^{+}\pi^{-}\pi^{0}]}{{\cal B}[X\to J/\psi\pi^{+}\pi^{-}]}=1.0\pm0.4\pm0.3\quad\text{Belle};\qquad
\frac{{\cal B}[X\to J/\psi\omega]}{{\cal B}[X\to
J/\psi\pi^{+}\pi^{-}]}=\begin{cases}
    1.6_{-0.3}^{+0.4}\pm0.2\quad\text{BESIII,}\\
    0.7\pm0.3\quad B^{+}\text{ events, BaBar,}\\
    1.7\pm1.3\quad B^{0}\text{ events, BaBar,}\label{eq:XisospinV}
\end{cases}
\end{eqnarray}
where the final states $J/\psi \pi^+\pi^-\pi^0$ and $J/\psi\pi^+\pi^-$ are mainly driven by the intermediate states $J/\psi \omega$ and $J/\psi \rho$, respectively. 

The theoretical interpretations of the $X(3872)$ include the $\chi_{c1}(2P)$ state, the $D^*\bar{D}/\bar{D}^*D$ molecular state, the mixing of the $c\bar{c}$ core with the $D^*\bar{D}/\bar{D}^*D$ component, the compact tetraquark state and so on. Obviously, in the pure $c\bar{c}$ picture and compact tetraquark picture, it is hard to understand the puzzles of its mass coincidence with the threshold and large isospin violation. In the $\bar{D}^{*0} D^0/\bar{D}^0
D^{*0}$ molecular picture, the proximity of the $X(3872)$ mass to the threshold is natural although such an exact coincidence is still confusing (see Sec.~\ref{sec:XandTccStates} for the discussion of the fine-tuning problem).  {The mass of the $X(3872)$ exactly coincides with the neutral
threshold $\xthn$ and is about $8$ MeV below the charged threshold
$D^{*-}D^+/D^-D^{*+}$.  In Sec.~\ref{sec:X-short}, we will review the theoretical interpretations of the large ratio in Eq.~\eqref{eq:XisospinV} based on the fine-tuning mass of the $X(3872)$. Meanwhile, we will see the kinematic mechanisms due to the threshold of the $J/\psi \rho(\omega)$ and the $\rho (\omega)$ widths will also contribute to the  large ratio in Eq.~\eqref{eq:XisospinV}.}  Another experimental evidence supporting the molecular picture is the dominant decay mode of $X(3872)$~\cite{ParticleDataGroup:2022pth},
\begin{equation}
    \Gamma(X\to D^0\bar{D}^0\pi^0)/\Gamma_{\mathrm{total}}=(49^{+18}_{-20})\%,\qquad\qquad \Gamma(X\to \bar{D}^{*0}{D}^0)/\Gamma_{\mathrm{total}}=(37\pm 9)\%,
\end{equation}
{where the branch fractions were extracted from  $\mathcal{B}(B^+\to D^0\bar{D}^0\pi^0 K^+)\Gamma(X\to D^0\bar{D}^0\pi^0)/\Gamma_{\mathrm{total}}$, $\mathcal{B}(B^+\to \bar{D}^{*0}{D}^0  K^+) \Gamma(X\to \bar{D}^{*0}{D}^0)/\Gamma_{\mathrm{total}}$ and $\mathcal{B}(B^+ \to X(3872)K^+)$.} The phase spaces of the above decay modes are severely suppressed. The large fractions indicate the strong coupling between the $X(3872)$ and the $\bar{D}^{*0} D^0/\bar{D}^0 D^{*0}$ channels.

The experimental progresses inspired the heated debates about the nature of $X(3872)$. The authors of Ref.~\cite{Bignamini:2009sk} argued that the experimental prompt production data of the $X(3872)$~\cite{CDF:2003cab,CMS:2013fpt,ATLAS:2016kwu,D0:2020nce,LHCb:2020sey,LHCb:2021ten} challenged the hadronic molecule interpretation. The prompt and non-prompt contributions are discriminated by analyzing displacement of the production vertex.  In Sec.~\ref{sec:X-short}, we will review the discussions ~\cite{Bignamini:2009sk,Artoisenet:2009wk,Albaladejo:2017blx,Braaten:2018eov}. In order to pin down the nature of the $X(3872)$, its heavy quark flavor partner was searched in CMS~\cite{CMS:2013ygz}, ATLAS~\cite{ATLAS:2014mka} and Belle~\cite{Belle:2014sys}. No evidence for the $X_b$ was observed. In Ref.~\cite{LHCb:2020xds}, the lineshape of the $X(3872)\to J/\psi \pi^+\pi^-$ was investigated in a Flatt\'e inspired model by LHCb Collaboration. The pole structure analysis indicated the $X(3872)$ is consistent with the $\xthn$ with the molecular component $\gtrsim$ 70\% in the context of Weinberg's compositeness criterion~\cite{Weinberg:1965zz}. The lineshape analysis results incited a debate on discerning the structure of the $X(3872)$ with the sign of the effective range in Ref.~\cite{Esposito:2021vhu,Baru:2021ldu}, which  will be discussed in Sec.~\ref{sec:1chanel_pionless}. In Ref.~\cite{LHCb:2020sey}, the LHCb Collaborations investigated the multiplicity-dependence of the prompt production of the $X(3872)$, $\psi(2S)$ and their ratio in $pp$ collisions and found that the ratio decreased with the multiplicity. Based on the Comover Interaction Model, the authors of Ref.~\cite{Esposito:2020ywk} argued that the above observation disfavored the molecular picture of the $X(3872)$, which was challenged by Ref.~\cite{Braaten:2020iqw}. The CMS Collaboration reported the evidence of the $X(3872)$ production in relativistic heavy ion collisions for the first time ~\cite{CMS:2021znk}, which provides a novel insight into the nature of $X(3872)$ ~\cite{Cho:2017dcy,Zhang:2020dwn,Chen:2021akx,Esposito:2021ptx}.

Very recently, the LHCb Collaboration observed the first doubly charmed tetraquark state $T^+_{cc}$ in the prompt production of the $pp$ collision~\cite{LHCb:2021vvq}. Its
mass with respect to the $D^{\ast+}D^0$ threshold and width are
\begin{equation}
    \delta m =( -273 \pm 61 \pm 5^ {+11}_{-14}) \text{ keV},\qquad\qquad \Gamma = (410 \pm 165 \pm 43 ^{+18}_{-38}) \text{ keV}. 
\end{equation}
In the fitting, the quantum number $J^P = 1^+$ is assumed. The LHCb Collaboration
also released the analysis in the unitarized Breit-Wigner profile including the $T_{cc}^+\to D^0D^0\pi^+,~D^0D^+\pi^0$ and $D^0D^+\gamma$ explicitly in the decay width~\cite{LHCb:2021auc}. The mass with respect to the $D^{*+}D^0$ threshold and width read
\begin{equation}
    \delta m_U=(-361 \pm 40) \text{ keV},\qquad\qquad \Gamma_U=(47.8\pm 1.9)\text{ keV}.
\end{equation}
{One should notice the discrepancy in the widths from the above two analyses. The conventional Breit-Wigner parameterization in the first analysis works well in the region far away from the thresholds. The $T_{cc}$ state lies very close to the $DD^*$ two-body thresholds and $DD\pi$ three-body thresholds. The second analysis constrained by the unitarity is more reasonable. In Sec.~\ref{sec:X_long_range}, we will review the related theoretical calculations (some of them were ahead of the second experimental analysis), which also support the narrower width.}

The observation of the $T_{cc}^+$ marks the striking progress in the search of the exotic states. It is the second doubly charmed hadron observed in experiments after the $\Xi_{cc}^{++}$~\cite{LHCb:2017iph}. Moreover, the $T_{cc}^+$ is manifestly exotic with four (anti)quarks ($cc\bar{q}\bar{q}$), which has been anticipated and debated for 40 years~\cite{Carlson:1987hh,Silvestre-Brac:1993zem,Semay:1994ht,Pepin:1996id,Gelman:2002wf,Vijande:2003ki,Janc:2004qn,Cui:2006mp,Navarra:2007yw,Vijande:2007rf,Ebert:2007rn,Lee:2009rt,Yang:2009zzp,Du:2012wp,Feng:2013kea,Ikeda:2013vwa,Luo:2017eub,Karliner:2017qjm,Eichten:2017ffp,Wang:2017uld,Cheung:2017tnt,Park:2018wjk,Francis:2018jyb,Junnarkar:2018twb,Deng:2018kly,Yang:2019itm,Liu:2019stu,Tan:2020ldi,Lu:2020rog,Braaten:2020nwp,Gao:2020ogo,Cheng:2020wxa,Noh:2021lqs,Faustov:2021hjs}. The $T_{cc}^+$ is also the second hadron with the mass almost coinciding with the threshold and a very narrow width after the $X(3872)$. In fact, the $T_{cc}^+$ and $X(3872)$ share the same fine-tuning problem. In the molecular scheme,
the one-pion-exchange interaction of the $\bar{D}^*D/\bar{D}D^*$ system
with the quantum numbers of $I(J^{PC})=0(1^{++})$ [corresponding to the
$X(3872)$] and that of the $D^*D$ system with $I(J^P)=0(1^+)$ are
exactly the same in the isospin symmetry
limit. The $D^*D$ molecular states had been predicted~\cite{Dias:2011mi,Li:2012cs,Li:2012ss,Xu:2017tsr} before the experimental observations. The discovery of the $T_{cc}$ also inspired huge amounts of investigations~\cite{Meng:2023iqj,Agaev:2021vur,Ling:2021bir,Chen:2021vhg,Dong:2021bvy,Feijoo:2021ppq,Yan:2021wdl,Xin:2021wcr,Huang:2021urd,Fleming:2021wmk,Azizi:2021aib,Hu:2021gdg,Chen:2021cfl,Albaladejo:2021vln,Du:2021zzh,Deng:2021gnb,Agaev:2022ast,Braaten:2022elw,He:2022rta,Abreu:2022lfy,Achasov:2022onn,Mikhasenko:2022rrl,Wang:2022jop,Lyu:2023xro}.

\subsubsection{$Z_c$, $Z_{cs}$ and $Z_b$} \label{sec1.3.2}

In addition to the isoscalar states $X(3872)$ and $T_{cc}^+$, there are several well-known isovector states $Z_c(3900)$ and $Z_c(4020)$ in the charmonium energy region, as well as the $Z_b(10610)$ and $Z_b(10650)$ in the bottomonium energy region~\cite{ParticleDataGroup:2022pth}. The charged charmoniumlike states $Z_c(3900)$ and $Z_c(4020)$ were observed by the BESIII Collaboration in the $J/\psi\pi^\pm$, $(D\bar{D}^\ast)^\pm$~\cite{BESIII:2013ris,BESIII:2013qmu,BESIII:2015pqw} and $h_c\pi^\pm$, $(D^\ast\bar{D}^\ast)^\pm$~\cite{BESIII:2013ouc,BESIII:2015tix} channels, respectively. The $Z_c(3900)$ was also confirmed by the Belle Collaboration~\cite{Belle:2013yex} and Xiao \etal~\cite{Xiao:2013iha}. The charged bottomoniumlike states $Z_b(10610)$ and $Z_b(10650)$ were observed by the Belle Collaboration in the $\Upsilon(nS)\pi^\pm~(n=1,2,3)$ and $h_b(mP)\pi^\pm~(m=1,2)$ invariant mass spectra~\cite{Belle:2011aa}, as well as in the $(B\bar{B}^\ast)^\pm$ and $(B^\ast\bar{B}^\ast)^\pm$ channels~\cite{Belle:2015upu}, respectively. {Besides, the Belle and LHCb Collaborations also observed several other $Z_c$ states with large widths in the $B$ meson decay processes, e.g., the $Z_c(4200)$~\cite{Belle:2014nuw,LHCb:2019maw} and $Z_c(4430)$~\cite{LHCb:2019maw,Belle:2007hrb,Belle:2013shl,LHCb:2014zfx}. In what follows, we will mainly focus on the $Z_c(3900)$, $Z_c(4020)$, $Z_b(10610)$ and $Z_b(10650)$. For the experimental and theoretical aspects of $Z_c(4200)$ and $Z_c(4430)$, we refer to reviews~\cite{Chen:2016qju,Guo:2017jvc,Brambilla:2019esw}.}

 The minimal quark content of these charged charmoniumlike and bottomoniumlike states should be $Q\bar{Q}q\bar{q}$ ($Q=c,b$, $q=u,d$), so they are good candidates of the manifestly exotic hadrons. There are many similarities among the $Z_c(3900)$, $Z_c(4020)$ and $Z_b(10610)$, $Z_b(10650)$ from their mass spectra and decay patterns, etc. For example, {(i) their masses from Breit-Wigner fits all lie about several MeVs above the corresponding $D\bar{D}^\ast$, $D^\ast\bar{D}^\ast$ and $B\bar{B}^\ast$, $B^\ast\bar{B}^\ast$ thresholds, respectively (see Refs.~\cite{Cleven:2011gp} and \cite{Ji:2022uie} for results with other parameterizations);} (ii) they both decay into the open and hidden heavy flavor channels, but the open heavy flavor channels are dominant; (iii) their $I^G(J^{PC})$ quantum numbers are measured to be $1^+(1^{+-})$ [the $J^P$ of $Z_c(4020)$ is not determined yet, but $1^+$ is presumed in most works]. There have been many efforts toward understanding the internal structures of these charged heavy quarkoniumlike states, such as the lattice QCD simulations, effective field theories (EFTs) (see the reviews in Sec.~\ref{sec:sec5.6}) and the phenomenological models, etc. The popular explanations include the molecular states (bound states, virtual states or resonances), compact tetraquarks, hadro-quarkonia and kinetic effects~(see reviews~~\cite{Chen:2016qju,Brambilla:2019esw,Guo:2017jvc,Esposito:2016noz,Hosaka:2016pey,Ali:2017jda,Liu:2019zoy,Lebed:2016hpi,Olsen:2017bmm,Olsen:2014qna}). The proximities to the $D^{(\ast)}\bar{D}^\ast/B^{(\ast)}\bar{B}^\ast$ thresholds indicate that their properties should be strongly correlated with the $D^{(\ast)}\bar{D}^\ast/B^{(\ast)}\bar{B}^\ast$ interactions. These states provide a very good platform to utilize the EFTs containing both the chiral and heavy quark symmetries.

The strange partner of the $Z_c$ states---the $Z_{cs}(3985)$ was recently observed by the BESIII Collaboration in the $D_s^-D^{\ast0}+D_s^{\ast-}D^0$ channel~\cite{BESIII:2020qkh}. {With the Breit-Wigner parameterization, its mass and width are measured to be} 
\begin{eqnarray}
m=3982.5^{+1.8}_{-2.6}\pm2.1~\mathrm{MeV},\qquad\qquad\Gamma=12.8^{+5.3}_{-4.4}\pm3.0~\mathrm{MeV}.
\end{eqnarray}
Similar to the $Z_c(3900)$, the $Z_{cs}(3985)$ lies about $6$ MeV above the $D_s^-D^{\ast0}/D_s^{\ast-}D^0$ thresholds. In addition, the mass difference is $m_{Z_{cs}(3985)}-m_{Z_c(3900)}\simeq m_{D_s^{(\ast)}}-m_{D^{(\ast)}}\simeq100$ MeV. These features inspired the interpretations of the $Z_{cs}(3985)$ as the $\SU(3)$ partner of the $Z_c(3900)$~\cite{Meng:2020ihj,Wang:2020htx,Yang:2020nrt}. The discovery of the $Z_{cs}(3985)$ has stimulated many works to explain its nature from various aspects~\cite{Meng:2020ihj,Wang:2020htx,Yang:2020nrt,Wan:2020oxt,Wang:2020kej,Sun:2020hjw,Cao:2020cfx,Chen:2020yvq,Du:2020vwb,Wang:2022ztm,Meng:2022xdf,Wang:2020rcx,Wang:2020iqt,Jin:2020yjn,Ikeno:2020mra,Xu:2020evn,Guo:2020vmu,Ozdem:2021yvo,Yan:2021tcp,Ortega:2021enc,Chen:2021erj,Maiani:2021tri,Shi:2021jyr,Faustov:2021hjs,Giron:2021sla,Karliner:2021qok,Ding:2021igr,Wu:2021ezz, Yang:2021zhe,Baru:2021ddn,Ferretti:2021zis,Wu:2021cyc,Du:2022jjv,Chen:2022yev,Han:2022fup}. Later, the LHCb Collaboration reported two $Z_{cs}$ states---the $Z_{cs}(4000)$ and $Z_{cs}(4220)$ in the $J/\psi K^+$ channel from the process $B^+\to J/\psi K^+\phi$~\cite{LHCb:2021uow}. The mass of the $Z_{cs}(4000)$ ($4003\pm6^{+4}_{-14}$ MeV) is close to that of $Z_{cs}(3985)$, but its width ($131\pm15\pm26$ MeV) is about ten times larger than that of $Z_{cs}(3985)$. Whether they are the same state~\cite{Ortega:2021enc} or totally different ones~\cite{Chen:2021erj,Maiani:2021tri,Meng:2021rdg} is still under debate. Some important implications for the $Z_{cs}(3985)$ and $Z_{cs}(4000)$ as two different states were derived in Ref.~\cite{Meng:2021rdg}. 

\subsubsection{$P_c$ and $P_{cs}$} \label{sec1.3.3}

In 2015, the LHCb Collaboration reported two pentaquark states, a {broader $P_{c}(4380)$ with a width $205\pm18\pm86$ MeV and narrower $P_{c}(4450)$ with a width $39\pm5\pm19$ MeV} in the $J / \psi p$ invariant mass distribution in the decay process $\Lambda_{b}^{0} \rightarrow J / \psi p K^{-}$~\cite{LHCb:2015yax}. In 2019, {in the same channel $J/\psi p$ but with larger statistics}, the LHCb Collaboration found that the $P_{c}(4450)^{+}$ should be dissolved into two substructures $P_{c}(4440)^{+}$ and $P_{c}(4457)^{+}$~\cite{LHCb:2019kea}. In addition, another narrow state $P_{c}(4312)^{+}$ was observed. Their resonance parameters are~\cite{LHCb:2019kea}
\begin{eqnarray}
&&P_{c}^{+}(4312): ~M_{P_{c}^{+}(4312)} =4311.9 \pm 0.7_{-0.6}^{+6.8}~\text{MeV},\qquad \Gamma_{P_{c}^{+}(4312)} =9.8 \pm 2.7_{-4.5}^{+3.7} ~\text{MeV},\\
&&P_{c}^{+}(4440): ~M_{P_{c}^{+}(4440)} =4440.3 \pm 1.3_{-4.7}^{+4.1}~\text{MeV}, \qquad \Gamma_{P_{c}^{+}(4440)} =20.6 \pm 4.9_{-10.1}^{+8.7} ~\text{MeV}, \\
&&P_{c}^{+}(4457): ~M_{P_{c}^{+}(4457)} =4457.3 \pm 1.3_{-4.1}^{+0.6} ~\text{MeV}, \qquad \Gamma_{P_{c}^{+}(4457)} =6.4 \pm 2.0_{-1.9}^{+5.7} ~\text{MeV}.
\end{eqnarray}
In the new analysis, the evidence of the broad $P_{c}(4380)${in Ref.~\cite{LHCb:2015yax}} was neither confirmed nor contradicted and awaits for a future complete amplitude analysis of the decay $\Lambda_{b}^{0} \rightarrow J / \psi p K^{-}$. 

The observation of the $P_c$ states confirmed the predictions of the hidden-charm molecular pentaquarks in Refs.~\cite{Yang:2011wz,Wu:2010jy,Wu:2010rv} and inspired the great enthusiasm about the hidden-charm pentaquark states. Various interpretations have been proposed such as the molecular states~\cite{Chen:2019bip,Chen:2019asm,Liu:2019tjn,Guo:2019kdc,Xiao:2019aya,Meng:2019ilv,Du:2019pij,Peng:2021hkr,Dong:2021juy,Xiao:2020frg,Wang:2019spc,Gutsche:2019mkg,Burns:2019iih,Yamaguchi:2019seo,Wang:2019ato,Lin:2019qiv,Voloshin:2019aut,Meng:2019ilv}, the compact pentaquark states~\cite{Kuang:2020bnk,Ali:2019npk,Zhu:2019iwm,Wang:2019got,Giron:2019bcs,Stancu:2019qga,Giron:2021fnl}, the hadro-charmonium states~\cite{Ferretti:2018ojb,Eides:2015dtr,Eides:2019tgv,Kubarovsky:2015aaa}, the triangle singularities~\cite{Guo:2015umn,Liu:2015fea,Bayar:2016ftu}, and the cusp effects~\cite{Kuang:2020bnk,Nakamura:2021dix}. The $P_{c}(4312) $ and $P_{c}(4440) / P_{c}(4457)$ are located only tens of MeV below the $ \Sigma_{c} \bar{D}$ and $\Sigma_{c} \bar{D}^{*}$ thresholds, respectively. The molecular scheme  is a more natural interpretation and attracted much attention. In the molecular picture, the $P_c(4312)$ is widely accepted as an $S$-wave  $\Sigma_{c} \bar{D}$ molecule with $J^P=\frac{1}{2}^-$. The $P_{c}(4440)$ and $P_{c}(4457)$ are treated as the $S$-wave $\Sigma_{c} \bar{D}^{*}$ molecules. However, there are two possible scenarios of the spin-parity assignments. In scenario I,  the $P_{c}(4440)$ and $ P_{c}(4457)$  are the $\Sigma_{c} \bar{D}^{*}$ molecules with the $ J^{P}= \frac{1}{2}^{-}$ and $\frac{3}{2}^{-}$ , respectively~\cite{He:2019ify,He:2019rva,Chen:2019asm,Wang:2019ato,Lin:2019qiv}. In scenario II, the $J^{P}$ quantum numbers of the  $P_{c}(4440)$ and $ P_{c}(4457)$ are the $\frac{3}{2}^{-} $ and $ \frac{1}{2}^{-}$, respectively~\cite{Liu:2019zvb,Du:2021fmf,Yalikun:2021bfm,PavonValderrama:2019nbk,Liu:2019zvb,Yamaguchi:2019seo}. Their decay patterns~\cite{Sakai:2019qph,Xiao:2019mvs,Voloshin:2019aut,Lin:2019qiv,Guo:2019fdo,Xu:2019zme,Cao:2019kst,Wang:2019spc,Ling:2021lmq,Dong:2020nwk} and  productions~\cite{Ling:2021sld,Park:2022nza,Winney:2019edt,Chen:2020opr,Yang:2020eye,GlueX:2019mkq,Du:2020bqj,Cao:2019kst,Kubarovsky:2015aaa,Karliner:2015voa,Wu:2019adv} were also investigated extensively. In Ref.~\cite{Rossi:2019szt}, the photo-production of $P_c$ was suggested to testing its multiquark nature in the framework of string-junction picture~\cite{Rossi:1977cy,Rossi:1977dp,Rossi:2004yr} unifying the baryons and multiquark states.

Recently, the LHCb Collaboration reported the evidence of a new pentaquark state $P_{c}(4337)^{+}$ in the $J / \psi p $ invariant mass distribution of the decay $B_{s} \rightarrow J / \psi p \bar{p}$. Its resonance parameters are~\cite{LHCb:2021chn}
\begin{eqnarray}
P_{c}(4337)^{+}:~M_{P_{c}(4337)^{+}} =4337_{-4 -2}^{+7+2}~\text{MeV}, \qquad\qquad\Gamma_{P_{c}(4337)^{+}}=29_{-12 -14}^{+26 +14}~\text{MeV}. 
\end{eqnarray}
The $P_{c}(4337)^{+}$ lies very close to the $P_{c}^{+}(4312)$ state and the $P_c(4380)$ signal. Its interpretations included the compact tetraquark state~\cite{Giron:2021fnl}, the cusp effect~\cite{Nakamura:2021dix}, the hadro-charmonium~\cite{Yan:2021nio}, coupled channel dynamics~\cite{Yan:2021nio}, the $\bar{D} \Sigma_{c}$ molecule~\cite{Yan:2021nio}, the $\bar{D} \Sigma_{c}^{*}$ molecule [$P_c(4380)^+$]~\cite{Wang:2021crr}, etc.

The partners of the $P_c$ states were predicted with the heavy quark spin symmetry, $[\Sigma_{c}^{*} \bar{D}]_{J=3 / 2}^{I=1 / 2}$ [candidate for the $P_c(4380)^+$] and $[\Sigma_{c}^{*} \bar{D}^{*}]^{I=1 / 2}_{J=\left(\frac{1}{2}, \frac{3}{2}, \frac{5}{2}\right)}$~\cite{Wang:2019ato,PavonValderrama:2019nbk,Xiao:2013yca,Pan:2019skd,Du:2021fmf,Xiao:2020frg,Du:2019pij,Liu:2019zvb,Liu:2019tjn}, 
which await the future experimental scrutiny.  
The studies of the $P_c$ states can be easily extended to the strange hidden-charm molecular pentaquarks ($P_{cs}$) under the SU(3) flavor symmetry and the $P_{cs}$ states were predicted in Refs.~\cite{Wu:2010jy,Chen:2015sxa,Chen:2016ryt,Shen:2019evi,Xiao:2019gjd,Wang:2019nvm,Xiao:2019gjd,Anisovich:2015zqa}. 

In 2020, the LHCb found the evidence of the $P_{c s}(4459)$ in  the $J / \psi \Lambda$ invariant mass distribution via the decay $\Xi_{b}^{-} \rightarrow J / \psi \Lambda K^{-}$~\cite{LHCb:2020jpq}, 
\begin{eqnarray}
P_{c s}(4459):~M_{P_{c s}(4459)}=4458.8 \pm 2.9_{-1.1}^{+4.7} ~\text{MeV}, \qquad\qquad \Gamma_{P_{c s}(4459)}=17.3 \pm 6.5_{-5.7}^{+8.0}~\text{MeV}.
\end{eqnarray}
which is only about $19$ MeV below the $\Xi_{c}^{0} \bar{D} ^{*0}$ threshold. The experiment did not confirm or contradict the existence of the two-peak hypothesis. 
The result inspired various theoretical works~\cite{Du:2021bgb,Ferretti:2021zis,Deng:2022vkv,Giron:2021fnl,Yang:2021pio,Peng:2020hql,Chen:2020uif,Chen:2020kco,Zhu:2021lhd,Lu:2021irg}. Most of them favored the ${P_{cs}}$ as the  $ \Xi_{c} \bar{D}^{*}$ molecule~\cite{Yang:2021pio,Du:2021bgb,Peng:2020hql,Chen:2020uif,Chen:2020kco,Zhu:2021lhd,Lu:2021irg,Liu:2020hcv}. Besides the molecular interpretation, there also exist other explanations such as the hadro-charmonium~\cite{Ferretti:2021zis}, compact pentaquark states~\cite{Deng:2022vkv,Giron:2021fnl}. There were intensive discussions of its production,  decays and other properties~\cite{Clymton:2021thh,Azizi:2021utt,Wu:2021caw}. Similar to the $P_c$ states, various partners of the $P_{cs}$ were also predicted in numerous theoretical works~\cite{Wang:2019nvm,Zhu:2021lhd,Xiao:2019gjd,Liu:2020hcv,Xiao:2021rgp,Peng:2020hql,Chen:2016ryt,Santopinto:2016pkp,Xiao:2019gjd}, which need to be examined by future experiments. 

{Very recently, the LHCb Collaboration observed the $P_{cs}(4338)$ (also named as the $P_{\psi s}^\Lambda(4338)^0$ according to the naming convention proposed in Ref.~\cite{Gershon:2022xnn}) in the decay $B^-\to J/\psi\Lambda \bar{p}$~\cite{LHCb:2022jad}. The amplitude analysis prefers the $J^P=\frac{1}{2}^-$ assignment for $P_{cs}(4338)$. Its mass and width in the relativistic Breit-Wigner fits read
\begin{eqnarray}
    P_{cs}(4338):~M_{P_{cs}(4338)}=4338.3\pm0.7\pm0.4~\text{MeV},\qquad\qquad \Gamma_{P_{cs}(4338)}=7.1\pm1.2\pm1.3~\text{MeV}.
\end{eqnarray} 
The $P_{cs}(4338)$ is very close to the $\Xi_c\bar{D}$ threshold (which is about $4336$ MeV). In Ref.~\cite{Karliner:2022erb}, the authors proposed it is the $\Xi_c\bar{D}$ molecule considering the near-threshold behavior, the spin-parity and the narrow width. In Ref.~\cite{Meng:2022wgl}, Meng \etal investigated the influence of the double-threshold ($\Xi_c^0\bar{D}^0$ and $\Xi_c^+D^-$) on the line shape (the invariant mass distribution of $J/\psi\Lambda$) of $P_{cs}(4338)$. A comparison between the $P_c$ and $P_{cs}$ states was given in Ref.~\cite{Chen:2022wkh}. Other approaches were also employed to  understand this state, such as the one-boson exchange model~\cite{Wang:2022mxy}, the triangle singularity~\cite{Burns:2022uha}, the effective field theory~\cite{Yan:2022wuz,Yang:2022ezl}, the coupled-channel dynamics~\cite{Nakamura:2022jpd,Giachino:2022pws}, the quark model~\cite{Ortega:2022uyu}, etc. Its electromagnetic properties were also studied in Refs.~\cite{Ozdem:2022kei,Wang:2022tib}.}

The discovery of the $P_c$ states and the evidence of the $P_{cs}$ also inspired the investigation of the other pentaquark states, such as $Q Q q q \bar{q}$~\cite{Zhou:2018bkn,Park:2018oib,Chen:2021kad,Yang:2020twg,Shimizu:2017xrg,Chen:2021htr,Chen:2021kad,Chen:2017vai}, $P_{c s s}$~\cite{Azizi:2021pbh}, the fully heavy pentaquark state~\cite{Yan:2021glh,An:2022fvs}, the $\bar{Q} q q q q$~\cite{An:2020vku,Richard:2019fms,Gignoux:1987cn,Lipkin:1987sk,Xing:2022aij}, $B \Xi_{c} $ and $B \Xi_{c}^{\prime} $ bound states~\cite{Shen:2022rpn}, etc.

\subsection{A short summary}

Quark model inherits some spirits and features of QCD and has been successfully employed to describe the conventional mesons and baryons at the quark level. The hadron spectroscopy had witnessed the success of quark model before 2003. However, the quark model was challenged by many near-threshold exotic states observed since 2003. More and more experimental evidences indicate the relevance and importance of the clustering effect of the quarks, e.g., the heavy hadronic molecules. {The loosely bound hadronic molecules are very well known as a building block of the matter, namely the nuclei}. The abundant hadron spectra are the opportunities and challenges for us to understand the nonperturbative behaviors of QCD at low energy domains. 

The chiral effective theories are built upon the chiral symmetry of QCD with the asymptotic hadronic d.o.f. These powerful tools have been developed and applied in the nuclear sectors with flying colors in the past decades. The heavy hadrons and exotic hadronic molecular states observed in recent years supply the fertile soil for the redevelopment of these effective theories. In this work, we will mainly review the developments and applications of chiral perturbation theory, chiral effective field theory, as well as their ramifications and variants in the heavy flavor sectors together with the heavy quark symmetry. We shall also cover the experimental measurements, lattice QCD simulations, and phenomenological model calculations briefly.
\section{Symmetries of low energy QCD and effective field theories} 

Effective field theory (EFT) is designed for systems with widely separated energy scales. One basic principle of EFT is that the low energy interaction does not depend on the details of dynamics at high energies. The most general Lagrangian contains all terms satisfying the requirements of underlying symmetries of the system, which are organized with increasing number of derivatives~\cite{Weinberg:1978kz}. Once the low energy scale $\mathcal{Q}$ and high energy scale $\Lambda$ for such a system are well identified, the amplitude of a soft process can be expanded in powers of $\mathcal{Q}/\Lambda$, i.e.,
\begin{eqnarray}\label{eq:EFTexpansion}
\mathcal{A}=\sum_\nu\mathcal{C}_\nu(\mathcal{Q}/\mu,c_i)\left(\frac{\mathcal{Q}}{\Lambda}\right)^\nu,
\end{eqnarray}
where the expansion coefficients $\mathcal{C}_\nu$ are the functions of regularization scale $\mu$ which arises from the loop diagrams and low energy constants (LECs). For a well-defined EFT, the expansion coefficients are of order unity (naturalness) {after separating out the common dimension of the amplitude}. The high energy dynamics above the $\Lambda$ are encoded in the LECs $c_i$ via a series of the local contact terms, which are the functions of $\Lambda$ in general, i.e., $c_i\equiv c_i(\Lambda)$. The expansion index $\nu$ is connected to the power counting given by the EFT. If the energy scale is largely separated in the system, the small value $\mathcal{Q}/\Lambda$ ensures the good convergence of the above expansion, thus only finite terms are needed for a given accuracy in practical calculations. {One can consult two recent books~\cite{meissner_rusetsky_2022,petrov_eft} for a more comprehensive introduction about EFT.}

QCD is one of the main ingredients of the Standard Model.
 The full Lagrangian of QCD is given in terms of the quark and gluon degrees of freedom, which reads
 \begin{eqnarray}\label{eq:QCDLagrangian}
 	\mathcal{L}_{\mathrm{QCD}}&=&\bar{q}(i\slashed{\mathcal{D}}-\mathcal{M})q-\frac{1}{4}\mathcal{G}_{\mu\nu,a}\mathcal{G}_{a}^{\mu\nu},
 \end{eqnarray}
 where the quark field is $q\equiv q_f^c$, and the summations over the flavor index $f$ and color index $c$ are implied. The covariant derivative is $\mathcal{D}_{\mu}=\partial_{\mu}-ig_s\lambda^{a}\mathcal{A}_{\mu}^{a}/2$, where $g_s$ is the strong coupling constant. $\lambda^a$ ($a=1,\dots,8$) represents the Gell-Mann matrix, and $\mathcal{A}_\mu^a$ denotes the gluon field. The quark mass matrix is defined as  $\mathcal{M}=\mathrm{diag}(m_u,m_d,m_s,m_c,m_b,m_t)$. The  field strength tensor reads $\mathcal{G}_{\mu\nu,a}=\partial_{\mu}\mathcal{A}_{\nu,a}-\partial_{\nu}\mathcal{A}_{\mu,a}+g_sf_{abc}\mathcal{A}_{\mu,b}\mathcal{A}_{\nu,c}$ [where $f_{abc}$ denotes the antisymmetric structure constant of $\SU(3)$ group], with $\tilde{\mathcal{G}}_{a}^{\mu\nu}=\epsilon^{\mu\nu\alpha\beta}\mathcal{G}_{\alpha\beta,a}/2$ being its dual.  QCD exhibits well separated scales. {In the following, we refer to the current quark mass in the $\overline{\rm MS}$ subtraction scheme at a renormalization scale $\mu =2$ GeV~\cite{ParticleDataGroup:2022pth}.  The $u$, $d$ and $s$ quarks are very light
 \begin{equation} (m_{u}=2.16_{-0.26}^{+0.49},~m_{d}=4.67_{-0.17}^{+0.48},~m_{s}=93_{-5}^{+11})~\mathrm{MeV}\ll\Lambda_\text{QCD},
 \end{equation}
  where  $\Lambda_{\mathrm{QCD}}\sim 200 \text{ MeV}$ is the nonperturbative scale of QCD. The $c$, $b$ and $t$ quarks are very heavy compared with $\Lambda_{\mathrm{QCD}}$,}
 	\begin{equation}
(m_{c}=1.27\pm0.02,~m_{b}=4.18_{-0.02}^{+0.03},~m_{t}=172.76\pm0.03)~\mathrm{GeV}\gg \Lambda_{\mathrm{QCD}}.
 	\end{equation}
Therefore, the $u$, $d$ and $s$ quarks are generally called as the light quarks, while $c$, $b$ and $t$ are the heavy quarks. The small mass of the light quarks and the large mass of the heavy quarks stimulate two extreme approximations, i.e., $m_u=m_d=m_s\to0$ and $m_c=m_b\to\infty$ (the top quarks are generally not considered in hadron physics due to its extremely unstable nature). In these two limits, QCD exhibits chiral symmetry and heavy quark symmetry, respectively.

Due to the nonperturbative nature of QCD at low energy domain, the quarks are confined in the color neutral hadrons with the not-fully-understood nonperturbative dynamics. It is hard to perform thorough analyses in terms of its fundamental d.o.f, i.e., the quarks and gluons. An alternative description for the physics occurring at $E\ll \Lambda_\chi$ is in terms of the asymptotic hadron states,	{where $\Lambda_\chi\sim 1\text{ GeV} $ is the chiral symmetry breaking scale~\footnote{$\Lambda_\chi$ can be chosen either as $4\pi f_\pi$, a factor appearing in the loop diagrams~\cite{Manohar:1983md} or the mass of the $\rho$-meson, a natural scale for chiral symmetry breaking.}}.  As a full-fledged EFT of QCD at low energies, the chiral perturbation theory ($\chi$PT) has been developed in the past decades. Now it has become the common language of nuclear physics and hadron physics when encountering the light quark dynamics. However, the richness of hadron spectrum also calls for combinations of chiral symmetry and HQS in the heavy-light systems (e.g., the singly heavy mesons $Q\bar{q}$, the singly heavy baryons $Qqq$, where $Q=c,b$ and $q=u,d,s$). The EFT based on the chiral symmetry and heavy quark symmetry for the heavy-light systems is denoted as the heavy hadron chiral perturbation theory (HH$\chi$PT). Besides, in recent years, an approximate symmetry, the so-called heavy diquark-antiquark symmetry (HDAS), is proposed for the heavy-heavy-light system (i.e., the doubly heavy baryons $QQq$), whereas the suitability is still an open question.

In the following, we outline the basic frameworks of $\chi$PT, heavy quark effective theory (HQET), HH$\chi$PT and chiral effective field theory ($\chi$EFT) with two matter fields, respectively. For more comprehensive details of these theories and related topics, we refer to some more specialized reviews~\cite{Ecker:1994gg,Bernard:1995dp,Cho:1992cf,Fettes:2000gb,Steininger:1998ya,Koch:1997ei,Neubert:1993mb,Epelbaum:2008ga,Machleidt:2011zz} and lecture notes~\cite{Scherer:2002tk,Manohar:2000dt,Pich:1998xt}, and the references therein. The readers who are familiar with these effective theories can skip this chapter and move on to the next one.

\subsection{Chiral perturbation theory}

\subsubsection{Chiral symmetry and its spontaneous breaking}\label{sec:CSSB}

Now we focus on the three flavor ($u$, $d$, and $s$) QCD with massless quarks (chiral limit), and pick out their left-handed ($q_L=P_L q$) and right-handed ($q_R=P_R q$) components with the projection operators
\begin{eqnarray}
P_{L}=\frac{1}{2}(1-\gamma^{5}),\quad P_{R}=\frac{1}{2}(1+\gamma^{5}),~\text{with }P_{L}+P_{R}=1,~~P_{R}^{2}=P_{R},~~P_{L}^{2}=P_{L},~~P_{R}P_{L}=P_{L}P_{R}=0.
\end{eqnarray}
The Lagrangian~\eqref{eq:QCDLagrangian} is expressed with the $q_L$ and $q_R$ as
\begin{eqnarray}\label{eq:QCDLagrangian_m0}
\mathcal{L}_{\mathrm{QCD}}^0=\bar{q}_{L}i\slashed{\mathcal{D}}q_{L}+\bar{q}_{R}i\slashed{\mathcal{D}}q_{R},
\end{eqnarray}
where the irrelevant gluon field strength tensor is omitted {for this discussion.} The Lagrangian~\eqref{eq:QCDLagrangian_m0} is invariant under the global $G=\SU(3)_{L}\otimes \SU(3)_{R}$ transformations\footnote{It is also invariant under the $U(1)_V\otimes U(1)_A$ transformations. The $U(1)_V$ trivially corresponds to the baryon number conservation, while the $U(1)_A$ is broken at quantum level due to the `$U(1)_A$ anomaly'.}
\begin{eqnarray}
q_{L}\overset{G} \longrightarrow g_{L}q_{L}=\exp(-i\vartheta_{L}^{a}\frac{\lambda^{a}}{2})q_{L},\qquad q_{R}\overset{G} \longrightarrow g_{R}q_{R}=\exp(-i\vartheta_{R}^{a}\frac{\lambda^{a}}{2})q_{R}.
\end{eqnarray}
According to the Noether's theorem, there are eight conserved left-handed currents $J_L^{\mu,a}$ and right-handed currents $J_R^{\mu,a}$, respectively, with $J_{L}^{\mu,a}=\bar{q}_{L}\gamma^{\mu}\frac{\lambda^{a}}{2}q_{L}$, $\partial_\mu J_{L}^{\mu,a}=0$, and $J_{R}^{\mu,a}=\bar{q}_{R}\gamma^{\mu}\frac{\lambda^{a}}{2}q_{R}$, $\partial_\mu J_{R}^{\mu,a}=0$. The vector currents and axial vector currents are the linear combinations of $J_{L}^{\mu,a}$ and $J_R^{\mu,a}$ 
\begin{eqnarray}
V^{\mu,a}&=&J_{R}^{\mu,a}+J_{L}^{\mu,a}=\bar{q}\gamma^{\mu}\frac{\lambda^{a}}{2}q,\quad\text{ with }\partial_{\mu}V^{\mu,a}=0,\nonumber\\
A^{\mu,a}&=&J_{R}^{\mu,a}-J_{L}^{\mu,a}=\bar{q}\gamma^{\mu}\gamma_{5}\frac{\lambda^{a}}{2}q,\quad\text{ with }\partial_{\mu}A^{\mu,a}=0.
\end{eqnarray}
The vector charges and axial-vector charges are given as 
\begin{eqnarray}
Q_{V}^{a}&=&\int d^{3}xV^{0,a}(\bm{x},t)=\int d^{3}xq^{\dagger}(\bm{x},t)\frac{\lambda^{a}}{2}q(\bm{x},t),\quad\text{ with }\frac{d}{dt}Q_{V}^{a}=0,\nonumber\\
Q_{A}^{a}&=&\int d^{3}xA^{0,a}(\bm{x},t)=\int d^{3}xq^{\dagger}(\bm{x},t)\gamma_{5}\frac{\lambda^{a}}{2}q(\bm{x},t),\quad\text{ with }\frac{d}{dt}Q_{A}^{a}=0.
\end{eqnarray}
They obey the following commutation relations,
\begin{eqnarray}
[Q_{V}^{a},Q_{V}^{b}]=if^{abc}Q_{V}^{c},\qquad [Q_{A}^{a},Q_{A}^{b}]=if^{abc}Q_{V}^{c},\qquad [Q_{V}^{a},Q_{A}^{b}]=if^{abc}Q_{A}^{c}.
\end{eqnarray}

Both the $Q_{V}^{a}$ and $Q_{A}^{a}$ commute with the QCD Hamiltonian $H_{\mathrm{QCD}}^0$, i.e.,  $[Q_{V}^{a},H_{\mathrm{QCD}}^0]=[Q_{A}^{a},H_{\mathrm{QCD}}^0]=0$, which implies the existence of degenerate light hadrons with opposite parities, i.e., the parity doublets, if the QCD vacuum satisfies the same symmetry as the Hamiltonian. From the experimental facts, the lightest vector hadron ($J^P=1^-$) is the $\rho$ meson and its mass is $m_\rho\simeq770$ MeV. In contrast, the lightest axial-vector state ($J^P=1^+$) is the $a_1$ meson and its mass $m_{a_1}\simeq1230$ MeV is much heavier than $m_\rho$. Therefore, it is far-fetched to regard the $a_1(1260)$ as the parity partner of the $\rho$. Moreover, there are eight light pseudoscalar mesons below $m_\rho$, and their masses are lighter than the scalar states in the hadron spectrum. These facts indicate that the vacuum (ground state) of QCD is not invariant under the continuous $\SU(3)_{R-L}$ transformation, i.e.,
\begin{eqnarray}\label{eq:SBCS}
Q_{A}^{a}|0\rangle\neq|0\rangle,
\end{eqnarray}
{where we keep having $Q_{V}^{a}|0\rangle=|0\rangle$.}

A continuous symmetry is spontaneously broken if this symmetry is not realized in its ground state. Therefore, Eq.~\eqref{eq:SBCS} implies the spontaneous breaking of chiral symmetry. Then the Goldstone's theorem~\cite{Goldstone:1961eq,Goldstone:1962es} demands that there should exist massless particles {with the same quantum numbers as the broken generators}, which are called Goldstone bosons. The number of Goldstone bosons is equal to the number of the broken generators, so there should exist eight Goldstone bosons according to Eq.~\eqref{eq:SBCS} {with $J^P=0^-$}. The eight lightest pseudoscalars ($\pi^{+},\pi^{-},\pi^{0},K^{+},K^{-},K^{0},\bar{K}^{0},\eta$) coincidentally correspond to the eight Goldstone bosons. {According to the Lorentz invariance, one can parameterize the following matrix elements,
\begin{equation}
    \langle0|A_{\mu}^{a}(0)|\varphi^{b}(p)\rangle=ip_{\mu}f_{0}\delta^{ab},\label{eq:goldstone}
\end{equation}
where $\varphi^a$ are the Goldstone bosons, and $f_0$ is the decay constant in the chiral limit~\footnote{In the weak decay element of pion, $\langle l^{-}\bar{v}|H_{W}|\pi^{-}\rangle=\frac{G_{F}\cos\theta_{C}}{\sqrt{2}}\langle l^{-}v|j^{-\mu}|0\rangle\text{\ensuremath{\langle0|J_{\mu}^{-}|\pi^{-}\rangle}}$, the leptonic current and the hadronic currents are $j_{\mu}^{-}=\bar{l}\gamma^{\mu}(1-\gamma_{5})\nu_{l}$ and $J_{\mu}^{-}=V_{\mu}^{+}-A_{\mu}^{+}$, respectively, where $G_F$ and $\theta_{C}$ are coupling constants and Cabibbo angle. Because of $\langle0|V_{\mu}(0)|\pi\rangle=0$, only the axial current has contribution. Thus, the matrix element of axial current is related to the leptonic decay constant $f_{0}$.}.}
The non-vanishing masses of the Goldstone bosons arise from light quark masses which break the chiral symmetry explicitly.

\subsubsection{Explicit breaking of chiral symmetry}\label{sec:Ebocs}

In deriving Eq.~\eqref{eq:QCDLagrangian_m0}, the approximation $\mathcal{M}\to0$ is assumed, whereas the light quarks have non-vanishing masses, which explicitly break the chiral symmetry. The mass term results in the mixing of left-handed and right-handed components of the quark fields, $\mathcal{L}_{\mathcal{M}}=-\bar{q}\mathcal{M}q=-(\bar{q}_{L}\mathcal{M}q_{R}+\bar{q}_{R}\mathcal{M}q_{L})$. An infinitesimal transformation on the $\mathcal{L}_{\mathcal{M}}$ under chiral group $G$ leads to the divergences of $V^{\mu,a}$ and $A^{\mu,a}$ as
\begin{eqnarray}\label{eq:divsVA}
\partial_{\mu}V^{\mu,a}=i\bar{q}[\mathcal{M},\lambda^{a}/2]q,\qquad \partial_{\mu}A^{\mu,a}=i\bar{q}\{\mathcal{M},\lambda^{a}/2\}\gamma_{5}q.
\end{eqnarray}
The three flavor light quark mass matrix can be expressed with the $\lambda_a$ matrix as
\begin{eqnarray}\label{eq:Minlambda}
\mathcal{M}=\left[\begin{array}{ccc}
m_{u} & 0 & 0\\
0 & m_{d} & 0\\
0 & 0 & m_{s}
\end{array}\right]=\frac{m_{u}+m_{d}+m_{s}}{\sqrt{6}}\lambda_{0}+\frac{(m_{u}+m_{d})/2-m_{s}}{\sqrt{3}}\lambda_{8}+\frac{m_{u}-m_{d}}{2}\lambda_{3}.
\end{eqnarray}
Inserting Eq.~\eqref{eq:Minlambda} into Eq.~\eqref{eq:divsVA} one obtains that:
\begin{itemize}
  \item[1.] When $m_u=m_d=m_s$, the eight vector currents $V^{\mu,a}$ are conserved due to $[1,\lambda^a]=0$, which corresponds to the emergence of the rigorous $\SU(3)_f$ (subscript $f$ denotes flavor) symmetry. In this case, we can define the $U$-spin and $V$-spin in the subgroup [$\SU(2)_f$] of $\SU(3)_f$, which are analogous to the well-known isospin $I$,
  \begin{eqnarray}
  u\overset{I} \longleftrightarrow d,\qquad d\overset{U} \longleftrightarrow s,\qquad s\overset{V} \longleftrightarrow u.
  \end{eqnarray}
  Similarly, we can further define the $G_U$-parity and $G_V$-parity as the $G$-parity for isospin symmetry, with the transformation operators
  \begin{eqnarray}
  \hat{G}=\hat{C}e^{i\pi \hat{I}_2},\qquad \hat{G}_U=\hat{C}e^{i\pi \hat{U}_2},\qquad \hat{G}_V=\hat{C}e^{i\pi \hat{V}_2}.
  \end{eqnarray}
  The eight axial-vector currents $A^{\mu,a}$ are not conserved anymore, and the chiral symmetry is explicitly broken [in comparison with the hidden (spontaneous) breaking in Sec.~\ref{sec:CSSB}]. The non-vanishing divergences of $A^{\mu,a}$ leads to the microscopic interpretation of the partially conserved
axial-vector current (PCAC) relation~\cite{Nambu:1961tp,Gell-Mann:1960mvl,chouKC:pcac}.
  \item[2.] When $m_u=m_d\neq m_s$, the $\SU(3)_f$ symmetry group breaks down to its subgroup $\SU(2)_f$. Now the isospin symmetry is still exact.
  \item[3.] When $m_u\neq m_d\neq m_s$, the isospin symmetry is also broken, which leads to the isospin breaking effect~\footnote{The other origin of the isospin symmetry breaking effect comes from the electromagnetic interaction.}.
\end{itemize}

The light masses of these pseudoscalar mesons [especially in the SU(2) sector] are deeply rooted in their nature as the pseudo Goldstone bosons. The quantum fluctuations of these pseudoscalar mesons are very important, which are  denoted as chiral dynamics. {The $\chi$PT is the low-energy effective field theory of the QCD.}

\subsubsection{Lowest order Lagrangians and power counting}

The phenomenological and experimental evidences all suggest that the chiral group $G=\SU(3)_{L}\otimes \SU(3)_{R}$ spontaneously breaks down to its vectorial subgroup $\SU(3)_V$ [or say $\SU(3)_{R+L}$], i.e.,
\begin{eqnarray}
G=\SU(3)_{L}\otimes \SU(3)_{R}\longrightarrow H=\SU(3)_V.
\end{eqnarray}
The interactions among light Goldstone bosons are described by the $\chi$PT~\cite{Gasser:1983yg,Gasser:1984gg}, in which the light octet are collected in terms of a $3\times3$ unitary matrix field $U(\varphi)$ transforming under $\SU(3)_L\otimes\SU(3)_R$ as
\begin{eqnarray}
U(\varphi)\overset{G} \longrightarrow g_R U(\varphi)g_L^{-1} \text{ or } g_L U(\varphi)g_R^{-1},\quad \text{ with } (g_L,g_R)\in G.
\end{eqnarray}
Different parameterizations of $U(\varphi)$ correspond to the different choices of coordinates in coset space $G/H$, where a commonly used representation is the exponential parametrization,
\begin{eqnarray}\label{eq:Upionmatrix}
U(\varphi)=\xi^2(\varphi)=\exp\left(i\frac{\varphi}{f_\varphi}\right),\qquad\varphi=\sum_a\lambda_{a}\varphi_{a}=\left[\begin{array}{ccc}
\pi^{0}+\frac{1}{\sqrt{3}}\eta_{8} & \sqrt{2}\pi^{+} & \sqrt{2}K^{+}\\
\sqrt{2}\pi^{-} & -\pi^{0}+\frac{1}{\sqrt{3}}\eta_{8} & \sqrt{2}K^{0}\\
\sqrt{2}K^{-} & \sqrt{2}\bar{K}^{0} & -\frac{2}{\sqrt{3}}\eta_{8}
\end{array}\right],
\end{eqnarray}
where the coset field $\xi(\varphi)$ is introduced as the square root of $U(\varphi)$, which transforms under the $\SU(3)_L\otimes\SU(3)_R$ as
\begin{eqnarray}
\xi(\varphi)\overset{G} \longrightarrow g_R \xi(\varphi) K^{-1}(\varphi,g)=K(\varphi,g)\xi(\varphi)g_L^{-1}, \text{ with } g= (g_L,g_R)\in G.
\end{eqnarray}
where the compensator field $K(\varphi,g)$ belongs to the unbroken subgroup $\SU(3)_V$. Introducing the coset field $\xi(\varphi)$ ensures the interactions of the Goldstone bosons {with themselves and} the matter fields (such as baryons, heavy mesons, etc.) can be described by the nonlinear representation theory on quantum fields~\cite{Coleman:1969sm,Callan:1969sn}.

The leading order (LO) Lagrangian for the interactions among the light Goldstone bosons without the external sources is given as
\begin{eqnarray}\label{eq:LagUU2}
\mathcal{L}_{2}=\frac{f_{\varphi}^{2}}{4}\mathrm{Tr}\left[\partial_{\mu}U\partial^{\mu}U^{\dagger}\right],
\end{eqnarray}
where the decay constant of the light Goldstone bosons $f_\varphi$ is introduced to yield the canonical kinetic terms (as well as the mass terms in the following).

In quantum field theory, the external fields can be either a computational technique or physical entities. The extended QCD Lagrangian with external fields reads
\begin{eqnarray}\label{eq:LQCD_ext}
\mathcal{L}_{\mathrm{QCD}}^{\mathrm{ext.}}=\mathcal{L}_{\mathrm{QCD}}+\bar{q}\gamma^{\mu}(v_{\mu}+\gamma_{5}a_{\mu})q-\bar{q}(s-i\gamma_{5}p)q,
\end{eqnarray}
where the $v_\mu$, $a_\mu$, $s$ and $p$ represent the external vector, axial-vector, scalar and pseudoscalar fields, respectively. The extended Lagrangian~\eqref{eq:LQCD_ext} is invariant under the local $G_l=\SU(3)_L\otimes\SU(3)_R$  transformations if the quark fields and external fields satisfy the following transformation rules,
\begin{eqnarray}
q_{L}\overset{G_l} \longrightarrow g_{L}q_{L},~ q_{R}\overset{G_l} \longrightarrow g_{R}q_{R},~ r_{\mu}\overset{G_l} \longrightarrow g_{R}r_{\mu}g_{R}^{-1}+ig_{R}\partial_{\mu}g_{R}^{-1},~ l_{\mu}\overset{G_l} \longrightarrow g_{L}l_{\mu}g_{L}^{-1}+ig_{L}\partial_{\mu}g_{L}^{-1},~ s+ip\overset{G_l} \longrightarrow g_{R}(s+ip)g_{L}^{-1},
\end{eqnarray}
where $g_{L/R}=g_{L/R}(x)$ is the function of $x$.

In practice, the quarks generally couple to the external sources, such as the electromagnetic field, the weak current, etc.  The quark mass matrix $\mathcal{M}$ is contained in the scalar field $s$. The photon, $W$ boson fields and quark mass terms are embedded in the $v_\mu$, $a_\mu$ and $s$ via
\begin{eqnarray}
r_{\mu}&=&v_{\mu}+a_{\mu}=-e\mathscr{Q}A_{\mu}+\dots,\nonumber\\
l_{\mu}&=&v_{\mu}-a_{\mu}=-e\mathscr{Q}A_{\mu}-\frac{e}{\sqrt{2}\sin\theta_{W}}(W_{\mu}^{\dagger}T^{+}+\mathrm{H.c.})+\dots,\nonumber\\
s&=&\mathcal{M}+\dots,
\end{eqnarray}
where $A_{\mu}$ and $W_{\mu}$ denote the photon and $W$ boson fields, respectively. The quark charge matrix $\mathscr{Q}$, CKM matrix $T^{+}$ and quark mass matrix $\cal{M}$  are given as, respectively,
\begin{eqnarray}
\mathscr{Q}=\frac{1}{3}\text{diag}(2,-1,-1),\qquad T^{+}=\left[\begin{array}{ccc}
0 & V_{ud} & V_{us}\\
0 & 0 & 0\\
0 & 0 & 0
\end{array}\right], \qquad \mathcal{M}=\text{diag}(m_u,m_d,m_s),\label{eq:external}
\end{eqnarray}
The electromagnetic field and the $W$ boson fields are gauge fields. 

The low energy effective field theory of the extended QCD contains the same external fields and has the same local $G_l=\SU(3)_L\otimes\SU(3)_R$ symmetry with extended QCD Lagrangians. The local transformation invariance requires that the fields $v_\mu$ and $a_\mu$ appear in the covariant derivatives of $U$,
\begin{eqnarray}
\nabla_{\mu}U=\partial_{\mu}U-ir_{\mu}U+iUl_{\mu},\qquad \nabla_{\mu}U^\dagger=\partial_{\mu}U^\dagger+i U^\dagger r_{\mu}-i l_{\mu} U^\dagger,\qquad \nabla_{\mu}U\overset{G_l} \longrightarrow g_{R}\nabla_{\mu}Ug_{L}^{-1}.
\end{eqnarray}
We can also introduce the field strength tensors as building blocks,
\begin{eqnarray}
f_{R}^{\mu\nu}=\partial^{\mu}r^{\nu}-\partial^{\nu}r^{\mu}-i[r^{\mu},r^{\nu}],\qquad f_{L}^{\mu\nu}=\partial^{\mu}l^{\nu}-\partial^{\nu}l^{\mu}-i[l^{\mu},l^{\nu}].
\end{eqnarray}
When the gauge field is introduced as the external field, the local group $G_l=\SU(3)_L\otimes\SU(3)_R$ will reduce to the corresponding gauge symmetry.  For example, if one introduces the external electromagnetic field, the $f_{R}^{\mu\nu}$ and $f_{L}^{\mu\nu}$ degrade into the usual electromagnetic field strength tensor $f_{R}^{\mu\nu}=f_{L}^{\mu\nu}=f^{\mu\nu}=-e\mathscr{Q}(\partial^\mu A^\nu-\partial^\nu A^\mu)$. When the external field is switched off, the above local symmetry will degenerate into the global chiral symmetry of QCD.

The most general LO Lagrangian which satisfies the Lorentz invariance and local chiral symmetry reads~\cite{Gasser:1984gg,Gasser:1984ux,Gasser:1984pr}
\begin{eqnarray}\label{eq:lagGB}
\mathcal{L}_{2}=\frac{f_{\varphi}^{2}}{4}\mathrm{Tr}\left[\nabla_{\mu}U\nabla^{\mu}U^{\dagger}+\chi^{\dagger}U+U^{\dagger}\chi\right],~\label{eq:piL0}
\end{eqnarray}
where the spurion $\chi=2B_0(s+ip)$, {and $B_0$ is a LEC. The above Lagrangian is the general LO Lagrangian with the external fields. If we focus on the effect induced by the light quark masses, we could set $s=\mathcal{M}=\text{diag}\{m_u,m_d,m_s\}, p=l_\mu=r_\mu=0$, where the scalar field $s$ is the same as that in the extended QCD Lagrangian in Eq.~\eqref{eq:external}. By comparing the vacuum expectation of the EFT Hamiltonian with that of the QCD Hamiltonian with the same external field, one can get $B_0=-\langle \bar{q} q\rangle/(3f_\varphi^2)$, where $\langle \bar{q} q\rangle$ is the quark condensate~\footnote{$\langle \mathcal{H}_{\mathrm{EFT}}\rangle=\mathcal{H}_{\mathrm{EFT}}|_{\varphi=0}=-f_\varphi ^2 B_0(m_u+m_d+m_s)$ and ${\partial \langle \mathcal{H}_{\mathrm{QCD}}\rangle\over \partial m_q}|_{m_u=m_d=m_s=0}={1\over 3} \langle \bar{q}q\rangle$.}. Expanding the Lagrangian containing the spurion $\chi$ at the hadronic level, one can see it corresponds to the mass of the pseudoscalar mesons, specifically,
	\begin{eqnarray}\label{eq:explicit_breaking}
		\chi=2B_0\mathrm{diag}(m_u,m_d,m_s)=\mathrm{diag}(m_\pi^2,m_\pi^2,2m_K^2-m_\pi^2).
	\end{eqnarray}
	The masses of the pseudoscalar mesons can be related to the quark masses.}

The $\chi$PT is an EFT of QCD at low energies. Its effective Lagrangians are organized in the same form as given in Eq.~\eqref{eq:EFTexpansion}, where the soft scale $\mathcal{Q}$ can be either the external momentum of the Goldstone bosons or their small masses. The expansion index $\nu$ (chiral order) is obtained from Weinberg's power counting based on the naive dimensional analysis (NDA)~\cite{Weinberg:1978kz}. The building blocks (fields and spurion) in the chiral Lagrangians are counted as
\begin{eqnarray}\label{eq:countingFields}
\mathcal{O}(p^0):U;\qquad \mathcal{O}(p^1):\nabla_\mu U,~v_\mu,~a_\mu;\qquad \mathcal{O}(p^2):f_{L}^{\mu\nu},~f_{R}^{\mu\nu},~s,~p.
\end{eqnarray}
In $\chi$PT without matter fields, the Lagrangians can be even order only. From Eq.~\eqref{eq:countingFields} one sees that the LO Lagrangian in Eq.~\eqref{eq:lagGB} is of order $\mathcal{O}(p^2)$. The chiral dimension (order) of a Feynmann diagram is
\begin{equation}
	D=4L-2I_{M}+\sum_{i}V_{i}d_{i},
\end{equation}
where $L$ and $I_M$ are the numbers of loops and Goldstone boson inner lines, respectively. $V_i$ and $d_i$ are the number of vertex $i$ and number of the derivatives in the vertex $i$, respectively. With the topological relation, $L=I_{M}-\sum_{i}V_{i}+1$,
one can obtain a more useful power counting,
\begin{equation}
	D=2L+2+\sum_{i}V_{i}(d_{i}-2).\label{eq:1.2:pwc}
\end{equation}
For the LO interaction ($d_i=2$), the amplitudes of Feynman diagrams with an extra loop is suppressed by two extra chiral orders. To achieve a certain precision, one needs to calculate the Feynman diagrams with a finite {numbers of} loops. With the power counting, one can calculate the amplitude perturbatively (in terms of $p/\Lambda_\chi$).

\subsection{Heavy quark effective theory}

In this part, we discuss the infinite heavy quark mass limit of low energy QCD~\cite{Politzer:1988wp,Isgur:1989vq,Eichten:1989zv,Georgi:1990um,Isgur:1990yhj,Grinstein:1990mj,Falk:1990yz,Mannel:1991mc}. The masses of the $c$ and $b$ quarks are much larger than the nonperturbative scale $\Lambda_{\mathrm{QCD}}$. We will see that the dynamics is largely simplified in the approximation $m_Q\to\infty$. For a hadron containing one single heavy quark, the typical transferred momentum $p^{\mathrm{ty}}$ between the heavy quark and light d.o.f is of order $\Lambda_{\mathrm{QCD}}$. The heavy quark is almost on-shell. The variation of the heavy quark velocity $\delta v= p^{\mathrm{ty}}/m_Q$ is very small due to $m_Q\gg p^{\mathrm{ty}}$. In the limit $m_Q\to \infty$, the heavy quark moves with a constant velocity. In the heavy quark limit, the heavy quark is at rest in the hadron rest frame and serves only as a static color source .

\subsubsection{Heavy quark flavor and spin symmetries}\label{sec:HQFSHQSS}

In the framework of heavy  quark effective {theory} (HQET), the four-momentum $p^\mu$ of a heavy quark is split into two parts,
\begin{eqnarray}
p^\mu=m_Qv^\mu+k^\mu,
\end{eqnarray}
where $v^\mu$ is the four velocity of the heavy quark with $v^2=1$, $k^\mu$ is called the residual momentum. $k\ll m_Q v$ since the heavy quark is almost on-shell. The heavy quark spinor field $Q(x)$ can also be separated into the large component $h_{v}^{(Q)}(x)$ and small component $H_{v}^{(Q)}(x)$
\begin{eqnarray}\label{eq:HQreduction}
Q(x)\equiv(P_{v+}+P_{v-})Q(x)=\exp(-im_{Q}v\cdot x)\left[h_{v}^{(Q)}(x)+H_{v}^{(Q)}(x)\right],
\end{eqnarray}
via introducing the projection operators $P_{v\pm}$,
\begin{eqnarray}
P_{v\pm}=\frac{1\pm \slashed{v}}{2},\quad \text{with } P_{v+}+P_{v-}=1,\quad P_{v\pm}^{2}=P_{v\pm},P_{v\pm}P_{v\mp}=0,\quad \slashed{v}h_v^{(Q)}=h_v^{(Q)},\quad \slashed{v}H_v^{(Q)}=-H_v^{(Q)}.
\end{eqnarray}
With $v_\mu=(1,\bm{0})$, one can see that the $h_{v}^{(Q)}(x)$ and $H_{v}^{(Q)}(x)$ correspond to the large component and small component of the spinor, respectively.  $H_v(x,t)$ is suppressed by the factor $1/m_Q$ in comparison with $h_v(x,t)$. The detailed derivation is presented in ~\ref{app:HFE}.
With Eq.~\eqref{eq:HQreduction}, the heavy quark Lagrangian becomes~\cite{Neubert:1993mb}
\begin{eqnarray}\label{HQexpansion1}
	\mathcal{L}_{\mathrm{QCD}}^{(Q)}=\bar{h}_{v}^{(Q)}(iv\cdot\mathcal{D})h_{v}^{(Q)}-\bar{H}_{v}^{(Q)}\left(iv\cdot\mathcal{D}+2m_{Q}\right)H_{v}^{(Q)}+\bar{h}_{v}^{(Q)}i\slashed{\mathcal{D}}_{\bot}H_{v}^{(Q)}+\bar{H}_{v}^{(Q)}i\slashed{\mathcal{D}}_{\bot}h_{v}^{(Q)},
\end{eqnarray}
where $\mathcal{D}_{\bot}^{\mu}=\mathcal{D}^{\mu}-v^{\mu}(v\cdot\mathcal{D}), ~v\cdot\mathcal{D}_{\bot}=0$. One can see that the $h_v$ and $H_v$ are light and heavy fields with the mass $0$ and $2m_Q$,  respectively. With the equation of motion or the path integral approach (see~\ref{app:HFE} for details), one can get the effective Lagrangian
\begin{eqnarray}\label{eq:LagHQET1}
	\mathcal{L}_{\mathrm{eff}}=\bar{h}_{v}^{(Q)}(iv\cdot\mathcal{D})h_{v}^{(Q)}+\bar{h}_{v}^{(Q)}i\slashed{\mathcal{D}}_{\bot}\frac{1}{(iv\cdot\mathcal{D}+2m_{Q}-i\epsilon)}i\slashed{\mathcal{D}}_{\bot}h_{v}^{(Q)}.
\end{eqnarray}
An expansion of Lagrangian in Eq.~\eqref{eq:LagHQET1} up to $\mathcal{O}(1/m_Q)$ reads
\begin{eqnarray}\label{eq:LagHQET2}
	\mathcal{L}_{\mathrm{HQET}}=\bar{h}_{v}^{(Q)}(iv\cdot\mathcal{D})h_{v}^{(Q)}+\frac{1}{2m_{Q}}\bar{h}_{v}^{(Q)}(i\slashed{\mathcal{D}}_{\bot})^{2}h_{v}^{(Q)}+\frac{g_s}{4m_{Q}}\bar{h}_{v}^{(Q)}\sigma^{\alpha\beta}\mathcal{G}_{\alpha\beta}h_{v}^{(Q)}+\mathcal{O}(1/m_{Q}^{2}),
\end{eqnarray}
where $\mathcal{G}_{\alpha\beta}=\lambda^{a}\mathcal{G}_{\alpha\beta}^{a}/2$.

 In the heavy quark limit, only the LO term survives. The original quark--gluon vertex $ig_s\gamma^\mu\lambda_a/2$ becomes $ig_s v^\mu\lambda_a/2$, i.e., the interaction is independent of the heavy quark spin. This is called the heavy quark spin symmetry (HQSS). In addition, the heavy quark mass is eliminated thus the reduced Lagrangian is invariant with a global transformation under heavy flavor $\mathrm{U}(2)$ group. This is the heavy quark flavor symmetry (HQFS). Therefore, the spin and flavor together form a larger $\SU(2)\otimes \mathrm{U}(2)\in \mathrm{U}(4)$ group.

The physical picture behind the HQSS and HQFS is that the soft gluon as a soft probe can only resolve the dynamics occurring at the scale of $\Lambda_{\mathrm{QCD}}^{-1}$. The light d.o.f can only perceive the color forces induced by the color charge of the heavy quark (chromoelectric interaction), while the chromomagnetic interaction that carries the spin and flavor information of the heavy quark vanishes when $m_Q\to\infty$.

The HQSS and HQFS lead to interesting spectroscopic implications for the hadrons containing a heavy quark. In the heavy quark limit, the heavy spin $j_h$ (spin of the heavy quark, $j_h={1\over 2}$) and light spin $j_\ell$ (the total angular momentum of the light d.o.f) are conserved separately. Thus, the singly heavy hadrons can be labeled by its light spin $j_\ell$. The states with total spin $J=j_\ell\pm {1\over 2}$ should be degenerate. An example is the masses of the ground state heavy mesons,
\begin{equation}\label{eq:masssplitBD1}
m_{D^\ast}-m_D\approx140 \text{ MeV}, \qquad m_{B^\ast}-m_B\approx45 \text{ MeV},
\end{equation}
where the mass splittings are relatively small compared with the masses of the heavy mesons. {The HQSS dictates the mass splittings are inversely proportional to heavy quark masses, namely $(m_{D^*}-m_D)/(m_{B^*}-m_B)=m_b/m_c$ in line with the above experimental results.} Meanwhile, HQFS requires that the mass difference between the mesons with different light d.o.f for the charm and bottom systems should be approximately the same. For example
\begin{equation}\label{eq:masssplitBD2}
	m_{B_s^{(\ast)}}-m_{B^{(\ast)}}\approx m_{D_s^{(\ast)}}-m_{D^{(\ast)}} \approx 100 \text{ MeV}.
\end{equation}

Apart from the mass spectrum, the HQSS and HQFS could be used to relate the different coupling constants and interactions. For example, in the HQFS, the vertices $BB^*\pi$ and $\bar{D}\bar{D}^*\pi$ can be related to each other. In the heavy quark limit, the interaction between two heavy hadrons is dominated by the interactions of the light d.o.f of the two particles. For example, the HQSS can bridge $D\bar{D}$ interaction and $D^*\bar{D}^*$ interaction. More examples will be discussed in Sec.~\ref{sec:HQSinHHM}.

In fact, the masses of the heavy quarks are finite and $m_c<m_b$, which indicates the HQS is explicitly broken. In Eq.~\eqref{eq:LagHQET2}, the second and third terms correspond to kinetic energy and chromomagnetic hyperfine interaction, respectively. The second term will break the HQFS while the third term will break both HQSS and HQFS. The chromomagnetic term gives the mass splittings in the spin doublets $(D,D^{\ast})$ and $(B,B^{\ast})$ at $\mathcal{O}(1/m_{Q})$, respectively. This leads to the refined relation $m_{B^*}^2-m_B^2\approx m_{D^*}^2-m_D^2$.

\subsubsection{An introduction to the heavy diquark-antiquark symmetry}\label{sec:HDAS}

The HDAS was proposed by Savage and Wise in Ref.~\cite{Savage:1990di}, which relates the doubly heavy baryons $QQq$ ($\bar{Q}\bar{Q}\bar{q}$) to the heavy mesons $\bar{Q}q$ ($Q\bar{q}$). The basic assumption (approximation) of HDAS is that the $QQ$ pair in color $\bar{\bm 3}$ in the doubly heavy baryons forms a compact object under the attractive color Coulomb interaction. The compact $QQ$ is called the diquark (for more general concepts of the diquark, we refer to~Ref. \cite{Anselmino:1992vg}). In the limit $m_Q\to \infty$, the heavy diquark becomes pointlike (the radius is inverse to the $m_Q$ for the Coulomb interaction) and only acts as a static color source in the $\bar{\bm 3}$ channel, which plays the same role as the heavy antiquark in the $\bar{Q}q$ meson.

If we denote the vector diquark as $\mathscr{D}_j$, the LO Lagrangian has the same form as that in Eq.~\eqref{eq:LagHQET2}
\begin{eqnarray}\label{eq:HDASLOexpansion}
\mathcal{L}_{\mathscr{D}}=\mathscr{D}^\dagger_j(iv\cdot \mathcal{D})\mathscr{D}_j,
\end{eqnarray}
with $j=1,2,3$ the spin index for the vector state. The Lagrangian~\eqref{eq:LagHQET2} plus~\eqref{eq:HDASLOexpansion} is invariant under the $\mathrm{U}(5)$ transformation, which is named as the superflavor symmetry~\cite{Georgi:1990ak}. The $S$-wave $QQq$ and $\bar{Q}q$ form the doublets $(\Xi_{QQ}^\ast,\Xi_{QQ})$ and $(\tilde{P}^\ast,\tilde{P})$ with spins-$(3/2,1/2)$ and -$(1,0)$, respectively. When the chromomagnetic interactions at $\mathcal{O}(1/m_Q)$ are considered [analogous to Eq.~\eqref{eq:LagHQET2}], they give the following relations of the mass splittings~\cite{Savage:1990di,Fleming:2005pd},
\begin{eqnarray}\label{eq:HDASmasssplitting}
m_{\Xi_{QQ}^\ast}-m_{\Xi_{QQ}}=\frac{3}{4}(m_{\tilde{P}^\ast}-m_{\tilde{P}}),
\end{eqnarray}
which is qualitatively supported by the calculations from the potential model~\cite{Ebert:2002ig} and lattice QCD~\cite{Lewis:2001iz,Brown:2014ena,Padmanath:2015jea}. A similar relation was extended to the singly heavy baryons $Qqq$ and doubly heavy tetraquarks $QQqq$~\cite{Mehen:2017nrh}. The corrections to Eq.~\eqref{eq:HDASmasssplitting} from the nonrelativistic QCD (NRQCD) calculation were given in~\cite{Mehen:2019cxn}.
Eq.~\eqref{eq:HDASmasssplitting} implies that the one-pion transition $\Xi_{QQ}^\ast\to \Xi_{QQ}\pi$ is inaccessible in experiments. Considering $m_{D^\ast}-m_D\approx m_\pi$, only the radiative and weak decays of the $\Xi_{QQ}^\ast$ state are allowed. The HDAS was also adopted to relate the axial coupling constant of the doubly heavy baryon to that of the heavy meson~\cite{Hu:2005gf} because the axial coupling $D^\ast D\pi$ can be directly extracted from the partial decay width of the $D^\ast$ in experiments. The details are given in Sec.~\ref{sec:1.5:combChandHQ}.

The HDAS is valid only if the spatial extent cannot be resolved from the perspective of the light d.o.f. If $m_Qv\gg\Lambda_{\mathrm{QCD}}$~\footnote{$m_Qv$ and $m_Qv^2$ are the typical transferred momentum and binding energy between heavy quarks in the NRQCD power counting.} and the diquark excitations are suppressed, the HDAS is a good approximation. However, for the doubly charmed baryons, $m_Qv\sim\Lambda_{\mathrm{QCD}}\sim m_Qv^2$, thus one expects that the HDAS breaking effect is sizable in the charmed sector. One can further consult the related discussions and calculations from NRQCD~\cite{Fleming:2005pd,Brambilla:2005yk,Cohen:2006jg}.

\subsection{Heavy baryon chiral perturbation theory}\label{sec:sec2.3}

When we extend the chiral perturbation theory to the heavy-light hadrons, we have to understand the meanings of ``heavy" in twofold ways. {First, the mass of the matter field is comparable to $\Lambda_\chi$, which is heavy compared to the pion mass.} In this sense, the nucleon mass is also heavy. At the hadronic level, the heavy baryon $\chi$PT (HB$\chi$PT) for the nucleon (or the matter field without heavy quarks) was invented to perform the chiral expansion. Meanwhile, for the system with a heavy quark, one can perform the heavy quark expansion and adopt the heavy quark symmetry. The heavy field expansion is performed at the quark level. In this subsection and Sec.~\ref{sec:1.5:combChandHQ}, we will focus on twofold meanings in order. In this subsection, we will take the nucleon as an example of the matter field with a large mass but without heavy quarks.

The Weinberg's power counting in Eq.~\eqref{eq:1.2:pwc} based on the NDA can be extended to the mater fields~\cite{Weinberg:1991um},
\begin{equation}
	D=4L-I_{N}-2I_{M}+\sum_{i}V_{i}d_{i},
\end{equation}
where $L$, $I_M$ and $I_N$ are the numbers of loops, Goldstone boson inner lines and matter field inner lines, respectively. $V_i$ and $d_i$ are the numbers of vertex $i$ and the derivatives in the vertex $i$, respectively. With two topological relations,
\begin{equation}
L=I_{N}+I_{M}-\sum_{i}V_{i}+1,\qquad 2I_{N}+E_{n}=\sum_{i}V_{i}n_{i},
\end{equation}
one can obtain a more useful power counting,
\begin{equation}
D=2L+2-\frac{E_{n}}{2}+\sum_{i}V_{i}\Delta_{i},\qquad\text{with }\Delta_{i}\equiv d_{i}+\frac{n_{i}}{2}-2~\label{eq:1.4:pwc},
\end{equation}
where $n_i$ is the number of matter fields connected by the vertex $i$. $E_n$ is the number of the external matter fields. The power counting can be generalized to the cases with more than one separately connected pieces and more than two matter fields~\cite{Weinberg:1992yk,Machleidt:2011zz,Epelbaum:2008ga}. In this section we focus on the process with one matter field and take $E_n=2$.

{In order to derive \eqref{eq:1.4:pwc}, we treat the ``momentum" of the matter field as a small scale, which is valid only when the mass of the matter field $M$ is removed properly.} However, in the practical calculations (e.g., in the dimensional regularization scheme), the mass of the matter field $M$ is comparable to or even larger than  $\Lambda_\chi$, which will make the naive power counting in Eq.~\eqref{eq:1.4:pwc} fail. In fact, compared to the matter field mass $M$, the chiral fluctuations involve very soft dynamics. Thus, one can integrate out the hard scale $M$ and recover the Weinberg's power counting. In the HB$\chi$PT~\cite{Jenkins:1990jv}, one can separate the momentum of the matter field into two parts, $p_\mu =Mv_\mu+q_\mu$, like the heavy quark expansion in HQET, where $v^2=1$. $Mv_\mu$ and $q_\mu$ are the mass term and residual momentum, respectively. The field is divided into the heavy field and massless light field. One can obtain the action and Lagrangian of the effective field theory by integrating out the heavy field with the assistance of the equation of motion~\cite{Neubert:1993mb} (in the classical sense) or the path integral approach~\cite{Bernard:1992qa} (in the quantum sense), see~\ref{app:HFE}.

We take the LO Lagrangian of the nucleon pion interaction as an example,{
\begin{equation}
	\ensuremath{\mathcal{L}^{(1)}=\bar{\Psi}(i\slashed{\mathcal{D}}-M+\frac{g_{A}}{2}\slashed u\gamma_{5})\Psi},\qquad\qquad   u_{\mu}= {i \over 2}\left[\xi^{\dagger}\partial_{\mu}\xi-\xi\partial_{\mu}\xi^{\dagger}\right],\label{eq:sec1.5:ncl_LO}
\end{equation}
where $\Psi=(p,n)^T$ and $\xi^2=U$. The complete form of $u_\mu$ with the external fields is given in Eq.~\eqref{eq:psef}.}  With the heavy field decomposition, the corresponding $\cal{A}$, $\cal{B}$  and $\cal{C}$ in Eq.~\eqref{eq:sec1.5:lightandheavy} read
\begin{eqnarray}
	{\cal A}&=&iv\cdot \mathcal{D}+g_{A}(u\cdot S),\\{\cal B}&=&i\slashed{\mathcal{D}}^{\perp}-\frac{g_{A}}{2}(v\cdot u)\gamma_{5},\\{\cal C}&=&i(v\cdot \mathcal{D})+2M+g_{A}(u\cdot S).
\end{eqnarray}
One can see that there is no mass term for the light field $H$, while the mass of the heavy field $h$ is $2M$.
Expanding the $\mathcal{C}^{-1}$ in powers of $1/M$ one obtains that
\begin{eqnarray}
	{\cal C}^{-1}=\frac{1}{2M}-\frac{i(v\cdot\mathcal{D})+g_{A}(u\cdot S)}{(2M)^{2}}+\dots.
\end{eqnarray}
The contribution of the heavy component will appear as the recoiling effect, which is suppressed by the power of $1/M$. In the HB$\chi$PT, one has to perform two expansions, the chiral expansion and heavy baryon expansion. The HB$\chi$PT has been widely used to investigate the chiral dynamics of the nucleon systems~(e.g., see Refs.~\cite{Bernard:1995dp,Bernard:2007zu} for reviews). For the heavy flavor systems, the recoiling effect is less important because of the much larger heavy quark mass $M$ as compared to the nucleon mass. However, the HQS breaking effect will appear as the $1/M$ correction for the heavy flavor system, which is particularly interesting for some systems. We will discuss this issue in Sec.~\ref{sec:HQSinHHM}.

\begin{figure}[hbtp]
	\begin{center}
		\includegraphics[width=1.0\textwidth]{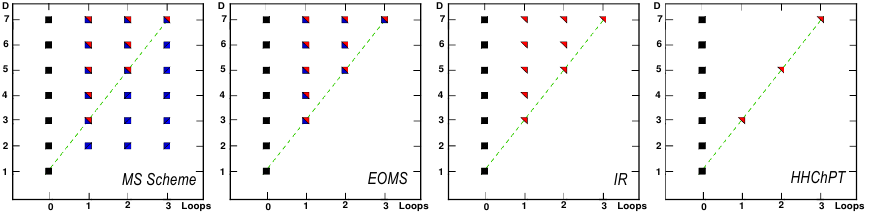}
		\caption{The comparisons of the different regularization schemes in LO Lagrangians~\cite{yaophd}.  $x$- and $y$-axis denote the number of loops and the corresponding chiral order, respectively. {The Green dashed lines stand for the Weinberg's power counting. The chiral orders increase linearly with the number of loops according to Eq.~\eqref{eq:1.4:pwc} if all the vertices are the LO ones}. The four subfigures (from left to right) correspond to MS, EOMS, infrared regularization and HH$\chi$PT schemes, respectively. The blue lower and red upper triangles represent the infrared regular and singular terms, respectively. The blue squares denote the PCB terms, while the black squares represent tree diagrams and counter terms. {A successful subtraction scheme should ensure that there is no contribution below the green dashed line.} } \label{Fig:1.4:bchpt}
	\end{center}
\end{figure}

Apart from the nonrelativistic expansion, the infrared regularization scheme~\cite{Becher:1999he,Becher:2001hv,Schindler:2003xv} and extended on-mass-shell scheme (EOMS)~\cite{Fuchs:2003qc} are two  Lorentz covariant approaches to performing chiral expansion for the matter fields. The motivation of the infrared regularization is to discern the soft dynamics (typically $m_\varphi$) and the hard dynamics (typically $M$). One can only focus on the soft part, which includes the chiral fluctuation effect. To this end, in the infrared regularization scheme, the loop integral is divided into the infrared singular and regular parts when the mass of the Goldstone boson approaches zero. The infrared singular part is typically the soft dynamics. It was shown that the infrared singular parts do not include the power counting breaking (PCB) effect~\cite{Becher:1999he} and thus are kept. The infrared regular terms can be expanded as polynomials of the $m_\varphi$ and thus are absorbed by the renormalization of the LECs. As shown in Fig.~\ref{Fig:1.4:bchpt}, the leading terms in the loop integrals of the infrared regularization are consistent with the Weinberg's power counting in Eq.~\eqref{eq:1.4:pwc}.

In the extended on-mass-shell scheme~\cite{Fuchs:2003qc}, all the PCB terms are polynomials of small scales (such as the $m_\varphi$), which can be absorbed by redefining the LECs. As shown in Fig.~\ref{Fig:1.4:bchpt}, in this scheme, some infrared regular terms which do not violate the power counting are kept. Like the infrared regularization scheme, the remaining terms in the loop integrals are at least at the order given by Eq.~\eqref{eq:1.4:pwc}. The terms with powers beyond the Weinberg's power counting only contribute at higher orders.

\subsection{Effective range expansion}~\label{sec:ere}
Effective range expansion (ERE) is a convenient parameterization scheme of the near-threshold partial wave $T$-matrix, which was first proposed by Schwinger in an unpublished note and reformulated by Bethe~\cite{Bethe:1949yr}. The ERE has extensive applications in both the experimental and theoretical aspects, such as fitting the scattering data~\cite{Blatt:1949zz}, Weinberg compositeness~\cite{Weinberg:1962hj}, determining the power counting of EFT~\cite{Kaplan:1998tg}, universality in low energy scattering~\cite{Braaten:2004rn}, $m_\pi$-dependence~\cite{Baru:2015ira} and finite volume effect~\cite{Morningstar:2017spu,CLQCD:2019npr} of the lattice QCD simulation and so on.

The partial wave $S$-matrix and $T$-matrix are parameterized as functions of the phase shift $\delta_l$,
\begin{equation}
S_{l}=1+2ikT_{l}(k)=e^{2i\delta_{l}(k)},\qquad T_{l}(k)=\frac{k^{2l}}{k^{2l+1}\cot\delta_{l}-ik^{2l+1}},
\end{equation}
where the unitarity $S_lS_l^\dagger=1$ is fulfilled. Here, we use the nonrelativistic formalism.  The power of the $k$ for $T_l$ is determined by the asymptotic behavior of the wave functions in the partial wave basis. One can see that the unitary cut of the $T$-matrix is ensured by introducing the $ik^{2l+1}$ in the denominator. $k^{2l+1}\cot\delta_{l}$ is a {meromorphic function, where we assume that the left-hand cuts are sufficiently far way}. Assuming there is no pole for $k^{2l+1}\cot\delta_{l}$ near the threshold, one can expand it with the Taylor series in powers of $k^2$,
\begin{equation}
k^{2l+1}\cot\delta_{l}=-\frac{1}{a_s}+\frac{1}{2}r_0k^{2}+v_{2}k^{4}+v_{3}k^{6}+...,~\label{eq:ere1}
\end{equation}
where the coefficients are named as shape parameters.  In particular, the leading two order coefficients $a_s$ and $r_0$ are scattering length and effective range, respectively. The Eq.~\eqref{eq:ere1} is called effective range expansion. In the derivation of Bethe's work~\cite{Bethe:1949yr}, the above expansion is formulated in the two body systems with a local interaction, namely $V(\bm{r},\bm{r'})=V(\bm{r})\delta^3(\bm{r}-\bm{r'})$. The $r_0$ gets its name since its value roughly equals to half of the range of the potential~\cite{newton2013scattering}. However, in the modern view, the potentials corresponding to the same scattering data are not unique and there is no good reason to exclude the non-local interactions. It was proven by Ekstein~\cite{Ekstein:1960xkd} that a large class of unitary transformations can relate the different Hamiltonians producing the same $S$-matrix. In the modern perspective, the ERE is not relevant to the locality of the interaction but an effective theory to depict the low-energy behavior of the two-body scattering with finite parameters truncated according to the expected precision. It was shown that the ERE is just equivalent to the pionless effective field theory~\cite{vanKolck:1998bw}. {But, once the one-pion exchange interaction is introduced, the accompanying left-hand cut will invalidate the ERE expansion.  }

\begin{figure}[htb]
    \centering
    \includegraphics[width=0.45\textwidth]{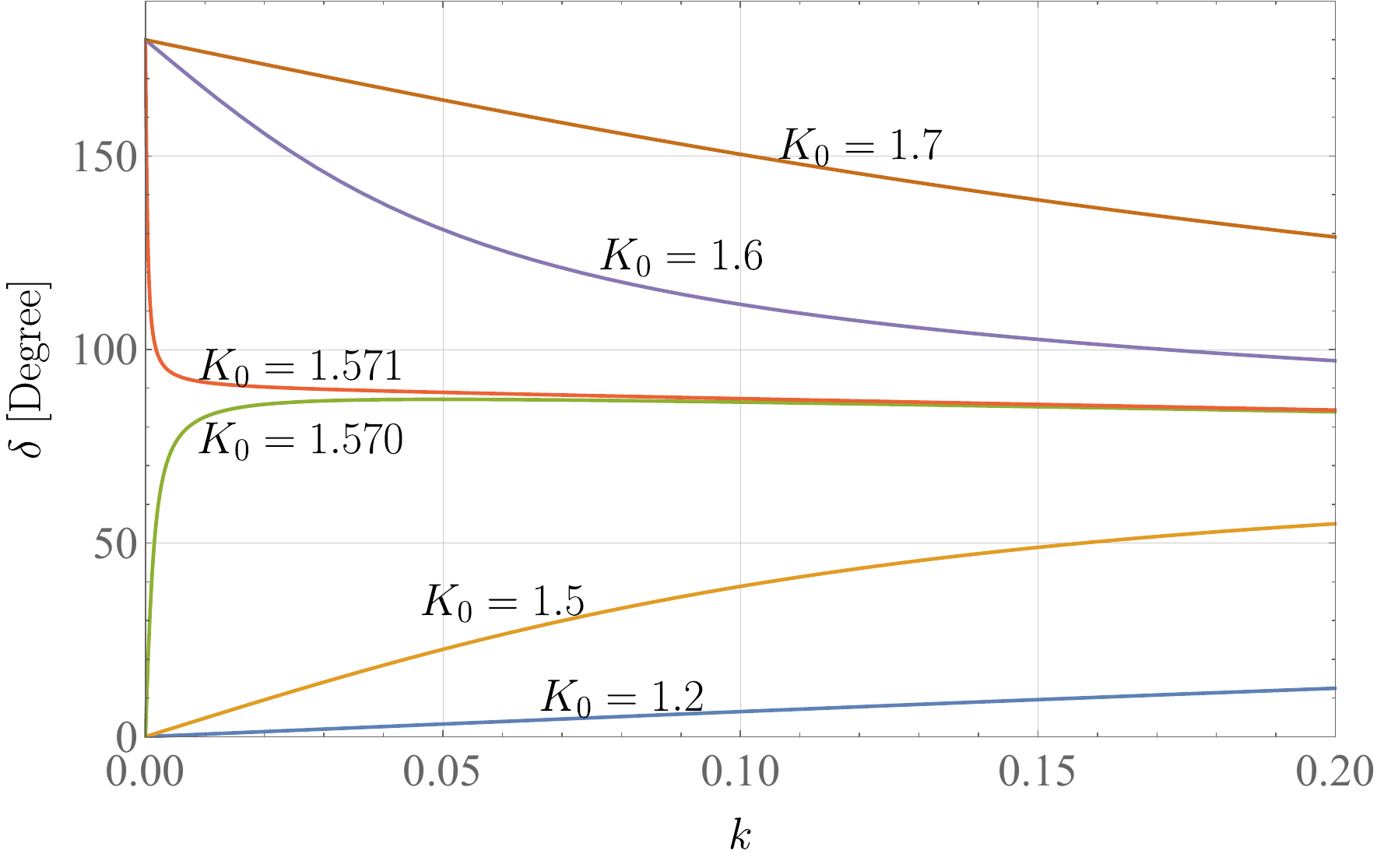}
        \includegraphics[width=0.45\textwidth]{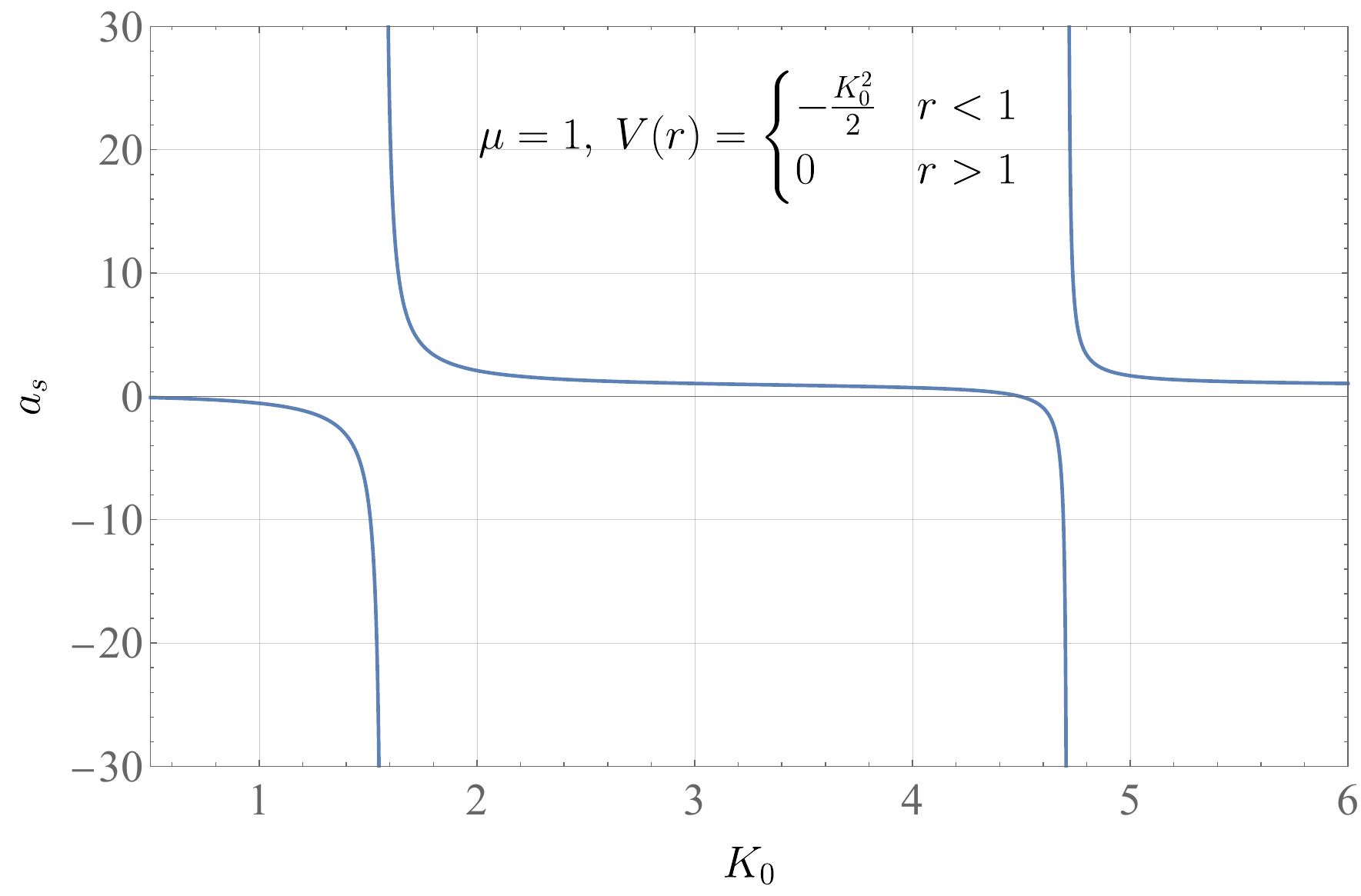}
    \caption{Levinson's theorem and scattering lengths with the square well potential  as an example (specifically in right panel). The left panel and the right panel illustrate the phase shifts near the threshold and scattering lengths respectively when the interaction becomes more attractive and then bound states appear.  }
    \label{fig:levinson}
\end{figure}

The scattering length encodes  important information. According to Levinson's theorem~\cite{levinson1949uniqueness,newton2013scattering}, if the potential admits $n$ bound states in the $S$-wave, the phase shift at zero energy is 
\begin{equation}
   \delta(0)=n\pi \quad [\text{with the convention }\delta(\infty)=0]. 
\end{equation}
 For the marginal potential that the $n_{\text{th}}$ bound state starts to appear, the phase shift is $\delta(0)=(n+\frac{1}{2})\pi$. The phase shifts within the square well potential are shown in Fig.~\ref{fig:levinson} when the potential becomes more attractive and the first bound state starts to appear at $K_0 \sim 1.57$. The scattering lengths will cross the infinity (from the negative one to the positive one or vise verse) when the new bound state appears (see the right panel of Fig.~\ref{fig:levinson}). When the interaction is repulsive everywhere or attractive everywhere but too weak to form the bound states, the sign of the scattering lengths can reflect the sign~\footnote{There are different conventions about the sign of the scattering length. Here, we follow the convention in Eq.~\eqref{eq:ere1}.} of the potential~\cite{scatteringlength},
\begin{eqnarray}
	V>0\text{ everywhere }	&\rightarrow& a_{s}>0,\nonumber\\
	V<0\text{ everywhere (but no bound states) }	&\rightarrow& a_{s}<0.
\end{eqnarray}
{For the system with the ``bare" poles, namely the discrete eigenstates of the free Hamiltonian $H_0$, the Levinson's theorem is generalized as~\cite{generallev1958,Vaughn:1961poz,Krivoruchenko:2020jtr,Li:2022aru} 
\begin{equation}
    \delta (0)=(n-n_{\rm bare})\pi,\label{eq:generalLevison}
\end{equation}
where $n_{\rm bare}$ is the number of the bare poles, and $n$ is the number of the bound states, namely the discrete eigenstates of the Hamiltonian $H$.
}

 Landau and Smorodinsky proved that the effective range is always positive $r_0>0$ if the local potential is attractive everywhere, i.e. $V(r)<0$ everywhere in their textbook~\cite{landau:text}. The arguments was cited in Ref.~\cite{Esposito:2021vhu} to claim that ``the molecular case gives always $r_0 > 0$". However, as we mentioned before, the potentials corresponding to the same observable are not unique. The local attractive everywhere potential is not the necessary condition of the hadronic bound states. The statement in Ref.~\cite{Esposito:2021vhu} extends the argument of Landau and Smorodinsky without an extra proof. In fact, according to the Wigner's theorem~\cite{Wigner:1955zz} (based on the unitary and causality), the effective range of the zero-range potential is non-positive~\cite{Phillips:1996ae}. {From the perspective of the inverse scattering problem, one can construct the non-local potential to permit any given scattering phase shift function (no matter the sign of the effective range) and bound states with given binding energies. For example, one can construct the rank-one separable potential to permit a loosely bound state and arbitrary phase shift~\cite{Tabakin:1969mr}. }In Sec.~\ref{sec:1chanel_pionless}, we will construct an example with the non-local interaction admitting the bound state but with negative effective range. See~\cite{xrange} for more detailed discussions on this issue.

The ERE in Eq.~\eqref{eq:ere1} is the simplest version, which has been extended to more complicated cases. The single-channel ERE was generalized to the multichannel cases~\cite{ROSS1960391,ROSS1961147,PavonValderrama:2005ku}. In fitting the pion-pion scattering, the effect of the inelastic thresholds are considered with ERE supplemented by a conformal expansion, which takes the unitarity and analyticity into account~\cite{Pelaez:2004vs}. Meanwhile, the convergence radius of Eq.~\eqref{eq:ere1} is determined by the appearance of the left-hand cuts. For example, for the $NN$ scattering systems, the ERE fails at the energy $|E_\text{lab}|\sim m_\pi^2/(2 m_N)=10.5$ MeV. If one includes the Coulomb interaction, the convergence radius of ERE is zero.  In Ref.~\cite{vanHaeringen:1981pb}, the modified ERE was proposed to overcome this problem by separating the long-range interaction and the short-range interaction, where the long-range interaction is known explicitly (e.g., OPE and Coulomb interaction). In the modified ERE, the contribution of the {long-range interaction} is calculated directly and the contribution of the {short-range} interaction can be expanded in the power of $k^2$, where the left-hand singularity is absent. It was shown that the effective-range function is actually metamorphic, where the poles might prevent performing the Taylor expansion in Eq.~\eqref{eq:ere1}. The problem can be solved by replacing the Taylor expansion with the Pad\'e approximation~\cite{Midya:2015eta}.

\subsection{Chiral unitary approach} \label{sec:2.4}

{The analyticity and unitarity are important features of the $S$-matrix, which are closely tied to the  causality and conservation of the probability current, respectively.} Considering these constraints, many theoretical tools have been invented to explore the nonperturbative dynamics. The chiral unitary approaches combine the $\chi$PT, unitarity and analyticity. Compared with the $\chi$PT, the 
chiral unitary formalism satisfies the exact unitary condition, while the $\chi$PT satisfies it perturbatively. A dazzling merit  of the unitary methods is that they can introduce the bound state or resonance poles in the amplitude, which is impossible in the $\chi$PT up to any finite order. In this subsection, we will introduce the basic concepts and frameworks. We refer to Refs.~\cite{Badalian:1981xj,Oller:2000ma,Oller:2019opk,Yao:2020bxx} for reviews.

For  an elastic scattering process $1+2\rightarrow1+2$, its partial wave amplitude is given by 
 \begin{eqnarray}
T_{\ell}(s)=\frac{1}{2(\sqrt{2})^{\alpha}}\int_{-1}^{1}dz T(s,z)P_{\ell}(z),
 \end{eqnarray}
where $(\sqrt{2})^{\alpha}$ is the symmetry factor. The $\alpha$ equals $1$ or $0$ when the two particles are identical and different, respectively. $\sqrt{s}$ is the total energy. $z=\text{cos}\theta$,  with $\theta$  the relative angle between the initial and final momenta in the center of mass system (c.m.s). $P_\ell$ is the Legendre polynomial.

The partial wave $S$-matrix is constrained by the unitary, 
\begin{eqnarray}
   S_{\ell}(s)=1+2i\rho(s)T_{\ell}(s),\qquad S_{\ell}^*S_\ell=1,
\end{eqnarray}
with
 \begin{eqnarray} \label{eq:rho}
\rho(s)=\frac{q}{8\pi\sqrt{s}},\qquad q=\frac{\sqrt{\left(s-\left(m_{1}+m_{2}\right)^{2}\right)\left(s-\left(m_{1}-m_{2}\right)^{2}\right)}}{2\sqrt{s}},\label{eq:qrelative}
 \end{eqnarray}
where $m_{1}$ and $m_{2}$ are the masses of the scattering particles. $q$ is the relative momentum in the c.m.s.
The optical theorem  $\operatorname{Im} T_{\ell}(s)=T_{\ell}^{\dagger}(s) \rho(s) T_{\ell}(s)$ leads to the unitary condition
\begin{eqnarray}\label{eq:righthc}
\operatorname{Im}{T}_{\ell}^{-1}(\sqrt{s})=-\rho(s), \qquad s \geq s_{\text{th}}=\left(m_{1}+m_{2}\right)^{2}. \end{eqnarray}
Based on Eq.~\eqref{eq:righthc}, one can define the $K$-matrix, $T_\ell=[K^{-1} - i\rho(s)]^{-1}$, where $K^{-1}=\text{Re}(T^{-1}_\ell)$. 

The $T_\ell$ is the function of $\sqrt{s}$, which is real and analytical along some intervals of real $\sqrt{s}$ axis. The $T_\ell$ can be analytically continued to the whole complex plane except for the poles and cuts. According to the Schwartz reflection principle (see Ref.~\cite{Oller:2019opk} for details), one gets
\begin{equation}
    T_\ell(\sqrt{s})=T_\ell(\sqrt{s}^*)^*.
\end{equation}  
With the unitary condition in~\eqref{eq:righthc}, we know the discontinuity appears across the real axis for $\sqrt{s}>m_1+m_2$,
 \begin{eqnarray} \label{eq:lefthc}
T_{\ell}(\sqrt{s}+i\epsilon)-T_{\ell}(\sqrt{s}-i\epsilon)=2i\operatorname{Im}T_{\ell}(\sqrt{s}),
 \end{eqnarray}
which corresponds to a right-hand cut (named after its approaching to the right), or unitary cut. The unitary cut is independent of the interaction and appears as the threshold opens, which is also classified as the kinetic cut.

Apart from the right-hand cut, there might exist the left-hand cuts (also called the dynamical cuts since they depend on the dynamical details, e.g., see Ref.~\cite{oller2019brief}). For the relativistic systems, the cross symmetry will transform the singularities of the $t$-channel or $u$-channel into the left-hand cuts in the $s$-channel. For the nonrelativistic system, the particle-exchange interaction such as the one-pion-exchange interaction in the $NN$ system will give rise to the left-hand cuts.

When the $T_\ell$ is continued to the complex plane of $\sqrt{s}$, the $q$ in Eq.~\eqref{eq:qrelative} is a multivalue function of $\sqrt{s}$. For simplicity, we neglect the existence of the left-hand cuts and focus on the single-channel problem. In this case, the Riemann surface has two sheets. Moving from  one sheet to the other one needs to cross the unitary cut and changes the sign of  $\text{Im }q$. The conventional definition is 
\begin{equation}
    \text{sheet I (physical): Im }q>0,\qquad \text{sheet II: Im }q<0. 
\end{equation}
The poles on the real axial of sheet I with $\sqrt{s}<m_1+m_2$ correspond to the bound states. {Actually, constrained by the causality, on the first sheet the poles can only appear on the real axis below the lowest threshold. } The poles in the sheet II with $\text{Im}(\sqrt{s})<0$ correspond to the resonances.
We refer to Ref.~\cite{Badalian:1981xj} for the topology of the Riemann surfaces for the coupled-channel problem.

 \subsubsection{Bethe-Salpeter equation}

One may incorporate the unitary condition through the Bethe-Salpeter equation (BSE). For simplicity, we start from the singe-channel scattering amplitude. 

The Bethe-Salpeter equation reads 
  \begin{eqnarray} \label{eq:BSE}
T&=&V+VGT,\\
T\left(q,q^{\prime},P\right)&=&V\left(q,q^{\prime},P\right)+\int \frac{d^{4}k}{(2\pi)^4}V\left(q,k,P\right)\frac{1}{k^{2}-m_{1}^{2}+i\epsilon}\frac{1}{\left(P-k\right)^{2}-m_{2}^{2}+i\epsilon}T(k,q^\prime,P), \label{eq:BS}
  \end{eqnarray}
where $P$, $q^{\prime}$, and $q$ are the total four-momentum, relative ones in the initial and final states, respectively. $G$ is the hadron-hadron loop function { with $k$ the four-momentum in the loop. $m_1$ and $m_2$ are the masses of the scattering hadrons.}  The Lippmann-Schwinger equation has a very similar form but in a nonrelativistic framework where the integral is performed for the three momentum. In Sec.~\ref{sec:noverd}, we will see that Eq.~\eqref{eq:BS} is equivalent to the dispersion relation of the $T$-matrix [c.f.~Eq.~\eqref{eq:N/Dg}] constrained by the unitary condition  if the left-hand cut is neglected~\cite{Oller:2000fj}.

\begin{figure}[hbtp]
	\begin{center}
		\includegraphics[width=1.0\textwidth]{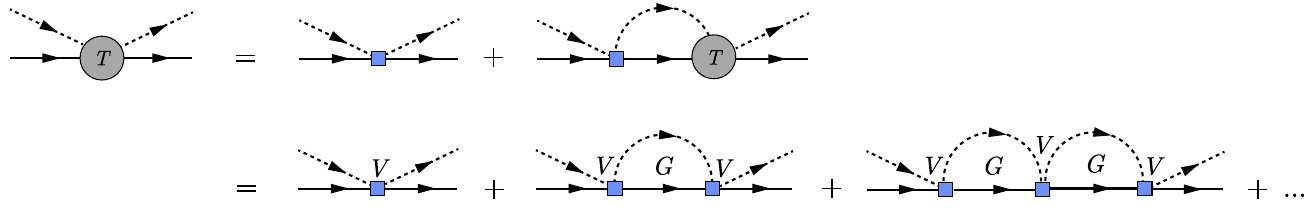}
		\caption{The resummation of the $s$-channel diagrams giving rise to the scattering amplitude.} \label{Fig:tmatrix}
	\end{center}
\end{figure}

 The BSE in Eq.~\eqref{eq:BS} can be understood as the nonperturbative resummation of the $s$-channel loops as shown in Fig.~\ref{Fig:tmatrix}. The BSE can generate the poles of the $S$-matrix corresponding to the bound states or resonances, which is impossible in $\chi$PT. Thus, the BSE was widely used to investigated the so-called ``dynamically generated" states. These poles originate from the hadron-hadron scattering dynamics and can be understood as the meson-meson/meson-baryon/baryon-baryon molecules to some extent.   
 
 One should note that the  unitary approach only considers the $s$-channel loops as shown in Fig.~\ref{Fig:tmatrix}, while the $\chi$PT has the  $t$-channel and $u$-channel ones as well, which are not included explicitly in the Bethe-Salpeter equation. If the singularities of the crossed loops locate far away from the concerned energy region, their energy dependency is expected to be smooth.  Their contributions can be absorbed by the subtraction terms in regularizing the Bethe-Salpeter equation.

Omitting the off-shell effect,  Eq.~\eqref{eq:BSE}  converts
into an algebraic equation 
 \begin{eqnarray}
&&T_{\text {on}}=V_{\text {on}}+V_{\text  {on}}GT_{\text  {on}}.
  \end{eqnarray}
The on-shell approximation $T_{\text {on}}$ still satisfies the unitary relation~\cite{Oller:1998hw,Oller:1997ng,Oller:1998zr}. It was shown that the off-shell contribution  could be absorbed by the renormalization of the couplings and masses in the kernel interaction~\cite{Oset:1997it,Oller:1998hw}. The loop function $G$ is given by
  \begin{eqnarray} \label{eq:loopfun}
G=i\int\frac{d^{4}k}{(2\pi)^{4}}\frac{1}{k^{2}-m_{1}^{2}+i\epsilon}\frac{1}{(P-k)^{2}-m_{2}^{2}+i\epsilon}.
  \end{eqnarray}
It is divergent and needs to be regularized. In literature, there are two common regularization methods. One is the dimensional regularization~\cite{Roca:2005nm}
  \begin{eqnarray}
  \label{eq:loopfdr}
&&G^{\text{DR}}(\sqrt{s})=\frac{1}{16\pi^{2}}\left\{ a(\mu)+\ln\frac{m_{2}^{2}}{\mu^{2}}+\frac{m_{1}^{2}-m_{2}^{2}+s}{2s}\ln\frac{m_{1}^{2}}{m_{2}^{2}}+\frac{q}{\sqrt{s}}\left[\ln\frac{\left(s+2q\sqrt{s}\right)^{2}-\left(m_{2}^{2}-m_{1}^{2}\right)^{2}}{\left(s-2q\sqrt{s}\right)^{2}-\left(m_{2}^{2}-m_{1}^{2}\right)^{2}}-2\pi i\right]\right\},   
  \end{eqnarray}
where $\mu$ is the regularization scale and $a(\mu)$ is the subtraction constant to cancel the $\mu$
dependence of the loop function. We may also introduce a hard cut-off parameter $\Lambda$ for the three momentum.
The $k_0$ component  in Eq.~\eqref{eq:loopfun} can be integrated out in the c.m.s, and the loop function reads~\cite{Oller:1997ti,Oller:1998hw} 
 \begin{eqnarray} \label{eq:loopfcut}
 G^{\text{CR}}(\sqrt{s}) =\int_{0}^{\Lambda}\frac{{k}^{2}d{k}}{4\pi^{2}}\frac{\left(\omega_{1}+\omega_{2}\right)}{\omega_{1}\omega_{2}\left[s-\left(\omega_{1}+\omega_{2}\right)^{2}+i\epsilon\right]}, \qquad \omega_{i}=\sqrt{{k}^{2}+m_i^{2}}.
  \end{eqnarray}

Constrained by the unitary relation, the two regularization schemes ensure the same imaginary parts. The analytical continuation to the second Riemann sheet is achieved by
 \begin{eqnarray}
G_\text{II}(\sqrt{s}+i \epsilon)=G_\text{I}(\sqrt{s}+i \epsilon)-2 i \operatorname{Im} G_\text{I}(\sqrt{s}+i \epsilon).
 \end{eqnarray}
However, their real parts depend on $a(\mu)$ or  $\Lambda$. For such a nonperturbative approach, the renormalization is not transparent. { For example, there are no free LECs in the LO Lagrangians contributing to the $\pi\pi$ scattering~\cite{Oller:1997ti} to absorb the  cutoff-dependence in $G(\sqrt{s})$. The $a(\mu)$ or $\Lambda$ can only be determined phenomenologically~\cite{Oller:1997ti}. However, in the spirit of EFT, one can expect that the regulator dependence will become weaker as one goes to the high orders.} In Ref.~\cite{Guo:2006fu}, the typical value of the $\Lambda$ was estimated as  $\Lambda\sim\sqrt{\Lambda_{\chi}^{2}-m_{\varphi}^{2}}\sim0.8\pm0.2~\text{GeV}
$, where $\Lambda_{\chi}\sim1$ GeV and $m_{\varphi}$ is the mass
of the light mesons such as $\pi$, {$K$ or $\eta$}.With the above estimation, the subtraction term in the dimensional regularization can be estimated by requiring the two schemes giving the same results in the low energy region. See Refs.~\cite{Oller:1998hw,Wu:2010rv} for the discussions on their relations. {One can get the regulator-independent results via some other unitary approaches, e.g. the inverse amplitude method in Sec.~\ref{sec:iam}.}

The above equations can be extended to the coupled-channel cases. The $T$-matrix, the potential $V$, and $G$ become the $n\times n$ matrix ($n$ is the number of the coupled channels), in which $G $ is a diagonal matrix. 

In the chiral unitary method, the  kernel interaction $V$ is obtained with the help of the $\chi$PT. We use the scattering of the pseduo-Goldstone boson off a matter field as an example. The relativistic $G$ in Eq.~\eqref{eq:loopfun} is superficially  $\mathcal{O}(p^0)$. However, the dynamics of the matter field is nonrelativistic. The $G$ function will reduce to $\mathcal{O}(p)$ if the propagator of the matter field is  nonrelativistic. The unitarized $T$-matrix within BSE is obtained by 
\begin{equation}
T=(1-VG)^{-1}V.
\end{equation}
By matching the $T$ with the chiral expansion amplitude $\cal A$ at low energy regions, the $V$ can be introduced order by order~\cite{Lutz:2001yb,Oller:2000fj,Borasoy:2005ie}, 
 \begin{eqnarray} \label{eq:uchpto}
V^{(1)}={\mathcal A^{(1)}}, \quad V^{(2)}=\mathcal{A}^{(2)}, \quad  V^{(3)}=\mathcal{A}^{(3)}-V^{(1)}GV^{(1)},~\label{eq:uch_match}
  \end{eqnarray}
  where the superscript denotes the chiral expansion order. With the kernel potential in Eq.~\eqref{eq:uch_match}, the unitarized $T$-matrix matches the chiral amplitude $\mathcal A$ precisely at a given order and satisfies the unitary condition.  One should note that the $s$-channel loop from the iteration of the lower chiral order amplitudes $\cal A$ has been subtracted to avoid the double counting. For instance, $V^{(3)}=\mathcal A^{(3)}-V^{(1)} G V^{(1)}$  at ${\cal O}(p^3)$.  In Ref.~\cite{Oller:2000fj}, the case with a resonance in the kernel potential was considered. The presence 
   of the resonance arises from the nonperturbative dynamics of chiral Lagrangians~\cite{Meissner:1999vr,Bernard:1991zc,Jamin:2000wn}. In this case, the power counting method in Eq.~\eqref{eq:uch_match} should be modified. {Recently, the authors in Refs.~\cite{Lutz:2022enz,Lutz:2018cqo} used an alternative regularization scheme of the HB$\chi$PT, infrared regularization scheme, and EOMS in Sec.~\ref{sec:sec2.3}. They introduced the renormalized scalar bubble-loop contributions independent of the renormalization scale and replaced the heavy-light field bubble loops subject to certain replacement rules. The renormalization scale dependence of the tadpole integral with only the light mesons is absorbed by the suitable counter terms. The left-hand cuts were also discussed and turned out to be significant.}
 
\subsubsection{N/D method }~\label{sec:noverd}

According to the analytic structures in Eq.~\eqref{eq:righthc} and Eq.~\eqref{eq:lefthc}, the N/D method is proposed to construct a general solution of the $T$-matrix~\cite{Chew:1960iv,Oller:1998zr}. In the following, the ``reduced''  amplitude $T_{\ell}^{\prime}={T_{\ell}(s)}/{q^{2\ell}}$ is used to remove the vanishing threshold behavior of
the partial wave.  In the N/D method, the $T_\ell^{\prime}$ has the form,
$T_{\ell}^{\prime}(s)={N_{\ell}^{\prime}(s)}/{D_{\ell}^{\prime}(s)}$,
 where $N'_{\ell}$ (numerator) and $D'_{\ell}$ (denominator) carry the analytic information of the left- and right-hand cuts, respectively. If the left-hand cut is neglected (see Ref.~\cite{Oller:2019opk} for the derivation of keeping the left-hand cut),  $N'_{\ell}$ becomes a polynomial. 
 
 Dividing the denominator by the numerator, one can set $N_{\ell}^{\prime}(s)=1$ and get~\cite{Oller:1998zr,Meissner:1999vr}
 \begin{equation}
     T_{\ell}^{\prime-1}(s)=  {D_{\ell}^{\prime}(s)}.
 \end{equation} 
 With the unitary condition, one has 
 \begin{eqnarray}
\text{Im}D_{\ell}^{\prime}=\begin{cases}
0 & s<s_{th}\\
-\rho(s)q^{2\ell} & s>s_{th}
\end{cases}.~\label{eq:ncut}
 \end{eqnarray}
{where $\rho$ is defined in Eq. \eqref{eq:rho}}. Using the dispersion relation, the  general forms of 
$D_{\ell}^{\prime}(s)\text{ and }T_{\ell}^{\prime}(s)$ are given by 
  \begin{eqnarray}
&& T_{\ell}^{\prime-1}(s)=D_{\ell}^{\prime}(s)=  -\frac{\left(s-s_{0}\right)^{\ell+1}}{\pi}\int_{s_{th}}^{\infty}ds^{\prime}\frac{q^{2\ell}\rho\left(s^{\prime}\right)}{\left(s^{\prime}-s\right)\left(s^{\prime}-s_{0}\right)^{\ell+1}}
 +\sum_{m=0}^{\ell}a_{m}s^{m}+\sum_{i}^{M_{\ell}}\frac{R_{i}}{s-s_{i}}, \label{eq:T_Lp}
 \end{eqnarray}
 where $s_0$ is the subtraction point and ${a}_{m}$ stands for the subtraction terms. According to Eq.~\eqref{eq:ncut}, one has the asymptotic behavior, 
 \begin{eqnarray}
\lim_{s\rightarrow\infty}\frac{\text{Im} D'_\ell}{s^\ell}= -\lim_{s\rightarrow\infty}\frac{q^{2\ell}\rho(s)}{s^{\ell}}=-\frac{1}{4^{\ell+2}\pi},
 \end{eqnarray}
 which requires the $\ell+1$ times subtraction. All the poles ($s_{i}$) in Eq.~\eqref{eq:T_Lp} are  called the  Castillejo--Dalitz--Dyson (CDD) poles~\cite{Castillejo:1955ed}. $M_\ell$ is the number of the CDD poles. When the CDD poles are located far away from the {energy region of interest}, they can be  absorbed into the subtraction terms. However, if they are close to the physical  state, one should be much cautious when dealing with the CDD poles. { In literature, there are various 
works where the interplay of the zeros of $T$-matrix (t-zeros) and poles of $T$-matrix (t-poles) was discussed in detail \cite{Hanhart:2011jz,Baru:2010ww,Braaten:2007dw,Artoisenet:2010va}. In Ref. \cite{Hanhart:2011jz}, the authors used the $X(3872)$ as a paradigm for a two-channel situation and compared the production line shapes in $B$-meson decays in the near-threshold region with and without the presence of the near-threshold t-zeros and important continuum channel interplay.
In the absence of these two factors, the production of the $X(3872)$ is regular and can be described by the simple Flatt\'e formulae. However, if there is a strong channel-entanglement or the presence of near-threshold t-zeros, the line shape can be dramatically distorted and exhibit an irregular behavior. The line shapes can be used to discriminate between the binding mechanisms for the $X(3872)$ as demonstrated in Ref. \cite{Artoisenet:2010va}. Additionally, the presence of the near-threshold t-zeros corresponds to more near-threshold t-poles, as shown in Refs. \cite{Baru:2010ww,Hanhart:2011jz}. }
 
 The physical resonances or bound states are the poles of $T'_{\ell}$ and satisfy $D_{\ell}^{\prime}(s_{\text{pole}})=0$.
 In contrast, the CDD poles are the poles  of the inverse amplitude  $T'^{-1}_\ell$ and the zeros of $T'_{\ell}(s)=0$.  In literature, the pole in the interaction potential was often called the CDD pole,  e.g.~\cite{Hyodo:2008xr,Chen:2012rp,Meng:2021uhz}. One should be cautious about the different meanings of the ``CDD pole" according to the context.

 If one insists on using the $T_\ell$ without eliminating the vanishing threshold effect, one obtains the form of the $T_\ell$~\cite{Meissner:1999vr} similar to Eq.~\eqref{eq:T_Lp},
 \begin{eqnarray} 
&&T^{-1}_\ell(s)=D_\ell(s) =\sum_{i} \frac{R_{i}}{s-s_{i}}+a\left(s_{0}\right)-\frac{s-s_{0}}{\pi} \int_{s_{{th }}}^{\infty} d s^{\prime} \frac{\rho\left(s^{\prime}\right)}{\left(s^{\prime}-s\right)\left(s^{\prime}-s_{0}\right)},\label{eq:T_L}
 \end{eqnarray}
 where the left-hand cut is neglected. The vanishing threshold behaviour of $T_\ell$ should be introduced as the CDD poles. The $a(s_0)$, $s_i$ and the residue $R_i$ are not known in advance. {Their values can be obtained in different ways such as using experimental data, matching to $\chi$PT, or fitting lattice data as discussed below. }

 The Eq.~\eqref{eq:T_L} can be written in the generalised form
 \begin{eqnarray} \label{eq:N/Dg}
&&T_\ell^{-1}(s)=\mathcal{V}^{-1}(s)-g(s),\quad \Rightarrow \quad T_\ell(s)=[1-\mathcal{V}(s)g(s)]^{-1}\mathcal{V}(s),
\end{eqnarray}
with
\begin{eqnarray}
g(s)=\tilde{a}\left(s_{0}\right)+\frac{s-s_{0}}{\pi} \int_{s_{th}}^{\infty} d s^{\prime} \frac{\rho\left(s^{\prime}\right)}{\left(s^{\prime}-s\right)\left(s^{\prime}-s_{0}\right)},
 \end{eqnarray}
 where  the $\tilde{a}\left(s_{0}\right)$ specifies the freedom of choosing subtraction terms. One can see the $\mathcal{V}$ and the $g(s)$ play the same roles as the kernel potential and loop function in BSE in Eq.~\eqref{eq:BS}. The explicit calculation shows that $g(s)$  is identical to Eq.~\eqref{eq:loopfun} up to a constant. Their right-hand cuts and imaginary parts along the cut are the same. In Ref.~\cite{Oller:2000fj}, the interaction kernel is determined by  the  $\chi$PT amplitude to any given chiral expansion order as well as the explicit resonance contributions, see Eq.~\eqref{eq:uch_match} and the context. The kernel potential $\mathcal{V}(s)$ contains all the contributions except the right-hand cut.

 The extension to the coupled-channel cases was given in Ref.~\cite{Oller:1998zr} with the matrix form, in which the $T$-matrix satisfies the unitary condition
\begin{eqnarray}
\left[\operatorname{Im}T_{\ell}^{-1}\right]_{ij}=-\rho_{ii}(s)\delta_{ij}.
 \end{eqnarray}

\subsubsection{Inverse amplitude method}~\label{sec:iam}

The inverse amplitude method (IAM) is another unitarization technique adopting the unitary condition in Eq.~\eqref{eq:righthc} for the inverse amplitude rather than the amplitude~\cite{Truong:1988zp,Truong:1991gv,Dobado:1989qm,Dobado:1996ps}, since they have the same analytic structures  except the possible pole contributions. This method has been used to study the light resonances, for instance {$f_0(500)$, $K^*_0(700)$, $f_0(980)$, $a_0(980)$, $\rho(770)$ and  $K^{*}(892)$} (see more discussions in Ref.~\cite{Pelaez:2015qba}).

Before deriving the formalism of the inverse amplitude method in the unitary perspective, it is instructive to adopt the Pad\'e expansion. The chiral partial wave amplitudes for the scattering of the light pseudoscalar mesons can be expanded as 
\begin{eqnarray}  
T(s)=T_{2}(s)+T_{4}(s)+T_{6}(s) +\ldots,
\end{eqnarray}
 where the $``\ldots"$ denotes the higher order amplitude $T_{2 k} \sim \mathcal{O}(p^{2 k})$.
 We can perform the $[1,1]$ and $[1,2]$ Pad\'e expansion for the $T$-matrix,
 \begin{equation}
     T^{[1,1]}=\frac{N^{(0)}+N^{(2)}}{1+D^{(2)}}=T_{2}+T_{4},\quad T^{[1,2]}=\frac{N^{(0)}+N^{(2)}}{1+D^{(2)}+D^{(4)}}=T_{2}+T_{4}+T_{6},
 \end{equation}
 where $[i,j]$ represent that the truncation chiral orders of the expansions of the numerator and denominator are $2i$ and $2j$,  respectively. In the above equation, we further match the $[1,1]$ order and $[1,2]$ order Pad\'e expansions to the chiral expansions of $T$ to $\mathcal{O}(p^4)$ and $\mathcal{O}(p^6)$, respectively, which can fix the  $N^{(2n)}$ and $D^{(2m)}$ terms. The final results are 
 \begin{equation}
     T^{[1,1]}=\frac{T_{2}^{2}}{T_{2}-T_{4}}+\mathcal{O}(p^{6}),\quad T^{[1,2]}=\frac{T_{2}^{2}}{T_{2}-T_{4}+T_{4}^{2}/T_{2}-T_{6}}+\mathcal{O}(p^{8}).~\label{eq:pade_iam}
 \end{equation}
 The Eq.~\eqref{eq:pade_iam} is just the conventional IAM formula to $\mathcal{O}(p^4)$ and $\mathcal{O}(p^6)$.
 
 Now, we inspect the above results from the unitarity. With the optical theorem, one has
\begin{eqnarray} \label{eq:iamta}
\operatorname{Im} T_{2}(s)=0, \qquad \operatorname{Im} T_{4}(s)= T_{2}(s)\rho(s)T_{2}(s), \qquad \operatorname{Im} T_{6}(s)=2 \rho(s)T_{2}(s) \operatorname{Re} T_{4}(s), \ldots.
\end{eqnarray}
The dispersion relations for the amplitudes are 
\begin{eqnarray} 
 T_{2}(s) =a_{0}+a_{1} s, \qquad
 T_{4}(s) =b_{0}+b_{1} s+b_{2} s^{2} +\frac{s^{3}}{\pi} \int_{s_{th}} d s^{\prime} \frac{\operatorname{Im} T_{4}\left(s^{\prime}\right)}{s^{\prime 3}\left(s^{\prime}-s-i \epsilon\right)}+\mathrm{LC} (T_4),  \label{eq:t4}
\end{eqnarray}
 where $\mathrm{LC}(T_4)$ represents the left-hand cut. {$a_i$ $ (i=0,1)$ and $b_j$ $(j=0,1,2)$} are the  subtraction constants. 

To implement unitarity, one defines $g(s)=T_{2}(s)^{2} / T(s)$. Since the $T_2(s)$ is real, the $g(s)$ has the same cuts as $T(s)$. One can obtain the integral equation for $g(s)$ with the dispersion relation. For instance, the  three times subtracted dispersion relation for $g(s)$ is 
 \begin{eqnarray} \label{eq:gsdp}
g(s)= g(0)+g^{\prime}(0) s+\frac{1}{2} g^{\prime \prime}(0) s^{2} +\frac{s^{3}}{\pi} \int_{s_{th}} d s^{\prime} \frac{\operatorname{Im} g\left(s^{\prime}\right)}{s^{\prime 3}\left(s^{\prime}-s-i \epsilon\right)}+\mathrm{LC}(g)+\mathrm{PC}(s),
\end{eqnarray}
 where $\mathrm{PC}(s)$ denotes the pole contribution of the $g(s)$. {Note that the correct pole contribution of $1/T$ (corresponding to Adler zero) cannot be obtained from the PC term of the $g(s)$ given the appearance of $T_2^2$ in the numerator.} With the unitarity relation in Eq.~\eqref{eq:righthc}, one obtains
  \begin{eqnarray}
 \operatorname{Im} g(s)=-T_{2}(s)\rho(s) T_{2}(s)=-\operatorname{Im} T_{4}(s),
 \end{eqnarray}
 on the right-hand cut. The subtraction constants in Eq.~\eqref{eq:gsdp} are related to the  chiral expansions. Up to $\mathcal {O}(p^4)$, the $g(s)$ can be expanded in the low energy region using Eq.~\eqref{eq:t4} after neglecting the pole contributions $\mathrm{PC}$, 
   \begin{eqnarray}
 g(s)=\frac{T_{2}^{2}}{T_{2}+T_{4}} \simeq a_{0}+a_{1} s-b_{0}-b_{1} s-b_{2} s^{2}
-\frac{s^{3}}{\pi} \int_{s_{\text {m }}} d s^{\prime} \frac{\operatorname{Im} T_{4}\left(s^{\prime}\right)}{s^{\prime 3}\left(s^{\prime}-s-i \epsilon\right)}-\mathrm{LC} (T_4)=T_{2}(s)-T_{4}(s),
  \end{eqnarray}
where one has approximated 
\begin{eqnarray}
 \operatorname{Im} g(s) \simeq-\operatorname{Im} T_{4}(s),\qquad \mathrm{LC}(g)=-\mathrm{LC}(T_4)+\ldots
   \end{eqnarray}
   on the left cut.  Thus, the elastic formula for IAM reads~\cite{Truong:1988zp,Truong:1991gv,Dobado:1989qm,Dobado:1996ps}
    \begin{eqnarray} \label{eq:iam}
   T(s) \simeq \frac{T_{2}^{2}(s)}{T_{2}(s)-T_{4}(s)},
   \end{eqnarray}
which recovers the $\chi$PT expansion by expanding Eq.~\eqref{eq:iam} as follows 
 \begin{eqnarray} 
   T(s) \simeq \frac{T_{2}^{2}(s)}{T_{2}(s)-T_{4}(s)}\simeq T_2+T_4+\mathcal O(p^6),
   \end{eqnarray}
The above discussion with the dispersion formalism can be extended to higher orders. For instance, up to the next-to-next-to-leading order (NNLO or N$^2$LO) one reads~\cite{Dobado:1996ps}
      \begin{eqnarray}
   T(s) \simeq \frac{T_{2}^{2}(s)}{T_{2}(s)-T_{4}(s)+T_{4}^{2}(s) / T_{2}(s)-T_{6}(s)}.
      \end{eqnarray}
  {It should be noticed that the above IAM method neglecting the pole contribution will give the imprecise Adler zeros of the $T$-matrix. Such a flaw casts some doubts about the robustness of this method~\cite{Boglione:1996uz} given the Adler zeros are zeros of the $T$-matrix arising from chiral symmetry and its spontaneous breaking~\cite{Adler:1965ga}. In Ref.~\cite{GomezNicola:2007qj}, the IAM is adjusted to incorporate the Alder zeros correctly. }

In Refs.~\cite{Oller:1997ng,Oller:1998hw,GomezNicola:2001as}, the IAM  is generalized to the coupled channels using the matrix formalism,
 \begin{eqnarray} 
 T(s)={T}^{(2)}(s) \cdot\left[{T}^{(2)}(s)-{T}^{(4)}(s)\right]^{-1} \cdot {T}^{(2)}(s).
  \end{eqnarray}
The IAM approach can be generalised to study the system with the matter field. In Ref.~\cite{Yao:2015qia}, the authors applied the IAM to derive the scattering amplitude of the pseudo-Goldstone boson off the heavy meson, 
    \begin{eqnarray} 
   T(s)={T}^{(1)}(s) \cdot\left[{T}^{(1)}(s)-{T}^{(2)}(s)-T^{(3)}(s)\right]^{-1} \cdot {T}^{(1)}(s),
   \end{eqnarray}
   where 
     \begin{eqnarray} 
    \text{Im}[{T}^{(1)}(s)]=\text{Im}[{T}^{(2)}(s)]=0, \qquad \text{Im}[{T}^{(3)}(s)]={T}^{(1)}(s) \tilde{\rho}(s){T}^{(1)}(s).
      \end{eqnarray}
   The loop function is $\mathcal O(p)$ since the propagator of the heavy meson is counted as ${\cal O} (p^{-1})$~\cite{Oller:2000fj} and its imaginary part  ${\tilde \rho}(s)$ is $\mathcal O(p)$ in this case.

 \subsubsection{Dynamically generated states} \label{sec2.5.4}
 
  The chiral unitary approaches combine the unitary condition with $\chi$PT and lead to the nonperturbative resummation of the infinite $s$-channel loop diagrams. The implementation of the unitary condition extends the applicable range of the $\chi$PT to higher energy region and generates the poles for the resonance/bound states in the unitarized scattering amplitudes. Such chiral unitary approaches have been successfully used to describe  a variety of scattering  processes of  the meson-baryon systems~\cite{Kaiser:1995eg,Oset:1997it,Krippa:1998us,Oller:2000fj,Garcia-Recio:2003ejq,Kolomeitsev:2003kt,Sarkar:2004jh,Hyodo:2008xr,Hyodo:2002pk,Oller:2005ig} and meson-meson systems~\cite{Dobado:1996ps,Oller:1997ti,Oller:1997ng,Oller:1998hw,Lutz:2003fm,Roca:2005nm}, which can also be extended to the  heavy flavor systems in Sec.~\ref{sec:sec4}.
 
In the chiral unitary methods, the  poles of the resonances or bound states can be generated by the dynamics of the hadron-hadron scattering. In this context, the  poles are called the ``dynamically generated" states. {These states arise from the hadron-hadron interactions ( which definitely happen at the hadron level). They are not the explicit d.o.fs in the free Hamiltonians or Lagrangians. Their properties such as the masses and decay widths are determined by the hadron-hadron scattering potentials.}

In contrast to the dynamically generated states, there are literally  ``preexisting" states, { which are assumed to exist as the explicit d.o.fs before the hadron interactions. The ``preexisting" states are often associated with the bound states or resonances composed of more fundamental d.o.fs than the scattering d.o.f, and their properties are determined by more fundamental theories. For example, in the meson-meson scattering, the compact quark states such as the $\bar q q$ or  $qq\bar q \bar q $ tetraquark state are governed by the direct QCD dynamics. They exist prior to the hadron-hadron interactions and are considered as the preexisting states. There is another potential origin of the ``preexisting" states. They can be the dynamically generated states in the other channels, which could be regarded as the preexisting ones in the relevant channels~\cite{Hyodo:2008xr,Gasparyan:2010xz,Jaffe:2004ph,Pelaez:2015qba}. 
 }   

In Ref.~\cite{Chew:1961cer}, the CDD pole was interpreted as an independent particle participating in the scattering.  In literature, the ``preexisting" states were associated with the CDD poles which act as the poles of the inverse $T$-matrix as shown in Eq.~\eqref{eq:T_Lp}. Thus, the  ``preexisting" states were often called the CDD poles directly,  e.g.~\cite{Hyodo:2008xr,Chen:2012rp}. One should be cautious about the different meanings of the ``CDD pole" according to the context. The ``preexisting" states are usually introduced explicitly through a bound state or resonance propagators in the effective field theory~\cite{Hyodo:2008xr,Jaffe:2004ph,MartinezTorres:2011pr,MartinezTorres:2011pr,Chen:2012rp}.

{  However, the above naive distinction (or classification) of the ``dynamically generated" and ``preexisting" states is model-dependent. From the perspective of the effective field theory, the parameterizations of the kernel potentials which generate the consistent $T$-matrix at the energy scale of interest are not unique. In some cases, the preexisting states are hidden in the low energy constants (LECs). Within a given energy range, it is possible to replace a theory with the non-perturbative two-particle interactions and no preexisting pole by an effective theory with a preexisting pole and perturbative two-particle interactions~\cite{Weinberg:1962hj}. This leads to ambiguity regarding the mechanisms of the resonance generation. Sometimes, it is difficult to distinguish the direct QCD dynamics at the quark level and the hadron-hadron interactions since both of them may contribute simultaneously. Even some well-accepted $ \bar qq$ mesons could be dynamically generated from the meson-meson interaction at least formally in some frameworks.  For example, the pole of the $\rho$ resonance can be reproduced well either from the $\chi$PT interaction without the bare pole in the chiral unitary approaches  (e.g.~\cite{Oller:1997ng,Nieves:1999bx}) or by introducing the bare pole in the $\pi\pi$ scattering, e.g.~\cite{Oller:1998zr,Chen:2012rp}. The $D^*$ was also dynamically generated from the $D\pi$ interaction in Ref.~\cite{Schmidt:2018vvl}. In Sec. \ref{sec:uchptothers}, we will illustrate that the dynamically generated mechanisms and its mixtures with the preexisting states can both reproduce the mass and width of the $D_{s0}^*(2317)$ and are indistinguishable within the current experimental uncertainties. }

\begin{figure}[hbtp]
	\begin{center}
		\includegraphics[width=0.9\textwidth]{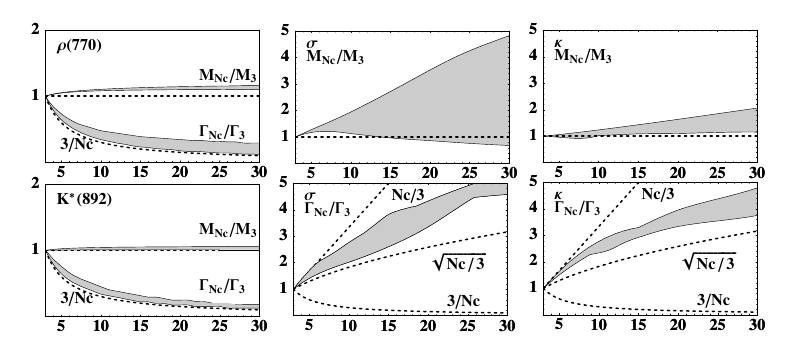}
		\caption{The trajectories of large $N_C$ scaling (dashed line) for the $ \rho$, $K^{*}$, $\sigma$ and $\kappa$ pole positions ($M_{N_C}-i\Gamma_{N_C}/2$).  The gray areas represent  the  uncertainties induced by  the  renormalization scale  $\mu \simeq 0.5-1$ GeV. The figure is taken from Ref.~\cite{Pelaez:2003dy}.} \label{Fig:largenclight}
	\end{center}
\end{figure}

For the system associated with the meson-meson scattering, to discern the genuine state and the dynamically generated ones, the studies of other properties, such as the trajectories in the large $N_C$ limit, are useful~\cite{Oller:1998zr}. The latter does not survive in the large $N_{C}$ limit, since the meson-meson scattering is $\mathcal O(N_{C}^{-1})$~\cite{Witten:1979kh}. In the large $N_C$ limit, the pole position of the genuine  $\bar q q$ state $M_{N_C}-i\Gamma_{N_C}/2$  should  satisfy $M_{N_C}/M_3$ being a constant and  $\Gamma_{N_C}/\Gamma_3\sim 3/N_C$, while the {dynamically generated} two-meson state will disappear~\cite{Ecker:1988te,Pelaez:2003dy}. As an example, the studies of the $N_C$ scaling  behaviors in Ref.~\cite{Pelaez:2003dy} favored  the $\rho$ and $K^*$  as the $q\bar q$ state, while their results supported  the $\sigma$ and $\kappa$ as the  ``dynamically generated" two-meson states as shown in Fig.~\ref{Fig:largenclight}. 
 
In fact, many theoretical methods are proposed to define or discern the above two concepts more framework-independently. {In principle, the generalized Levinson's theorem~\cite{generallev1958,Vaughn:1961poz,Li:2022aru} in Eq.~\eqref{eq:generalLevison} can tell the number of the ``preexisting" states. However, in practice, the phase shifts at the infinite energy are not accessible observables. In the 1960s, Weinberg exploited the relations of the elementary particle and the composite particle~\cite{Weinberg:1962hj,Weinberg:1965zz}, which are similar to ``preexisting" state and ``dynamically generated" state, respectively. For an $S$-wave shallow  bound state, the probability  of the hadron-hadron  molecule component $(1-Z)$ (where $Z$ is the wave function renormalization constant) is related to the $S$-wave scattering length $a_s$ and effective range $r_e$~\cite{Weinberg:1965zz},
\begin{equation}
    a_{s}=\frac{2}{\kappa}\frac{1-Z}{2-Z},\qquad\qquad r_{e}=-\frac{1}{\kappa}\frac{Z}{1-Z},
\end{equation}
where $\kappa$ is the binding momentum. The above relations provide a criterion of compositeness from the low-energy scattering observables. This relation was used to identify the deuteron as a molecule composed of two nucleons instead of an elementary state~\cite{Weinberg:1965zz}. Later, this relation has been generalized to describe the unstable resonances in many theoretical works~\cite{Baru:2003qq,Baru:2010ww,Hyodo:2011qc,Aceti:2012dd,Sekihara:2014kya,Hyodo:2013iga,Chen:2013upa,Kamiya:2015aea,Guo:2015daa}. It should be noticed that the Weinberg compositeness criterion, even the version for the bound states, has its validity range. By construction, one can always get a pure two-body bound state with any given small binding energy with any given phase shifts~\cite{Tabakin:1969mr}, which means any $Z$ from Weinberg's criterion. The conditions to ensure and validate the Weinberg's criterion include: $\kappa \ll \Lambda$ (where $\Lambda$ is $m_\pi$ for the $NN$ system), the non-pole term in the Low equation can be neglected, the form factor $\langle p|V|B\rangle $ is a constant, and so on. In Ref.~\cite{Li:2021cue}, Li \etal investigated the effect of the non-constant form factor. At the same time, the authors demonstrated that the low energy observables (phase shifts) in Weinberg's criterion cannot reveal the short-range structure information of a state.  It was pointed that the Weinberg “compositeness” is a measure for the probability to find the constituents separated by a distance greater than the interaction range~\cite{Bruns:2019xgo,Bruns:2022hmb}. In Ref.~\cite{Song:2022yvz}, Song \etal also considered the role of the interaction range (the cutoff of the interaction) in the Weinberg's criterion.  }

\subsection{Superfield representations combining the chiral and heavy quark symmetries} ~\label{sec:1.5:combChandHQ}

For the heavy-light systems, their dynamics are constrained by HQS and chiral symmetry, thus the combination of these two symmetries is necessary. At higher order, the corrections come from two sides: the $1/m_Q$ correction and the chiral expansion. Thus, the HH$\chi$PT has a double expansion in $ \Lambda_{\mathrm{QCD}}/m_Q$   and   $p/\Lambda_{\chi}$.  The $\Lambda_{\chi}$ and $m_Q$ are  identified as 
the high-energy scale, while the $\Lambda_{\mathrm{QCD}}$ and the $ p_{\pi,K,\eta}\sim m_{\pi,K,\eta} $ are identified as the low-energy scale. In literature, the two expansions were often combined (treating $\Lambda_{\mathrm{QCD}}/m_Q$ as the same order as $p/\Lambda_{\chi}$). In this review, we will discuss these corrections separately.

In Sec.~\ref{sec:HQFSHQSS}, we have discussed the HQSS, which is manifested in heavy hadron spectrum. In the heavy quark limit, these degenerate states are collected in the superfield representation owing to the development of $\chi$PT and HQET in 1990s~\cite{Georgi:1990cx,Mannel:1990vg,Falk:1991nq,Cho:1992cf,Wise:1992hn}. In the superfield representations, the Lagrangians can be written with a compact form. The LO terms satisfy the HQS and chiral symmetry, while the explicit breaking terms from finite light quark masses and heavy quark masses can be systematically included order by order.

In the following, we outline some basic properties of the superfields. For a heavy-light system, if one assumes the heavy d.o.f is ultra nonrelativistic and decouples from the light d.o.f, the field of this system can be expressed with the product of the heavy and light ones, i.e., $\psi_{hl}\sim\psi_h\psi_l$~\cite{Neubert:1993mb}. Accordingly, we take the $S$-wave heavy-light hadrons as an example,
\begin{eqnarray}
\text{heavy mesons: } \mathcal{H}&\sim& u_h \bar{v}_l,\\
\text{singly heavy baryons: } \psi_Q^\mu&\sim& u_h A_l^\mu,\\
\text{doubly heavy baryons: } \psi_{QQ}^\mu&\sim& u_l A_h^\mu,
\end{eqnarray}
where $u_h$ and $v_l$ ($h$ and $l$ for heavy and light d.o.f respectively) denote the quark and antiquark. $A_{h/l}^\mu$ represents the vector diquark. The $\mathcal{H}$, $\psi_Q^\mu$ and $\psi_{QQ}^\mu$ transform as the antitriplet, sextet and triplet under the $\SU(3)_L\otimes\SU(3)_R$ rotations. Meanwhile, they are the linear combinations of the degenerate states in the heavy (di)quark symmetry. {Here, we aim to explain the superfield technique to satisfy the heavy (di)quark spin symmetry.
However, for the scalar diquarks, there are no heavy (di)quark spin doublets. There is no need to construct the superfields.} For the rules to construct the Lagrangians keeping or breaking heavy (di)quark symmetry, see the details in~\ref{app:1} and Ref.~\cite{Falk:1991nq}.

In the following, we present the matrix representations of the heavy hadrons as the flavor $\SU(3)$ multiplet, and define the superfields of degenerate states under the HQSS. We give the low order Lagrangians to illustrate how to combine the two symmetries for different classes. One can find the definition of building blocks and some technical details in~\ref{app:1}.

\subsubsection{$S$-wave heavy mesons}\label{sec:SwaveHM}

For the ground-state heavy mesons $Q\bar{q}$, $S_Q=\frac{1}{2}$, $j_\ell=\frac{1}{2}$, $J=0$ and $J=1$ correspond to the pseudoscalar $P$ and vector meson $P^*$, respectively.  In flavor space, the doublet $P$ with $J^P=0^-$ and $P^*$ with $J^P=1^-$ are
\begin{eqnarray}
P=(D^{0},D^{+},D_{s}^{+})\text{ or }(B^{-},\bar{B}^{0},\bar{B}_{s}^{0}),\qquad
 P^*=(D^{\ast0},D^{\ast+},D_{s}^{\ast+})\text{ or }(B^{\ast-},\bar{B}^{\ast0},\bar{B}_{s}^{\ast0}),
\end{eqnarray}
where the corresponding antiparticle's doublet is denoted as $\tilde{P}$ and $\tilde{P}^*$ (column vector in flavor space), respectively. The degenerate $P$ and $P^*$ ($\tilde{P}$ and $\tilde{P}^*$) are combined into the superfield $\mathcal{H}$ ($\tilde{\mathcal{H}}$) with the form
\begin{equation}~\label{eq:spfd:smeson}
{\cal H}=\Lambda_{+}(P_{\mu}^*\gamma^{\mu}+iP\gamma_{5}),\qquad\tilde{{\cal H}}=(\tilde{P}_{\mu}^*\gamma^{\mu}+i\tilde{P}\gamma_{5})\Lambda_{-},
\end{equation}
where $\Lambda_{\pm}=(1\pm\slashed{v})/2$,  $v^2=1$. The conjugation of $\cal{H}$ and $\tilde{\mathcal{H}}$ is defined as
$\bar{\mathcal{H}}=\gamma_0\mathcal{H}^\dagger\gamma_0$ and $\bar{\tilde{\mathcal{H}}}=\gamma_0\tilde{\mathcal{H}}^\dagger\gamma_0$. The properties of the $\mathcal{H}$ field under the Lorentz, chiral and heavy quark spin transformations can be found in Ref.~\cite{Manohar:2000dt}.

With the superfields, the low order Lagrangians read~\cite{Wise:1992hn,Manohar:2000dt}
\begin{eqnarray}
{\cal L}_{{\cal H}\varphi}&=&-i\langle{\cal H}v\cdot \mathcal{D}\bar{{\cal H}}\rangle+g_{b}\langle{\cal H}\slashed{u}\gamma_{5}\bar{{\cal H}}\rangle-\frac{\delta_{b}}{8}\langle{\cal H}\sigma^{\mu\nu}\bar{{\cal H}}\sigma_{\mu\nu}\rangle,\label{eq:app1:lagD}\\
{\cal L}_{\tilde{{\cal H}}\varphi}&=&-i\langle\bar{\tilde{{\cal H}}}v\cdot \mathcal{D}\tilde{{\cal H}}\rangle+g_{b}\langle\bar{\tilde{{\cal H}}}\slashed{u}\gamma_{5}\tilde{{\cal H}}\rangle-\frac{\delta_{b}}{8}\langle\bar{\tilde{{\cal H}}}\sigma^{\mu\nu}\tilde{{\cal H}}\sigma_{\mu\nu}\rangle,\label{eq:app1:lagDbar}
\end{eqnarray}
where $\langle\dots\rangle$ denotes the trace in spinor space, i.e., for the gamma matrix. {If we consider the heavy quark flavor symmetry, the $g_b$ is the same for the charmed and bottom sector.} For the mesons, the chiral covariant derivative is defined as $\mathcal{D}_{\mu}=\partial_{\mu}+\Gamma_{\mu}$, and the complete form of chiral connection $\Gamma_\mu$ with the external fields is given in Eq.~\eqref{eq:connection}. $g_b$ is the axial coupling constant, which can be determined from the decay width of $D^{\ast+}\to D^0\pi^+$~\cite{ParticleDataGroup:2022pth} for the charmed ones  ($|g_b|\simeq0.59$), or the lattice QCD calculations for the bottom ones~\cite{Ohki:2008py}. The $\delta_b$ terms originate from the chromomagnetic interaction, e.g., see Eq.~\eqref{eq:LagHQET2}, which contribute to the mass splittings between $P$ and $P^\ast$,
\begin{equation}
\delta_b=m_{P^\ast}-m_P.\label{eq:mss:meson0}
\end{equation}
For the general higher order chiral Lagrangians of the heavy mesons, we refer to Ref.~\cite{Jiang:2019hgs}.

\subsubsection{$P$-wave heavy mesons}

For the $P$-wave excited heavy mesons, combining the light spin $j_\ell=\frac{1}{2},\frac{3}{2}$ with the $S_Q=\frac{1}{2}$ yields two spin doublets $(0^+,1^+)$ and $(1^+,2^+)$. These two doublets are described by the superfields $\mathcal{S}$ and $\mathcal{T}^\mu$~\cite{Falk:1991nq}, respectively,
\begin{eqnarray}
{\cal S}&=&\Lambda_{+}\left[R^{*\mu}\gamma_{\mu}\gamma_{5}-R\right],~\label{eq:spfd:pmeson1}\\
{\cal T^{\mu}}&=&\Lambda_{+}\left\{ Y^{*\mu\nu}\gamma_{\nu}-\sqrt{\frac{3}{2}}Y_{\nu}\gamma_{5}\left[g^{\mu\nu}-\frac{1}{3}(\gamma^{\mu}-v^{\mu})\gamma^{\nu}\right]\right\}.~\label{eq:spfd:pmeson2}
\end{eqnarray}
The corresponding low order chiral Lagrangians are given by~\cite{Kilian:1992hq,Casalbuoni:1996pg}
\begin{eqnarray}
	{\cal L}_{{\cal S}\varphi}&=&\langle{\cal S}\left(iv\cdot \mathcal{D}-\delta_{{\cal S}}\right)\bar{{\cal S}}\rangle+g'_{b}\langle{\cal S}\slashed{u}\gamma_{5}\bar{{\cal S}}\rangle+\frac{\delta'_{b}}{8}\langle{\cal S}\sigma^{\mu\nu}\bar{{\cal S}}\sigma_{\mu\nu}\rangle,\label{eq:app1:lagS}\\
	{\cal L}_{{\cal T}\varphi}&=&\langle{\cal T}^{\mu}\left(iv\cdot \mathcal{D}-\text{\ensuremath{\delta_{{\cal T}}}}\right)\bar{{\cal T}}_{\mu}\rangle+g''_{b}\langle{\cal T}^{\nu}\slashed{u}\gamma_{5}\bar{{\cal T}}_{\nu}\rangle+\frac{3\delta''_{b}}{16}\langle{\cal T}^{\rho}\sigma^{\mu\nu}\bar{{\cal T}}_{\rho}\sigma_{\mu\nu}\rangle, \label{eq:app1:lagTT}\\
	{\cal L}_{{\cal S}{\cal H}\varphi}&=&h\langle{\cal H}\slashed{u}\gamma_{5}\bar{{\cal S}}\rangle+\mathrm{H.c.},\label{eq:app1:lagSH}\\
\mathcal{L}_{\mathcal{T}\mathcal{H}\varphi}&=&k_1\langle\mathcal{T}^\mu\gamma_\lambda\gamma_5(\mathcal{D}_\mu u^\lambda)\bar{\mathcal{H}}\rangle+k_2\langle\mathcal{T}^\mu\gamma_\lambda\gamma_5(\mathcal{D}^\lambda u_\mu)\bar{\mathcal{H}}\rangle+\mathrm{H.c.},\\
\mathcal{L}_{\mathcal{S}\mathcal{T}\varphi}&=&\tilde{h}\langle{\cal T}^\mu u_\mu\gamma_{5}\bar{{\cal S}}\rangle+\mathrm{H.c.},\label{eq:app1:lagST}
\end{eqnarray}
where $\delta_{\mathcal{S}}$ and $\delta_{\mathcal{T}}$ contribute to the mass difference with respect to the ground states,
\begin{equation}
\delta_{\mathcal{S/T}}=\bar{m}_{\mathcal{S/T}}-\bar{m}_{\mathcal{H}},\quad\bar{m}_{\mathcal{H/S}}\equiv{3m_{P^*/R^*}+m_{P/R}\over 4},\quad \bar{m}_{\mathcal{T}}\equiv {5m_{Y^*}+3m_{Y}\over 8}.~\label{eq:mss_meson1}
\end{equation}
 The $\delta_b^\prime$ and $\delta_b^{\prime\prime}$ terms produce the mass splittings in the doublets $\mathcal{S}$ and $\mathcal{T}$, respectively,
 \begin{equation}
\delta_b^\prime=m_{R^*}-m_{R},\quad \delta_b^{\prime\prime}=m_{Y^*}-m_{Y}.~\label{eq:mss_meson2}
 \end{equation}

\subsubsection{Singly heavy baryons}\label{sec:singlyHB}

The two light quarks $qq$ in the ground-state singly heavy baryons can form $\bm 3\otimes\bm 3=\bar{\bm 3}\oplus\bm 6$ representations in $\SU(3)$ flavor space. The antitriplet have $j_\ell=0$, while the sextet have $j_\ell=1$. Therefore, the spin of the antitriplet baryon is $\frac{1}{2}$, while the spins of the sextet are $\frac{1}{2}$ and $\frac{3}{2}$, respectively.  In the SU(3) flavor space, the conventional representations of the singly heavy baryons are~\cite{Yan:1992gz}
\begin{equation}\label{eq:shbaryons36}
	B_{\bar{\bm3}}=\left[\begin{array}{ccc}
		0 & \Lambda_{c}^{+} & \Xi_{c}^{+}\\
		-\Lambda_{c}^{+} & 0 & \Xi_{c}^{0}\\
		-\Xi_{c}^{+} & -\Xi_{c}^{0} & 0
	\end{array}\right],\qquad
   B_{\bm6}=\left[\begin{array}{ccc}
		\Sigma_{c}^{++} & \frac{\Sigma_{c}^{+}}{\sqrt{2}} & \frac{\Xi_{c}^{\prime+}}{\sqrt{2}}\\
		\frac{\Sigma_{c}^{+}}{\sqrt{2}} & \Sigma_{c}^{0} & \frac{\Xi_{c}^{\prime0}}{\sqrt{2}}\\
		\frac{\Xi_{c}^{\prime+}}{\sqrt{2}} & \frac{\Xi_{c}^{\prime0}}{\sqrt{2}} & \Omega_{c}^{0}
	\end{array}\right],\qquad
   B_{\bm6}^{*}=\left[\begin{array}{ccc}
		\Sigma_{c}^{*++} & \frac{\Sigma_{c}^{*+}}{\sqrt{2}} & \frac{\Xi_{c}^{*+}}{\sqrt{2}}\\
		\frac{\Sigma_{c}^{*+}}{\sqrt{2}} & \Sigma_{c}^{*0} & \frac{\Xi_{c}^{*0}}{\sqrt{2}}\\
		\frac{\Xi_{c}^{*+}}{\sqrt{2}} & \frac{\Xi_{c}^{*0}}{\sqrt{2}} & \Omega_{c}^{*0}
	\end{array}\right],
\end{equation}
where we use the $B_{\bar{\bm3}}$, $B_{\bm6}$ and $B_{\bm6}^{*}$ to denote the spin-$\frac{1}{2}$ antitriplet, spin-$\frac{1}{2}$ and spin-$\frac{3}{2}$ sextet, respectively.

Without the HQS, the low order Lagrangians are constructed as~\cite{Yan:1992gz}
\begin{eqnarray}
{\cal L}_{B_Q\varphi}&=&\text{Tr}\left[\bar{B}_{\bm6}(i\slashed{\mathcal{D}}-m_{\bm6})B_{\bm6}\right]+\frac{1}{2}\text{Tr}\left[\bar{B}_{\bar{\bm3}}(i\slashed{\mathcal{D}}-m_{\bar{\bm3}})B_{\bar{\bm3}}\right]\nonumber\\
&&+\text{Tr}\left\{\bar{B}_{\bm6}^{*\mu}\left[-g_{\mu\nu}(i\slashed{\mathcal{D}}-m_{\bm6}^{*})i(\gamma_{\mu}\mathcal{D}_{\nu}+\gamma_{\nu}\mathcal{D}_{\mu})-\gamma_{\mu}(i\slashed{\mathcal{D}}+m_{\bm6}^{*})\gamma_{\nu}\right]B_{\bm6}^{*\nu}\right\}\nonumber\\
&&+g_{1}\text{Tr}\left[\bar{B}_{\bm6}\slashed{u}\gamma_{5}B_{\bm6}\right]+g_{2}\text{Tr}\left[\bar{\text{\ensuremath{B}}}_{\bm6}\slashed{u}\gamma_{5}B_{\bar{\bm3}}\right]+\text{H.c}.+g_{3}\text{Tr}\left[\bar{B}_{\bm6}^{*\mu}u_{\mu}B_{\bm6}\right]+\text{H.c.}\nonumber\\
&&+g_{4}\text{Tr}\left[\bar{B}_{\bm6}^{*\mu}u_{\mu}B_{\bar{\bm3}}\right]+\text{H.c.}+g_{5}\text{Tr}\left[\bar{B}_{\bm6}^{*\nu}\slashed{u}\gamma_{5}B_{\bm6\nu}^{*}\right]+g_{6}\text{Tr}\left[\bar{B}_{\bar{\bm3}}\slashed{u}\gamma_{5}B_{\bar{\bm3}}\right],\label{eq:app1:lagBc1}
\end{eqnarray}
where the chiral covariant derivative is
\begin{equation}
\mathcal{D}_{\mu}B=\partial_{\mu}B+\Gamma_{\mu}B+B\Gamma_{\mu}^{T}.~\label{eq:covB6}
\end{equation}
The Lagrangian~\eqref{eq:app1:lagBc1} is reduced to a compact form with the superfield representation, where the $B_{\bm 6}$ and $B_{\bm 6}^\ast$ are described by the superfield $\psi_Q^\mu$ via
\begin{equation}
	\psi_{Q}^{\mu}={\cal B}_{\bm6}^{\ast\mu}+\sqrt{\frac{1}{3}}(\gamma^{\mu}+v^{\mu})\gamma^{5}{\cal B}_{\bm6}, \text{ and its conjugate }\bar{\psi}_{Q}^{\mu}=\bar{{\cal B}}_{\bm6}^{\ast\mu}-\sqrt{\frac{1}{3}}\bar{\cal B}_{\bm6}\gamma^{5}(\gamma^{\mu}+v^{\mu}),\label{eq:sec1.5:supBc}
\end{equation}
with ${\cal B}_{\bm6}$ ($\equiv\Lambda_+B_{\bm6}$) and ${\cal B}_{\bm6}^{\ast}$ ($\equiv\Lambda_+B_{\bm6}^\ast$) the nonrelativistic reduced fields of $B_{\bm6}$ and $B_{\bm6}^\ast$, respectively, see ~\ref{app:HFE}.
Then the Lagrangian~\eqref{eq:app1:lagBc1} is reexpressed as
\begin{eqnarray}
	{\cal L}_{{\psi_Q}\varphi}&=&-\text{Tr}\left[\bar{\psi}_{Q}^{\mu}iv\cdot \mathcal{D}\psi_{Q\mu}\right]+ig_{a}\epsilon_{\mu\nu\rho\sigma}\text{Tr}\left[\bar{\psi}_{Q}^{\mu}u^{\rho}v^{\sigma}\psi_{Q}^{\nu}\right]+i\frac{\delta_{a}}{2}\text{Tr}\left[\bar{\psi}_{Q}^{\mu}\sigma_{\mu\nu}\psi_{Q}^{\nu}\right]\nonumber\\
&&+\frac{1}{2}\mathrm{Tr}\left[\bar{\mathcal{B}}_{\bar{\bm3}}(iv\cdot \mathcal{D}+\delta_{c})\mathcal{B}_{\bar{\bm3}}\right]+g_{c}\mathrm{Tr}\left(\bar{\psi}_Q^{\mu}u_{\mu}\mathcal{B}_{\bar{\bm3}}+\mathrm{H.c.}\right),\label{eq:app1:lagBc2}
\end{eqnarray}
where $g_a$ and $g_c$ are two independent axial couplings, and the $\delta_a$ term is proportional to the mass splitting between the spin-$3\over 2$ sextet and spin-$1\over 2$ sextet. The $\delta_c$ contributes to the mass splitting between the flavor antitriplet and sextet.
Unfolding Eq.~\eqref{eq:app1:lagBc2} one obtains the following relations
\begin{eqnarray}
	g_{1}=-\frac{2}{3}g_{a},\quad g_{3}=-\frac{1}{\sqrt{3}}g_{a},\quad g_{5}=g_{a};\quad g_{2}=-\frac{1}{\sqrt{3}}g_{c},\quad g_{4}=g_{c},\quad g_{6}=0,
\end{eqnarray}
in which the $g_2$ and $g_4$ can be extracted from the partial decay widths of $\Sigma_c\to\Lambda_c\pi$ and $\Sigma_c^\ast\to\Lambda_c\pi$, respectively. $g_6=0$ is due to the fact that the interactions between the  pseduscalar meson and the spin-$0$ diquark are forbidden  because of the conservation of the parity and angular momentum. Meanwhile, the couplings $g_1$, $g_3$ and $g_5$ are determined with the quark model~\cite{Meguro:2011nr,Liu:2011xc,Meng:2018gan},
\begin{equation}
\begin{array}{ll}
g_{2}=-0.60, & g_{4}=-\sqrt{3} g_{2}=1.04; \\
g_{1}=-\sqrt{\frac{8}{3}} g_{2}=0.98, & g_{3}=\frac{\sqrt{3}}{2} g_{1}=0.85, \quad g_{5}=-\frac{3}{2} g_{1}=-1.47.
\end{array}
\end{equation}

It should be noticed that, for the SU(2) case, there are two conventional representations for the isospin triplet $\Sigma_{c}$,
\begin{equation}
	\Sigma_{c}^{\text{I}}=\left[\begin{array}{cc}
		\Sigma_{c}^{++} & \frac{\Sigma_{c}^{+}}{\sqrt{2}}\\
		\frac{\Sigma_{c}^{+}}{\sqrt{2}} & \Sigma_{c}^{0}
	\end{array}\right],\quad
	\Sigma_{c}^{\text{II}}=\bm{\Sigma}_{c}\cdot\bm{\tau}=\sqrt{2}\left[\begin{array}{cc}
		\frac{\Sigma_{c}^{+}}{\sqrt{2}} & \Sigma_{c}^{++}\\
		\Sigma_{c}^{0} & -\frac{\Sigma_{c}^{+}}{\sqrt{2}}
	\end{array}\right].
\end{equation}
The different representations will not change the physical results so long as they are used consistently. The first one is the direct reduction of $B_{\bm6}$ and the covariant derivative is similar to Eq.~\eqref{eq:covB6} but transformed into the SU(2) case. The second representation is like the $\Sigma$ in the nucleon octet. The corresponding covariant derivative in the Lagrangian becomes
$\mathcal{D}_\mu \Sigma_{c}^\text{II}=\partial_\mu +[\Gamma_{\mu},\Sigma_{c}^\text{II}]$.
The isospin triplet with specific $I_3=(+1,0,-1)$ is also different in the two representations, e.g., $(\Sigma_{c}^{++},\Sigma_{c}^{+},\Sigma_{c}^{0})^\text{I}, (-\Sigma_{c}^{++},\Sigma_{c}^{+},\Sigma_{c}^{0})^\text{II}$.

\subsubsection{Doubly heavy baryons}

For the ground-state doubly heavy baryons $QQq$, $j_\ell=\frac{1}{2}$, and $S_{QQ}=1$,  there exist two multiplets, the $B_{QQ}$ with spin-$1\over 2$ and the $B^{*}_{QQ}$ with spin-$3\over 2$,
\begin{equation}
	B_{QQ}=\left[\begin{array}{c}
		\Xi_{cc}^{++}\\
		\Xi_{cc}^{+}\\
		\Omega_{cc}^{+}
	\end{array}\right],\qquad\qquad
    B^{*}_{QQ}=\left[\begin{array}{c}
		\Xi_{cc}^{*++}\\
		\Xi_{cc}^{*+}\\
		\Omega_{cc}^{*+}
	\end{array}\right].
\end{equation}

The low order Lagrangians after heavy baryon reduction reads (see Ref.~\cite{Qiu:2020omj} for higher order Lagrangians)
\begin{eqnarray}
	\mathcal{L}_{{\mathcal{B}_{QQ}}\varphi}&=&\bar{\mathcal{B}}_{QQ}iv\cdot \mathcal{D}{\cal B}_{QQ}-\bar{{\cal B}}^{*\mu}_{QQ}(iv\cdot \mathcal{D}-{\delta}_4){\cal B}_{{QQ}\mu}^{*}+2\tilde{g}_{1}\bar{{\cal B}}_{QQ}(S\cdot u){\cal B}_{QQ}\nonumber\\
	&&+2\tilde{g}_{2}\bar{{\cal B}}^{*\mu}_{QQ}(u\cdot S){\cal B}_{{QQ}\mu}^{*}+\tilde{g}_{3}(\bar{{\cal B}}^{*\mu}_{QQ}u_{\mu}{\cal B}_{QQ}+\bar{{\cal B}}_{QQ}u_{\mu}{\cal B}^{*\mu}_{QQ}),\label{eq:app1:lagBcc1}
\end{eqnarray}
where the covariant derivative is the same as that in Eqs.~\eqref{eq:app1:lagD} and~\eqref{eq:app1:lagDbar}, $\mathcal{D}_{\mu}=\partial_{\mu}+\Gamma_{\mu}$. The mass splitting $\delta_4$ is given as ${\delta}_4=m_{B^\ast_{QQ}}-m_{B_{QQ}}$. $S^\mu=\frac{i}{2}\gamma_5\sigma^{\mu\nu}v_\nu$ is the covariant spin operator.

If one assumes the heavy diquark as a compact and heavy object, one can adopt the HDAS to treat the $(B_{QQ},B^\ast_{QQ})$ pair as a spin doublet. They can be uniformly depicted by the superfield
\begin{equation}
	\psi^{\mu}_{QQ}={\mathcal{B}^{\ast\mu}_{QQ}}+\sqrt{\frac{1}{3}}(\gamma^{\mu}+v^{\mu})\gamma^{5}{\cal B}_{QQ}, \text{ and its conjugate }\bar{\psi}^{\mu}_{QQ}={\bar{\mathcal{B}}^{\ast\mu}_{QQ}}-\sqrt{\frac{1}{3}}{\bar{\cal B}}_{QQ}\gamma^{5}(\gamma^{\mu}+v^{\mu}).\label{eq:sec1.5:supBcc}
\end{equation}
The Lagrangian~\eqref{eq:app1:lagBcc1} is expressed with the $\psi^{\mu}_{QQ}$ as
\begin{eqnarray}
	{\cal L}_{\psi_{QQ}\varphi}=-\bar{\psi}^{\mu}iv\cdot \mathcal{D}\psi_{\mu}+\tilde{g}_{b}\bar{\psi}^{\mu}\slashed{u}\gamma_{5}\psi_{\mu}+\frac{i\tilde{\delta}_{b}}{4}\bar{\psi}^{\mu}\sigma_{\mu\nu}\psi^{\nu},\label{eq:app1:lagBcc2}
\end{eqnarray}
where there exists only one coupling $\tilde{g}_{b}$ within the HDAS, and the $\tilde{\delta}_{b}$ term breaks the heavy diquark symmetry and contributes to the mass splitting of spin-$1\over 2$ and spin-$3\over 2$ multiplets. The couplings $\tilde{g}_i~(i=1,2,3)$ and mass splitting $\delta_4$ in Eq.~\eqref{eq:app1:lagBcc1} can be related to $\tilde{g}_{b}$ and $\tilde{\delta}_b$ via
\begin{equation}
	\tilde{g}_{1}=\frac{1}{3}\tilde{g}_{b},\qquad\tilde{g}_{2}=\tilde{g}_{b},\qquad\tilde{g}_{3}=2\sqrt{\frac{1}{3}}\tilde{g}_{b},\qquad \delta_4={3\over4}\tilde{\delta}_b.
\end{equation}

As discussed in Sec.~\ref{sec:HDAS}, the Lagrangians of the doubly heavy baryons can be related to those of the singly heavy mesons with the HDAS. In the superifield formalism, we have $\psi_{QQ}^{\mu}\sim u_{l}A_{h}^{\mu}$ and $\tilde{{\cal H}}\sim u_{l}\bar{v}_{h}$. For the Lagrangians with the heavy d.o.f as the spectator, we have
\begin{eqnarray}
    {\cal L}_{\tilde{{\cal H}}}&=&\xi\left(\bar{u}_{l}\Gamma u_{l}\right)\left(A_{h}^{*\mu}A_{h\mu}\right)=\xi\bar{\psi}^{\mu}\Gamma\psi_{\mu},\nonumber\\{\cal L}_{\psi_{QQ}}&=&\xi\left(\bar{u}_{l}\Gamma u_{l}\right)\left(\bar{v}_{h}v_{h}\right)=\xi\bar{\tilde{{\cal H}}}\Gamma\tilde{{\cal H}},~\label{eq:hdas1}
    \end{eqnarray}
    where the Lagrangians for the singly heavy meson and doubly heavy baryons share the same coupling constant $\xi$ in HDAS once the heavy parts are normalized properly. Therefore, comparing  Eqs.~\eqref{eq:app1:lagDbar} and~\eqref{eq:app1:lagBcc2}, one can obtain the relation~\cite{Hu:2005gf},
\begin{equation}
    \tilde{g}_b=g_b.~\label{eq:hdas2}
\end{equation}
Moreover, assuming only the chromomagnetic interaction [the third term in Eq.~\eqref{eq:LagHQET2}] breaks the heavy (di)quark symmetry, the $\delta_b$ and $\tilde{\delta}_b$ terms in Eqs.\eqref{eq:app1:lagDbar} and~\eqref{eq:app1:lagBcc2} can be related to each other as well~\cite{Savage:1990di},
\begin{equation}
	\delta_b=\tilde{\delta}_b.~\label{eq:hdas3}
\end{equation}

\subsection{KSW and Weinberg schemes for two matter field systems}\label{sec:chiEFT}

For the system with one matter field, one can perform the perturbative expansion with the guideline of the Weinberg's power counting and heavy hadron expansion. However, when one adopts the same scheme in the two matter field system, one has to overcome two obstacles at least.

\begin{figure}[htbp]
 	\centering
 	\includegraphics[width = 0.6\textwidth]{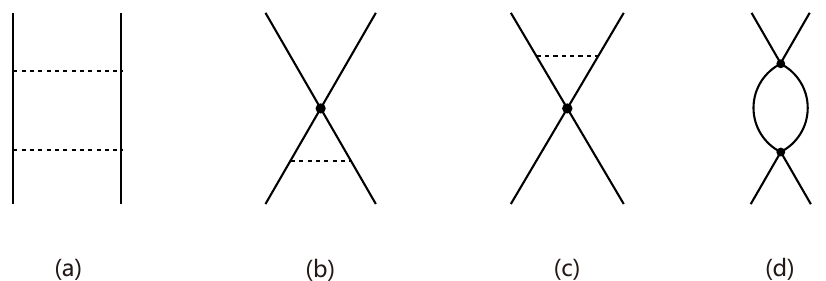}
 	\caption{Three types of Feynmann diagrams that contain the pinched singularities for the $NN$ scattering at the leading order of the heavy baryon expansion. The (a), (b)/(c) and (d) represent the box, triangle and bubble diagrams, respectively. The solid and dashed lines denote the nucleon and pion, respectively.}\label{fig:PinchSingularity}
 \end{figure}

We use the two nucleon system to illustrate the first obstacle---the pinched singularity~\cite{Weinberg:1991um}. For the box diagram as shown in Fig.~\ref{fig:PinchSingularity}(a), if we take the leading order of the heavy baryon expansion, the amplitude reads
\begin{equation}
{\cal A}\sim\int dl^{0}\frac{1}{v\cdot l+i\varepsilon}\frac{1}{-v\cdot l+i\varepsilon},
\end{equation}
where only the two propagators of the intermediate nucleons are presented. The contour of the integration is pinched by the two poles, $l_0=\pm i\varepsilon$. It is impossible to avoid the singularities by distorting the contour. In order to eliminate the pinched singularity, one has to keep the kinetic terms in the Lagrangian: ${\cal L}=\bar{{\cal N}}(i\partial_{0}+\frac{\bm{\bm{\partial}}^{2}}{2M}){\cal N}$. The amplitude becomes
\begin{eqnarray}
	{\cal A}\sim\int dl^{0}\frac{1}{v\cdot l+\frac{\bm{p}_{1}^{\prime2}}{2M}+i\varepsilon}\frac{1}{-v\cdot l+\frac{\bm{p}_{2}^{\prime2}}{2M}+i\varepsilon} \sim\frac{1}{\frac{\bm{p}_{2}^{\prime2}}{2M}+\frac{\bm{p}_{1}^{\prime2}}{2M}} \sim\frac{1}{|\bm{p}|}\frac{M}{|\bm{p}|}.\label{eq:sec1.5:delpinch}
\end{eqnarray}
Although the pinched singularity is cured by including the kinetic energy of the nucleon in the leading terms, the amplitude is enhanced by a large factor $M/|\bm p|$. This strong enhancement explicitly breaks the naive power counting with which the $l_0$ integral should be of $\mathcal{O}(1/|\bm p|)$.
It is worthwhile to notice that the pinched singularity will appear not only in the box diagrams but also in the triangle diagrams [Fig.~\ref{fig:PinchSingularity}(b), (c)] and bubble diagram [Fig.~\ref{fig:PinchSingularity}(d)] in the two nucleon scattering.

The pinched singularity is a typical feature of the nonrelativistic two-body systems. An analog is the heavy quarkonium in the NRQCD ~\cite{Bodwin:1994jh,Caswell:1985ui}. One can first separate the nonrelativistic kinetic energy $q^0$ and momentum $\bm{q}$ from the mass in the relativistic momentum $p^\mu=(m_Q,\bm{0})+(q^0,\bm{q})$.  For the system with two heavy quarks, a different power counting is adopted in NRQCD . In the NRQCD, $q^0$ and $\bm{q}^2/m_Q$ is counted as the same order, while in the HQET the $q^0$ and $\bm{q}$ are regarded as the same order. In the NRQCD, the calculation is organized in power of the velocity $v\equiv|\bm q |/m_Q$~\footnote{One should not confuse the velocity here with that in HQET and heavy field expansion.  }. In the nonrelativistic theory, the kinetic energy is at the order of $m_Qv^2$ and the three momentum is at the order of $m_Q v$. One can see that keeping the kinetic term in Eq.~\eqref{eq:sec1.5:delpinch} essentially takes the similar power counting with NRQCD.

 \begin{figure}[htbp]
 	\centering
 	\includegraphics[width = 0.5\textwidth]{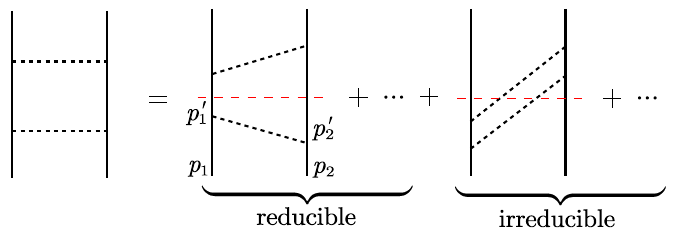}
 	\caption{Box diagrams in the time-ordered perturbation theory.}\label{fig:box}
 \end{figure}

 The unnaturalness is the second obstacle preventing the two matter field systems from suiting Weinberg's power counting. The Weinberg's power counting is based on the NDA. An important prerequisite of its validity is the naturalness, which implies all the dimensionless parameters of the expansion is at order of unity. The dimensionless parameter is obtained by factoring out $\Lambda^n$, where $\Lambda$ is the breaking scale of the effective field theory. The opposite of the naturalness is the unnaturalness or the fine tuning. A typical example of the fine tuning is the nuclear force. At the scale $p\ll m_\pi$, one can integrate out the pion by treating it as a hard scale. It is expected that the pionless effective field theory ($\slashed{\pi}$EFT) is valid, which is equivalent to the ERE~\cite{vanKolck:1999mw}. For example, the ERE for the $S$-wave $NN$ system is
 \begin{equation}
p\cot\delta=-\frac{1}{a_s}+\frac{1}{2}r_{0}p^{2}+\dots,
 \end{equation}
where $\delta$ is the phase shift and $p$ is the magnitude of three momentum. $a_s$ and $r_0$ are the scattering length and effective range, respectively. In the expansion, the large scale is the pion mass $m_\pi$. By assuming naturalness, the scattering length $a_s$ should be of order $1/ m_\pi \approx$1.4 fm. However, the experimental {neutron-proton} scattering lengths for the $^1 S_0$ and $^3S_1$ channels are 
 \begin{eqnarray}
 	a_{^1S_0}\simeq-23.76\text{ fm},\qquad a_{^3 S_1}\simeq5.42\text{ fm}.
 \end{eqnarray}
Thus, the loss of the naturalness indicates the failure of Weinberg's power counting for the $NN$ systems.

The unnaturalness of the $NN$ interaction {can} be attributed to the appearance of the bound state or virtual state, like the deuteron as the $^3S_1$ bound state. The bound state, virtual state and the resonance are all nonperturbative phenomena, which indicate the failure of the perturbative framework conducted by Weinberg's power counting. One has to face the same obstacle of unnaturalness in the heavy hadron systems. The main motivation of investigating the two matter filed system in the heavy flavor sector is to decode the nature of the exotic hadrons in experiments. In the molecular scheme, the exotic hadrons are interpreted as the bound state, virtual state or resonance of two heavy hadrons. Therefore, the same unnaturalness will appear. For the $NN$ systems, there are two types of frameworks to perform the effective field theory properly, the Kaplan-Savage-Wise (KSW) scheme~\cite{Kaplan:1998tg,Kaplan:1998we} and the Weinberg scheme~\cite{Weinberg:1990rz,Weinberg:1991um}~\footnote{One should not confuse with the Weinberg scheme in calculating the nuclear force with the Weinberg's power counting.}. They were widely used for the heavy hadron systems as well.

\subsubsection{KSW scheme}~\label{sec:ksw}

Kaplan, Savage and Wise proposed an elegant approach to reformulating the power counting of the $NN$ systems considering the unnatural large scattering length~\cite{Kaplan:1998tg,Kaplan:1998we}. The scheme starts from the scale less than the pion mass in the $^1S_0$ channel. Then the tree level amplitude from the contact interaction reads
\begin{equation}
	i\mathcal{A}_{\mathrm{tree}}=-i(\mu/2)^{4-d}\sum_{n=0}^{\infty}C_{2n}(\mu)p^{2n}=-i(\mu/2)^{4-d}C(p^{2},\mu),~\label{eq:sec1.6:ctc}
\end{equation}
where $p$ is the nucleon momentum in the center of mass frame. $M$ and $\mu$ are the nucleon mass and subtraction scale. $d$ is the dimension of space-time. $C_{2n}$ are the LECs and $C$ is the polynomial of $C_{2n}$. In the Table~\ref{tab:sec1.5:ksw_pc}, we present the dimension of the corresponding LECs. For such a nonrelativistic system, the double expansions in power of $1/M$ and $p/\Lambda$ are performed. In Eqs.~\eqref{eq:sec1.5:delpinch}, every loop will contribute a factor $M$. In order to ensure the diagrams with  arbitrary loops have the same counting of $1/M$ as the tree-level diagrams, we count $C_{2n}$ as order of $1/ M$. With the common factor of $1/ M$, we can focus on the $p/ \Lambda$ expansion. For example, the dimension of $C_0$ is $-2$. Thus, one expects $C_0 \sim 1/(M\Lambda)$ from NDA.  With the vertices, the amplitude of the one-loop bubble diagram reads
\begin{eqnarray}
I_{n}&=&-i(\mu/2)^{4-d}\int\frac{d^{d}q}{(2\pi)^{d}}\bm{q}^{2n}\frac{i}{(E/2-q^{0}-\bm{q}^{2}/2M+i\varepsilon)}\frac{i}{(E/2+q^{0}-\bm{q}^{2}/2M+i\varepsilon)}\nonumber\\
&=&-M(ME)^{n}(-ME-i\varepsilon)^{\frac{d-3}{2}}\Gamma\left(\frac{3-d}{2}\right)\frac{(\mu/2)^{4-d}}{(4\pi)^{(d-1)/2}}.
\end{eqnarray}
In the calculation, the kinetic energy terms are kept to eliminate the pinched singularity. In the conventional dimensional regularization, the pole at $d=4$ is subtracted to track the logarithmic ultraviolet divergence $\ln \mu$. However, if one calculates the above integration in a hard cutoff regularization, one can find the linear ultraviolet divergence $\mu$. To track the power ultraviolet divergence, an extra subtraction at $d=3$ was made by the counter term
\begin{equation}
\delta I_{n}=-\frac{M(ME)^{n}\mu}{4\pi(d-3)}.
\end{equation}
The above procedure was called the power divergence substation (PDS) scheme. The regularized amplitude becomes
\begin{equation}
I_{n}^{\mathrm{PDS}}=I_{n}+\delta I_{n}=-(ME)^{n}\left(\frac{M}{4\pi}\right)(\mu+ip).\label{eq:sec1.5:int}
\end{equation}
The amplitude considering bubble diagrams nonperturbatively then becomes
\begin{equation}
i{\cal A}=\frac{-iC(p^{2},\mu)}{1+MC(p^{2},\mu)(\mu+ip)/4\pi}.
\end{equation}
In order to investigate the effect of the unnatural scattering length, one can match the above amplitude to the effective range expansion,
\begin{equation}
p\cot\delta=-\frac{4\pi}{MC(p^{2},\mu)}-\mu=-\frac{1}{a_s}+\frac{1}{2}r_{0}p^{2}+\dots.
\end{equation}
Then the first two LECs read
\begin{equation}
C_{0}(\mu)=\frac{4\pi}{M}\left(\frac{1}{-\mu+1/a_s}\right),\qquad C_{2}(\mu)=\frac{2\pi}{M}\left(\frac{1}{-\mu+1/a_s}\right)^{2}r_{0},\dots.
\end{equation}

\begin{figure}[htbp]
	\centering
	\includegraphics[width = 0.4\textwidth]{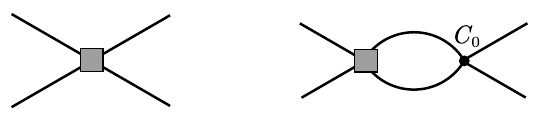}
	\caption{The effect of an extra loop with the LO vertex. In the natural case, the right diagram is suppressed by an extra factor $Mp$ compared with the left one. In the unnatural case, the right one has the same order with the left one.}\label{fig:KSW_Order}
\end{figure}

Taking $\mu\sim p$, one can obtain the power counting either in the natural or unnatural cases. In the natural case $1/a_s\sim \Lambda$, the powers of $C_{2n}$ satisfy the naive counting in Table~\ref{tab:sec1.5:ksw_pc}. In the unnatural case, $a\gg 1/\Lambda$, the power counting is presented in Table~\ref{tab:sec1.5:ksw_pc}. The powers of the LECs are increased by the unnaturalness. From  Eq.~\eqref{eq:sec1.5:int}, one can see that adding an extra loop (see Fig.~\ref{fig:KSW_Order}) will introduce a factor of order $Mp$ from the integration. For the natural case, taking the extra vertex in the loop into consideration, the extra loop will introduce a factor of order $p/\Lambda$ at most (for the LO vertex, see Fig.~\ref{fig:KSW_Order}). Thus, one can perform the calculation perturbatively for the natural case. For the unnatural case, introducing an extra loop with the LO vertex will introduce a factor of order $1$. Thus, one has to include the $C_0$ nonperturbatively. Although the higher order vertices ($C_2$, $C_4$,...) were enhanced by the unnaturalness, one can still deal with them perturbatively.

\begin{table}
\renewcommand{\arraystretch}{1.5}
	\centering
	\caption{Power counting and dimensions of the contact terms in Eq.~\eqref{eq:sec1.6:ctc} for both natural and unnatural cases.}~\label{tab:sec1.5:ksw_pc}
\setlength{\tabcolsep}{5.0mm}
{
	\begin{tabular}{c|cccccc}
		\hline
		& $C_{0}$ & $C_{2}$ & $C_{4}$ &  & $C_{2n}$ & $C_{2n}p^{2n}$\tabularnewline
		\hline
		Dimension & $-2$ & $-4$ & $-6$ & $\cdots$ & $-2(n+1)$ & $-2$\tabularnewline
		Natural & $\frac{1}{M}\frac{1}{\Lambda^{1}}$ & $\frac{1}{M}\frac{1}{\Lambda^{3}}$ & $\frac{1}{M}\frac{1}{\Lambda^{5}}$ & $\cdots$ & $\frac{1}{M}\frac{1}{\Lambda^{2n+1}}$ & $\frac{1}{M}\frac{p^{2n}}{\Lambda^{2n+1}}$\tabularnewline
		Unnatural & $\frac{1}{M}\frac{1}{p}$ & $\frac{1}{M}\frac{1}{p^{2}\Lambda^{1}}$ & $\frac{1}{M}\frac{1}{p^{3}\Lambda^{2}}$ & $\cdots$ & $\frac{1}{M}\frac{1}{p^{n+1}\Lambda^{n}}$ & $\frac{1}{M}\frac{p^{n-1}}{\Lambda^{n}}$\tabularnewline
		\hline
	\end{tabular}
}
\end{table}

It is worthwhile to stress that the validity of the above power counting is independent of the specific regularization scheme---PDS. In Refs.~\cite{Mehen:1998zz,Mehen:1998tp,Gegelia:1998xr,Gegelia:1998iu,Cohen:1998bv}, the alternative but equivalent regularization schemes were adopted to obtain the same power counting. The key point of choosing the regularization scheme is to track the power divergence. In Ref.~\cite{vanKolck:1998bw}, the author illustrated the same power counting without specific regularization scheme. In Ref.~\cite{Birse:1998dk}, the equivalent power counting was obtained by performing the Wilsonian renormalization group equation to the Lippmann-Schwinger equation.

 The power counting can be extended to $p\sim m_\pi$. In the $^1S_0$ channel, the one-pion exchange (OPE) interaction is
 \begin{equation}
V_\pi(\bm{p},\bm{p'})=-{g_A^2 \over 2f^2}\left({m_\pi^2 \over \bm{q}^2+m_\pi^2}-1 \right),
 \end{equation}
where $\bm{q}=\bm{p'}-\bm{p}$.  The counting of the OPE interaction is $p^0$, which is of the same order as the $C_2$ term, thus can be treated perturbatively.

 \begin{figure}[htbp]
	\centering
	\includegraphics[width = \textwidth]{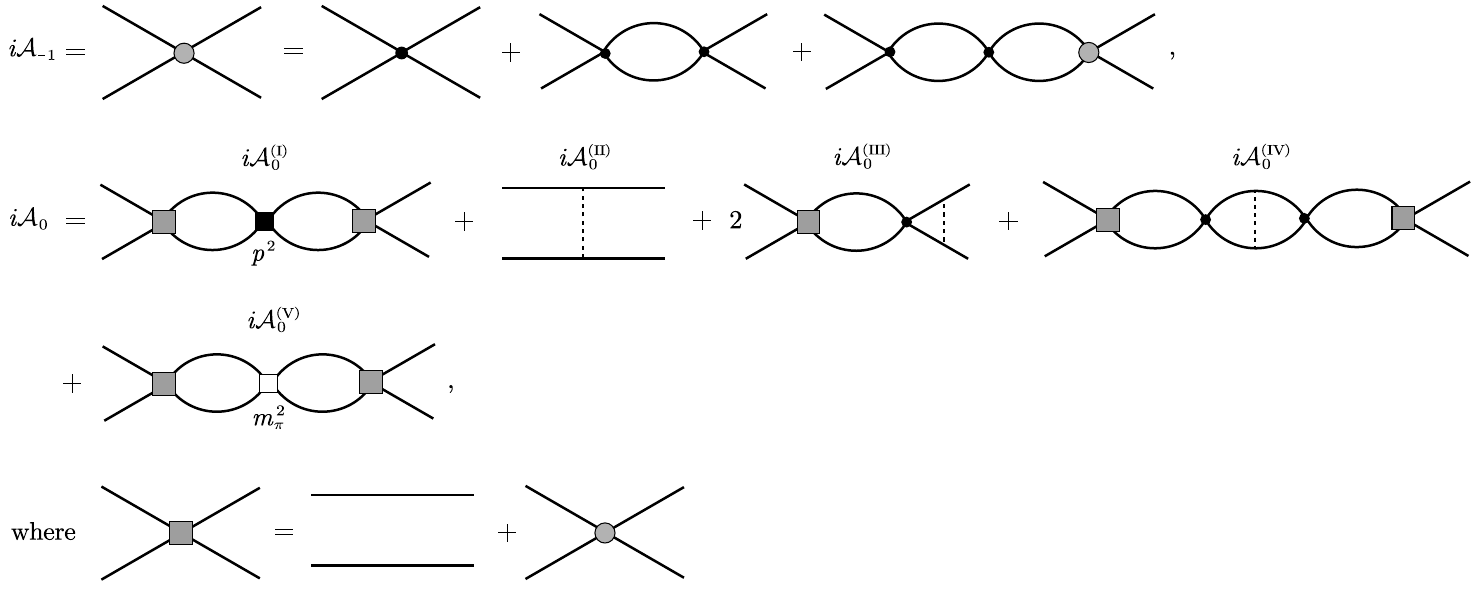}
	\caption{Feynman diagrams for KSW. The LO vertices are represented by the solid dot. The NLO vertices are labeled by solid square and empty square. }\label{fig:KSE}
\end{figure}

The KSW scheme is an elegant framework to treat the unnatural $NN$ systems with very clear power counting. In this framework, the renormalization is very transparent. However, the resulting scattering amplitudes didn't converge in certain spin-triplet channels~\cite{Fleming:1999ee,Cohen:1999iaa} for the nucleon
momenta around the pion mass (the subsequent improvements seemed to yield a convergent
expansion~\cite{Beane:2008bt,Kaplan:2019znu}). It was shown that the breakdown of the KSW
expansion arises from the perturbative treatment
of the pion exchange contributions~\cite{Gegelia:1998ee,Gegelia:1999ja,Cohen:1999iaa,Cohen:1998jr}. The proper scale where the OPE is perturbative was obtained in Ref.~\cite{Birse:2005um}.

The KSW type scheme was also applied in the heavy hadronic systems. Using the XEFT to investigate the $X(3872)$ inherits the characters of the KSW scheme---the nonperturbative LO contact interaction and the perturbative OPE and higher contact interactions. We will discuss the relevant EFT in Sec.~\ref{sec:XEFT}  {and explain why the perturbative OPE works for the $X(3872)$ although it
failed for the NN system.}

\subsubsection{Weinberg scheme}
The state-of-the-art nuclear forces~(see Refs.~\cite{Reinert:2017usi,Epelbaum:2014efa,Epelbaum:2014sza,Entem:2017gor} for recent progresses and see Refs.~\cite{Machleidt:2011zz,Epelbaum:2008ga,Epelbaum:2019kcf,Hammer:2019poc} for reviews) are based on the Weinberg's seminal works~\cite{Weinberg:1990rz,Weinberg:1991um}. Weinberg proposed
to adopt the power counting law to calculate the $NN$ potentials perturbatively. The $NN$ potentials are regarded as the effective potential of the  Schr\"odinger equation or the kernel of the Lippmann-Schwinger equations (LSEs). From the perspective of the time-ordered perturbation theory as shown in Fig.~\ref{fig:box}, the contribution of the loop diagrams (e.g. box diagrams) is divided into the two-particle-reducible (2PR) part and two-particle-irreducible (2PIR) part. The 2PR part is subtracted in the $NN$ potential. The remaining 2PIR part is iterated to all orders by solving the Schr\"odinger equation or
Lippmann-Schwinger equation. The 2PR part is recovered by iterating the tree diagrams automatically.

In the practical calculation, one can obtain the 2PIR contribution by subtracting the contribution of the poles of the nucleons~\cite{Kaiser:1997mw,Kaiser:1998wa} rather than the time-ordered perturbation theory~\cite{Ordonez:1995rz}. For the heavy hadron systems, the subtraction becomes more subtle due to the appearance of the mass splitting, see the Appendix of Ref.~\cite{Wang:2019ato}.

Compared with the KSW scheme, the renormalization of Weinberg scheme is less transparent. The amplitudes are nonperturbative expressions obtained by solving the integral equations with ultraviolet (UV) divergences. For example, one usually introduces regulators to the $NN$ potential to render the solutions of LSEs or Schr\"odinger equation finite. Sometimes the Gauss form regulator is used in the calculation
\begin{equation}
	V(\bm{p},\bm{p'})\to V(\bm{p},\bm{p})e^{-(p^n+p'^n)\over \Lambda^n}.
\end{equation}
In such a nonperturbative renormalization, the cutoff independence is implicit. In Refs.~\cite{Epelbaum:2018zli,Epelbaum:2009sd,Gasparyan:2021edy}, several conceptional ambiguities in the renormalization of the nuclear forces in Weinberg scheme were clarified. Though the Weinberg scheme is not as elegant as KSW scheme, it turns out to be convergent and practical in calculating high-precision nuclear forces~\cite{Machleidt:2011zz,Epelbaum:2008ga,Epelbaum:2019kcf}. In the Weinberg scheme, the OPE interaction is iterated to all orders. In the heavy hadron systems, the Weinberg scheme was also used to identity the hadronic molecules by calculating the mass spectrum (e.g.~\cite{Meng:2019ilv,Wang:2019ato}) and fitting the experimental line shapes~\cite{Du:2019pij,Du:2021zzh}. We will discuss this formalism in Sec.~\ref{sec:chap5}.

\section{Masses, axial charges, strong and electromagnetic decays of heavy hadrons}\label{sec.3}

The HH$\chi$PT was developed  to investigate the heavy-light hadrons incorporating the chiral symmetry and the heavy quark symmetry consistently. It has successfully described the properties of the heavy-light systems, including the spectra, form factors, decay patterns, etc. In this section, we will review the chiral corrections to the masses, axial currents, magnetic moments, electromagnetic decays and strong decays of the heavy mesons, singly heavy baryons and doubly heavy baryons within the framework of HH$\chi$PT. The cross-talk between HH$\chi$PT and other frameworks will also be discussed.

\subsection{Heavy hadron masses}\label{sec:heavy hadron mass}

In this subsection, we briefly introduce the application of the HH$\chi$PT in the study of the mass spectroscopy of the heavy mesons and baryons with the help of the experimental data and lattice QCD simulations.
In the  HH$\chi$PT,  the  propagator of the meson is proportional to
\begin{eqnarray} \label{eq:GM}
\frac{i}{2 (v \cdot k-\delta^{\text{tree}})-\Sigma(v \cdot k)},
\end{eqnarray}
where the factor $2$  results from the normalization of the heavy meson field (see~\ref{app:HFE}). {{The heavy meson momentum $p$ is split into the large component with velocity $v$ and small residue  momentum $k$ with $p=m_0 v+k$ and $m_0$ being the bare mass of the reference heavy meson. $\delta^{\text{tree}}$ denotes the tree-level mass difference of the particle relative $m_0$.}}  The $-i\Sigma(v \cdot k)$ is the sum of all the one-particle irreducible (1PIR) self-energy contribution. In general, it is  complex and the imaginary part is related to the decay width. {The on-shell renormalization condition  is given by,}
\begin{eqnarray}\label{eq:solktild}
2 (v \cdot \tilde k-\delta^{\text{tree}})-\Sigma(v \cdot \tilde k)=0,
\end{eqnarray}
The physical mass { corresponds the position of the pole of the propagator in Eq. \eqref{eq:GM}} and is given by,
\begin{eqnarray}\label{eq:hmm}
m^{\text{phy}}_H=m_0+v\cdot \tilde k=m_0+\delta^{\text{tree}} +{1\over2} \mathrm{Re}\left[\Sigma(v \cdot  \tilde k)\right],
\end{eqnarray}

The propagators of the singly and doubly heavy  baryons are similar to those of the mesons after the heavy field expansion.
Their mass expressions  can be written as
 \begin{eqnarray} \label{eq:bm}
M={M}_0+\delta^{\text{tree}}+ \mathrm{Re}\left[\Sigma(v \cdot  \tilde k)\right],
\end{eqnarray}
where $M_0$ is the bare mass of the heavy baryon.

\subsubsection{Heavy mesons}\label{Sect.2.1.1}

As shown in Sec.~\ref{sec1.1}, the heavy mesons can be labeled with the light and heavy d.o.f in the heavy quark spin basis. In the heavy quark limit, the mesons with the same $j^P_\ell$ are degenerate and they can be embedded into the superfields as shown in Eqs.~\eqref{eq:spfd:smeson},~\eqref{eq:spfd:pmeson1} and~\eqref{eq:spfd:pmeson2}.

Constrained by the heavy quark symmetry and the chiral symmetry, the low order Lagrangians for the heavy mesons $Q\bar q$ are given in Eqs.~\eqref{eq:app1:lagD},~\eqref{eq:app1:lagS}-\eqref{eq:app1:lagST}. Among them, we have included the $1/m_Q$ corrections which break the heavy quark symmetry, 
\begin{eqnarray}\label{eq:Lag-mesonhm}
\mathcal{L}_{1 / m_{Q}}=-\frac{\delta_{b}}{8}\langle{\cal H}\sigma^{\mu\nu}\bar{{\cal H}}\sigma_{\mu\nu}\rangle
+\frac{\delta'_{b}}{8}\langle{\cal S}\sigma^{\mu\nu}\bar{{\cal S}}\sigma_{\mu\nu}\rangle+\frac{3\delta''_{b}}{16}\langle{\cal T}^{\rho}\sigma^{\mu\nu}\bar{{\cal T}}_{\rho}\sigma_{\mu\nu}\rangle,
\end{eqnarray}
where the constant $\delta_b/\delta^{\prime}_b/\delta^{\prime \prime}_b$ is responsible for the mass splitting,  $\delta_b/\delta_b^{\prime}/\delta_b^{\prime\prime}=\left(M_{P^{*}/R^{*}/Y^*}-M_{P/R/Y} \right)$ in the $\mathcal H$/$\mathcal S$/$\mathcal T$ multiplets~\cite{Mehen:2004uj}, which is suppressed by $1/m_Q$ and vanishes in the heavy quark limit. 

\begin{figure*}[htbp]
\centering
\includegraphics[width=1.0\textwidth]{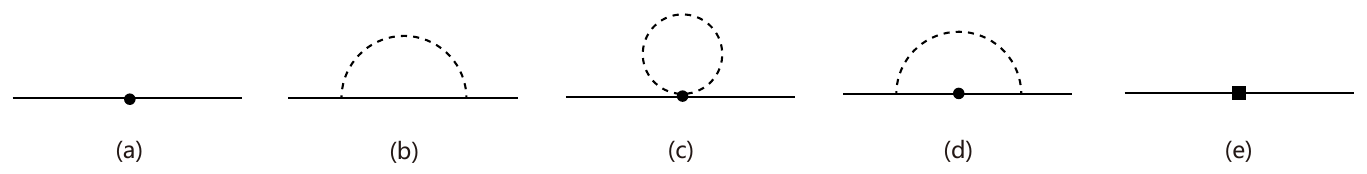}
\caption{The Feynman diagrams contributing to the mass corrections of the heavy hadrons. The solid and dashed lines represent  the
heavy hadrons and  the Goldstone bosons, respectively. In the loop diagrams, the external and the internal heavy hadrons can be the same or different.  The solid dot and black square denote the vertices at $ \mathcal O(p^2)$ and $\mathcal O(p^4)$, respectively. The one-loop diagram (b) is at $\mathcal O(p^3)$ while the other one-loop diagrams are at $\mathcal O(p^4)$.}
\label{fig:mass}
\end{figure*}

At $\mathcal{O}(p^2)$ of the chiral expansion,  the chiral Lagrangians with the light quark masses read~\cite{Boyd:1994pa,Cheng:1993gc,Cheng:2017oqh,Jenkins:1992hx,Yan:1992gz,Kilian:1992hq,Mehen:2005hc,Alhakami:2016zqx,Ananthanarayan:2007sa,Alhakami:2019ait,Jiang:2019hgs}
\begin{eqnarray} \label{eq:mm_chiral}
 \mathcal{L}^{(2)}_{\text{chiral }}&=&a_{H}\left\langle  {\mathcal{H}}m^{\text{\ensuremath{\xi}}}\bar{{\mathcal{H}}}\right \rangle +\sigma_{H}\left\langle  {\mathcal{H}}  \bar{{\mathcal{H}}}\right\rangle \text{Tr}( m^{\text{\ensuremath{\xi}}})-a_{S}\left \langle  \mathcal{S}m^{\text{\ensuremath{\xi}}}\bar{\mathcal{S}} \right \rangle  -\sigma_{S} \left \langle {\mathcal{S}} \bar{\mathcal{S}} \right \rangle  \text{Tr}(  m^{\text{\ensuremath{\xi}}})\nonumber \\
 &&-a_{T} \left \langle \mathcal{T}_{\rho}m^{\text{\ensuremath{\xi}}}\bar{\mathcal{T}}^{\rho}\right \rangle -\sigma_T \left \langle \mathcal{T}_{\rho} \bar{\mathcal{T}}^{\rho} \right \rangle \text{Tr}( m^{\text{\ensuremath{\xi}}}) \nonumber \\
 &&-\frac{\Delta_{H}^{(a)}}{8}\left\langle {\mathcal{H}}\sigma_{\mu\nu}m^{\text{\ensuremath{\xi}}}\bar{\mathcal{H}}\sigma^{\mu\nu}\right\rangle -\frac{\Delta_{H}^{(\sigma)}}{8}\left\langle \mathcal H\sigma_{\mu\nu} \bar{\mathcal H}\sigma^{\mu\nu} \right\rangle   \text{Tr}( m^{\text{\ensuremath{\xi}}})  \nonumber \\
 &&+\frac{\Delta_{S}^{(a)}}{8}\left\langle \mathcal{S}\sigma_{\mu\nu}m^{\text{\ensuremath{\xi}}}  \bar{\mathcal{S}}\sigma^{\mu\nu} \right\rangle +\frac{\Delta_{S}^{(\sigma)}}{8}\left\langle {\mathcal{S}}\sigma_{\mu\nu} \bar{\mathcal{S}}\sigma^{\mu\nu}\right\rangle  \text{Tr}(  m^{\text{\ensuremath{\xi}}})  \nonumber \\
 &&+\frac{3}{16}\Delta_{T}^{(a)}\left\langle {\mathcal{T}}_{\rho}\sigma_{\mu\nu}m^{\text{\ensuremath{\xi}}}  \bar{\mathcal{T}}^{\rho}\sigma^{\mu\nu} \right\rangle +\frac{3}{16}\Delta_{T}^{(\sigma)}\left\langle{\mathcal{T}}_{\rho}\sigma_{\mu\nu}  \bar{\mathcal{T}}^{\rho}\sigma^{\mu\nu}\right\rangle \text{Tr}( m^{\text{\ensuremath{\xi}}}),
\end{eqnarray}
where $m^{\xi}={\chi^+ / 4B_0}=\text{diag}\{m_u,m_d,m_s\}+\mathcal O(p^4)$. In the above interaction, the $\Delta^{(a)}_{H/S/T}$ and $a_{H/S/T}$ terms contribute to the SU(3) flavor breaking effect, while the $\Delta^{(\sigma)}_{H/S/T}$ and $\sigma_{H/S/T}$ terms keep the SU(3) flavor symmetry. The terms in $\mathcal{L}^{(2)}_{\text{chiral }}$ will contribute to the mass spectrum of the heavy mesons  through  Fig.~\ref{fig:mass}(a) at  $\mathcal O(p^2)$ of the chiral expansion. The contributions from  $\Delta_{H / S / T}^{(a / \sigma)}$   are further suppressed by $1/m_Q$.

Up to the  order of $1/m_Q$ and  up to the NLO chiral expansion [$\mathcal O(p^2)$], the mass corrections come from  the tree diagrams~\cite{Mehen:2005hc,Alhakami:2016zqx,Ananthanarayan:2007sa,Alhakami:2019ait},
\begin{eqnarray}
	 \delta^{\text{tree}}_{P_q} &=&\sigma_{H} \bar{m}+a_{H} m_{q}-\frac{3}{4} \left( \delta_{b}+ \Delta_{H}^{(\sigma)} \bar{m}+\Delta_{H}^{(a)} m_{q}\right), \label{eq:ms_meson1} \\
	  \delta^{\text{tree}}_{P_q^*} &=&\sigma_{H} \bar{m}+a_{H} m_{q}+\frac{1}{4} \left(\delta_{b}+ \Delta_{H}^{(\sigma)} \bar{m}+\Delta_{H}^{(a)} m_{q}\right), \label{eq:ms_meson2}  \\
	 \delta^{\text{tree}}_{R_q}&=&\delta_{\mathcal S}+\sigma_{S} \bar{m}+a_{S} m_{q}-\frac{3}{4} \left(\delta^{\prime}_b+\Delta_{S}^{(\sigma)} \bar{m}+\Delta_{S}^{(a)} m_{q}\right), \label{eq:ms_meson3} \\
	  \delta^{\text{tree}}_{R_q^{*}}&=&\delta_{\mathcal S}+\sigma_{S} \bar{m}+a_{S} m_{q}+\frac{1}{4} \left( \delta^{\prime}_b+ \Delta_{S}^{(\sigma)} \bar{m}+ \Delta_{S}^{(a)} m_{q}\right), \label{eq:ms_meson4} \\
	 \delta^{\text{tree}}_{Y_q}&=&\delta_{\mathcal T}+a_{T} m_{q}+\sigma_{T} \bar{m}-\frac{5}{8}\left(\delta^{\prime \prime}_b+\Delta_{T}^{(a)} m_{q}+\Delta_{T}^{(\sigma)} \bar{m}\right), \label{eq:ms_meson5}\\
	 \delta^{\text{tree}}_{Y_q^{*}}&=&\delta_{\mathcal T}+a_{T} m_{q}+\sigma_{T} \bar{m}+\frac{3}{8}\left(\delta^{\prime \prime}_b+\Delta_{T}^{(a)} m_{q}+\Delta_{T}^{(\sigma)} \bar{m}\right), \label{eq:ms_meson6}
	\label{eq:mesonmass}
\end{eqnarray}
where $\bar{m}=m_{u}+m_{d}+m_{s}$. The subscript $q$ denotes different light quarks $u$, $d$ and $s$.

The one-loop diagrams start to contribute at $\mathcal O(p^3)$ of the chiral expansion. The $\mathcal O (p^3)$ contributions come from the wave function renormalization diagram in Fig.~\ref{fig:mass}(b). The axial coupling vertices in the loops arise from the  Lagrangians in Eqs.~\eqref{eq:app1:lagD},~\eqref{eq:app1:lagS}-\eqref{eq:app1:lagST}. The intermediate states in  the loop can be the same as or different from the external states. The different intermediate states will include the mass corrections from the other kinds of heavy mesons. The other one-loops in Fig.~\ref{fig:mass} will contribute at $\mathcal O(p^4)$.
In Fig. ~\ref{fig:mass}(c) and (d) diagrams, there are three types of the $\mathcal O(p^2)$  vertices which stem from  $a_{H/S/T}$, $\sigma_{H/S/T}$ and $\Delta_{H / S / T}^{(a / \sigma)}$ terms. The $a_{H/S/T}$ terms keep the HQSS  but break the $\SU(3)_V$ symmetry.  
The $\sigma_{H/S/T}$ terms keep the HQSS and yield the flavor-independent contributions to the  masses of the pseudoscalar and vector mesons in the same heavy quark spin doublet.  The $\Delta_{H / S / T}^{(a / \sigma)}$ terms breaks the HQSS and their loop corrections  will be further suppressed by $1/m_Q$.~At $\mathcal O(p^4)$, the tree diagram in Fig.~\ref{fig:mass}(e) is governed by the Lagrangians with two insertions of the light quark mass matrix~\cite{Jenkins:1992hx}. As an example, the Lagrangians of the $\mathcal H$ multiplet at $\mathcal O(p^4)$ are given by
\begin{eqnarray}
\mathcal {L}^{(4)}_{\text{chiral}}&=&b\left\langle \mathcal H m_{\xi} m_{\xi}  \bar{\mathcal H} \right \rangle+c \operatorname{Tr} (m_{\xi}) \left\langle \mathcal H m_{\xi} \bar{\mathcal H} \right\rangle + d\operatorname{Tr} (m_{\xi} m_{\xi}) \left\langle {\mathcal H } \bar{\mathcal H}\right\rangle \nonumber\\
&&-\frac{1}{8} \Delta^{(b)} \left\langle \mathcal H \sigma_{\mu v} m_{\xi} m_{\xi} \bar{\mathcal H} \sigma^{\mu \nu}\right\rangle
-\frac{1}{8} \Delta^{(c)} \operatorname{Tr}( m_{\xi}) \left\langle \mathcal H \sigma_{\mu \nu} m_{\xi} \bar{\mathcal H} \sigma^{\mu \nu}\right\rangle  \nonumber \\
&&-\frac{1}{8} \Delta^{(d)} \operatorname{Tr}( m_{\xi} m_{\xi}) \left\langle \mathcal H \sigma_{\mu \nu}\bar{\mathcal H} \sigma^{\mu \nu}\right\rangle+\dots,
\end{eqnarray}
where $b$, $c$, $d$, and $\Delta^{(b/c/d)}$  are the LECs. The Lagrangians for the $\mathcal S$ and $\mathcal T$ multiplets are similar.

So far, there exist many theoretical predictions for the heavy mesons using the HH$\chi$PT. The early studies only considered the ground state $D$ and $B$ mesons. The mass splittings of these  heavy mesons have been studied in the heavy quark limit and SU(3) flavor symmetry~\cite{Isgur:1989vq,Isgur:1990yhj}, and receive  the $1/m_Q$ and the light quark mass corrections~\cite{Falk:1990pz,Jenkins:1992hx,Yan:1992gz,Cheng:1993gc,Casalbuoni:1996pg, Rosner:1992qw,Randall:1992pb,DiBartolomeo:1994ir,Amoros:1997rx,Goity:1992tp,Blok:1996iz,Guo:2001ph} (see Ref.~\cite{Casalbuoni:1996pg} for a review). For instance, the leading mass splittings ${m_{D^{*}}-m_D}$ and $m_{B^*}-m_B$ are proportional to $\delta_b$ as shown in Eq.~\eqref{eq:ms_meson1} and  Eq.~\eqref{eq:ms_meson2}. Since $\delta_b$ is proportional to $1/m_Q$, one has
\begin{eqnarray}\label{eq:dmq}
\frac{m_{D^{*}}-m_D}{m_{B^*}-m_B}=\frac{m_b}{m_c}.
\end{eqnarray}
Its leading corrections come from the mass corrections of the heavy quarks and have been studied in HQET~\cite{ Falk:1990pz,Amoros:1997rx,Neubert:1993mb} 
\begin{eqnarray}
\frac{m_{D^{*}}-m_D}{m_{B^*}-m_B}=\frac{m_{\mathrm{b}}}{m_{\mathrm{c}}}\left(\frac{\alpha_{\mathrm{s}}\left(m_{\mathrm{b}}\right)}{\alpha_{\mathrm{s}}\left(m_{\mathrm{c}}\right)}\right)^{-9/25}\simeq\frac{m_{\mathrm{B}}}{m_{\mathrm{D}}}\left(\frac{\alpha_{\mathrm{s}}\left(m_{\mathrm{B}}\right)}{\alpha_{\mathrm{s}}\left(m_{\mathrm{D}}\right)}\right)^{-9/25},
\end{eqnarray}
with the $\alpha_s$ the running coupling constant.

Another popular hyperfine mass splitting is~\cite{Jenkins:1992hx,Cheng:1993gc,Casalbuoni:1996pg,Rosner:1992qw,Randall:1992pb,DiBartolomeo:1994ir}
 \begin{eqnarray}
&&\Delta M_{D}=\left(M_{D_{s}^{*}}-M_{D_{s}}\right)-\left( M_{D^{*+}}-M_{D^{+}} \right),\\
&&\Delta M_{B}=\left(M_{B_{s}^{*}}-M_{B_{s}}\right)-\left( M_{B^{*0}}-M_{B^{0}} \right).
\end{eqnarray}
{The leading order electromagnetic corrections from the $s $ and $d$ quarks are the same due to their equal electric charges. Hence, both $\Delta M_D$ and $\Delta M_B$ vanish in the chiral and heavy quark limit.}

In the HH$\chi$PT formalism, the lowest order contribution  arises from the $\Delta_{H}^{(a)}$ term in $\mathcal L ^{(2)}_{\text {chiral}}$ through the diagram in Fig.~\ref{fig:mass}(a) at the order $1/m_{Q}$. The mass splittings then  satisfy  the following relation~\cite{ Rosner:1992qw},
\begin{eqnarray}\label{eq:mssdl}
\frac{\Delta M_{B}}{\Delta M_{D}}=\frac{m_{c}}{m_{b}},
\end{eqnarray}
{which is a direct consequence of Eq. \eqref{eq:dmq}.}

 The one-loop diagram in Fig.~\ref{fig:mass} will  provide corrections to $\Delta M_D$ and $\Delta M_B$. In the loop diagrams, the different masses of the pseudoscalar mesons, $m_{\pi}$/$m_K$/$m_{\eta}$, the mass splittings between the heavy spin doublets $\delta_b$/$\delta^{\prime}_b$/$\delta^{\prime\prime}_b$, the mass splittings between the non-strange and strange mesons $\Delta_s$=$\Delta_{H}^{(a)}(m_s-m_{u/d})$, as well as the different axial couplings for the $P^*P^*\pi$ and $P^*P\pi$ vertex may contribute to the $\Delta M_D$ and $\Delta M_B$.  The one-loop correction was given by~\cite{Jenkins:1992hx,Casalbuoni:1996pg, Randall:1992pb,DiBartolomeo:1994ir,Cheng:1993gc}
\begin{eqnarray}
\Delta{M_{D/B}}^{(b)}&=&\frac{g_{b}^{2}\delta_{b}}{16\pi^{2}f_{\varphi}^{2}}\left[m_{K}^{2}\log\left(\Lambda_{\chi}^{2}/m_{K}^{2}\right)+\frac{1}{2}m_{\eta}^{2}\log\left(\Lambda_{x}^{2}/m_{\eta}^{2}\right)-\frac{3}{2}m_{\pi}^{2}\log\left(\Lambda_{\chi}^{2}/m_{\pi}^{2}\right)\right]\nonumber\\&&+\frac{g_{b}^{2}\delta_{b}}{16\pi^{2}f_{\varphi}^{2}}\left(6\pi m_{K}\Delta_{s}\right)-\frac{g_{b}^{2}}{24\pi f_{\varphi}^{2}}\frac{\Delta_{g_b}}{g_{b}}\left(m_{K}^{3}+\frac{1}{2}m_{\eta}^{3}-\frac{3}{2}m_{\pi}^{3}\right),
\end{eqnarray}
where the terms in the first line is the so-called chiral logarithm. The terms in the second line are  both of order $m_{q}^{3/2}$.  $\Delta_{g_b}$ represents the difference between the axial coupling in the $\mathcal H$ multiplet and stems from the Lagrangian at the order $1 / m_Q$ ~\cite{DiBartolomeo:1994ir},
\begin{eqnarray} \label{app:coupling}
	\mathcal{L}^{g_b}_{1 / m_Q}&=&i \frac{\xi_{1}}{m_Q}\left \langle {\mathcal  H} \gamma_{\mu} \gamma_{5} u^{\mu} \bar{ {\mathcal H }} \right\rangle +i \frac{\xi_{2}}{m_Q}\left \langle\mathcal {H} \bar{ \mathcal H}u^\mu \gamma_{\mu} \gamma_{5}   \right \rangle ,
\end{eqnarray}

{
The $\xi_1$ term is the recoil correction to the leading order interaction term $g_b$ in Eq. (\ref{eq:app1:lagD}) and leads
to a deviation of the $DD^*\pi$ and $BB^*\pi$ couplings. However, the $\xi_1$ term keeps the HQSS and does not contribute to the $\Delta_{g_b}$. On the other hand, 
 the $\xi_2$ term breaks the HQSS, hence is forbidden in the leading order, which contributes to the coupling difference as}
\begin{eqnarray}
	\Delta_{g} =2 \frac{\xi_{2}}{m_Q}.
\end{eqnarray}

The calculation can be generalized to the analogous $d-u$ type hyperfine splitting  $\left(D^{*+}-D^{+}\right)-\left(D^{*0}-D^{0}\right)$.  
{Both the electromagnetic interaction and the current quark mass difference contribute to the isospin violation. Their effects are quantitatively of the same order. Fortunately, the dominant contribution to the isospin violation from $m_d-m_u$ can be easily accounted for through the spurion $\chi$ so long as one keeps $m_d\neq m_u$ from the beginning. In the leading order, the spurion $\chi$ leads to the mass splitting of the heavy mesons and baryons, and also the mass splitting of the eight pseudoscalar mesons. The higher order isospin violating corrections to the heavy hadron masses may arise from the chiral loop contributions where the intermediate isospin multiplets of the pions or kaons have different masses with $m_{\pi^+}\neq m_{\pi^0}$ etc. Another source of higher order correction is the multiple insertions of the spurion $\chi$ in the construction of higher order Lagrangians. However, both types of higher order corrections are much smaller than the leading order term. Therefore, one generally considers the leading order correction due to $m_d-m_u$ in literature.} 

Apart from the light quark mass difference , the electromagnetic effect was also important~\cite{Jenkins:1992hx}. The Lagrangians contributing to  the isospin hyperfine splittings read 
\begin{eqnarray}
	\mathcal{L}_{\mathrm{QED}}^{\mathscr{Q}_{h} \mathscr{Q}_{h} \slashed S}&=&-\frac{1}{8}\Delta_{\text {em}}\alpha_{\text {em}}\left\langle \mathscr{Q}_{h}^{2}\mathcal{H}\sigma_{\mu\nu}\bar{\mathcal{H}}\sigma^{\mu\nu}\right\rangle , \label{eq:emm1}\\
	\mathcal{L}_{\mathrm{QED}}^{\mathscr{Q}_{h} Q}&=& a_{\text {em}}\alpha_{\text {em}}\left\langle \mathscr{Q}_{h}\mathcal{H}{Q}_{+}\bar{\mathcal{H}}\right\rangle , \label{eq:emm2}\\
	\mathcal{L}_{\mathrm{QED}}^{\mathscr{Q}_{h} {Q} \slashed S}&=&-\frac{1}{8}\Delta_{\text {em}}^{(a)}\alpha_{\text {em}}\left\langle \mathscr{Q}_{h}{\mathcal{H}}{Q}_{+}\sigma_{\mu\nu}\bar{\mathcal{H}}\sigma^{\mu\nu}\right\rangle, \label{eq:emm3}\\
	\mathcal{L}_{\mathrm{QED}}^{{Q} {Q}}&=&b_{\text {em}}\alpha_{\text {em}}\left\langle \mathcal{H}Q_{+}Q_{+}\bar{\mathcal{H}}\right\rangle +d_{\text {em}}\alpha_{\text {em}}\text{Tr}(Q_{+}Q_{+})\left\langle \mathcal{H}\bar{\mathcal{H}}\right\rangle, \label{eq:emm4}\\
	\mathcal{L}_{\mathrm{QED}}^{Q Q \slashed S}&=&-\frac{1}{8}\Delta_{\text {em}}^{(b)}\alpha_{\text {em}}\left\langle \mathcal{H}\sigma_{\mu\nu}Q_{+}Q_{+}\bar{\mathcal{H}}\sigma^{\mu\nu}\right\rangle -\frac{1}{8}\Delta_{\text {em}}^{(d)}\alpha_{\text {em}}\text{Tr}(Q_{+}Q_{+})\left\langle \mathcal{H}\sigma_{\mu\nu}\bar{\mathcal{H}}\sigma^{\mu\nu}\right\rangle, \label{eq:emm5}
\end{eqnarray}
where $\mathscr{Q}_{h}$ is the heavy quark electric charge and equal to $\frac{2}{3}$ and $-\frac{1}{3}$ for the charmed and bottom quarks, respectively. The $Q_+=\xi^{\dagger}\mathscr{Q}_l \xi+ \xi\mathscr{Q}_l\xi^{\dagger}$ with $\mathscr{Q}_l=\rm{diag}(2/3,-1/3,-1/3)$ is the light quark electric charge. The charge factor $e^2$ has been absorbed in $\alpha_{\text {em}}$. The above Lagrangians are suppressed by $\alpha_{\text {em}}$, which are counted as $\mathcal O(p^2)$ in Refs.~\cite{Meissner:1997fa,Meissner:1997ii}.   The Lagrangians with the heavy quark charge $\mathscr{Q}_{h}$ will contribute to the masses of the charmed and bottom hadrons due to their different electric charges. The  Lagrangians $\mathcal{L}_{\mathrm{QED}}^{\mathscr{Q}_{h} Q}$, $\mathcal{L}_{\mathrm{QED}}^{\mathscr{Q}_{h} {Q}\slashed{S}}$, the $b_{\text {em}}$ term in $\mathcal{L}_{\mathrm{QED}}^{{Q} {Q}}$, as well as the $\Delta_{\text {em}}^{(b)} $ term in $\mathcal{L}_{\mathrm{QED}}^{Q Q \slashed S}$ will contribute to the isospin breaking effect. The Lagrangians with the superscript $ \slashed S$ will break the HQSS and other ones without the $ \slashed S$, i.e., $\mathcal{L}_{\mathrm{QED}}^{\mathscr{Q}_{h} Q}$ and $ \mathcal{L}_{\mathrm{QED}}^{Q Q} $, conserve the HQSS. 

\begin{figure*}[htbp]
\centering
\includegraphics[width=0.3\textwidth]{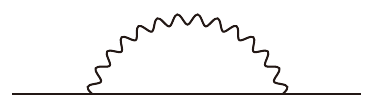}
\caption{The QED one-loop self-energy diagram at $\mathcal O(p^3)$. }
\label{fig:mass_QEDloop}
\end{figure*}

The Lagrangians in Eqs.~\eqref{eq:emm1}-\eqref{eq:emm5} will contribute to the  $d-u$ type hyperfine splitting at the tree level through the Fig.~\ref{fig:mass}(a)  at $\mathcal O(p^2)$ and up to $1/m_Q$.
The $\mathcal{L}_{\mathrm{QED}}^{\mathscr{Q}_{h} Q}$ and $ \mathcal{L}_{\mathrm{QED}}^{Q {Q}} $  contribute to the electromagnetic mass splittings at $\mathcal O(p^2)$, which are  of order $2-5$ MeV for the charmed mesons. The contributions from other Lagrangians $	\mathcal{L}_{\mathrm{QED}}^{\mathscr{Q}_{h} \mathscr{Q}_{h} \slashed S}$, $\mathcal{L}_{\mathrm{QED}}^{\mathscr{Q}_{h} {Q} \slashed S}$, and $\mathcal{L}_{\mathrm{QED}}^{Q Q \slashed S}$ will be further suppressed by the $1/ m_{Q}$. The electromagnetic correction may also appear at $\mathcal O(p^3)$ through the photon loop as illustrated in Fig.~\ref{fig:mass_QEDloop}, where the meson-photon interaction vertex  arises from the chiral connection $\Gamma_\mu$ [given in Eq.~\eqref{eq:connection}] in the chiral covariant derivative of the LO Lagrangians. This loop correction vanishes at $\mathcal O(p^3)$ in heavy hadron regularization and infrared regularization schemes~\cite{Guo:2008ns}. Up to $\mathcal O(p^3)$ and up to the order $1/m_Q$, the electromagnetic contributions to the isospin mass splitting arise from the Lagrangians in Eqs.~\eqref{eq:emm2}-\eqref{eq:emm5} at the tree level.

After the observation of the $D_{s0}^{*}(2317)$ and ${D_{s 1}(2460)}$, the HH$\chi$PT has been extended to discuss the mass corrections of the excited heavy mesons with spin-parity  $0^+$ and $1^+$~\cite{Jiang:2019hgs,Mehen:2005hc,Alhakami:2016zqx,Guo:2007up,Cheng:2014bca,Alhakami:2020vil,Becirevic:2004uv,Fajfer:2006hi,Fajfer:2016xkk,Ananthanarayan:2007sa}.
In these works, the two even-parity states $[{D_{s0}^{*}(2317)}, {D_{s 1}(2460)}]$ and their non-strange partner states $[{D_{0}^{*}(2300)},{D_{1}(2430)}]$ were treated as the spin doublets $\left(0^{+}, 1^{+}\right)$ with $j_{\ell}^{P}=\frac{1}{2}^{+}(L=1)$, which were described by the superfield $\mathcal S$.
As mentioned in Sec.\ref{sec1.1}, one mystery of the excited charmed mesons is their surprising mass degeneracy,
\begin{eqnarray}
&&m_{D_{s0}^{*}(2317)}-m_{D_{0}^{*}(2300)}=-25.2 \, \mathrm{MeV}, \\
&&m_{D_{s 1}(2460)}-m_{D_{1}(2430)}=47.5  \, \mathrm{MeV},
\label{eq:massdegeneracy}
\end{eqnarray}
while the typical splittings around $100$ MeV are naively expected due to the mass of the strange quark.

This mystery was first investigated in Ref.~\cite{Becirevic:2004uv}. The authors calculated the mass difference $(m_{D^{*}_0}-m_{D})-(m_{D_{s 0}^{*}}-m_{D_{s}})$ in the heavy quark limit. Only the Lagrangians preserving the heavy quark symmetry were considered and the power counting  was determined by the chiral expansion only.  The chiral corrections were calculated up to $\mathcal O(p^3)$. The LO Lagrangians in Eqs.~\eqref{eq:app1:lagD},~\eqref{eq:app1:lagS}-\eqref{eq:app1:lagST} and the $a_{H/S}$ as well as $\sigma_{H/S}$ terms in  the next-to-leading order $\mathcal{L}_{\text {chiral }}$ [in Eq.\eqref{eq:mm_chiral}] contribute up to $\mathcal O(p^2)$ at the tree level. The $ \mathcal O(p^3)$ corrections come from the one-loop diagram in Fig.~\ref{fig:mass}(b). The mass difference only depends on the SU(3) breaking counterterm $a_S-a_H$, which was determined by the lattice calculation. The result was obtained as $(m_{D^{*}_0}-m_{D})-(m_{D_{s 0}^{*}}-m_{D_{s}})\sim -100$ MeV, which is inconsistent with the experimental value $\sim 100$ MeV.

Later,  the authors in Ref.~\cite{Mehen:2005hc} included both the heavy quark symmetry breaking effect up to $1/m_Q$  and the SU(3) breaking effects with  the Lagrangians  $\mathcal L_{1/m_Q}$ in Eq.~\eqref{eq:Lag-mesonhm} and $\mathcal{L}^{(2)}_{\text {chiral }}$ in Eq.~\eqref{eq:mm_chiral}, respectively. They derived  the mass spectrum of  the $\mathcal H$ and $\mathcal S$ multiplets up to $\mathcal O(p^3)$ and to  $\mathcal O(1/m_Q)$ in the HH$\chi$PT. There are totally eleven LECs: three coupling constant $g_b, g_b^{\prime}, h$ and eight  combinations of the other LECs which appear in the  $\mathcal{L}^{(2)}_{\text {chiral }}$ [Eq.~\eqref{eq:mm_chiral}] and $\mathcal L_{1/m_Q}$ [Eq.~\eqref{eq:Lag-mesonhm}]. There are different combinations of the LECs as shown in Ref.~\cite{Mehen:2005hc} and  Refs.~\cite{Alhakami:2016zqx,Alhakami:2018jgb,Alhakami:2019ait}. Since the number of the LECs  exceeds that of the experimental data (eight charmed mesons),  there are multiple fitting solutions~\cite{Mehen:2005hc,Ananthanarayan:2007sa}. Therefore, it is impossible to draw firm conclusions. The two fits  in Ref.~\cite{Mehen:2005hc} either underestimate the excited non-strange meson or have unnaturally large LECs at $\mathcal O(1/m_Q)$. Thus the mass difference  $(m_{D^{*}_0}-m_{D})-(m_{D_{s 0}^{*}}-m_{D_{s}})$ still remains a big puzzle.

In Ref.~\cite{Guo:2007up}, Guo {\it et al} pointed out that the bare mass of the heavy-light scalar mesons obtained in the quark model was  pulled down significantly by the hadron loop with the HH$\chi$PT formalism.  The authors in Ref.~\cite{Cheng:2014bca} calculated the self-energy corrections  arising from the hadronic loop in Fig.~\ref{fig:mass}(b) for both the charmed-strange mesons and their nonstrange partners. 
They considered the LO Lagrangians in Eqs.~\eqref{eq:app1:lagD},~\eqref{eq:app1:lagS}-\eqref{eq:app1:lagST} and the $1/m_Q$ correction $\mathcal L_{1/m_Q}$ in Eq.~\eqref{eq:Lag-mesonhm}. However, their results showed that the mass degeneracy and the physical masses of $ D_{s 0}^{*}$ and $ D_{0}^{*}$ cannot be achieved simultaneously in the HH$\chi$PT formalism.

In Refs.~\cite{Alhakami:2016zqx,Alhakami:2019ait}, the authors used the similar framework as in Ref.~\cite{Mehen:2005hc} and calculated the expressions for the heavy meson masses up to $\mathcal O(p^3)$. Compared with Ref.~\cite{Mehen:2005hc}, the additional loop corrections with the intermediate spin-$3/2$ $\mathcal T$ mesons were considered in calculating the $\mathcal S$ multiplet masses due to their small mass splitting ($\lesssim 130$ MeV) and the considerable LO couplings.    With the masses of the odd-parity and even-party charmed mesons as inputs, they  predicted the near degeneracy of the nonstrange and strange scalar $B$ mesons. The authors also pointed out that the calculations in Refs.~\cite{Guo:2007up,Cheng:2014bca} were not complete since they only included the hadronic loops with the ground  mesons in the $\mathcal H$ doublet  but missed the contribution from the $\mathcal S$ doublets to the self-energy corrections of the scalar $D_{(s) 0}^{*}$. In addition, they criticized that the physical masses rather than the bare masses should be used in evaluating loop functions.
Later, the authors in Ref.~\cite{Cheng:2014bca} revisited their calculations in Ref.~\cite{Cheng:2017oqh}. Their results showed that the unsatisfactory  mass degeneracy in Ref.~\cite{Cheng:2014bca}  can be implemented  by adjusting  the $g^{\prime}_b$, $h$, and the renormalization scale $\Lambda$.  The mass  degeneracy of the $D_{s 0}^{*}$ and $  D_{0}^{*}$ can be quantitatively understood as a consequence of the self-energy corrections from the hadronic loop.  More technical differences in the approach of the HH$\chi$PT  among Refs.~\cite{Casalbuoni:1996pg,Mehen:2005hc,Guo:2007up,Boyd:1994pa,Cheng:2014bca,Cheng:2017oqh,Alhakami:2016zqx,Ananthanarayan:2007sa} were given in Refs.~\cite{Alhakami:2016zqx,Cheng:2017oqh}.

Another mystery of the $D_{s 0}^{*}(2317)$ and $ D_{s 1}^{*}(2460)$ is the fine-tuning problem
\begin{eqnarray}
\left(m_{D_{s 1}(2460)}-m_{D_{s0}^{*}(2317)}\right)-\left(m_{D^{*}}-m_{D}\right)\leq 2~\text{MeV},
\label{eq:finetuning}
\end{eqnarray}
which cannot be dictated by the QCD symmetries alone.  With the parity doubling model---an analog of the linear $\Sigma$-model for the heavy mesons, the equality was obtained at the tree level~\cite{Bardeen:2003kt,Bardeen:1993ae,Nowak:1992um,Nowak:2003ra}.  The parity doubling model embeds the chiral symmetry, its spontaneous breaking and heavy quark symmetry. The  heavy  spin multiplet $\left(0^{+}, 1^{+}\right)$ with $j_{\ell}=1 / 2$ is assumed to be the chiral partner of the ground doublet  $\left(0^{-}, 1^{-}\right)$. Their linear combinations $D\left(0^{+}, 1^{+}\right)+D\left(0^{-}, 1^{-}\right) $ and $D\left(0^{+}, 1^{+}\right)-D\left(0^{-}, 1^{-}\right)$ transform  as (approximately) pure $(1,3)$ and $(3,1)$, respectively, under SU(3)$_{L}$ $\otimes$ SU(3)$_{R}$ chiral symmetry. The $\Sigma$ transforms as $(\bar{3}, 3)$. With the spontaneously broken chiral symmetry, the vacuum expectation of the $\Sigma$ is nonzero leading to the similar effective Lagrangians to those in HH$\chi$PT. Compared with HH$\chi$PT, the parity doubling model predicts the same hyperfine splittings in the $\mathcal H$ and $\mathcal S$ doublets, for example, $m_{D^{*}}-m_{D}= m_{D^*_{1}}-m_{D_{0}^{*}}$, at the tree level, i.e., $\delta^{\prime}_b=\delta_b$.  In Ref.~\cite{Mehen:2005hc}, Mehen \textit{et al} proved that the equality will survive in the one-loop level with $|g_b|=|g'_b|$. It seems that this model explained the relation in Eq.~\eqref{eq:finetuning}.

In the section, all the discussions are performed assuming the  $D^*_{s0}(2317)/D^*_{s1}(2460)$ states as the $P$-wave quark-antiquark mesons. In the same picture, their analogous partner states in the bottom sector, especially the not yet observed heavy bottom mesons $B^*_{(s)0} $ and $B^{\prime}_{(s)1} $, were discussed with heavy quark flavor symmetry in many theoretical works, such as Refs.~\cite{Mehen:2005hc,Cheng:2017oqh,Cheng:2014bca,Bardeen:2003kt,Colangelo:2012xi,Colangelo:2005gb,Alhakami:2020vil}.  Since the two positive parity $D_s$ reside very close to the $DK/D^*K$ thresholds, other explanations, such as the  molecular states composed of $D^{(*)}$ and $K$ mesons, are also quite popular and will be discussed in Sec.~\ref{sec:sec4}.

\subsubsection{Singly heavy baryons}

The $\chi$PT was first employed to investigate the heavy baryons containing a heavy quark in  Refs.~\cite{Wise:1992hn,Yan:1992gz,Burdman:1992gh, Cheng:1993kp,Cho:1992gg, Cho:1992cf, Pirjol:1997nh}. The one-loop chiral corrections to their masses were given in Refs.~\cite{Savage:1995dw,Guo:2002tg,Guo:2008ns,Jiang:2014ena}.
In Ref.~\cite{Jiang:2014ena}, Jiang {\it et al} systematically calculated the chiral corrections to the heavy baryon masses as well as the axial charges up to $\mathcal O(p^3)$ using the HH$\chi$PT formalism. In this section, we follow their framework to illustrate the calculation of the heavy baryon masses.

The mass corrections at $\mathcal O (p)$ come from the Lagrangians in Eq.~\eqref{eq:app1:lagBc1} and Eq.~\eqref{eq:app1:lagBcc1} [or Eq.~\eqref{eq:app1:lagBc2} in the superfield formalism]. At  $\mathcal O (p^2)$, the mass splittings arise from the light quark mass difference as well as the electromagnetic effects  due to  different electric charges. The Lagrangians at  $\mathcal O (p^2)$ contain both the SU(3) flavor breaking Lagrangian $\mathcal{L}_{bc}^{(2)}$ and the QED Lagrangian $ \mathcal{L}_{\mathrm{QED}}^{(2)}$ as follows~\cite{Cheng:1993kp,Guo:2008ns,Jiang:2014ena,Savage:1995dw}
 \begin{eqnarray}
\mathcal{L}_{bc}^{(2)}&= & b_{1}\text{Tr}\left(\bar{B}_{\bm6}\chi_{+}B_{\bm6}\right)+b_{5}g_{\mu\nu}\text{Tr}\left(\bar{B}_{\bm6}^{*\mu}\chi_{+}B_{\bm6}^{*\nu}\right) +b_{6}\text{Tr}\left(\bar{B}_{\bar{\bm3}}\chi_{+}B_{\bar{\bm3}}\right)\nonumber \\
 &&+c_{1}\text{Tr}\left(\bar{B}_{\bm6}B_{\bm6}\right)\text{Tr}(\chi_{+}) +c_{5}g_{\mu\nu}\text{Tr}\left(\bar{B}_{\bm6}^{\mu}B_{\bm6}^{*\nu}\right)\text{Tr}(\chi_{+})+c_{6}\text{Tr}\left(\bar{B}_{\bar{\bm3}}B_{\bar{\bm3}}\right)\text{Tr}(\chi_{+}), \label{eq:lbc}\\
 \mathcal{L}_{\mathrm{QED}}^{(2)66}&= & e_{1}^{66}\text{Tr}\left(\bar{B}_{\bm6}Q_{+}^{2}B_{\bm6}\right)+e_{2}^{66}\text{Tr}\left(\bar{B}_{\bm6}Q_{+}B_{\bm6}\right)\text{Tr}(Q_{+}) +e_{3}^{66}\text{Tr}\left(\bar{B}_{\bm6}B_{\bm6}\right)\text{Tr}(Q_{+}^{2})+e_{4}^{66}\text{Tr}\left(\bar{B}_{\bm6}Q_{+}B_{\bm6}Q_{+}^{T}\right),\\
\mathcal{L}_{\mathrm{QED}}^{(2)\bar{3}\bar{3}}&= & e_{1}^{\bar{3}\bar{3}}\text{Tr}\left(\bar{B}_{\bar{\bm3}}Q_{+}^{2}B_{\bar{\bm3}}\right)+e_{2}^{\bar{3}\bar{3}}\text{Tr}\left(\bar{B}_{\bar{\bm3}}Q_{+}B_{\bar{\bm3}}\right)\text{Tr}(Q_{+}) +e_{3}^{\bar{3}\bar{3}}\text{Tr}\left(\bar{B}_{\bar{\bm3}}B_{\bar{\bm3}}\right)\text{Tr}(Q_{+}^{2}),\\
\mathcal{L}_{\mathrm{QED}}^{(2) 6^{*} 6^{*}}&=& e_{1}^{6^{*} 6^{*}} g_{\rho \sigma} \text{Tr}\left(\bar{B}_{\bm6}^{* \rho} Q_{+}^{2} B_{\bm6}^{* \sigma}\right) +e_{2}^{6^{*} 6^{*}} g_{\rho \sigma} \text{Tr}\left(\bar{B}_{\bm6}^{* \rho} Q_{+} B_{\bm6}^{* \sigma}\right) \text{Tr} (Q_{+}) ,\nonumber \\
&&+e_{3}^{6^{*} 6^{*}} g_{\rho \sigma} \text{Tr}\left(\bar{B}_{\bm6}^{* \rho} B_{\bm6}^{* \sigma}\right) \text{Tr} (Q_{+}^{2}) +e_{4}^{6^{*} 6^{*}} g_{\rho \sigma} \text{Tr}\left(\bar{B}_{\bm6}^{* \rho} Q_{+} B_{\bm6}^{* \sigma} Q_{+}^{T}\right).
\end{eqnarray}
Here, the $Q_+=\xi^{\dagger}\mathscr{Q} \xi+ \xi\mathscr{Q}\xi^{\dagger}$ with $\mathscr{Q}=\mathscr{Q}_{h}+\mathscr{Q}_{l}=e \operatorname{diag}(2,0,0)$ or $e \operatorname{diag}(1,-1,-1)$  the charge matrix of the charmed or bottom baryons. These Lagrangians  will contribute to the mass corrections at $\mathcal O(p^2)$ through diagram in Fig.~\ref{fig:mass}(a). One should note that the above Lagrangians are constructed in the heavy quark limit and no recoil effects are considered.  

The chiral one-loop corrections start to contribute at $\mathcal O(p^3)$ through the diagram in Fig.~\ref{fig:mass}(b). The tree-level QED contributions $\Sigma_{\mathrm{QED}}$ are at $\mathcal O(p^2)$. The one-loop QED corrections as shown in Fig.~\ref{fig:mass_QEDloop} stem from the  chiral connection term $\Gamma_\mu$ in the leading order Lagrangians and  vanish as in the heavy meson case~\cite{Guo:2008ns}. All the QED effects may be very small due to the double expansion in the chiral order and fine-structure constant. 

In Ref.~\cite{Jiang:2014ena}, the authors derived the chiral corrections up to $\mathcal O(p^3)$ with the contributions from the tree diagrams in Fig.~\ref{fig:mass}(a) and (b). They determined the values of the involved LECs with the experimental masses as input in five cases to investigate the role of the different corrections. The LEC values agreed with those in the study of the light-meson-heavy-baryon scattering lengths~\cite{Liu:2012uw,Liu:2012sw}. The results for the mass spectrum as well as the decay widths were consistent with the experimental data.

In Ref.~\cite{Guo:2008ns}, the authors calculated the isospin mass splitting  of the spin-$\frac{1}{2}$ heavy baryons in the  isospin multiplets $\Sigma_{c(b)} $ and $ \Xi_{c(b)}^{\prime}$. They calculated the corrections  to $\mathcal{O}(p^{3})$ with the chiral  perturbation theory using the infrared regularization. They focused on the isospin splitting and used the experimental splitting values as input. The results showed that the electromagnetic interaction played an important role in turning the mass order of the $\Sigma_{c}$ isotriplet into an unnatural pattern ({compared with the naive expectation that follows from $ m_d > m_u$}). For the $\Sigma_b$ states, the natural mass order is restored.

In Ref.~\cite{Savage:1995dw}, the authors studied the following equal spacing rule between the sextet heavy baryon with the LO Lagrangian in Eq.~\eqref{eq:app1:lagBc1} and the $\mathcal{L}_{bc}^{(2)}$ in Eq.~\eqref{eq:lbc},
 \begin{eqnarray} \label{eq:hbsr}
\frac{1}{3}\left(M_{\Sigma_{c}^{++}}+M_{\Sigma_{c}^{+}}+M_{\Sigma_{c}^{0}}\right)+M_{\Omega_{c}^{0}}-\left(M_{\Xi_{c }^{\prime+}}+M_{\Xi_{c }^{\prime 0}}\right)=0.
\end{eqnarray}
Up to $\mathcal O(p^2)$, the relation survives and the $\mathcal O(p^3)$ corrections from the loop in Fig.~\ref{fig:mass}(b) are small. They also studied the mass splitting between the spin-$\frac{3}{2}$ and spin-$\frac{1}{2}$ sextet heavy baryons. In this case, apart from the above Lagrangians, the $\delta_a$ term at $\mathcal O({1/m_Q})$ as shown in Eq.~\eqref{eq:app1:lagBc2} as well as its chiral corrections at $\mathcal{O}(p^2)$ with the insertion of one light quark mass matrix needs to be taken into account~\cite{Savage:1994ti,Savage:1994zw,Savage:1995dw}. Similar to  Eq.~\eqref{eq:hbsr}, the hyperfine splitting $\delta_{\Sigma_{c}}+\delta_{\Omega_{c}}-2 \delta_{\Xi_{c}}$  with $\delta=m_{B^*}-m_B$ equals zero up to $\mathcal O(p^2)$ and $\mathcal O(1/m_Q)$, and receives small one-loop corrections from  Fig.~\ref{fig:mass}(b) at $\mathcal O (p^3)$.

In Ref.~\cite{Guo:2002tg}, the authors calculated the masses of the  $\Sigma_{b}$,  $\Sigma_{b}^{*}$, $\Lambda_{b}$ as well as  the mass splitting between the $\Sigma_{b}$ and  $\Sigma_{b}^{*}$ by extrapolating the lattice QCD data at the unphysical pion mass to the physical region. The  Lagrangian in Eq.~\eqref{eq:app1:lagBc2} including the $1/m_Q$ correction and the $\mathcal{L}_{bc}^{(2)}$ in Eq.~\eqref{eq:lbc} were considered. Up to $\mathcal O(p^3)$, the tree diagram (a) and the one-loop diagram (b) in Fig.~\ref{fig:mass}  were calculated  and the phenomenological function for the extrapolation of lattice QCD was proposed. They found that the extrapolation was quite different with the naive
linear extrapolation when $m_\pi<500$ MeV.

 The parity doubling model mentioned in the heavy meson sector has also been  developed to study the heavy baryons (see Refs.~\cite{Harada:2019udr,Kim:2020imk,Kim:2021ywp,Suenaga:2021qri,Kawakami:2020sxd,Dmitrasinovic:2020wye,Suenaga:2021ikr,Chen:2008qv} and references therein). With the diquarks $qq$  as the building blocks, the chiral diquark effective field theory is built  based on the SU(3)$_R$ $\otimes$ SU(3)$_{L}$ chiral symmetry and its  spontaneously breaking induced by the nonvanishing  vacuum expectation value of the $\Sigma$ field. The $0^{+}({}^{1}S_{0})$ and  $0^{-}({}^{3}P_{0})$ diquarks are assumed to be the chiral partners and belong to the   $(\bar{3}, 1)$ and $ (1, \bar{3})$ representations of the SU(3)$_R$ $\otimes$ SU(3)$_{L}$  symmetry, respectively. Analogous to the heavy mesons, the  linear-$\Sigma$-model effective Lagrangian is constructed with the $\Sigma$ field belonging to $(\bar{3},3)$ representation. The $1^-$ diquark  and $1^+$ diquark form another chiral partner doublet.  With the chiral effective model, the mass spectra of the singly heavy baryons have been studied.

In addition to the HH$\chi$PT, the investigations of the heavy baryon mass spectrum with other effective field theory methods have been reviewed in Refs.~\cite{Chen:2016spr,Cheng:2021qpd}, such as the  Large $N_c$, the QCD sum rule  in the framework of HQET, and the non-linear chiral SU(3) Lagrangian, etc.

\subsubsection{Doubly heavy baryons}

\begin{figure*}[htbp]
\centering
\includegraphics[width=0.7\textwidth]{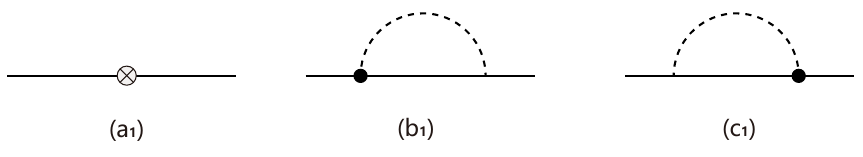}
\caption{The additional  Feynman diagrams  contributing to the mass spectrum of the heavy hadrons arising from the recoil effect in the HH$\chi$PT formalism. The solid and dashed lines represent  the
heavy hadrons and  the Goldstone bosons, respectively.  The solid dot and circled cross denote the vertices from the vertex at $ \mathcal O(p^2)$ and $\mathcal O(p^3)$, respectively. The two one-loop diagrams are at $\mathcal O(p^4)$.}\
\label{fig:hhmwr}
\end{figure*}

In the doubly heavy baryon sector,  the first attempt of calculating their masses up to ${\mathcal O}(p^4)$ in the baryon chiral perturbation theory was made in Ref.~\cite{ Sun:2014aya} with the HH$\chi$PT formalism.  Later, the masses of the  doubly charmed baryons was also calculated with the EOMS formalism up to $\mathcal{O}(p^{3})$~\cite{Sun:2016wzh} and up to ${\mathcal O}(p^4)$~\cite{Yao:2018ifh}. Recently, the authors of Ref.~\cite{Tong:2021raz} performed a similar calculation up to $ \mathcal{O}(p^{3})$ in HH$\chi$PT  with new lattice results. The leading order Lagrangian is given in Eq.~\eqref{eq:app1:lagBcc1}. The higher order Lagrangians  up to  ${\mathcal O}(p^4)$ after the non-relativistic projections read~\cite{Sun:2014aya,Yao:2018ifh,Qiu:2020omj} \begin{eqnarray}
\mathcal{L}^{(2)}&=&  c_{1} \bar{\mathcal{B}}_{QQ} {\cal B}_{QQ}\text{Tr}(\chi_{+}) +\frac{c_{2}}{2}\bar{\mathcal{B}}_{QQ} {\cal B}_{QQ} \text{Tr}(v \cdot u)^{2}+c_{3}\bar{\mathcal{B}}_{QQ}(v \cdot u)^{2}{\cal B}_{QQ}+\frac{c_{4}}{2}\bar{\mathcal{B}}_{QQ} {\cal B}_{QQ} \text{Tr} (u^{2}) \nonumber \label{eq:dhbl}\\
&& +\frac{c_{5}}{2}  \bar{\mathcal{B}}_{QQ} u^{2}{\cal B}_{QQ}+\frac{c_{6}}{2}\bar{\mathcal{B}}_{QQ} \left[S^{\mu}, S^{v}\right]\left[u_{\mu}, u_{v}\right]{\cal B}_{QQ}+c_{7} \bar{\mathcal{B}}_{QQ} \hat{\chi}_{+} {\cal B}_{QQ}\nonumber \\
&&+\frac{2}{m}\bar{\mathcal{B}}_{QQ} \left(S \cdot D\right)^{2} {\cal B}_{QQ}-\frac{i \tilde g_{1}}{ m}\bar{\mathcal{B}}_{QQ} \left\{S \cdot D, v \cdot u\right\} {\cal B}_{QQ}-\frac{\tilde g_{1}^{2}}{2 m}\bar{\mathcal{B}}_{QQ} (v \cdot u)^{2}   {\cal B}_{QQ}+\dots, \\
\mathcal{L}^{(3)}&=&h_{1}  \bar{\mathcal{B}}_{QQ} S\cdot u {\cal B}_{QQ}\text{Tr}(\chi_{+})+h_{2} \bar{\mathcal{B}}_{QQ} S^{\mu}\left\{\hat{\chi}_{+}, u_{\mu}\right\}  {\cal B}_{QQ}+h_{3} \bar{\mathcal{B}}_{QQ} S^{\mu} \text{Tr} (  \hat{\chi}_{+} u_{\mu}) {\cal B}_{QQ}\nonumber \\
&&-\frac{i}{m^{2}}  \bar{\mathcal{B}}_{QQ} S \cdot D v \cdot D S \cdot D{\cal B}_{QQ}+\dots,\\
  \mathcal{L}^{(4)}&= & e_1\bar{\mathcal{B}}_{QQ}  {\cal B}_{QQ}\text{Tr}(\chi_{+}) \text{Tr}(\chi_{+}) +e_{2} \bar{\mathcal{B}}_{QQ}  \hat{\chi}_{+} {\cal B}_{QQ}\text{Tr}(\chi_{+}) +e_{3}\bar{\mathcal{B}}_{QQ}  {\cal B}_{QQ}\text{Tr}(\hat{\chi}_{{+}} \hat{\chi}_{+})\nonumber\\
  &&+e_{4}\bar{\mathcal{B}}_{QQ} \hat{\chi}_{+} \hat{\chi}_{+} {\cal B}_{QQ} + e_{5}\bar{\mathcal{B}}_{QQ}  {\cal B}_{QQ}\text{Tr}(\chi_{-})\text{Tr}(\chi_{-})+ e_{6} \bar{\mathcal{B}}_{QQ} \hat{\chi}_{-}{\cal B}_{QQ}\text{Tr}(\chi_{-})\nonumber\\
  &&+e_{7}\bar{\mathcal{B}}_{QQ}  {\cal B}_{QQ}\text{Tr}(\hat{\chi}_{-} \hat{\chi}_{-})+ e_{8} \bar{\mathcal{B}}_{QQ}\hat{\chi}_{-}  \hat{\chi}_{-} {\cal B}_{QQ}+\dots.
 \end{eqnarray}
The relativistic forms are referred to Refs.~\cite{ Sun:2014aya,Yao:2018ifh,Qiu:2020omj,Yao:2018ifh}. The last three terms in $\mathcal{L}^{(2)}$ and the last one in $\mathcal{L}^{(3)}$  come from the  second  term in  Eq.~\eqref{eq:sec1.5:lightandheavy} which is suppressed at least by $1/m$ in the nonrelativistic projections, while the others are the non-relativistic forms of their corresponding relativistic operators. 
 Apart from the contributions in the heavy quark limit,  the recoil terms will contribute through the diagrams in Fig.~\ref{fig:hhmwr}.  The last term in $\mathcal{L}^{(3)}$ will lead to an additional tree-level diagram at $\mathcal O(p^3)$ as illustrated in Fig.~\ref{fig:hhmwr}$\mathrm{(a_1)}$. The $\tilde g_1/m$ term in $\mathcal{L}^{(2)}$ leads to two additional loop diagrams at $\mathcal O(p^4)$ as shown in Fig.~\ref{fig:hhmwr}$\mathrm{(b_1)}$ and $\mathrm{(c_1)}$. The physical masses can be obtained with Eq.~\eqref{eq:bm}. Up to $\mathcal O(p^3)$, there are four undetermined LECs involved: the bare mass $M_0$ in Eq.\eqref{eq:bm}, the $\tilde g_{1}$ in Eq.~\eqref{eq:app1:lagBcc1}, and the 
$c_{1}$, $c_{7}$ in Eq.~\eqref{eq:dhbl}. In Refs.~\cite{Sun:2014aya,Sun:2016wzh,Yao:2018ifh}, $\tilde g_{1}$ was related to the axial coupling of the singly heavy mesons in the HDAS as shown in Eqs.~\eqref{eq:hdas1},~\eqref{eq:hdas2} and~\eqref{eq:hdas3}.

In Ref.~\cite{Hu:2005gf}, a superfiled formalism of HDAS was proposed
   \begin{eqnarray}
 \mathcal{L}=\left\langle\mathscr{H}\left(iD_{0}\right)\mathscr{H}^{\dagger}\right\rangle-g\left\langle \mathscr{H}\bm\sigma\cdot\bm u\mathscr{H}^{\dagger}\right\rangle+\frac{\Delta_{H}}{4}\left\langle \mathscr{H}_{a}^{\dagger}\Sigma^{i}\mathscr{H}_{a}\sigma^{i}\right\rangle,
\label{eq:diquarkL}
    \end{eqnarray}
   where $\mathscr{H}$ is  a $5\times2$ matrix field formed by the  $2\times 2$ matrix field for the heavy baryon $H_{a}=\bm P_{a}^{*}\cdot\bm\sigma+P_{a}$  and the $3\times 2$ matrix field for the doubly heavy baryon  $T_{a,i\beta}=\sqrt{2}\left(\Xi_{a,i\beta}^{*}+\frac{1}{\sqrt{3}}\Xi_{a,\gamma}\sigma_{\gamma\beta}^{i}\right)$. The $H$ and $T$ fields are the nonrelativisitc forms of our $\mathcal{H}$ and $\psi_{QQ}^\mu$ superfields in Sec.~\ref{sec:1.5:combChandHQ}. The $g$ and $\Delta_H$ can be related to the $g_b$, $\tilde{g}_b$ and  $\delta_b$, $\tilde{\delta}_b$ in Sec.~\ref{sec:1.5:combChandHQ}. The $\Sigma^i$ is an extended Pauli matrix which is determined by assuming the violation  of heavy (di)quark spin symmetry only arises from the chromomagnetic interaction in Ref.~\cite{Hu:2005gf}.

   The LECs $c_1$, $c_7$ and $M_0$ are determined from the lattice QCD data~\cite{Alexandrou:2012xk,Na:2008hz,Lewis:2001iz,Liu:2009jc,Namekawa:2012mp,Bahtiyar:2020uuj,Brown:2014ena}. In Refs.~\cite{Sun:2014aya,Sun:2016wzh,Yao:2018ifh}, the authors used the same lattice QCD data for the $\Xi_{cc}$~\cite{Alexandrou:2012xk}, which were calculated with different $m_\pi$ and $m_c$.  They introduced  the $m_c$ dependence of the doubly heavy baryon masses through the bare mass $M_0$ in Eq.\eqref{eq:bm} as follows~\cite{Sun:2014aya,Sun:2016wzh,Yao:2018ifh}
   \begin{eqnarray}
M_{0}=\tilde{m}_{0}+2 m_{c}+\alpha / m_{c}+O\left(1 / m_{c}^{2}\right),
    \end{eqnarray}
where two new unknown parameters $\tilde{m}_{0}$ and $\alpha$ appear to replace $M_{0}$. $\tilde{m}_{0}$ is the mass of the light d.o.f in the doubly charmed baryons. $\alpha$ contains two kinds of $1/m_ c$ corrections stemming from the  Lagrangian terms at $\mathcal O (1/m_Q)$ in the HQET as shown in Eq.~\eqref{eq:LagHQET2}.

In  the HH$\chi $PT formalism, the contribution from the $c_1$ term was the same for all the doubly charmed baryons so that it can be absorbed into $M_0$. In Ref.~\cite{Sun:2014aya}, the authors used the lattice results  for the $\Xi_{cc}$~\cite{Alexandrou:2012xk} and  $\Omega_{cc}$~\cite{Alexandrou:2012xk,Na:2008hz,Lewis:2001iz,Liu:2009jc,Namekawa:2012mp} to determine the    LECs  $\tilde{m}_{0}$,  $\alpha$  and $c_7$  and obtained the following mass spectrum
  \begin{eqnarray}
m_{\Xi_{cc}}=3.665_{-0.097}^{+0.093}~\mathrm{GeV},\qquad m_{\Omega_{cc}}=3.726_{-0.097}^{+0.093}~\mathrm{GeV}.
     \end{eqnarray}
 In the EOMS formalism, the authors in Refs.~\cite{Sun:2016wzh,Yao:2018ifh}  also fitted four LECs  with the same lattice QCD data~\cite{Alexandrou:2012xk}.
In Ref.~\cite{Yao:2018ifh}, the authors also took the finite volume effect into consideration and obtained the masses
   \begin{eqnarray}
 m_{\Xi_{cc}}=3.591\pm0.067~\mathrm{GeV},\qquad m_{\Omega_{cc}}=3.657\pm0.100~\mathrm{GeV}.
         \end{eqnarray}
The authors of Ref.~\cite{Tong:2021raz} calculated  the masses of the doubly bottom  and the charmed-bottom baryons up to $\mathcal O(p^3)$ with the HH$\chi$PT formalism. The authors used the  quark model value for $\tilde{g}_{1}$ from Ref.~\cite{Li:2017cfz} and determined the LECs $M_{0}$,  $c_{7}$ by fitting the  lattice QCD data~\cite{Bahtiyar:2020uuj,Brown:2014ena} without considering the mass dependence on the  heavy quark.  The expressions of the mass corrections up to $\mathcal O(p^4) $ were also derived in Refs.~\cite{Sun:2014aya,Yao:2018ifh}. Unfortunately,  there are too many LECs involved and cannot be determined with the current lattice QCD and experimental data.

 The mass splittings of the doubly heavy baryons  have been studied with the help of the HDAS~\cite{Hu:2005gf,Mehen:2006vv,Brodsky:2011zs}. The hyperfine splittings $m_{\Xi^{*}_{cc}}-m_{\Xi_{cc}}-\frac{3}{4}\left(m_{P^{*}}-m_{P}\right)$ vanishes at the leading order, which is guaranteed  by the Lagrangian constructed in HDAS as illustrated in  Eq.~\eqref{eq:diquarkL}. In Ref.~\cite{Hu:2005gf}, the authors  calculated the  one-loop corrections from Fig.~\ref{fig:mass}(b) at $\mathcal O(p^3)$ and kept the nonanalytic parts. The deviation from the HDAS symmetry was small ($<10$ MeV) and changed slightly with different subtraction scale. In Ref.~\cite{Brodsky:2011zs}, the authors investigated isospin splittings of the doubly charmed baryons up to $\mathcal O(p^2)$ with the following Lagrangians,
   \begin{eqnarray}
\mathcal{L}_{\mathrm{ISV}}=&-c \left\langle\mathscr{H} \chi_{+}\mathscr{H}^{\dagger} \right\rangle -d_{0} \left\langle \hat{Q}_h \mathscr{H}Q_{l+} \mathscr{H}^{\dagger} \right\rangle -d_{1}\left\langle  \mathscr{H}\left(Q_{l+}^{2}-Q_{l-}^{2} \right)\mathscr{H}^{\dagger}  \right\rangle -d_{2}\left\langle  \mathscr{H}Q_{l+} \mathscr{H}^{\dagger}\right\rangle  \text {Tr}( Q_{l+}),
  \end{eqnarray}
  where $Q_{l\pm}$ is defined with the light quark electric matrix. $\hat{Q}_h$ is the heavy quark charge operator and its projections on the $H_a$ and $T_a$ fields are $\hat{Q}_hH_a=\mathscr{Q}_hH_a$ and $\hat{Q}_h T_{a}=-2 \mathscr{Q}_{h} T_{a}$.
   No loop corrections were considered. They predicted the isospin splittings for the  spin-$3/2$ and -$1/2$ doubly heavy baryons in the heavy quark spin symmetry
    \begin{eqnarray}
  M_{\Xi_{c c}^{++}}-M_{\Xi_{c c}^{+}}= 1.5 \pm 2.7~\text{MeV},\quad
  M_{\Xi_{b b}^{-}}-M_{\Xi_{b b}^{0}}=6.3 \pm 1.7,~\text{MeV},\quad
  M_{\Xi_{b c}^{+}}-M_{\Xi_{b c}^{0}}=-0.9 \pm 1.8~\text{MeV}.
     \end{eqnarray}
 In addition to the above formalisms, some other effective field theory, such as the HQET~\cite{Hwang:2008dj},  the parity doubling model~\cite{ Ma:2015lba,Ma:2015cfa} were also used to study the mass spectroscopy of the doubly heavy baryons.

\subsection{Axial vector currents, axial couplings and strong decays}\label{sec:2.3}

The matrix elements of the axial vector currents involving heavy hadrons are very important in flavor physics and in search for physics beyond the Standard Model. In the weak interaction, the axial charges of the heavy hadrons are associated with many  observables at low energies such as the weak transition form factors~\cite{Isgur:1989qw,Burdman:1992gh,Burdman:1993es,Falk:1993fr,Lee:1992ih,Randall:1993qg,Goity:1994xn,Boyd:1995pq,Ligeti:1997aq}, decay constants~\cite{Grinstein:1992qt,Neubert:1992fk,Boyd:1994pa} and so on.  Meanwhile, the coupling constants of the heavy hadrons and pseudoscalar Goldstone bosons (axial coupling for short) can be related to the axial charges through the Goldberger-Treiman relation, {which is a natural consequence of the spontaneous breaking of the chiral symmetry}. The axial couplings also play important roles in the strong interaction. The pionic strong decays of the heavy hadrons are related to the axial coupling constants. In addition, more and more exotic hadrons were observed as the candidates of the heavy hadronic molecules in the past decades, e.g. the $X(3872)$ as the $\bar{D}^*D/\bar{D}D^*$ molecule~\cite{Belle:2003nnu}, the $P_c$ states as the $\Sigma_c^{(*)}\bar{D}^{(*)}$ molecules~\cite{LHCb:2015yax,LHCb:2019kea}, and the $T_{cc}$ state as the $D^*D$ molecule~\cite{LHCb:2021auc}. In the molecular picture, the exotic hadrons are formed by exchanging the light mesons. Therefore, the precise determination of the axial coupling constants is crucial to understand these molecular structures. The investigations of the molecular hadrons within $\chi$EFT will be reviewed in Sec.~\ref{sec:chap5}. One can refer to~Refs.~\cite{Chen:2016qju,Guo:2017jvc,Brambilla:2019esw} for general reviews of hadronic molecules. In short, the matrix elements of the axial currents and axial coupling constants are essential for understanding both strong and weak interactions of the heavy hadrons.

\begin{figure}[hbtp]
	\begin{center}
		\includegraphics[width=0.6\textwidth]{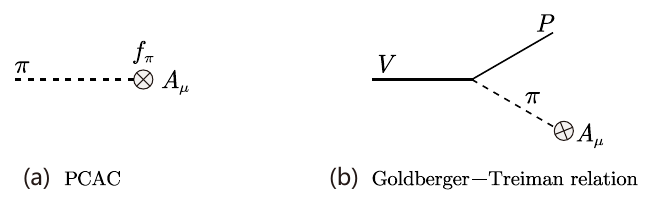}
		\caption{The partially conserved axial current and Goldberger-Treiman relations. The circled cross represents an insertion of the axial current.} \label{Fig:2.3:PCAC_GT}
	\end{center}
\end{figure}

We take the axial vector current $A_\mu$ between of the pseudoscalar $P$ and vector meson $V$ as an example to illustrate the Goldberger-Treiman relation~\cite{deDivitiis:1998kj}. In general, the axial current matrix can be parameterized as follows,
\begin{equation}
\langle P(p')|A^a_{\mu}|V(p)\rangle=\varepsilon_{\mu}F_{1}(q^{2})+2(\varepsilon\cdot q)k_{\mu}F_{2}(q^{2})+(\varepsilon\cdot q)q_{\mu}F_{3}(q^{2}),
\end{equation}
where $q=p'-p$ and $k=(p'+p)/2$. $\varepsilon$ is the polarization vector of $V$. $F_1$, $F_2$ and $F_3$ are the scalar functions. {From the Goldstone theorem in Sec.~\ref{sec:CSSB}, we know the axial current can annihilate the Goldstone bosons in Eq.~\eqref{eq:goldstone}, which is a natural consequence of spontaneous broken chiral symmetry. If we introduce the explicit breaking of chiral symmetry, we can relate the axial current to the pion field with the partially conserved axial current (PCAC) relation $\partial^{\mu}A_{\mu}^{a}=f_{\pi}m_{\pi}^{2}\pi^{a}$ as shown in Fig.~\ref{Fig:2.3:PCAC_GT}(a).} Thus, we can get
\begin{equation}
	\langle P(p')|\partial^{\mu}A_{\mu}^{a}|V(p)\rangle=\langle P(p')|\pi^{a}|V(p)\rangle f_{\pi}m_{\pi}^{2}+q^{\mu}N_{\mu}=(\varepsilon\cdot q)g_{VP\pi}(q^{2})\frac{1}{q^{2}-m_{\pi}^{2}}f_{\pi}m_{\pi}^{2}+q^{\mu}N_{\mu},~\label{eq:gt-derive}
\end{equation}
where the axial coupling constant $g_{VP\pi}$ is introduced to depict the $VP\pi$ vertex {and $N_\mu$ term represents the non-pole contribution. In the chiral perturbation theory, the non-pole contribution starts to appear from the $\mathcal{O}(p^3)$ Lagrangian~\cite{Goity:1999by}. Therefore, in the following derivation, we neglect the non-pole term as shown in Fig.~\ref{Fig:2.3:PCAC_GT}(b).}Thus, one can relate the axial coupling constant $g_{VP\pi}$ to the axial current matrix elements,
\begin{equation}
	g_{VP\pi}(q^{2})\frac{1}{q^{2}-m_{\pi}^{2}}f_{\pi}m_{\pi}^{2}=\left[F_{1}(q^{2})+2(q\cdot k)F_{2}(q^{2})+q^{2}F_{3}(q^{2})\right].
\end{equation}
When $q^2\to 0$, one can obtain the Goldberger-Treiman relation,
\begin{equation}
g_{VP\pi}(0)=-\frac{1}{f_{\pi}}F_{1}(0)\approx g_{VP\pi}(m_{\pi}^{2}),
\end{equation}
where the $F_1(0)$ is the axial charge. The PCAC and Goldberger-Treiman relations are illustrated in Fig.~\ref{Fig:2.3:PCAC_GT}. In lattice QCD, the axial coupling constants and strong decays of the ground heavy hadrons were extracted by calculating the matrix element of the light quark axial vector current~\cite{Detmold:2012ge,Becirevic:2009xp,Becirevic:2009yb,Abada:2003un,deDivitiis:1998kj,Becirevic:2012pf}.

Since the Goldstone bosons are very light (massless in chiral limit), the chiral corrections cannot be neglected in the calculation of the axial vector current at low energy (with small $q^2$) or the axial charges. The $\chi$PT was first used to calculate the chiral corrections of the axial vector current of the nucleon systems~\cite{Jenkins:1990jv,Jenkins:1991es,Zhu:2000zf,Zhu:2002tn,Hemmert:2003cb,Beane:2004rf,Smigielski:2007pe,Flores-Mendieta:2012fxp,Yao:2017fym,Sauerwein:2021jxb}. The similar formalism combining heavy quark symmetries has been extended to  calculate the axial currents of the heavy hadron systems. As for the strong decays of the ground heavy hadrons such as $D^*\to D \pi$, the small phase space for the decays ensures a valid chiral expansion. Therefore, we will review the chiral corrections to the axial vector current and strong decays within HH$\chi$PT. In order to avoid the confusions arising from different conventions in literature, we will construct the general Lagrangians contributing to the axial vector currents of the heavy baryons first. 

\subsubsection{Lagrangians and Feynman diagrams}~\label{subII:2.3:Lag_feyn}

\begin{figure}[hbtp]
	\begin{center}
		\includegraphics[width=0.9\textwidth]{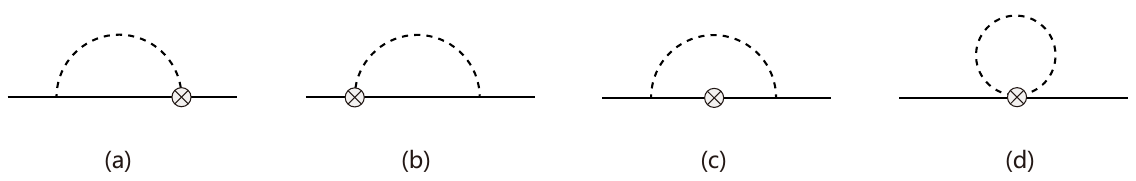}
		\caption{The topological loop diagrams contributing to the axial vector current. The circled cross represents an insertion of the axial vector current.} \label{Fig:2.3:loop_axial}
	\end{center}
\end{figure}

For the heavy hadrons, the axial coupling terms in the $\mathcal{O}(p)$ Lagrangians in Eqs.~\eqref{eq:app1:lagDbar},~\eqref{eq:app1:lagBc2} and~\eqref{eq:app1:lagBcc2} will contribute to the LO the axial vector currents of the heavy mesons, singly heavy baryons and doubly heavy baryons, respectively. The NLO [$\mathcal{O}(p^3)$] axial vector currents arise from both the tree diagrams and one-loop diagrams. The LO axial vector current multiplied by the wave function renormalization factor of the heavy hadrons will contribute to the $\mathcal{O}(p^3)$ axial vector current. In addition, the loop diagrams in Fig.~\ref{Fig:2.3:loop_axial} will contribute to the $\mathcal{O}(p^3)$ axial current if the vertices are all LO ones in Eqs.~\eqref{eq:lagGB}, ~\eqref{eq:app1:lagDbar},~\eqref{eq:app1:lagBc2} and~\eqref{eq:app1:lagBcc2}.  The chiral correction from the two seagull diagrams contain a factor $v\cdot q$, which is suppressed by $1/M_H$~\cite{Zhu:2000zf} ($M_H$ is the heavy hadron mass). In literature, this contribution was usually neglected if the recoiling effect was not included.

\begin{figure}[hbtp]
	\begin{center}
		\includegraphics[width=0.9\textwidth]{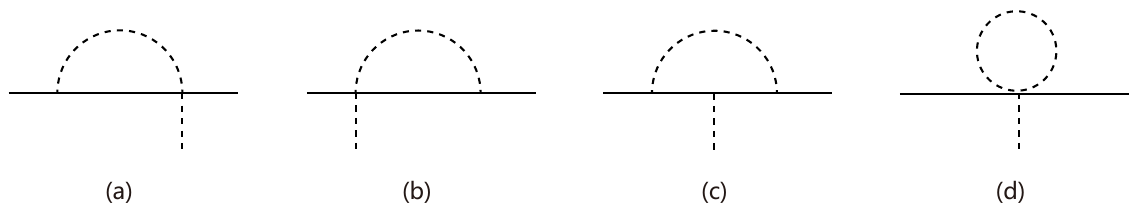}
		\caption{The topological loop diagrams contributing to strong decays. } \label{Fig:2.3:loop_strong_decay}
	\end{center}
\end{figure}

The $\mathcal{O}(p^3)$ Lagrangians will contribute to the $\mathcal{O}(p^3)$ axial vector currents at the tree level, which are constructed with the building blocks $\chi_+$ and $u_\mu$. With the knowledge of representations of the SU(3) group as shown in Table~\ref{tab:app1:g_irp}, we can construct the Lagrangians as follows,
\begin{eqnarray}
{\cal L}_{\tilde{{\cal H}}\varphi}^{(3)}&=&h_{1}\langle\bar{\tilde{{\cal H}}}\slashed{u}\gamma_{5}\tilde{{\cal H}}\rangle\text{\text{Tr}(\ensuremath{\chi_{+}})}+h_{2}\langle\bar{\tilde{{\cal H}}}\{\hat{\chi}_{+},\slashed{u}\}\gamma_{5}\tilde{{\cal H}}\rangle+h_{3}\langle\bar{\tilde{{\cal H}}}\gamma_{\mu}\gamma_{5}\tilde{{\cal H}}\rangle\text{Tr}(\hat{\chi}_{+}u^{\mu})+ih_{4}\langle\bar{\tilde{{\cal H}}}[\hat{\chi}_{+},\slashed{u}]\gamma_{5}\tilde{{\cal H}}\rangle,\\
{\cal L}_{\psi_{QQ} \varphi}^{(3)}&=&\tilde{h}_{1}\bar{\psi}_{QQ}^{\nu}\slashed{u}\gamma_{5}\psi_{QQ\nu}\text{\text{Tr}(\ensuremath{\chi_{+}})}+ \tilde{h}_{2}\bar{\psi}_{QQ}^{\nu}\{\hat{\chi}_{+},\slashed{u}\}\gamma_{5}\psi_{QQ\nu}+ \tilde{h}_{3}\bar{\psi}_{QQ}^{\nu}\gamma_{\mu}\gamma_{5}\psi_{QQ\nu}\text{Tr}(\hat{\chi}_{+}u^{\mu})\nonumber\\&&+ i\tilde{h}_{4}\bar{\psi}_{QQ}^{\nu}[\hat{\chi}_{+},\slashed{u}]\gamma_{5}\psi_{QQ\nu},\label{eq:nlo_psiQQ}\\
{\cal L}_{\psi_Q\varphi}^{(3)} &=& ih_{5}\epsilon_{\mu\nu\rho\sigma}\text{Tr}\left(\bar{\psi}_{Q}^{\mu}u^{\rho}v^{\sigma}\psi_{Q}^{\nu}\right)\text{Tr}(\ensuremath{\chi_{+}})+ ih_{6}\epsilon_{\mu\nu\rho\sigma}\text{Tr}\left(\bar{\psi}_{Q}^{\mu}\{u^{\rho},\hat{\chi}_{+}\}v^{\sigma}\psi_{Q}^{\nu}\right)+ h_{7}\epsilon_{\mu\nu\rho\sigma}\text{Tr}\left(\bar{\psi}_{Q}^{\mu}[u^{\rho},\hat{\chi}_{+}]v^{\sigma}\psi_{Q}^{\nu}\right) \nonumber\\
&&+ih_{8}\epsilon_{\mu\nu\rho\sigma}\text{Tr}\left(\bar{\psi}_{Q}^{\mu}v^{\sigma}\psi_{Q}^{\nu}\right)\text{Tr}(u^{\rho}\hat{\chi}_{+})+ ih_{9}\epsilon_{\mu\nu\rho\sigma}\text{Tr}\left(\bar{\psi}_{Q}^{\mu}v^{\sigma}u^{\rho}\psi_{Q}^{\nu}\hat{\chi}_{+}^{T}\right),\\
{\cal L}_{\mathcal{B}_{\bar{\bm 3}}\varphi}^{(3)}&=&h_{10}\text{Tr}\left(\bar{{\cal B}}_{\bar{\bm3}}\slashed{u}\gamma_{5}{\cal B}_{\bar{\bm3}}\right)\text{Tr}(\ensuremath{\chi_{+}})+ h_{11}\text{Tr}\left(\bar{{\cal B}}_{\bar{\bm3}}\{\hat{\chi}_{+},\slashed{u}\}\gamma_{5}{\cal B}_{\bar{\bm3}}\right)+h_{12}\text{Tr}\left(\bar{{\cal B}}_{\bar{\bm3}}\gamma^{\mu}\gamma_{5}{\cal B}_{\bar{\bm3}}\right)\text{Tr}(\hat{\chi}_{+}u_{\mu}),\label{eq:lagnlo33phi}\\
{\cal L}_{\psi_Q\mathcal{B}_{\bar{\bm 3}}\varphi}^{(3)} &=&h_{13}\mathrm{Tr}\left(\bar{\psi}_Q^{\mu}u_{\mu}\mathcal{B}_{\bar{\bm3}}\right)\text{Tr}(\ensuremath{\chi_{+}})+ h_{14}\mathrm{Tr}\left(\bar{\psi}_{Q}^{\mu}\{\ensuremath{\hat{\chi}_{+}},u_{\mu}\}\mathcal{B}_{\bar{\bm 3}}\right)+h_{15}\mathrm{Tr}\left(\bar{\psi}_{Q}^{\mu}[\ensuremath{\hat{\chi}_{+}},u_{\mu}]\mathcal{B}_{\bar{\bm 3}}\right)+\mathrm{H.c.}.~\label{eq:2.3:nlolag}
\end{eqnarray}
The $\hat{\chi}_+$ in the above Lagrangians will introduce the SU(3) flavor symmetry violation in the vertices. According to the representation reduction of SU(3) group $\bm8\otimes\bm8 \to \bm8_1(\bm8_2)$, {where $\bm8_1$ and $\bm8_2$ are {two different reduction ways of the octet representation} (see ~\ref{app:1} for details),} we introduce $\{\hat{\chi}_{+},u_{\mu}\}$ and $[\hat{\chi}_{+},u_{\mu}]$ as the building blocks. However, in the relativistic Lagrangian with two identical matter fields, e.g., Eq.~\eqref{eq:lagnlo33phi}, one of them will be eliminated by the constraint of charge conjugation. Note that the charge conjugation will introduce the transpose of the building blocks $\hat{\chi}_+$ and $u_\mu$ as shown in Table~\ref{tab:building}. The transpose will introduce an extra sign for $[\hat{\chi}_+,u^\mu]$ but not for $\{\hat{\chi}_+, u^\mu\}$. Therefore, only one of them will survive. However, for the Lagrangians with different matter fields, e.g., Eq.~\eqref{eq:2.3:nlolag}, both $\{\hat{\chi}_{+},u_{\mu}\}$ and $[\hat{\chi}_{+},u_{\mu}]$ terms will contribute. In the superfield formalism, we keep both of them in the Lagrangians to include the two different matter field case. However, one should be cautious about these terms for the processes with identical matter fields by checking the C-parity conservation. In the practical calculations, {if one does not care about the pion mass dependence}, the $h_1$, $\tilde{h}_1$, $h_5$, $h_{10}$ and $h_{14}$ terms can be absorbed by the LO Lagrangians, since the building block $\text{Tr}(\chi_+)$ only contributes a constant at the leading order.

The Lagrangians and Feynman diagrams which contribute to the strong decays are very similar to those of the axial vector currents. The leading order strong decays arise from the LO axial coupling terms in Eqs.~\eqref{eq:app1:lagD},~\eqref{eq:app1:lagBc2} and~\eqref{eq:app1:lagBcc2}. At  $\mathcal{O}(p^3)$, the tree level contribution to the strong decays stems from the same Lagrangians of the axial vector currents in Eq.~\eqref{eq:2.3:nlolag}. The LO results multiplied by the wave function renormalization factors will contribute to the  $\mathcal{O}(p^3)$  strong decay amplitudes. The wave function renormalization of the Goldstone bosons will also give rise to the $\mathcal{O}(p^3)$ corrections apart from those of the heavy hadrons. The remaining $\mathcal{O}(p^3)$ loop corrections arise from the diagrams in Fig.~\ref{Fig:2.3:loop_strong_decay}. Among them, the contributions of Figs.~\ref{Fig:2.3:loop_strong_decay}(a) and~\ref{Fig:2.3:loop_strong_decay}(b) are suppressed by $1/M_H$. The contribution of the diagram~\ref{Fig:2.3:loop_strong_decay}(d) will be canceled out by the renormalization of the Goldstone bosons~\cite{Cheng:1993kp}. Thus, if one neglects the recoiling effect, only diagram~\ref{Fig:2.3:loop_strong_decay}(c) and the wave function renormalization of the heavy hadrons contribute to the strong decays at the NLO.

\subsubsection{Axial couplings and strong decays of the ground state heavy hadrons}~\label{sec:2.3.1}

In Ref.~\cite{Yan:1992gz}, the authors constructed the chiral Lagrangians embedding the HQSS, which is in fact the LO Lagrangian of the HH$\chi$PT. The axial coupling constants were estimated by relating them to that of the nucleon within the quark model. With the nucleon axial charge $g_{A}=1.25$ as input, the coupling constants in Eqs.~\eqref{eq:app1:lagD} and~\eqref{eq:app1:lagBc1} were,
\begin{equation}
	|g_{b}|=0.75,\quad g_{1}=\frac{4}{3}\times0.75,\quad g_{2}=-\sqrt{\frac{2}{3}}\times0.75,
	\end{equation}
and the strong decay widths were estimated to be
\begin{eqnarray}
 \Gamma(D^{*-}\to D^{0}\pi^{-})=0.10\text{ MeV},\quad\Gamma(\Sigma_{c}^{0}\to\Lambda^{+}\pi^{-})=2.45\text{ MeV}.	
\end{eqnarray}

The above calculations were improved by including SU(3) symmetry breaking effect~\cite{Cheng:1993kp} and $1/M_H$ correction~\cite{Cheng:1993gc}, respectively. In Ref.~\cite{Cheng:1993kp}, the strong decays of the heavy mesons were calculated to the NLO of the chiral expansion including the SU(3) symmetry breaking effect. 	With the $|g_b|=0.75$, the strong decay widths of $D^{*+}$ read
\begin{equation}
\Gamma(D^{*+}\to D^{0}\pi^{+})=0.276\text{ MeV},\quad\Gamma(D^{*+}\to D^{+}\pi^{0})=0.125\text{ MeV}.
\end{equation}
In Ref.~\cite{Cheng:1993gc}, the authors discussed the $1/M_H$ effect by considering the reparametrization invariance of the heavy hadron fields and dynamical effect from the $1/m_Q$ terms in HQET.

In Ref.~\cite{Stewart:1998ke}, the author calculated the $D^*\to D\pi$ decay within HH$\chi$PT. The chiral expansion is truncated at NLO, which includes the tree and one-loop diagrams. In the analytical calculation, the SU(3) flavor and HQSS breaking effects were included.  Apart from the Lagrangians and Feynman diagrams in Sec.~\ref{subII:2.3:Lag_feyn}, the author also constructed the following $\mathcal{O}(p^2)$ terms,
\begin{equation}
\langle\bar{\tilde{{\cal H}}}(iv\cdot\nabla \slashed{u})\gamma_{5}\tilde{{\cal H}}\rangle,\quad \langle\bar{\tilde{{\cal H}}}(i\slashed{\nabla} v\cdot u)\gamma_{5}\tilde{{\cal H}}\rangle.
\end{equation}
The two terms give rise to the amplitudes proportional to $v\cdot k_\pi$, which are expected to be tiny, considering the small phase space of $D^*\to D\pi$.  In the numerical analysis, the contributions from these two terms were neglected.  In addition, the effect of the heavy quark spin and flavor breaking operators in Eq.~\eqref{eq:LagHQET2} were included by two operators at the hadron level,
\begin{eqnarray}
	\frac{1}{m_{Q}}\langle\bar{\tilde{{\cal H}}}\slashed{u}\gamma_{5}\tilde{{\cal H}}\rangle,\quad \frac{1}{m_{Q}}\langle\bar{\tilde{{\cal H}}}u^{\mu}\tilde{{\cal H}}\gamma_{\mu}\gamma_{5}\rangle.
\end{eqnarray}
The first operator only breaks the HQFS due to the different $m_Q$ for the charm and bottom quarks. Apart from the violation of the HQFS, the second term can break the HQSS by flipping the heavy quark spin. With the ratios of $\mathcal{B}(D^{*0}\rightarrow D^{0}\gamma)/\mathcal{B}(D^{*0}\rightarrow D^{0}\pi^{0})$, $\mathcal{B}(D^{*+}\rightarrow D^{+}\gamma)/\mathcal{B}(D^{*+}\rightarrow D^{+}\pi^{0})$ and $\mathcal{B}(D_{s}^{*}\rightarrow D_{s}\pi^{0})/\mathcal{B}(D_{s}^{*}\rightarrow D_{s}\gamma)$~\cite{CLEO:1997rew}, the author obtained two solutions, $g_b=0.27$ or $g_b=0.76$.

In Ref.~\cite{Guetta:1999vb}, the authors studied the decays $B^*\to B\gamma \gamma $ and $D^*\to D\gamma \gamma $ to determine the  $g_{B^*(D^*)B(D)\pi}$ and $g_{B^*(D^*)B(D)\gamma}$ with the LO chiral Lagrangians, In the processes, they considered Feynman diagrams including the $B^*(D^*)\to B(D) \gamma$, $B^*(D^*)\to B(D) \pi$, and $\pi^0\to 2\gamma$ vertices which was fixed by the chiral anomaly.

In Ref.~\cite{Detmold:2011rb}, the axial vector current matrix elements of the heavy mesons were calculated to NLO. In order to perform the chiral extrapolation for the partially-quenched lattice QCD simulations, the partially-quenched chiral perturbation theory was adopted for both SU$(4|2)$ and SU$(6|3)$ symmetries~\cite{Bernard:1993sv}. The finite volume effect was also incorporated. This calculation was used to extrapolate the lattice QCD simulations of the axial couplings of the $B^*B\pi$, $\Sigma_b^*\Sigma_b\pi$ and $\Sigma_b^{(*)}\Lambda_b\pi$~\cite{Detmold:2011bp,Detmold:2012ge}. The numerical results expressed in our conventions in Eqs.~\eqref{eq:app1:lagBc2} and~\eqref{eq:app1:lagD} were
\begin{equation}
	g_{b}=0.449\pm0.051,\quad g_{a}=0.84\pm0.20,\quad g_{3}=0.71\pm0.13.
\end{equation}

In Ref.~\cite{Jiang:2014ena}, the axial charges of the ground state heavy baryons were calculated up to NLO within HH$\chi$PT in the isospin symmetry limit but with explicit SU(3) breaking. In the calculations, the recoiling corrections were discarded. In the numerical analysis, the authors took the values of the axial charges in the chiral quark model~\cite{Li:2012bt} as the pseudo-experimental data. With the pseudo-experimental data as input, the authors determined the LECs and gave the numerical contributions order by order. The results showed that the convergence of the chiral expansion is quite good.

In Ref.~\cite{Hu:2005gf}, the authors constructed the LO chiral Lagrangians for the doubly charmed baryons with the HDAS. The axial coupling constant is related to those of the heavy mesons. Apart from the ground state heavy hadrons, the authors also constructed the chiral Lagrangians for the excited states. The quenched and partially quenched chiral Lagrangians were proposed in Ref.~\cite{Mehen:2006vv} to suit the chiral extrapolation in lattice QCD simulations. Later, Mehen constructed the chiral Lagrangians for the excited doubly heavy baryons and doubly heavy tetarquarks~\cite{Mehen:2017nrh}.

In Ref.~\cite{Sun:2016wzh}, the chiral corrections to the axial vector currents and axial charges of the doubly heavy baryons were investigated to the NLO ($\mathcal{O}(p^3)$) of the chiral expansion. The LO axial coupling constant is determined by the HDAS, i.e.,  $\tilde{g}_b=g_b$ in Eqs.~\eqref{eq:app1:lagD} and ~\eqref{eq:app1:lagBcc2}. At the NLO, the loop diagrams and corrections are similar to the discussions in Sec.~\ref{subII:2.3:Lag_feyn}. At the NLO, the authors also constructed the Lagrangians equivalent to the $\tilde{h}_{1,2,3}$ terms in Eq.~\eqref{eq:nlo_psiQQ} to consider the tree-level diagram contributions. If one only considers the spin-$1\over 2$ doubly charmed baryons, the $\tilde{h}_4$ term vanishes due to the charge conjugation symmetry. In the numerical analysis, the $\tilde{h}_1$ term can be absorbed by the LO axial coupling constant. The remaining LECs $\tilde{h}_{2}$ and $\tilde{h}_3$ were estimated by the naturalness assumption.

\subsubsection{$D_s^*\to D_s \pi^0$ isospin violating decays}

In the above discussions of the strong decays, the SU(3) breaking effect is kept but the isospin violating effect is neglected, because it is tiny in general. However, for the specific process $D_s^*\to D_s \pi^0$, the isospin violating effect is significant.

The mass difference between the $D_s^*$ and $D_s$ is slightly larger than the neutral pion mass by about 2 MeV, which makes the $D_s^*\to D_s \pi^0$ and $D_s^*\to D_s \gamma$ the dominant decay modes of the $D_s^*$. The strong decay process $D_s^*\to D_s \pi^0$ violates the isospin symmetry. The double suppressions from the phase space and the
isospin violation make this strong decay width very tiny, around several eVs. The branching fraction of this strong decay
mode is $(5.8 \pm 0.7)$\%. For comparison, the branching fraction of the electromagnetic decay mode $D_s^* \to D_s\gamma$ is about $(93.5\pm0.7)$\%~\cite{CLEO:1995ewe,BaBar:2005wmf,ParticleDataGroup:2022pth}. The magnitude of the strong decay mode challenges our physical intuitions about the strong decays and deserves a refined investigation.

\begin{figure}[hbtp]
	\begin{center}
		\includegraphics[width=0.6\textwidth]{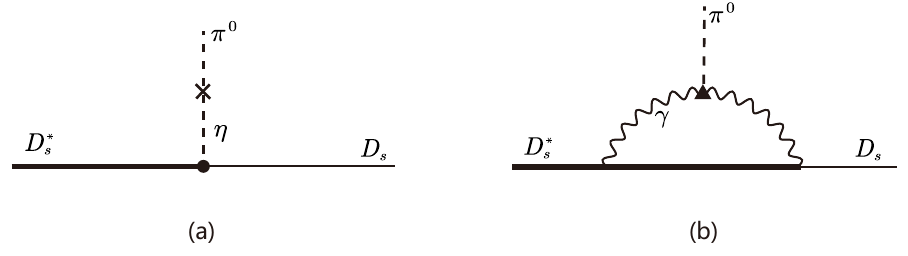}
		\caption{The Feynman diagrams of $D_s^*\to D_s\pi^0$. LO contribution (a) and axial current anomaly contribution (b). } \label{Fig:2.3:isoV_decay}
	\end{center}
\end{figure}

In Ref.~\cite{Cho:1994zu}, the author calculated the $D_s^*\to D_s\pi^0$ decay rate using HH$\chi$PT. The isospin violating effect is attributed to the $\pi^0-\eta$ mixing as shown in Fig.~\ref{Fig:2.3:isoV_decay}. The $D_s^*D_s \eta$ vertex stems from the LO axial coupling term in Eq.~\eqref{eq:app1:lagD}. The $\pi^0-\eta$ mixing effect arises from the mass term in Eq.~\eqref{eq:lagGB}, which reads,
\begin{equation}
	\mathcal{L}_\text{mixing}=-{B_0 \over \sqrt{3}}(m_u-m_d)\eta \pi^0.
\end{equation}
One can see the mass difference of the up and down quarks drives the $\pi^0-\eta$ mixing and leads to the isospin violating decays. Another possible origin of the isospin violation is the electromagnetic interaction, which corresponds to the graph (b) in Fig.~\ref{Fig:2.3:isoV_decay}. The $\pi^0\gamma\gamma$ vertex is induced by the axial vector current anomaly in QED. However, the diagram is suppressed by $\alpha_\text{em}^2$ and can be neglected safely. In Ref.~\cite{Cho:1994zu}, the author got the ratio
\begin{equation}\label{eq:ratiopig}
{\Gamma(D_s^*\to D_s\pi^0) \over \Gamma(D^{*+}\to D^+\gamma)} \approx 8\times 10^{-5}.
\end{equation}
A similar calculation considering the LO contribution was performed in Ref.~\cite{Terasaki:2015eao} as well. The result reads
\begin{equation}
    	\mathcal{R}_{\pi/\gamma}={\Gamma(D_s^*\to D_s\pi^0) \over \Gamma(D_s^{*}\to D_s\gamma)} \approx 1.8 \%.~\label{eq:ratio}
\end{equation}

The calculation of the $D_s^*\to D_s \pi^0$ decay was performed to the NLO in Ref.~\cite{Stewart:1998ke}. The NLO loop diagrams include graphs which allow the direct emission of the $\pi^0$ in Fig.~\ref{Fig:2.3:loop_strong_decay}. If the mass splittings of the intermediate isospin multiplets are neglected, the amplitudes will cancel with each other. In the calculation, the mass splittings of the isospin doublet or triplet in the loops are kept, which give rise to the nonvanishing amplitude. In addition, the loop corrections to the $\pi^0-\eta$ mixing effect contribute to the decay width as well, which correspond to diagrams replacing the external Goldstone lines in Fig.~\ref{Fig:2.3:loop_strong_decay} with the $\pi^0-\eta$ mixing.

In Ref.~\cite{Yang:2019cat}, the authors calculated the $D_s^*\to D_s \pi^0$ decay to the NLO. Apart from all the diagrams in Ref.~\cite{Stewart:1998ke}, the authors also constructed the NLO Lagrangians,
\begin{equation}
\langle\bar{\tilde{\mathcal{H}}}\slashed{u}\hat{\chi}_{+}\gamma_{5}\tilde{\mathcal{H}}\rangle,\;\langle\bar{\tilde{\mathcal{H}}}\slashed{u}\gamma_{5}\tilde{\mathcal{H}}\rangle\mathrm{Tr}\left[\chi_{+}\right],\; i\langle\bar{\tilde{\mathcal{H}}}\slashed{\nabla}\hat{\chi}_{-}\gamma_{5}\tilde{\mathcal{H}}\rangle,\;i\langle\bar{\tilde{\mathcal{H}}}\gamma^{\mu}\gamma_{5}\tilde{\mathcal{H}}\rangle\nabla_{\mu}\mathrm{Tr}\left[\chi_{-}\right],\;\langle\bar{\tilde{\mathcal{H}}}\nabla_{\nu}\slashed{\nabla}u^{\nu}\gamma_{5}\tilde{\mathcal{H}}\rangle.
\end{equation}
In the numerical analysis, two strategies were adopted to estimate the unknown LECs. In the strategy A, the nonanalytic dominant approximation is made. Thus, the NLO result only contains the nonanalytic loop corrections and no contributions from the unknown LECs, which yielded,
\begin{equation}\label{eq:wDsDspi}
	\Gamma(D_s^*\to D_s\pi^0)=8.1_{-2.6}^{+3.0} \text{ eV},\quad \Gamma[D_s^*]=139.0_{-54.6}^{+77.9}\text{ eV}.
\end{equation}
 {
Combing the radiative decay $ \Gamma(D_s^{*}\to D_s\gamma)=0.32^{+0.3}_{-0.3}$ keV in Ref.~\cite{Wang:2019mhm}, one can get the ratio defined in Eq.~\eqref{eq:ratiopig} as
\begin{equation}
   \mathcal{R}_{\pi/\gamma} \approx 2.5\%.\label{eq:ratio2}
\end{equation}
Here we give the value only using the central value and do not give the rigorous uncertainties because the uncertainties of the radiative decay width of $D_s^*$ are too large to use the conventional error propagating formula.} In the strategy B, the authors made use of the naturalness assumption. The $\Gamma(D_s^*\to D_s\pi^0)$ was estimated in the range of 5-12 eV. {It is worth noticing that the recent experimental value from BESIII of the ratio is $ \mathcal{R}_{\pi/\gamma}=(6.16 \pm 0.40 \pm 0.17)\%$~\cite{BESIII:2022kbd}. Apparently, the theoretical results in Eq.~\eqref{eq:ratio} and \eqref{eq:ratio2} are of the same order as the experimental value. We cannot give more information without the rigorous uncertainties in theoretical calculation. }

In Ref.~\cite{Cheung:2015rya}, the chiral Lagrangian for the strong and radiative decays of the strange heavy mesons was formulated based on HQET, in which the $1/m_Q$ correction and SU(3) symmetry breaking effect were included. The $\pi^0-\eta$ mixing vertex was estimated with the new data of $\Gamma(\eta\to 3\pi^0)$~\cite{ParticleDataGroup:2022pth}.

\subsubsection{Strong decays of the excited heavy hadrons}

In principle, the $\chi$PT cannot be used to investigate the excited heavy hadrons consistently.
The mass splittings between the excited and ground heavy mesons do not vanish in the chiral limit, which are numerically much larger than the pion mass. 
 For example $m_{D_0^*}-m_D\approx 470 \text{ MeV}\gg m_\pi$. We cannot expect good convergence in such a high energy scale. Meanwhile, the excited heavy hadrons appear as the resonances in the scattering of the ground state heavy hadrons and pions, which implies the nonperturbative dynamics. However, from the phenomenological perspective, one can still construct the chiral Lagrangians to depict the chiral dynamics of the excited heavy hadrons, discarding the rigorous power counting. For example, in Refs.~\cite{Cheng:2006dk,Cheng:2015naa}, the authors exploited the chiral Lagrangians to discuss the strong decays of the $S$-, $P$-, and higher wave singly charmed baryons. In Refs.~\cite{Gupta:2018zlg,Gandhi:2019lta}, the authors investigated the strong decays of the excited charmed mesons with the chiral Lagrangian and heavy quark expansion. Among the excited heavy hadrons, the nature of the $\DsI$ and $\DsII$ remains unsettled decades after their discovery. Whether they are  molecule states or conventional $P$-wave heavy mesons continues to be a heated topic.

In Ref.~\cite{Mehen:2004uj}, the authors constructed the chiral Lagrangians in Eqs.~\eqref{eq:app1:lagD},~\eqref{eq:app1:lagS} and~\eqref{eq:app1:lagSH}. The coupling constant $h$ is extracted from the experimental decay widths of the $D_1(1^+)$ and $D_0(0^+)$~\cite{CLEO:1999cmp,Belle:2003nsh,FOCUS:2003gru}, $h^2=0.44\pm 0.11$. With the coupling constants, the authors calculated the isospin violating decays of the $\DsI$ and $\DsII$ with the tree level $\pi^0-\eta$ mixing effect. The results showed that the relevant branching ratios $\text{Br}[D_{s1}(2460)\rightarrow D_{s}^{*}\gamma]/\text{Br}[D_{s1}(2460)\rightarrow D_{s}^{*}\pi^{0}]$ and $\text{Br}\left[D_{s0}(2317)\rightarrow D_{s}^{*}\gamma\right]/\text{Br}\left[D_{s0}(2317)\rightarrow D_{s}\pi^{0}\right]$ deviate significantly from the experimental results~\cite{CLEO:2003ggt}. After considering the HQSS violating operators (with the LECs determined by the naturalness) and one-loop chiral corrections to the electromagnetic decays, the theoretical results became consistent with the experimental data. In Ref. ~\cite{Mehen:2005hc}, the authors calculated the loop corrections to the coupling constants $g_b$ and $g'_b$ and derived their one-loop renormalization group equation as well. This work also investigated the decay patterns in the molecular scheme, see Sec.~\ref{Sect.2.4.1} for details.

\begin{figure}[hbtp]
	\begin{center}
		\includegraphics[width=0.9\textwidth]{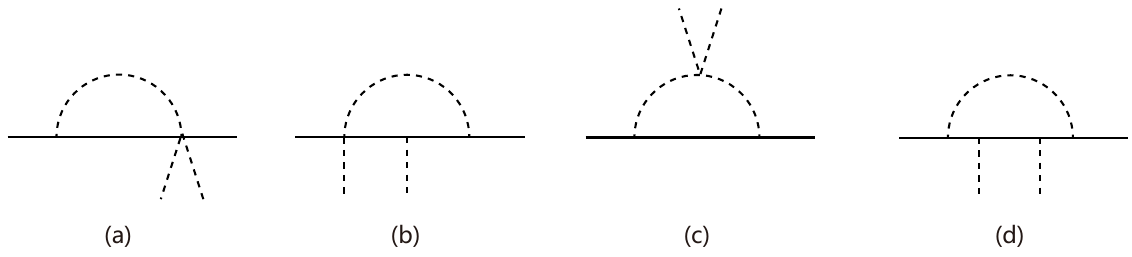}
		\caption{The one-loop diagrams for the $\DsII\to D_s\pi\pi$ decay.} \label{Fig:2.3:loop_isoV2}
	\end{center}
\end{figure}

In Ref.~\cite{Fajfer:2006hi}, the authors considered the one-loop correction to the $g_b$, $g_b^\prime$ and $h$ in Eqs.~\eqref{eq:app1:lagD},~\eqref{eq:app1:lagS} and~\eqref{eq:app1:lagSH} and included the positive parity mesons in the loop diagrams. Although the author also constructed the counter terms, their contributions were neglected due to lacking of input in the numerical analysis. Thus, the results are regularization-scale-dependent. With the regularization scale $\mu\simeq 1$ GeV, they got
\begin{eqnarray}
	\text{LO:}\; &&g_b=0.61, \qquad h=0.52,\qquad g'_b=-0.15,\\
	\text{One-loop:}\;&& g_b=0.66, \qquad h=0.47,\qquad g'_b=-0.06.
\end{eqnarray}
In Ref.~\cite{Fajfer:2015zma}, the isospin violating decays $\DsII\to D_s\pi\pi$, $\DsII\to D_s^*\pi$ and $\DsI\to D_s\pi$ were investigated. The $g_b$, $g_b^\prime$ and $h$ terms in Eqs.~\eqref{eq:app1:lagD},~\eqref{eq:app1:lagS} and~\eqref{eq:app1:lagSH} contribute to the $\DsII\to D_s^*\pi$ and $\DsI\to D_s\pi$ tree diagrams via the $\pi^0-\eta$ mixing effect. The one-loop diagrams for the two decays were similar to those for $D_s^*\to D_s\pi^0$. There is no tree-level contribution for the $\DsII\to D_s\pi\pi$ decays. The related one-loop diagrams were presented in Fig.~\ref{Fig:2.3:loop_isoV2}. In the calculation, only the diagrams (a), (b) and (c) were taken into consideration. The counter terms were introduced to absorb the divergence of the loop diagrams and estimated with the experimental ratios of the decay widths roughly. The numerical results read
\begin{eqnarray}
 &	\Gamma(\DsII^+\to D_s^+\pi^+\pi^-) \simeq 0.25 \text{ keV},\qquad\Gamma(\DsII^+\to D_s^+\pi^0\pi^0) \simeq 0.15 \text{ keV},\\
 & \Gamma(\DsII^+\to D_s^{*+}\pi^0)=2.7\sim 3.4\text{ keV},\quad\Gamma(\DsI^+\to D_s^+\pi^0)=2.4\sim 4.7\text{ keV}.
\end{eqnarray}
{The above results are significantly smaller than the hadronic widths predicted in the molecular pictures, see Refs.~\cite{Lutz:2007sk,Guo:2008gp} for examples.}The same framework was used to investigate the isospin violating decays of the positive parity $B_s$ mesons~\cite{Fajfer:2016xkk}. The numerical results read
\begin{equation}
\Gamma\left(B_{s 1}^{0} \rightarrow B_{s}^{0} \pi \pi\right) \sim 10^{-3} \mathrm{keV},\quad\Gamma\left(B_{s 0}^{* 0} \rightarrow B_{s}^{0} \pi^{0}\right) \leq 55 \mathrm{keV}, \quad\Gamma\left(B_{s 1}^{0} \rightarrow B_{s}^{* 0} \pi^{0}\right) \leq 50 \mathrm{ keV}.
\end{equation}

\subsection{Magnetic moments and radiative transitions of the heavy hadrons}\label{Sect:radiative_transitions}

The mass, charge, spin and magnetic moment are intrinsic characteristics of a particle. The magnetic moment $\boldsymbol{\mu}$ of a particle (e.g., the electron) is related to its charge $e$, mass $m$ and spin $\boldsymbol{S}$ via
\begin{eqnarray}\label{eq:mm}
\boldsymbol{\mu}=g\frac{e}{2m}\boldsymbol{S},
\end{eqnarray}
where $g$ is the Land\'e $g$-factor, which usually indicates whether a particle is elementary or composite. A renowned example is the neutron's magnetic moment, which is expected to vanish if it is an elementary particle due to its electro-neutral property [see Eq.~\eqref{eq:mm}]. But the experimental measurement told us that the neutron carries a non-zero, large and negative magnetic moment ~\cite{Frisch:1933,Esterman:1933,Breit:1934},
which strongly indicated that the neutron is not an elementary particle. Its inner structure was understood until the quark model was developed in 1960s. Now we know that the neutron is composed of three constituent quarks $udd$. The sum of their magnetic moments leads to the neutron magnetic moment.


The magnetic moment of a hadron is related to its magnetic form factor $\mathscr{G}_M(q^2)$ at $q^2=0$. Generally, the electromagnetic properties of a hadron are encoded in the following matrix element,
\begin{eqnarray}\label{eq:emcurrent}
\left\langle \mathscr{H}(p^\prime)\left|\mathscr{J}_{\text{em}}^\mu(q^2)\right|\mathscr{H}(p)\right\rangle,
\end{eqnarray}
where $\mathscr{J}_{\text{em}}^\mu$ denotes the electromagnetic current and $\mathscr{H}$ is the hadron state. The transferred momentum is $q=p-p^\prime$, with $p$ and $p^\prime$ the 4-momenta of the initial and final states, respectively. The covariant expansion of Eq.~\eqref{eq:emcurrent} depends on the spins of the specified hadrons and is illustrated in~\ref{app:EM_FF}.

The proton and neutron are stable against the strong and electromagnetic decays. Their long lifetimes allow the direct measurements of their magnetic moments and charge radius ~\cite{Gao:2003ag,Arrington:2006zm,Perdrisat:2006hj,Pacetti:2014jai,Carlson:2015jba,Chupp:2017rkp,Bezginov:2019mdi,Xiong:2019umf}. But for the other hadron states, it is generally hard to measure their magnetic moments directly in experiments due to their very short lifetimes. For example, the magnetic moment of the lightest vector meson $\rho$ was extracted from the $e^+e^-\to \pi^+\pi^-\pi^0\pi^0$ process via an indirect way~\cite{GarciaGudino:2013alv}, in which the vector meson dominance model was adopted to describe the $\gamma^\ast\to 4\pi$ vertex. The charge radius of the pion was extracted from the analysis of the $e^+e^-\to\pi^+\pi^-$ data with the help of the dispersion relations~\cite{Colangelo:2018mtw}. In contrast to the magnetic moment, the radiative decays of the heavy hadrons are more accessible in experiments. For instance, the branching fraction of $D^{\ast0}\to D^0\gamma$ can reach up to $(35.3\pm0.9)\%$. For the $D_s^\ast$, $B^\ast$ and $B_s^\ast$ mesons, the radiative decays even dominate their decay modes.

For a lepton (e.g., electron and muon), the QED quantum fluctuations can introduce sizable and detectable corrections to its magnetic moment. If a hadron contains the light quarks $u$, $d$, $s$ as its components, it will naturally couple to the light Goldstone bosons (such as $\pi$, $K$ and $\eta$) due to the spontaneous breaking of chiral symmetry in low energy QCD. Apart from the QED, the chiral corrections (fluctuations of the light Goldstone bosons) will make considerable contributions to the magnetic moments.

\begin{figure}[!hbtp]
	\begin{center}
		\includegraphics[width=\linewidth]{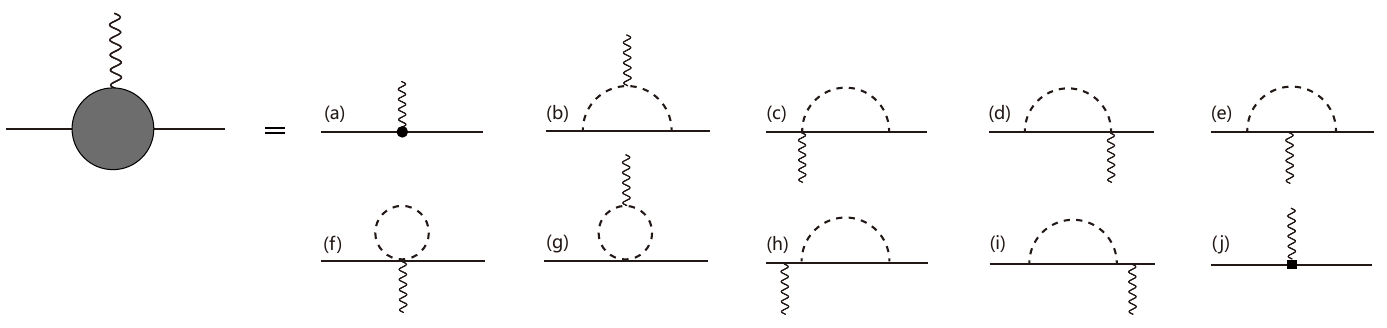}
		\caption{The tree level and one-loop level diagrams for the magnetic moments and radiative transitions of the heavy hadrons. Each one represents a set of diagrams that have the same topological structure. The solid, dashed and wiggly lines denote the heavy hadrons, light Goldstones and photon, respectively. The solid dot and solid square stand for the $\mathcal{O}(p^2)$ and $\mathcal{O}(p^4)$ vertices, respectively.} \label{Fig:2.4.RadiativeDecays}
	\end{center}
\end{figure}

In this section, we focus on the chiral corrections to the low energy electromagnetic properties (e.g. magnetic moments and electromagnetic decays) of the heavy hadrons within the HH$\chi$PT. The tree level and one-loop level Feynman diagrams are given in Fig.~\ref{Fig:2.4.RadiativeDecays}. The chiral order $\mathcal{O}$ of the diagrams in Fig.~\ref{Fig:2.4.RadiativeDecays} is organized by the Weinberg power counting scheme in Eq.~\eqref{eq:1.4:pwc}.
The order of the (transition) magnetic moment is usually defined as
\begin{eqnarray}
	\mathcal{O}_\mu=\mathcal{O}-1.
\end{eqnarray}

\subsubsection{Heavy mesons}\label{Sect.2.4.1}

The magnetic moments of the charmed and bottom vector mesons were first calculated by Bose and Singh with the Bag model in 1980~\cite{Bose:1980vy}. Later, the extended Nambu-Jona-Lasinio model and Bag model were used to study the magnetic moments and radiative transitions of the ground-state vector mesons by Deng {\it et al}~\cite{Deng:2013uca,Luan:2015goa} and \v{S}imonis~\cite{Simonis:2016pnh,Simonis:2018rld}, respectively. A systematic study in the framework of $\chi$PT up to $\mathcal{O}(p^4)$ was performed by Wang {\it et al} in 2019~\cite{Wang:2019mhm}. Very recently, Aliev {\it et al} also investigated the same entities with the light-cone QCD sum rules~\cite{Aliev:2019lsd}. 

The radiative decays of the heavy vector meson $V\to P\gamma$ ($V$ and $P$ are the vector and pseudoscalar mesons, respectively.) were intensively studied in various quark models~\cite{Hackman:1977am,Eichten:1979ms,Pham:1981ru,Thews:1984tn,Miller:1988tz,Kamal:1992uv,Ivanov:1994ji,Jaus:1996np,Deandrea:1998uz,Choi:2007se,Priyadarsini:2016tiu}, quark potential models~\cite{Bose:1980vy,Godfrey:1985xj,Goity:2000dk,Ebert:2002xz,Simonis:2016pnh,Simonis:2018rld}, synthesis of heavy quark effective theory and vector meson dominance model~\cite{Colangelo:1993zq}, QCD sum rules~\cite{Aliev:1994nq,Dosch:1995kw,Zhu:1996qy,Li:2020rcg}, extended Nambu-Jona-Lasinio model~\cite{Deng:2013uca,Luan:2015goa}, lattice QCD simulations~\cite{Becirevic:2009xp} and $\chi$PT~\cite{Cho:1992nt,Amundson:1992yp,Cheng:1992xi,Casalbuoni:1996pg,Stewart:1998ke,Singer:1999ak,Savage:2001jw,Wang:2019mhm}. In $\chi$PT, the magnetic moments and radiative transitions are deduced from the same Lagrangians with the same set of parameters. In what follows, we review the results obtained with $\chi$PT and summarize the results from other models in Table~\ref{tab:radwidth}.

\begin{table*}
\renewcommand{\arraystretch}{1.4}
 \tabcolsep=1.5pt
\caption{The radiative decay widths of the charmed and bottom vector mesons calculated in various approaches, where $\Gamma_V^\gamma$ denotes the decay width of $V\to P\gamma$, while $\Gamma_V$ represents the total width of meson $V$. The QM, VMD, QCDSR, NJL, LQCD, and $\chi$PT are abbreviations of quark model, vector meson dominance model, QCD sum rule, extended Nambu-Jona-Lasinio model, lattice QCD, and chiral perturbation theory, respectively. The averaged value in the last row is given as an average of the corresponding values calculated in the listed references. Since the pion emission decay for the bottom vector meson is kinematically forbidden, their widths are saturated by the radiative decays, i.e., we approximately have $\Gamma_{B^{\ast-}}\simeq\Gamma_{B^{\ast-}}^\gamma$, $\Gamma_{\bar{B}^{\ast0}}\simeq\Gamma_{\bar{B}^{\ast0}}^\gamma$ and $\Gamma_{\bar{B}_s^{\ast0}}\simeq\Gamma_{\bar{B}_s^{\ast0}}^\gamma$. Although the $D_s^\ast$ can decay into $D_s\pi^0$ via isospin violating decay, its minor branching ratio~\cite{ParticleDataGroup:2022pth} makes the approximation $\Gamma_{D_s^{\ast+}}\approx\Gamma_{D_s^{\ast+}}^\gamma$ reasonable. The values in this table are all given in units of keV.}\label{tab:radwidth}
\begin{threeparttable}
\setlength{\tabcolsep}{0.94mm}
{
\begin{tabular}{ccccccccccc}
\hline
\multirow{2}{*}{Models}&\multicolumn{6}{c}{Charmed vector mesons}&\multicolumn{3}{c}{Bottom vector mesons}&\multirow{2}{*}{Refs.}\\
&$\Gamma_{D^{\ast0}}^\gamma$&$\Gamma_{D^{\ast+}}^\gamma$&$\Gamma_{D_s^{\ast+}}^\gamma$&$\Gamma_{D^{\ast0}}$&$\Gamma_{D^{\ast+}}$&$\Gamma_{D_s^{\ast+}}$&$\Gamma_{B^{\ast-}}^\gamma$&$\Gamma_{\bar{B}^{\ast0}}^\gamma$&$\Gamma_{\bar{B}_s^{\ast0}}^\gamma$\\
\hline
\multirow{12}{*}{QM}& $\sim23$ & $\sim0.9$  & $\sim0.1$  & $-$ & $-$  & $\sim0.1$  & $-$  & $-$ & $-$  &\cite{Hackman:1977am}\\
                                     & $35.2$ & $2.4$  & $0.32$  & $78.6$ & $78.0$  & $\sim0.32$  & $1.7$  & $0.5$ & $0.2$  &\cite{Eichten:1979ms}\\
                                     & $12.2$& $4.3$  & $1.1$  & $59.4$ & $79.0$  & $\sim1.1$  & $-$  & $-$ & $-$  &\cite{Miller:1988tz}$^{\rm a}$\\
                                     & $21.8$& $1.7$  & $-$  & $64.2$ & $86.4$ & $-$  & $-$  & $-$ & $-$  &\cite{Kamal:1992uv}$^{\rm b}$\\
                                     & $17.9$& $0.36$  & $0.1$  & $56.0$ & $80.5$ & $\sim0.1$  & $0.40$  & $0.13$ & $-$  &\cite{Ivanov:1994ji}$^{\rm c}$\\
                                     & $21.7$& $0.56$  & $-$  & $65.1$ & $92.0$ & $-$  & $0.43$  & $0.14$ & $-$  &\cite{Jaus:1996np}\\
                                     & $13.1$& $0.25$  & $-$  & $38.0$ & $62.0$ & $-$  & $-$  & $0.05$ & $-$  &\cite{Deandrea:1998uz}\\
                                     & $20.0^{+0.3}_{-0.3}$& $0.9^{+0.02}_{-0.02}$  & $0.18^{+0.01}_{-0.01}$  & $55^{+6}_{-6}$ & Input~\cite{ParticleDataGroup:2006fqo} & $0.19^{+0.01}_{-0.01}$  & $0.4^{+0.03}_{-0.03}$  & $0.13^{+0.01}_{-0.01}$ & $0.068^{+0.017}_{-0.017}$  &\cite{Choi:2007se}\\
                                     & $26.5$& $0.93$  & $0.21$  & $99.8^{+19.5}_{-19.5}$ & Input~\cite{ParticleDataGroup:2014cgo} & $0.32^{+0.06}_{-0.06}$  & $0.57$  & $0.18$ & $0.12$  &\cite{Priyadarsini:2016tiu}\\
                                     & $26.0$ & $0.94$  & $0.2$  & $67.6$ & $94.3$  & $\sim0.2$  & $0.6$  & $0.2$ & $0.1$  &\cite{Goity:2000dk}$^{\rm d}$\\
                                     & $11.5$ & $1.04$  & $0.19$  & $-$ & $-$  & $-$  & $0.19$  & $0.07$ & $0.054$  &\cite{Ebert:2002xz}\\
                                     & $19.7$ & $1.1$  & $0.4$  & $-$ & $-$  & $-$  & $0.46$  & $0.15$ & $0.10$  &\cite{Simonis:2016pnh}\\
VMD                           & $16.0^{+12.5}_{-9.0}$ & $0.51^{+0.69}_{-0.44}$  & $0.24^{+0.24}_{-0.24}$  & $36.7^{+9.7}_{-9.7}$ & $46.1^{+14.2}_{-14.2}$  & $\sim0.24^{+0.24}_{-0.24}$  & $0.22^{+0.09}_{-0.09}$  & $0.075^{+0.027}_{-0.027}$ & $-$  &\cite{Colangelo:1993zq}\\
\multirow{3}{*}{QCDSR}& $2.43^{+0.21}_{-0.21}$ & $0.22^{+0.06}_{-0.06}$  & $0.25^{+0.08}_{-0.08}$  & $\sim8$ & $\sim12$  & $\sim0.25$  & $-$  & $-$ & $-$  &\cite{Aliev:1994nq}\\
                                           & $3.7^{+1.2}_{-1.2}$ & $0.09^{+0.40}_{-0.07}$  & $-$  & $11^{+4}_{-4}$ & $12^{+7}_{-7}$  & $-$  & $0.10^{+0.03}_{-0.03}$  & $\sim0.04$ & $-$  &\cite{Dosch:1995kw}\\
                                           & $12.9^{+2}_{-2}$ & $0.23^{+0.1}_{-0.1}$  & $0.13^{+0.05}_{-0.05}$  & $\sim36$ & $\sim46$  & $\sim0.13$  & $0.38^{+0.06}_{-0.06}$  & $0.13^{+0.03}_{-0.03}$ & $0.22^{+0.04}_{-0.04}$  &\cite{Zhu:1996qy}$^{\rm e}$\\
NJL                             & $19.4$ & $0.7$  & $0.09$  & $65.9$ & $124.8$  & $\sim0.09$  & $0.25$  & $0.22$ & $0.10$  &\cite{Deng:2013uca}\\
LQCD                         & $27^{+14}_{-14}$ & $0.8^{+7}_{-7}$  & $-$  & $68^{+17}_{-17}$ & Input~\cite{CLEO:2001sxb} & $-$  & $-$  & $-$ & $-$  &\cite{Becirevic:2009xp}\\
\multirow{4}{*}{$\chi$PT} & $8.8^{+17.1}_{-17.1}$ & $8.3^{+8.1}_{-8.1}$  & $-$  & $50.6^{+61.9}_{-61.9}$ & $97.0^{+95.6}_{-95.6}$  & $-$  & $0.66^{+0.93}_{-0.93}$  & $0.13^{+0.20}_{-0.20}$ & $-$  &\cite{Cho:1992nt}$^{\rm f}$\\
                                             & $34$ & $2$ & $0.3$ & $102$ & $141$ & $\sim0.3$ & $0.84$ & $0.28$ & $-$ &\cite{Cheng:1992xi}\\
                                             & $-$  & $-$  & $-$  & $18$ & $26$ & $0.06$ & $-$ & $-$ & $-$ &\cite{Stewart:1998ke}$^{\rm g}$\\
                                             & $16.2^{+6.5}_{-6.0}$ & $0.73^{+0.7}_{-0.3}$ & $0.32^{+0.3}_{-0.3}$ & $77.7^{+26.7}_{-20.5}$ & Input~\cite{ParticleDataGroup:2018ovx} & $0.62^{+0.45}_{-0.50}$ & $0.58^{+0.2}_{-0.2}$  & $0.23^{+0.06}_{-0.06}$ & $0.04^{+0.03}_{-0.03}$ &\cite{Wang:2019mhm}$^{\rm h}$\\
\hline
Average$^{\rm i}$                      &$18.5$ & $1.4$ & $0.3$ & $55.6$ & $71.8$ & $\sim0.3$ & $0.5$ & $0.2$ & $0.1$ &\\
\hline
\end{tabular}
}
\begin{tablenotes}
       \footnotesize
       \item[a] Radiative decays with zero anomalous moment for the charmed quark are quoted.
       \item[b] The results in model (c) of Ref.~\cite{Kamal:1992uv} are quoted.
       \item[c] The results calculated with constituent heavy quark masses $(m_c,m_b)=(1.6,5.0)$ GeV are quoted.
       \item[d] The results with $\kappa^q=0.45$ are quoted.
       \item[e] The $D^{\ast0}$ and $D^{\ast+}$ widths are inferred by combining the strong decays of those calculated in Ref.~\cite{Belyaev:1994zk}.
       \item[f] We adopt the results calculated with constituent heavy quark masses $(m_c,m_b)=(1.7,5.0)$ GeV.
       \item[g] The results with uncertainties from experiment and counter terms are not quoted here.
       \item[g] The results in SU(3) case with $\Delta\neq0$ are quoted.
       \item[i] Only the central values are used.
     \end{tablenotes}
\end{threeparttable}
\end{table*}

Owing to the synthetical development of $\chi$PT and HQET in 1990s~\cite{Wise:1992hn,Burdman:1992gh,Yan:1992gz,Cho:1992gg,Luke:1992cs}, the electromagnetic interactions are gauged into the Lagrangians of HH$\chi$PT by Cho and Georgi~\cite{Cho:1992nt}, in which the LO [$\mathcal{O}(p^2)$] contribution [corresponding to Fig.~\ref{Fig:2.4.RadiativeDecays}(a)] to the $\bar{D}^\ast\to \bar{D}\gamma$ decay is considered. It is described equivalently by the following Lagrangian,
\begin{eqnarray}\label{eq:LHHgamma2}
\mathcal{L}_{H\gamma}^{(2)}=a\langle\tilde{\mathcal{H}}\sigma^{\mu\nu}\bar{\tilde{\mathcal{H}}}\rangle\operatorname{Tr}(f_{\mu\nu}^+)+\hat{a}\langle\bar{\tilde{\mathcal{H}}}\sigma^{\mu\nu}\hat{f}_{\mu\nu}^+\tilde{\mathcal{H}}\rangle,
\end{eqnarray}
in which the formulations of $f_{\mu\nu}^+$ and $\hat{f}_{\mu\nu}^+$ are given in Eqs.~\eqref{eq:fmnpm} and~\eqref{eq:fmnt}, respectively. $a$ and $\hat a$ are two independent LECs, standing for the coupling strengths of the photon with the heavy and light quark parts, respectively. Therefore, the first and second terms in Eq.~\eqref{eq:LHHgamma2} represent the heavy quark current and light quark current contributions, respectively. From the properties of the superfields in Eq.~\eqref{eq:app1:hlHH}, the first term can flip the heavy spin which breaks the heavy quark spin symmetry and is suppressed by $1/m_Q$.

{In the quark model, the transition M1 form factor of $V\to P\gamma$ can be parameterized as~\cite{Wang:2019mhm}
\begin{eqnarray}\label{eq:muff}
    \mu^\prime_{\bar{Q}q}=\mathscr{Q}_{\bar{Q}} \frac{1}{\Lambda_{\bar{Q}}}-\mathscr{Q}_q \frac{1}{\Lambda_q},
\end{eqnarray}
where $\mathscr{Q}_{\bar{Q}}$ and $\mathscr{Q}_q$ are the charge matrices of the heavy anti-quark $\bar{Q}$ and light quark $q$, respectively. $\Lambda_{\bar{Q}}$ and $\Lambda_q$ are the mass parameters that can be understood as the masses of the constituent quarks in the quark model. Heavy quark symmetry guarantees $\Lambda_{\bar{Q}}\approx m_{\bar{Q}}$ (see the discussions in Ref.~\cite{Manohar:2000dt}). In Ref.~\cite{Wang:2019mhm}, Wang \etal adopted the vector meson dominance (VMD) model~\cite{Pacetti:2014jai,Colangelo:1993zq} to estimate the value of $\Lambda_q$. It turns out that $\Lambda_u=\Lambda_d=0.366$ GeV, $\Lambda_s=0.596$ GeV. These values are very close to the constituent quark masses $m_u$, $m_d$, and $m_s$ which are usually adopted in the quark models. } 

{Expanding Eq.~\eqref{eq:LHHgamma2} and comparing with Eq.~\eqref{eq:muff} one obtains $a\sim 1/m_Q$ and $\hat{a}\sim 1/m_q$}. In Ref.~\cite{Cho:1992nt}, Cho {\it et al} fixed the masses of the $c$ and $b$ quarks, while they left the light quark mass and axial coupling $g_b$ to be determined from the branching ratios of $V\to P\pi$ and $V\to P \gamma$. Cheng {\it et al} also studied  similar processes with the same framework~\cite{Cheng:1992xi}, in which the $a$ and $\hat{a}$ were determined from the constituent quark model. Along this line, Wang {\it et al} obtained~\cite{Wang:2019mhm}
\begin{eqnarray}
a=\frac{1}{24m_{Q}},\quad\quad\quad \hat{a}=-\frac{1}{8m_q}.
\end{eqnarray}

The NLO contribution comes from the loop diagram in Fig.~\ref{Fig:2.4.RadiativeDecays}(b) and was first incorporated by Amundson {\it et al}~\cite{Amundson:1992yp}. {In diagram \ref{Fig:2.4.RadiativeDecays}(b), the relevant vertices are the $\mathcal{H}\varphi$ coupling from Eq.~\eqref{eq:app1:lagDbar}  and $\varphi\gamma$ coupling from Eq.~\eqref{eq:piL0}, respectively}. In principle, the diagrams in Fig.~\ref{Fig:2.4.RadiativeDecays}(c) and (d) also contribute at this order, but their contributions vanish in the heavy quark limit.  {We take the Fig.~\ref{Fig:2.4.RadiativeDecays}(c) as an example. If we are considering the $D^\ast\to D\gamma$ decay, then the intermediate heavy state is $D^\ast$. In this case, its amplitude
\begin{eqnarray}\label{eq:vanishamp}
\mathcal{A}\propto\epsilon_{\mu\nu\alpha\beta}\varepsilon_{D^\ast}^\mu\varepsilon_\gamma^{\ast\alpha}v^\beta\int\frac{d^d \ell}{(2\pi)^d}\frac{i\ell_\rho}{\ell^2-m_\varphi^2+i\epsilon}\frac{-i(g^{\rho\nu}-v^\rho v^\nu)}{v\cdot \ell+i\epsilon}\sim\epsilon_{\mu\nu\alpha\beta}\varepsilon_{D^\ast}^\mu\varepsilon_\gamma^{\ast\alpha}v^\beta v_\rho(g^{\rho\nu}-v^\rho v^\nu)=0,
\end{eqnarray}
where $\varepsilon_{D^\ast}$ and $\varepsilon_\gamma$ denote the polarization vector of $D^\ast$ and photon, respectively. $v$ and $\ell$ represent the four-velocity in the superfield notation and the momentum in the loop, in order. One can consult Ref.~\cite{Wang:2018atz} for the Lorentz structure of the loop integral in Eq.~\eqref{eq:vanishamp}. Similar relation holds for the amplitude of Fig.~\ref{Fig:2.4.RadiativeDecays}(d).
}

In order to absorb the divergence of the loop correction for Fig.~\ref{Fig:2.4.RadiativeDecays}(b), Wang {\it et al} constructed the $\mathcal{O}(p^3)$ Lagrangian~\cite{Wang:2019mhm},
\begin{eqnarray}\label{eq:LHgamma3}
\mathcal{L}_{H\gamma}^{(3)}=-ic\langle\tilde{\mathcal{H}}\sigma^{\mu\nu}\bar{\tilde{\mathcal{H}}}\rangle v\cdot\nabla\mathrm{Tr}(f_{\mu\nu}^+)-i\hat{c}\langle\bar{\tilde{\mathcal{H}}}\sigma^{\mu\nu}v\cdot\nabla\hat{f}_{\mu\nu}^+\tilde{\mathcal{H}}\rangle.
\end{eqnarray}
But their contributions can be absorbed by renormalizing the LO LECs $a$ and $\hat{a}$, i.e., $a\to a+c v\cdot q$ and $\hat{a}\to\hat{a}+\hat{c}v\cdot q$.

The Feynman diagrams (e)-(j) in Fig.~\ref{Fig:2.4.RadiativeDecays} give the $\mathcal{O}(p^4)$ contributions. The vertices in these diagrams are given above except those in diagrams~\ref{Fig:2.4.RadiativeDecays}(g) and (j). For the diagram~\ref{Fig:2.4.RadiativeDecays}(g), the $\mathcal{O}(p^2)$ two-pion coupling vertex is given by~\cite{Wang:2019mhm}
\begin{eqnarray}\label{eq:Hphiphi2}
\mathcal{L}_{H\varphi\varphi}^{(2)}=ib\langle\bar{\tilde{\mathcal{H}}}\sigma^{\mu\nu}[u_\mu,u_\nu]\tilde{\mathcal{H}}\rangle,
\end{eqnarray}
where $b$ is the coupling constant, and its value was phenomenologically determined through the VMD model in Ref.~\cite{Wang:2019mhm}. As the $u_\mu$ transforms as an adjoint representation, one may have three terms $\mathrm{Tr}(u_\mu u_\nu)$, $[u_\mu,u_\nu]$ and $\{u_\mu,u_\nu\}$ (they belong to the $\bm1$, $\bm8_1$ and $\bm8_2$ flavor representations, respectively) sandwiched in the paired $\bar{\tilde{\mathcal{H}}}$ and $\tilde{\mathcal{H}}$. However, $\mathrm{Tr}(u_\mu u_\nu)$ and $\{u_\mu,u_\nu\}$ vanish when contracted with $\sigma^{\mu\nu}$ due to their symmetrized Lorentz index. Therefore, only one single term remains in Eq.~\eqref{eq:Hphiphi2}. The general principles to construct the independent Lagrangians are discussed in~\ref{app:1}.

The tree-level diagram at $\mathcal{O}(p^4)$ in Fig.~\ref{Fig:2.4.RadiativeDecays}(j) is governed by the following Lagrangians~\cite{Wang:2019mhm},
\begin{eqnarray}\label{eq:Hgamma4}
\mathcal{L}_{H\gamma}^{(4)}=\hat{d}\langle\tilde{\mathcal{H}}\sigma^{\mu\nu}\hat{\chi}_+\bar{\tilde{\mathcal{H}}}\rangle \mathrm{Tr}(f_{\mu\nu}^{+})+\bar{d}\langle\bar{\tilde{\mathcal{H}}}\sigma^{\mu\nu}\tilde{\mathcal{H}}\rangle\mathrm{Tr}(\hat{f}_{\mu\nu}^{+}\hat{\chi}_+)
+d\langle\bar{\tilde{\mathcal{H}}}\sigma^{\mu\nu}\{\hat{\chi}_+,\hat{f}_{\mu\nu}^+\}\tilde{\mathcal{H}}\rangle,
\end{eqnarray}
in which the spurion $\hat{\chi}_{+}$ introduces the SU(3) flavor breaking effect.
There are six types of flavor structures constructed with the $\chi_+$ and $f_{\mu\nu}^+$ as listed in Table~\ref{tab:app1:g_irp}, but not all of them survive in the Lagrangians. For example, $\mathrm{Tr}(\chi_+) \hat{f}_{\mu\nu}^+$ and $\mathrm{Tr}(\chi_+)\mathrm{Tr}(f_{\mu\nu}^+)$ are assimilated into the $\mathcal{O}(p^2)$ Lagrangians~\eqref{eq:LHHgamma2}, thus can be replaced by renormalizing $\hat{a}$ and $a$, respectively. Besides, both $\hat{\chi}_+$ and $\hat{f}_{\mu\nu}^+$ are
diagonal matrices at the LO, which makes the leading term of the $[\hat{\chi}_+,\hat{f}_{\mu\nu}^+]$ vanish after the expansion. Therefore, only three terms survive in Eq.~\eqref{eq:Hgamma4} eventually.

Summing up the contributions of all the diagrams in Fig.~\ref{Fig:2.4.RadiativeDecays} with all possible intermediate states, one obtains the (transition) magnetic moments of the corresponding vector states (see the analytical expressions in Ref.~\cite{Wang:2019mhm}). In Refs.~\cite{Cho:1992nt,Cheng:1992xi}, Cho {\it et al} and Cheng {\it et al} calculated the decay widths of $D^\ast\to D\gamma$ and $B^\ast\to B\gamma$ at the tree level in HH$\chi$PT, respectively. The Lagrangians of Ref.~\cite{Wang:2019mhm} are the same as those in Refs.~\cite{Cho:1992nt,Cheng:1992xi} at the LO. In Ref.~\cite{Amundson:1992yp}, Amundson {\it et al} investigated the same process with the same framework to the NLO, but the heavy quark spin symmetry breaking effect was ignored. In heavy quark limit ($a\to0$), they found $\mu_V^u:\mu_V^d:\mu_V^s=2:-1:-1$ for the LO results. This ratio relation still holds when the loop corrections are included in the strict SU(3) symmetry [neglecting the mass splitting of SU(3) multiplets]~\cite{Wang:2019mhm,Amundson:1992yp,Savage:2001jw}. In particular, the authors of Ref.~\cite{Wang:2019mhm} noticed $|\mu_V|=|\mu_{V\to P\gamma}|$ in the heavy quark limit by taking mass splittings $\delta_b=0$ in the loops. Because the heavy quark {spin} completely decouples in the heavy quark limit, the $VV\gamma$ and $VP\gamma$ three point Green functions depict the same light quark dynamics. In a more realistic calculation, the authors of Ref.~\cite{Wang:2019mhm} further showed that the HQSS breaking effect (the physical value of $\delta_b$ is used in loop integrals) can induce sizable corrections, especially for the charmed vector mesons, see Table~\ref{tab:MagneticMomentsDB}.

\begin{table*}[htbp]
\renewcommand{\arraystretch}{1.2}
 \tabcolsep=1.5pt
\caption{The magnetic moments of the charmed and bottom vector mesons (in units of nucleon magnetons $\mu_N$) calculated with the $\chi$PT, Bag model (Bag), extended Bag model (eBag), extended Nambu-Jona-Lasinio model (NJL) and light-cone QCD sum rules (LCSR), respectively. The results from $\chi$PT are given in SU(2) and SU(3) symmetries as well as in the cases of the mass splittings $\Delta=0$ and $\Delta\neq 0$, respectively.}\label{tab:MagneticMomentsDB}
\setlength{\tabcolsep}{2.05mm}
{
\begin{tabular}{ccccccccc}
\hline
\multirow{2}{*}{States}&\multicolumn{2}{c}{SU(2)}&\multicolumn{2}{c}{SU(3)}&\multicolumn{4}{c}{The results from other theoretical works}\\
&$\Delta=0$&$\Delta\neq0$&$\Delta=0$&$\Delta\neq0$&Bag~\cite{Bose:1980vy} &eBag~\cite{Simonis:2018rld} &NJL~\cite{Luan:2015goa}&LCSR~\cite{Aliev:2019lsd}\\
\hline
$D^{\ast0}$&$-1.38^{+0.25}_{-0.25}$&$-1.60^{+0.25}_{-0.25}$&$-1.18^{+0.25}_{-0.25}$&$-1.48^{+0.38}_{-0.22}$&$-0.89$&$-1.28$&$-$&$0.30^{+0.04}_{-0.04}$\\
$D^{\ast+}$&$1.14_{-0.15}^{+0.15}$&$1.39_{-0.15}^{+0.15}$&$1.31_{-0.20}^{+0.15}$&$1.62_{-0.24}^{+0.08}$&$1.17$&$1.13$&$1.16$&$1.16^{+0.08}_{-0.08}$\\
$D_s^{\ast+}$&$-$&$-$&$0.62_{-0.15}^{+0.15}$&$0.69_{-0.22}^{+0.10}$&$1.03$&$0.93$&$0.98$&$1.00^{+0.14}_{-0.14}$\\
$B^{\ast-}$&$-1.86^{+0.25}_{-0.25}$&$-1.90^{+0.20}_{-0.20}$&$-1.71^{+0.25}_{-0.25}$&$-1.77^{+0.30}_{-0.25}$&$-1.54$&$-1.56$&$-1.47$&$-0.90^{+0.19}_{-0.19}$\\
$\bar{B}^{\ast0}$&$0.75_{-0.11}^{+0.11}$&$0.78_{-0.11}^{+0.11}$&$0.87_{-0.13}^{+0.11}$&$0.92_{-0.15}^{+0.11}$&$0.64$&$0.69$&$-$&$0.21^{+0.04}_{-0.04}$\\
$\bar{B}_s^{\ast0}$&$-$&$-$&$0.25_{-0.11}^{+0.11}$&$0.27_{-0.13}^{+0.10}$&$0.47$&$0.51$&$-$&$0.17^{+0.02}_{-0.02}$\\
\hline
\end{tabular}
}
\end{table*}

In Ref.~\cite{Stewart:1998ke}, Stewart calculated the decays $D^\ast\to D\pi$ and $D^\ast\to D\gamma$ up to the one-loop level within HH$\chi$PT, in which the $\mathcal{B}(D^{\ast+}\to D^+\gamma)$ as well as the ratios of the $D\gamma$ and $D\pi^0$ branching fractions were used to extract the $D^\ast D\pi$ and $D^\ast D\gamma$ couplings. The $D^\ast D\pi$ coupling $g_b$ is determined to be around $0.3$, which is close to the current value from the $D^{\ast+}\to (D\pi)^+$ partial widths.

In addition to the ground state vector states, the excited heavy mesons were also systematically studied in the HH$\chi$PT~\cite{Falk:1991nq,Kilian:1992hq,Falk:1992cx,Falk:1993iu,Falk:1995th,Bardeen:2003kt,Colangelo:2003vg}.  The radiative decay $D_{s0}^\ast(2317)\rightarrow D_s^\ast\gamma$ was firstly studied by Colangelo {\it et al} with the HQS and VMD ansatz~\cite{Colangelo:2003vg}. They obtained $\Gamma(D_{s0}^\ast\to D_s^\ast\gamma)\simeq1$ keV (decaying to $D_s\gamma$ is forbidden due to the angular momentum and parity conservation). In Ref.~\cite{Mehen:2004uj}, Mehen and Springer studied the electromagnetic decays of the $D_{s0}^\ast (2317)$ and $D_{s1}(2460)$ in HH$\chi$PT considering the chiral loop corrections and HQSS breaking effect. In the compact meson scenario, the tree level contribution for $\{0^+,1^+\}\to\{0^-,1^-\}\gamma$ is mediated by the light quark current coupling with the photon, the Lagrangian reads
\begin{eqnarray}\label{eq:lagsjg}
\mathcal{L}_{\mathcal{S}\mathcal{H}\gamma}=\hat{\beta} \langle \mathcal{S} \sigma^{\mu \nu}\hat{f}_{\mu \nu}^+\bar{\mathcal{H}}\rangle.
\end{eqnarray}
The HQSS violating operators at $\mathcal{O}(1/m_c)$ are constructed for the $\mathcal{S}\to \mathcal{H}\gamma$ decay as
\begin{equation}\label{eq:HQSSvio}
\mathcal{L}=i\beta^\prime\epsilon^{\mu \nu \alpha \beta}\langle\bar{\mathcal{H}} \sigma_{\mu \nu} \mathcal{S} \gamma_{5}\rangle \Tr(f_{\alpha \beta}^+) +i\beta^{\prime \prime} \langle\bar{\mathcal{H}} \sigma^{\mu \nu} \mathcal{S} \gamma^{\alpha}\rangle \partial_{\alpha} \Tr(f_{\mu \nu}^+)+\text{H.c.},
\end{equation}
where the $\beta^\prime$ and $\beta^{\prime\prime}$ are proportional to $1/m_c$. The NLO corrections receive contributions from Eq.~\eqref{eq:HQSSvio} and the loop diagrams in Fig.~\ref{Fig:2.4.RadiativeDecays}(b)-(d) with the possible intermediate states in the loops (see Ref.~\cite{Mehen:2004uj}).

{For convenience, we define three ratios of the branching fractions for the $\{0^+,1^+\}\to\{0^-,1^-\}\gamma$ and $\{0^+,1^+\}\to\{0^-,1^-\}\pi$ decays:
\begin{eqnarray}\label{eq:bfratios}
 \mathcal{R}_1=\frac{\operatorname{Br}\left[D_{s 1}(2460) \rightarrow D_{s}^{*} \gamma\right]}{\operatorname{Br}\left[D_{s 1}(2460) \rightarrow D_{s}^{*} \pi^{0}\right]},\quad
 \mathcal{R}_2=\frac{\operatorname{Br}\left[D_{s 1}(2460) \rightarrow D_{s} \gamma\right]}{\operatorname{Br}\left[D_{s 1}(2460) \rightarrow D_{s}^{*} \pi^{0}\right]},\quad
 \mathcal{R}_3=\frac{\operatorname{Br}\left[D_{s0}^\ast(2317) \rightarrow D_{s}^\ast \gamma\right]}{\operatorname{Br}\left[D_{s 0}^\ast(2317) \rightarrow D_{s} \pi^{0}\right]},
\end{eqnarray}
in which only the $\mathcal{R}_2$ was measured to certain ranges by the Belle Collaboration
\begin{eqnarray}
    \mathcal{R}_2=
    \begin{cases}
    0.38\pm0.11\pm0.04~\text{\cite{Belle:2003guh}}\\
    0.55\pm0.13\pm0.08~\text{\cite{Belle:2003kup}}
\end{cases},
\end{eqnarray}
while for the $\mathcal{R}_1$ and $\mathcal{R}_3$ only the upper limits were estimated with given confidence level~\cite{ParticleDataGroup:2022pth}. For example, the bounds quoted by the CLEO Collaboration~\cite{CLEO:2003ggt}
\begin{eqnarray}
  \text{CLEO bounds}~\text{\cite{CLEO:2003ggt}}:~\mathcal{R}_1<0.16,\qquad \mathcal{R}_3<0.059.
\end{eqnarray}
}
In the compact-meson scenario, Mehen and Springer~\cite{Mehen:2004uj} calculated the $\mathcal{R}_1$ and $\mathcal{R}_3$ within the LO HH$\chi$PT [the $\mathcal{R}_2$ was used as input to determine the $\hat{\beta}$ in Eq.~\eqref{eq:lagsjg}], but the results exceed the CLEO bounds significantly. The ratios can be made consistent with the CLEO bounds when the NLO contributions are included (in which the corresponding LECs are of natural size).
They also investigated the electromagnetic decays of the $D_{s0}^\ast(2317)$ and $D_{s1}(2460)$ and assumed these states are the $DK$ and $D^\ast K$ hadronic molecules with $I=0$, respectively. In the molecular scenario, they obtained~\cite{Mehen:2004uj}
\begin{eqnarray}\label{eq:bfmolecules}
\mathcal{R}_1=3.23,\qquad\qquad
\mathcal{R}_2=2.21,\qquad\qquad
\mathcal{R}_3=2.96.
\end{eqnarray}
In a recent study~\cite{Fu:2021wde}, Fu \etal studied the ratios in the molecular scenario and obtained 
\begin{eqnarray}\label{eq:bfmolecules1}
\mathcal{R}_1=0.12\pm0.02,\qquad\qquad
\mathcal{R}_2=0.38(\rm fixed)\pm0.08,\qquad\qquad
\mathcal{R}_3=0.028\pm0.009,
\end{eqnarray}
in which the central value of $\mathcal{R}_2$~\cite{Belle:2003guh} was also used as input to fix the LO counter term in the radiative decays. {One can see the ratios obtained from Ref.~\cite{Fu:2021wde} are about one order of magnitude smaller than those from Ref.~\cite{Mehen:2004uj}. The authors of Ref.~\cite{Fu:2021wde} considered the extra NLO isospin violating vertices which enhance the strong decay widths of the charmed-strange mesons significantly.}  A similar investigation was performed by Lutz~\cite{Lutz:2007sk} and Guo~\cite{Guo:2008gp} to study the decays of the $D_{s0}^\ast(2317)$ and $D_{s1}(2460)$, as well as the effective Lagrangian approach in Refs.~\cite{Faessler:2007gv,Cleven:2014oka}.


\subsubsection{Singly heavy baryons}\label{Sect.2.4.2}

{As the heavy flavor siblings of the nucleons, the electromagnetic properties of the singly heavy baryons at low energies are {sensitive to} their inner constituents, structures and the corresponding chiral dynamics of the light diquarks}. The experimentalists are trying to measure the charm baryon dipole moments (particularly $\Lambda_c^+,\Xi_c^+$) at LHC~\cite{Aiola:2020yam}. Before giving the systematic applications of $\chi$PT to the electromagnetic properties of the singly heavy baryons, we first briefly review the works from other frameworks in this field.
The electromagnetic properties of the singly heavy baryons have been studied with the bag models~\cite{Bose:1980vy,Simonis:2018rld,Bernotas:2012nz,Bernotas:2013eia,Zhang:2021yul}, the various quark models~\cite{Barik:1983ics,Ivanov:1996fj,Ivanov:1999bk,Tawfiq:1999cf,Julia-Diaz:2004yqv,Kumar:2005ei,Faessler:2006ft,Sharma:2010vv,Majethiya:2011ry,Wang:2017kfr,Hazra:2021lpa}, the QCD sum rules~\cite{Zhu:1997as,Zhu:1998ih,Aliev:2008ay,Aliev:2008sk,Wang:2010xfj,Aliev:2011bm,Agamaliev:2016fou}, the hyper central model~\cite{Patel:2007gx}, the Skyrme model~\cite{Oh:1991ws,Oh:1995eu}, the pion mean-field approach~\cite{Yang:2018uoj,Yang:2019tst,Kim:2021xpp}, and the bound-state approach~\cite{Scholl:2003ip}. Meanwhile, the {\it ab initio} calculations in lattice QCD have been performed to investigate the electromagnetic properties of the singly heavy baryons. For example, in Refs.~\cite{Bahtiyar:2015sga,Bahtiyar:2016dom}, Bahtiyar {\it et al} studied the $\Omega_c^\ast\to\Omega_c\gamma$ and $\Xi_c^\prime\to\Xi_c\gamma$ transitions in $2+1$-flavor lattice QCD, where the radiative decay widths, electric and magnetic form factors, as well as the magnetic moments are computed (see also~\cite{Can:2013tna,Can:2015exa} for the charmed and charmed-strange baryons).

The $\chi$PT was systematically applied to the radiative decays and magnetic moments of the singly heavy baryons in a series of works~\cite{Cheng:1992xi,Cho:1994vg,Savage:1994wa,Banuls:1999br,Tiburzi:2004mv,Jiang:2015xqa,Wang:2018gpl,Meng:2018gan,Wang:2018cre,Shi:2018rhk}. The LO calculations were done by Cheng {\it et al}~\cite{Cheng:1992xi}, in which the decay rates of $\Xi_c^\prime\to\Xi_c\gamma$, $\Sigma_c\to\Lambda_c\gamma$ and $\Sigma_c\to\Lambda_c\pi\gamma$ were calculated. The LECs in the chiral Lagrangians were estimated with the constituent quark model, which paved a way for latter works to determine the LECs ~\cite{Wang:2019mhm,Jiang:2015xqa,Wang:2018gpl,Meng:2018gan,Wang:2018cre}. Cho extended the chiral Lagrangians to the $P$-wave $\bar{\bm 3}$ baryons (with the orbital excitation between the heavy quark and light diquark pair, i.e., $j_\ell=1$), in which the $J^P=\frac{1}{2}^-$ and $\frac{3}{2}^-$ doublets are described with the superfield $\mathscr{R}_{\mu i}=1/\sqrt{3}(\gamma^\mu+v^\mu)\gamma^5 \mathcal{R}_i+\mathcal{R}_{\mu i}^\ast$~\cite{Cho:1994vg}. The LO expressions for the one-pion, di-pion and one-photon transitions are presented. Savage investigated the magnetic dipole (M1) and electric quadrupole (E2) contributions to the $\Sigma_c^\ast\to\Lambda_c\gamma$ process~\cite{Savage:1994wa}. The E2 contribution in the process of $j_\ell=1$ to $j_\ell=0$ is $1/m_c$ suppressed and vanishes in the heavy quark limit. Savage found that the E2 contribution is actually enhanced by a small energy denominator arising from the infrared behaviour of the pion loop graphs, which compensates the suppression and yields a few percent amplitude ratio for $\mathcal{A}_{\text{E2}}/\mathcal{A}_{\text{M1}}$ (this ratio is also affected by the mass splitting of the $\Sigma_c^\ast$ and $\Sigma_c$). Ba\~{n}uls {\it et al} calculated the $B_{\bm 6}\to B_{\bar{\bm 3}}\gamma$, $B_{\bm 6^\ast}\to B_{\bar{\bm 3}}\gamma$ and $B_{\bm 6^\ast}\to B_{\bm 6}\gamma$ decays up to NLO~\cite{Banuls:1999br}, where the magnetic dipole and electric quadrupole contributions were separately computed. Tiburzi studied the same process in a partially quenched approach~\cite{Tiburzi:2004mv}, which allows the pion mass extrapolation and the zero-momentum extrapolation in lattice QCD. The updated calculations to higher orders were performed in Refs.~\cite{Jiang:2015xqa,Wang:2018gpl,Meng:2018gan,Wang:2018cre}. Here we follow the notations in these references to review how the $\chi$PT is employed to study the electromagnetic properties of the heavy baryons.

 The chiral Lagrangians contributing to the electromagnetic properties of the singly heavy baryons were constructed in Ref.~\cite{Jiang:2015xqa} with the field $B_{\bar{\bm3}}$, $B_{\bm6}$ and $B_{\bm6^*}$ fields.  Wang {\it el al} found the number of LECs can be largely reduced in the superfield formalism with the HQSS~\cite{Wang:2018gpl,Meng:2018gan,Wang:2018cre}. The superfield $\psi_Q^\mu$ for the singly heavy baryons is given in Eq.~\eqref{eq:sfofshb}. A similar form was obtained by Cheng {\it et al}~\cite{Cheng:1992xi} in heavy quark symmetry and quark model calculation. The LO Lagrangians that contribute to  Fig.~\ref{Fig:2.4.RadiativeDecays}(a) read
\begin{eqnarray}\label{eq:Lag_psigamma}
	\mathcal{L}_{\psi_{Q}\gamma}^{(2)}&=&i\kappa_{1}\text{Tr}(\bar{\psi}_{Q}^{\mu}\hat{f}_{\mu\nu}^{+}\psi_{Q}^{\nu})+\kappa'_{1}\text{Tr}(\bar{\psi}_{Q}^{\lambda}\sigma^{\mu\nu}\psi_{Q\lambda})\text{Tr}(f_{\mu\nu}^{+})\nonumber+\kappa_{2}\epsilon^{\mu\nu\alpha\beta}\text{Tr}(\bar{\psi}_{Q\mu}\hat{f}_{\alpha\beta}^{+}v_{\nu}B_{\bar{\bm{3}}})+\text{H.c.}\nonumber\\
	&&+\kappa'_{2}i\text{Tr}(\bar{B}_{\bar{\bm{3}}}\gamma^{\nu}\gamma^{5}\psi_{Q}^{\mu})\text{Tr}(f_{\mu\nu}^{+})+\text{H.c.}+\kappa'_{3}\text{Tr(}\bar{B}_{\bar{\bm{3}}}\sigma^{\mu\nu}B_{\bar{\bm{3}}})\text{Tr}(f_{\mu\nu}^{+}),
\end{eqnarray}
where the $\kappa_{1,2}$ and $\kappa'_{1,2,3}$ terms denote the contributions from the light quark current and heavy quark current, respectively. The $\kappa'_{1,2,3}$ terms can flip the spin of the heavy quark and break heavy quark spin symmetry. There is no contribution from the light quarks for the $B_{\bar{\bm{3}}}B_{\bar{\bm{3}}}\gamma$,  because the spin of the light diquark in the flavor anti-triplet is zero. The $\mathcal{O}(p^2)$ di-pion coupling vertices of the heavy baryons in Fig.~\ref{Fig:2.4.RadiativeDecays}(g) are depicted by
\begin{eqnarray}
\mathcal{L}_{\psi_Q \varphi \varphi}^{(2)}&=& \kappa_4 \operatorname{Tr}(\bar{\psi}_Q^{\mu}[u_{\mu}, u_{\nu}] \psi^{\nu}_Q)+i\kappa'_{4}\text{Tr(}\bar{B}_{\bar{\bm{3}}}\sigma^{\mu\nu}[u_{\mu},u_{\nu}]B_{\bar{\bm{3}}})
+i \kappa_{5}\epsilon^{\sigma \mu \nu \rho} \operatorname{Tr}(\bar{B}_{\bar{\bm3}}[u_{\mu}, u_{\nu}] v_{\rho} \psi_{Q\sigma})+\text{H.c.}\nonumber\\
&&+i \kappa_{6} \operatorname{Tr}(\bar{B}_{\bar{\bm3} a b} \epsilon^{\sigma \mu \nu \rho} u_{i \mu}^{b} u_{j \nu}^{a} v_{\rho} \psi_{Q\sigma}^{i j})+\text{H.c.}.
\end{eqnarray}
Most of the $\mathcal{O}(p^3)$ Lagrangians can be absorbed by renormalizing the $\mathcal{O}(p^2)$ ones. The remaining terms may contribute to the lowest-order E2 transitions (see the discussions in Ref.~\cite{Wang:2018cre}), which read,
\begin{equation}
\mathcal{L}_{\psi_Q \gamma}^{(3)}=n_1 \operatorname{Tr}(\bar{B}_{\bm3} \nabla_{\lambda} \tilde{f}_{\mu \nu}^{+} S^{\lambda} v^{\mu} B_{\bm6^{*}}^{\nu})+m_1 \operatorname{Tr}(\bar{B}_{\bm6} \nabla_{\lambda} \tilde{f}_{\mu \nu}^{+} S^{\lambda} v^{\mu} B_{\bm6^{*}}^{\nu})+\tilde{m}_{1} \operatorname{Tr}[\bar{B}_{\bm6} S^{\lambda} v^{\mu} B_{\bm6^{*}}^{\nu}]\nabla_{\lambda} \operatorname{Tr}(f_{\mu \nu}^{+})+\text {H.c.}+\dots.
\end{equation}
 At N$^2$LO, the Lagrangians for the tree-level transitions in Fig.~\ref{Fig:2.4.RadiativeDecays}(j) were constructed as
\begin{equation}
\mathcal{L}_{\psi_Q \gamma}^{(4)}= i \kappa_{7} \operatorname{Tr}(\bar{\psi}_{Q\mu}^{a b}\{\chi_{+}, \hat{f}_{\mu \nu}^{+}\}_{a b}^{i j} \psi_{Qi j}^{\nu})
+\kappa_{8} \operatorname{Tr}(\bar{\psi}^{\lambda}_Q \chi_{+} \sigma^{\mu \nu} \psi_{Q\lambda}) \operatorname{Tr}(f_{\mu \nu}^{+}).
\end{equation}

In Refs.~\cite{Banuls:1999br,Wang:2018gpl}, the Coleman-Glashow relations~\cite{Coleman:1961jn} were obtained at $\mathcal{O}(p^2)$ for the singly heavy baryons, which is analogous to that of the octet baryons~\cite{Jenkins:1992pi,Meissner:1997hn,Li:2017vmq}. But these relations are broken when higher order contributions are involved~\cite{Wang:2018gpl}. The $\mathcal{O}(p^4)$ calculations were performed in Refs.~\cite{Wang:2018gpl,Meng:2018gan,Wang:2018cre}, in which the LECs were estimated either from the quark model or fitting the results of lattice QCD simulations (one can also consult Refs.~\cite{Wang:2018gpl,Meng:2018gan,Wang:2018cre} for the summarized results from their calculations and the other models). Besides, they also discussed how the inclusion of the $\bar{\bm 3}$ baryons as the intermediate states in the loops affects the convergence of the chiral expansion ~\cite{Wang:2018gpl,Meng:2018gan,Wang:2018cre}.

\subsubsection{Doubly heavy baryons}\label{Sect.2.4.3}

Among the doubly heavy baryons, only the $\Xi_{cc}^{++}$ was observed in experiments~\cite{LHCb:2017iph}. The electromagnetic properties of the doubly heavy baryons have been investigated in various quark models~\cite{Julia-Diaz:2004yqv,Faessler:2006ft,Hazra:2021lpa,Lichtenberg:1976fi,Jena:1986xs,Silvestre-Brac:1996myf,Albertus:2006ya,Patel:2008xs,Branz:2010pq}, the bag model~\cite{Bose:1980vy,Bernotas:2012nz}, the Skyrme model~\cite{Oh:1991ws}, the light-cone QCD sum rule~\cite{Ozdem:2018uue,Ozdem:2019zis}, and lattice QCD~\cite{Can:2013tna,Can:2013zpa}.

The HH$\chi$PT was used to study the magnetic moments of the spin-$\frac{1}{2}$~\cite{Li:2017cfz,Li:2020uok} and spin-$\frac{3}{2}$~\cite{Meng:2017dni} doubly charmed and bottom baryons, as well as the radiative transitions from the spin-$\frac{3}{2}$ to spin-$\frac{1}{2}$ states~\cite{Li:2017pxa}. The transition amplitudes were calculated up to N$^2$LO in these works, while the numerical results were presented up to NLO, because the LECs at N$^2$LO cannot be fixed well. As for the heavy mesons, the spin-$\frac{1}{2}$ and spin-$\frac{3}{2}$ states can be incorporated in the superfield notations [see the superfield $\psi_{QQ}^\mu$ of the doubly heavy baryons in Eq.~\eqref{eq:sfofdhb}]. The LO Lagrangians for the electromagnetic interaction read
\begin{equation}\label{eq:psiQQgamma2}
\mathcal{L}_{\psi_{QQ}\gamma}^{(2)}=ia^{\prime}\bar{\psi}_{QQ}^{\mu}\psi_{QQ}^{\nu}\text{Tr}(f_{\mu\nu}^{+})+\hat{a}^{\prime}\bar{\psi}_{QQ}^{\rho}\sigma^{\mu\nu}\hat{f}_{\mu\nu}^{+}\psi_{QQ}^{\rho},
\end{equation}
while the NLO dipion Lagrangian reads
\begin{equation}\label{eq:psiQQphiphi}
\mathcal{L}_{\psi_{QQ}\varphi\varphi}^{(2)}=i b^\prime \bar{\psi}_{QQ}^{\mu}[u_{\rho}, u_{\sigma}] \sigma^{\rho \sigma} g_{\mu \nu} \psi^{\nu}_{QQ}.
\end{equation}
The N$^2$LO electromagnetic coupling Lagrangians read
\begin{equation}\label{eq:psiQQgamma4}
\mathcal{L}_{\psi_{QQ}\gamma}^{(4)}=i\hat{d}^{\prime}\bar{\psi}_{QQ}^{\mu}\hat{\chi}_{+}\psi_{QQ}^{\nu}\Tr(f_{\mu\nu}^{+})+\bar{d}^{\prime}\bar{\psi}_{QQ}^{\rho}\sigma^{\mu\nu}\psi_{QQ\rho}\Tr(\hat{f}_{\mu\nu}^{+}\hat{\chi}_{+})+d^{\prime}\bar{\psi}_{QQ}^{\rho}\sigma^{\mu\nu}\{\hat{f}_{\mu\nu}^{+},\hat{\chi}_{+}\}\psi_{QQ\rho}.
\end{equation}
One can see that the structures in Eqs.~\eqref{eq:psiQQgamma2}-\eqref{eq:psiQQgamma4} are analogous to those for the heavy mesons in Sec.~\ref{Sect.2.4.1}.

In Ref.~\cite{Li:2017cfz}, Li {\it et al} calculated the  magnetic moments of the spin-$\frac{1}{2}$ doubly heavy baryons. The LO tree level contributions as well as the axial coupling $\tilde{g}_b$ are estimated from the quark model. However, the spin-$\frac{3}{2}$ states were not included as the intermediate states of the loops. An improved calculation was performed in Ref.~\cite{Li:2020uok}, in which the spin-$\frac{3}{2}$ baryons as the intermediate states were considered. They found that the magnetic moment of the $\Xi_{cc}^{++}$ changes acutely from $-0.25\mu_N$ to $0.35\mu_N$ (but note that the positive value is contradictory with the results in other works), and the changes for the magnetic moments of the $\Xi_{cc}^{+}$ and $\Omega_{cc}^+$ were also very significant. The magnetic moments of the spin-$\frac{3}{2}$ doubly heavy baryons were calculated in Ref.~\cite{Meng:2017dni}. The mass splittings of the spin-$\frac{1}{2}$ and spin-$\frac{3}{2}$ doubly heavy baryons are not fixed in experiments, thus a model dependent value was adopted in Refs.~\cite{Li:2020uok,Meng:2017dni}. The results in Ref.~\cite{Meng:2017dni} are compatible with most of the other approaches. The radiative transitions from the spin-$\frac{3}{2}$ to spin-$\frac{1}{2}$ doubly heavy baryons were calculated in Ref.~\cite{Li:2017pxa}, in which the radiative decay widths of the spin-$\frac{3}{2}$ states were estimated to be
\begin{eqnarray}
\Gamma[\Xi_{cc}^{\ast++}\to\Xi_{cc}^{++}\gamma]=22.0~\text{keV},\qquad\Gamma[\Xi_{cc}^{\ast+}\to\Xi_{cc}^{+}\gamma]=9.57~\text{keV},\qquad\Gamma[\Omega_{cc}^{\ast+}\to\Omega_{cc}^{+}\gamma]=9.45~\text{keV}.
\end{eqnarray}

A covariant version of $\chi$PT with the extended-on-mass-shell renormalization scheme~\cite{Becher:1999he} was employed for the doubly heavy baryons in Refs.~\cite{Liu:2018euh,HillerBlin:2018gjw,Shi:2021kmm}. In Refs.~\cite{Liu:2018euh,HillerBlin:2018gjw}, the authors calculated the magnetic moments of the spin-$\frac{1}{2}$ ones via fitting the lattice data~\cite{Can:2013tna,Can:2015exa,Can:2013zpa}. Since the lattice data were only available for the $\Xi_{cc}^+$ and $\Omega_{cc}^+$, the two LECs at LO cannot be fixed simultaneously. In Ref.~\cite{Liu:2018euh}, the axial coupling was fixed either by quark model or heavy diquark-antiquark symmetry~\cite{Hu:2005gf}, or fitting the lattice data. By fitting the lattice data, they obtained a relatively smaller axial coupling than that predicted by either the quark model~\cite{Li:2017cfz} or the heavy diquark-antiquark  symmetry~\cite{HillerBlin:2018gjw}. The finite volume effect was not considered in these two works~\cite{Li:2017cfz,HillerBlin:2018gjw}. The follow-up calculations for the spin-$\frac{3}{2}$ ones were provided in Ref.~\cite{Shi:2021kmm}, in which the magnetic moments of the spin-$\frac{3}{2}$ doubly charmed baryons as a function of $m_\pi^2$ were predicted. The spin-$\frac{3}{2}$ states have not been observed yet. Therefore, the results in all above mentioned works strongly depend on the phenomenological model calculations, such as the mass splittings between the spin-$\frac{3}{2}$ and spin-$\frac{1}{2}$ states appearing in the loops, or the axial coupling constant.

\section{Scattering of the Goldstone bosons and heavy hadrons} \label{sec:sec4}

In this section, we review  the theoretical progresses about the scattering of the light Goldstone mesons off  the  heavy flavor  hadrons. In Sec.~\ref{sec:4.1}, we first review the perturbative calculations of the scattering amplitude  for the scattering of the light Goldstone boson off the heavy mesons in $\chi$PT. Then we will 
review the application of the chiral unitary approaches (discussed in  Sec.~\ref{sec:2.4})  to uncover the nature of the $D^*_{s0}(2317)$ and $D_{s 1}(2460)$ and related states in Sec.~\ref{sec:uchpthm}.  We focus on the nonpertubative calculations with the chiral unitary approaches, which incorporate both the low energy dynamics governed by the $\chi$PT and the unitary condition in a coupled-channel framework. In Sec.~\ref{sec:uchpt_Ds}, the scattering amplitudes in $\chi$PT are employed as the kernel interactions. The $D^*_{s0}(2317)$ and $D_{s 1}(2460)$ are identified as the dynamically generated  poles originating from the resummations of the $\cal H \varphi$ interactions, and are interpreted as  the $D^{(*)}K$ molecules.
In Sec.~\ref{sec:uchptDscore}, the $c\bar s$ cores are included in addition to the $D^{(*)}K$ scattering potential. In Sec.~\ref{sec:uchpt_D}, the two-pole structures of the non-strange charmed mesons in the chiral unitary approaches with the SU(3) flavor symmetry are reviewed. With the HDAS symmetry, the scatterings of the light Goldstone boson off the heavy mesons and doubly heavy baryons are related to each others.  In Sec.~\ref{sec:uchptothers},  we review the scattering of the light Goldstone mesons off the doubly heavy baryons with the similar frameworks. 

\subsection{Perturbative scattering amplitude in $\chi$PT} \label{sec:4.1} 

We choose the scattering ${\cal H} \left(p_{1}\right) \varphi \left(p_{2}\right) \rightarrow {\cal H} \left(p_{3}\right) \varphi \left(p_{4}\right)$ (${\cal H} $ and $\varphi$ are the heavy and light mesons, respectively) as an example. The  Mandelstam variables are  defined as 
  \begin{eqnarray}
s=\left(p_{1}+p_{2}\right)^{2}, \qquad t=\left(p_{1}-p_{3}\right)^{2}, \qquad u=\left(p_{1}-p_{4}\right)^{2},
  \end{eqnarray}
where $p_i$ is the momentum of the scattering meson. The power counting rules are 
  \begin{eqnarray} \label{eq:hmpcr}
m_{\cal H} \sim O\left(p^{0}\right), \quad m_{\varphi} \sim O\left(p^{1}\right), \quad t \sim O\left(p^{2}\right),\quad s-m_{{\cal H}}^{2} \sim O\left(p^{1}\right), \quad u-m_{{\cal H}}^{2} \sim O\left(p^{1}\right),
  \end{eqnarray}
with $p$ the typical small momentum in the chiral expansion. $m_{\cal H}=m_{P/P^*}$ and $m_{\varphi}$ denote the masses of the heavy and light mesons, respectively.
In the $\chi$PT, the naive power counting rule is listed in Eq.~\eqref{eq:1.2:pwc}. As discussed in Sec.~\ref{sec:sec2.3}, the $m_{\cal H}$ breaks the naive power counting rule in calculating the ${\cal H}  \varphi$ scattering amplitude.  If one adopts Eq.~\eqref{eq:hmpcr}, the loop integrals in the relativistic formalism contain the power counting breaking terms which have the lower chiral order than those given by Eq.~\eqref{eq:1.2:pwc}. To deal with the PCB terms, many regularization approaches have been developed. The most well-known ones are the HH$\chi$PT, extended on-mass-shell and infrared regularization schemes as discussed in Sec.~\ref{sec:sec2.3}.

\begin{figure}[btp]
	\begin{center}
		\includegraphics[width=1.0\textwidth]{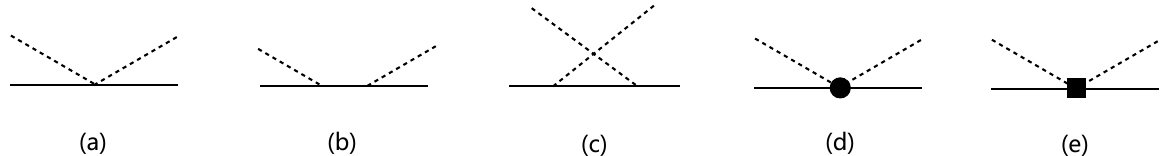}
		\caption{The tree diagrams for the scattering of the pesudoscalar light mesons  (dashed line) off the  heavy mesons (solid line). Each one represents a set of diagrams which have the same topological structure. The solid dot and square stand for the $\mathcal O(p^2)$ and $\mathcal O(p^3)$ vertices. The diagrams (a)-(c) are at LO, while (d) and (e) are at NLO and N$^2$LO, respectively.} \label{Fig:HMPTree}
	\end{center}
\end{figure}

\begin{figure}[btp]
	\begin{center}
		\includegraphics[width=1.0\textwidth]{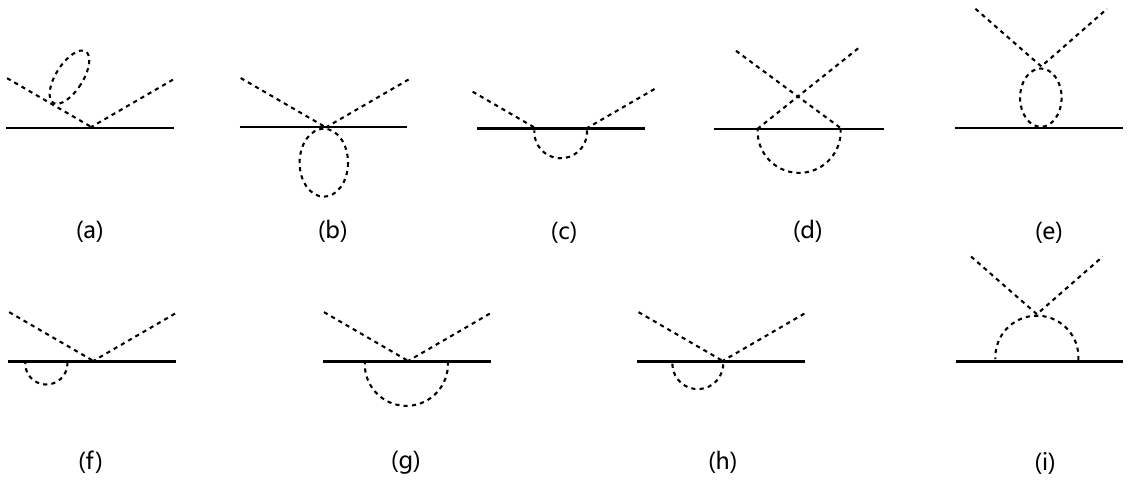}
		\caption{The loop diagrams which survive in the heavy quark limit $m_{\cal H}\rightarrow \infty $ for the scattering of the pesudoscalar light mesons (dashed line) off the heavy mesons (solid line). Each one represents a set of diagrams which have the same topological structure. } \label{Fig:HMPlooph}
	\end{center}
\end{figure}

\begin{figure}[hbtp]
	\begin{center}
		\includegraphics[width=1.0\textwidth]{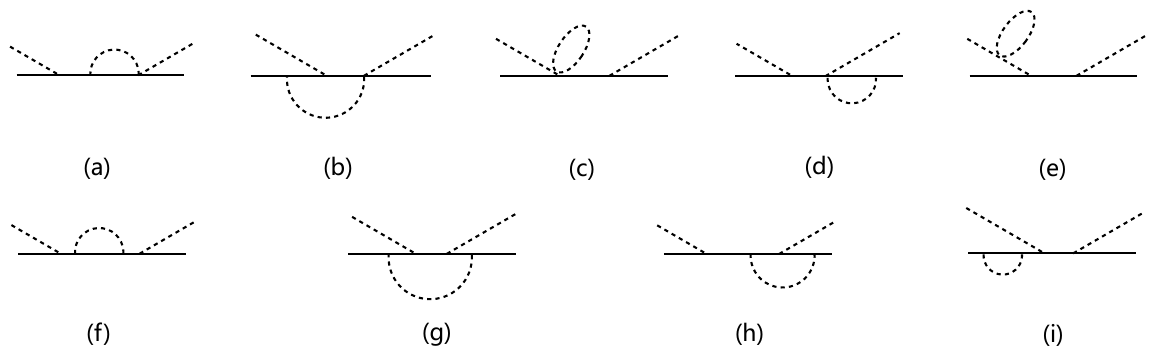}
		\caption{The topological loop diagrams for the scattering of the pesudoscalar light mesons (dashed line) and  heavy mesons (solid line) that vanish in the heavy quark limit $m_{\cal H}\rightarrow \infty $.} \label{Fig:HMPloopv}
	\end{center}
\end{figure}

In the HH$\chi$PT, the heavy components of the matter fields are integrated out and only the masses of the light Goldstone mesons appear in the loop calculation. Thus, the PCB terms disappear in the chiral expansion when the heavy hadron mass approaches infinity. At the same time, the Lagrangians in HH$\chi$PT   are also organized by the inverse of the heavy hadron mass ($1/m_{\cal H}$) and  the contributions of the heavy components are at least  $\mathcal O (1/m_{\cal H})$, which are usually suppressed and can be included as the recoiling corrections. In the heavy 
meson limit, the pseudoscalar and the vector heavy mesons are degenerate and can be treated as the same spin doublet as shown in Sec.~\ref{sec:1.5:combChandHQ}. The Lagrangians for the ${\cal H} \varphi$ scattering read
  \begin{eqnarray}
\mathcal{L}_{{\cal H} \varphi}=\mathcal{L}_{\varphi \varphi}^{(2)}+\mathcal{L}_{{\cal H} \varphi}^{(1)}+\mathcal{L}_{{\cal H} \varphi}^{(2)}+\mathcal{L}_{{\cal H} \varphi}^{(3)}+\dots,
  \end{eqnarray}
  where the numbers in the superscripts denote the chiral orders. The  ellipsis represents the higher order Lagrangians.  The LO Lagrangian for the pseudoscalar mesons $\mathcal{L}_{\varphi \varphi}^{(2)}$ is given in Eq.~\eqref{eq:LagUU2}. 
 In the  HH$\chi$PT, the LO Lagrangian    $\mathcal{L}_{{\cal H} \varphi}^{(1)}$   is given in Eq.~\eqref{eq:lagGB}. The NLO and N$^2$LO Lagrangians  read~\cite{Hofmann:2003je,Liu:2009uz,Huang:2021fdt,Liu:2011mi}~\footnote{The LECs $c_0$ and $c_1$  are equivalent to  $\sigma_H$  and $a_H$ in Eq.~\eqref{eq:mm_chiral} up to a constant $4B_0$, respectively. }
  \begin{eqnarray} 
\mathcal{L}_{{\cal H} \varphi}^{(2)}&=& c_{0} \left \langle {\cal H} {\bar{\cal H}}\right \rangle \text{Tr}\left(\chi_{+}\right)+c_{1}\left\langle {\cal H} \chi_{+} {\bar{\cal H}}\right\rangle-c_{2}\left \langle {\cal H}{ \bar{ \cal H}} \right \rangle \text{Tr}\left(u^{\mu} u_{\mu}\right)-c_{3}\left\langle {\cal H} u^{\mu} u_{\mu} {\bar{\cal H}}\right\rangle \nonumber \\ \label{lhhchpt} 
&&-c_{4}\left \langle {\cal H} {\bar{\cal H}}\right\rangle \text{Tr}\left(v \cdot u v \cdot u\right)-c_{5}\left \langle {\cal H} v \cdot u v \cdot u {\cal \bar{H}} \right \rangle,\\
\mathcal{L}_{{\cal H} \varphi}^{(3)}&=&k_1\left\langle {\cal H}\left[\chi_{-}, v \cdot u\right] {\bar{\cal H}}\right\rangle+i k_{2}\left\langle {\cal H}\left[u^{u},\left[v\cdot\partial, u_{u}\right]\right] {\bar{\cal H}}\right\rangle+i k_{3}\left\langle {\cal H}\left[v\cdot \mu,\left[v\cdot\partial, v\cdot u\right]\right] {\bar{\cal H}}\right\rangle, \label{lhhchpt3} 
  \end{eqnarray}
where the $k_2$ and $k_3$ terms are not given explicitly in Refs.~\cite{Liu:2009uz,Huang:2021fdt} since they can be absorbed by the $k_1$ term at the threshold. The ${\cal H} \varphi$ scattering  amplitude  up to N$^2$LO is given by  
  \begin{eqnarray}\label{eq:Dphcha}
\mathcal{A}(s, t)=\mathcal{A}_{\text {LO }}^{(\text{WT)}}(s, t)+\mathcal{A}_{\text {LO }}^{(\text {EX})}+\mathcal{A}_{\text {NLO }}^{(\text{Tree})}+\mathcal{A}^{(\text {Tree})}_{\text{N$^2$LO}}+\mathcal{A}^{(\text {Loop)}}_{\text{N$^2$LO}}.
  \end{eqnarray}
The corresponding tree and loop Feynman diagrams are shown in  Fig.~\ref{Fig:HMPTree}, and Figs.~\ref{Fig:HMPlooph}-\ref{Fig:HMPloopv}, respectively. 
  
At LO $[\mathcal  O (p)]$, the diagram (a) in  Fig.~\ref{Fig:HMPTree} stems from the chiral connection term in  $  \mathcal{L}_{\cal H \varphi}^{(1)}$ and yields the famous Weinberg–Tomozawa (WT)  term~\cite{Weinberg:1966kf,Tomozawa:1966jm}. 
  \begin{eqnarray} \label{eq:wta}
\mathcal{A}_{\text {LO }}^{(\text{WT})}=\mathcal{C}_{\mathrm{LO}} \frac{s-u}{4 f_{\varphi}^{2}},
  \end{eqnarray}
where $C_{\text{LO}}$ is the flavor coefficient for the different scattering process.  The tree diagrams (b) and (c) in Fig.~\ref{Fig:HMPTree}  arise from the axial-vector coupling term  in $  \mathcal{L}_{\cal H \varphi}^{(1)}$. They contribute to the $s$- and $u$-channel exchanging amplitude $\mathcal{A}_{\text {LO }}^{(\text {EX})}$. Compared to $\mathcal{A}_{\text {LO }}^{(\text{WT})}$, the  $\mathcal{A}_{\text {LO }}^{(\text {EX})}$ is  suppressed by  $1/m_{\cal H}$ and vanishes in the heavy meson limit. At the threshold, the  $\mathcal{A}_{\text {LO }}^{(\text {EX})}$ is actually $\mathcal O(p^2)$ and can be encoded into the LECs in $\mathcal{L}_{{\cal H} \varphi}^{(2)}$~\cite{Altenbuchinger:2013vwa,Geng:2010vw,Yao:2015qia}. 

At   NLO  $[\mathcal  O (p^2)]$, the scattering amplitude $\mathcal{A}_{\text {NLO }}^{(\text{Tree})}$ comes from the  tree diagram (d)  in Fig.~\ref{Fig:HMPTree} , which stems from the NLO Lagrangian $\mathcal{L}_{{\cal H} \varphi}^{(2)}$.  The N$^2$LO amplitudes arise from both the tree and loop diagrams. The  $\mathcal{L}_{{\cal H} \varphi}^{(3)}$ contributes to $\mathcal{A}^{(\text {Tree})}_{\text{N$^2$LO}}$ through the tree diagram \ref{Fig:HMPTree}(e). The loop diagrams in Fig.~\ref{Fig:HMPlooph}  and  Fig.~\ref{Fig:HMPloopv} contribute to the $\mathcal{A}^{(\text {Loop)}}_{\text{N$^2$LO}}$. The chiral corrections from  Fig.~\ref{Fig:HMPlooph}  survive in the heavy meson limit,  while the corrections from Fig.~\ref{Fig:HMPloopv}  are proportional to $1/m_{\cal H}$ and vanish in the heavy meson limit. All the vertices arise from $\mathcal{L}^{(1)}_{{\cal H} \varphi}$ and $\mathcal{L}_{\varphi \varphi}^{(2)}$. One can  obtain  the $\cal H \bar \varphi$ scattering amplitude using  the cross symmetry. 

Within the HH$\chi$PT scheme,  the authors of Ref.~\cite{Liu:2009uz} calculated the $S$-wave scattering lengths of the heavy pseudoscalar and Goldstone bosons ($P\varphi$) up to N$^2$LO. They considered only the tree diagrams in Fig.~\ref{Fig:HMPTree} and the loop diagrams in Fig.~\ref{Fig:HMPlooph} in the heavy meson limit. The scattering length $a$ was related to the perturbative amplitude at the threshold ($ T_{\mathrm{th}}$) by~\footnote{There are different definitions, for instance, $T_{\mathrm{th}}=-{8 \pi\left(m_{\mathcal H}+m_{\varphi}\right)} a$ in Refs.~\cite{Guo:2015dha,Liu:2012zya}.}
  \begin{eqnarray} \label{eq:slp}
 T_{\mathrm{th}}=8\pi\left(1+\frac{M_\varphi}{m_{\cal H}}\right)a.
  \end{eqnarray}
 {With the scattering lengths from lattice QCD simulations \cite{Liu:2008rza}, they extracted the LECs and predicted the attractive isoscalar $DK$ interactions.} In this work, the scattering lengths were extracted from the perturbative amplitudes. If there exists the bound state or resonance, the perturbative calculation will fail. In this case, the perturbative amplitude can be regarded as the kernel of the nonperturbative calculation. The sign and magnitude of the  kernel at threshold~\footnote{In this case, the experimental scattering lengths cannot be extracted from the  perturbative amplitude at the threshold. One can extract the $a_\text{Born}$ as shown in Eq.~\eqref{eq:slu}.} still reflect the properties of the interaction and provide the hints for the existence of the possible bound states.

In Ref.~\cite{Liu:2011mi}, the authors applied a similar formalism to study the scattering lengths of the heavy vector and light Goldstone bosons  ($P^*\varphi$) up to N$^2$LO. Besides the HH$\chi$PT  scheme, they also calculated the scattering lengths in the framework of the infrared regularization. The results for the pion scattering channels are similar in the two frameworks and the chiral expansions converge well. The loop contributions for the  $\eta$ and $K$ scattering are large, but canceled by the N$^2$LO tree diagrams, which leads to 
 the convergent results. In particular, they introduced the contribution of the $D_{s 1}(2460)$ in the $D^{*} K$ scattering through the LEC $c_3$ in $\mathcal{L}_{{\cal H} \varphi}^{(2)}$. 
  
\begin{figure}[hbtp]
	\begin{center}
		\includegraphics[width=1.0\textwidth]{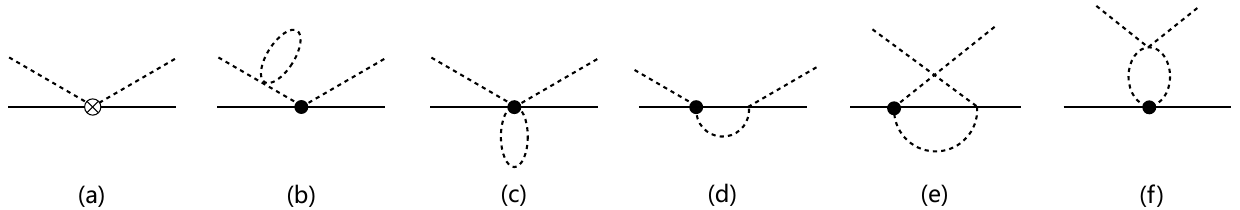}
		\caption{The N$^3$LO topological loop diagrams contributing to the ${\cal H}\varphi$ scattering  amplitude in the heavy quark limit without the vector $P^*$ meson as the intermediate state in the loop. The solid dot and the crossed circle stand for the $\mathcal O(p^2)$ and $O(p^4)$ vertices stemming from $\mathcal L ^{(2)}_{{\cal H} \varphi}$ and $\mathcal L ^{(4)}_{{\cal H} \varphi}$, respectively.} \label{Fig:HMn3lo}
	\end{center}
\end{figure}

In Ref.~\cite{Huang:2021fdt}, the authors calculated the $P\varphi $ scattering to N$^3$LO $[\mathcal{O}(p^{4}) ]$ with the HH$\chi$PT scheme without considering  the vector $P^{*}$ meson.  The Lagrangian  at N$^3$LO reads, 
  \begin{eqnarray}
  \mathcal{L}_{\cal H \varphi}^{(4)}&=&e_{1}\left \langle{ \cal H} {\bar{\cal H}}\right \rangle \text{Tr}\left [(v \cdot \partial v \cdot u)(v \cdot \partial v \cdot u)\right]+e_{2}\left \langle {\cal H} (v \cdot \partial v \cdot u)(v \cdot \partial v \cdot u) {\bar{\cal H}} \right \rangle,
  \end{eqnarray}
where the terms with the quark mass matrix $\chi_{\pm}$ in $\mathcal{L}_{\mathcal H \varphi}^{(4)}$ are not given explicitly since they can be absorbed by the $c_0$ and $c_1$ terms in  $\mathcal{L}_{\mathcal H \varphi}^{(2)}$ {if one is not interested in the light quark mass dependence of the observables.} Without the contribution of the vector mesons, the LO interaction comes from the Weinberg-Tomozawa interaction through the tree diagram (a) in Fig.~\ref{Fig:HMPTree}, while the diagrams (b) and (c) do not exist. At N$^2$LO,  the loop diagrams with the same topological structures as the (a), (b) and (e) diagrams in Fig.~\ref{Fig:HMPlooph} contribute and the relevant vertices originate from the kinetic terms of $\mathcal{L}_{\cal H \varphi}^{(1)}$ and $\mathcal{L}_{\varphi \varphi}^{(2)}$.  
At N$^3$LO, there are additional diagrams as shown in Fig.~\ref{Fig:HMn3lo}. The tree diagram stems from $\mathcal{L}_{\cal H \varphi}^{(4)}$, and the $P^{(\ast)}P^{(\ast)}\varphi$ vertices in the loops are from both  the $\mathcal{L}_{\mathcal H \varphi}^{(1)}$ and  $\mathcal{L}_{\mathcal H \varphi}^{(2)}$. 
  
The authors of Ref.~\cite{Huang:2021fdt}  calculated the scattering amplitude up to N$^3$LO and extracted the scattering lengths with two formalisms: the perturbative one as shown in Eq.~\eqref{eq:slp} and iterated method~\cite{Kaiser:1995eg}
  \begin{eqnarray} \label{eq:slu}
a=a_{\mathrm{Born}}\left(1-\frac{1}{2}\mu a_{\mathrm{Born}}\right)^{-1},
 \end{eqnarray}
where the $\mu$ is a cutoff scale. $a_{\mathrm{Born}}$ contains the contributions of the diagrams except the $s$-channel loop diagram (c) in Fig.~\ref{Fig:HMPlooph} as well as diagram (d) in Fig.~\ref{Fig:HMn3lo}, which can be generated through the iteration of the tree diagrams (a) and (d) in Fig.~\ref{Fig:HMPTree}.  Their isoscalar scattering lengths with Eq.~\eqref{eq:slu} are consistent with the lattice QCD results~\cite{Liu:2012zya,Mohler:2013rwa} as well as the results in Refs.~\cite{Liu:2009uz,Guo:2018tjx}. They also obtained the scattering length of the meson and doubly charmed baryon using the heavy diquark-antiquark symmetry. Their results support the existence of the isoscalar $\bar{K}\Xi_{QQ}~(Q=c,b)$ bound state.

Another approach to eliminate the PCBs is the EOMS scheme. Instead of performing the nonrelativistic projections and integrating out the heavy filed  in HH$\chi$PT, the EOMS performs the  chiral expansion with the covariant Lagrangians. Apart from the substraction of the UV divergences, the extra regularization removing the PCB terms has to be performed. The relevant Lagrangians  responsible for the $P \varphi$ and $P^*\varphi$ interactions up to N$^2$LO read~\cite{Guo:2009ct,Yao:2015qia,Du:2016ntw,Geng:2010vw,Altenbuchinger:2013vwa}~\footnote{ In the heavy quark limit, the Lagrangians recover the formalisms constructed with the superfield as shown in Eq.~\eqref{eq:lagGB},  Eq.~\eqref{lhhchpt} and Eq.~\eqref{lhhchpt3}. One has $m_P=m_P^*$, $h_i=\tilde h_i$ for ($i=1,\dots,5$), $g_k=\tilde g_k$ for $k=0,\dots,3$. The LECs in these two kinds of Lagrangians are related to each other, for instance $h_{0}=2 m_{\cal H} c_{0}$ and $h_{1}=2 m_{\cal H} c_{1}$, with $m_{\cal H}$ the mass of the heavy meson.}
    \begin{eqnarray} 
    \mathcal{L}_{PP^*\varphi}^{(1)}&=&\mathcal{D}_{\mu}P\mathcal{D}^{\mu}P^{\dagger}-m_{P}^{2}PP^{\dagger}-\frac{1}{2}\mathcal{F}^{\mu v}\mathcal{F}_{\mu\nu}^{\dagger}\nonumber \\
    &&+m_{P^*}^{ 2} P^{* v} P_{v}^{* \dagger}+i g_{0}\left(P_{\mu}^{*} u^{\mu} P^{\dagger}-P u^{\mu} P_{\mu}^{* \dagger}\right)+\frac{\tilde {g}{_0}}{2}\left(P_{\mu}^{*} u_{\alpha} \partial_{\beta} P_{\nu}^{* \dagger}-\partial_{\beta} P_{\mu}^{*} u_{\alpha} P_{\nu}^{* \dagger}\right) \epsilon^{\mu \nu \alpha \beta}, \label{eq:dphisl1}\\
 \mathcal{L}_{PP^*\varphi}^{(2)}&=&P\left[-h_{0}\text{Tr}(\chi_{+})-h_{1} \chi_{+}+h_{2}\text{Tr}(u_{\mu} u^{\mu})-h_{3} u_{\mu} u^{\mu}\right] P^{\dagger}  +{\mathcal D}_{\mu} P\left[h_{4}\text{Tr}( u_{\mu} u^{\nu})-h_{5}\left\{u^{\mu}, u^{\nu}\right\}\right] {\mathcal D}_{v} P^{\dagger} \nonumber \\
&&+P^{*\nu}\left[-{\tilde h}_{0}\text{Tr}(\chi_{+})-{\tilde h}_{1} \chi_{+}+{\tilde h}_{2}\text{Tr}(u_{\mu} u^{\mu})-{\tilde h}_{3} u_{\mu} u^{\mu}\right] P_{\nu}^{*\dagger}  +{\mathcal D}_{\mu} P^{*\alpha}\left[{\tilde h}_{4}\text{Tr}( u_{\mu} u^{\nu})-{\tilde h}_{5}\left\{u^{\mu}, u^{\nu}\right\}\right] {\mathcal D}_{v} P_\alpha^{*\dagger} , \label{eq:dphisl2}\\
\mathcal{L}_{PP^*\varphi}^{(3)}&=&g_{1}P[\chi_{-},u_{\nu}]{\mathcal D}^{v}P^{\dagger}+g_{2}P[u^{\mu},\text{\ensuremath{\nabla}}_{\mu}u_{v}+\nabla_{v}u_{\mu}]{\mathcal D}^{v}P^{\dagger}+g_{3}P[u_{\mu},\nabla_{v}u_{\rho}]{\mathcal D}^{\mu v\rho}P^{\dagger}\nonumber\\
&&+ i{\tilde{g}}_{1}P^{*\alpha}[\chi_{-},u_{\nu}]{\mathcal D}^{v}P_{\alpha}^{*\dagger}+{\tilde{g}}_{2}P^{*\alpha}[u^{\mu},\nabla_{\mu}u_{v}+\nabla_{v}u_{\mu}]{\mathcal D}^{v}P_{\alpha}^{*\dagger}+{\tilde{g}}_{3}P^{*\alpha}[u_{\mu},\nabla_{v}u_{\rho}]{\mathcal D}^{\mu v\rho}P_{\alpha}^{*\dagger},\label{eq:dphisl3}
    \end{eqnarray}
where the ${\mathcal D}^{\mu v \rho}=\left\{{\mathcal D}_{\mu},\left\{{\mathcal D}_{\nu}, {\mathcal D}_{\rho}\right\}\right\}$ and  $\mathcal{F}_{\mu \nu}=\left({{\mathcal D}}_{\mu} P_{v}^{*}-{{\mathcal D}}_{v} P_{\mu}^{*}\right)$. 
$m_P$ and $m_{P^*}$ are the masses of the pseudoscalar and vector heavy mesons, respectively. According to the SU(3) group representation, one has $\bm8\otimes\bm8 \to \bm8_1(\bm8_2)$ as shown in Table~\ref{tab:building}. Apart from the $\{u^\mu,u^\nu\}$ building block, there is an additional term with the building block  $[u^{\mu}, u^{\nu} ] $ in $\mathcal{L}_{PP^*\varphi}$ ~\cite{Guo:2009ct}, which was shown to be $\mathcal O(p^3)$ and can be absorbed by the $\mathcal{L}_{PP^*\varphi}^{(3)}$~\cite{Yao:2015qia,Du:2016ntw}.  

 {The calculation of the tree diagrams is independent of the renormalization scheme, and the corresponding tree diagrams are listed in Fig.~\ref{Fig:HMPTree}.} In the EOMS scheme, at the N$^2$LO, the loop diagrams will generate both the UV divergent and the PCB terms. Taking the $P\varphi$ scattering as an example, the $\mathcal {L}_{PP^*\varphi}^{(1)}$ and the terms with the $P$ field in  $\mathcal{L}_{PP^*\varphi}^{(2)}$  and $\mathcal{L}_{PP^*\varphi}^{(3)}$  will contribute to  the $P\varphi\rightarrow P\varphi$ scattering process up to N$^2$LO, while the terms with the vector meson $P^*$ at the NLO and N$^2$LO Lagrangians do not contribute at this order. The bare LECs  can be decomposed into the finite and divergent parts~\cite{Yao:2015qia,Du:2017ttu}
      \begin{eqnarray} 
m_P^{2}=m_P^{r 2}(\mu)+\beta_{m_P^{2}}\lambda, \qquad m_{P^*}^2=m_{P^*}^{r 2}(\mu)+\beta_{m_{P^*}^{2}}\lambda, \qquad h_{i}=h_{i}^{r}(\mu)+{h^0_{i} }\lambda, \qquad g_{j}=g_{j}^{r}(\mu)+g^0_{j} \lambda,
      \end{eqnarray}
      where $\lambda=\mu^{d-4}\left[(4 \pi)^{d / 2}(d-4)\right]^{-1}$ with $\mu$ the regularization scale and $d$ the dimension.  The UV parts will be absorbed by the divergent term proportional to $\lambda$.
  Up to N$^2$LO, the PCB terms originating from the loop integrals are  $\mathcal O(p^2)$ and can be subtracted by redefining the LECs $h^r_i$ and  $g_{0}^{r}$ as follows 
        \begin{eqnarray} 
        h_{i}^{r}(\mu)=\tilde h_i+h^{\text{ PCB}}_i,\qquad\qquad g_{0}^{r}(\mu)=\tilde {g}_{0}+g_0^{\text{ PCB}},
      \end{eqnarray}
while the other $g^r_j$s do not change. The explicit forms of the coefficients  $\beta_{m^2}/\beta_{m^{*2}}/h^0/g^0$ and $h^{\text{ PCB}}_i/g_0^{\text{ PCB}}$ are referred to Refs.~\cite{Yao:2015qia,Du:2017ttu}.  Eventually, the UV and PCB terms  were absorbed by the LECs. 
With the Lagrangians in Eqs.~\eqref{eq:dphisl1}-~\eqref{eq:dphisl3}, the scattering amplitudes of the heavy and light mesons have been calculated up to LO~\cite{Burdman:1992gh,Wise:1992hn,Yan:1992gz}, NLO~\cite{Hofmann:2003je,Guo:2008gp,Guo:2009ct,Cleven:2010aw,Altenbuchinger:2013gaa,Guo:2015dha}, N$^2$LO with the EOMS schemes~\cite{Geng:2010vw,Yao:2015qia,Du:2017ttu}.  
In Refs.~\cite{Yao:2015qia,Du:2017ttu}, the authors performed the calculations with and without the contribution of the vector $P^*$ mesons and found the differences were negligible in the vicinity of the thresholds. In the above works, the perturbative scattering amplitudes were used as the {kernel} interactions in the chiral unitary method and will be discussed in  Sec.~\ref{sec:uchpthm}.

The Goldstone meson and charmed meson scattering was also investigated in Ref.~\cite{Gamermann:2006nm}, where the authors studied both the open and the hidden charmed scalar meson resonances. 
They extended the chiral Lagrangian to study the flavor breaking effect in the Skyrme models. They also considered the SU(4) flavor symmetry and its breaking effect.  

The Goldstone meson and heavy hadron interaction Lagrangians may contain many unknown LECs. In general, the LECs should be determined by fitting the experimental data. However, the experimental information of the $\mathcal H \varphi$ scattering is still scarce nowadays. {{One may reduce the number of LECs in the effective field theory by taking approximate limits. In the large $N_C$ limit, the LECs may have different orders}}. For instance, the $c_1(h_1)$ in Eq.~\eqref{lhhchpt}~[Eq.~\eqref{eq:dphisl1}] can be obtained by the mass differences between the charmed-strange and charmed non-strange mesons.  Furthermore, in Eq.~\eqref{eq:dphisl1}, the  $h_1/h_3/h_5$ terms at the NLO Lagrangians are $\mathcal O (N_C^{0})$. The $h_0/h_2/h_4$ terms are $\mathcal O (N_C^{-1})$ and suppressed in the large $N_C$ limit. Their contributions have been neglected in Refs.~\cite{Lutz:2007sk,Cleven:2010aw}. In Ref.~\cite{Geng:2010vw}, the authors did not include the $h_4$ and $h_5$ terms since their contributions have the same structures as those from the $h_2$ and $h_3$ terms in the vicinity of the thresholds. An alternative way is using the lattice QCD data~\cite{Liu:2012zya,Mohler:2013rwa,Lang:2014yfa,Bali:2017pdv,Moir:2016srx,Mohler:2012na}, which have been extensively employed to determine the LECs in the perturbative or unitarized $\chi$PT calculations~\cite{Guo:2009ct,Liu:2012zya,Wang:2012bu,Altenbuchinger:2013vwa,Liu:2009uz,Geng:2010vw,Altenbuchinger:2013gaa,Du:2017ttu,Guo:2018tjx,Yang:2021tvc,Guo:2018kno}. The LECs are also estimated using phenomenological models such as the resonance saturation method~\cite{Ecker:1988te,Bernard:1993fp,Liu:2011mi,Du:2016tgp,Liu:2009uz,Liu:2011mi,Du:2016tgp}, where the effective Lagrangians are first constructed with the resonances and then the resonances are integrated out to estimate the LECs for the mesonic Lagrangians. In Ref.~\cite{Du:2016tgp}, the authors  constructed the effective Lagrangians with the resonances including the scalar charmed mesons, the light vector mesons, and the tensor mesons. Then they estimated the LECs for the mesonic Lagrangian by integrating out the resonances. The results at NLO were consistent with those obtained by fitting the lattice QCD data, while those at N$^2$LO had a sizable deviation.

The $\chi$PT amplitude is valid and useful in the low energy region where the higher resonance/bound states are integrated out. The implementation of the unitarity properties will help to extend the $\chi$PT amplitude to the higher regions where the resonances/bound states appear.

\subsection{Scatterings of Goldstone bosons off the heavy mesons} \label{sec:uchpthm}

As discussed in Sec.~\ref{sec1.1}, the $D^*_{s0}(2317)$ and $D_{s 1}(2460)$ are two narrow charm strange mesons with the spin-parity $J^P=0^+$ and $J^P=1^+$, respectively. They are located around $45$ MeV below the $DK$ and $D^*K$ thresholds, respectively. {Since} their observation, their inner structures remain puzzling. The two parity-even $D_s$ states cannot  be simply  categorized  as the $ c \bar s$ mesons in the quark model, since their masses are  much smaller than the quark model predictions. Moreover, the mass splittings between the SU(3) flavor partners are unnatural with $m_{D^{*}_{s0}(2317)}-m_{D_0^{*}(2300)}=-25.2$ MeV and $m_{D_{s1}(2460)}-m_{D_{1}(2430)}=47.5 $ MeV, which are predicted to be around $100$ MeV due to the strange quark mass in the conventional quark model. {Various explanations have been proposed, including the quenched and unquenched $c\bar{s}$ quark model, the $D^{(*)}K$ hadronic molecules, the $cq\bar s\bar q$ compact tetraquark states, the mixing of the $c\bar{s}$ state and tetraquark state, and other possibilities (see Sec.~\ref{sec1.1} for more discussions). The proximity of the two $D_s^{(*)}$ states to the $D^{(*)}K$ thresholds has led to the popularity of the $D^{(*)}K$ molecular explanation, which suggests that the dominant component of these mesons is a $D^{(*)}K$ molecule. The molecular explanation was first proposed in Refs.~\cite{Barnes:2003dj,vanBeveren:2003kd} and has been further studied in various works (for more details, see the reviews~\cite{Chen:2016spr,Dong:2017gaw,Guo:2017jvc,Yao:2020bxx}). The molecular framework provides a consistent explanation of the mass puzzles and the fine-tuning problem in Eq. (\ref{eq:finetuning}). In this subsection, we will review research works on the formation of the molecular state through two widely used interaction mechanisms and the unitary approach, resulting in the molecular state as a pole of the $T$ matrix.}

In Sec.~\ref{sec:uchpt_Ds}, the $D^*_{s0}(2317)$ and $D_{s 1}(2460)$ are identified as the dynamically generated poles originating from the resummations of the $\cal H \varphi$ interactions, and are interpreted as the $D^{(*)}K$ molecules. {{The $\cal H \varphi$ 
 scattering amplitude expanded in $\chi$PT (see Sec.~\ref{sec:4.1}) was used as the kernel interactions in the chiral unitary approaches.}} The molecular picture seemed to successfully describe the experimental data on the $\mathcal H \varphi$ invariant mass distribution as well as the lattice QCD simulations.

In Sec.~\ref{sec:uchptDscore}, the kernel potentials contain both the $\cal H \varphi$ hadronic potentials and the contributions from the $c\bar s$ cores. In this case, the physical states are the mixtures of both the $c\bar s$ and the $D^{(*)}K$ components. The contents of the components are vital to identify the nature of the resonance. If the molecular components are dominant in the $D^*_{s0}(2317)$ and $D_{s 1}(2460)$ states, they could still owe their origin to the $\cal H\varphi$ scattering and be understood as the hadronic molecules. 

\begin{figure}[btp]
	\begin{center}
		\includegraphics[width=0.70\textwidth]{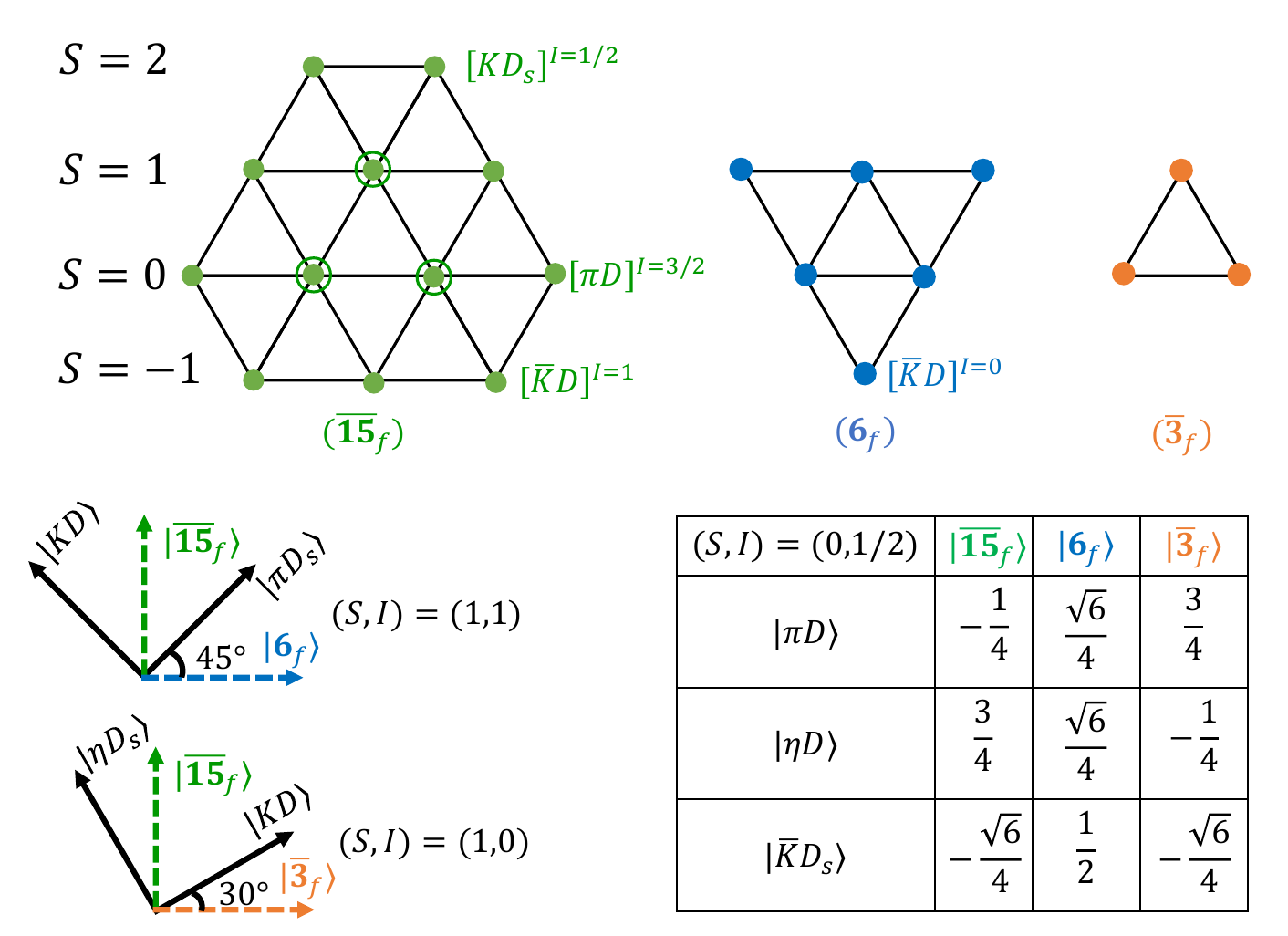}
		\caption{The flavor representations of the ${\cal H} \varphi$ system in the SU(3) flavor symmetry and relations with di-hadron basis. The heavy ($\cal H$) and light ($\varphi$) mesons are in the $\bar {\bm 3}_f$ and ${\bm 8}_f$ representations, respectively. Thus, the possible flavor representations of the ${\cal H} \varphi$ system are $\bar{\bm 3}_f \otimes {\bm 8}_f=\overline{\bm{1 5}}_f \oplus {\bm 6}_f \oplus \bar {\bm 3}_f$~\cite{Albaladejo:2016lbb,Gamermann:2006nm}. ``$S$" is the strange number. The upper panel is the weight diagrams of representation taken from Ref.~\cite{Albaladejo:2016lbb}. For the states in the SU(3) symmetry corresponding to the specific di-hadron system, we label them in the upper panel. For the states corresponding to the mixture of different di-hadron states, we illustrate their mixing angles in the lower panel, including the $(S,I)=(1,1),(1,0)$ and $(0,1/2)$ ones.} \label{Fig:hmsu3}
	\end{center}
\end{figure}

Apart from the successful explanations of the two positive $D_s$ states, the chiral unitary approaches also provide important insights into the non-strange charmed mesons with positive parity. In the SU(3) symmetry limit, the ${\cal H} \varphi$ systems can be categorized into the $\bar {\bm 3}_f$, ${\bm 6}_f$ and $\overline{\bm 15}_f$ flavor representations as illustrated in Fig.~\ref{Fig:hmsu3}. {In the non-strange sector, the coupled channel effects of $D\pi$, $D_s\bar K$ and $D\eta$  are considered. }
The chiral unitary methods predict the two-pole structures for both the scalar $D^*_0$ and axial-vector $D^*_1$ charmed mesons.  {Within the two-pole picture, the lower and higher $D^*_0$ mesons are located around $2100$ MeV and $2450$ MeV, respectively. They mainly couple with the $D\pi$ and $D_s\bar K$ channels, respectively. The two-pole picture could accommodate the broad  $D^*_0(2300)$ signal and explain the mass splitting puzzle in the quark model.} 

In contrast, there are only one scalar and axial-vector charmed mesons listed in the Review of Particle Physics (RPP)~\cite{ParticleDataGroup:2022pth}, the $D_{0}^{*}(2300)$ {[was called as $D_{0}^{*}(2400)$ before]} and $D_{1}(2430)$, which are the two lightest scalar and axial-vertor charmed mesons, respectively. The lattice QCD simulations reported only one resonance, but did not exclude the second one~\cite{Moir:2016srx,Gayer:2021xzv}. There are many theoretical works to understand the ``controversial" results in literature and we will review them in Sec.~\ref{sec:uchpt_D}. 
 With the heavy quark spin symmetry, the similar calculations were generalized to the bottom sectors, especially the unobserved $P$-wave $B^*_s$ states. 
 
\subsubsection{$D^\ast_{s0}(2317)$ and $D_{s 1}(2460)$ in the molecular picture} \label{sec:uchpt_Ds}

The LO chiral amplitude of the $\cal H \varphi$ scattering is parameter-free and used as the kernel interactions in the chiral unitary approaches~\cite{Kolomeitsev:2003ac,Guo:2006fu,Guo:2006rp,Gamermann:2006nm,Gamermann:2007fi,Flynn:2007ki}. 
Kolomeitsev \etal~\cite{Kolomeitsev:2003ac} and Guo \etal~\cite{Guo:2006fu,Guo:2006rp} studied the $J^P=0^{+}$ and $J^P=1^{+}$ heavy mesons with the similar SU(3) chiral Lagrangians.  In the charm-strange sector, two bound states were found in the isosinglet $DK$ and $D^{*} K$ channels and identified as the $D_{s0}^{*}(2317)$ and $D_{s1}(2460)$ states, respectively. For the heavy non-strange mesons with isospin $I=1/2$, the two-pole structures were predicted for the scalar $D^*_0/B^*_0$ and axial vector $D_1/B_1$ mesons, respectively. The lower pole is broad and the higher one is narrow. The two poles belong to the $\bar {\bm 3}_f$ and ${\bm 6}_f$ representations up to some mixing effects. The partner mesons in the bottom sector were also predicted. Moreover, the decay widths of the isospin violating decays $D_{s0}^{*+}(2317)\rightarrow D_{s}^{+}\pi^{0}$ and $B_{s0}^{*0}(5729)\rightarrow B_{s}^{0}\pi^{0}$ were calculated to be $8.69$ keV and $1.54 $ keV, respectively~\cite{Guo:2006fu}. 

In Ref.~\cite{Gamermann:2006nm} and Ref.~\cite{Gamermann:2007fi}, the authors studied the dynamical generations of the open and hidden charm $0^+$ and $1^+$ resonances with the SU(4) flavor symmetric Lagrangians. {However, the SU(4) flavor symmetry { is just an assumption and not a real symmetry of QCD}. In their calculations, the exchange of heavy mesons was suppressed, the SU(4) flavor symmetry was broken into the SU(3) symmetry by suppressing the exchange of the heavy mesons.} The  $D_{s0}^{*}(2317)$ and $D_{s1}(2460)$ were identified as the $DK$  and $D^*K$ bound states. The two-pole structures were also found for the charmed mesons with the strangeness-isospin number  $(S,I)=(0,1/2)$,  similar to Refs.~\cite{Guo:2006fu,Kolomeitsev:2003ac,Guo:2006rp}. The lower scalar and the axial vector poles in the $\bar 3_f$ representation were identified as the  {$D_{0}^{*}(2300)$} and $D_{1}(2420)$ in PDG~\cite{ParticleDataGroup:2022pth}, respectively.  The other higher poles for the charmed mesons with $J^P=0^{+}$ and $J^P=1^{+}$ were broader than the predictions in Refs.~\cite{Guo:2006fu,Kolomeitsev:2003ac,Guo:2006rp}. 
   
To achieve  better accuracy, the NLO chiral amplitude was considered in Refs.~\cite{Hofmann:2003je,Guo:2009ct,Cleven:2010aw,Wang:2012bu}. With the scattering amplitude up to NLO, the mass dependence of the $P$-wave heavy mesons on the light Goldstone meson masses can be investigated. In  Ref.~\cite{Guo:2009ct}, the authors calculated the $S$-wave scattering lengths of the $P\varphi$ system with the NLO chiral amplitude as well as its unitarized version. The unitarized one reproduced the quark mass dependence of the scattering lengths from lattice QCD~\cite{Liu:2008rza}. In Ref.~\cite{Cleven:2010aw}, the authors studied the masses of $D_{\mathrm{s} 0}^{*}(2317)$ and $D_{s 1}(2460)$ as well as their bottom partner states. They  proposed the linear dependence of the heavy meson masses on  $m_K$ as a specific character  of  a molecule, which can be a criterion for  investigating the molecular nature. The difference between the $S$-wave $DK$ and $D^*K$ scattering potentials first arises from the $u$-channel exchange tree diagram as shown in Fig.~\ref{Fig:HMPTree}(c). Compared with the mass splitting of the $D$ and $D^*$ mesons, the difference of the scattering potentials is $ {\cal O} (M_K^{2}/\Lambda_{\chi}^{2})$ . Thus, the hyperfine mass splitting between the  $D^*_{s0}(2317)$ and ${D_{s 1}(2460)}$  almost equals the $D$ and $D^*$ mass difference. Later, the authors of Ref.~\cite{Wang:2012bu}  also calculated the $D_{\mathrm{s} 0}^{*}(2317)$ scattering length and its mass trajectories of $m_{\pi/K}$ as in Ref.~\cite{Guo:2009ct}. They determined the LECs with the help of the scattering lengths from lattice QCD simulations~\cite{Liu:2008rza} rather than applying the large $N_{c}$ approximation as in Ref.~\cite{Guo:2009ct}.  

In 2012, the authors of Ref.~\cite{Liu:2012zya} studied the $P \varphi$ scattering in lattice QCD simulations and obtained the $S$-wave scattering lengths of five channels with several $m_\pi$  without the disconnected contribution: $D \bar{K}~(I=0,1), ~D_{s} K, ~D \pi~(I=3/2), ~D_{s} \pi$. The chiral  extrapolation was incorporated into the unitarized scattering lengths, which were obtained in their previous works~\cite{Guo:2008gp,Guo:2009ct}. 
By fitting the lattice QCD data, they determined the unknown LECs and the subtraction constant.   Their results favored the $DK$ molecular interpretation of the $D_{s 0}^{*}(2317)$. Moreover, the $DK$ was shown to be dominant in the $D_{s 0}^{*}(2317)$ with the probability $70\%$ by using the Weinberg's compositeness relation. The decay width was $\Gamma[D_{s 0}^{*}(2317) \rightarrow D_{s} \pi]=(133 \pm 19) $ {keV}.

With the lattice scattering lengths in Ref.~\cite{Liu:2012zya} and the data from another group~\cite{Mohler:2012na,Mohler:2013rwa,Lang:2014yfa} as input, the  light-quark mass and the $N_{C}$ dependencies of the pole positions in the $P\varphi$ channels were  analyzed~\cite{Guo:2015dha}. The pole positions of the  $D_{s0}^{*}(2317)$ as well as the charmed mesons with the strangeness-isospin number $(S, I)=(0,1 / 2)$ were found to be not like that of the $c \bar s/ c \bar q$ mesons. They did not tend to fall down to the real axis at large $N_C$. In contrast, the imaginary part of a genuine $c \bar s/c \bar q$ states tends to vanish in the large $N_C$ limit. 

\begin{figure}[btp] 
	\begin{center} 
	\includegraphics[width=0.73\textwidth]{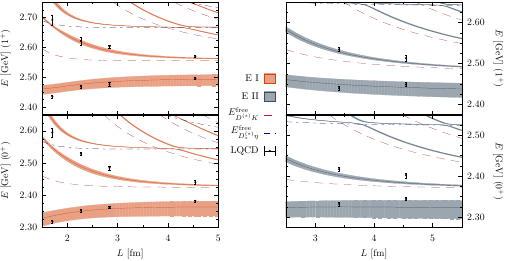}
\caption{\label{fig:levels-NLODs} Comparison of the energy levels using the unitarized NLO $P\varphi$ amplitudes of Refs.~\cite{Liu:2012zya} with the lattice QCD results~\cite{Bali:2017pdv}. The two panels correspond to two sets of lattice QCD data obtained with $m_{\pi}=290$ {MeV} (Ensemble I) and $150$ MeV (Ensemble II), respectively. The 68\% CL uncertainty bands originate from  the errors of the LECs in Ref.~\cite{Liu:2012zya}. Figures taken from Ref.~\cite{Albaladejo:2018mhb}}
	\end{center} 
\end{figure}

The fixed LECs and the unitarized NLO $P \varphi$ amplitude in Ref.~\cite{Liu:2012zya} have been used in the subsequent works~\cite{Albaladejo:2018mhb,Albaladejo:2016lbb,Du:2017zvv}. In Ref.~\cite{Du:2017zvv}, the authors summarized the predictions of the pole positions. The results of  the lowest positive-parity $D_{s 0}^{*}/D_{s 1}$ and  $B_{s 0}^{*}/B_{s 1}$ were consistent with the available experimental data and the lattice QCD results~\cite{Bali:2017pdv,Lang:2015hza}, respectively. For the heavy non-strange ones, two-pole structures were predicted: one flavor antitriplet and a  nontrivial sextet meson as shown in Fig.~\ref{Fig:hmsu3}. The authors of Ref.~\cite{Albaladejo:2018mhb} calculated the energy levels in the finite volume and compared their results with the lattice QCD simulations. The finite volume effect was taken into consideration. In a finite box with the size $L$, the finite-volume effects are induced by discretizing the three momentum $\vec{q} \in \mathbb{R}^{3}$ and its integral as follows
  \begin{eqnarray}  \label{eq:finiteq}
\vec{q}\rightarrow \frac{2 \pi}{L} \vec{n}, ~(\vec{n} \in \mathbb{Z}^{3}),\qquad\qquad
\int_{\mathbb{R}^{3}} \frac{d^{3} q}{(2 \pi)^{3}} \rightarrow \frac{1}{L^{3}} \sum_{n \in \mathbb{Z}^{3}}.
      \end{eqnarray} 
The loop function ${G}_{i i}(s) $ then transforms into its discrete form $\tilde{{G}}_{i i}(s, L)$. Correspondingly, the $T$-matrix $T(s)$ in the discrete form (in the cutoff regularization) reads~\cite{Doring:2011vk},
  \begin{eqnarray}  \label{eq:fvg}
 {G}_{i i}(s) & \rightarrow& \tilde{{G}}_{i i}(s, L)={G}_{i i}(s)+\lim _{\Lambda \rightarrow \infty}\left(\frac{1}{L^{3}} \sum_{\vec{n}}^{|\vec{q}|<\Lambda} I_{i}(\vec{q})-\int_{0}^{\Lambda} \frac{q^{2} d q}{2 \pi^{2}} I_{i}(\vec{q})\right), \\ 
 V(s) & \rightarrow& \tilde{V}(s, L)=V(s), \\ 
 T^{-1}(s) & \rightarrow& \widetilde{T}^{-1}(s, L)=V^{-1}(s)-\tilde{{G}}(s, L), 
      \end{eqnarray} 
where the integrand $I_i(\vec{q})$ can be read from Eq.~\eqref{eq:loopfcut}. This formalism gave a nice description of the energy levels of the $0^{+}$ and $1^+$ charmed-strange states in lattice QCD simulations~\cite{Bali:2017pdv} as shown in Fig.~\ref{fig:levels-NLODs}.
  
One should note the all the parameters of the unitarized amplitudes have been determined using the scattering lengths in  lattice QCD simulations~\cite{Liu:2012zya} prior to Ref.~\cite{Bali:2017pdv}. The  $DK$ ($I=0$) scattering lengths in the two lattice works are   
   \begin{eqnarray} 
a_s^{\text{I}}=-0.86(3)~\text{fm}~ 
\qquad \qquad 
a_s^{\text{II}}=-1.49(0.13)(-0.30)~\text{fm},
   \end{eqnarray} 
   where $a_s^{\text{I}}$ and $a_s^{\text{II}}$ are the results from Refs.~\cite{Liu:2012zya,Bali:2017pdv}, respectively.

In Ref.~\cite{Guo:2018tjx}, the authors fitted the finite-volume energy levels and scattering lengths from lattice calculations and  successfully described the lattice data~\cite{Liu:2012zya,Lang:2015hza,Moir:2016srx}. After performing the chiral extrapolation, they made predictions of the resonance parameters at the physical $m_\pi$, including the phase shifts, the inelasticities, the pole position, and the residues of the pole. They reproduced the $D_{s 0}^{*}(2317)$ mass and found two possible solutions corresponding to the bound and virtual states, respectively. The two poles with the strange-isospin number $(S, I)=(0,1 / 2)$ persist. The lower one has the mass around $2100$ MeV and width more than $200$ MeV, while the higher one has the mass $2300 \sim 2500$ {MeV} and width $70\sim 200$ MeV. 
   
The chiral amplitudes were also extended to N$^2$LO~\cite{Yao:2015qia,Du:2017ttu,Guo:2018kno,Guo:2018gyd}. In Ref.~\cite{Yao:2015qia}, the authors calculated the  ${\cal H} \varphi$ scattering amplitude up to N$^2$LO with the $\chi$PT in the EOMS scheme without the contribution of the $D^*$.  With both BSE and the IAM approaches,
they obtained the unitarized amplitude and 
the scattering length. With the chiral extrapolation of the meson masses and the decay constants, they got a good fit to the lattice QCD data of the $S$-wave scattering lengths~\cite{Liu:2012zya,Mohler:2013rwa}. However, most of the parameters suffered from large uncertainties. They also calculated the diagrams with the $D^*$ and found its contribution to be negligible. Later in Ref.~\cite{Du:2017ttu}, they performed a complete calculation with the contribution of the explicit vector charmed meson and used a similar fit procedure as in Ref.~\cite{Yao:2015qia}. Their results showed that the $D^{*}$ contribution was negligible for the $S$-wave scattering near the threshold. They also searched for the $J^{P}=0^{+}$ poles on the Riemann sheets and presented their trajectories with the varying pion mass. The LECs still bear large uncertainties. More lattice data are required to obtain more solid conclusions.

In Refs.~\cite{Guo:2018kno}, the authors used the lattice results of both the heavy meson masses and the  $S$-wave scattering lengths to determine the LECs of the Lagrangians up to $\mathcal O(p^4)$. They calculated the isospin-violating strong decay width $\Gamma[D_{s 0}^{*}(2317)]=(104-106)$ keV and found a clear signal of the exotic sextet charmed meson in the $D\pi$~\cite{Guo:2018kno} and $D^*\pi$~\cite{Guo:2018gyd} $S$-wave coupled systems, which tended to support the two-pole structure for the charmed meson with $J^P=0^+$ and $J^P=1^+$. A similar calculation was extended to the bottom sectors~\cite{Guo:2021rjv}.
 
Besides the lattice data, the $B$ decay process also provided a platform to test the molecular explanation. In Ref.~\cite{Albaladejo:2016hae}, the authors studied the $B$ decays $B^{+} \rightarrow \bar{D}^{0} D^{0} K^{+}$, $B^{0} \rightarrow D^{-} D^{0} K^{+}$~\cite{BaBar:2014jjr} and $B_{s} \rightarrow \pi^{+} D^{0} K $~\cite{LHCb:2014ioa}. The  unitarized LO amplitude for the $DK$ scattering~\cite{Guo:2006fu,Flynn:2007ki} was used, which contains an unknown subtraction constant $a(\mu)$ in Eq.~\eqref{eq:loopfdr}. The $D_{s 0}^{*}(2317)$ appeared as the $DK$ bound state and their presence led to the enhancement in the $D K$ invariant mass spectra. 
With the Weinberg's compositeness condition, they extracted the $DK$ component to be $P_{D K}=70_{-6~-8}^{+4~ +4} \%$.  In Refs.~\cite{Du:2017zvv,Du:2019oki}, the authors fitted the LHCb data  of $B^{-} \rightarrow D^{+} \pi^{-} \pi^{-}$~\cite{LHCb:2016lxy} and $B_{s}^{0} \rightarrow \bar{D}^{0} K^{-} \pi^{+}$~\cite{LHCb:2014ioa} with the unitarized NLO amplitudes, which will be explained in Sec.~\ref{sec:uchpt_D}. 

The chiral unitary methods were also applied to the other open-charm system. For instance, the $X(2900)$ was explained as the dynamically generated $D K_{1}$ resonance. More discussions can be found in Refs.~\cite{Chen:2016spr,Dong:2017gaw,Guo:2017jvc,Yao:2020bxx}.

 \subsubsection{$D^\ast_{s0}(2317)$ and $D_{s 1}(2460)$ with the $(c\bar s)$ component} \label{sec:uchptDscore}
 
{{In this subsection, we review the works discussing the interplay between the $c\bar{s}$ and $D^*K$ channels in the study of the $D^\ast_{s}$ states. Since the quark model almost successfully explains the spectrum of the conventional ground state hadrons, one naturally expects the existence of the $P$-wave $c\bar s$ mesons. If the masses of the $c\bar{s}$ mesons and $D^{(*)}K$ components are close, their coupled-channel effects may be significant and should be taken into account. The physical state is a mixture of both the conventional $c\bar{s}$ meson and the $D^{(*)}K$~component~\cite{vanBeveren:2003kd,vanBeveren:2003jv,Ortega:2016mms,Browder:2003fk,Mohler:2011ke,MartinezTorres:2011pr,Mohler:2013rwa,Lang:2014yfa,Albaladejo:2018mhb,Bali:2017pdv,Yang:2021tvc}. The proportion of each component will be related to the strength of their coupling. So far, the interplay of the molecular and compact components is not yet well understood. In literature, there are a lot of works discussing their  interplay~\cite{Hanhart:2011jz,Hanhart:2022qxq,Sekihara:2014kya,Sazdjian:2022kaf,Hyodo:2011qc,Guo:2015daa,Guo:2016bjq}. If the coupling is weak, the physical state will be dominated by the compact $c\bar{s}$ component, with the $D^{*}K$ component dressing the $c\bar{s}$ core. Recently, an interesting mechanism was proposed which allows the molecular states to decouple from the compact states in the strong coupling regime \cite{Hanhart:2022qxq}. The authors considered $N$ compact resonances and a scattering channel. When these compact resonances strongly interact with the scattering state, $N-1$ compact states are formed as the dressed cores while a single molecular state emerges as a bound or virtual state. In this molecular state, the compact components are suppressed compared to the molecular component, leading to the decoupling of the molecular and compact states.}}

To understand the structure of a particle and pin down whether it is elementary or composite, the probability of different components is of special importance.  { As outlined in Sec.~\ref{sec2.5.4}, the probability of finding the compact component of the bound states can be quantified using the field renormalization constant $Z$, which is equivalent to the overlap of the bound state wave function with the bare $c\bar s$ core. $1-Z$ represents the molecular (two-body scattering) component. $Z$ ranges from zero to one. A value close to $Z\sim 0$ indicates that the physical state is primarily molecular, while a value close to $Z\sim 1$ implies that it is predominantly compact. Moreover, the couplings of the $T$ matrix poles to different components are related to $Z$, which conveys the information of the strength of the coupling.}

In some works, the $D_{s 0}^{*}(2317)$ and $D_{s 1}(2460)$ states are dominated by the genuine $c\bar s$ states but the masses of the bare $c\bar s$ core are lowered through their coupling to the $D^{(*)} K$ channels as discussed in Sec.~\ref{Sect.2.1.1}. In Ref.~\cite{Ortega:2016mms}, the authors studied the $c\bar s$ meson with the quark model ~\cite{Vijande:2004he,Valcarce:2005em,Segovia:2013wma} and derived its coupling with the $D K\left(D^{*} K\right)$ channels using the $^3P_0$ model~\cite{LeYaouanc:1972vsx}. With both the $c\bar s $ and $D^*K$ degrees of freedom, they obtained the Hamiltonian with the Resonating Group Method (RGM)~\cite{Tang:1978zz}. By solving the coupled-channel Schrödinger-type equation, they found that the coupling with the $D^{(*)}K$ scattering states will dress the $c\bar s$ state and lead to its mass shift. The dressed state $D^*_{s0}(2317)$  contained around the $34 \%$ $D K$ component. The percentages of the $D^*K$ component in the  $D_{s 1}(2460)$ and $D_{s 1}(2536)$ are around $50 \%$. 

In Ref.~\cite{Alexandrou:2019tmk}, the authors investigated the $D^*_{s0}(2317)$ state using lattice QCD simulations and found that the $D^*_{s0}(2317)$ is mainly the $c\bar s$ state in the quark model with a small $DK$ component. 
However, the $DK$ component was found to be dominant in the $D_{s 0}^{*}(2317)$ from other lattice QCD calculations~\cite{Bali:2017pdv,Mohler:2013rwa,Cheung:2020mql} and in some theoretical works~\cite{Liu:2012zya,Albaladejo:2016hae,MartinezTorres:2014kpc,Albaladejo:2018mhb,Yang:2021tvc} with the probability larger than $60\%$. In Ref.~\cite{MartinezTorres:2014kpc}, the extracted percentages of the $ DK$ components were $70\%$ in the $D_{s0}^{*}(2317)$ and $(57\pm 21\pm 6)\%$ in the $D_{s 1}(2460)$ state, respectively. The authors reanalyzed the three energy levels in the lattice QCD simulations~\cite{Mohler:2013rwa,Lang:2014yfa} beyond the effective range expansion. They used the auxiliary potentials to construct the unitarized $T$-matrix in two schemes. In the first scheme, they considered the one-single channel $D^{(*)}K $ with two energy dependent potentials. One potential was the linear function of $s$ and the other one included an additional pole to account for a genuine $c\bar s$ contribution explicitly. In the second scheme, the coupled-channel unitarized $T$-matrix with the $D^{(*)}K$, $D_{s}^{(*)}\eta$ (energy independent potential) was considered.  With an extended Lüscher method as shown in Eq.~\eqref{eq:fvg}~\cite{Doring:2011vk,MartinezTorres:2011pr}, they determined the two poles of the $T$-matrix for the $D_{s 0}^{*}(2317)$ and $D_{s 1}(2460)$ and identified them as the $DK$ and $ D^{*}K$ bound state. The reformulation of the Weinberg's compositeness condition ~\cite{Weinberg:1965zz,Baru:2003qq} was used to extract the probability of the meson-meson component and its generalization form reads~\cite{Hyodo:2013nka,Aceti:2014ala},
\begin{eqnarray} \label{eq:wcg}
-\sum_{i} g_{i}^{2} \frac{\partial G_{i}}{\partial s}-\sum_{i, j} g_{i} g_{j} G_{i} \frac{\partial V_{ij}}{\partial s} G_{j}=1,
\end{eqnarray}
where $G_i$ is the loop function in the $i$th two-meson channel. $V_{ij}$ is the potential between the $i$th and $j$th two-meson channels and $g_i$ is the coupling of the physical state to the $i$th two-meson channel. The first term stands for the probabilities of the meson-meson  components.  

In Ref.~\cite{Albaladejo:2018mhb}, the authors considered both the $c\bar s $ and two-meson channels such as the $D^{(*)}K$ and $D^{(*)}\eta$ to describe the $0^{+}$ and $1^{+}$ charm-strange energy levels from the lattice QCD calculation~\cite{Bali:2017pdv}. The ${\cal H} \varphi$ interactions were obtained by the LO scattering amplitude with the HH$\chi$PT. Two sets of the $c\bar s$ bare masses were employed~\cite{Segovia:2012yh}. The Lagrangian for the coupling of the $\cal H \varphi$ channel with the bare $c\bar s$ mesons is given in Eq.~\eqref{eq:app1:lagSH} which contains a dimensionless LEC $h$. One should note that Eq.~\eqref{eq:app1:lagSH} has employed the heavy quark spin symmetry in the system. The authors constructed the unitarized $T$-matrix with the chiral unitary method and  extended it to the finite volume with Eq.~\eqref{eq:fvg}. They successfully described the energy levels in the $0^+$ and $1^+$ channels in lattice QCD simulations. With the Weinberg's compositeness condition in Eq.~\eqref{eq:wcg}, their results showed that the $D_{s 0}^{*}(2317)$ and $D_{s 1}(2460)$ resonances have the predominantly hadronic molecular $DK$ and $D^* K$ components, with the probabilities being $65 \%$ and $56 \%$, respectively. Moreover, one additional broad resonance was found in both the  $0^{+}$ and $1^{+}$  channels. The similar formalism was also applied to study the analogue resonances of the two $B^*_s$ states in the bottom sector~\cite{Albaladejo:2016ztm} with the help of the lattice data~\cite{Lang:2015hza}. They found two poles in the $S$-wave $\bar{B}^{(*)} K$ amplitude with the masses $5709 \pm 8 \mathrm{MeV}\left(0^+\right)$  and $5755 \pm 8 \mathrm{MeV}\left(1^+\right)$, corresponding to the $\bar{B}_{s 0}^{*} $ and $ \bar{B}_{s 1}$. The molecular $\bar{B}^{(*)} K$ components were around $50 \%$.

In Ref.~\cite{Yang:2021tvc}, the authors included both the $c\bar s$ and the $D^{(*)}K$ components explicitly and 
systematically studied the four positive parity $D_{s}$ states: $D_{s0}^{*}(2317)$, $D_{s1}(2460)$, $D_{s1}^{*}(2536)$ and $ D_{s2}^{*}(2573)$ with the chiral unitary approach. The authors employed the
Godfrey-Isgur (GI) quark model~\cite{Godfrey:1985xj} to specify the $c\bar s$ core explicitly, and refitted the masses of the well-established mesons. The $P$-wave bare $c\bar s$ cores were almost automatically accommodated into the heavy quark
spin basis without the pre-implementation of HQSS, which implies that the HQSS is a good approximation in the $D_s$ sectors. The couple-channel effects $(g)$ between the bare $c\bar s$ core and the $D^{(*)}K$ channels were described by the quark pair creation (QPC, aka, the $^3P_0$) model~\cite{LeYaouanc:1977fsz,Kokoski:1985is,Page:1995rh,Blundell:1996as,Ackleh:1996yt,Morel:2002vk, Ortega:2016mms,Morel:2002vk, Ortega:2016mms}. The interactions ($v$) between the two-meson channels $D^{(*)}K$ were derived with the vector meson exchange Lagrangians~\cite{Lin:1999ad,Oset:2009vf,Zhao:2014gqa}.

\begin{figure}[!htp]
\centering
\includegraphics[width=0.8\linewidth]{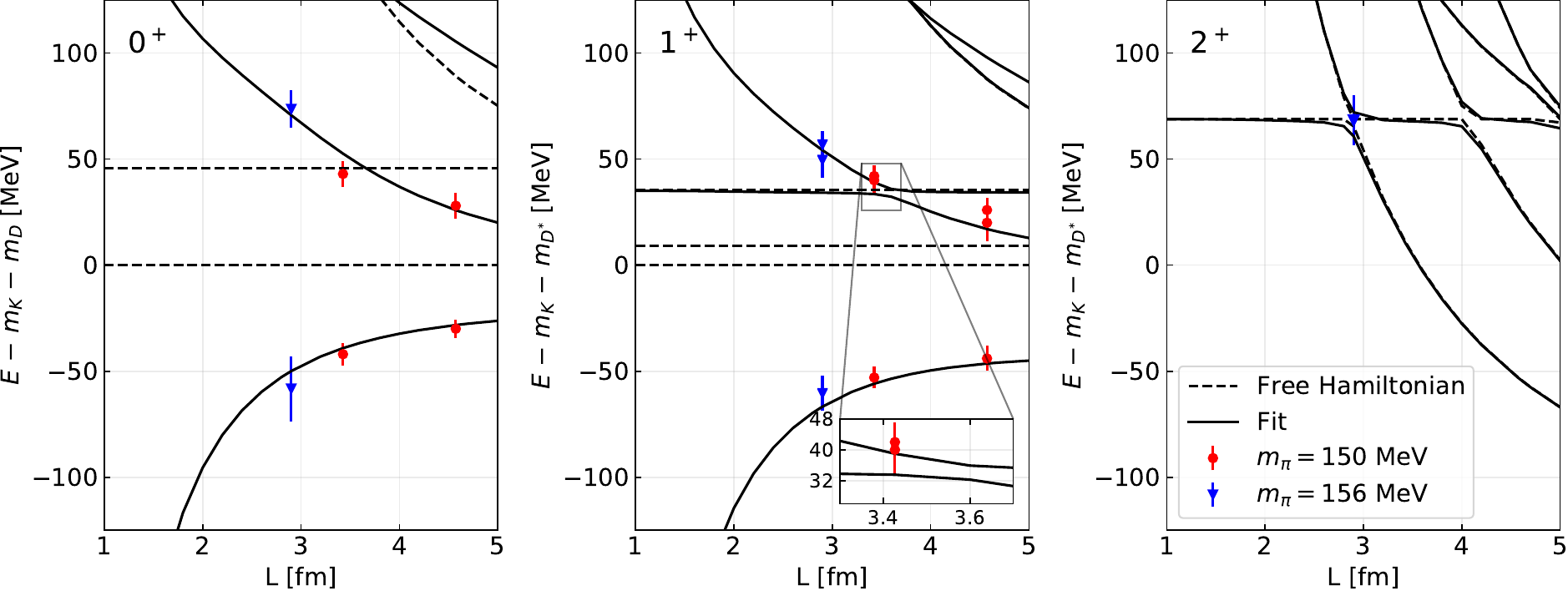}
\caption{The comparison of the dependence of lattice binding energies on length $L$ for the $D^*_{s0}(2317)$ (left), the $D_{s1}(2460)/D_{s1}^\ast(2536)$ (middle), and $D^*_{s2}(2573)$ (right) states with the pion mass $m_{\pi}=150$ MeV~\cite{Bali:2017pdv} and $m_{\pi}=156$ MeV~\cite{Lang:2014yfa}. The black curves stand for the eigenvalues using the finite-volume Hamiltonian, while the dashed lines represent those of the bare $c\bar s$ cores and $D^{(*)}K$ thresholds obtained with the free Hamiltonian $H_0$. Figures taken from Ref.~\cite{Yang:2021tvc}.}
\label{fig:fitspec}
\end{figure}

To use the lattice QCD data, they adopted the Hamiltonian effective field theory (HEFT)~\cite{Hall:2013qba, Wu:2014vma, Hall:2014uca, Liu:2015ktc}. The momentum is discretized according to Eq.~\eqref{eq:finiteq} in a cubic box with the size $L$. However, instead of the loop function, the Hamiltonian matrix is discretized in the HEFT as follows  
\begin{eqnarray}
H_{0}&=&\sum_{i=1,n}|B_{i}\rangle m_{i}\langle B_{i}|+\sum_{\alpha,k}|\vec{q}_{k},-\vec{q}_{k}\rangle_{\alpha}\left(\sqrt{m_{\alpha_{B}}^{2}+q_{\alpha}^{2}}+\sqrt{m_{\alpha_{M}}^{2}+q_{\alpha}^{2}}\right)_{\alpha}\left\langle \vec{q}_{k},-\vec{q}_{k}\right|,\\
H_{I}&=&\sum_{k}\left(\frac{2\pi}{L}\right)^{3/2}\sum_{\alpha}\sum_{i=1,n}\left(|\vec{q}_{k},-\vec{q}_{k}\rangle_{\alpha}g_{i,\alpha}^{\dagger}(s,\vec{q}_{k})\left\langle B_{i}|+|B_{i}\right\rangle g_{i,\alpha}(s,\vec{q}_{k})\langle\vec{q}_{k},-\vec{q}_{k}|\right)\\ \nonumber
&&+\sum_{k,l}\left(\frac{2\pi}{L}\right)^{3/2}\sum_{\alpha,\beta}\left|\vec{q}_{k},-\vec{q}_{k}\right\rangle _{\alpha}v_{\alpha,\beta}(s,\vec{q}_{k}-\vec{q}_{l}){}_{\beta}\left\langle \vec{q}_{l},-\vec{q}_{l}\right|,
\end{eqnarray}
where $H_0$ and $H_I$ are the free and interacting Hamiltonian. The $i,j$ and $\alpha, \beta$ represent the bare $c\bar s$ core and the $D^{(*)} {K}$ channels, respectively. $k,l$ are the indices of the {discretized} momentum. The discretized energy levels were obtained from the eigenvalues of the Hamiltonian matrix for each $L$, which are used to compare with the ones in the lattice QCD as shown in Fig.~\ref{fig:fitspec}. In the figure, the data in the $0^+$ and $1^+$ sectors was used as input. In the $0^+$ channel, the $0^+$ $c\bar s$ core and the $S$-wave $DK$ channels were involved. In the $1^+$ one, two $c\bar s$ cores with $J^P=1^+$ as well as the $S$-wave and $D$-wave $D^*K$ channels were considered. The lowest eigenvalue corresponded to the $D_{s 1}(2460)$ and its $c\bar s$ core had a significant mass shift due to the coupling with the $S$-wave $D^*K$ channel. In contrast, the $c\bar s$ core for the $D_{s 1}^{*}(2536)$ mainly couples with the $D$-wave $D^*K$ channel and its energy levels tended to be stable. With the increasing length, the kinetic energy of the $D^*K$ channel  decreased. When its eigenvalue approached that of the $D_{s 1}^{*}(2536)$, an interesting crossing appeared around $L=3.5$ fm. In the $2^+$ sector, the results for the $D_{s 2}^{*}(2573)$ are the predictions from the Hamiltonian matrix with the $2^+$ $c\bar s$ core and the $D$-wave $DK$ as well as the $D^*K$ channels. The authors determined the parameters and got the unitarized $T$-matrix through the
Lippmann-Schwinger equation in the infinite volume. The physical $D^*_{s0}(2317)$ and $D_{s1}(2460)$ are the mixtures
of the bare $c\bar s$ core and $D^{(*)}K$ component with the probabilities  $P_{D^*_{s0}(2317)}(DK)\approx68.0 \%$ and $P_{D_{s1}(2460)}(D^*K)\approx47.6\%$, respectively. The
$D^*_{s1}(2536)$ and $D^*_{s2}(2573)$ are almost dominated by the
bare $c\bar{s}$ core with $P_{D^*_{s1}(2536)}(c\bar s)\approx98.2\%$ and $P_{D^*_{s2}(2573}(c\bar{s})\approx 95.9\%$, respectively. The different mass shifts and mixing patterns of the four $D_s$ states are governed by the heavy quark symmetry. In the heavy quark limit, the $c\bar s$ cores in the $D^*_{s0}(2317)$ and $D_{s1}(2460)$ can couple with the $S$-wave $DK$ and $D^*K$ channels, respectively. However, the coupling of  
the $c\bar s$ cores in the $D^*_{s1}(2536)$ and $D^*_{s2}(2573)$ with the $S$-wave two-meson channels are forbidden by HQSS. They can only couple with the $D$-wave ones which are significantly suppressed in the vicinity of the thresholds. Another prediction was that the mass of the $D_{s 0}^{*}(2317)$ tended to be stable with increasing $m_\pi$ which can be checked in the future lattice calculation.

{{With the interplay of the $c\bar s$ and $D^*K$ components, 
the positive parity $D_s$ states can take two forms: either a compact $c\bar s$ state dressed by the $D^{(*)}K$ or a molecular state dominated by the $D^{(*)}K$ components. Both scenarios can explain the mass shifts observed in comparison to the prediction from the conventional quark model. To distinguish these two scenarios, the experimental observations of the electromagnetic decays into $D_s^{(*)}\gamma$ or strong decays into the isospin-violating $D^{(*)}\pi^0$ channels will be very helpful. In Sec.~\ref{Sect.2.4.1}, the significant differences in the decays of $D^*_{s0}(2317)$ and $D_{s1}(2460)$ in the compact and molecular scenarios were discussed. The decay widths for the $D^{(*)}\pi$ modes are less than 10 keV in the compact scenario \cite{Colangelo:2003vg,Fajfer:2015zma}, while the molecular scenario leads to the enhanced decay widths due to the isospin-breaking effects in the $K$ and $D^{(*)}$ meson masses via loops. For examples, the widths were around 100 keV in Refs. \cite{Fu:2021wde,Lutz:2007sk,Guo:2008gp}.}}

 {{One may also compare the formation mechanisms of the molecular $D^{*}K$ states. In Sec.~\ref{sec:uchpt_Ds}, the molecular state is formed via the $D^{(*)}K$ interactions in $\chi$PT, which are then used as the kernel in the chiral unitary method. This subsection considers the interplay between the $c\bar s$ component and the $D^{(*)}K$. Besides the interactions in the $D^{*}K$ channels, the $D^{(*)}K$ can also interact through exchanging the compact $c\bar s$ state. Both mechanisms without and with the $c\bar s$ component can describe the physical $D^*_{s0}(2317)$ and $D_{s1}(2460)$ masses as well as lattice QCD data and yield the consistent probabilities of the $D^{(*)}K$ components (over $60\%$). These probabilities were extracted using either the Weinberg's compositness relation or the wave function of the physical state. The contribution of the $c\bar s$ component to the $D^{(*)}K$ interactions in the latter case is short-range, which may be incorporated into the LECs in the chiral Lagrangian. These LECs can be determined from experiments or lattice simulations. Then both mechanisms contain physically similar information and can be considered more or less equivalent in this context as the results are consistent. }}

\subsubsection{Two-pole structures of $D_{0}^{*}(2300) $ and $D_{1}(2430)$} \label{sec:uchpt_D}

The structures of the $D^*_{s0}(2317)$ and the $D_{s1}(2460)$ have been successfully disentangled with the chiral unitary methods as discussed in Sec.~\ref{sec:uchpt_Ds}. Under the SU(3) flavor symmetry, the unitarized scattering amplitude can be directly extended to study the other $\cal H \varphi$ scattering channels as shown in Fig.~\ref{Fig:hmsu3} and Table~\ref{tab:SI}. The possible strangeness-isospin quantum numbers of the $\cal H \varphi$ channels are
  \begin{eqnarray}
(S,I)=(2,\frac{1}{2}),~(1,0),~(1,1),~(0,\frac{1}{2}),~( 0,\frac{3}{2}),~(-1,0),~(-1,1).
  \end{eqnarray}

\begin{table}
\centering
\renewcommand{\arraystretch}{1.5}
 \tabcolsep=1.5pt
\caption{The possible strange-isospin quantum numbers $(S,I)$  for the $\mathcal H \varphi$ system. The symbols ``$\times$", ``$\otimes$", and ``$\medcirc$" represent that the pole does not exist, exists perhaps, does exist, respectively. The script ``Irreps." denotes the irreducible representations in the SU(3) flavor symmetry.}\label{tab:SI}
\setlength{\tabcolsep}{1.45mm}
{
\begin{tabular}{cccccc}
\hline 
$(S,I)$ & Irreps. & channels & $\langle  \boldsymbol{F}_{\mathcal{H}} \cdot \boldsymbol{F}_{\varphi}\rangle _{{\cal H} \varphi}$ & Poles & References \tabularnewline
\hline 
$(2,\frac{1}{2})$ & $\overline{\bm{15}}_{f}$ &  $D_{s}K$ & $\frac{1}{2}$ &$\times$ & \tabularnewline
\multirow{2}{*}{$(1,1)$} & \multirow{2}{*}{$\overline{\bm{15}}_{f}$,$\bm{6}_{f}$} & \multirow{2}{*}{  $\left(\begin{array}{c}
DK\\
D_{s}\pi
\end{array}\right)$} & \multirow{2}{*}{$\left(\begin{array}{cc}
0 & -\frac{1}{2}\\
-\frac{1}{2} & 0
\end{array}\right)$ } & \multirow{2}{*}{  $\left(\begin{array}{c}
\times\\
\times
\end{array}\right)$} & cusp effect~\cite{Kolomeitsev:2003ac,Guo:2018tjx}, \tabularnewline
& & & & & no experimental relavant pole~\cite{Albaladejo:2016lbb,Hofmann:2003je,Gamermann:2006nm}.
\tabularnewline
$(1,0)$ & $\overline{\bm{15}}_{f}$,$\bar{\bm{3}}_{f}$ &  $\left(\begin{array}{c}
DK\\
D_{s}\eta
\end{array}\right)$ & $\left(\begin{array}{cc}
-1 & -\frac{\sqrt{3}}{2}\\
-\frac{\sqrt{3}}{2} & 0
\end{array}\right)$ & { $\left(\begin{array}{c}
\medcirc\\
\times
\end{array}\right)$}
 & $D^*_{s0}(2317)$ (see Sec.~\ref{sec:uchpt_Ds})\tabularnewline
$(0,\frac{3}{2})$ & $\overline{\bm{15}}_{f}$$ $ &  $D\pi$ & $\frac{1}{2}$ & $\times$ &\tabularnewline
$(0,\frac{1}{2})$ & $\overline{\bm{15}}_{f}$,$\bm{6}_{f}$,$\bar{\bm{3}}_{f}$ &  $\left(\begin{array}{c}
D\pi\\
D\eta\\
D_{s}\bar{K}
\end{array}\right)$ & $\left(\begin{array}{ccc}
-1 & 0 & -\frac{\sqrt{6}}{4}\\
0 & 0 & \frac{\sqrt{6}}{4}\\
-\frac{\sqrt{6}}{4} & \frac{\sqrt{6}}{4} & -\frac{1}{2}
\end{array}\right)$ &  {  $\left(\begin{array}{c}
\medcirc\\
\times\\
\medcirc
\end{array}\right)$}  & two-pole structures (see Sec.~\ref{sec:uchpt_D})\tabularnewline
$(-1,1)$ & $\overline{\bm{15}}_{f}$$ $ & $D\bar{K}$ & $\frac{1}{2}$ & $\times$ &\tabularnewline
$(-1,0)$ & $\bm{6}_{f}$ & $D\bar{K}$ & $-\frac{1}{2}$ &$\otimes$ &  cusp effect~\cite{Kolomeitsev:2003ac,Guo:2018tjx}, 
virtual~\cite{Albaladejo:2016lbb}, bound states~\cite{Hofmann:2003je}\tabularnewline
\hline 
\end{tabular}}
\end{table}

\begin{figure}[htp]
\centering
\includegraphics[width=0.5\linewidth]{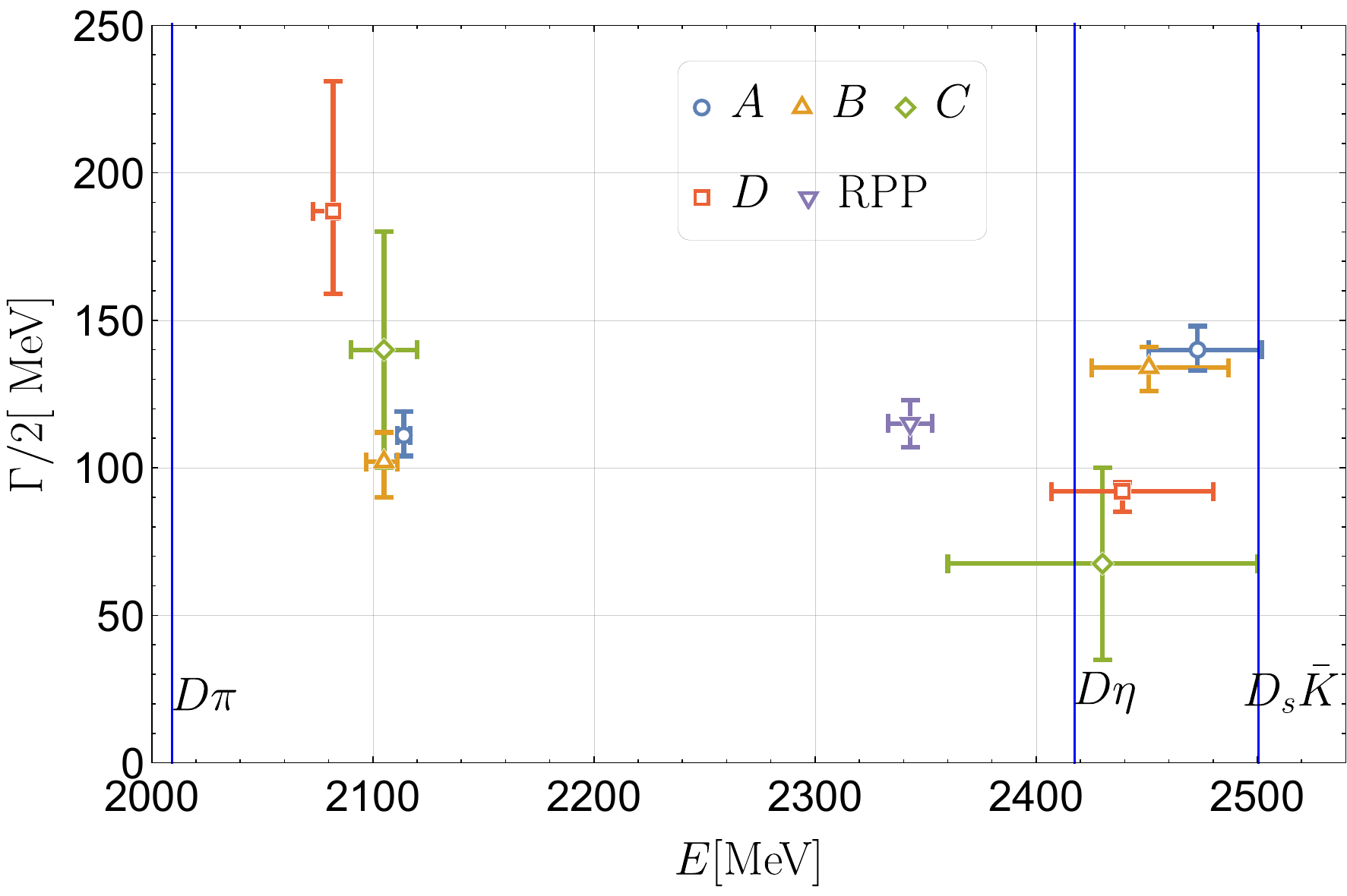}
\caption{The comparison of the two-pole structures in the chiral unitary approaches for the channels with the strangeness-isospin number $(S,I)=(1,\frac{1}{2})$ and RPP~\cite{ParticleDataGroup:2022pth}. The $A,~B,~C,~D$ represent the results in Ref.~\cite{Guo:2015dha}, Refs.~\cite{Albaladejo:2016lbb,Du:2017zvv}, Ref.~\cite{Guo:2018tjx}, Refs.~\cite{Guo:2018kno,Guo:2018gyd}, respectively.}
\label{fig:two-pole}
\end{figure}

One striking observation is that there are two poles  in the channels with the strangeness-isospin quantum numbers $(S,I)=(0,1/2)$ where the   $D^{(*)} \pi$, $D^{(*)}\eta$ and $D^{(*)}_{s} \bar{K}$  are involved in the $0^+$ ($1^+$) sector as illustrated in Fig.~\ref{fig:two-pole}.

The presence of the two-pole structures originated from the Weinberg-Tomozawa terms at LO, which read~\cite{Weinberg:1966kf,Tomozawa:1966jm}~\footnote{The potential is obtained with the scattering amplitude in Eq.~\eqref{eq:wta}. In some theoretical works, the $C_{\alpha, \mathcal{H} \varphi}$  is related to $C_{\text{LO}}$ up to the additional form factors due to the renormalization, for instance $\sqrt{\left(m_{{\cal H}_i}+E_{{\cal H}_i}\right)\left(m_{{\cal H}_j}+E_{{\cal H}_j}\right) /\left(4 m_{{\cal H}_i} m_{{\cal H}_j}\right)}$.}
  \begin{eqnarray} \label{eq:Casimiro}
V^{\text{(WT)}}(s, u)\sim~-C_{\alpha,\cal H\varphi}\frac{(s-u)}{4 f_\varphi^{2}}, \qquad C_{\alpha,\cal H\varphi}=-\left\langle 2 \boldsymbol{F}_{\cal H} \cdot \boldsymbol{F}_{\varphi}\right\rangle_{\alpha}=C_{2}({\cal H})-C_{2}(\alpha)+3,
 \end{eqnarray}
 with $\alpha$ the flavor representation of the $\cal H\varphi$ system. $\boldsymbol{F}_{\cal H}$ and $\boldsymbol{F}_{\varphi}$ are the SU(3) generators for the heavy meson and the light Goldstone boson. $C_2$ is the Casimir operator. For the $\overline{\bm{15}} _f\oplus {\bm 6}_f \oplus \bar {\bm 3}_f$ flavor representations as shown in Fig.~\ref{Fig:hmsu3}, the LO potentials are $V^{\text{WT}}\sim(1,-1,-3)w$ where the $w$ is positive. Thus, the $\cal H\varphi$ potentials are  attractive in the ${\bm 3}_f$ and ${\bm 6}_f$ representation, while they are repulsive  in the $\overline{\bm {15}}_f$ one. Furthermore, the attractive potentials in the anti-triplet channel  are much stronger than that in the sextet one. In the SU(3) limit, the lower pole therefore belongs to the $\bar {\bm 3}_f$ which is the partner state of the $D^*_{s0}(2317)$ [$D_{s1}(2460)$] state with isospin $I=0$, while the higher pole belongs to the sextet. The two-pole structures have been predicted using the LO unitarized scattering amplitude in the chiral unitary method~\cite{Kolomeitsev:2003ac,Guo:2006fu,Gamermann:2006nm,Guo:2006rp,Gamermann:2007fi,Flynn:2007ki} as mentioned in  Sec.~\ref{sec:uchpt_Ds}. In literature, there were extensive discussions of the  similar two-pole structures ~\cite{Meissner:2020khl} in the $K_{1}(1270)$~\cite{Roca:2005nm,Geng:2007ix,Garcia-Recio:2013uva} and $\Lambda(1405)$ systems~\cite{Oller:2000fj,Jido:2003cb,Garcia-Recio:2003ejq,Gubser:2009sn}.  
 
\begin{figure}[hbt] 
	\begin{center} 
\includegraphics[width=0.48\textwidth]{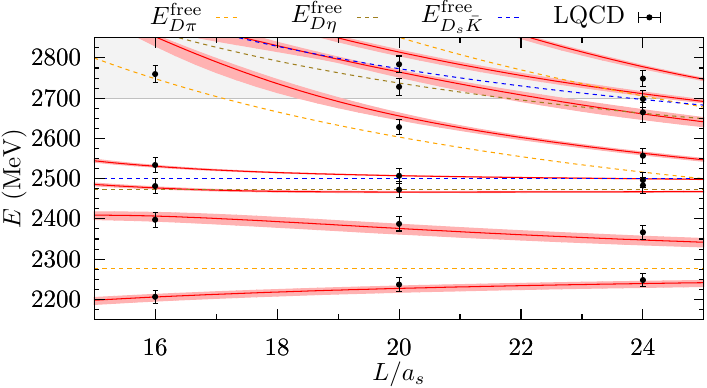}~~
\includegraphics[width=0.4\textwidth]{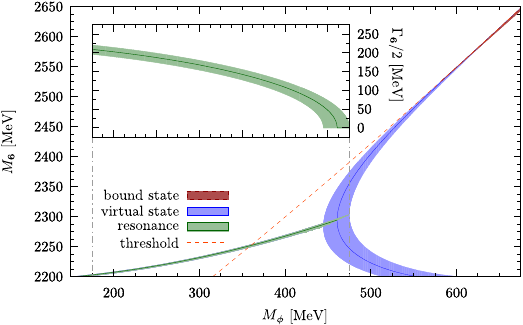}~~
\caption{Left panel: Comparison of the predictions from Ref.~\cite{Albaladejo:2016lbb} for the energy levels in the $(S,I)=(0,1/2)$ channel with the lattice QCD data~\cite{Moir:2016srx} (black dots) using $m_\pi \simeq 391$ MeV (red lines and bands). Right panel: Mass of the predicted sextet state $M_6$ at the SU(3) symmetric point as a function of the Goldstone boson mass $M_\varphi$.
The left and right panel are taken from Ref.~\cite{Albaladejo:2016lbb} and Ref.~\cite{Du:2017zvv}, respectively.
  \label{fig:D0poles}} 
	\end{center} 
\end{figure}

 The chiral amplitudes up to NLO ~\cite{Guo:2009ct,Liu:2012zya,Guo:2015dha,Albaladejo:2016lbb,Guo:2018tjx} and N$^2$LO~\cite{Guo:2018kno,Guo:2018poy} do not change the predictions. As mentioned in Sec.~\ref{sec:uchpt_Ds}, the authors of Ref.~\cite{Albaladejo:2018mhb} used the same NLO unitarized amplitude and the fixed LECs from the lattice scattering lengths in Ref.~\cite{Liu:2012zya} to provide a good description of the  energy levels in the $0^+$ and $1^+$ charm-strange channels from the later lattice QCD calculation~\cite{Bali:2017pdv}.  In Ref.~\cite{Albaladejo:2016lbb}, the authors extended the same formalism to predict the finite volume energy levels of the strangeness-isospin $(S, I)=(0,1/2)$ channel with $J^{{P}}=0^{+}$. They considered the $S$-wave $D \pi$, $D \eta$ and $D_{s} \bar{K}$ coupled-channel scatterings and found two poles at $(2105_{-8}^{+6}-i 102_{-12}^{+10}) ~\mathrm{MeV}$ and $(2451_{-26}^{+36}-i 134_{-8}^{+7}) ~\mathrm{MeV}$, which dominantly couples with the $D \pi$ and $D_{s} \bar{K}$ channels, respectively. Their results successfully described the finite volume energy levels of the coupled-channel $D \pi, D \eta$ and $D_{s} K$ scattering from the later lattice QCD calculations~\cite{Moir:2016srx} in the c.m.s as shown in Fig.~\ref{fig:D0poles}. The consistence seemed to support the existence of the two pole structures. The same resonance patterns also occurred to their partners in the bottom sector. The authors also investigated the trajectories of the two poles from the physical to the SU(3) symmetric cases by varying the meson masses. In Ref.~\cite{Guo:2018tjx}, the authors fitted the lattice data~\cite{Moir:2016srx} in the  strangeness-isospin $(S, I)=(0,{1\over 2})$ channels obtained in the c.m.s and the moving frames. The extracted parameters are similar to those in Ref.~\cite{Albaladejo:2016lbb}.

 The successful descriptions of the lattice QCD results, for instance, the scattering lengths~\cite{Liu:2012zya,Guo:2018tjx,Guo:2009ct,Guo:2015dha,Guo:2018kno,Guo:2018poy}, the phase shift~\cite{Guo:2018tjx,Guo:2018kno,Guo:2018poy}, the finite energy levels in the charmed strange channels~\cite{Albaladejo:2018mhb,Guo:2018tjx} as well as the  charmed channels with  $(S,I)=(0,1/2)$ the Refs.~\cite{Albaladejo:2016lbb,Guo:2018tjx},  tended to support the existence of the two-pole structures for the positive $D^{*}_0$ and $D^*_1$ charmed mesons. 
  
Up to now, only one pole was reported in lattice QCD simulations~\cite{Moir:2016srx,Gayer:2021xzv}. In Ref.~\cite{Moir:2016srx}, the Hadron Spectrum Collaboration reported a scalar $D_{0}^{*}$ meson with $m_{\pi} \simeq 390$ MeV. Very recently, the Hadron Spectrum Collaboration computed the isospin-$1\over2$ $D \pi$ scattering amplitudes using lattice QCD at $m_{\pi} \approx 239 \mathrm{MeV}$ and found a broad $D_{0}^{*}$ resonance around $2200-i400$ MeV~\cite{Gayer:2021xzv}, which was lighter than $D_{s 0}^{*}(2317)$. Its mass was consistent with the predictions for the lower $D^*_0$ state in the chiral unitary approaches. 

In Ref.~\cite{Du:2017zvv}, the authors argued that the amplitudes in lattice QCD simulations~\cite{Moir:2016srx} contained an additional pole, but the pole position depended strongly on the parametrization. The pole might be located too deep in the complex plane to be captured at $m_{\pi} \simeq 390 \mathrm{MeV}$~\cite{Albaladejo:2016lbb,Du:2017zvv,Guo:2018tjx}. They concluded the existence of the additional pole was not ruled out. In Ref.~\cite{Du:2017zvv}, the authors studied the mass trajectory of the sextet pole with different Goldstone boson mass $m_\varphi$ by varying the quark masses in the SU(3) symmetric case. The result is displayed in the right panel in Fig.~\ref{fig:D0poles}. With the increasing $m_\varphi$, the energy-dependent LO WT terms become larger and lead to the stronger $\mathcal H \varphi$ potentials. The sextet pole changes from the resonance to the virtual state and even the bound state with a large enough $m_\varphi$~\cite{Du:2017zvv}. The authors of Ref.~\cite{Gregory:2021rgy} studied the lightest $D_{s 0}^{*}(2317)$ and $D_{0}^{*}$ as well as their axial vector partners at the SU(3) symmetric point using lattice QCD. They found the evidence for the nontrivial bound sextet state at $M_{\pi}=612(90)$ MeV. More lattice calculations are needed to check the extra sextet pole.

\begin{figure}[hbt] 
	\begin{center} 
\includegraphics[width=0.85\textwidth]{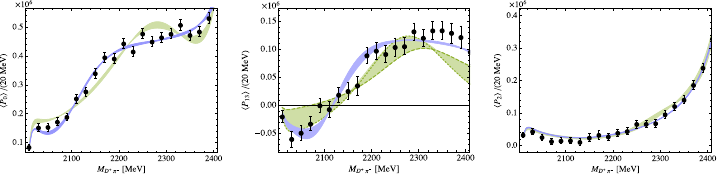}
\caption{The best fits of the LHCb angular moments (denoted with the black dot)~\cite{LHCb:2016lxy} using the Breit-Wigner parametrization (green band, $\chi^2/\text{d.o.f}=2.0$) and the unitarized amplitude in the  chiral unitary approach (blue band, $\chi^2/\text{d.o.f}=1.2$).
  \label{fig:Lhcbam}} 
	\end{center} 
\end{figure}

There is only one scalar and one axial-vector charmed meson listed in the Review of Particle Physics~\cite{ParticleDataGroup:2022pth}, the $D_{0}^{*}(2300)$ and $D_{1}(2430)$. Their resonance parameters were extracted with a simple Breit-Wigner (BW) parametrization in the $D^{(*)}\pi$ invariant mass spectrum, { which correspond to the constant interaction vertices. The pionic couplings contain the additional  $E_\pi$ dependence from the chiral symmetry (Goldstone theorem). The inclusion of the energy-dependent interaction results in a shift of the pole position from the simple Breit-Wigner parametrization~\cite{Du:2019oki}}. Moreover, the $D_{s 0}^{*}(2300)$ signal spreads in a wide range and the high energy region overlaps with the $S$-wave $D^{(*)} \eta$ and the  $D_{s}^{(*)} \bar{K}$ thresholds. Thus, the coupled-channel effects should be taken into consideration ~\cite{Du:2019oki,Du:2020pui}. The $B^-\to  D^+\pi^-\pi^-$ decay might help to confirm the molecular explanation of the $D^*_{s0}$ states as well as the two-pole structures~\cite{Du:2020pui}. The unitarized $D \pi$ scattering amplitude seemed to describe the LHCb data quite well as shown in Fig.~\ref{fig:Lhcbam}. 
{ In the vicinity of the $D^{*}_0(2300)$ and $D_1(2430)$, there are two broad $D^\ast_0$ and two  $D^*_1$  states produced.  The wide $D^*(2300)$ state in RPP is actually a combination of two $D^*_0$ poles.} The mass of the lowest $D_0^{*}$ was predicted to be around $2.1$ GeV. 
  
The existence of the $D_{s 0}^{*}(2317)\left[D_{s 1}(2460)\right]$, and the two-pole structures of the charmed meson with $(S,I)=(0,{1\over 2})$ are closely related to the chiral dynamics through the WT terms. 
After the spontaneous breaking of the chiral symmetry, the remaining SU(3) flavor symmetry constrains the $\mathcal{H}\varphi$ molecules as the multiplet in Table~\ref{tab:SI}. The flavor representation can be directly used to determine the existence of the pole in the SU(3) limit. According to Eq.~\eqref{eq:Casimiro}, the potentials for the $\overline{\bm{15}}_f$, $\bm{6}_f$ and $\bar{\bm 3}_f$ are repulsive, attractive, and more attractive, respectively. There are no poles in the $(S,I)=(2, \frac{1}{2})$, $(S,I)=(0, \frac{3}{2})$, and $(S,I)=(-1,1)$ channels in the $\overline{\bm{15}}_f$ representation. The potential in the $(S,I)=(-1,0)$ channel is attractive but weaker than that in the $\bar {\bm 3}_f$  channel, hence the existence of the pole is uncertain and still in discussion.

With the SU(3) symmetry breaking, the states with the other $(S,I)$ are composed of the $\mathcal {H} \varphi$ channels with different thresholds as shown in the lower panel of Fig.~\ref{Fig:hmsu3}. The mass splittings between different channels break the SU(3) flavor symmetry and play an important role in the formation of the unnatural states. Especially, their underlying structures are sensitive to these thresholds. For instance, in the $(0,{1\over2})$ sector, the $D_s\bar K$ is located around $490$ MeV higher than the $D\pi$ one. The lower $D^*_0$ mainly couples to the $D\pi$ channel while the higher one couples to the $D_{s} \bar{K}$ channels. 
Thus, it may be more reasonable to categorize the states using the channels instead of the SU(3) flavor representations.

{
In the quark model, there only exists one non-strange scalar charmed meson below 2.4 GeV. In the molecular scenario, the positive parity $D$-states can be dynamically generated through the scattering of the pseudo-scalar meson and the ground state charm meson. Such an interaction arises from the Weinberg-Tomozawa terms at LO. In other words, the presence of the positive parity molecular $D$-states is closely related to the chiral symmetry. Moreover, there exist two broad molecular states. The higher state is around 2.45 GeV while the lower one is around 2.1 GeV. Both of them have a large width around $200\sim 400$ MeV. The large widths of the $D^{*}_0(2300)$ and $D_1(2430)$ suggest that they may correspond to two distinct $D^*_0$ and two $D^*_1$ poles respectively. Up to now, the two-pole structures with a large decay width seemed to describe the experimental data successfully. More lattice QCD calculations from different groups are needed to examine the two-pole structures of the $D^{*}_0$ and $D^*_1$ mesons carefully.}

\subsection{Scatterings of Goldstone bosons off the heavy baryons}\label{sec:uchptothers}

The frameworks developed to investigate the $D_{s0}^*(2317)$ and $D_{s1}(2460)$ were also used to explore the scattering of the light pseudoscalar mesons ($\varphi$) off the heavy baryons ($B_Q$ or $B_{QQ}$) and predict the exotic states~\cite{Liu:2010bw,Liu:2012uw,Guo:2017vcf,Meng:2018zbl,Yan:2018zdt}. Among them, the $\varphi B_{QQ}$ systems are the most interesting ones~\cite{Guo:2017vcf,Meng:2018zbl,Yan:2018zdt} because the $B_{QQ}$ can be related to the heavy-light mesons in the HDAS as discussed in Sec.~\ref{sec:HDAS}. The same chiral dynamics between $\varphi B_{QQ}$ and $\varphi D(\bar{B})$ in the HDAS limit leads to the possible existence of the analogs of $D_{s0}^*(2317)$ and $D_{s1}(2460)$ in the doubly heavy sector. We review the calculation within the EFT frameworks. The above systems were also explored in other approaches (e.g. Refs.~\cite{Dias:2018qhp,Yu:2019yfr}). 

In Ref.~\cite{Liu:2010bw}, Liu \etal investigated the pseudoscalar meson and decuplet baryon scattering lengths to N$^2$LO within HB$\chi$PT. The same framework was extended to calculate the scattering of the pseudoscalar meson off the singly charmed baryons in Ref.~\cite{Liu:2012uw}. In their calculation, the mass splittings of the different multiplets were kept in the small scale expansion scheme~\cite{Hemmert:1996xg,Hemmert:1997ye}. The chiral expansion was performed to N$^2$LO including the diagrams similar to Figs.~\ref{Fig:HMPTree},~\ref{Fig:HMPlooph} and~\ref{Fig:HMn3lo}. In the numerical analysis, the authors adopted the SU(4) flavor symmetry to determine the LECs. From the signs of the scattering lengths, one can get a rough idea about the possible existence of the loosely bound state in each channel.

In Ref.~\cite{Meng:2018zbl}, Meng and Zhu investigated the scattering lengths of $\varphi B_{cc}^{(*)}$ to $\mathcal{O}(p^3)$ in the HB$\chi$PT. Within the HDAS, the LECs are related to those of the $\varphi D^{(*)}$ scattering in Ref.~\cite{Liu:2009uz,Liu:2011mi}. The authors proved that the analytical results for the scattering lengths in the HDAS satisfy
\begin{equation}
	a_{\varphi B_{cc}}=a_{\varphi B_{cc}^*}=a_{\bar{\varphi}D}=a_{\bar{\varphi} D^*}.
\end{equation}
The corrections from the recoil effect and the mass splitting between spin-$1\over 2$ and spin-$3\over 2$ doubly charmed baryons were tiny. The LECs determined from fitting lattice QCD results~\cite{Liu:2012zya} and using the resonance saturation model are consistent with each other. The interactions for the $[\pi \Xi_{c c}^{(*)}]^{(1 / 2)}$, $[K \Xi_{c c}^{(*)}]^{(0)}$, $[K \Omega_{c c}^{(*)}t]^{(1 / 2)}$, $[\eta \Xi_{c c}^{(*)}]^{(1 / 2)}$, $[\eta \Omega_{c c}^{(*)}]^{(0)}$, and $[\bar{K} \Xi_{c c}^{(*)}]^{(0)}$ channels are attractive, where the superscripts represent the isospin. Among them, the most attractive channel $[\bar{K} \Xi_{c c}^{(*)}]^{(0)}$ is likely to form the analogs of the $D_{s 0}^{*}(2317)\left[D_{s 1}(2460)\right]$ in the doubly heavy sector.

In Ref.~\cite{Guo:2017vcf}, Guo predicted several states by investigating the $S$-wave scattering of the doubly charmed baryons $(\Xi_{cc}^{++},\Xi_{cc}^+,\Omega_{cc}^+)$ and the light pseudoscalar mesons $(\pi, K, \eta)$ in the unitarized LO chiral effective field theory. They predicted a $\Xi_{cc}\bar{K}$ bound state in the $(S,I)=(-1,0)$ channel and two resonance structures in the $\Xi_{cc}\pi$, $\Xi_{cc}\eta$ and $\Omega_{cc} K$ coupled channels with $(S,I)=(0,1/2)$. They are the analogs of the $D^*_{s0}(2317)$ and the two-pole structures for the $D_0^*$ in the doubly charmed sector.

In Ref.~\cite{Yan:2018zdt}, the authors predicted the negative-parity doubly charmed baryons by investigating the scattering of the light pseudoscalar off the ground doubly charmed baryons using the unitarized version of $\chi$PT up to NLO. The authors considered the mixing effect between the $\varphi B_{cc}$ molecules and the $P$-wave doubly charmed baryons $B_{cc}^P$. The $B_{cc}^P$ is the conventional baryon with the $P$-wave excitation within the $(cc)$ diquark, which is expected to be the lowest excitation of the doubly charmed baryons~\cite{Gershtein:2000nx,Ebert:2002ig,Mehen:2017nrh}. They predicted two narrow states in  $(S,I)=(-1,1/2)$ channel and one narrow state in the $(S,I)=(0,1/2)$ channel, respectively.

\section{Chiral effective field theory for heavy hadronic molecules}~\label{sec:chap5}

In this section, we will review the applications of the EFT in the hadronic molecules composed of two matter fields.  An important analog is the deuteron in the $NN$ systems. In order to investigate the deuteron-like states in the heavy flavor sector, numerous EFTs are constructed. In this section, we will first review these EFTs from the simple to complex ones, specifically from the single-channel EFTs to coupled-channel EFTs, from the pionless EFTs [$\slashed{\pi}$EFTs (contact EFTs)] to the chiral EFT, from the perturbative pion EFTs to nonperturbative pion EFTs. After that, we will review applications of these EFTs to specific hadronic molecular candidates. One may also consult the review in Ref.~\cite{Guo:2017jvc}. Since we are only interested in the energy regions close to two hadron thresholds, the EFTs for the hadronic molecules are usually nonrelativistic.

 Unlike the $\chi$EFT for the $NN$ systems which is constrained by the precise $NN$ phase shifts from the experiments, the EFTs constructed for the hadronic molecules are still preliminary. In this section, we will also review some frameworks that are not rigorous EFTs but incorporate the spirits of EFT. From our perspective, an EFT is the quantum field theory which should include quantum fluctuations and renormalizations. Therefore, we will pay more attentions to the frameworks which include quantum fluctuations (either by calculating loop corrections or iterating nonperturbatively) and scale-independent renormalizations. After all, the tree-level results or the cutoff-sensitive results from the so-called EFTs are usually equivalent to some classical frameworks. Meanwhile, the EFTs are based on separated scales and specific symmetries. In this section, we will discuss some frameworks making use of separated scales and symmetries. In principle, the EFTs are performed according to the power counting which originates from the scale analysis and should manifest convergence in high order calculations.  According to the present experimental status, it is usually hard to fix the LECs in higher order calculations. Therefore, in this section, we have to pay some attentions to the validity of the scale analysis.

\subsection{Pionless EFT}~\label{sec:pionlessEFT}
\subsubsection{Single-channel systems}~\label{sec:1chanel_pionless}
We start from the LO pionless EFT in the single-channel for the $S$-wave systems. The interaction between particles $A$ and $B$ is $V(p,p')=v\Theta(\Lambda-p)\Theta(\Lambda-p^\prime)$. We choose a cutoff UV regulator $\Theta(\Lambda-p)$ in this section. One can also use other regularization schemes such as the PDS in Sec.~\ref{sec:ksw}, which will not change the results qualitatively. In order to reproduce a bound state or virtual state, we resum the interaction nonperturbatively via LSEs,
\begin{equation}
T(p,p^{\prime})=V(p,p^{\prime})+\int\frac{d^{3}q}{(2\pi)^{3}}V(p,q)G(E,q)T(q,p^{\prime}),\qquad G(E,q)=\frac{1}{E-\frac{q^{2}}{2\mu}+i\epsilon}.
\end{equation}
For such a separable interaction, we can assume the $T$-matrix is also separable, i.e., $T(p,p')=t\Theta(\Lambda-p)\Theta(\Lambda-p^\prime)$. Then the LSEs can be solved as
\begin{equation}
t=v+vFt,\quad\Longrightarrow\quad t^{-1}=v^{-1}-F, \label{eq:4.1_LSE_sl}
\end{equation}
where the $F$ is
\begin{equation}
	F(E)=\int^{\Lambda}\frac{d^{3}q}{(2\pi)^{3}}\frac{1}{E-\frac{q^{2}}{2\mu}+i\epsilon}\approx\begin{cases}
		-\frac{\mu}{2\pi}\left(\frac{2}{\pi}\Lambda+ik\right) & \text{with }E>0,\:k\equiv\sqrt{2\mu E}\\
		-\frac{\mu}{2\pi}\left(\frac{2}{\pi}\Lambda-\kappa\right) & \text{with }E\leq0,\:\kappa\equiv\sqrt{-2\mu E}
	\end{cases}.~\label{eq:4.1F}
\end{equation}
In the cutoff regularization, the $F$ has the linear UV behavior which is similar to that of PDS in Eq.~\eqref{eq:sec1.5:int} up to a factor for $\Lambda$. One can see the relation of the cutoff regularization and dimensional regularization for the nonrelativistic system in Ref.~\cite{Phillips:1998uy}. Meanwhile, the $(v^{-1}-F)$ should be cutoff-independent to make the $T$-matrix renormalization group invariant. The pole of the $T$-matrix obtained in the real axis from $v^{-1}-F(E)=0$ corresponds to the bound state (in the first Riemann sheet) or virtual state (in the second Riemann sheet) of the $AB$ system.

It was shown that the $\slashed{\pi}$EFT is equivalent to the ERE~\cite{vanKolck:1999mw}. We will illustrate the equivalence at the LO. One can perform the ERE as
\begin{eqnarray}
t^{-1}&=&-\frac{\mu}{2\pi}(k\cot\delta-ik)=-\frac{\mu}{2\pi}\left(-\frac{1}{a_{s}}-ik+\frac{1}{2}r_{0}k^{2}...\right)\nonumber\\
&=&v^{-1}-F=-\frac{\mu}{2\pi}\left(-\frac{2\pi}{\mu}v^{-1}-\frac{2}{\pi}\Lambda-ik\right).
\end{eqnarray}
One can obtain the $k^0$ and $k^1$ terms from the LO $\slashed{\pi}$EFT, which matches to the ERE truncated at the LO. The scattering length term and $t^{-1}$ read, respectively,
\begin{equation}
	\frac{1}{a_{s}}=\left(\frac{2\pi}{\mu}v^{-1}+\frac{2}{\pi}\Lambda\right),\qquad t^{-1}\sim \left(-\frac{1}{a_{s}}-ik\right).
\end{equation}
The $k^1$ term is required by the unitarity. The approximation of using the contact interaction without derivative is equivalent to keeping the LO ERE (only the scattering length term). The effective range and higher order terms are neglected.   For $a_s>0$, one can see the $E=-\gamma^2/(2\mu)=-1/(2\mu a_s^2)$ corresponding to a bound state pole with the binding momentum $\gamma=1/a_s$. For $a_s<0$, one can see that $E=-1/(2\mu a_s^2)$ at the second Riemann sheet corresponds to a virtual state.

\begin{figure}[hbtp]
	\begin{center}
		\includegraphics[width=0.98\textwidth]{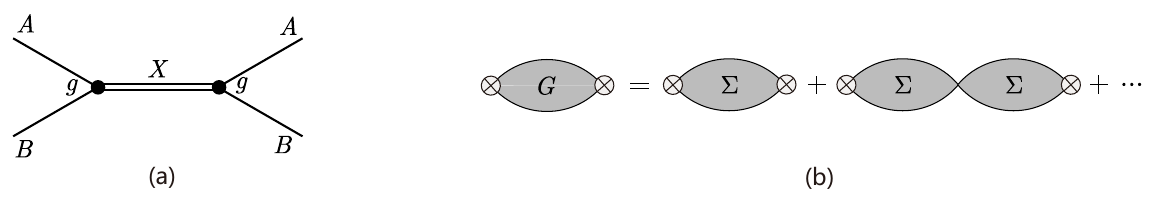}
		\caption{(a)The scattering of $A$ and $B$ is saturated by their bound state $X$.  (b) The two point Green's function of $\hat{X}(x)\equiv \hat{A}(x) \hat{B}(x)$.} \label{Fig:4.1:bound_state}
	\end{center}
\end{figure}

 For the system with a bound state, the $T$-matrix can be reexpressed as
\begin{eqnarray}
	t^{-1}\sim (-\gamma-ik).~\label{eq:4.1_tMtx_biding}
\end{eqnarray}
 Another interesting quantity is the coupling constant $g$ of the bound state $X$ and its components $A$ and $B$. As shown in Fig.~\ref{Fig:4.1:bound_state}(a), the vertex of $XAB$ is $g\Theta(\Lambda-p)$. The $T$-matrix of the $A$ and $B$ scattering is saturated by the bound state. The coupling constant is obtained from the residue of $t$ at its pole position~\cite{Gamermann:2009fv,Gamermann:2009uq}, i.e.,
\begin{equation}
t\sim {g^2 \over E-E_X},\quad\Longrightarrow\quad	g=\frac{4M_{X}\sqrt{\pi\gamma}}{\sqrt{\mu}}~\label{eq:t_matrix_x_pole}.
\end{equation}
The coupling constant only depends on the binding energy which is determined by the scattering length. {The relation between the coupling constant and $\sqrt{\gamma}$ is also derived from the Weinberg compositeness criterion~\cite{Weinberg:1965zz} in the case of a pure molecule.}

One can construct the composite operator $\hat{X}(x)\equiv \hat{A}(x) \hat{B}(x)$, where $\hat{A}$ and $\hat{B}$ are the field operators of $A$ and $B$, respectively. The bound state can also be determined by the pole of the two-point Green's function of $\hat{X}$~\cite{Kaplan:1998sz},
\begin{equation}
	G(E)=\int d^{4}xe^{-iEt}\langle0|T\{X^{\dagger}(x)X(0)\}|0\rangle=i\frac{Z(E_{X})}{E-E_{X}+i\epsilon},\label{eq:4.1_twopointG}
\end{equation}
where $T\{...\}$ represents the time-ordered product. The Green's function can be calculated from the diagrams in Fig.~\ref{Fig:4.1:bound_state}(b) as
\begin{equation}
	G	=\Sigma+\Sigma\left(-iv\Sigma\right)+\Sigma\left(-iv\Sigma\right)^{2}+...=\frac{\Sigma(E)}{1+iv\Sigma(E)}=\frac{iF}{1-vF},
\end{equation}
where the $\Sigma=iF$ is used. We obtain the same equation of the binding energy $1-vF=0$ as that from the LSEs. The wave function renormalization factor $Z$ is determined as
\begin{equation}
Z(E_X)=\frac{1}{2\pi}\left(\frac{2}{\pi}\Lambda-\gamma\right)^{2}\gamma.
\end{equation}
{One notes that the dimension of $Z$ is $M^3$. In our calculation, we adopt the fields of the $\hat{A}$ and $\hat{B}$ with the dimension $M^{3/2}$ as in most of the NREFT calculations. Thus, the composite operator $\hat{X}(x)\equiv \hat{A}(x) \hat{B}(x)$ carries the dimension $M^3$ and finally gives rise to the $Z$ with $M^3$.  In the present discussion, we focus on the relations of $Z$ with the cutoff and binding momentum. }

The above results about the bound state can also be obtained in the language of quantum mechanics. We investigate the same bound state in Schr\"odinger equation,
\begin{equation}
\frac{q^{2}}{2\mu}\phi(q)+\int\frac{ d^{3}p}{(2\pi)^{3}}V(q,p)\phi(p)=-\frac{\gamma^{2}}{2\mu}\phi(q).
\end{equation}
The wave function is
\begin{equation}
\phi(p)=\xi\frac{\Theta(\Lambda-p)}{-\frac{\gamma^{2}}{2\mu}-\frac{p^{2}}{2\mu}},\qquad \xi^{2}\approx\frac{\gamma}{4\pi^{2}\mu^{2}},
\end{equation}
where the normalization constant $\xi$ is determined by $\int|\phi(p)|^{2}d^{3}p=1$. The wave function can be related to the scattering length as
\begin{equation}
	\phi(p)=\frac{1}{\pi\sqrt{a_{s}}}\frac{1}{1/a_{s}^{2}+p^{2}},\qquad\varphi(r)=\frac{1}{\sqrt{2\pi a_s}}\frac{e^{-r/a_s}}{r},~\label{eq:wavefunction_universal}
\end{equation}
where we have taken the cutoff to infinity. $\varphi(r)$ is the wave function in the coordinate space. Another interesting quantity is the wave function at the origin,
\begin{equation}
	\varphi(0)=-\frac{\gamma^{1/2}}{(2\pi)^{1/2}}\left(\frac{2}{\pi}\Lambda-\gamma\right),\label{eq.4.1:wv_r=0}
\end{equation}
which is cutoff-dependent and linear divergent.

One can set up two corresponding relations between the languages of quantum mechanics and quantum field theory,
\begin{equation}
	\langle p|\hat{V}|\phi\rangle \sim g \Theta(\Lambda-p) ,\qquad \sqrt{Z}\sim\varphi(0).
\end{equation}
 One can take the approximation of the LSEs near the pole,
\begin{equation}
\hat{T}=\hat{V}+\hat{V}\frac{1}{E-\hat{H}+i\epsilon}\hat{V},\quad\Longrightarrow\quad T(p,p')\sim\frac{\langle p|\hat{V}|\phi\rangle\langle\phi|\hat{V}|p'\rangle}{E-E_{X}},
\end{equation}
with
\begin{equation}
	\langle p|\hat{V}|\phi\rangle=\langle p|H-H_{0}|\phi\rangle=\left(E_{0}-\frac{p^{2}}{2\mu}\right)\langle p|\phi\rangle=\xi\Theta(\Lambda-p).
\end{equation}
One can see that the $\langle p|\hat{V}|\phi\rangle$ corresponds to the $XAB$ vertex $g\Theta(\Lambda-p)$ in quantum field theory language. The bound state can be represented as
\begin{equation}
	|X(\bm{P})\rangle\equiv\int\frac{d^{3}\bm p}{(2\pi)^{3}}\phi(\bm{p})|A(\bm{p}_{1})B(\bm{p}_{2})\rangle,\qquad\bm{P}\equiv\bm{p}_{1}+\bm{p}_{2},\quad\bm{p}\equiv\frac{\bm{p}_{1}-\bm{p}_{2}}{2}.\label{eq:4.1_bdstate}
\end{equation}
From Eq.~\eqref{eq:4.1_twopointG}, we know the wave function renormalization factor $Z$ can be extracted from
\begin{equation}
	\langle X(P)|\hat{X}(x)|\Omega\rangle=\sqrt{Z}e^{iP\cdot x}.\label{eq:4.1_Z}
\end{equation}
With Eqs.~\eqref{eq:4.1_bdstate} and~\eqref{eq:4.1_twopointG}, one notices that $\sqrt{Z}\sim\varphi(0)$ up to some constant factors. Both the quantum mechanics (e.g.,~\cite{Voloshin:2003nt,Voloshin:2005rt}) and quantum field theory (e.g.,~\cite{Fleming:2007rp}) languages were used in literature.

From the above calculation, one can see that the binding energy, the wave function of the bound state and the low energy scattering properties are determined by one parameter $a_s$. For the two body system which admits a shallow bound state with an unnatural large scattering length, the low-energy behavior is universal and depends on the scattering length only. For such a system, the LO ERE and LO contact interaction will be a good approximation (see Ref.~\cite{Braaten:2004rn} for a review of the universality).

In the LO single-channel contact interaction, {since $a_s$} is real, the pole of the $T$-matrix can only appear in the real axis corresponding to either the bound state or the virtual state.  In order to interpret the resonance with a width, one can introduce the LO coupled-channel contact interaction or resort to the NLO contact interaction (e.g.,~\cite{Albaladejo:2015lob,Meng:2020ihj,Yang:2020nrt}). In the LO coupled-channel contact scheme, one may obtain the effective $a_s$ for the elastic channel with an imaginary part~(e.g.,~\cite{Braaten:2005ai}) considering the inelastic channel. It will be discussed in Sec.~\ref{sec:nopi_cc}.  Meanwhile, using the NLO contact interaction is equivalent to keeping the effective range term in ERE~(e.g., see Refs.~\cite{Meng:2021rdg,Guo:2020vmu,Hyodo:2013iga}).

We take the $S$-wave interaction to the NLO as an example,
\begin{equation}
	V(p,p')=\frac{c_{a}}{\Lambda}+\frac{c_{b}}{\Lambda^{3}}(\bm{p}^{2}+\bm{p}'^{2}),\label{eq:contactnlo}
\end{equation}
where $c_a$ and $c_b$ are the LECs. For simplicity, we assume the interaction is independent on energy. For such a separable interaction, we adopt the techniques in Ref.~\cite{Epelbaum:2017byx} and obtain,
\begin{eqnarray}
	{1\over T(k)}=\frac{-G_{0}\Lambda^{5}c_{a}-2G_{2}\Lambda^{3}c_{b}+\left(G_{2}^{2}-G_{0}G_{4}\right)c_{b}^{2}+\Lambda^{6}}{\Lambda^{5}c_{a}+c_{b}\left[c_{b}\left(G_{0}k^{4}-2G_{2}k^{2}+G_{4}\right)+2k^{2}\Lambda^{3}\right]},\quad\text{ with } 	G_{n}=\int\frac{d^{3}{q}}{(2\pi)^{3}}\frac{q^{n}}{E-\frac{q^{2}}{2\mu}+i\epsilon}. \nonumber\\
\end{eqnarray}
In the hard cutoff regularization scheme, we obtain the analytical expressions of $G_n$ as
\begin{eqnarray}
	G_{0}=\frac{4\pi}{(2\pi)^{3}}2\mu\left[k\tanh^{-1}\left(\frac{k}{\Lambda}\right)-\Lambda-i\frac{\pi}{2}k\right],	\qquad G_{n}
	=k^{2}G_{n-2}-\frac{\mu}{\pi^{2}}\frac{\Lambda^{n+1}}{n+1}.\label{eq:G0}
\end{eqnarray}
Performing the effective range expansion, $T^{-1}(k)=-\frac{\mu}{2\pi}\left(-\frac{1}{a_{s}}-ik+\frac{1}{2}r_{0}k^{2}+...\right)$, we obtain that
\begin{eqnarray}
	a_{s}&=&\frac{9\pi\mu\left(5\pi^{2}c_{a}-\mu c_{b}^{2}\right)}{\Lambda\left(30\pi^{2}\mu\left(3c_{a}+2c_{b}\right)-8\mu^{2}c_{b}^{2}+90\pi^{4}\right)},\\ r_{0}&=&\frac{4\left(30\pi^{2}\mu^{2}c_{b}^{2}\left(10c_{b}-9c_{a}\right)+225\pi^{4}\mu\left(3c_{a}^{2}+5c_{b}^{2}\right)+52\mu^{3}c_{b}^{4}+1350\pi^{6}c_{b}\right)}{27\pi\Lambda\mu\left(\mu c_{b}^{2}-5\pi^{2}c_{a}\right){}^{2}}.
\end{eqnarray}
Now, we obtain the non-vanishing $r_0$. The resonance poles can be obtained from $T^{-1}=0$ in the NLO ERE formula~\cite{Hyodo:2013iga}.

\begin{figure}[hbtp]
	\begin{center}
		\includegraphics[width=0.30\textwidth]{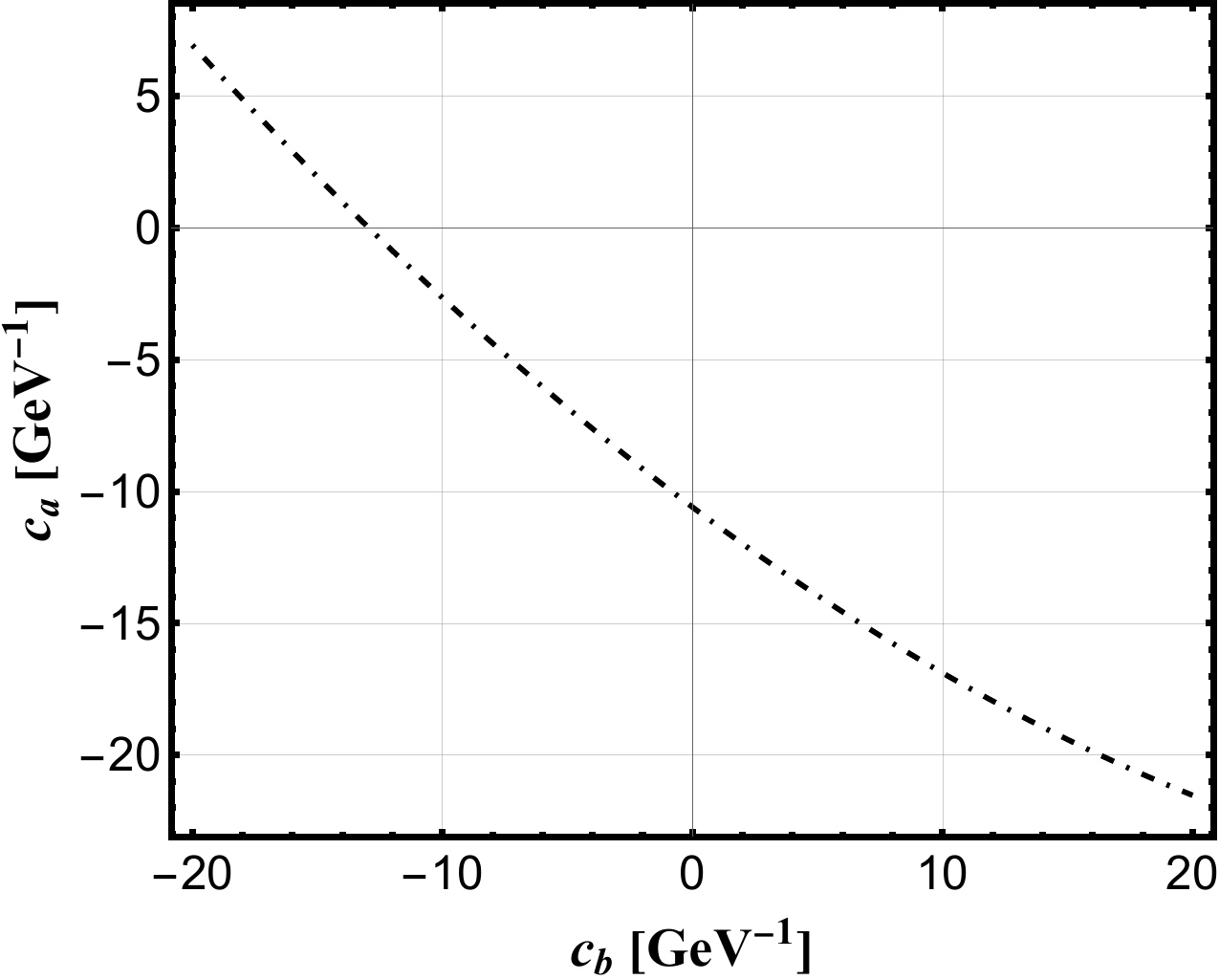}~~~~~
		\includegraphics[width=0.30\textwidth]{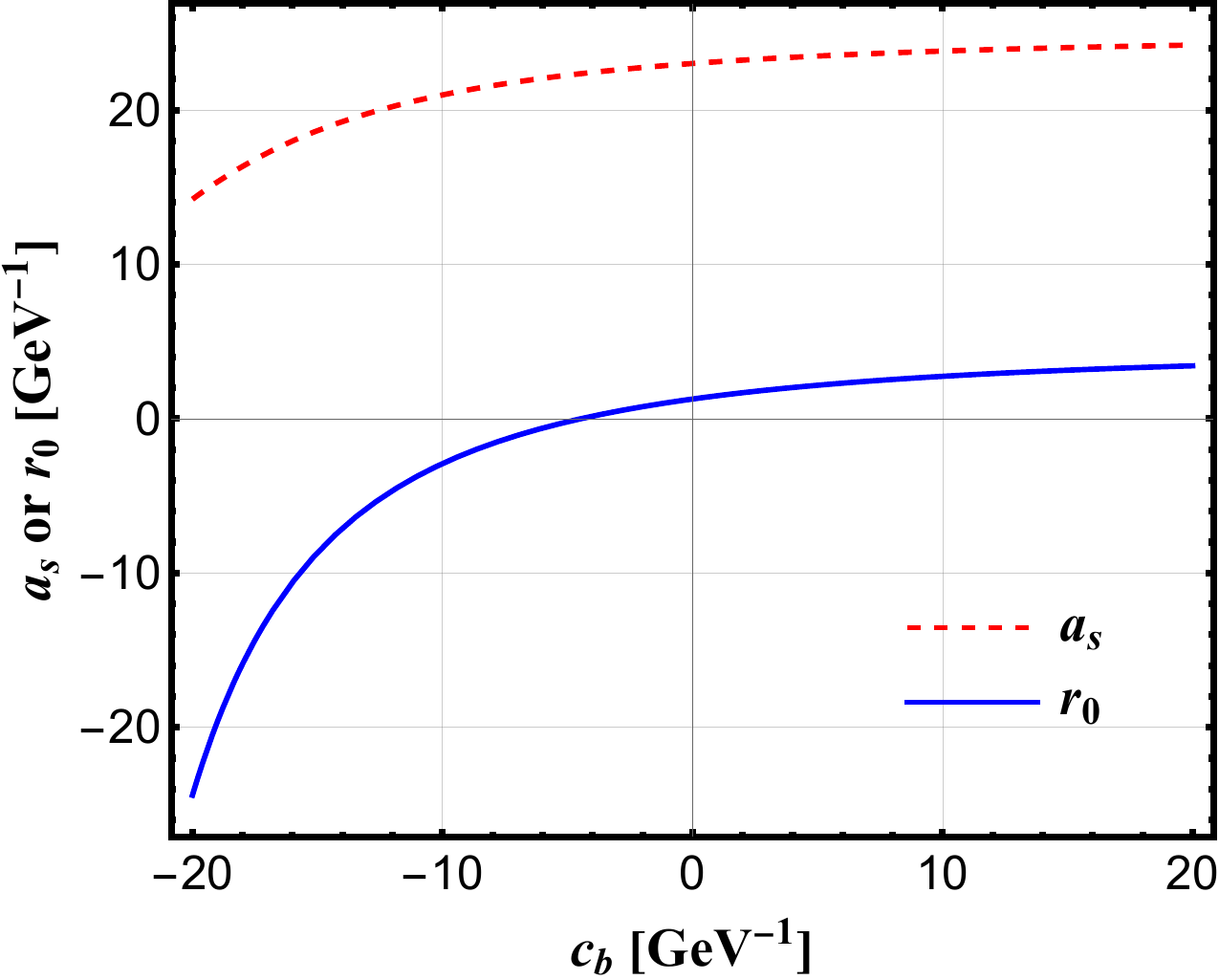}~~~~~
  \includegraphics[width=0.30\textwidth]{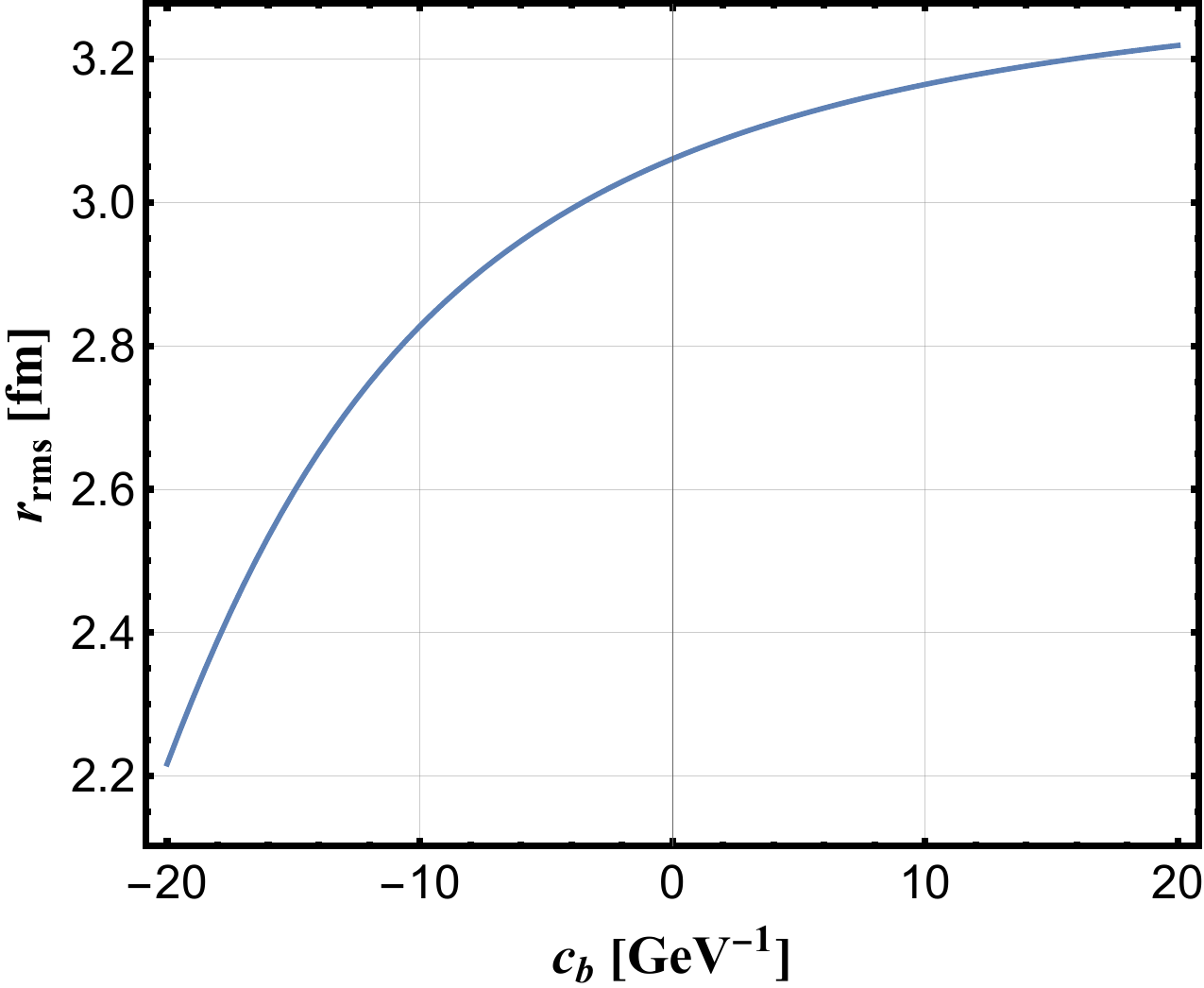}
		\caption{The scattering length and effective range for the system in the non-local interaction  in Eq.~\eqref{eq:contactnlo} with a binding energy $E_b=1$ MeV. {The relation of $c_a$ and $c_b$ is constrained by permitting a bound state with the binding energy $E_b=1$ MeV (left panel). The $a_s$ and $r_0$ depend on the coupling constants (middle panel). The root-mean-square radii $r_{\rm rms}$ vary with the coupling constants (right panel). } } \label{Fig:4.1:ere_sign}
	\end{center}
\end{figure}

With the above expressions, we can check the validity of of the extension of the argument of Landau and Smorodinsky in the non-local potential~\cite{landau:text}. It was proved that the effective range is always positive ($r_0>0$) if the local potential is attractive everywhere, i.e., $V(r)<0$ everywhere in their textbook~\cite{landau:text}. The argument was cited in Ref.~\cite{Esposito:2021vhu} to claim that ``the molecular case gives always $r_0 > 0$". We use Eq.~\eqref{eq:contactnlo} as an example of the non-local potential case. For simplicity, we take $\Lambda = 1$ GeV and $\mu=1$ GeV. We constrain the interaction in Eq.~\eqref{eq:contactnlo} via admitting a bound state with the binding energy $E_b=1$ MeV, which will constrain the relation of the two LECs as shown in the left panel of Fig.~\ref{Fig:4.1:ere_sign}. In the middle panel, we present the $a_s$ and $r_0$ numerically. One can see that the negative $r_0$ is allowed for such a non-local interaction. {In the right panel, we present the variation of the root-mean-square radii $r_{\rm rms}$ with the coupling constants. One can see these states are all loosely bound states with $r_{\rm rms}$ around 2$-$3 fm. } Therefore, it is biased to conclude that the molecule solution implies the always positive effective range. See~\cite{xrange} for more detailed discussions on this issue.

\subsubsection{Coupled-channel systems}~\label{sec:nopi_cc}
In Ref.~\cite{Cohen:2004kf}, Cohen, Galman and van Kolck proposed a coupled-channel EFT for two channels with different thresholds. For the LO contact interaction, the renormalization is given explicitly. Within the hard cutoff regularization, the interaction reads
\begin{equation}
V(p,p^\prime)=\left[\begin{array}{cc}
	v_{11} & v_{12}\\
	v_{12} & v_{22}
\end{array}\right]\Theta(\Lambda-p)\Theta(\Lambda-p^\prime).~\label{eq:v_couplechannel}
\end{equation}
For the separable interaction, one can transform the LSEs into the algebraic equation of the coefficient matrices $v$ and $t$, i.e.,
\begin{equation}
T=V+VGT,\quad\Longrightarrow\quad t=(1-vF)^{-1}v,
\end{equation}
where the $F=\text{diag}\{F_a,F_b\}$ with
\begin{equation}
F_{a}(E)=\int^{\Lambda}\frac{d^{3}{p''}}{(2\pi)^{3}}\frac{1}{E-E_{a,p''}},\qquad E_{a,p''}=\delta_{a}+\frac{p''^{2}}{2\mu_a}.
\end{equation}
We can choose the $\delta_1=0$ and $\delta_2=\delta$.  We use the $\mu_i$ to represent the reduced masses of two channels. With the definitions $\kappa_{1}\equiv\sqrt{-2\mu_1 E}$ and $\kappa_{2}\equiv\sqrt{-2\mu_2(E-\delta)}$, the solution of the LSEs is
\begin{equation}
t^{-1}=\frac{1}{D}\left(\begin{array}{cc}
	\frac{1}{\mu_{1}}b_{11}b_{12}^{2}\left(1-b_{22}\kappa_{2}\right) & \frac{1}{\sqrt{\mu_{1}\mu_{2}}}b_{11}b_{12}b_{22}\\
	\frac{1}{\sqrt{\mu_{1}\mu_{2}}}b_{11}b_{12}b_{22} & \frac{1}{\mu_{1}}b_{12}^{2}b_{22}\left(1-b_{11}\kappa_{1}\right)
\end{array}\right),
\end{equation}
where $D=\frac{1}{2\pi}\left[b_{12}^{2}\left(b_{11}\kappa_{1}-1\right)\left(b_{22}\kappa_{2}-1\right)-b_{11}b_{22}\right]$. The $b_{11}$, $b_{22}$ and $b_{12}$ are cutoff-independent parameters defined by
\begin{equation}
\begin{cases}
	\frac{1}{b_{11}} & =\frac{2\pi}{\mu_{1}}\left(\frac{v_{22}}{v_{11}v_{22}-v_{12}^{2}}-F_{1}\right)+\kappa_{1}\\
	\frac{1}{b_{22}} & =\frac{2\pi}{\mu_{2}}\left(\frac{v_{11}}{v_{11}v_{22}-v_{12}^{2}}-F_{2}\right)+\kappa_{2}\\
	\frac{1}{b_{12}} & =\frac{2\pi}{\sqrt{\mu_{1}\mu_{2}}}\frac{v_{12}}{v_{11}v_{22}-v_{12}^{2}}
\end{cases}.~\label{eq:inverse-bij}
\end{equation}
The cutoff dependence in the $F_1$ and $F_2$ is canceled out by the cutoff-dependent coupling constants $v_{ij}$. Thus, the renormalization of the EFT is given explicitly. In literature, when a similar contact interaction as in Eq.~\eqref{eq:v_couplechannel} was used to depict the neutral and charged channels of $X(3872)$, it was often assumed that $v_{11}=v_{22}=v_{12}=v_{21}$, which is equivalent to the vanishing isovector interaction. However, we can see that the $v_{12}$ and $v_{11}$, $v_{22}$ in Eq.~\eqref{eq:inverse-bij} have to cancel out the quite different divergent behavior.  In such an EFT framework, one can set $v_{11}=v_{22}$. However, it is illegitimate to introduce the $v_{ii}=v_{12}$ or $v_{21}$ in order to meet the renormalization group invariance.

For the coupled-channel system, if one expands the $1/t_{11}$ according to the ERE as
\begin{equation}
	\frac{1}{t_{11}}=-\frac{\mu_1}{2\pi}\left[-\frac{1}{a_\text{eff}}+\frac{1}{2}r_\text{eff}k^{2}-ik+...\right],
\end{equation}
one can get the expression of the parameters $a_{\text{eff}}$ and $r_{\text{eff}}$ as
\begin{equation}
\frac{1}{a_{\text{eff}}}=\frac{1}{b_{11}}+\frac{1}{b_{12}^{2}}\left[\sqrt{2\delta\mu_{2}}-\frac{1}{b_{22}}\right]^{-1},\qquad r_{\text{eff}}=-\frac{1}{\sqrt{2\delta\mu_{2}}}\frac{b_{22}^{2}}{\left(b_{12}\sqrt{2\delta\mu_{2}}-1\right){}^{2}},
\end{equation}
where $a_\text{eff}$ and $r_\text{eff}$ represent the effective scattering length and effective range, respectively. If the second threshold is below the first one, i.e., $\delta<0$, the inelastic channel (the second channel) renders $a_\text{eff}$ a complex number. Another interesting conclusion is $r_\text{eff}<0$ if the second threshold is larger than the first one, i.e., $\delta>0$. In such a coupled-channel system, the bound state is allowable by tuning the parameters, but with the negative effective range, which is another example challenging the statement in Ref.~\cite{Esposito:2021vhu}.

Meanwhile, if $a_\text{eff}$ is unnaturally large, the long-range behavior only depends on  $a_\text{eff}$, which is the universality for the coupled-channel system~\cite{Braaten:2005ai}. From the expression of $a_\text{eff}$, one can see that the universality can be obtained by fine-tuning either the interaction or the threshold difference of the two channels.

If there exists the bound state solution with $E_0<0$, one can obtain the residues of the $T$-matrix (see Ref.~\cite{Meng:2021jnw} for details),
\begin{equation}
\lim_{E\to E_{0}}(E-E_{0})t=\frac{2\pi}{\mu^{2}}\left[\begin{array}{cc}
	\gamma_{1}\cos^{2}\theta & \sqrt{\gamma_{1}\gamma_{2}}\sin\theta\cos\theta\\
	\sqrt{\gamma_{1}\gamma_{2}}\sin\theta\cos\theta & \gamma_{2}\sin^{2}\theta
\end{array}\right],
\end{equation}
with $\lim_{E\to E_{0}}\kappa_{i}\equiv\gamma_{i}$, and $\tan^{2}\theta\equiv\frac{b_{22}\gamma_{1}\left(b_{11}\gamma_{1}-1\right)}{b_{11}\gamma_{2}\left(b_{22}\gamma_{2}-1\right)}$. The $b_{12}$ is eliminated by setting $D=0$. For convenience, we omit the difference of two reduced masses and set $\mu_1=\mu_2=\mu$. The coupling constants between the bound state and two channels are
\begin{equation}
	g_{1}	=\frac{4M_{T}\sqrt{\pi\gamma_{1}}}{\sqrt{\mu}}\cos\theta,\qquad g_{2}=\frac{4M_{T}\sqrt{\pi\gamma_{2}}}{\sqrt{\mu}}\sin\theta.~\label{eq:coupling_two_channel}
\end{equation}
One can obtain similar results in quantum mechanics. The wave function of the bound states is
\begin{equation}
	\langle\bm{p}|\psi\rangle=c_{1}\phi_{1}(p)|1\rangle+c_{2}\phi_{2}(p)|2\rangle,\qquad 	\phi_{i}(p)=\xi_{i}\frac{\Theta(\Lambda-p)}{E_{0}-\frac{p^{2}}{2\mu}-\delta_{i}},\qquad\xi_{i}^{2}\approx\frac{\gamma_{i}}{4\pi^{2}\mu^{2}}.
\end{equation}
Similar to the single-channel case, one gets
\begin{equation}
	\langle\bm{p},i|\hat{V}|\psi\rangle	=\langle p,i|H-H_{0}|\psi\rangle=\left(E_{0}-\frac{p^{2}}{2\mu}-\delta_{i}\right)\langle\bm{p},i|\psi\rangle
	=c_{i}\phi_{i}(p)=c_{i}\xi_{i}\Theta(\Lambda-p).
\end{equation}
When one relates $\langle\bm{p},i|\hat{V}|\psi\rangle$ with $g_i\Theta(\Lambda-p)$, one can see that $c_1=\cos\theta$ and $c_2=\sin\theta$. Therefore, the $\theta$ angle defined in the quantum field theory language is just the mixing angle of the two channels.

\subsection{XEFT}~\label{sec:XEFT}

The XEFT is a nonrelativistic EFT to depict the long-range properties of the $X(3872)$, in which the $D^{*0}$, $\bar{D}^{0}$ and $\pi^{0}$ are explicit degrees of freedom~\cite{Fleming:2007rp}. It is very similar to the KSW scheme for the $NN$ system in Sec.~\ref{sec:ksw}. In this section, we will take the $X\to D^{0}\bar{D}^{0}\pi^{0} $ and $\xthn$ scattering as examples to illustrate this formalism.

 \begin{figure}[htbp]
	\centering
	\includegraphics[width = 0.1\textwidth]{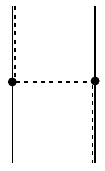}
	\caption{The one-pion exchange interaction for the $\xthn$ scattering. The double (solid plus dashed) line denotes the $D^{\ast0}/\bar{D}^{\ast0}$, while the single solid and dashed lines denote the $D^0/\bar{D}^0$ and $\pi^0$, respectively.}\label{fig:4.1X_OPE}
\end{figure}

For the $D^0\bar{D}^{\ast0}/\bar{D}^0D^{\ast0}$ scattering, the $D^{\ast0}D^0\pi^0/\bar{D}^{\ast0}\bar{D}^0\pi^0$ vertices contribute to the OPE interaction (see Fig.~\ref{fig:4.1X_OPE}). The mass splitting $\Delta=M_{D^{\ast0}}-M_{D^0}$ will enter the pion propagator. The {static} OPE potential is~\footnote{The $f_\pi$ in Ref.~\cite{Fleming:2007rp} is different from that in this review by a factor $\sqrt{2}$.}

\begin{equation}\label{eq:OPEinXEFT}
\frac{g_{b}^{2}}{4f_{\pi}^{2}}\frac{(\bm{q}\cdot\bm{\bm{\varepsilon}}^{\prime\dagger})(\bm{q}\cdot\bm{\bm{\varepsilon}})}{\bm{q}^{2}-u^{2}},
\end{equation}
where $u^2$ is defined as $u^2=\Delta^2-m_{\pi^0}^2$. {The above static propagator is an economy choice in most studies, but the cost is the analytic structure of OPE is changed due to the omission of the energy term, see Sec.~\ref{sec:x_methodology} for details.}

\begin{table*}
\centering
\renewcommand{\arraystretch}{1.5}
	\caption{Some scales associated with the $X(3872)$. The $T^{n}=M_{D^{*0}}+M_{D^{0}}$ and $T^{c}=M_{D^{*+}}+M_{D^{+}}$ are the neutral and charged thresholds. $E_{\pi^0}^\text{max}=(M_X^2-4M_{D^0}^2+m_{\pi^0}^2)/(2M_X)$ is the maximum energy of the pion in the $X\to D^0\bar{D}^0\pi^0$ decay.}~\label{tab:4.1xscale}
\setlength{\tabcolsep}{3.1mm}
{
	\begin{tabular}{c|c|c|c|c}
		\hline
		& \multicolumn{4}{c}{$\Delta=M_{D^{*0}}-M_{D^{0}}=142$ MeV, ~~~~~~~~~~$m_{\pi^{0}}=135$ MeV}\tabularnewline
		\cline{1-5} 	\cline{2-5} \cline{3-5} \cline{4-5} \cline{5-5}
		\multirow{2}{*}{Energy scale}	& $E_{X}^{c}\equiv T^{c}-M_{X}$ & $E_{X}\equiv T^{n}-M_{X}$ & $\delta=\Delta-m_{\pi^{0}}$ & $E_{\pi^{0}}^{max}-m_{\pi^{0}}$\tabularnewline
		\cline{2-5} \cline{3-5} \cline{4-5} \cline{5-5}
		& 8.4 MeV & 0.2 MeV & 7 MeV & $6.6$ MeV\tabularnewline
		\hline
		\multirow{2}{*}{Momentum} &
		$\begin{array}{rl}
			\gamma^{c} & =\sqrt{2\mu E_{X}^{c}}\\
			& \sim p_{D^{*}}\sim p_{D}
		\end{array}$
		&
		$\begin{array}{rl}
			\gamma & =\sqrt{2\mu E_{X}}\\
			& \sim  p_{D^{*}}\sim p_{D}
		\end{array}$
		&
		$\begin{array}{rl}
			u & =\sqrt{\Delta^{2}-m_{\pi^{0}}^{2}}\\
			& \sim\sqrt{2m_{\pi^{0}}\delta}\\
			& \sim p_{\pi}
		\end{array}$
		& 	
		$p_{\pi}  =\sqrt{2m_{\pi}(E_{\pi}^{max}-m_{\pi})}$
		\tabularnewline
		\cline{2-5} \cline{3-5} \cline{4-5} \cline{5-5}
		& 128 MeV & 20 MeV & 45 MeV & 42 MeV\tabularnewline
		\hline
		\multirow{2}{*}{Velocity} & $v_{D^{(*)}}=\frac{p_{D^{(*)}}}{M_{D^{(*)}}}$ & $v_{D^{(*)}}=\frac{p_{D^{(*)}}}{M_{D^{(*)}}}$ & $v_{\pi}=\frac{p_{\pi}}{m_{\pi}}$ & $v_{\pi}=\frac{p_{\pi}}{m_{\pi}}$\tabularnewline
		\cline{2-5} \cline{3-5} \cline{4-5} \cline{5-5}
		& 0.06 & 0.01 & 0.32 & 0.31\tabularnewline
		\hline
		Process & $-$ & $-$ & OPE & $X\to D^{0}\bar{D}^{0}\pi^{0}$\tabularnewline
		\hline
	\end{tabular}
}
\end{table*}

In Table~\ref{tab:4.1xscale}, we list the estimation of different scales associated with $X(3872)$.  One can see the momentum of the $D^{(*)0}$ estimated by the binding momentum of the $X(3872)$, the momentum of the $\pi^0$ in the OPE and the $X\to D^{0}\bar{D}^{0}\pi^{0} $ are all small scale. In comparison, the momentum associated with the charged channel is about $3-6$ times larger. Therefore, in the XEFT, the charged channel is integrated out. {In fact, some KSW-type EFTs containing the charged channel have been developed . In this review, we use the terminology ``XEFT" to denote the EFT without the charged channel as in the original paper~\cite{Fleming:2007rp}.} Meanwhile, the $\Delta$ and $m_{\pi^0}$ are also large scales compared with $u\sim p_\pi$. Thus, in XEFT, the expansion is performed in the powers of $p_{D}\sim p_{D^{\ast}}\sim p_{\pi}\sim u\sim\gamma$. We use $Q$ to label these small scales. The velocities of the explicit degrees of freedom, $D^0$, $\bar{D}^{*0}$ and $\pi^0$ are much smaller than $1$, thus they are all treated nonrelativistically. The Lagrangians are constructed as follows (\cvII in Table~\ref{tab:two-convention}),
\begin{eqnarray}
	\mathcal{L}&=&\bm{D}^{\ast\dagger}\cdot\left(i\partial_{t}+\frac{{\vec{\bm\nabla}}^{2}}{2m_{D^{\ast}}}\right)\bm{D}^{\ast}+D^{\dagger}\left(i\partial_{t}+\frac{{\vec{\bm\nabla}}^{2}}{2m_{D}}\right)D+\bar{\bm{D}}^{\ast\dagger}\cdot\left(i\partial_{t} +\frac{{\vec{\bm\nabla}}^{2}}{2m_{D^{\ast}}}\right)\bar{\bm{D}}^{\ast}+\bar{D}^{\dagger}\left(i\partial_{t}+\frac{{\vec{\bm\nabla}}^{2}}{2m_{D}}\right)\bar{D} \nonumber\\
&&+\pi^{\dagger}\left(i\partial_{t}+\frac{{\vec{\bm\nabla}}^{2}}{2m_{\pi}}+\delta\right)\pi+\frac{g_{b}}{2f_{\pi}}\frac{1}{\sqrt{2m_{\pi}}}\left(D\bm{D}^{\ast\dagger}\cdot\vec{\bm{\nabla}}\pi +\bar{D}^{\dagger}\bar{\bm{D}}^{\ast}\cdot\vec{\bm{\nabla}}\pi^{\dagger}\right)+\mathrm{H.c.}\nonumber\\
&&-\frac{C_{0}}{2}\left(\bar{\bm{D}}^{\ast}D+\bm{D}^{\ast}\bar{D}\right)^{\dagger}\cdot\left(\bar{\bm{D}}^{\ast}D+\bm{D}^{\ast}\bar{D}\right) +\frac{C_{2}}{16}\left(\bar{\bm{D}}^{\ast}D+\bm{D}^{\ast}\bar{D}\right)^{\dagger}\cdot\left(\bar{\bm{D}}^{\ast}(\overleftrightarrow{\bm{\nabla}})^{2}D+\bm{D}^{\ast}(\overleftrightarrow{\bm{\nabla}})^{2}\bar{D}\right)+\mathrm{H.c.} \nonumber\\ &&+\frac{B_{1}}{\sqrt{2}}\frac{1}{\sqrt{2m_{\pi}}}\left(\bar{\bm{D}}^{\ast}D+\bm{D}^{\ast}\bar{D}\right)^{\dagger}\cdot D\bar{D}{\vec{\bm\nabla}}\pi+\mathrm{H.c.}+\dots, \label{eq:XEFT_lag}
\end{eqnarray}
where $\overleftrightarrow{\bm{\nabla}}=\overleftarrow{\bm{\nabla}}-\overrightarrow{\bm{\nabla}}$.
The power counting can be summarized as follows,
\begin{gather}
p_{D}\sim p_{D^{\ast}}\sim p_{\pi}\sim u\sim\gamma\sim Q,\\
\text{Propagators:}\sim\frac{1}{E_{D}}/\frac{1}{E_{D^{\ast}}}/\frac{1}{E_{\pi}}\sim Q^{-2},\\
\text{Loop integrals:}\sim Q^{5},\\
D^{\ast0}D^{0}\pi^{0}\text{ vertex:}\sim Q,\quad\text{OPE interaction: }\frac{g_{b}^{2}}{4f_{\pi}^{2}}\frac{\bm{q}\cdot\bm{\bm{\varepsilon}}^{\prime\dagger}\bm{q}\cdot\bm{\bm{\varepsilon}}}{\bm{q}^{2}-u^{2}}\sim Q^{0}.
\end{gather}
Like the LO contact interaction in the KSW scheme, the NDA of $C_0$ is $1/(M_D \Lambda)$. However, for a system with unnaturally large scattering length, the scale of $C_0$ is $1/(M_DQ)$ as shown in Table~\ref{tab:sec1.5:ksw_pc}. One can find that
\begin{equation}
	C_0={2\pi \over  \mu_{DD^*}}{1\over  {1\over a_s}-\mu},
\end{equation}
where $\mu$ is the renormalization scale in PDS, and $\mu_{DD^*}$ is the reduced mass of the $D^0$ and $\bar{D}^{*0}$ pair. The corresponding vertex is at the order of $Q^{-1}$. An extra loop with an extra vertex $C_0$ will introduce an extra factor at order $1$ [$Q^5\times (Q^{-2})^2\times Q^{-1}\sim Q^0$] as shown in Sec.~\ref{sec:ksw} (see Fig.~\ref{fig:KSW_Order}). Therefore,  the $C_0$ should be resummed nonperturbatively. {In this unnatural case, the LEC $C_2$ is at the order of $Q^{-2}$ and the Lagrangian with the $C_2$ term is at the order of $Q^0$, which is the NLO contribution as the OPE.} The NLO and higher order vertices can be included perturbatively. It seems that the XEFT should be different from KSW because {the pion mass is treated as a large scale in XEFT.} However, the $u$ is another small scale and serves as the similar role as the pion mass in the KSW framework. {There also exists the KSW-type EFTs containing the $D^*\bar{D}^*$ system, where the pion mass is treated as a small scale as in Refs.~\cite{Kaplan:1998sz,Kaplan:1998tg}. However, in this review, we restrict the terminology ``XEFT" for the neutral $D^*\bar{D}$ system without the $D^*\bar{D}^*$ channel as in the original paper of Fleming \etal~\cite{Fleming:2007rp}, which is extended to investigate the $D^*D$ system at most.}

\begin{figure}
	\centering
	\includegraphics[width = 1.0\textwidth]{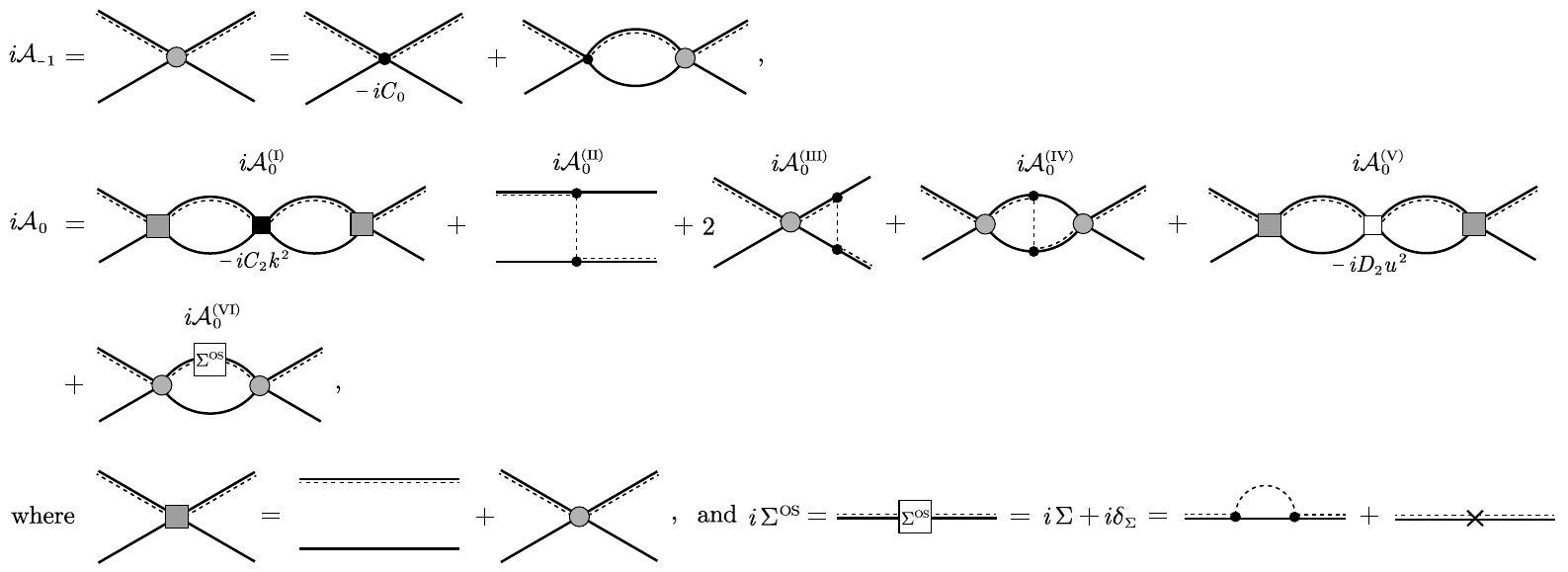}
	\caption{Feynman diagrams in Ref.~\cite{Jansen:2013cba} for the $\xthn$ scattering in the $J^{PC}=1^{++}$ channel to NLO.}\label{fig:4.x_xmassXEFT}
\end{figure}

With XEFT, one can calculate the $\xthn$ scattering process (e.g.,~\cite{Jansen:2013cba}). The Feynman diagrams up to NLO are listed in Fig.~\ref{fig:4.x_xmassXEFT}. One can see that the $C_0$ term is treated nonperturbatively, while the OPE and NLO contact terms are calculated perturbatively. With the scattering amplitude, one can extract the coupling constants of the $X(3872)$ with $\xthn$ by calculating the residue of the $T$-matrix as shown in Fig.~\ref{Fig:4.1:bound_state}(a).

\begin{figure}
	\centering
	\includegraphics[width = 1.0\textwidth]{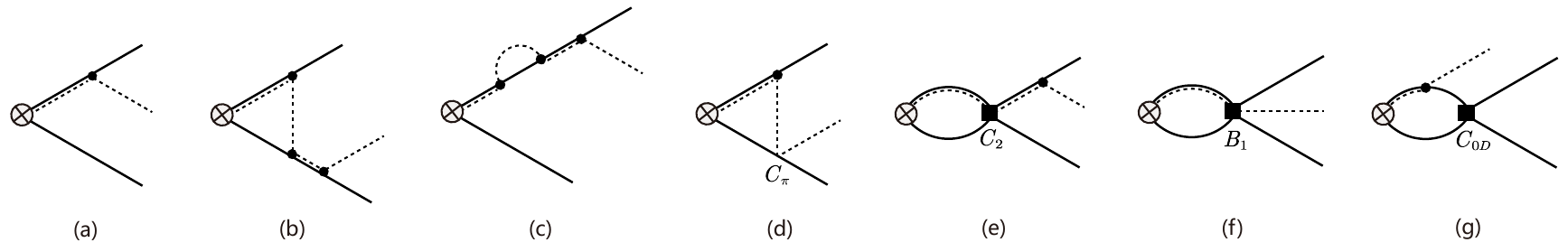}
	\caption{Feynman diagrams for $X\to D^0\bar{D}^0\pi^0$ up to NLO in XEFT. The circled cross represents the vertex (form factor)$ XD^*\bar{D}$ and its charge conjugations.}\label{fig:4.x_xtoDDpi}
\end{figure}

The Feynman diagrams contributing to $X\to D^{0}\bar{D}^{0}\pi^{0}$ is presented {in Fig.~\ref{fig:4.x_xtoDDpi}}.  The diagram (a) contributes to the LO decay widths. The circled cross represents the vertex (form factor) of $ XD^*\bar{D}$ and its charge conjugations, which is obtained by iterating the $C_0$ terms. The diagrams (b)-(g) contributing to the NLO. The pion vertices, $C_2$ and $B_1$ are included as perturbation. The LEC $B_1$ is at the same order of $Q^{-2}$ as $C_2$. The $C_\pi$ and $C_{0D}$ terms are omitted in the original work of XEFT~\cite{Fleming:2007rp} and first pointed out in Ref.~\cite{Guo:2017jvc}, which will be discussed in details in Sec.~\ref{sec:X_long_range}.

\subsection{Chiral effective field theory}

In Sec.~\ref{sec:pionlessEFT}, we have discussed the $\slashed{\pi}$EFT, in which the pions are totally integrated out. In Sec.~\ref{sec:XEFT}, we elucidated the XEFT, in which the pions are treated perturbatively. Modern theory of nuclear forces ($\chi$EFT) is built upon the Weinberg scheme (see Sec.~\ref{sec:chiEFT}), in which the pion is an explicit d.o.f and treated nonperturbatively. The EFTs with different treatment of the pion are valid in different low energy scales~\cite{Barford:2002je,Birse:2005um}, see Sec.~\ref{sec:x_methodology} for details. In this section, $\chi$EFT  is generalized to heavy hadron systems combining both the chiral symmetry and the heavy quark symmetry. The expansion parameter in this case is $\mathcal{Q}/\Lambda$, where the corresponding low energy scale $\mathcal{Q}$ can be  either the pion mass $m_\pi$, the momentum of pion, or the residual momenta of heavy hadrons, while the high energy scale $\Lambda$ can  be either the chiral symmetry breaking scale $\Lambda_\chi$ or the heavy hadron masses. As stated in Sec.~\ref{sec:chiEFT}, Weinberg's proposal is to extract the effective potentials from the two particle irreducible diagrams. The importance of the corresponding irreducible diagrams is measured via the power counting given in Eq.~\eqref{eq:1.4:pwc}. In this part, we take the $D\bar{D}^\ast/\bar{D}D^\ast$ system as an example to illustrate the generalizations of $\chi$EFT in the heavy hadron systems.

 \begin{figure*}
 	\centering
 	\includegraphics[width = 1.0\textwidth]{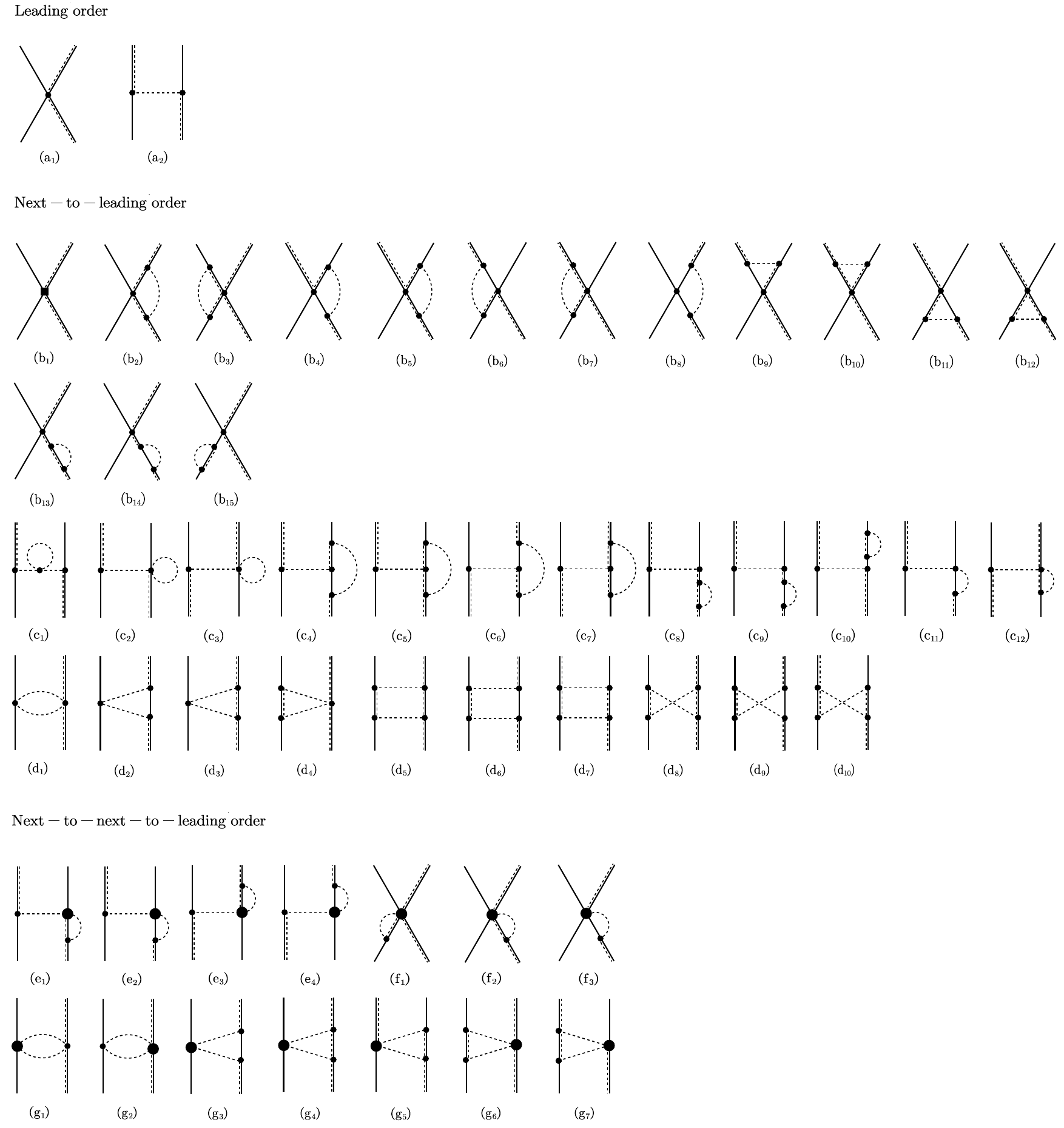}
 	\caption{The LO, NLO and N$^2$LO Feynmann diagrams of the $D\bar{D}^\ast/\bar{D}D^\ast$ scatterings in $\chi$EFT, where the double (solid plus dashed) line denotes the $D^\ast/\bar{D}^\ast$, while the single solid line and dashed line denote the $D/\bar{D}$ and pion, respectively. The small dot (`$\bullet$') denotes the LO vertex, such as the vertices in Lagrangias~\eqref{eq:LOctLagHH},~\eqref{eq:app1:lagD} and~\eqref{eq:app1:lagDbar}. The vertex denoted by black square (`{\tiny$\blacksquare$}') in diagram $\mathrm{(b_1)}$ comes from the NLO contact Lagrangian~\eqref{eq:NLOcontactDDast}. The large dot (`{\Large$\bullet$}') denotes the NLO vertex, such as the two-pion coupling vertices in $(\mathrm{e_1})-(\mathrm{e_4})$ and $(\mathrm{g_1})-(\mathrm{g_7})$ from the NLO Lagrangian~\eqref{eq:HHppNLO}, and vertices in diagrams $(\mathrm{f_1})-(\mathrm{f_3})$ from the Lagrangian~\eqref{eq:HHHHpNLO}. The diagrams $(\mathrm{b_2})-(\mathrm{b_{15}})$ represent the one-loop corrections to LO contact term, while the diagrams $(\mathrm{c_1})-(\mathrm{c_{12}})$ are the one-loop corrections to OPE.\label{fig:chiEFT_graphs}}
 \end{figure*}

\subsubsection{Leading order interactions}\label{sec:LOinteractions}

According to the power counting in Eq.~\eqref{eq:1.4:pwc}, the LO effective potentials of $D\bar{D}^\ast/\bar{D}D^\ast$ receive contributions from the contact terms and one-pion exchange (OPE) interaction, see the diagrams in the first row of Fig.~\ref{fig:chiEFT_graphs}. For a more concrete case, we consider the $D\bar{D}^\ast/\bar{D}D^\ast$ system with $I^G(J^{PC})=1^+(1^{+-})$ ($C$-parity for the neutral systems only), which corresponds to the $Z_c(3900)$ with flavor wave function
\begin{eqnarray}\label{eq:fwZc3900}
|1^+(1^{+-})\big\rangle=\frac{1}{\sqrt{2}}(D\bar{D}^{\ast}+D^{\ast}\bar{D}),
\end{eqnarray}
where the relative sign is determined by the convention $\hat{C}D^\ast\to-\bar{D}^\ast$ (\cvI in Table~\ref{tab:two-convention}).
One can derive the static OPE potential of this state with the LO chiral Lagrangians in Eqs.~\eqref{eq:app1:lagD} and~\eqref{eq:app1:lagDbar}, which reads
\begin{eqnarray}
V_{\text{OPE}}=-\frac{g_b^2}{4f_\pi^2}\frac{(\bm q\cdot\bm{\varepsilon})(\bm q\cdot\bm{\varepsilon}^{\prime\dagger})}{\bm q^2+m_\pi^2},\label{eq:VOPE1}
\end{eqnarray}
with $\bm q=\bm p-\bm p^\prime$ the transferred momentum.
$\bm p~(\bm p^\prime)$ denotes the momentum of the initial (final) states in the center of mass system (c.m.s). $\bm{\varepsilon}$ ($\bm{\varepsilon}^{\prime\dagger}$) is the polarization vector of the initial (final) $D^\ast$ or $\bar{D}^\ast$ meson. Besides, the Breit
approximation $\mathcal{V}=-\mathcal{M}/\sqrt{\Pi_i2m_i\Pi_f2m_f}$ [$m_i$ ($m_f$) stands for the mass of initial (final) state] has been used to relate the effective potential $\mathcal{V}$ to the scattering amplitude $\mathcal{M}$~\cite{Berestetsky:1982bf}. Note that the term $p_0-p_0^\prime\approx m_{D^\ast}-m_D=\delta_b$ is ignored in the denominator of Eq.~\eqref{eq:VOPE1}. One recovers the structure of Eq.~\eqref{eq:OPEinXEFT} once the mass difference is considered.
\begin{figure}[!hptb]
\begin{centering}
    \scalebox{1.0}{\includegraphics[width=0.35\linewidth]{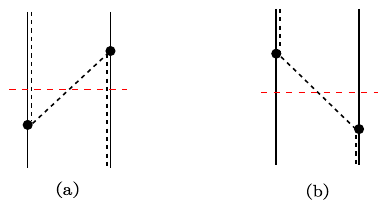}}
    \caption{The OPE diagrams for the $D\bar{D}^\ast/\bar{D}D^\ast$ scattering in time-ordered perturbation theory. The notations are the same as those in Fig.~\ref{fig:chiEFT_graphs}. The red-dashed horizontal line indicates the time at which the intermediate state is evaluated. In diagrams (a) and (b), the propagators are retarded and advanced ones, respectively. The corresponding intermediate states in diagrams (a) and (b) are $D^\ast\bar{D}^\ast\pi$ and $D\bar{D}\pi$, respectively.\label{fig:TOPT}}
\end{centering}
\end{figure}

In Refs.~\cite{Baru:2011rs,Wang:2018jlv}, the time-ordered perturbation theory (TOPT) (see~\cite{Schwartz:2014sze,Sterman:1993hfp} for the details of the Feynman rules in TOPT) is adopted to derive the nonstatic (energy dependent) OPE potential. Two typical contributions are shown in Figs.~\ref{fig:TOPT}(a) and~\ref{fig:TOPT}(b), which read
\begin{eqnarray}
V_{\text{OPE}}^{\text{(a)}}&=&-\frac{g_b^2}{4f_\pi^2}\frac{(\bm q\cdot\bm{\varepsilon})(\bm q\cdot\bm{\varepsilon}^{\prime\dagger})}{2E_\pi(E_\pi+E_{D^\ast}+E_{\bar{D}^\ast}-E-i\epsilon)},\label{OPEel}\\
V_{\text{OPE}}^{\text{(b)}}&=&-\frac{g_b^2}{4f_\pi^2}\frac{(\bm q\cdot\bm{\varepsilon})(\bm q\cdot\bm{\varepsilon}^{\prime\dagger})}{2E_\pi(E_\pi+E_{D}+E_{\bar{D}}-E-i\epsilon)},\label{OPEin}
\end{eqnarray}
where
\begin{eqnarray}
E_\pi=\sqrt{\bm q^2+m_\pi^2},\qquad
 E_{i}=m_{i}+\frac{\bm p_i^2}{2m_{i}},\quad (i=D^\ast,\bar{D}^\ast,D,\bar{D}),
\end{eqnarray}
and $E$ represents the total energy of the system.

If we only focus on the $\SU(2)$ case [see Ref.~\cite{Liu:2012vd} for the $\SU(3)$ case], the LO contact potential is obtained from the following Lagrangian without derivatives~\cite{Liu:2012vd,AlFiky:2005jd,Valderrama:2012jv},
\begin{eqnarray}\label{eq:LOctLagHH}
\mathcal{L}_{\mathrm{ct}}^{(0)}&=&D_a\big\langle\bar{\tilde{\mathcal{H}}}\tilde{\mathcal{H}}\big\rangle\big\langle\mathcal{H}\bar{\mathcal{H}}\big\rangle+D_b\big\langle\bar{\tilde{\mathcal{H}}}\gamma^{\mu}\gamma_{5}\tilde{\mathcal{H}}\big\rangle\big\langle\mathcal{H}\gamma_{\mu}\gamma_{5}\bar{\mathcal{H}}\big\rangle\nonumber\\
&&+E_a\big\langle\bar{\tilde{\mathcal{H}}}\tau_{i}\tilde{\mathcal{H}}\big\rangle\big\langle \mathcal{H}\tau_{i}\bar{\mathcal{H}}\big\rangle+ E_b\big\langle\bar{\tilde{\mathcal{H}}}\gamma^{\mu}\gamma_{5}\tau_{i}\tilde{\mathcal{H}}\big\rangle\big\langle\mathcal{H}\gamma_{\mu}\gamma_{5}\tau_{i}\bar{\mathcal{H}}\big\rangle,
\end{eqnarray}
where $D_a$, $D_b$, $E_a$ and $E_b$ are four LECs. The $\tau_i$ is the isospin Pauli matrix. There are other forms of couplings such as $\big\langle\bar{\tilde{\mathcal{H}}}\gamma_\mu\tilde{\mathcal{H}}\big\rangle\big\langle\mathcal{H}\gamma^\mu\bar{\mathcal{H}}\big\rangle$, which is equal to $\big\langle\bar{\tilde{\mathcal{H}}}v_\mu\tilde{\mathcal{H}}\big\rangle\big\langle\mathcal{H}v^\mu\bar{\mathcal{H}}\big\rangle$ with heavy field reduction and is absorbed by adjusting the $D_a$. The term $\big\langle\bar{\tilde{\mathcal{H}}}\gamma_5\tilde{\mathcal{H}}\big\rangle\big\langle\mathcal{H}\gamma_5\bar{\mathcal{H}}\big\rangle$ vanishes in the heavy quark limit. The remaining terms $\big\langle\bar{\tilde{\mathcal{H}}}\sigma_{\mu\nu}\tilde{\mathcal{H}}\big\rangle\big\langle\mathcal{H}\sigma^{\mu\nu}\bar{\mathcal{H}}\big\rangle$ and $\big\langle\bar{\tilde{\mathcal{H}}}\sigma_{\mu\nu}\gamma_5\tilde{\mathcal{H}}\big\rangle\big\langle\mathcal{H}\sigma^{\mu\nu}\gamma_5\bar{\mathcal{H}}\big\rangle$ can also be absorbed by adjusting the $D_b$ (see the properties of the gamma matrices under the heavy field reduction in Ref.~\cite{Scherer:2002tk}).

It is worth mentioning that the contact Lagrangians introduced here as well as those in Secs.~\ref{sec:pionlessEFT} and~\ref{sec:XEFT} should be regarded as the parameterization of the dynamics that occur at the scale which is much shorter than the scale we are working, but they are not the true zero-range interaction, e.g., see the discussions in Ref.~\cite{Cohen:1996my}, which implies that the regularization is necessary from the outset.

With the Lagrangian~\eqref{eq:LOctLagHH}, the LO contact potential of the state~\eqref{eq:fwZc3900} is
\begin{eqnarray}\label{eq:LOcontactDDast}
V_{\mathrm{ct}}^{(0)}=(-D_a-D_b+E_a+E_b)\bm{\varepsilon}\cdot\bm{\varepsilon}^{\prime\dagger},
\end{eqnarray}
where the LECs may be separately determined either by fitting the experimental data or from the phenomenological meson exchange model such as the resonance saturation model (RSM)~\cite{Ecker:1988te,Epelbaum:2001fm}.

\begin{figure}[!hptb]
\begin{centering}
    \scalebox{1.0}{\includegraphics[width=0.6\linewidth]{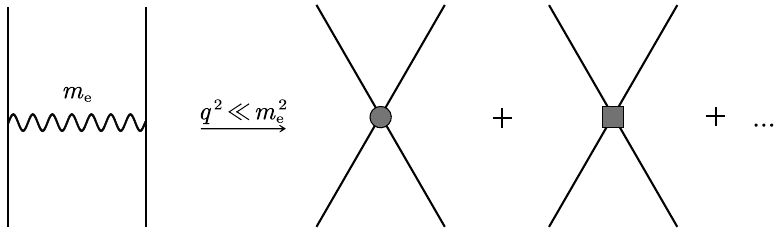}}
    \caption{{An illustration of the resonance saturation model. The wiggly line denotes the exchanged resonances. The circle and square blobs represent the contact vertices with zero and two derivatives, respectively. The ellipsis stands for the higher order terms (with more derivatives). } \label{fig:rsm}}
\end{centering}
\end{figure}

{Here, we append a brief introduction to the RSM. The basic idea of RSM is to localize the resonance-exchange contribution if one is interested in the region $q^2\ll m_{\rm e}^2$, e.g., see the illustration in Fig.~\ref{fig:rsm}. The exchanged resonances may contain many types with different quantum numbers, such as the scalar (s), pseudoscalar (p), vector (v), axial-vector (a), and tensor (t), etc. We take the $NN$ interaction as an example. Within the framework of the one-boson exchange (OBE) model, the effective potential of $NN$ can be written as
\begin{eqnarray}
    V_{NN}=V_\pi+\sum_{\rm e=s,p,v,a,t}V_{\rm e},
\end{eqnarray}
where $V_\pi$ denotes the one-pion exchange contribution, which is usually regarded as the long-range force both in the OBE model and the $\chi$EFT. In the region $q^2\ll m_{\rm e}^2$, one can make the following expansion for $V_{\rm e}$,
\begin{eqnarray}\label{eq:nonlotolo}
    V_{\rm e}=(\bar{N}\Gamma_i N)\left(\frac{g_{\rm e}^2\delta^{ij}}{q^2-m_{\rm e}^2}\right)(\bar{N}\Gamma_j N)=-\frac{g_{\rm e}^2}{m_{\rm e}^2}\left[(\bar{N}\Gamma_i N)(\bar{N}\Gamma^i N)+\frac{q^2}{m_{\rm e}^2}(\bar{N}\Gamma_i N)(\bar{N}\Gamma^i N)+\dots\right],
\end{eqnarray}
where $g_{\rm e}$, $q$ and $m_{\rm e}$ denote the coupling constant, the transferred momentum and the mass of the exchanged resonance in order. $\Gamma_i$ are the projectors on the appropriate quantum numbers for a given resonance exchange. From Eq.~\eqref{eq:nonlotolo}, the interaction is changed into the contact form in the soft-momentum approximation. One then can estimate the LECs via comparing the structures in Eq.~\eqref{eq:nonlotolo} and those from the contact chiral Lagrangians. The LECs estimated with the RSM in the meson-meson~\cite{Ecker:1988te}, meson-baryon~\cite{Bernard:1996gq}, and baryon-baryon~\cite{Epelbaum:2001fm} sectors are close to the values determined from the experimental data. The RSM is also applied to the heavy-heavy sectors~\cite{Wang:2022jop,Yan:2021tcp,Du:2016tgp}. One should note that the RSM can work only for a certain range of regulators. For example, the momentum cutoff for the OBE potentials in the $NN$ case is in the range $1-2$ GeV, while it is commonly around $0.5$ GeV for the chiral $NN$ potentials ~\cite{Epelbaum:2001fm}. Therefore, though one may use the RSM to estimate the LECs in EFTs, one should consistently use the cutoff constrained by the validity of the EFTs rather than that of the OBE model.}

\subsubsection{Next-to-leading order interactions}\label{sec:NLOinteractions}

The NLO interactions can be divided into four parts. The first part is the one-loop corrections to the LO OPE [see Figs.~\ref{fig:chiEFT_graphs}($\mathrm{c}_1$)-($\mathrm{c}_{12}$)], which include the corrections to the $D^\ast D\pi$ vertex, the renormalizations of the wave functions of the pion and $D$ mesons. The second part arises from the one-loop corrections to the LO contact diagrams [see Figs.~\ref{fig:chiEFT_graphs}($\mathrm{b}_2$)-($\mathrm{b}_{15}$)]. These two contributions have been systematically considered for the $DD^\ast$~\cite{Xu:2017tsr}, $BB^\ast$\cite{Wang:2018atz} and $\Sigma_c N$~\cite{Meng:2019nzy} systems. Partial diagrams were also calculated for the $X(3872)$~\cite{Wang:2013kva} (see the criticisms in Ref.~\cite{Baru:2015nea}). However, these two parts do not induce new structures. Thus, if one does not care about the $m_\pi$ dependence of the $D\bar{D}^\ast$ forces, their contributions can be included by using the physical values of $m_\pi$, $g_b$, $f_\pi$, $m_D$, $m_{D^\ast}$ and the LO contact LECs (see also the discussions on the one-loop corrections in the $NN$ cases~\cite{Epelbaum:2005pn,Epelbaum:2008ga,Machleidt:2011zz}, and the possible appearance of Goldberger-Treiman discrepancy at this order~\cite{Epelbaum:2002gb}).

The third part comes from the NLO contact Lagrangian [see diagram~\ref{fig:chiEFT_graphs}($\mathrm{b}_1$)], which carries two derivatives or has an insertion of the light quark mass terms,
\begin{eqnarray}\label{eq:NLOcontactDDast}
\mathcal{L}_{\mathrm{ct}}^{(2)}	=	\sum_{i=1}^{4}\mathcal{D}_{\mu}\mathcal{D}^{\mu}\langle\bar{\tilde{\mathcal{H}}}\mathbb{O}_{i}\tilde{\mathcal{H}}\rangle\langle\mathcal{H}\mathbb{O}_{i}\bar{\mathcal{H}}\rangle+\sum_{i=5}^{6}\mathcal{D}_{\mu}\mathcal{D}^{\nu}\langle\bar{\tilde{\mathcal{H}}}\mathbb{O}_{i}^{\mu} \tilde{\mathcal{H}}\rangle\langle\mathcal{H}\mathbb{O}_{i\nu}\bar{\mathcal{H}}\rangle
+\sum_{i=1}^{4}\chi\langle\bar{\tilde{\mathcal{H}}}\mathbb{O}_{i}\tilde{\mathcal{H}}\rangle\langle\mathcal{H}\mathbb{O}_{i}\bar{\mathcal{H}}\rangle,
\end{eqnarray}
where the $\mathbb{O}_i$ are defined as $\mathbb{O}_{1}=\bm{\mathrm{I}},\mathbb{O}_{2}=\gamma^{\mu}\gamma_{5},\mathbb{O}_{3}=\tau_{i},\mathbb{O}_{4}=\gamma^{\mu}\gamma_{5}\tau_{i},\mathbb{O}_{5}^{\mu}=\gamma^{\mu}\gamma_{5},\mathbb{O}_{6}^{\mu}=\gamma^{\mu}\gamma_{5}\tau_{i}$. The two covariant derivatives terms stand for the following allocations,
\begin{eqnarray}
\mathcal{D}_{\mu}\mathcal{D}^{\mu}\langle\bar{\tilde{\mathcal{H}}}\mathbb{O}_{i}\tilde{\mathcal{H}}\rangle\langle\mathcal{H}\mathbb{O}_{i}\bar{\mathcal{H}}\rangle&\equiv& C_{i1}^{d}\left(\langle\mathcal{D}_{\mu}\bar{\tilde{\mathcal{H}}}\tilde{\mathcal{H}}\rangle\langle\mathcal{H}\mathcal{D}^{\mu}\bar{\mathcal{H}}\rangle+\langle\bar{\tilde{\mathcal{H}}}\mathcal{D}_{\mu}\tilde{\mathcal{H}}\rangle\langle\mathcal{D}^{\mu}\mathcal{H}\bar{\mathcal{H}}\rangle\right)\nonumber\\ &&+C_{i2}^{d}\left(\langle\mathcal{D}_{\mu}\bar{\tilde{\mathcal{H}}}\tilde{\mathcal{H}}\rangle\langle\mathcal{D}^{\mu}\mathcal{H}\bar{\mathcal{H}}\rangle+\langle\bar{\tilde{\mathcal{H}}}\mathcal{D}_{\mu}\tilde{\mathcal{H}}\rangle\langle\mathcal{H}\mathcal{D}^{\mu}\bar{\mathcal{H}}\rangle\right)\nonumber\\
		&&+C_{i3}^{d}\left(\langle\mathcal{D}^{2}\bar{\tilde{\mathcal{H}}}\tilde{\mathcal{H}}\rangle\langle\mathcal{H}\bar{\mathcal{H}}\rangle+\langle\bar{\tilde{\mathcal{H}}}\mathcal{D}^{2}\tilde{\mathcal{H}}\rangle\langle\mathcal{H}\bar{\mathcal{H}}\rangle\right)\nonumber\\
		&&+C_{i4}^{d}\left(\langle\bar{\tilde{\mathcal{H}}}\tilde{\mathcal{H}}\rangle\langle\mathcal{D}^{2}\mathcal{H}\bar{\mathcal{H}}\rangle+\langle\bar{\tilde{\mathcal{H}}}\tilde{\mathcal{H}}\rangle\langle\mathcal{H}\mathcal{D}^{2}\bar{\mathcal{H}}\rangle\right)\nonumber\\
		&&+C_{i5}^{d}\left(\langle\mathcal{D}_{\mu}\bar{\tilde{\mathcal{H}}}\mathcal{D}^{\mu}\tilde{\mathcal{H}}\rangle\langle\mathcal{H}\bar{\mathcal{H}}\rangle+\langle\bar{\tilde{\mathcal{H}}}\tilde{\mathcal{H}}\rangle\langle\mathcal{D}_{\mu}\mathcal{H}\mathcal{D}^{\mu}\bar{\mathcal{H}}\rangle\right).
\end{eqnarray}
The $\chi\langle\bar{\tilde{\mathcal{H}}}\mathbb{O}_{i}\tilde{\mathcal{H}}\rangle\langle\mathcal{H}\mathbb{O}_{i}\bar{\mathcal{H}}\rangle$ terms are the $m_\pi^2$ related terms, allowing the following forms
\begin{eqnarray}
\chi\langle\bar{\tilde{\mathcal{H}}}\mathbb{O}_{i}\tilde{\mathcal{H}}\rangle\langle\mathcal{H}\mathbb{O}_{i}\bar{\mathcal{H}}\rangle\equiv C_{i1}^{\chi}\langle\bar{\tilde{\mathcal{H}}}\mathbb{O}_{i}\tilde{\mathcal{H}}\rangle\langle\mathcal{H}\mathbb{O}_{i}\bar{\mathcal{H}}\rangle\mathrm{Tr}(\chi_{+})
+C_{i2}^{\chi}\langle\bar{\tilde{\mathcal{H}}}\hat{\chi}_{+}\mathbb{O}_{i}\tilde{\mathcal{H}}\rangle\langle\mathcal{H}\mathbb{O}_{i}\bar{\mathcal{H}}\rangle
+C_{i3}^{\chi}\langle\bar{\tilde{\mathcal{H}}}\mathbb{O}_{i}\tilde{\mathcal{H}}\rangle\langle\mathcal{H}\hat{\chi}_{+}\mathbb{O}_{i}\bar{\mathcal{H}}\rangle,
\end{eqnarray}
where $\hat{\chi}_+=\chi_+-\frac{1}{2}\Tr(\chi_+)$ in the $\SU(2)$ case. These terms can be absorbed into the LO contact interaction, if one does not care about the pion mass dependence.

Similar to the $NN$ case~\cite{Erkelenz:1971caz}, it is more convenient to extract the operator structures of the $D\bar{D}^\ast/\bar{D}D^\ast$ interactions in momentum space. There are four independent operators for the $D\bar{D}^\ast/\bar{D}D^\ast$ system~\cite{Wang:2020dko},
\begin{align}\label{eq:OperatorsOp2}
\mathcal{O}_1&=\bm{\varepsilon}^{\prime\dagger}\cdot\bm{\varepsilon},&\mathcal{O}_2&=(\bm{\varepsilon}^{\prime\dagger}\times\bm{\varepsilon})\cdot(\bm q\times\bm k),\nonumber\\
\mathcal{O}_3&=(\bm q\cdot\bm{\varepsilon}^{\prime\dagger})(\bm q\cdot\bm{\varepsilon}),&\mathcal{O}_4&=(\bm k\cdot\bm{\varepsilon}^{\prime\dagger})(\bm k\cdot\bm{\varepsilon}),
\end{align}
with $\bm k=(\bm p^\prime+\bm p)/2$ the average momentum. A similar term $[\bm{\varepsilon}^{\prime\dagger}\cdot(\bm q\times \bm k)][\bm{\varepsilon}\cdot(\bm q\times \bm k)]$ contributes to the N$^3$LO (or even higher order) potentials~\cite{Erkelenz:1971caz}. Note that the remaining two terms $(\bm q\times\bm{\varepsilon}^{\prime\dagger})\cdot(\bm q\times\bm{\varepsilon})$ and $(\bm k\times\bm{\varepsilon}^{\prime\dagger})\cdot(\bm k\times\bm{\varepsilon})$ in Ref.~\cite{Wang:2020dko} are not independent since $(\bm q\times\bm{\varepsilon}^{\prime\dagger})\cdot(\bm q\times\bm{\varepsilon})=\bm q^2(\bm{\varepsilon}^{\prime\dagger}\cdot\bm{\varepsilon})-(\bm q\cdot\bm{\varepsilon}^{\prime\dagger})(\bm q\cdot\bm{\varepsilon})$. With the operators in Eq.~\eqref{eq:OperatorsOp2}, the NLO contact potential of the state~\eqref{eq:fwZc3900} can be parameterized as follows,
\begin{eqnarray}\label{eq:Vct2DDast}
V_{\text{ct}}^{(2)}=(C_1\bm q^2+C_2\bm k^2)\mathcal{O}_1+\sum_{i=2}^{4}C_{i+1}\mathcal{O}_i,
\end{eqnarray} 
in which the $m_\pi^2$ related terms are omitted. The $C_is$ are the linear combinations of the LECs in Eq.~\eqref{eq:NLOcontactDDast}. An additional term $C_6m_\pi^2\mathcal{O}_1$ should be added into Eq.~\eqref{eq:Vct2DDast} if one is interested in the $m_\pi$ dependence, {see e.g.,~\cite{Jansen:2013cba,Baru:2013rta,Baru:2015tfa,Meng:2019nzy,Haidenbauer:2020uci}.} The $V_{\text{ct}}^{(2)}$ only contributes to the $S$- and $P$-wave interactions. The $D$-wave interaction needs contact terms at least at $\mathcal{O}(p^4)$~\cite{Machleidt:2011zz}. Projecting the $V_{\text{ct}}^{(2)}$ into $S$-wave (considering the $S$- and $D$-wave mixing) with the partial wave decomposition (PWD)~\cite{Golak:2009ri}, one obtains
\begin{eqnarray}\label{eq:VctPWD}
[V_{\text{ct}}^{(2)}]_{LL^\prime}=\left[
\begin{array}{cc}
V_{^3S_1}&V_{^3S_1-^3D_1}\\
V_{^3D_1-^3S_1}&V_{^3D_1}
\end{array} \right]=\left[
\begin{array}{cc}
\tilde{C}_\text{s}+C_\text{s}(p^2+p^{\prime2})&C_\text{sd}p^2\\
C_\text{sd}p^{\prime2}&0
\end{array} \right],
\end{eqnarray}
where
\begin{eqnarray}\label{eq:LECsPWD}
\tilde{C}_\text{s}&=&4\pi(-D_a-D_b+E_a+E_b),\nonumber\\
C_\text{s}&=&\pi(4C_1+C_2+\frac{4}{3}C_4+\frac{1}{3}C_5),\nonumber\\
C_\text{sd}&=&-\frac{\sqrt{2}}{3}\pi(4 C_4+C_5).
\end{eqnarray}
The $C_3$ does not appear in Eq.~\eqref{eq:LECsPWD}. Because the related operator $\mathcal{O}_2$ is responsible for the spin-orbit ($S$-$L$) coupling, which vanishes in the $S$-wave ($L=0$) case. In performing PWD, the polarization vectors $\bm{\varepsilon}$ and $\bm{\varepsilon}^{\prime\dagger}$ are related to the conventional spin operators of the vector particles with the spin transition operators (see the Appendix~C of Ref.~\cite{Wang:2019ato}).

The fourth part of the NLO effective potential originates from the two-pion exchange (TPE) diagrams [see Figs.~\ref{fig:chiEFT_graphs}($\mathrm{d}_1$)-($\mathrm{d}_{10}$)], in which the one-pion and two-pion coupling vertices are governed by the axial coupling terms and chiral connections of the LO Langrangians~\eqref{eq:app1:lagD} and~\eqref{eq:app1:lagDbar}, respectively. The TPE contributions up to NLO have been considered for heavy hadron systems in a series of works~\cite{Liu:2012vd,Wang:2013kva,Xu:2017tsr,Wang:2018atz,Meng:2019ilv,Wang:2019ato,Meng:2019dba,Liu:2019ruv,Meng:2019nzy,Wang:2019nvm,Wang:2020dhf,Wang:2020dko,Wang:2020htx,Chen:2021htr,Xu:2021vsi}. These systems will be discussed later. 
 The planar box diagram~\ref{fig:chiEFT_graphs}($\mathrm{d_6}$) contains the 2PR component (see the definitions about 2PR and 2PIR in Sec.~\ref{sec:chiEFT}), which can also be generated from the iterated OPE when the OPE potential is fed into the nonrelativistic Lippmann-Schwinger equation via
\begin{eqnarray}\label{eq:V2pit}
V_{d_6}^{\mathrm{it}}(\bm p,\bm p^\prime)&=&\int\frac{d^3\bm q}{(2\pi)^3}V_{\mathrm{OPE}}(\bm p,\bm q)\frac{2\mu_{DD^\ast}}{\bm p^2-\bm q^2+i\epsilon}V_{\mathrm{OPE}}(\bm q,\bm p^\prime).
\end{eqnarray}
The $V_{d_6}^{\mathrm{it}}$ is proportional to the reduced mass of $D\bar{D}^\ast$, thus it breaks the naive power counting in Eq.~\eqref{eq:1.4:pwc}. Therefore, one needs to subtract the 2PR contribution in $V_{d_6}$, which can be done with the old-fashioned TOPT~\cite{Ordonez:1993tn,Ordonez:1995rz} or the
covariant perturbation theory~\cite{Kaiser:1997mw}. In the Appendix~B of Ref.~\cite{Wang:2019ato}, the authors demonstrated another trick to make the 2PR subtraction with the mass splitting $\delta_b$ kept, which is based on the principle-value integral within the framework of covariant perturbation theory. Notably, the 2PR contribution also emerges in the one-loop corrections of the LO contact term, e.g., diagrams~\ref{fig:chiEFT_graphs}$(\mathrm{b_{9}})$ and~\ref{fig:chiEFT_graphs}$(\mathrm{b_{11}})$, which can be easily seen via replacing one of the $V_{\mathrm{OPE}}$s in Eq.~\eqref{eq:V2pit} with the LO contact potential~\eqref{eq:LOcontactDDast}.

 For the graphs~\ref{fig:chiEFT_graphs}($\mathrm{d_5}$),~\ref{fig:chiEFT_graphs}($\mathrm{b_{10}}$) and~\ref{fig:chiEFT_graphs}($\mathrm{b_{12}}$), there is no pinched singularity if one keeps the mass splittings of the intermediate states and outer legs.  One can choose either to not perform 2PR subtraction or perform the 2PR subtraction with the inclusion of the inelastic channel $D\bar{D}^\ast\leftrightarrow D^\ast\bar{D}^\ast$ in the LO tree diagrams. In the first choice, the coupled-channel effect of $D^\ast\bar{D}^\ast$ is included via loop diagrams while in the second choice the coupled-channel effect is included by iterating the tree diagrams.

In strict heavy quark symmetry limit, i.e., $m_Q\to\infty$, the exact HQSS guarantees $\delta_b\to 0$. Then the 2PIR TPE potential at NLO can be formulated in a compact form~\cite{Wang:2020dko},
\begin{eqnarray}\label{VTPEform}
V_{\mathrm{TPE}}&=&-\mathcal{O}_1\frac{24(4g_b^2+1)m_\pi^2+(38g_b^2+5)\bm q^2}{2304\pi^2f_\pi^4}
+\mathcal{O}_1\frac{6(6g_b^2+1)m_\pi^2+(10g_b^2+1)\bm q^2}{768\pi^2f_\pi^4}\ln\frac{m_\pi^2}{(4\pi f_\pi)^2}\nonumber\\
&&+\mathcal{O}_1\frac{4(4g_b^2+1)m_\pi^2+(10g_b^2+1)\bm q^2}{384\pi^2f_\pi^4y}\varpi\arctan\frac{y}{\varpi},
\end{eqnarray}
where $\varpi=\sqrt{\bm q^2+4m_\pi^2}$, and $y=\sqrt{2pp^\prime\cos\vartheta-p^2-p^{\prime2}}$ (with $p^{(\prime)}=|\bm p^{(\prime)}|$, and $\vartheta$ the scattering angle in the c.m.s of $D\bar{D}^\ast$).
The $\mathcal{O}_{2,\dots,4}$ terms vanish and only the central potential, i.e., the $\mathcal{O}_1$ term survives~\cite{Meng:2019ilv}.

\subsubsection{Next-to-next-to-leading order interactions}

The N$^2$LO interactions of the $D\bar{D}^\ast/\bar{D}D^\ast$ system are composed of three types of diagrams [see Figs.~\ref{fig:chiEFT_graphs}($\mathrm{e}_1$)-($\mathrm{e}_4$), ($\mathrm{f}_1$)-($\mathrm{f}_3$) and ($\mathrm{g}_1$)-($\mathrm{g}_7$)] due to the absence of the four-body contact Lagrangians at this order. In these diagrams, the following subleading vertices~\cite{Gasser:1987rb,Fettes:1998ud,Fettes:2000gb,Jiang:2019hgs} are inserted,
\begin{eqnarray}
\mathcal{L}_{\mathcal{H}\varphi}^{(2)}&=&\tilde{c}_1\big\langle\mathcal{H}\bar{\mathcal{H}}\big\rangle\Tr(\chi_+)+\tilde{c}_2\big\langle\mathcal{H}u^\mu u^\nu v_\mu v_\nu\bar{\mathcal{H}}\big\rangle+\tilde{c}_3\big\langle\mathcal{H}u^\mu u_\mu\bar{\mathcal{H}}\big\rangle\nonumber\\
&&+i\tilde{c}_4\big\langle\mathcal{H}[u^\mu,u^\nu]\sigma_{\mu\nu}\bar{\mathcal{H}}\big\rangle+\tilde{c}_5\big\langle\mathcal{H}\hat{\chi}_+\bar{\mathcal{H}}\big\rangle,\label{eq:HHppNLO}\\
\mathcal{L}_{\mathcal{H}\mathcal{H}\varphi}^{(1)}&=&\tilde{d}_1\big\langle \bar{\tilde{\mathcal{H}}}\tilde{\mathcal{H}}\big\rangle\big\langle \mathcal{H}\slashed{u}\gamma_5\bar{\mathcal{H}}\big\rangle+\tilde{d}_2\big\langle \bar{\tilde{\mathcal{H}}}\slashed{u}\gamma_5\tilde{\mathcal{H}}\big\rangle\big\langle \mathcal{H}\bar{\mathcal{H}}\big\rangle,\label{eq:HHHHpNLO}
\end{eqnarray}
where the $\tilde{c}_i$ and $\tilde{d}_i$ are corresponding LECs.
One can accordingly obtain the $\mathcal{L}_{\tilde{\mathcal{H}}\varphi}^{(2)}$ for the anticharmed ones (see Sec.~\ref{sec:SwaveHM}). Eq.~\eqref{eq:HHppNLO} contributes to diagrams~\ref{fig:chiEFT_graphs}($\mathrm{e}_1$)-($\mathrm{e}_4$), ($\mathrm{g}_1$)-($\mathrm{g}_7$), and Eq.~\eqref{eq:HHHHpNLO} contributes to diagrams~\ref{fig:chiEFT_graphs}($\mathrm{f}_1$)-($\mathrm{f}_3$), respectively.

The subleading $NN\pi\pi$ couplings can be extracted~\cite{Fettes:1998ud,Bernard:1996gq,Fettes:2000bb,Krebs:2012yv} with the benefit of abundant $\pi N$ scattering data~\cite{Koch:1985bn,Arndt:2006bf}, while for the $\mathcal{H}\mathcal{H}\pi\pi$ ($\tilde{\mathcal{H}}\tilde{\mathcal{H}}\pi\pi$) couplings, only a few terms in Eq.~\eqref{eq:HHppNLO} were estimated for specific problems~\cite{Hofmann:2003je,Kolomeitsev:2003ac,Guo:2008gp,Liu:2009uz,Liu:2011mi,Altenbuchinger:2013vwa,Yao:2015qia,Guo:2015dha,Du:2017ttu,Du:2017zvv,Guo:2018kno}, or see discussions in Sec.~\ref{sec:uchpthm}.

It is pointed out that calculating the TPE loop diagrams with dimensional regularization induces unphysically attractive forces in the isospin scalar $NN$ channel~\cite{Epelbaum:2003gr,Epelbaum:2003xx}, which results from the heavily involved higher momentum modes of the intermediate states. An alternative approach, the spectral function regularization (SFR), was adopted to calculate the TPE contributions~\cite{Epelbaum:2003gr,Epelbaum:2003xx}, in which a cutoff $\tilde{\Lambda}$ (usually chosen to be $0.5<\tilde{\Lambda}<1.0$ GeV) is used in the dispersion relation to suppress the high momentum modes and restrict the potentials in the low-energy region where $\chi$EFT works healthily (see more details in~\cite{Epelbaum:2005pn}). Recently, in the new generation of nuclear force, the local or semi-local regularization is adopted for the pion exchange contributions~\cite{Epelbaum:2014efa,Reinert:2017usi}, which resolves all the issues mentioned above.

At present, the N$^2$LO TPE potentials for the $D D^\ast$ system have been attained in Ref.~\cite{Wang:2022jop}. The higher order contributions, such as the three-pion exchange interactions are still absent. However, they deserve the systematical studies in the future since they are crucial for us to establish a uniform picture between the nuclear forces and their analogues in the heavy hadron sectors.

\subsubsection{Nonperturbative renormalizations}\label{sec:NonperturbativeRenor}

We have handled the regularization of the loop diagrams in deriving the effective potentials above. The potentials will be iterated into the LSEs to generate the possible molecular states in the $D\bar{D}^\ast/\bar{D}D^\ast$ system. Once again, one needs to impose regularization on LSEs to suppress the higher momentum contributions~\cite{Epelbaum:2009sd}. In addition, regularizing the LSEs is also required from the outset---the contact interaction is used to mimic (model) the short-distance physics but not the incarnation of the idealized zero-range forces~\cite{Cohen:1996my}. The renormalization of the Weinberg scheme is not as transparent as that of the KSW scheme but has been proved more practical in the $NN$ systems (see discussions in Sec.~\ref{sec:chiEFT}).  In practical calculations, the LSEs is regularized by multiplying a form factor on the potentials, i.e.,
\begin{eqnarray}
V(p,p^\prime)\to V(p,p^\prime)\mathcal{F}(p,p^\prime,\Lambda),
\end{eqnarray}
where the regulator $\mathcal{F}(p,p^\prime,\Lambda)$ is usually chosen to be the Gaussian form
\begin{eqnarray}
\mathcal{F}(p,p^\prime,\Lambda)=\exp\bigg[-\bigg(\frac{p}{\Lambda}\bigg)^{2n}-\bigg(\frac{p^{\prime}}{\Lambda}\bigg)^{2n}\bigg],\text{ with }n\geq1,
\end{eqnarray}
or the hard cutoff form
\begin{eqnarray}\label{eq:hardcutoff}
\mathcal{F}(p,p^\prime,\Lambda)=\Theta(\Lambda-p)\Theta(\Lambda-p^\prime),\text{ with }\Theta\text{ the step function},
\end{eqnarray}
in which $\Lambda$ is the cutoff parameter.

The value of $\Lambda$ is constrained in the applicable region of $\chi$EFT (such as $\Lambda\sim0.5$ GeV in the $NN$ case). Putting the $\Lambda$ beyond the breaking scale of $\chi$EFT may render the results uncontrollable and lose its predictive power. Here, we adopt the elucidations of nonperturbative renormalizations in Refs.~\cite{Epelbaum:2009sd,Machleidt:2011zz}, which are based on the spirit of Lepage~\cite{Lepage:1997cs}: As long as the cutoff pertains to the working region of $\chi$EFT and the associated errors induced by its finite value are within the theoretical uncertainty at the given order, the $\chi$EFT gradually achieves `nonperturbative renormalization' when the calculation is put to higher orders and the resulting physical observables become more and more insensitive to the different values of the cutoff, e.g., see the performance of nuclear forces~\cite{Entem:2003ft,Entem:2017gor}.

\subsection{Heavy quark symmetry and SU(3) flavor symmetry in heavy hadronic molecules}\label{sec:HQSinHHM}

One prominent feature of the interactions in the heavy hadron systems is the embedding of heavy quark symmetry and light quark flavor symmetry simultaneously. In the light-meson-exchange picture, the interactions between the heavy hadrons are mediated by exchanging the soft light mesons [roughly the SU(3) flavor multiplets], which do not depend on the heavy quark masses and spins in the heavy quark limit. The interactions constrained by HQS and SU(3) flavor symmetry are reflected in the spectrum of the heavy hadronic molecules. With these symmetries, one can make abundant predictions with a few inputs. In the following, we take the $S$-wave $B^{(\ast)}\bar{B}^{(\ast)}$ system as an example to illustrate the HQS and SU(3) light flavor symmetry in the heavy hadronic molecules. The possible symmetry breaking effects are also discussed.

\begin{figure}
	\centering
	\includegraphics[width = 0.5\textwidth]{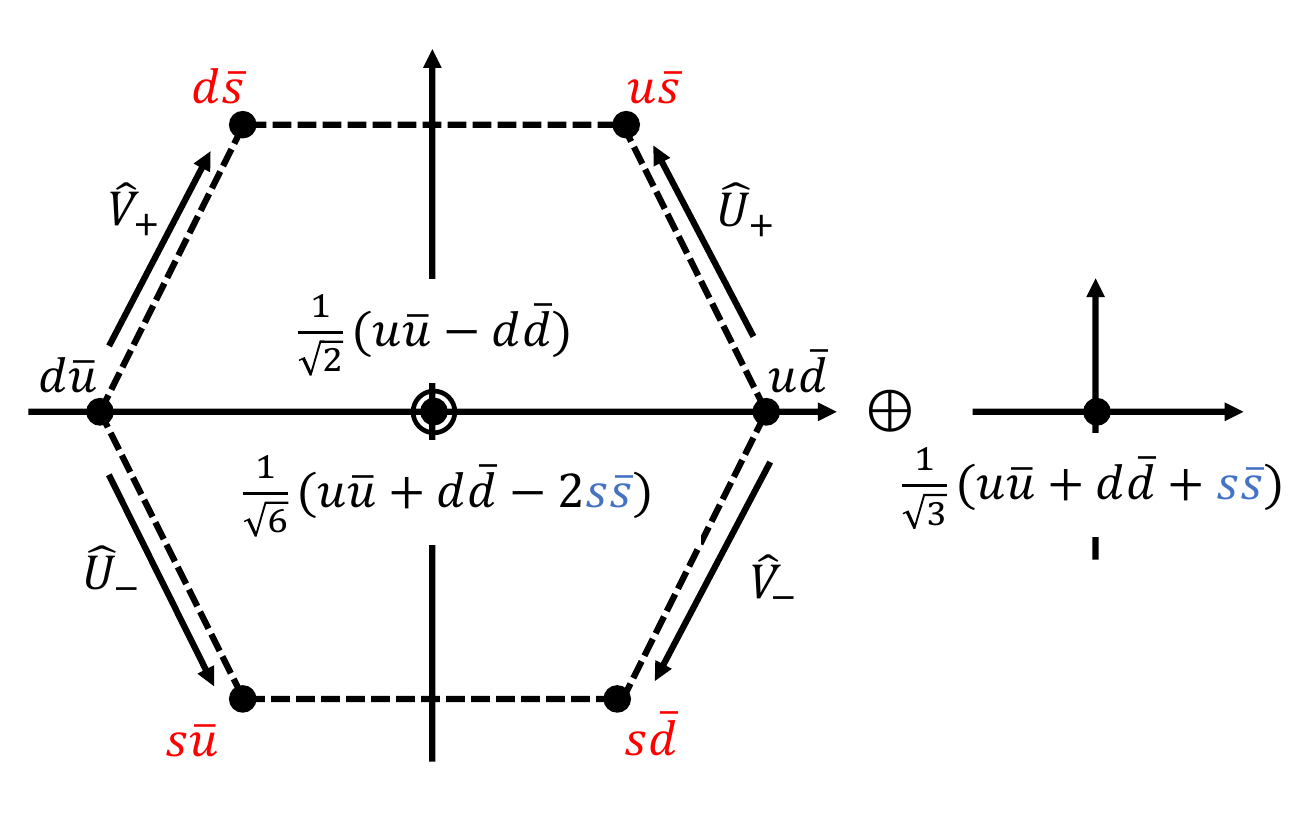}
	\caption{The SU(3) flavor multiplets with the strict flavor symmetry. We use red and blue to label the state with one and two (anti)strange quark, respectively. }\label{fig:4.x_SU3_multi}
\end{figure}

In the SU(3) flavor symmetry limit and the heavy quark limit, the di-meson spectrum is classified by the irreducible representation of SU(3) group and the total spin of the light d.o.fs. In the spin space, the total angular momentum of the $B^{(\ast)}\bar{B}^{(\ast)}$ system can be $J=0,1,2$ {if the S-wave orbital angular momentum is presumed}. One can make the following decomposition for $J$,
\begin{eqnarray}\label{eq:twoparticlebasis}
\begin{cases}
J=0, & B\bar{B},B^{\ast}\bar{B}^{\ast}\\
J=1, & B\bar{B}^{\ast}(B^{\ast}\bar{B}),B^{\ast}\bar{B}^{\ast}\\
J=2 & B^{\ast}\bar{B}^{\ast}
\end{cases}.
\end{eqnarray}  
For the $B\bar{B}^{\ast}(B^{\ast}\bar{B})$ system, one can define the $C$ parity even and $C$ parity odd basis as [$C$ parity for the neutral ones in the following]
\begin{eqnarray}
C\text{ parity even}:&&\frac{1}{\sqrt{2}}(B\bar{B}^{\ast}-B^{\ast}\bar{B}),~\label{eq:cparty1}\\
C\text{ parity odd}:&&\frac{1}{\sqrt{2}}(B\bar{B}^{\ast}+B^{\ast}\bar{B}),~\label{eq:cparty2}
\end{eqnarray}
in which the convention $\hat{C}B^{\ast}\to-\bar{B}^{\ast}$ is implied (\cvI in Table~\ref{tab:two-convention}).  For the $\bar{B}^{(\ast)}$ and ${B}^{(\ast)}$ mesons, we denote the spins of the mesons as the sum of the internal heavy quark and light quark as $\bm j_1 (\bar{B}^{(\ast)})= \bm s_1(b)+\bm\ell_1(\bar{q})$, $\bm j_2({B}^{(\ast)})=\bm s_2(\bar{b})+\bm\ell_2({q})$. In the strict HQSS, it is more convenient to use the basis $|\ell_{1}\ell_{2}S_{\ell},s_{1}s_{2}S_{h};JM\rangle$, where the heavy spin, light spin and total spin are defined as $\bm S_h=\bm s_1+\bm s_2$, $\bm S_\ell=\bm \ell_1+\bm \ell_2$, $\bm J=\bm j_1+\bm j_2=\bm S_h+\bm S_\ell$. Since the interaction does not depend on the $\bm S_h$, the  hadronic molecules form two multiples labeled by the $S_\ell$,
\begin{eqnarray}
	S_{\ell}=0:&&\left\{|0_{h}^{-+}\otimes0_{\ell}^{-+},0^{++}\rangle,|1_{h}^{--}\otimes0_{\ell}^{-+},1^{+-}\rangle\right\},\nonumber\\
	S_{\ell}=1:&&\left\{|0_{h}^{-+}\otimes1_{\ell}^{--},1^{+-}\rangle,|1_{h}^{--}\otimes1_{\ell}^{--},0^{++}\rangle,|1_{h}^{--}\otimes1_{\ell}^{--},1^{++}\rangle,|1_{h}^{--}\otimes1_{\ell}^{--},2^{++}\rangle\right\}.
\end{eqnarray}
The above discussion in the spin space can be extended to the system with strange (anti)quark in the strict SU(3) flavor symmetry. The Eqs.~\eqref{eq:cparty1} and~\eqref{eq:cparty2} are also the states with fixed $G$ parity once their isospin were determined. For the systems with (anti)strange quark, one can introduce $G_{U/V}$ parities to label the $B^*_{s}\bar{B}/B_s \bar{B}^*$ states~\cite{Meng:2020ihj}.
With the strict SU(3) flavor symmetry, the $B^{(\ast)}_{(s)}\bar{B}_{(s)}^{(\ast)}$ will form two multplets $\mathbf{8}_F$ and $\mathbf{1}_F$ as shown in Fig.~\ref{fig:4.x_SU3_multi}.
With the strict HQSS and SU(3) flavor symmetry, the interaction of $q\bar{q}$ can be parameterized as
\begin{equation}
	V_{q\bar{q}}=c_{1}+c_{2}\bm{\ell}_{1}\cdot\bm{\ell}_{2}+c_{3}\mathbb{C}_{2}+c_{4}(\bm{\ell}_{1}\cdot\bm{\ell}_{2})\mathbb{C}_{2}.\label{eq:vqq}
\end{equation}
$\mathbb{C}_2=-\sum_{i=1}^8 \lambda_F^i\lambda_F^{i\ast}$ is the Casimir operator in the flavor space. There are four independent operators in spin and flavor space in the symmetry limits.

Apart from the interaction terms, the mass terms will violate the symmetries. There are two related mass splittings [see also Eqs.~\eqref{eq:masssplitBD1} and~\eqref{eq:masssplitBD2}],
\begin{eqnarray}
	m_{B_{s}^{(*)}}-m_{B^{(*)}}\simeq90\text{ MeV},\qquad
	m_{B_{(s)}^{*}}-m_{B_{(s)}}\simeq45\text{ MeV},\label{eq:masssplitting}
\end{eqnarray}
which break the $\SU(3)$ symmetry and HQSS, respectively. For the near-threshold states, the interaction is weak compared to the above mass splittings. Although we can still assume the interactions satisfy the symmetries, in Eq.~\eqref{eq:vqq}, the mass splitting in the Eq.~\eqref{eq:masssplitting} will separate the di-meson systems into several blocks.
\begin{eqnarray}
	\text{spin space:}&&\{B_{(s)}\bar{B}_{(s)}\},\{B_{(s)}\bar{B}_{(s)}^{*}/B_{(s)}^{*}\bar{B}_{(s)}\},\{B^{*}_{(s)}\bar{B}_{(s)}^{*}\},\nonumber\\
	\text{flavor space:}&&\{\bar{B}^{(\ast)}B^{(\ast)}\},\{B_{s}^{(\ast)}\bar{B}^{(\ast)}\},\{\bar{B}_{s}^{(\ast)}B^{(\ast)}\},\{B_{s}^{(\ast)}\bar{B}_{s}^{(\ast)}\}.
\end{eqnarray}
The mixing between states in different blocks is suppressed by the mass splittings in Eq.~\eqref{eq:masssplitting}. In fact, it is shown to be suppressed by at least two orders in EFT power counting in Ref.~\cite{Valderrama:2012jv}. Thus, we need only focus on the mixing effect inside each block.

The two particle basis in Eq.~\eqref{eq:twoparticlebasis} can be expanded with the HQSS basis $|S_h\otimes S_\ell\rangle$ via the $9j$ symbols, {if the orbital angular momentum $L=0$ is presumed.} 
Then we have~\cite{Ozpineci:2013zas}
\begin{eqnarray}
	\left[\begin{array}{c}
		|B\bar{B}\rangle\\
		|B^{\ast}\bar{B}^{\ast}\rangle
	\end{array}\right]_{0^{++}}	&=&	\left[\begin{array}{cc}
		\frac{1}{2} & \frac{\sqrt{3}}{2}\\
		-\frac{\sqrt{3}}{2} & \frac{1}{2}
	\end{array}\right]\left[\begin{array}{c}
		|0_{h}^{-+}\otimes0_{\ell}^{-+}\rangle\\
		|1_{h}^{--}\otimes1_{\ell}^{--}\rangle
	\end{array}\right],\\
	|\frac{1}{\sqrt{2}}(B\bar{B}^{\ast}-B^{\ast}\bar{B})\rangle_{1^{++}}	&=&	-|1_{h}^{--}\otimes1_{\ell}^{--}\rangle,\label{eq:BB1plusplus}\\
	\left[\begin{array}{c}
		|\frac{1}{\sqrt{2}}(B\bar{B}^{\ast}+B^{\ast}\bar{B})\rangle\\
		|B^{\ast}\bar{B}^{\ast}\rangle
	\end{array}\right]_{1^{+-}}	&=&	\left[\begin{array}{cc}
		\frac{1}{\sqrt{2}} & -\frac{1}{\sqrt{2}}\\
		-\frac{1}{\sqrt{2}} & -\frac{1}{\sqrt{2}}
	\end{array}\right]\left[\begin{array}{c}
		|1_{h}^{--}\otimes0_{\ell}^{-+}\rangle\\
		|0_{h}^{-+}\otimes1_{\ell}^{--}\rangle
	\end{array}\right],\\
	|B^{\ast}\bar{B}^{\ast}\rangle_{2^{++}}	&=&	-|1_{h}^{--}\otimes1_{\ell}^{--}\rangle,
\end{eqnarray} 
where the subscripts at the left hand side denote the $J^{PC}$ of the $B^{(\ast)}\bar{B}^{(\ast)}$ system, which are expressed by $S_h^{P_h C_h}\otimes S_\ell^{P_\ell C_\ell}$. One can also include the $G$ parity, such as $|B^{\ast}\bar{B}^{\ast}\rangle_{0^{+}(2^{++})}	=	-|0^{-}(1_{h}^{--})\otimes0^{-}(1_{\ell}^{--})\rangle$, and $|B^{\ast}\bar{B}^{\ast}\rangle_{1^{-}(2^{++})}	=	-|0^{-}(1_{h}^{--})\otimes1^{+}(1_{\ell}^{--})\rangle$.

The simultaneous conservation of $\bm J$ and $\bm S_h$ means that the $\bm S_\ell$ is also a conserved quantity. Therefore, the interaction is largely simplified for an effective Hamiltonian $\hat{H}_{\mathrm{eff}}\equiv \tilde{c}_1+\tilde{c}_2 \bm{\ell}_1 \cdot \bm{\ell}_2$ that satisfies the HQS. The conservation of $\bm S_\ell$ requires
\begin{eqnarray}
\mathcal{C}_{0}^{\alpha}&=&\langle S_{h}\otimes0;\alpha|\hat{H}_{\mathrm{eff}}|S_{h}\otimes0;\alpha\rangle=\tilde{c}_1-{3\over 4}\tilde{c}_2,\label{eq:C0alpha}\\
\mathcal{C}_{1}^{\alpha}&=&\langle S_{h}\otimes1;\alpha|\hat{H}_{\mathrm{eff}}|S_{h}\otimes1;\alpha\rangle=\tilde{c}_1+{1\over 4}\tilde{c}_2,\label{eq:C1alpha}
\end{eqnarray}
where $\mathcal{C}_{0}^{\alpha}$ and $\mathcal{C}_{1}^{\alpha}$ ($\alpha$ denotes the other conserved quantities, such as the isospin) are the corresponding LECs in the $S_\ell=0$ and $S_\ell=1$ sectors, respectively. Then the effective potentials read
\begin{eqnarray}
V_{0^{++}}^{\alpha}&=&\left[\begin{array}{cc}
	\frac{1}{4}(\mathcal{C}_{0}^{\alpha}+3\mathcal{C}_{1}^{\alpha}) & \frac{\sqrt{3}}{4}(-\mathcal{C}_{0}^{\alpha}+\mathcal{C}_{1}^{\alpha})\\
	\frac{\sqrt{3}}{4}(-\mathcal{C}_{0}^{\alpha}+\mathcal{C}_{1}^{\alpha}) & \frac{1}{4}(3\mathcal{C}_{0}^{\alpha}+\mathcal{C}_{1}^{\alpha})
\end{array}\right]=\left[\begin{array}{cc}
	\tilde{c}_{1} & \frac{\sqrt{3}}{4}\tilde{c}_{2}\\
	\frac{\sqrt{3}}{4}\tilde{c}_{2} & \tilde{c}_{1}-\frac{1}{2}\tilde{c}_{2}
\end{array}\right],\label{eq:V0pp}\\
V_{1^{+-}}^{\alpha}&=&\left[\begin{array}{cc}
	\frac{1}{2}(\mathcal{C}_{0}^{\alpha}+\mathcal{C}_{1}^{\alpha}) & \frac{1}{2}(-\mathcal{C}_{0}^{\alpha}+\mathcal{C}_{1}^{\alpha})\\
	\frac{1}{2}(-\mathcal{C}_{0}^{\alpha}+\mathcal{C}_{1}^{\alpha}) & \frac{1}{2}(\mathcal{C}_{0}^{\alpha}+\mathcal{C}_{1}^{\alpha})
\end{array}\right]=\left[\begin{array}{cc}
	\tilde{c}_{1}-\frac{1}{4}\tilde{c}_{2} & \frac{1}{2}\tilde{c}_{2}\\
	\frac{1}{2}\tilde{c}_{2} & \tilde{c}_{1}-\frac{1}{4}\tilde{c}_{2}
\end{array}\right],\label{eq:V1pmepsV2pm}\\
V_{1^{++}}^{\alpha}&=&\mathcal{C}_{1}^{\alpha}=\tilde{c}_{1}+\frac{1}{4}\tilde{c}_{2},\label{eq:V1ppepsV2pp1}\\
V_{2^{++}}^{\alpha}&=&\mathcal{C}_{1}^{\alpha}=\tilde{c}_{1}+\frac{1}{4}\tilde{c}_{2}.\label{eq:V1ppepsV2pp}
\end{eqnarray}
An equivalent description is performed through the superfield representations, e.g., see Refs.~\cite{Mehen:2011yh,Nieves:2012tt,Hidalgo-Duque:2012rqv}. Here we demonstrate Eqs.~\eqref{eq:V1ppepsV2pp1} and~\eqref{eq:V1ppepsV2pp} by expanding the LO Lagrangian~\eqref{eq:LOctLagHH},
\begin{eqnarray}
V_{1^{++}}^\alpha&=&(-D_{a}+D_{b}+E_{a}-E_{b})\bm{\varepsilon}^{\prime\dagger}\cdot\bm{\varepsilon},\\
V_{2^{++}}^\alpha&=&(E_{a}-D_{a})(\bm{\varepsilon}^{\dagger}\cdot\bm{\varepsilon})(\bm{\varepsilon}^{\prime\dagger}\cdot\bm{\varepsilon}^{\prime})+(E_{b}-D_{b})\left[(\bm{\varepsilon}^{\prime}\cdot\bm{\varepsilon})(\bm{\varepsilon}^{\prime\dagger} \cdot\bm{\varepsilon}^{\dagger})-(\bm{\varepsilon}^{\prime\dagger}\cdot\bm{\varepsilon})(\bm{\varepsilon}^{\dagger}\cdot\bm{\varepsilon}^{\prime})\right],
\end{eqnarray}
where $\alpha=1$. One can easily verify that $V_{1^{++}}^\alpha=V_{2^{++}}^\alpha$ with the substitution $\bm{\varepsilon}^{\prime\dagger}\cdot\bm{\varepsilon}\to1$, $(\bm{\varepsilon}^{\dagger}\cdot\bm{\varepsilon})(\bm{\varepsilon}^{\prime\dagger}\cdot\bm{\varepsilon}^{\prime})\to1$, $(\bm{\varepsilon}^{\prime}\cdot\bm{\varepsilon})(\bm{\varepsilon}^{\prime\dagger}\cdot\bm{\varepsilon}^{\dagger})\to(\bm{S}_{1}\cdot\bm{S}_{2})^{2}-1$, $(\bm{\varepsilon}^{\prime\dagger}\cdot\bm{\varepsilon})(\bm{\varepsilon}^{\dagger}\cdot\bm{\varepsilon}^{\prime})\to\bm{S}_{1}\cdot\bm{S}_{2}+(\bm{S}_{1}\cdot\bm{S}_{2})^{2}-1$, where $\bm S_1$ and $\bm S_2$ denote the spin operators of the $B^\ast$ and $\bar{B}^\ast$, respectively.

\begin{figure}
	\centering
	\includegraphics[width = 0.7\textwidth]{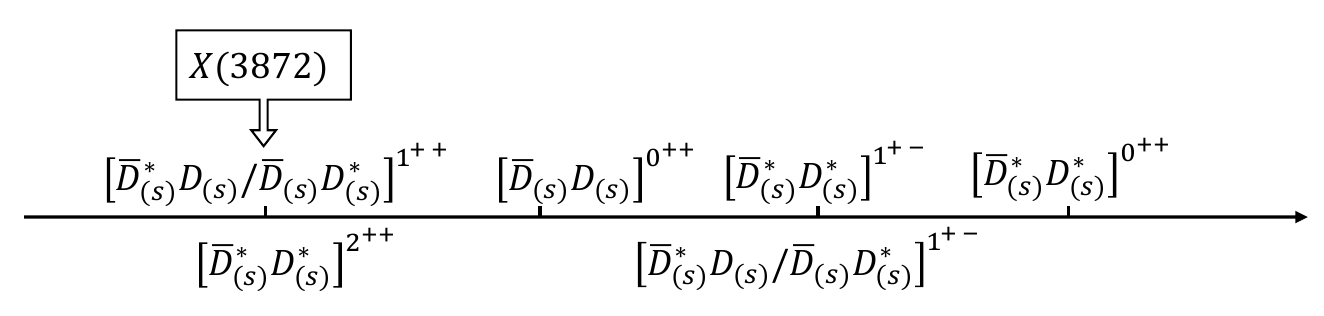}
	\caption{The order of the six $\bar{D}_{(s)^{(*)}}D_{(s)}^{(*)}$ states according to interaction in Eqs.~\eqref{eq:C0alpha}-\eqref{eq:V1ppepsV2pp}. The arrows indicates that the interaction becomes more attractive or repulsive according to the sign of $\tilde{c}_2$. }\label{fig:4.x_inter_order}
\end{figure}

Since the mixing effect induced by the off-diagonal terms in Eqs.~\eqref{eq:V0pp}-\eqref{eq:V1ppepsV2pp} is suppressed, we only focus on the diagonal terms. In the parameterization with $\tilde{c}_1$ and $\tilde{c}_2$, the interactions for the six states have the same coefficient of $\tilde{c}_1$ and are arranged in order of $\tilde{c}_2$ in Fig.~\ref{fig:4.x_inter_order}. The trivial prediction in the heavy quark symmetry is the existence of the $[D^*\bar{D}^*]^{2^{++}}$ molecule state whose interaction is the same as that of the $X(3872)$. In addition, if there exists any other bound state with the binding energy larger than that of the $X(3872)$ among the other four channels, one can naturally expect the existence of all the six molecular states~\cite{Nieves:2012tt,Meng:2020cbk}. The arrow in  Fig.~\ref{fig:4.x_inter_order} indicates the more attractive trend. See more related discussions in Refs.~\cite{Bondar:2011ev,Voloshin:2011qa,Guo:2009id,Albaladejo:2015dsa,Liu:2019stu,Voloshin:2016cgm,Voloshin:2004mh,Voloshin:2013dpa,Voloshin:2018tav,Nieves:2011zz,Baru:2016iwj,Cleven:2015era,Guo:2013xga}.

The above formalism can be extended to the other systems, such as the $\Sigma_c^{(\ast)}\bar{D}^{(\ast)}$. In this case,
\begin{eqnarray}
I=\alpha:\begin{cases}
J=\frac{1}{2}, & \Sigma_{c}\bar{D},\Sigma_{c}\bar{D}^{\ast},\Sigma_{c}^{\ast}\bar{D}^{\ast}\\
J=\frac{3}{2}, & \Sigma_{c}\bar{D}^{\ast},\Sigma_{c}^{\ast}\bar{D},\Sigma_{c}^{\ast}\bar{D}^{\ast}\\
J=\frac{5}{2}, & \Sigma_{c}^{\ast}\bar{D}^{\ast}
\end{cases},
\end{eqnarray}
where $\alpha=1/2,3/2$. The two particle basis can then be expanded with the HQSS basis as
\begin{eqnarray}
\left[\begin{array}{c}
|\Sigma_{c}\bar{D}\rangle\\
|\Sigma_{c}\bar{D}^{\ast}\rangle\\
|\Sigma_{c}^{\ast}\bar{D}^{\ast}\rangle
\end{array}\right]_{J=\frac{1}{2}}	&=&	\left[\begin{array}{ccc}
\frac{1}{2} & -\frac{1}{2\sqrt{3}} & \sqrt{\frac{2}{3}}\\
-\frac{1}{2\sqrt{3}} & \frac{5}{6} & \frac{\sqrt{2}}{3}\\
\sqrt{\frac{2}{3}} & \frac{\sqrt{2}}{3} & -\frac{1}{3}
\end{array}\right]\left[\begin{array}{c}
|0_{h}\otimes\frac{1}{2}_{\ell}\rangle\\
|1_{h}\otimes\frac{1}{2}_{\ell}\rangle\\
|1_{h}\otimes\frac{3}{2}_{\ell}\rangle
\end{array}\right],\label{eq:LOcontactJ12}\\
\left[\begin{array}{c}
|\Sigma_{c}\bar{D}^{\ast}\rangle\\
|\Sigma_{c}^{\ast}\bar{D}\rangle\\
|\Sigma_{c}^{\ast}\bar{D}^{\ast}\rangle
\end{array}\right]_{J=\frac{3}{2}}	&=&	\left[\begin{array}{ccc}
\frac{1}{3} & -\frac{1}{\sqrt{3}} & \frac{\sqrt{5}}{3}\\
-\frac{1}{\sqrt{3}} & \frac{1}{2} & \frac{1}{2}\sqrt{\frac{5}{3}}\\
\frac{\sqrt{5}}{3} & \frac{1}{2}\sqrt{\frac{5}{3}} & \frac{1}{6}
\end{array}\right]\left[\begin{array}{c}
|1_{h}\otimes\frac{1}{2}_{\ell}\rangle\\
|0_{h}\otimes\frac{3}{2}_{\ell}\rangle\\
|1_{h}\otimes\frac{3}{2}_{\ell}\rangle
\end{array}\right],\\
|\Sigma_{c}^{\ast}\bar{D}^{\ast}\rangle_{J=\frac{5}{2}}	&=&	|1_{h}\otimes\frac{3}{2}_{\ell}\rangle.
\end{eqnarray}
Similarly, there are two independent LECs,
\begin{eqnarray}
\mathcal{C}_{1/2}^{\alpha}&=&\langle S_{h}\otimes\frac{1}{2};\alpha|\hat{H}_{\mathrm{eff}}|S_{h}\otimes\frac{1}{2};\alpha\rangle,\label{eq:C12alpha}\\
\mathcal{C}_{3/2}^{\alpha}&=&\langle S_{h}\otimes\frac{3}{2};\alpha|\hat{H}_{\mathrm{eff}}|S_{h}\otimes\frac{3}{2};\alpha\rangle.\label{eq:C32alpha}
\end{eqnarray}
It is straightforward to obtain the effective potentials $V_{1/2}^\alpha$, $V_{3/2}^\alpha$ and $V_{5/2}^\alpha$, see details in Ref.~\cite{Xiao:2013yca}. The related SU(3) flavor symmetry was investigated in Ref.~\cite{Peng:2019wys}.

\subsection{$X(3872)$ and $T_{cc}^+$ states}\label{sec:XandTccStates}

There is no doubt that $X(3872)$ is the superstar in the exotic hadron family. In this section, we will review the theoretical progress of the $X(3872)$ and its analogue $T_{cc}^+$ as the hadronic molecules in EFTs.

\subsubsection{Methodology}~\label{sec:x_methodology}
	\begin{figure}[htbp]
		\centering
		\includegraphics[width = 0.7\textwidth]{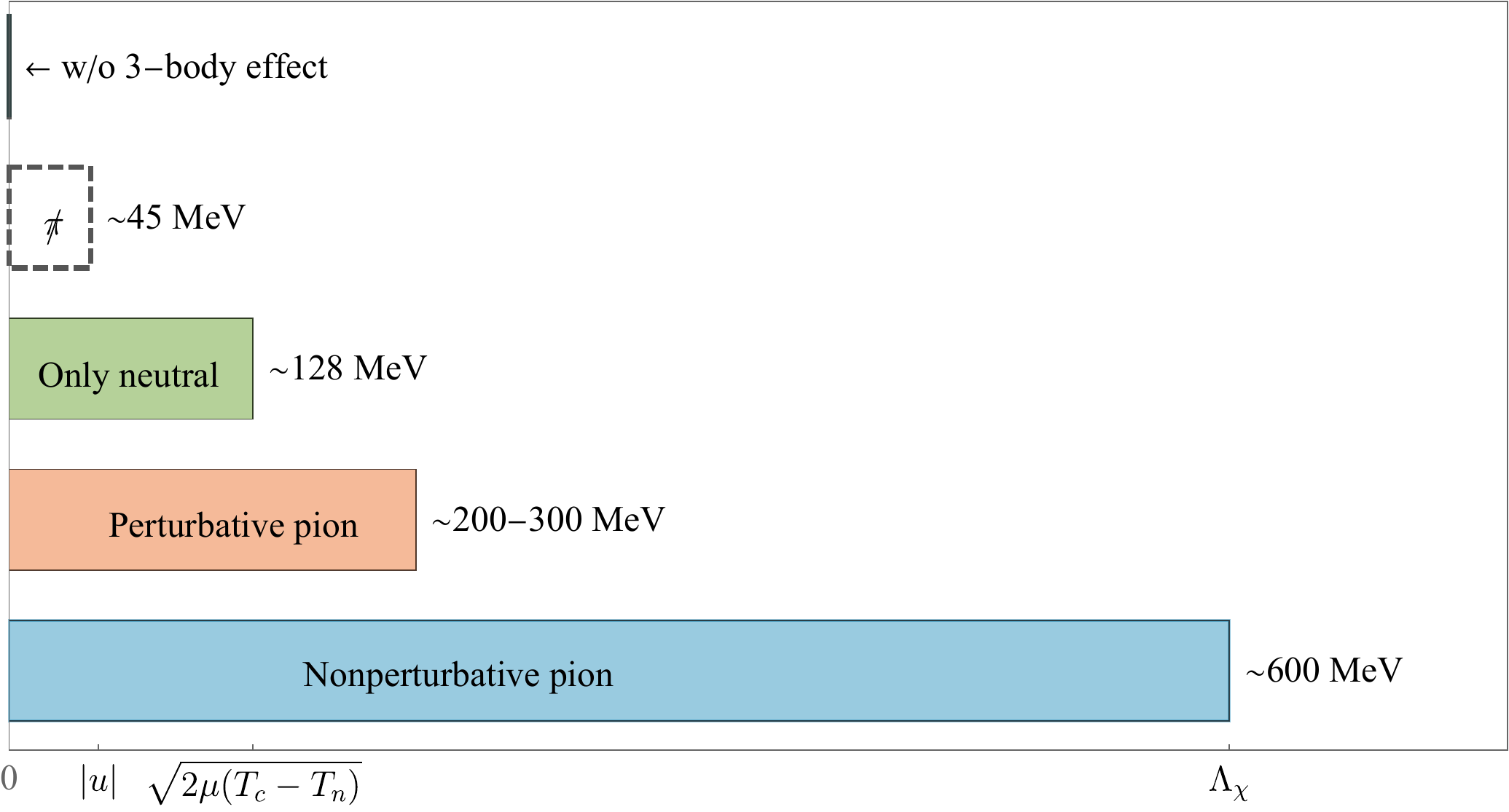}
		\caption{{The validity range of the EFTs for the $X(3872)$. We here choose the $\xthn$ threshold as the baseline. For simplicity, we use the ``invalid energy range or scale" to denote the scale where the EFT breaks down. For the $X(3872)$, the invalid energy range of the EFTs without considering the three-body dynamics, pionless EFTs (conditionally valid), EFTs with only the neutral channel, EFTs with the perturbative pion and EFTs with the nonperturbative pion are given in order. The three-body dynamics should be considered from the $\xthn$ threshold, because the $D^0\bar{D}^0\pi^0$ threshold is below it. Therefore, the pionless EFTs cannot work rigorously without considering the three-body effect. If one treats the $D^{\ast 0}$ as a stable state, the pionless EFTs will be valid conditionally up to the scale $|u|= \sqrt{|m_\pi^2-(M_{D^{*}}-M_D)^2|}$. The invalid scale of the EFTs with only the neutral channel is estimated by the momentum corresponding to the threshold difference. In the figure, $T_c$, $T_n$ and $\mu$ are the charged threshold, neutral threshold and reduced mass, respectively. The invalid scale of the perturbative pion is from Ref.~\cite{Valderrama:2012jv}. The EFTs with the nonperturbative pion (e.g., the $\chi$EFT) will be valid to the chiral breaking scale $\Lambda_\chi$.} }\label{fig:4.x_eft_validity}
	\end{figure}
\begin{figure}[htbp]
		\centering
		\includegraphics[width = 1.0\textwidth]{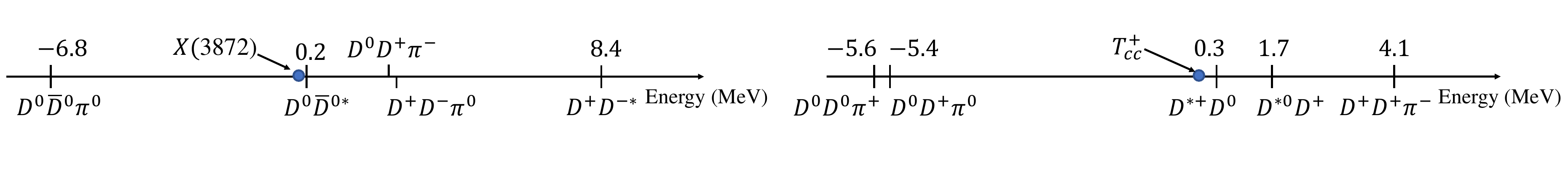}
		\caption{$X(3872)$ (left one) and $T_{cc}^+$ (right one) related two-body and three-body thresholds.}\label{fig:4.x_xmass}
	\end{figure}

We will first review some characters of the $X(3872)$ and $T_{cc}^+$ from the theoretical perspectives. These characters will hint which kind of EFT is suitable for the $X(3872)$ and $T_{cc}^+$ states. {In Fig.~\ref{fig:4.x_eft_validity}, we summarize the valid scales and relevant d.o.fs of different EFTs for the $X(3872)$. The details are as follows.}
\begin{itemize}
	\item {\it Molecular interpretation.} There are different interpretations for the $X(3872)$, e.g.,~the molecular picture~\cite{Tornqvist:2004qy,Liu:2008fh,Liu:2009qhy,Li:2012cs}, compact tetraquark state~\cite{Maiani:2004vq,Maiani:2007vr}, radial excitation of the charmonium~\cite{Barnes:2003vb,Kalashnikova:2010hv}, etc.. Naturally, one can blend the above mentioned pictures, e.g., the coupled-channel picture of the $c\bar{c}$ and di-meson d.o.fs~\cite{Braaten:2003he,Kalashnikova:2005ui,Barnes:2007xu,Ortega:2009hj,Li:2009ad,Baru:2010ww,Yamaguchi:2019vea}. We refer to Refs.~\cite{Chen:2016qju,Esposito:2016noz,Brambilla:2019esw} for general reviews and Ref.~\cite{Kalashnikova:2018vkv} for the specialized review of the $X(3872)$ as a molecule. The evidences supporting the molecular interpretation of $X(3872)$ include its mass coinciding with the $\xthn$ threshold~$m_{D^0}+m_{D^{*0}}-m_{X(3872)}=(0.00\pm 0.18 )$ MeV and large branching ratio of $\mathscr{R}[X(3872)\to (D^0\bar{D}^0\pi^0,\bar{D}^{*0}D^0)]\approx80\%$~\cite{ParticleDataGroup:2022pth}. In this section, we focus on the molecular picture, which suits to be depicted in the EFT at the hadronic level. For the EFT at the quark-gluon level, e.g., Born-Oppenheimer EFT, we refer to Ref.~\cite{Brambilla:2019esw} for reviews.

	\item {\it Universality.}	As shown in Fig.~\ref{fig:4.x_xmass}, the $X(3872)$ is below the $\xthn$ threshold about 0.2 MeV. The $T_{cc}^+$ is below the $\tth$ threshold about 0.3 MeV. If we regard the $X(3872)$ or $T_{cc}^+$ as the bound state of the related particles, the binding energy is unnaturally small as compared to the typical scale $m_\pi^2/(2\mu_{DD^*})\sim 10$ MeV. The shallow $S$-wave bound state indicates the unnaturally large scattering length as compared to $1/m_\pi$, and the low-energy universality. The universality motivated numerous works based on LO contact interaction to depict the long-range dynamics of the $X(3872)$, e.g.,~	\cite{Voloshin:2003nt,Braaten:2003he,Braaten:2005ai}. However, the universality is a two-edged sword, which makes the long-range dynamics only depend on one parameter (binding-energy or scattering length), and also makes the mechanism of forming the $X(3872)$ hard to be detected. As pointed out in Ref.~\cite{Braaten:2003he}, there are two fine-tuning mechanisms to form the $X(3872)$, the fine-tuning of the $\xthn$ interaction and reduced mass to form a loosely bound state, and the fine-tuning of the $P$-wave charmonium, i.e., $\chi_{c1}(2P)$, to the proximity of the $\xthn$ threshold. It is worthwhile to stress that in the latter mechanism, the $c\bar{c}$ component plays a crucial role to form the $X(3872)$ but its proportion could be very small (suppressed by $1/a_s$). In order to discriminate the two mechanisms, one has to detect the dynamics beyond the ``universal" region---tens of MeV from the threshold. In Ref.~\cite{Artoisenet:2010va}, Artoisenet \etal  proposed an EFT mixing the zero-range amplitude and Flatt\'e scattering amplitude~\cite{Hanhart:2007yq,Baru:2010ww} to discern the two different mechanisms from the line shape.

	\item {\it Pion-exchange interaction.} The OPE interaction in the $\DastDbar$ and $D^\ast D$ systems only allows the $D^\ast D\pi/\bar{D}^\ast\bar{D}\pi$ vertex as shown in Fig.~\ref{fig:4.1X_OPE}. The mass difference of the $D^*$ and $D$ will appear in the propagator of the pion as
	\begin{equation}
		(q^{2}-m_{\pi}^{2})^{-1}=\left[(E_{D^{*}}-E_{D})^{2}-(\bm{q}^{2}+m_{\pi}^{2})\right]^{-1}\approx\left[(M_{D^{*}}-M_{D})^{2}-(\bm{q}^{2}+m_{\pi}^{2})\right]^{-1}\equiv-(\bm{q}^{2}+\tilde{u}^{2})^{-1},~\label{eq:4.x-ope-prop}
	\end{equation}
	where the energy dependence other than the mass splitting of the $D^*$ and $D$ is neglected (static approximation). $\tilde{u}\equiv \sqrt{m_\pi^2-(M_{D^*}-M_D)^2}$ is introduced as the effective mass in the propagator. The values $\tilde{u}$ for the states with different charges are given in Table~\ref{tab:4.xmass}. If we use $\tilde{u}$ as a natural scale to estimate the natural binding energy, the result is $0.3-1.0$ MeV. Among them, the $D^{*0}D^0\pi^0$ one is about 1.0 MeV. The estimation hints that the convergent range of the \nopi is $p<|\tilde{u}|\sim 44.12$ MeV rather than $p<m_{\pi}$. The universality is valid in a smaller range. Therefore, the OPE interaction is needed to exploit the larger-range dynamics, e.g., in~\cite{Fleming:2007rp,Schmidt:2018vvl}. {It should be stressed that for the $T_{cc}^+$ and $X(3872)$, the lowest relevant two-body thresholds are above some $DD\pi$ or $D\bar{D}\pi$ three-body thresholds as shown in Fig.~\ref{fig:4.x_xmass}. The pionless EFTs cannot work rigorously without considering the three-body effect. If one treats the $D^{\ast}$ as a stable state, the pionless EFTs will be valid conditionally up to the scale $|\tilde{u}|$.}

	\item {\it Perturbative pion versus nonperturbative pion.} In the $^3S_1$ $NN$ system, the perturbative treatment of the OPE interaction fails in the expansion in the KSW scheme~\cite{Fleming:1999ee}. However, the relative ratios of the two-pion and one-pion exchange interaction for the $NN$ system and $\bar{D}^*D$ system are estimated as
	\begin{equation}
		\bar{D}^*D:\;	\frac{g_b^{2}\mu_{DD^{\ast}}|\tilde{u}|}{8\pi f_{\pi}^{2}}\approx\frac{1}{16},\quad NN:\; \frac{g_{A}^{2}\mu_{NN}m_{\pi}}{8\pi f_{\pi}^{2}}\approx\frac{1}{2}.
	\end{equation}
Compared with the $NN$ system, the expansion parameter for the $\bar{D}^*D$ system is relatively small. Therefore, one expects that the validity range of the perturbative pion for the $\bar{D}^*D$ system is larger. The OPE interaction is included perturbatively in XEFT~\cite{Fleming:2007rp}. In the Weinberg scheme, the OPE interaction is resummed nonperturbatively ~\cite{Baru:2011rs,Baru:2013rta,Baru:2015tfa,Du:2021zzh,Xu:2017tsr,Wang:2018atz,Schmidt:2018vvl}. {In Ref.~\cite{Valderrama:2012jv}, the breaking-down scale of the perturbative pions is estimated to be around $200-300$ MeV for the $X(3872)$ via a technique to analyze the scattering in the power law singular potentials based on the renormalization group~\cite{gao1999repulsive,Birse:1998dk}. }
	\item {\it Three-body dynamics.}
	One may notice that the effective mass $\tilde{u}$ defined in Eq.~\eqref{eq:4.x-ope-prop} can be imaginary as shown explicitly in Table~\ref{tab:4.xmass}. The corresponding potential from Eq.~\eqref{eq:4.x-ope-prop} in coordinate space is an oscillating function rather than the conventional Yukawa potential~\cite{Suzuki:2005ha}. In other words, the exchanged pion can be on mass-shell. The imaginary part of the potential arises from the opening of the three-body threshold. The $X(3872)$ is above the $D^0\bar{D}^0\pi^0$ threshold about $6.8$ MeV and the $T_{cc}^+$ state is above the $D^0D^0\pi^+$ and $D^0D^+\pi^0$ thresholds about $5.5$ MeV, e.g., see Fig.~\ref{fig:4.x_xmass}. In order to consider this effect, the three-body dynamics should be incorporated properly. There are two types of three-body cuts (see Fig.~\ref{fig:4.x_3body_cut}) that were studied in Refs.~\cite{Braaten:2007ct,Baru:2011rs,Jansen:2013cba,Schmidt:2018vvl,Du:2021zzh}. {In the previous discussions, the pion propagators in the OPE interactions are mentioned many times, see Eqs.~\eqref{eq:OPEinXEFT}, \eqref{eq:VOPE1}, \eqref{OPEel}, \eqref{OPEin} and \eqref{eq:4.x-ope-prop}. However, among them, only Eqs.~\eqref{OPEel} and \eqref{OPEin} keep the three-body cuts properly, which should be a better choice in the calculation. }
	
	\item {\it Factorization versus non-factorization.} The productions and decays of the $X(3872)$ should be associated with the short-range dynamics unavoidably. For example, in the $X\to J/\psi \pi^+\pi^-$, the formation of the $J/\psi$ would involve the short-distance interaction around its size, which is much shorter than the size of the $X(3872)$ as a hadronic molecule. In the process $B\to XK$, the production of the $c$ and $\bar{c}$ quark is a high-energy dynamics compared with the binding momentum of the $X(3872)$. With the separated scale, Braaten \etal proposed the factorization formula~\cite{Braaten:2005jj}, which can be derived from the operator product expansion in Ref~\cite{Braaten:2006sy}. In the literature, the factorization formula was used either implicitly or explicitly~\cite{Braaten:2004fk,Braaten:2004ai,Braaten:2004jg,Braaten:2007dw,Braaten:2007ft,Stapleton:2009ey,Braaten:2013poa,Meng:2021kmi,Mehen:2011ds,Fleming:2011xa}. We take the decays involving the short-range dynamics $X\to H$ [where $H$ denotes the final states produced through short-range dynamics, such as $J/\psi\rho(\omega)$] as an example.  The amplitude reads
	\begin{eqnarray}
{\cal T}[X\to H]&\sim&\int\frac{d^{3}q}{(2\pi)^{3}}\psi(\bm{q})\times{\cal A}_{\text{short}}\big[\{D(\bm{q})\bar{D}^{*}(-\bm{q})\}^{+}\to H\big]\nonumber\\
&\sim&{\cal A}_{\text{long}}(E)\times{\cal A}_{\text{short}}\big[\{D\bar{D}^{*}\}^{+}\to H\big],~\label{eq:4.x_decay_factorization}
	\end{eqnarray}
where $\psi(\bm{q})$ is the wave function of the $X(3872)$. The superscript of $\{D\bar{D}^{*}\}^{+}$ represents the even $C$ parity state. Strictly speaking, the short-range part will (insensitively) depend on the momentum of $D(\bar{D}^*)$. One can ignore the $\bm{q}$-dependence and factorize out the long- and short-range parts. The long-range part is the molecular wave function at origin. The factorization formula can simplify the calculation by circumventing the complicated loop integrals sometimes. What is more important, it is easy to establish the relation between two processes with the same short- or long-range dynamics, within the factorization approach (this will be illustrated later). Meanwhile, one can depict the same dynamics by calculating the loop diagrams (non-factorization approach), which incorporates the long- and short-range dynamics simultaneously.  In general, it aims to include more dynamical details than the factorization approach. However, apart from the tedious calculations, a caveat is that the dynamics at different energy scales should be incorporated properly in a unified framework, which is a subtle work. We will compare the factorization and non-factorization frameworks in detail later.

\end{itemize}

	 \begin{figure}[htbp]
		\centering
		\includegraphics[width = 0.5\textwidth]{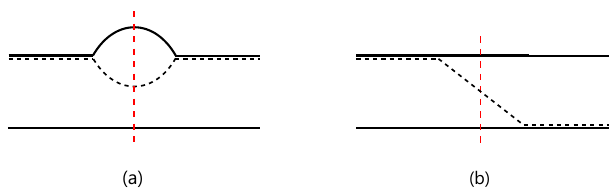}
		\caption{Two types of three-body cuts for the three-body dynamics of $X(3872)$ or $T_{cc}^+$.}\label{fig:4.x_3body_cut}
	\end{figure}

\begin{table}
	\centering
\renewcommand{\arraystretch}{1.5}
	\caption{Mass scales in the OPE interactions for the $\DastDbar$ and $D^*D$ systems (in units of MeV). The $\Delta^i\equiv D^{*i}-D^i$, $m_\pi^i$ and $\tilde{u}_i\equiv\sqrt{m_\pi^{i2}-\Delta^{i2}}$ are listed in order, where $i$ represents different charges.}~\label{tab:4.xmass}
\setlength{\tabcolsep}{5.1mm}
{	
\begin{tabular}{c|c|c}
		\hline
		$\Delta^{i}|m_{\pi}^{i}|\tilde{u}_i$ & $D^{*0}$ & $D^{*+}$\tabularnewline
		\hline
		$D^{0}$ & $142.02\; |\; 134.98 \;|\; i44.12$ & $145.43 \;|\; 139.57 \;|\; i40.83$\tabularnewline
		\hline
		$D^{+}$ & $137.20\; |\; 139.57\; |\; 25.61$ & $140.61 \;|\; 134.98 \;|\; 39.39$\tabularnewline
		\hline
	\end{tabular}
}
\end{table}

In order to fit these characters of the $X(3872)$ or $T_{cc}^+$, many attempts have been made to include new ingredients in EFTs. We will review the progresses in EFT frameworks to investigate the $X(3872)$ or $T_{cc}^+$. We will discuss the XEFT with revised power counting, the Galilean invariant XEFT, the perturbative scale of the pion-exchange interaction, the formulation of the three-body dynamics, factorization and non-factorization schemes for the processes with short-distance dynamics.

In the original version of XEFT~\cite{Fleming:2007rp}, the $m_\pi$ and $M_{D^{(*)}}$ are integrated out as the large scales. However, they will appear in the kinetic energy terms in the calculation. In Ref.~\cite{Fleming:2007rp}, the $m_\pi/M_D$ expansion is performed to simplify the calculation, though it is not implemented systematically in the power counting scheme. In Ref.~\cite{Alhakami:2015uea}, a modified power counting for XEFT is proposed. In this power counting, $\Delta$ and $m_\pi$ is treated as $\mathcal{O}(Q)$ and $\delta$ is treated as $\mathcal{O}(Q^2)$. Accordingly, the power counting of other quantities is
\begin{eqnarray}
\Delta,~m_{\pi},~p_D&\sim& Q,\nonumber\\
u,~p_{\pi}&\sim& Q^{3/2},\nonumber\\
\delta,~{p_{\pi}^{2}}/({2m_{\pi}}),~{p_{D}^{2}}/({2M_{D}})&\sim& Q^{2}.
\end{eqnarray}
This power counting is very convenient to match with the HH$\chi$PT.  In Ref.~~\cite{Alhakami:2015uea}, it was employed to simplify the loop diagrams rather than constructing the Lagrangians order by order.

It was pointed out by Braaten~\cite{Braaten:2015tga} that there are several problems which prevented the original XEFT~\cite{Fleming:2007rp} from obtaining accurate predictions. First, in the original XEFT, the results are not frame-independent, which was specified, e.g., in the center-of-mass frame. Meanwhile, the UV divergence in the NLO calculation can be canceled out only when the low order $m_\pi/M_D$ expansions are kept. In addition, the renormalization scheme in the original XEFT made it hard to include the decays with large momentum in the final state, such as $D^{*0}\to D^0\gamma$. In order to overcome these shortcomings, Braaten proposed the Galilean-invariant XEFT based on the feature that $M_{D^*}\approx M_D+m_\pi$~\cite{Braaten:2015tga}. In a nonrelativistic theory, the Galilean boost with velocity $\bm{v}$ will change the momentum and energy as
\begin{equation}
\bm{p}\to\bm{p}+m\bm{v},\qquad E\to E+m\bm{v\cdot p}+\frac{1}{2}mv^{2}.
\end{equation}
The combination $E-\frac{\bm p^{2}}{2m}$ is Galilean invariant. Thus, the kinetic terms of the $\pi^0$ and $D^0$ in Eq.~\eqref{eq:XEFT_lag} satisfy the Galilean invariance. In Ref.~\cite{Braaten:2015tga}, the author chose a different baseline of the mass and eliminate the $\delta$ term in the $\pi^0$ kinetic term. In the Galilean-invariant XEFT, the $D^*$ is regarded as the combination of $D$ and $\pi$, thus its kinetic mass is $M+m$ (where $M$ and $m$ are the masses of $D^0$ and $\pi^0$, respectively). The kinetic term of the $D^{\ast0}$ reads
\begin{equation}
	\mathcal{L}_{D^{*0}}=\bm{D}^\dagger \cdot \left[i\partial_0+{\bm\nabla^2 \over 2(M+m)}-\left(\delta-i{\Gamma_{*0}\over 2}\right)\right]\bm{D},
\end{equation}
where $\Gamma_{*0}$ denotes the $D^{\ast0}$ width. The partial width of $D^{*0}\to D^0\gamma$ can be included in the $\Gamma_{*0}$ term. For the interaction terms, the Galilean invariant Lagrangian can be obtained from modifying those in Eq.~\eqref{eq:XEFT_lag} as follows
\begin{itemize}
	\item In the $\bm{D}^\dagger \cdot D \rnb \pi$ term, the operator $\rnb$ is replaced by $(M\rnb-m\lnb)/(M+m)$;
	\item In the $(\bar{D}\bm{D})^\dagger \cdot \bar{D}(\lrnb)^2\bm{D}$, the $\lrnb$ should be replaced by $4(M\rnb-(M+m)\lnb)^2/(2M+m)^2$;
	\item In the interaction $(\bar{D}\bm{D})^\dagger \cdot \bar{D}D\rnb \pi$, the $\rnb$ should be replaced by $(2M\rnb-m\lnb)/(2M+m)$.
\end{itemize}
The above results are obtained by satisfying two conditions---the Galilean invariance and recovery of the original XEFT in the center-of-mass frame. We can see the Galilean invariance easily in the form, e.g., $(m^{-1}\rnb -M^{-1}\lnb)\mu_{M,m}$, where $m^{-1}\rnb$ and $M^{-1}\lnb$ are proportional to the velocity operators. In the Galilean-invariant XEFT, apart from the conservation of $N_c\equiv N_{D^0}+N_{D^{*0}}$ and $N_{\bar{c}}\equiv N_{\bar{D}^0}+N_{\bar{D}^{*0}}$, an extra conservation of the pion number $N_\pi\equiv N_{\pi^0}+N_{D^{*0}}+N_{\bar{D}^{*0}}$ is guaranteed by the Galilean symmetry. {Constrained by the Galilean symmetry, it is hard to include the $D^*D^*\pi$ vertex. The heavy quark spin symmetry has to be given up. } In the calculation, the complex on-shell renormalization scheme was used, which introduces an extra counter term,
\begin{equation}
	\mathcal{L}=-\frac{\delta D_{0}}{2}(\bar{D} \boldsymbol{D})^{\dagger} \cdot\left[i \partial_{0}+\frac{\bm\nabla^{2}}{2(2 M+m)}\right](\bar{D} \boldsymbol{D}).
\end{equation}
With the constraints of the Galilean symmetry, the cutoff-dependence is eliminated to all orders of $m/M$. In Ref.~\cite{Braaten:2020nmc}, the Galilean-invariant XEFT was reformulated in a more simple way by introducing the auxiliary field and adopting a new complex threshold renormalization scheme.

In Ref.~\cite{Valderrama:2012jv}, Valderrama investigated the power counting in the heavy meson molecules, such as the $X(3872)$, $Z_b(10610)$ and $Z_b(10650)$. They argued that the coupled-channel effect from the HQSS partner channels (e.g., $D\bar{D}$, $\DastDbar$ and ${D}^*\bar{D}^*$) is suppressed by at least two orders, which can be safely neglected in the low order calculation. Meanwhile, the energy scale that the pion can be treated perturbatively was explored within a framework developed in atomic physics to handle the singular power-law  potential~\cite{gao1999repulsive} (which was employed successfully in the $NN$ system~\cite{Birse:2005um}). It was shown that the validity of the perturbative pion is related to the $1/r^3$ singularity in the short range part of the tensor force in the OPE interaction~\cite{Birse:2005um}. Expanding the Schr\"odinger equation with the power-law potential in the Bessel functions makes it possible to obtain the analytical solutions nonperturbatively and determine the radius of convergence~\cite{Birse:2005um}. In the Ref.~\cite{Valderrama:2012jv}, the approach was extended to the heavy meson systems. For the $I^{G}(J^{PC})=0^{\pm}(1^{+\pm})$, the critical momentum of the perturbative pion is about $290^{+120}_{-80}$ MeV, which corresponds to the critical binding energy $42_{-19}^{+46}$ MeV. Therefore, it is reasonable to adopt the perturbative pion in XEFT. One can see that the $\gamma_c$ in Table~\ref{tab:4.1xscale} is below the critical momentum. Thus, in the EFT,  considering the charged channel $\xthc$, one can treat the OPE interaction as the NLO contribution perturbatively as well.

 \begin{figure}[htbp]
	\centering
	\includegraphics[width = 1.0\textwidth]{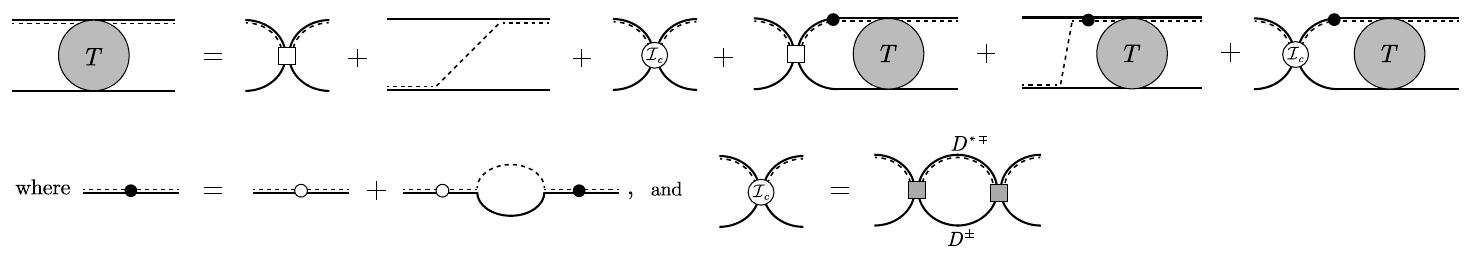}
	\caption{Nonperturbative $D\bar{D}^*$ amplitude in Ref.~\cite{Schmidt:2018vvl}.}\label{fig:4.x_three_body}
\end{figure}

In Ref.~\cite{Braaten:2007ct}, the author proposed a toy model to investigate the three-body dynamics of the $X(3872)$ with an equivalently perturbative pion-exchange interaction, which is the first work to discuss the three-body dynamics of the $X(3872)$.  In this toy model, three spin-zero mesons $D_1$, $D_2$ and $\phi$ are used to mimic the $D$, $D^\ast$ and pion, respectively. The Lagrangians are
\begin{eqnarray}
{\cal L}_{{\rm free}}&=&\sum_{i=1,\,2}D_{i}^{\dagger}\left(i\frac{\partial}{\partial t}-M_{i}+\frac{1}{2M_{i}}\nabla^{2}\right)D_{i}+\phi^{\dagger}\left(i\frac{\partial}{\partial t}-m+\frac{1}{2m}\nabla^{2}\right)\phi,\\
{\cal L}_{{\rm int}}&=&-\lambda_{0}D_{1}^{\dagger}D_{2}^{\dagger}D_{1}D_{2}-g\big(D_{2}^{\dagger}D_{1}\phi+D_{1}^{\dagger}\phi^{\dagger}D_{2}\big)-\delta MD_{2}^{\dagger}D_{2}.
\end{eqnarray}
In the calculation, the vertex $\lambda_0$ is summed to all orders but the vertex $g$ is treated perturbatively. The cutoff dependence was removed by renormalizing the $D_2$ mass, the $\lambda_0$ and short-range coefficients in the operator product expansion. In the NLO, the nonzero width of the $D_2$ mesons was introduced by either summing its self-energy correction to all orders or using the complex mass scheme to remove the nonphysical IR divergence. In Ref.~\cite{Jansen:2013cba}, the binding energy of the $X(3872)$ is investigated up to NLO in XEFT, where the one-pion-exchange interaction is treated perturbatively. Two types of three-body cuts in Fig.~\ref{fig:4.x_3body_cut} were considered, which will be reviewed in detail in Sec.~\ref{sec:x-mass-crrection}.

In Ref.~\cite{Baru:2011rs}, the three-body dynamics of the $X(3872)$ was investigated in a Faddeev-type framework with nonperturbative pion dynamics for the first time. In Ref.~\cite{Schmidt:2018vvl}, Schmidt \etal proposed an EFT with $D^0$, $\bar{D}^0$ and $\pi^0$ as the basic degrees of freedom for the $X(3872)$ with the exact Galilean invariance. The $D^{\ast0}$ is dynamically generated as the $P$-wave resonance of $D^0 \pi^0$. The LO Lagrangians are (\cvII in Table~\ref{tab:two-convention})
\begin{eqnarray}
	\mathcal{L}&=&\mathcal{L}_{\text{kin}}+(\mathcal{L}_{D\pi}+\mathcal{L}_{\bar{D}\pi})+\mathcal{L}_{D\bar{D}\pi},\\
\mathcal{L}_{\text{kin}}&=&D^{\dagger}\left[i\,\partial_{0}+\frac{\nabla^{2}}{2M_{D}}\right]D+\bar{D}^{\dagger}\left[i\,\partial_{0}+\frac{\nabla^{2}}{2M_{D}}\right]\bar{D}+\pi^{\dagger}\left[i\,\partial_{0}+\frac{\nabla^{2}}{2m_{\pi}}\right]\pi,\\
\mathcal{L}_{D\pi}&=&\bm{D}^{\dagger}\left[\Delta_{0}+\Delta_{1}\,i\,\partial_{\text{cm}}+\sum_{n\geq2}\Delta_{n}\,(i\,\partial_{\text{cm}})^{n}\right]\bm{D}+g\left[\bm{D}^{\dagger}\cdot\big(\pi\overleftrightarrow{\bm{\nabla}}D\big)+\text{H.c.}\right],\\
\mathcal{L}_{D\bar{D}\pi}&=&-C_{0}\,\frac{1}{2}\left[\bar{D}\bm{D}+D\bar{\bm{D}}\right]^{\dagger}\cdot\left[\bar{D}\bm{D}+D\bar{\bm{D}}\right],
\end{eqnarray}
where $i\partial_{\text{cm}}\equiv i\partial_0+\nabla^2/(2M)$ (with $M\equiv M_D+m_\pi$) and $\overleftrightarrow{\bm{\nabla}}\equiv\ \mu_{D\pi}\,\big(m_{\pi}^{-1}\,\overleftarrow{\bm{\nabla}}-M_{D}^{-1}\,\overrightarrow{\bm{\nabla}}\big)$ are Galilean-invariant derivatives~\cite{Braaten:2015tga,Braaten:2020nmc}. $\mu_{D\pi}$ is the reduced mass of the  $D$ and $\pi$. The $\bm{D}~(\bar{\bm{D}})$ is the auxiliary vector field to embed the $P$-wave interaction. The auxiliary field can be eliminated by the path integral or the equations of  motion. In general, the value $\Delta_1$ can be taken as $\Delta_1= \pm 1$, in which the $\Delta_1= + 1$ makes the $\bm{D}$ a physical field. The Feynman diagrams are presented in Fig.~\ref{fig:4.x_three_body}. The three-body cut associated with the self-energy of the $D^*$ is included by incorporating the full propagator of the $D^*$. The pion exchange potential is given by
\begin{equation}
	i\,V^{ij}\left(\bm{p},\,\bm{q};\,E\right)\equiv i\,g^{2}\frac{\left(\alpha\,\bm{p}+\bm{q}\right)^{i}\left(\alpha\,\bm{q}+\bm{p}\right)^{j}}{E-\frac{\bm{p}^{2}}{2\mu}-\frac{\bm{q}^{2}}{2\mu}-\frac{\bm{p}\cdot\bm{q}}{m_{\pi}}+i\epsilon},
\end{equation}
where $i$ and $j$ are the indices of the polarization vector. The $\alpha$ is defined as $\alpha=M_{D}/M$. The three-body dynamics for the pion-exchange interaction is included as well. Meanwhile, at NLO the charged channel is included in {the kernel $\mathcal{I}_c$ as shown in Fig.~\ref{fig:4.x_three_body}.} 

\begin{figure}[htbp]
	\centering
	\includegraphics[width = 1.0\textwidth]{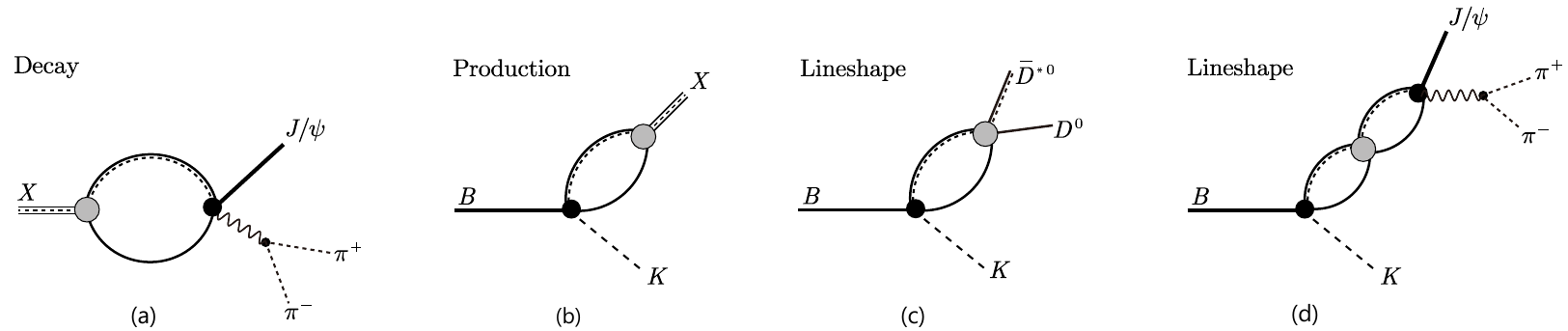}
	\caption{Examples to use the factorization formula in Ref.~\cite{Braaten:2005jj}.}\label{fig:4.x_factation}
\end{figure}

With the separated scale, Braaten \etal adopted the factorization formula~\cite{Braaten:2005jj} to explore the decays and productions of the $X(3872)$ when the short-range dynamics is involved. The amplitude of the decays involving short-rang dynamics is given in Eq.~\eqref{eq:4.x_decay_factorization}. The long-range part involves the wave function at origin. Another equivalent approach to writing down the long-range dynamics is
\begin{equation}
	\frac{d^{3}q}{(2\pi)^{3}}\psi(\bm{q})\sim \int\frac{d^{3}q}{(2\pi)^{3}}\frac{g}{E-\frac{\bm{q}^{2}}{2\mu}},
\end{equation}
where the coupling constant and  two-body nonrelativistic propagator are introduced. The equivalence has been shown in Sec.~\ref{sec:1chanel_pionless}. According to Eq.~\eqref{eq.4.1:wv_r=0}, the long-range and short-range parts are both cutoff-dependent. One can define the cutoff-independent factorization formula by absorbing the linear divergent part in the wave function at origin into the short-range part,
\begin{equation}
	{\cal T}[X\to H]\sim{\cal \tilde{A}}_{\text{long}}(E)\times{\cal \tilde{A}}_{\text{short}}[\{D\bar{D}^{*}\}^{+}\to H],
\end{equation}
where we use $\tilde{\mathcal{A}}$ to represent the cutoff-independent amplitude.
Similarly, the production of the $X(3872)$ as shown in Fig.~\ref{fig:4.x_factation}(b) can be formulated as
\begin{equation}
	{\cal T}[B\to XK]\sim{\cal \tilde{A}}_{\text{short}}[B\to\{D\bar{D}^{*}\}^{+}K]\times{\cal \tilde{A}}_{\text{long}}[\{D\bar{D}^{*}\}^{+}\to X].
\end{equation}
If one wants to investigate the line shape of $B\to D\bar{D}^{*}K$ as shown in Fig.~\ref{fig:4.x_factation}(c), the factorization reads
\begin{equation}
	{\cal T}[B\to D\bar{D}^{*}K]\sim{\cal \tilde{A}}_{\text{short}}[B\to\{D\bar{D}^{*}\}^{+}K]\times{\cal \tilde{A}}_{\text{long}}[\{D\bar{D}^{*}\}^{+}\to\{D\bar{D}^{*}\}^{+}].
\end{equation}
For the line shape $B\to K J/\psi \pi \pi$ in Fig.~\ref{fig:4.x_factation}(d), the factorization is
\begin{equation}
	{\cal T}[B\to HK]\sim{\cal \tilde{A}}_{\text{short}}[B\to\{D\bar{D}^{*}\}^{+}K]\times{\cal \tilde{A}}_{\text{long}}[\{D\bar{D}^{*}\}^{+}\to\{D\bar{D}^{*}\}^{+}]\times{\cal \tilde{A}}_{\text{short}}[\{D\bar{D}^{*}\}^{+}\to H].
\end{equation}
It is worthwhile to mention that the charged $\xthc$ channel can be regarded as either the short-range dynamics~\cite{Braaten:2005ai} or long-range dynamics~\cite{Meng:2021kmi} depending on the specific framework.

\begin{figure}[htbp]
	\centering
	\includegraphics[width = 1.0\textwidth]{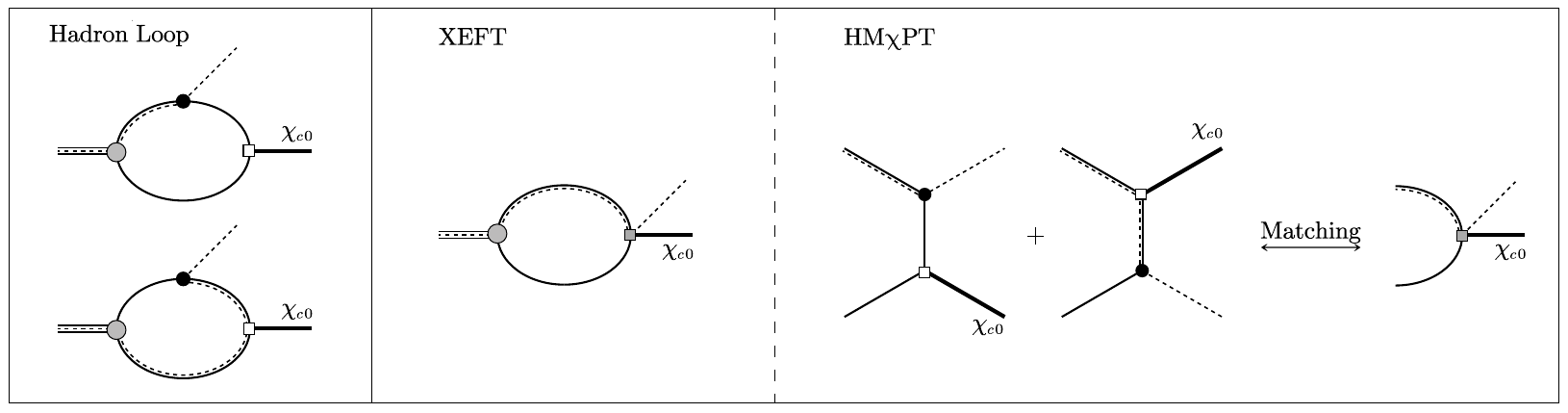}
	\caption{$X(3872) \to  \chi_{c0} \pi$ in the non-factorization scheme (hadron loop) and factorization scheme (matching HM$\chi$PT to the XEFT).}\label{fig:4.x_hadron_loop_fraction}
\end{figure}

Apart from the factorization framework, the non-factorization approach is also used when calculating these short-range involved dynamics. An example is the NREFT expanded with the velocity of particles~\cite{Guo:2009wr,Guo:2010zk,Guo:2010ak,Guo:2013zbw,Guo:2014taa,Guo:2017jvc}. We take the hadron loop diagrams in Fig.~\ref{fig:4.x_hadron_loop_fraction} as an example. In the loop diagrams, the momentum and energy of the three nonrelativistic intermediate states are counted as $\mathcal{O}(v)$ and $\mathcal{O}(v^2)$, respectively. Thus, the propagators are counted as $\mathcal{O}(v^{-2})$ and the loop integral $\int d^4 l$ is counted as $\mathcal{O}(v^5)$. The general amplitude is counted as
\begin{equation}
	\mathcal{M}\sim {v^5q^n \over (v^2)^3}.~\label{eq:guoNREFT_pw}
\end{equation}
where $q^n$ is the scale arising from the derivative coupling vertices.

The relation of these two approaches was investigated by Mehen~\cite{Mehen:2015efa} with the $X(3872)\to\chi_{cJ}\pi^0$ process as an example. In the non-factorization approach (also known as the hadronic loop approach) the diagrams are listed in the Fig.~\ref{fig:4.x_hadron_loop_fraction}. The $D\bar{D}\chi_{c0}$ and $D^*\bar{D}^*\chi_{c0}$ vertices were constructed in HM$\chi$PT. The loop integral in the non-factorization approach is finite (an extra regulator is not needed). In the factorization approach, the short-distance dynamics is depicted by the $D\bar{D}^*\chi_{c0} \pi$ vertex in XEFT, which is matched to the HM$\chi$PT diagrams. In the factorization approach, the charged channel $\xthc$ is integrated out. The short-distance dynamics in these two approaches arise from the same $D\bar{D}\chi_{c0}$ and $D^*\bar{D}^*\chi_{c0}$ vertices in HM$\chi$PT. However, one cannot expect these two approaches to give the same results. The two approaches do include the same dynamics in the region $p<\gamma_c$, where $\gamma_c$ is the binding momentum with respect to the charged threshold. When the cutoff $\Lambda \sim \gamma_c$ is introduced, the non-factorization approach will give the same results as the factorization approach~\cite{Mehen:2015efa}. However, in the region $p>\gamma_c$, these two approaches contain different UV behaviors, which will lead to different results. Mehen pointed out ~\cite{Mehen:2015efa} that the $X(3872)\to \chi_{cJ}\pi^0$ rate in the non-factorization approach including the charged channels will be similar to the results from the factorization approach because of the cancellations of the contribution from the UV region in the neutral and charged channels for the specific process. However, this is just a specific example. In Ref.~\cite{Guo:2017jvc}, it was found that the contributions of the charged and neutral channels for $X(3872)\to\chi_{cJ}\pi\pi$ are constructive. Thus, these two approaches (even with the charged channel in the non-factorization approach) will give different results.

{We can analyze the scales appearing in the above example following Ref.~\cite{Hanhart:2007wa}. The binding momenta for the neutral and charged channels are $\gamma_n\sim 20$ MeV and $\gamma_c\sim 120$ MeV, respectively (see Table~\ref{tab:4.1xscale} for details). The vertex $X\bar{D}^*D$ is dressed by the one-pion exchange interaction with the typical scale $|u|~\sim 45$ MeV. The typical scale of the vertex $D^*D\pi$ is estimated as $\sqrt{2m_{\pi}|m_{D^{*}}-m_{D}-m_{\pi}|} \sim |u|\sim 45$ MeV in the nonrelativistic approximation. The vertex $D^{(*)}\bar{D}^{(*)}\chi_{c0}$ is relevant to the rearrangement of the charmed quarks into a bound state. Here we use the binding momentum of the charmonium, $m_cv\sim $ 900 MeV to estimate the typical scale, where the $v\sim 0.3$ is the typical velocity of charm quark. One can see that there are at least three well-separated scales in total, $|u|\sim \gamma_n$, $\gamma_c$ and $m_c v$.} In the above examples, one should be cautious of taking the cutoff $\Lambda $ to infinity, because one cannot expect the hadronic vertices still work properly at very high energy scale. One cannot even expect that one single cutoff is enough to mimic all the form factors of the vertices at (widely) separated scale. Meanwhile, in the NREFT, there exist two velocities as pointed out in Ref.~\cite{Guo:2017jvc}. The power counting might fail due to the two separated velocities and the appearance of triangle singularity (see Ref.~\cite{Guo:2019twa} for reviews). However, in order to include the effect of triangle singularity, one has to resort to the non-factorization approach~\cite{Guo:2019qcn,Braaten:2019gfj,Braaten:2019yua,Braaten:2019sxh,Braaten:2019gwc}. {Therefore, the non-factorization approach is trying to embed details of higher energy scales than those in the factorization approach. The cost for the non-factorization approach is the arbitrariness of choosing UV behaviors. Moreover, the appearance of multiple scales would destroy the convergence of EFT.} In the specific process, the two approaches might give different results, which need to be checked by experiments.

\subsubsection{Mass corrections}~\label{sec:x-mass-crrection}

\begin{figure}[htbp]
	\centering
	\includegraphics[width = 1.0\textwidth]{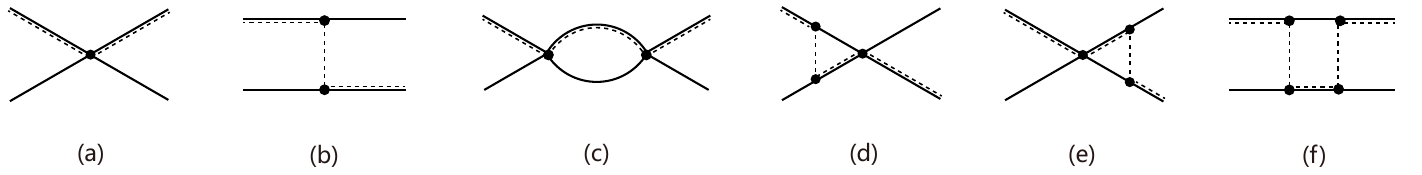}
	\caption{Feynman diagrams contribute to $T^{(0)}$ and $T^{(2)}$ in Ref.~\cite{Wang:2013kva}.}\label{fig:4.x_Pingwang}
\end{figure}

In Ref.~\cite{Wang:2013kva}, the authors investigated the $\DastDbar$ scattering in the unitarized HH$\chi$PT with the pion exchange and a contact interaction. In the calculation, the $\mathcal{O}(p^0)$ amplitudes come from the (a) and (b) diagrams in Fig.~\ref{fig:4.x_Pingwang}. The diagrams (c)-(f) are counted as $\mathcal{O}(p^2)$ from Weinberg's power counting. However, the amplitudes of such two-particle reducible diagrams are enhanced by a factor $M/p$ if one keeps the kinetic energy terms to remove the pinched singularity as shown in Eq.~\eqref{eq:sec1.5:delpinch}. The amplitudes in diagrams (c)-(f) counted as $\mathcal{O}(p)$ can be generated by iterating the tree-level amplitudes in (a) and (b). The authors obtained a unitary amplitude $T^{\text{phy}}=T^{(0)2}/(T^{(0)}-T^{(1)})$ on the basis of the leading order Pad\'e approximation (see Ref.~\cite{Glockle:99109} for Pad\'e approximation and see Ref.~\cite{Dobado:1996ps} for the similar unitary method for the $\pi\pi$ scattering). The authors concluded that (i) the pion-exchange interaction is sufficient to form a bound state without the contact interaction, (ii) the result is not sensitive to the strength of the contact interaction, and (iii) the $X(3872)$ pole disappears when the pion mass is larger than $142$ MeV.

\begin{figure}
	\centering
\includegraphics[width = 0.3\textwidth]{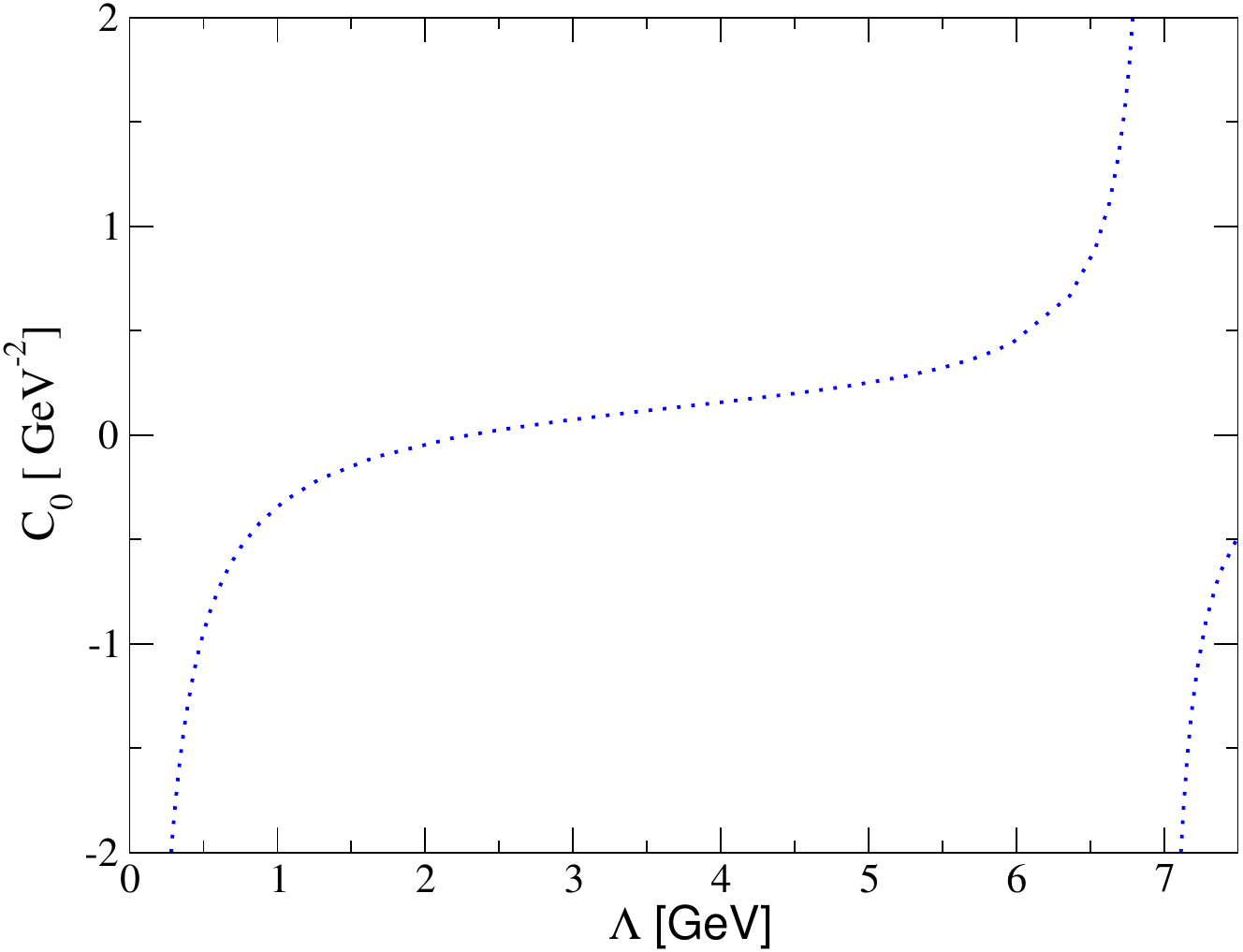}
\caption{The cutoff dependence of the contact coupling constant obtained in Ref.~\cite{Baru:2011rs} by imposing a bound state with a binding energy $0.5$ MeV in an interaction with the dynamical pion.}\label{fig:4.x_c0_cut}
\end{figure}

However, the authors of  Ref.~\cite{Baru:2015nea} illustrated in an explicit cutoff-regularization scheme that the OPE interaction is well defined only in the context of the definite contact interaction because the divergent part of the iterating pion-exchange interaction should be canceled out by the contact counter term. Thus, the separation of the pion-exchange and contact contributions is scheme-dependent. In Ref.~\cite{Baru:2011rs}, the cutoff dependence of the contact coupling constants was obtained (see Fig.~\ref{fig:4.x_c0_cut}), in which a bound state was imposed with a binding energy $0.5$ MeV in an interaction with the dynamical pion. One notices that the $C_0=0$ corresponds to the cutoff $\Lambda\approx 1.0$ GeV. Thus, the bound state was naively attributed to the OPE interaction when the cutoff is around $1.0$ GeV  (e.g.,~\cite{Tornqvist:2004qy,Thomas:2008ja}), which seems a scheme-dependent conclusion. 

\begin{figure}
	\centering
	\includegraphics[width = 0.32\textwidth]{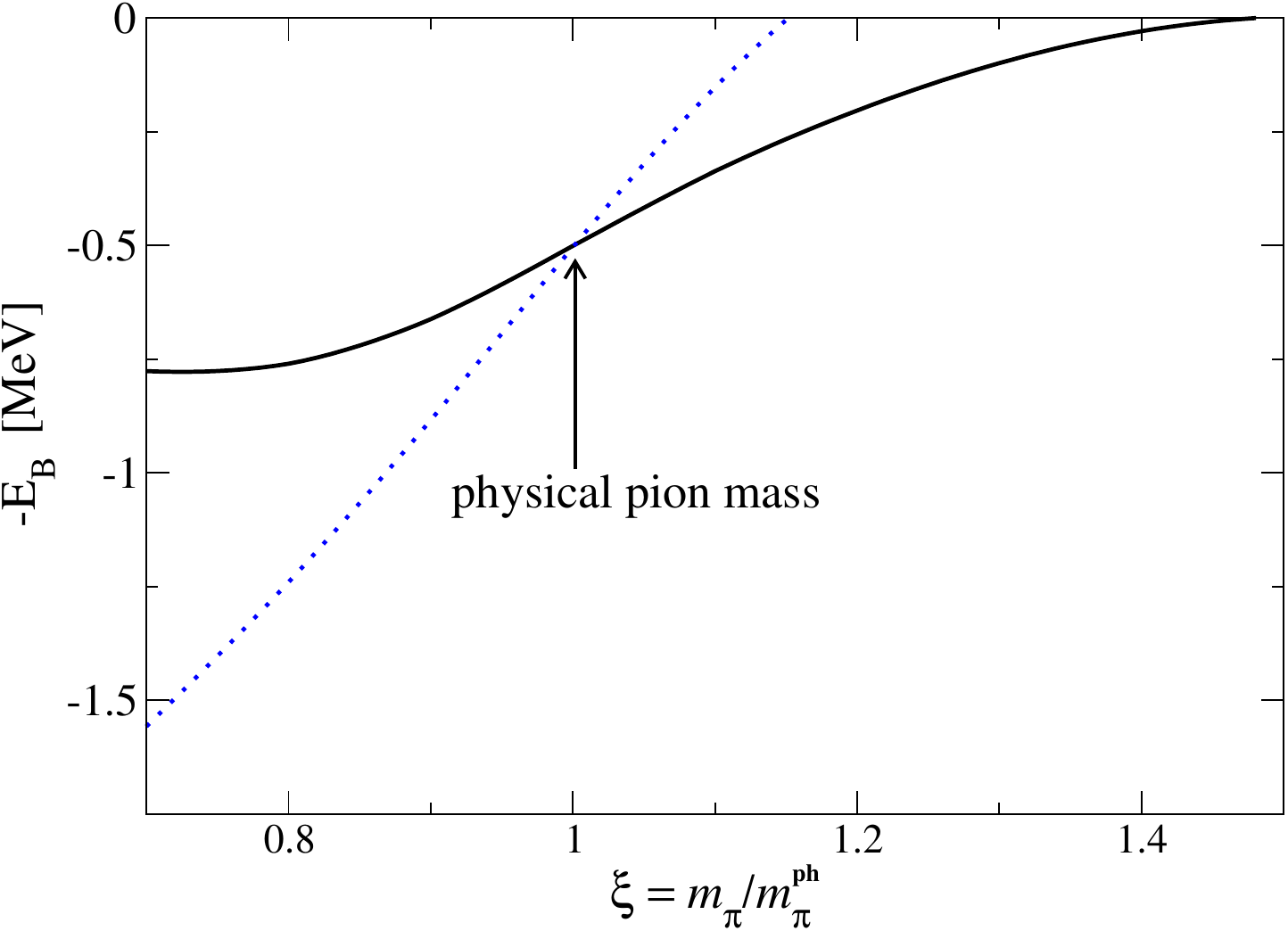}
	\includegraphics[width = 0.32\textwidth]{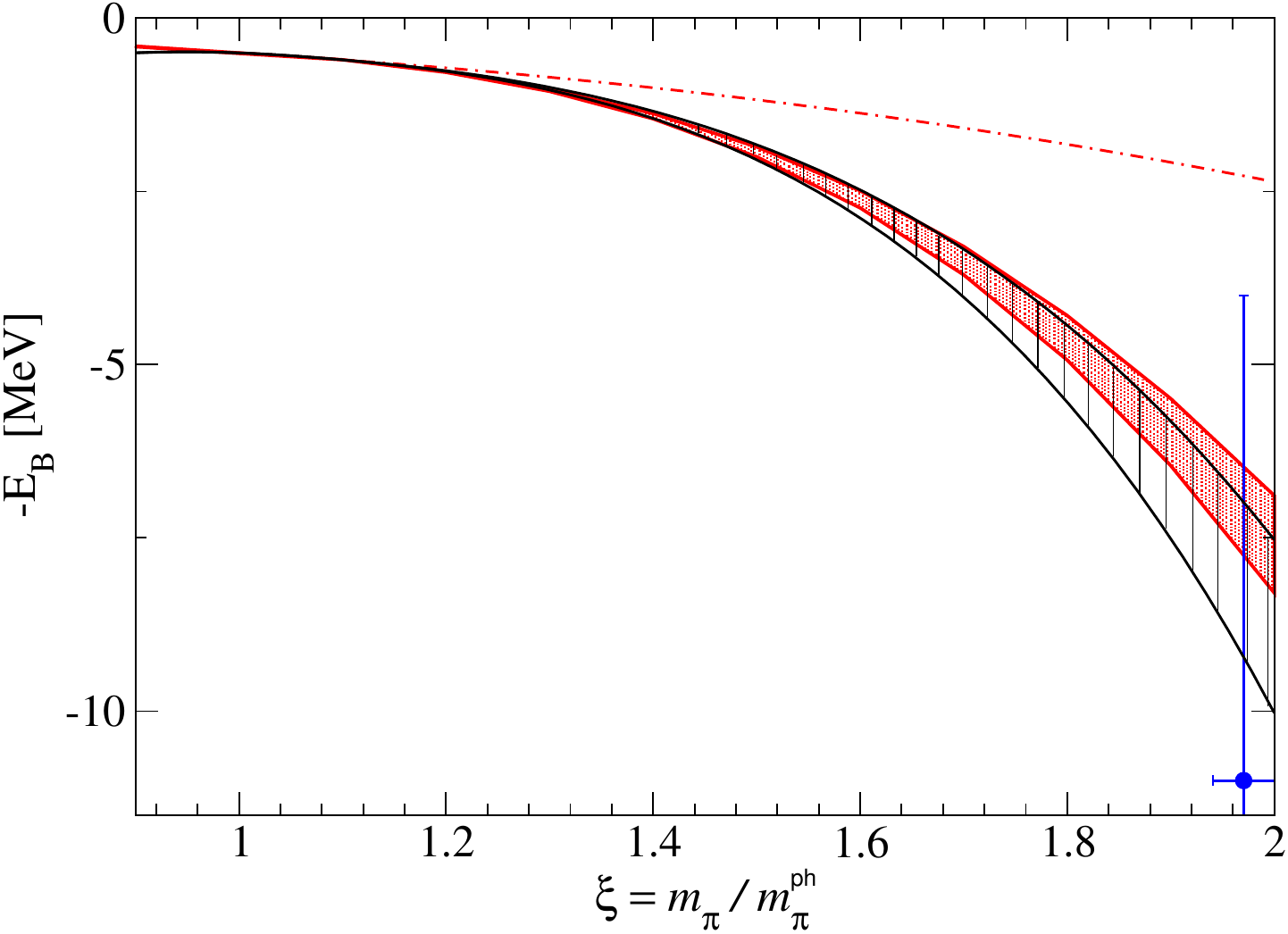}
	\includegraphics[width = 0.33\textwidth]{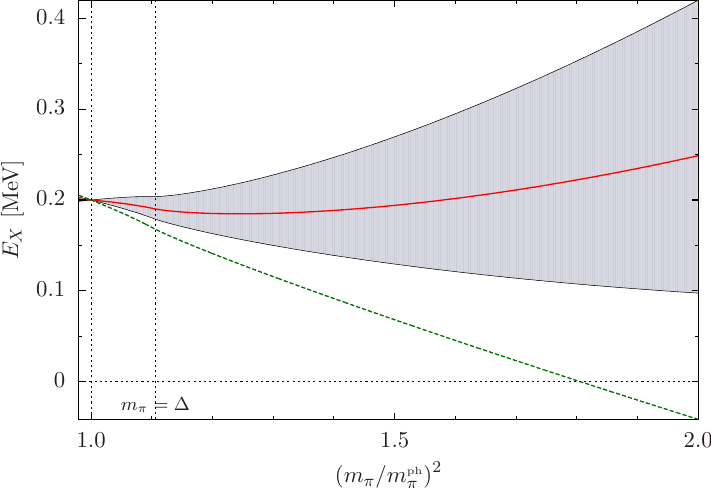}
	\caption{Left panel: The LO results including the fully three-body dynamics (black solid curve) and in the simple static OPE (blue dotted curve) in Ref.~\cite{Baru:2015tfa} within the modified Weinberg formulation. Middle panel: The NLO results in the conventional Weinberg formulation (black hatched band) in Ref.~\cite{Baru:2013rta} and modified Weinberg formulation (red dotted band) in Ref.~\cite{Baru:2015tfa}. The dot-dashed curve comes from the calculation without pions. The dot with error bars represents the results from lattice QCD simulations~\cite{Padmanath:2015era}. Right panel: The NLO results (gray band) in XEFT in Ref.~\cite{Jansen:2013cba}, where the unknown $d_2$ and $r_0$ are estimated with naturalness. The red solid curve represents the result considering the LO contact and OPE interaction. The green dashed curve represents an example of unnatural parameters.}\label{fig:4.x_chiral_bid}
\end{figure}

In Ref.~\cite{Baru:2013rta}, the authors investigated the $m_\pi$-dependence of the binding energy of the $X(3872)$ within a nonrelativistic Faddeev-type three-body equation for the $D\bar{D}\pi$ system in the $J^{PC} = 1^{++}$ channel. In the calculation, the dynamical pion was included nonperturbatively, and the $m_{\pi}$-dependence of the $D^{(\ast)}$ mass~\cite{Cleven:2010aw}, pion decay constant $f_{\pi}$~\cite{Gasser:1983yg} as well as the $D^\ast D\pi$ coupling constant $g$~\cite{Becirevic:2012pf} were determined by $\chi$PT or lattice QCD. However, the $m_\pi$-dependence of the contact interaction is hard to determine, which is conjectured on the basis of the $X(3872)$ as a bound state and the cutoff-independent binding energy $E_B$. The form of the contact interaction is assumed as
\begin{equation}
C_{0}(\Lambda,\xi)=C_{0}^{\text{ph}}(\Lambda)\left[1+f(\Lambda)\frac{m_{\pi}^{\text{ph}2}}{M^{2}}(\xi^{2}-1)\right].
\end{equation}
The superscript ``ph" represents the quantities at the physical pion mass. The $\xi$ is defined as $\xi\equiv m_{\pi}/m_{\pi}^{\text{ph}}$. $M$ is a large scale around $m_\rho\approx 800$ MeV. The unknown $f(\Lambda)$ is evaluated with the requirement that the binding energy $E_B^\text{ph}$ and its slope $S_{E_B}\equiv(\partial E_B/\partial m_\pi)|_{m_\pi=m_\pi^\text{ph}}$ should be cutoff-independent at the physical point. The final results depend on the specific value of $S_{E_B}$. A two-pion-exchange diagram was used to estimate the natural results of $f(\Lambda)$ and $S_{E_B}$. The final result is given with varying $S_{E_B}$ in a natural scale as shown in the middle subfigure of Fig.~\ref{fig:4.x_chiral_bid}. The results indicated that the behavior of the quark-mass dependence of the $X(3872)$ strongly depends on the contact interaction. In other words, ignoring the contact interaction might yield misleading results. Meanwhile, the $X(3872)$ can turn into a virtual state only when the $S_{E_B}$ is in a unnatural range, which is different from the results in Ref.~\cite{Wang:2013kva}.

In Ref.~\cite{Baru:2015tfa}, the chiral extrapolation of the binding energy of the $X(3872)$ was investigated in a similar framework with that in Ref.~\cite{Baru:2013rta}, but in the modified Weinberg formulation~\cite{Epelbaum:2012ua}. The modified Weinberg formulation was proposed to solve the problem which was referred to as the ``inconsistency of Weinberg's approach"~\cite{Savage:1998vh}. The idea is to make the UV behavior milder by interchanging the expansion order of $1/\Lambda$ and $1/m$. For example, within the cutoff regularization, the relativistic loop function reads
\begin{equation}
	I=\frac{4\,i}{(2\,\pi)^{4}}\int\frac{d^{4}k\,\theta(\Lambda-|\bm{k}|)}{\left[k^{2}-m^{2}+i\,0^{+}\right]\left[(P-k)^{2}-m^{2}+i\,0^{+}\right]},
\end{equation}
where $P=(2\sqrt{m^{2}+\bm{p}\,^{2}},\bm{0}\,)$. The integral is logarithmically divergent. Our conventional nonrelativistic expansion based on $p\ll \Lambda\ll m$ will give rise to a linear divergence. However, if one first performs the $1/\Lambda$ expansion and then performs the $1/m$ expansion based on $p \ll m\ll\Lambda$, one can retrieve the logarithmic divergence and solve the inconsistency of Weinberg's approach. With this modified Weinberg formulation, the theory is perturbatively renormalizable and all divergence can be removed by renormalizing the LECs of the LO contact interaction. Unlike the approach of Ref.~\cite{Baru:2013rta}, where additional input is needed to maintain the renormalization, the modified Weinberg approach can predict the $m_\pi$-dependence of the binding energy without extra parameters.  At the leading order the contact coupling constant $C_0$ is $m_\pi$-independent. The LO results are shown in the left subfigure of Fig.~\ref{fig:4.x_chiral_bid}, in which the binding energy of the $X(3872)$ will decrease and finally become unbound with the increasing of $m_\pi$. However, the NLO results shown in the middle subfigure of Fig.~\ref{fig:4.x_chiral_bid} will change the tendency and become consistent with those in Ref.~\cite{Baru:2013rta}. It is worthwhile to stress that the NLO results were based on some assumptions and naive dimensional analysis.

In Ref.~\cite{Jansen:2013cba}, the binding energy of the $X(3872)$ was investigated in XEFT, where the $\xthn$ scattering in the $J^{PC}=1^{++}$ channel was calculated up to NLO as shown in the diagrams of Fig.~\ref{fig:4.x_xmassXEFT}. Apart from the Lagrangian~\eqref{eq:XEFT_lag}, an extra NLO Lagrangian was introduced to include the $m_\pi$-dependence of the contact term (\cvII) in Table~\ref{tab:two-convention},
\begin{equation}
{\cal L}^{\text{NLO}}=-\frac{D_{2}u^{2}}{2}(\bar{\boldsymbol{D}}D+\boldsymbol{D}\bar{D})^{\dagger}\cdot(\bar{\boldsymbol{D}}D+\boldsymbol{D}\bar{D}),
\end{equation}
where $u^2$ is defined as $u^{2}=\Delta^{2}-m_{\pi}^{2}$. Compared with the KSW scheme for the $NN$ system, a novel feature is the $\mathcal{A}_0^{(\text{VI})}$. The contribution of the diagram for the $NN$ system is canceled out by the counter term in the on-shell renormalization scheme because the pions are always off-shell which only contribute to the real part of the pole of the nucleon propagator. For the XEFT, the pions could be on-shell and contribute to the imaginary part of the pole of the $D^\ast$ which cannot be removed in the on-shell renormalization scheme.  In the calculation, the authors showed that the $\mathcal{A}_0^{(\text{VI})}$ is IR divergent if one dresses the $D^{*0}$ propagator with one loop. Instead, the author used the full propagator of the $D^{*0}$ that is dressed to all orders. If the full propagator is used in $\mathcal{A}_{-1}$ and $\mathcal{A}_0^{(\text{I},\dots,\text{V})}$, the contribution of $\mathcal{A}_0^{(\text{VI})}$ is included automatically. To the NLO, the cutoff-dependence of the $C_2$ and $D_2$ is obtained by the renormalization up to two unknown parameters, $r_0$ and $d_2$. With the naturalness estimation of the two parameters, the $m_\pi$-dependence of the binding energy is given in the right subfigure of Fig.~\ref{fig:4.x_chiral_bid}. In the numerical analysis, the $m_\pi$-dependence of the pion decay constant, $D$ meson axial coupling constant and masses of the $D^{(*)}$ were taken from Refs.~\cite{Gasser:1983yg,Colangelo:2001df,Becirevic:2012pf,Guo:2009ct}. One can see that the binding energy of the $X(3872)$ depends on the pion mass moderately. In Ref.~\cite{Jansen:2015lha}, the calculation was extended to the $X(3872)$ in the finite volume. The finite volume correction to the binding energy was obtained explicitly and fully determined by the infinite volume parameters. The numerical results showed that the finite volume effect is significant even for a large box with the length around $20$ fm.

In Ref.~\cite{Xu:2017tsr}, Xu \etal calculated the $DD^*$ interaction in $\chi$EFT up to NLO with the Weinberg scheme. In their calculation, the LO contact term, OPE and the NLO TPE interactions were investigated. The LECs of the LO contact terms were estimated in the resonance saturation model~\cite{Ecker:1988te,Donoghue:1988ed,Bernard:1996gq,Epelbaum:2001fm,Du:2016tgp}. The numerical results showed that there is no bound state in the isovector channel. In the isoscalar channel, there exists a bound state with the binding energy $17.5_{-3.9-14.0}^{+4.1+18.3}$ MeV, where the first and second uncertainties arise from the inclusion of the $f_0$, $a_0$ and $a_1$, $f_1$ mesons in the resonance saturation model, respectively. The prediction was confirmed by the observation of the $T_{cc}^+$~\cite{LHCb:2021auc,LHCb:2021vvq}. The similar $\chi$EFT was adopted to investigate the $\bar{B}^{(*)}\bar{B}^{(*)}$ systems in Ref.~\cite{Liu:2012vd,Wang:2018atz}. It was shown that there exist the $\bar{B}\bar{B}^*$ and $\bar{B}^*\bar{B}^*$ bound states with $I(J^P)=0(1^+)$. The former one is naturally the heavy quark flavor symmetry partner of the $T_{cc}^+$ state.

\subsubsection{Partners of $X(3872)$}

The $X(3872)$ as a hadronic molecule implies the existence of  other states which are related to the $X(3872)$ through various symmetries such as the {heavy quark spin symmetry, heavy quark flavor symmetry,}  SU(3) flavor symmetry and so on. In this section, we review the partner states of the $X(3872)$ as predicted in various EFT frameworks. We will also pay attention to the caveats of these predictions.

In Ref.~\cite{AlFiky:2005jd}, the author adopted the $\slashed{\pi}$EFT with heavy quark symmetry to explore the partners of the $X(3872)$. It was found that the existence of the $\xthn$ bound state does not exclude and support the existence of the $D^0\bar{D}^0$ bound state. As shown in Eqs.~\eqref{eq:V0pp},~\eqref{eq:V1pmepsV2pm} and~\eqref{eq:V1ppepsV2pp} there are two independent coupling constants in the $D^{(*)}\bar{D}^{(*)}$ systems. The interactions for the $\DastDbar$ and $D\bar{D}$ systems are independent in the heavy quark spin symmetry limit.

In Ref.~\cite{Nieves:2011zz}, Nieves \etal discussed the $B\bar{B}^*$ bound states deduced from the weakly bound $X(3872)$ within EFT framework including the contact and OPE interactions. The heavy quark symmetry was used to relate the cutoffs in the charmed and bottom systems by $\Lambda_B=\Lambda_X+\mathcal{O}({1\over m_Q})$. The contact dynamics is determined from the phenomenological model in Refs.~\cite{Gamermann:2006nm,Gamermann:2007fi}. Within the phenomenological model, the contact interaction in the neutral-charged basis can be expressed as
\begin{equation}
	\langle\bm{k}|V_{C}|\bm{k}'\rangle_{C=+1}=\langle\bm{k}|V_{C}|\bm{k}'\rangle_{C=-1}=C_{0}^{D\bar{D}^{*}}\left(\begin{array}{cc}
		1 & 1\\
		1 & 1
	\end{array}\right).
\end{equation}
One can see that the model exerts two constrains. First, the diagonal and off-diagonal terms are constrained to be equal, which implies the vanishing interaction in the isovector channel. Meanwhile, the states with odd and even parities have the same interaction. With the relations, the authors predicted the $I^G(J^{PC})=0^{\pm}(1^{+\pm})$ $B\bar{B}^* /B^*\bar{B}$ bound states in the $^3S_1 - ^3D_1$ waves and an even $C$ parity $^3P_0$ states. The conclusions in Ref.~\cite{Nieves:2011zz} strongly rely on the validity of the phenomenological model~\cite{Gamermann:2006nm,Gamermann:2007fi}.

Nieves \etal investigated the heavy quark spin symmetry partners of the $X(3872)$ within the pionless EFT~\cite{Nieves:2012tt}. Apart from the trivial prediction of the $J^{PC}=2^{++}$ state [see Eq.~\eqref{eq:V1ppepsV2pp}], the authors predicted a total six $D^{(*)}\bar{D}^{(*)}$ molecular states with the extra assignment of the $X(3915)$ as the $0^{++}$ $D^*\bar{D}^*$ molecule. The result is very natural, which has been explained in Sec.~\ref{sec:HQSinHHM} according to  Fig.~\ref{fig:4.x_inter_order}. In Ref.~\cite{Nieves:2012tt}, the author also considered the subleading effect, like the HQSS breaking effect, OPE interaction and coupled-channel effect, which brought uncertainties of about $40-50$ MeV to the binding energy of the most bound cases.

In Ref.~\cite{Hidalgo-Duque:2012rqv}, Hidalgo-Duque investigated the partner states of the $X(3872)$ in HQSS and light flavor symmetry. In this work, the coupled-channel effects related to the mass splittings $M_{D^*}-M_D$ and $M_{D_s}-M_D$ are neglected. As shown in Eq.~\eqref{eq:vqq}, there are four independent coupling constants to depict the general $D^{(*)}\bar{D}^{(*)}$ systems in the HQSS and SU(3) flavor symmetry. The four independent coupling constants are determined by treating the $X(3872)$ as the $\DastDbar$ state, $X(3915)$ as the $D^*\bar{D}^*$ state, $Y(4140)$ as the $D^*_s\bar{D}_s^*$ state and the isospin violating decay ratio of $X(3872)$ (the isospin violating decay ratio depends on the binding energy and mixing angle of the neutral and charged components, see Sec.~\ref{sec:X-short}). In the numerical analysis of Ref.~\cite{Hidalgo-Duque:2012rqv}, the isospin partner of the $X(3872)$ is ruled out. Meanwhile, the author predicted the  $D^{(*)}_{(s)}\bar{D}^{(*)}_{(s)}$ full molecular spectrum.

In Ref.~\cite{Meng:2020cbk}, Meng \etal predicted the $[\bar{D}_{s}^{*}D_{s}^{*}]^{0^{++}}$,
$[\bar{D}_{s}^{*}D_{s}/\bar{D}_{s}^{}D_{s}^*]^{1^{+-}}$, and
$[\bar{D}_{s}^{*}D_{s}^{*}]^{1^{+-}}$ bound states as the partners of the $X(3872)$ in the HQSS and SU(3) flavor symmetry, which is the consequence of  the
existence of the $[\bar{D}_{s}D_{s}]^{0^{++}}$ bound state
supported by the lattice QCD calculation~\cite{Prelovsek:2020eiw}
and the observation of $\chi_{c0}(3930)$ by the LHCb Collaboration~\cite{LHCb:2020pxc,LHCb:2020bls}. In the single-channel scheme, the authors obtained the results by assuming $X(3872)$ as the weakly bound $\xthn$ state, which can be related to the $\bar{D}_s^*D_s/D_s^*\bar{D}_s$ state in the SU(3) flavor symmetry. Then, according to Fig.~\ref{fig:4.x_inter_order}, one can infer that the arrow in Fig.~\ref{fig:4.x_inter_order} represents the more attractive interaction if the $[\bar{D}_{s}D_{s}]^{0^{++}}$ is the deeper bound state. In the Ref.~\cite{Meng:2020cbk}, the authors also considered the coupled-channel effect for the $X(3872)$. Within a cutoff-independent framework, it was shown that treating $X(3872)$ as a coupled-channel molecule will not change the results qualitatively, instead will make it become a deeper bound state.

In Ref.~\cite{Baru:2016iwj}, the authors explored the consistence of the strict heavy quark limit and the OPE interaction. The calculation showed that the implication of HQSS, degeneration of $|0_{H}^{-+}\otimes1_{L}^{--},1^{+-}\rangle$, $|1_{H}^{--}\otimes1_{L}^{--},0^{++}\rangle$, $|1_{H}^{--}\otimes1_{L}^{--},1^{++}\rangle$ [$X(3872)$], and $|1_{H}^{--}\otimes1_{L}^{--},2^{++}\rangle$ are still robust considering the OPE interaction when all the partial waves and channels are included. The similar consistences of chiral dynamics and the heavy quark symmetry were also explored in Ref.~\cite{Meng:2018zbl}. In Ref.~\cite{Baru:2016iwj}, the $2^{++}$ channels were investigated with the nonperturbative pions considering the heavy quark symmetry breaking effect. The results showed that the nonperturbative pion approach will lead to significant shifts of the mass and width about $50$ MeV as compared to the perturbative pion approach.

Recently, Xu explored the $S$-wave interaction for the $D\bar{D}^*$ system with $I^G(J^{PC})=0^+(1^{++})$, $0^-({1^{+-}})$, $1^+(1^{+-})$ and $1^-({1^{++}})$ within the $\chi$EFT~\cite{Xu:2021vsi}. The calculation was performed up to NLO, including the contact term, OPE and TPE interactions with the Weinberg scheme. The results indicated the existence of the $0^-(1^{+-})$ and $1^+(1^{+-})$ molecular states in addition to the $X(3872)$ as the $0^+(1^{++})$ molecule.

In Ref.~\cite{Guo:2013xga}, the HDAS was used to investigate the triply heavy partners of $X(3872)$ in contact EFT. With this symmetry, $X(3872)$ as the molecule implies several isoscalar baryonic molecules composed of $\Xi_{QQ'}^{(*)}D^{*}$ and $\Xi_{QQ'}^{(*)}\bar{B}^{*}$.

In Ref.~\cite{Zhang:2020mpi}, the hadronic atom $D^{\pm}D^{*\mp}$ was investigated, which is formed mainly by the Coulomb interaction. Unlike the conventional hadronic atoms~\cite{Holstein:1999nq,Gasser:2007zt}, the strong interaction correction to the Coulomb interaction is treated nonperturbatively due to the pole of the $X(3872)$. In their calculation, the strong interaction in the neutral and charged bases was introduced as
\begin{equation}
	V=C_0 \left(\begin{array}{cc}
		1 & 1\\
		1 & 1
	\end{array}\right),
\end{equation}
where the vanishing isovector interaction is presumed. With this interaction, the correction to the binding energy of the hadronic atom is
\begin{equation}
\Delta E_{n}=\frac{2 \alpha^{3} \mu_{c}^{2}}{n^{3} \sqrt{2 \mu_{c} \Delta}}\left[-1-i+\mathcal{O}\left(\alpha \sqrt{\frac{\mu_{c}}{\Delta}}\right)\right]^{-1},
\end{equation}
where $\mu_c$ is the reduced mass and $n$ is the principal quantum number.  $\Delta$ is the threshold difference of the charged and neutral channels. The correction is independent of the binding energy of the $X(3872)$ and depends on the threshold difference. The authors evaluated the ratio of the production rate of the $X$ atom with respective to the $X(3872)$ in $B$ decays and at hadron collider. The null observing of the $X$ atom will give lower limit of the binding energy of the $X(3872)$.

Canham studied the scatterings of the $D$ and $D^*$ mesons off the $X(3872)$ as a weakly bound state of the $\xthn$ in a contact EFT~\cite{Canham:2009zq}. In Ref.~\cite{Braaten:2010mg}, the scattering of the ultrasoft pion and $X(3872)$ was investigated in XEFT. The breakup cross section of $\pi^+ X(3872)\to D^{*+}\bar{D}^{*0}$ was calculated. In Ref.~\cite{Contessi:2020jqa}, the hadronic systems composed of three and four $X(3872)$ were investigated. A $4X(4^{++})$ octamer and a bound $3X(3^{++})$ were predicted with strict HQSS.

\subsubsection{Long-range dynamics}~\label{sec:X_long_range}

In Ref.~\cite{Voloshin:2003nt}, Voloshin investigated the $X(3872)\to D^0\bar{D}^0\pi^0$ and $X(3872)\to D^0\bar{D}^0\gamma$ decays driven by $D^{*0}\to D^0\pi^0(\gamma)$ and its charge conjugation $\bar{D}^{*0}\to \bar{D}^0\pi^0(\gamma)$. The Feynman diagram of the $X(3872)\to D^0\bar{D}^0\pi^0$ decay corresponds to  Fig.~\ref{fig:4.x_xtoDDpi}(a). The universal wave function in Eq.~\eqref{eq:wavefunction_universal} up to a different normalization was adopted. The decay patterns for different $C$ parities of the $X(3872)$ are distinct due to the different interference of the contributions from the $D^{*0}$ and $\bar{D}^{*0}$ decays. Although the calculation was not performed in an EFT framework, it is equivalent to the results from the LO contact EFT as shown in Sec.~\ref{sec:1chanel_pionless}.

\begin{figure}
	\centering
	\includegraphics[width = 0.9\textwidth]{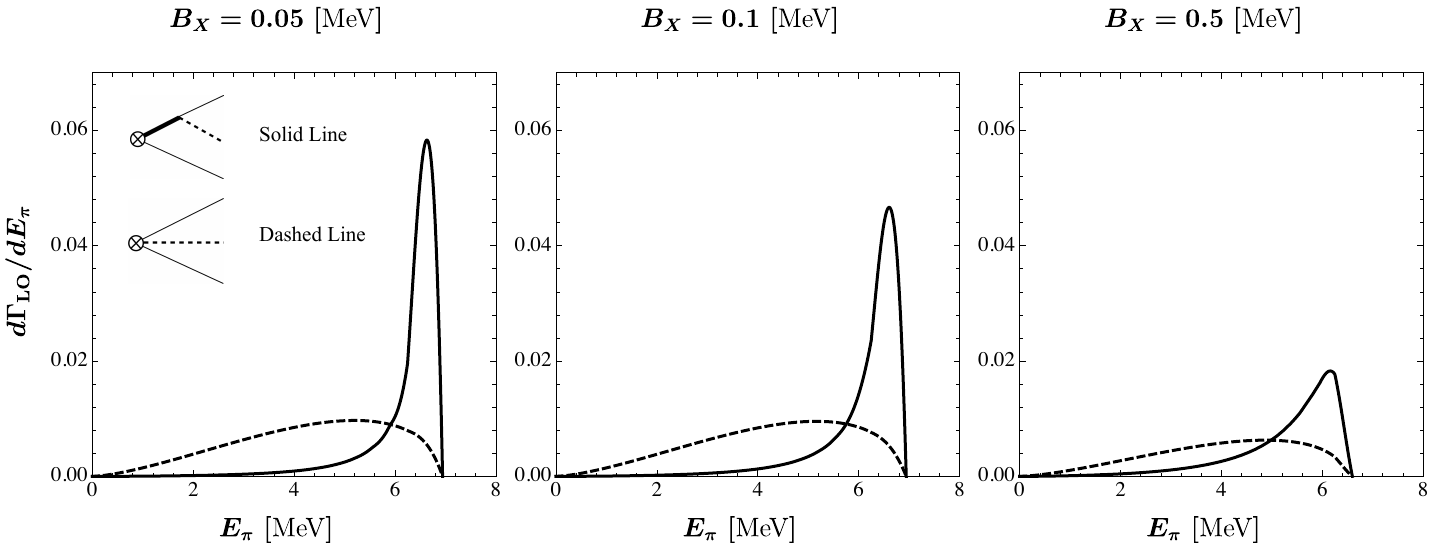}
	\caption{The differential distribution in the pion energy. The solid and dashed curves correspond to the decays driven by $D^{*0}\to D^0\pi^0$ (or its charge conjugation) and contracting the $D^{*0}$ propagator into a contact vertex, respectively.}\label{fig:4.x_pDD}
\end{figure}

In the original work of XEFT~\cite{Fleming:2007rp}, the $X\to D^0\bar{D}^0\pi^0$ was calculated up to NLO.  The contributions of Feynman diagrams (a), (b), (c), (e) and (f) in Fig.~\ref{fig:4.x_xtoDDpi} were investigated. The decay rate for $X\to D^0\bar{D}^0\pi^0$ was given as a function of the binding energy. The results showed that the nonanalytical correction from the perturbative pion exchange is very small, about $1\%$ of the decay rate. The calculation was updated in Ref.~\cite{Dai:2019hrf} by including the $\pi^0 D^0$, $\pi^0\bar{D}^0$ and $D^0\bar{D}^0$ rescattering effect. Apart from the Lagrangians in Eq.~\eqref{eq:XEFT_lag}, two extra terms were included, which were first pointed out in Ref.~\cite{Guo:2017jvc},
\begin{equation}
	{\cal L}^{\text{NLO}}=\frac{C_{\pi}}{2m_{\pi}}(D^{\dagger}\pi^{\dagger}D\pi+\bar{D}^{\dagger}\pi^{\dagger}\bar{D}\pi)+C_{0D}D^{\dagger}\bar{D}^{\dagger}D\bar{D},
\end{equation}
where the first and second terms correspond to the $C_\pi$ and $C_{0D}$ vertices in diagrams (d) and (g) of Fig.~\ref{fig:4.x_xtoDDpi}, respectively. The two new interactions change the uncertainties of the decay rate and the binding energy. In the calculation, the interaction of $D^0\bar{D}^0$ was resummed to include the rescattering effect from the bubble diagram to all orders. The results showed that the rescattering effect of $D^0\bar{D}^0$ only leads to a small modification. The differential distribution in the pion energy, i.e., $d\Gamma/dE_\pi$ was studied, which is sensitive to the binding energy and the correct inclusion of the virtual $D^{*0}/\bar{D}^{*0}$ propagator as shown in Fig.~\ref{fig:4.x_pDD}. The differential decay width becomes very different when one contracts the virtual $D^{*0}/\bar{D}^{*0}$ propagator into a contact vertex.

In Ref.~\cite{Guo:2014hqa}, Guo \etal  investigated the $X(3872)\to D^0\bar{D}^0\pi^0$ in the coupled-channel contact ($\slashed{\pi}$) EFT (see Sec.~\ref{sec:nopi_cc}). The neutral ($\xthn$) and charged ($\xthc$) channels were both considered. The diagrams (a) and (g) in Fig.~\ref{fig:4.x_xtoDDpi} were calculated considering the charged channel. The two coupling constants were fixed by fitting the $X\to J/\psi \rho$ and $X\to J/\psi \omega$ line shapes~\cite{Hanhart:2011tn} and those of $Z_b(10610)$~\cite{Cleven:2011gp}. The cutoff is specified as $\Lambda=0.5$ GeV and $1.0$ GeV. In the calculation, the final state interaction of $D\bar{D}$ was induced by solving the LSEs, which is important numerically.

In Ref.~\cite{Meng:2021kmi}, Meng \etal obtained the decay widths of $X(3872)\to D^0 \bar{D}^0\pi^0$ and $X(3872)\to D^0 \bar{D}^0\gamma $ as a by-product of investigating its isospin violating decays in the coupled-channel $\slashed{\pi}$EFT. The cutoff-dependence was removed in such an EFT. The LECs were determined by the isospin violating decay ratio and the binding energy of the $X(3872)$. The decay widths are presented in Fig.~\ref{fig:4.x_X_decay_lmeng}. The strong and radiative decay widths are about $30$ keV and $10$ keV, respectively, for the binding energy from $-300$ keV to $-50$ keV. The calculation details will be given in Sec.~\ref{sec:X-short}.

\begin{figure}
	\centering
	\includegraphics[width = 0.3\textwidth]{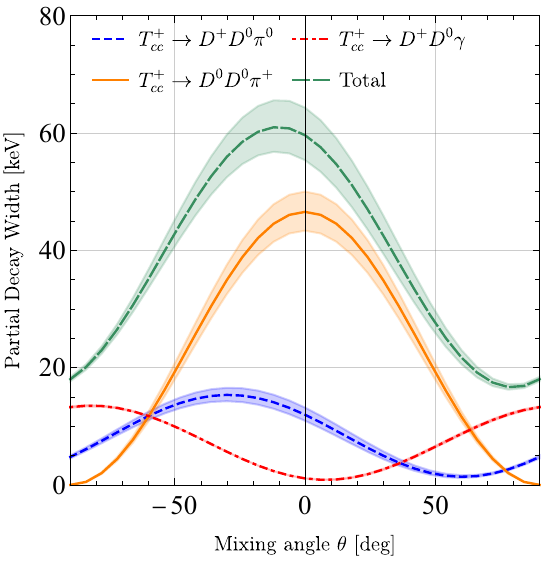}
	\includegraphics[width = 0.3\textwidth]{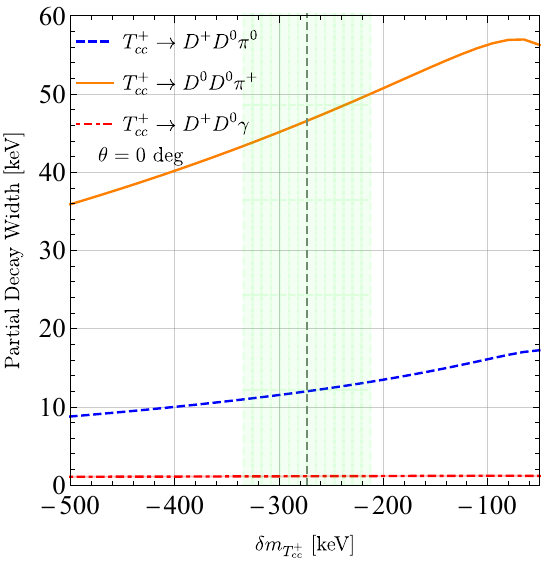}
		\includegraphics[width = 0.3\textwidth]{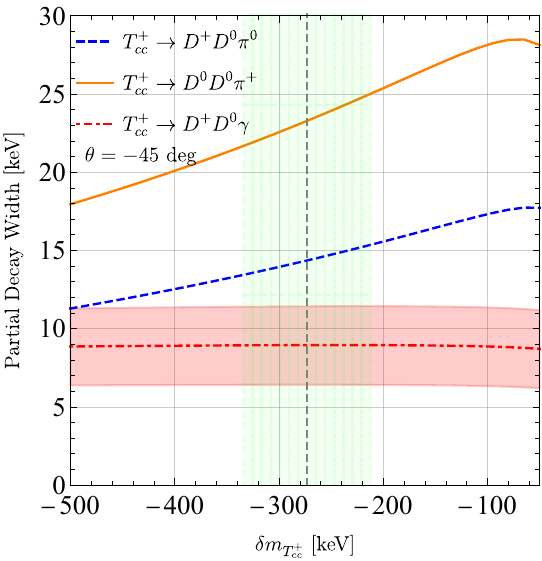}
	\caption{The strong and radiative decay widths of the $T_{cc}^+$ obtained in Ref.~\cite{Meng:2021jnw}. Left panel: the mixing angle dependence of the partial widths, where the colored shadow represents the uncertainties stemming from the $T_{cc}^+$ mass. Middle and right panels: the dependence of the partial widths on the binding energy of the $T_{cc}^+$ in single-channel limit ($\theta=0^\circ$) and isospin limit ($\theta=-45^\circ$), respectively.}\label{fig:4.x_Tcc_lmeng}
\end{figure}

In Ref.~\cite{Meng:2021jnw}, Meng \etal investigated the kinetically allowed $T_{cc}^+\to D^0D^0\pi^+$, $T_{cc}^+\to D^+D^0\pi^0$ and $T_{cc}^+\to D^+D^0\gamma$ decays within the coupled-channel EFT (see Sec.~\ref{sec:nopi_cc}). The $T_{cc}^+$ was treated as the bound state of $D^{*0}D^+/D^{*+}D^0$. In the coupled-channel EFT, the coupling constants of the $T_{cc}^+$ with the two channels were related to the binding energy and mixing angle of two components. This framework includes the possible isospin violating effect and satisfies the renormalization group invariance. The numerical results were given in Fig.~\ref{fig:4.x_Tcc_lmeng}. The results showed that the largest decay mode is the $T_{cc}^+\to D^0D^0\pi^+$, which is just the experimental discovery channel. The total width (strong plus radiative decays) in the single-channel and isospin singlet limits are $59.7_{-4.4}^{+4.6}$ keV and $46.7_{-2.9}^{+2.7}$ keV, respectively, which is much smaller than the width first reported (about $410$ keV) in experiment~\cite{LHCb:2021vvq}. However, the theoretical calculation was supported by the subsequent experimental analysis in a unitarized profile, which gives $\Gamma=47.8\pm 1.9$ keV~\cite{LHCb:2021auc}. In Ref.~\cite{Fleming:2021wmk}, Fleming \etal obtained similar results with the similar two-channel contact EFT but in a relativistic form. Apart from the decay widths, the differential spectra as a function of the invariant mass of the final state $DD$ pair was given. The effect of the possible bound state from the $DD$ system was discussed. In Ref.~\cite{Ling:2021bir}, the authors obtained consistent results with those in Ref.~\cite{Meng:2021jnw} within a single-channel phenomenological approach in the presumed isoscalar channel. 

In Ref.~\cite{Yan:2021wdl}, the subleading contributions to the strong and electromagnetic decays of the $T_{cc}^+$ state were investigated within the framework of EFT, including the contribution of the seagull two-body operator, the $DD$ final state interaction, the isospin violating effect of the molecular wave function and the contribution of the compact tetraquark component. Although it is hard to determine all the parameters in calculation, the authors provided some reasonable estimation of these effects.

Du \etal ~\cite{Du:2021zzh} adopted a coupled-channel EFT including the three-body dynamics (e.g., see Fig.~\ref{fig:4.x_3body_cut}) to fit the observed line shape in the $D^0D^0\pi^+$ channel~\cite{LHCb:2021vvq}. With the analysis, the scattering length and effective range of the $D^*D$ scattering was extracted and the compositeness parameter of the $T_{cc}^+$ was calculated, which is close to the unity indicating the molecular nature of the $T_{cc}^+$. Employing the HQSS, a $D^*D^*$ molecular partner of the $T_{cc}^+$ with $I(J^P)=0(1^+)$ was predicted. {In Ref.~\cite{Albaladejo:2021vln}, the partners of the $T_{cc}^+$ state in the HQSS were investigated.}

\subsubsection{Short-range dynamics}~\label{sec:X-short}

\begin{figure}
	\centering
	\includegraphics[width = 0.9\textwidth]{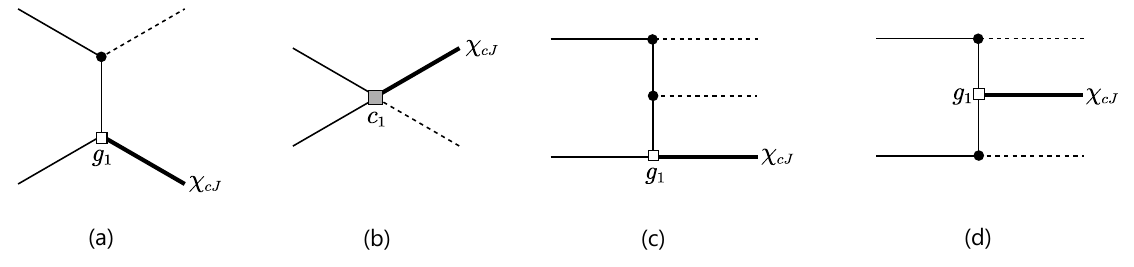}
	\caption{The topological diagrams contributing to $D\bar{D}^*\to \chi_{cJ} \pi$ and $D\bar{D}^* \to \chi_{cJ}\pi \pi$ in HM$\chi$PT~\cite{Fleming:2008yn}. The thin solid line represents $D$ or $\bar{D}^*$. }\label{fig:4.x_XtoChicJ}
\end{figure}

In Ref.~\cite{Fleming:2008yn}, Fleming \etal calculated the $X(3872)\to \chi_{cJ}(\pi^0,\pi\pi)$ processes with the factorization formula. The short-range dynamics is depicted by the local operators [for the $D^0\bar{D}^0\chi_{cJ}(\pi^0,\pi\pi)$ vertices] in XEFT. The coupling constants of the local operators are determined by matching to the HM$\chi$PT as shown in Fig.~\ref{fig:4.x_XtoChicJ}, where $g_1$ and $c_1$ are unknown couplings. The results showed that the two pion transitions are highly suppressed compared with the one pion transition, except the $X\to\chi_{c1}\pi^0\pi^0$, which is  almost at the same order as the $X\to\chi_{c1}\pi^0$. The enhancement of $X\to\chi_{c1}\pi^0\pi^0$ arises from regulating the IR divergence (when the $D$ meson is on-shell) with the $D^{*0}$ width. In Ref.~\cite{Fleming:2011xa}, the author corrected the calculation with the operator product expansion by regulating the IR divergence with the binding energy of the $X(3872)$. From the updated results , the $X\to \chi_{c1}\pi\pi$ process does not receive the large enhancement.

\begin{figure}
	\centering
	\includegraphics[width = 0.5\textwidth]{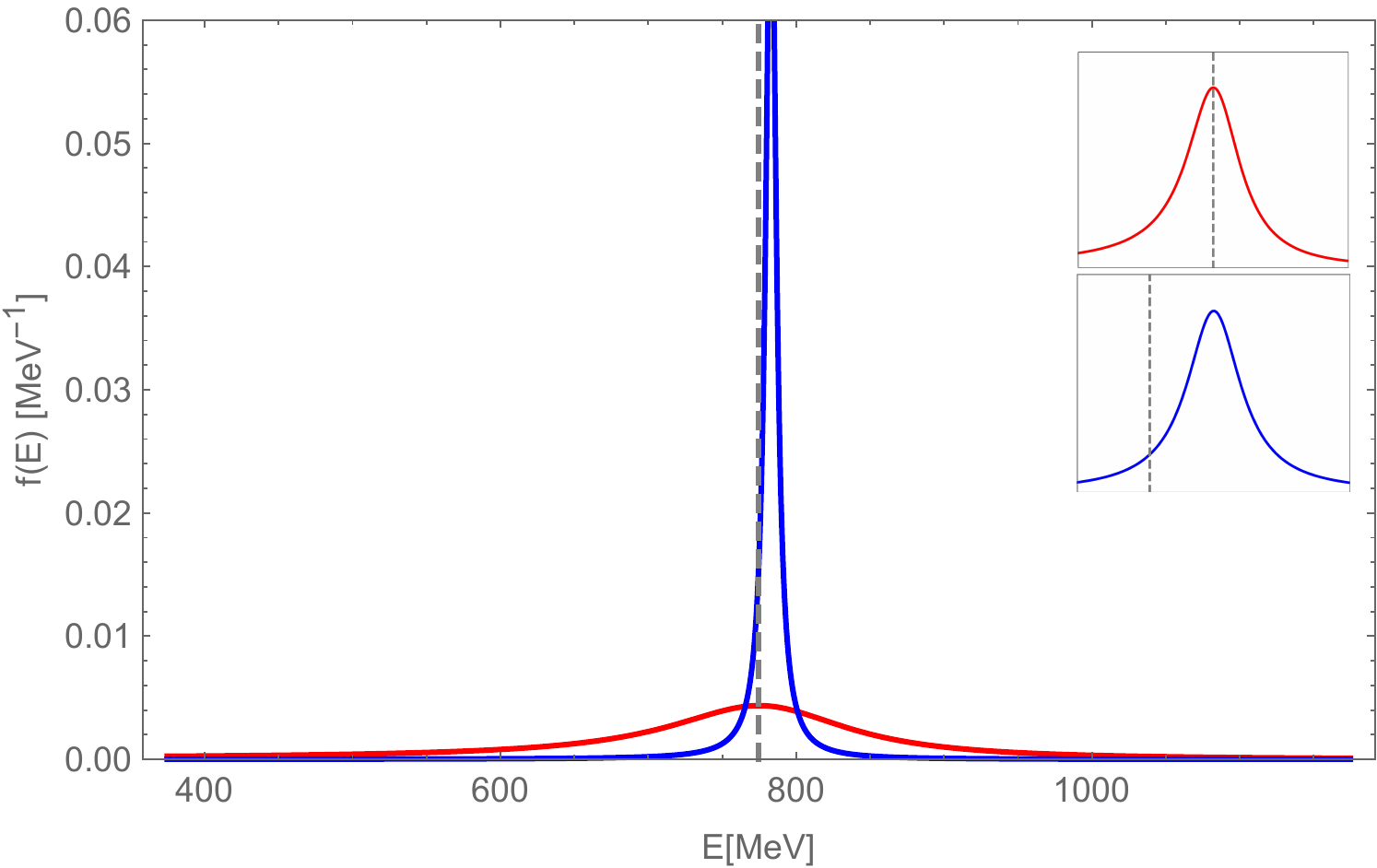}
	\caption{Schematic illustration of the kinematic mechanism of the large isospin violating decay of the $X(3872)$. The red and blue curves are the Breit-Wigner distribution of $\rho$ and $\omega$ mesons.  The dashed vertical line corresponds to the $M_{X(3872)}-M_{J/\psi}$. The two small subfigures are the rescaled plots. }\label{fig:4.x_X_isospin_violating_phss}
\end{figure}

We move on to the $X\to J/\psi h$ decays ($h$ denotes the light hadrons or photon). An interesting issue is the isospin violating decay of the $X(3872)$ as reviewed in Sec.~\ref{sec1.3.1}. The isospin conserving decay mode driven by $X\to J/\psi\omega$ is strongly suppressed kinetically as shown in Fig.~\ref{fig:4.x_X_isospin_violating_phss}.
Only a small portion of the Breit-Winger distribution of the $\omega$ meson is in the kinetically allowed region. In contrast, about half of the Breit-Winger distribution of the $\rho$ meson is allowed by kinetics. In addition to the kinetic restriction, there are also dynamical reasons for the large isospin violating decays of $X(3872)$.

\begin{figure}
	\centering
	\includegraphics[width = 0.3\textwidth]{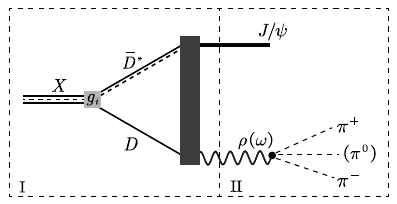}
	\caption{Isospin violating decays of $X(3872)$. The $X\to J/\psi \pi^+\pi^- (\pi^0) $ decays are divided into two pieces, the $X\to J/\psi V$ and $V\to \pi^+\pi^- (\pi^0)$ with $V\equiv\rho$ and $\omega$.  }\label{fig:4.x_X_isospin_decay}
\end{figure}

The $X\to J/\psi h$ was investigated by Braaten \etal with the factorization formula~\cite{Braaten:2005ai}, where $h$ stands for $\pi^+\pi^-\pi^0$, $\pi^+\pi^-$, $\pi^0\gamma$ or $\gamma$. It was assumed that these decay modes are driven by the $X\to J/\psi\rho$ and $X\to J/\psi \omega$ (see Fig.~\ref{fig:4.x_X_isospin_decay}). The amplitudes $V\to h$ ($V\equiv\rho$ or $\omega$) were determined by the effective Lagrangian approaches with the partial decay widths of the $\rho$ and $\omega$ as inputs. The isospin violating decay ratio was determined as
\begin{equation}
\frac{\Gamma\left[X \rightarrow J / \psi \pi^{+} \pi^{-} \pi^{0}\right]}{\Gamma\left[X \rightarrow J / \psi \pi^{+} \pi^{-}\right]}=0.087 \frac{\left|G_{X \psi \omega}\right|^{2}}{\left|G_{X \psi \rho}\right|^{2}}.
\end{equation}
The part-II ($V\to h$) of Fig.~\ref{fig:4.x_X_isospin_decay} contributes a factor about $0.087$, which incorporates the kinetic effect. In the part-I ($X\to J/\psi V$) of  Fig.~\ref{fig:4.x_X_isospin_decay}, the author adopted the factorization formula. The long-range part was obtained by the universality, which is equivalent to the single-channel ($\xthn$) $\slashed{\pi}$EFT, while the short-range part was fixed by a model proposed by Swanson~\cite{Swanson:2003tb,Swanson:2004pp}. The partial width of $X\to J/\psi \pi^+\pi^-\pi^0$ was predicted as a function of the binding energy and total width of the $X(3872)$.

\begin{figure}
	\centering
	\includegraphics[width = 0.3\textwidth]{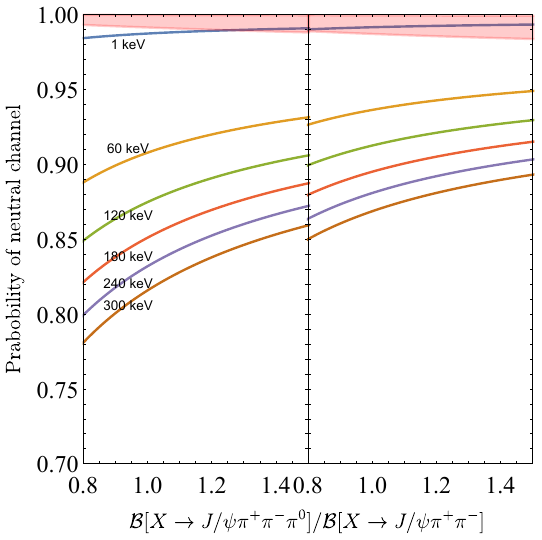}~~~~~~~~~~~~~~~~~~~
	\includegraphics[width = 0.3\textwidth]{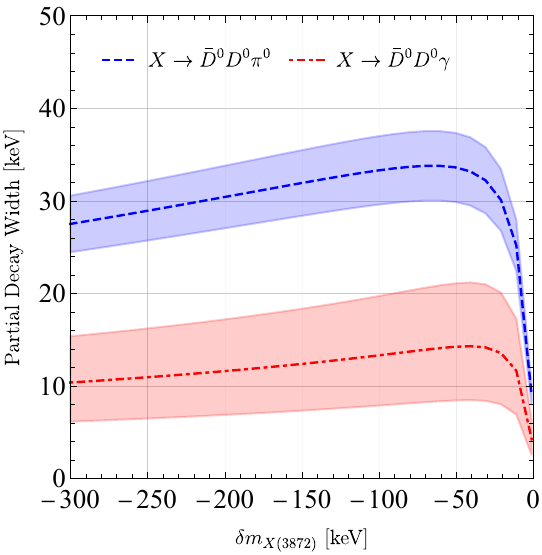}	
		\caption{Numerical results in Ref.~\cite{Meng:2021kmi}. Left panel: probability of the neutral channel in the wave function of the $X(3872)$. The binding energies are labeled near the related curves. Right panel: the partial widths of the $X(3872)$ with the mixing angle of neutral and charged channels extracted from the left subfigure. }\label{fig:4.x_X_decay_lmeng}
\end{figure}

In Ref.~\cite{Meng:2021kmi}, Meng \etal also investigated the isospin violating decays, in which the ratio is divided into two parts.
\begin{equation}
		R\equiv \frac{{\cal B}^{I=0}(X\to J/\psi\pi^{+}\pi^{-}\pi^{0})}{{\cal B}^{I=1}(X\to J/\psi\pi^{+}\pi^{-})}=R_{1}\times R_{2},
\end{equation} 
where the $R_2$ includes the kinetic mechanism, and it was estimated in two approaches from Refs.~\cite{Gamermann:2009uq,Braaten:2005ai} (which give $R_2=0.147$ and $0.087$, respectively). The $R_1$ was handled in the factorization formula. The long-range part is investigated in the coupled-channel EFT (see Sec.~\ref{sec:nopi_cc}), in which the isospin violating effect in the wave function was considered. In the short-range part, the isospin symmetry is exact and the unknown matrix element cancels out in the ratio while the isospin Clebsch-Gordon coefficients survive. The $R_1$ reads
\begin{equation}
	  R_{1}=\left(\frac{{g}_{1}{F}_{1}-{g}_{2}{F}_{2}}{{g}_{1}{F}_{1}+{g}_{2}{F}_{2}}\right)^{2}\approx\left(\frac{{g}_{1}-{g}_{2}}{{g}_{1}+{g}_{2}}\right)^{2}
	=\left(\frac{{c}_{1}{\gamma}_{1}^{1/2}-{c}_{2}{\gamma}_{2}^{1/2}}{{c}_{1}{\gamma}^{1/2}+{c}_{2}{\gamma}^{1/2}}\right)^{2},~\label{eq:ratio_x3872}
\end{equation} 
where the $F_i$, $g_i$ and $\gamma_i$ are defined in Sec.~\ref{sec:nopi_cc}. $c_1$ and $c_2$ are the coefficients of the neutral and charged channels in the wave function of the $X(3872)$, respectively. The proportion of the neutral channel was determined using the experimental $R$ as shown in the left panel of Fig.~\ref{fig:4.x_X_decay_lmeng}. With both values of $R_2$, the proportion $R$ is over $80\%$. With the proportion as input, the partial widths of the $X\to \bar{D}^0D^0\pi^0$ and $X\to \bar{D}^0D^0\gamma$ were calculated (see the right panel of Fig.~\ref{fig:4.x_X_decay_lmeng}). {To derive Eq.~\eqref{eq:ratio_x3872}, the difference of $F_1$ and $ F_2$ is neglected, which is different from  Ref.~\cite{Gamermann:2009fv}. In Ref.~\cite{Gamermann:2009fv}, the authors presumed $g_1 = g_2$ and kept the difference of $F_1$ and $F_2$ to make $R_1$ non-vanishing. The isospin violation effect of the conventional hadrons can stem from the not fully offset loop integrals due to the displaced charged and neutral thresholds. However, such effects are usually tiny. For the case of the $X(3872)$, the isospin violation effects from the difference of $F_1$ and $F_2$ depend on the cutoff and are too small to explain the large isospin violation effect with a reasonable cutoff. Thus, the authors made a different assumption that $g_1$ and $g_2$ are different while $F_1$ and $F_2$ are the same as in Ref.~\cite{Meng:2021kmi}. In fact, the relation $g_1=g_2$ is by no means guaranteed. In Ref.~\cite{Gamermann:2009fv}, the validation of $g_1 = g_2$ is based on the constraint $v_{11} = v_{22} = v_{12}$ in Eq.~\eqref{eq:v_couplechannel}, namely, the vanishing interaction in the isovector channel. However, the LECs $v_{ij}$ are cutoff-dependent, which cannot be equal to each other with the varying cutoff [see Eq.~\eqref{eq:inverse-bij}]. On the other hand, the couplings $g_i$
are related to the mixing angle of the two channels, binding energy and mass difference of the two thresholds [see Eq.~\eqref{eq:coupling_two_channel}], which are all physical observables. Therefore,
it is reasonable to infer the more important effect resulting from different coupling constants. One can find comparisons of these two works in details in Ref.~\cite{Meng:2021kmi}.} 

Mehen \etal  calculated the radiative decays of $X(3872)\to \psi(2S)\gamma$ and $\psi(4040)\to X(3872)\gamma$ with the factorization formula by matching the short-range dynamics in HH$\chi$PT to the operators of XEFT~\cite{Mehen:2011ds}. In these two decays, the photon energy is about $181$ MeV and $165$ MeV. In the long-range dynamics, the $X(3872)$ is regarded as a bound state of the $\xthn$. In the $X\to J/\psi \gamma$ decay, the energy of the photon is about $700$ MeV, which lies beyond the working range of HH$\chi$PT. Thus, in this framework, one cannot obtain the branching fraction
\begin{equation}
R=\Gamma[X\to \psi(2S)\gamma]/\Gamma[X\to J/\psi \gamma],\label{eq:Rxtopsigamma}
\end{equation}
which has been measured in experiments~\cite{BaBar:2008flx,Belle:2011wdj,LHCb:2014jvf}. However, it was shown that the polarization of $\psi(2S)$ in $X\to \psi(2S)\gamma$ and the angular distribution of $X$ in $\psi(4040)$ decays can be used to determine the quantum number of $X(3872)$. Similar framework was also used to study the $\psi(4160)\to X(3872)\gamma $~\cite{Margaryan:2013tta}, where the correlation between the polarization of $\psi(4160)$ and angular distribution of the final states can be utilized to explore the structure of the $X(3872)$.

\begin{figure}
	\centering
\includegraphics[width = 0.99\textwidth]{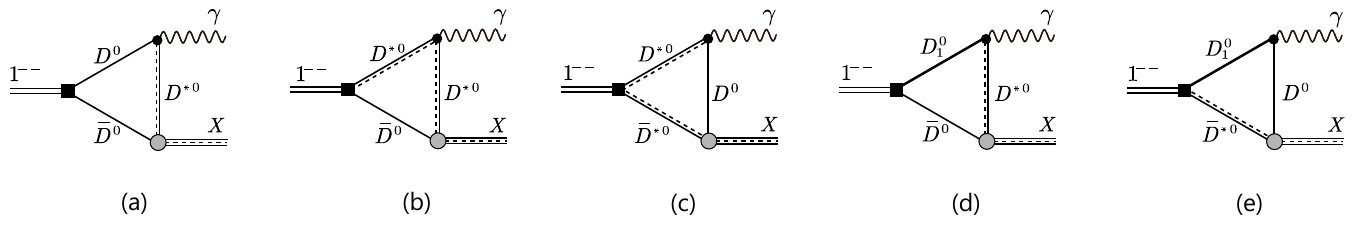}
\caption{Feynman diagrams for the production of the $X(3872)$ in the radiative transitions of the vector charmonium(-like) states, including $\psi(4040)$, $\psi(4160)$, $\psi(4415)$ and the $Y(4260)$ in Ref.~\cite{Guo:2013zbw}. }\label{fig:4.x_psitoXgamma}
\end{figure}

Guo \etal investigated the production of the $X(3872)$ in the radiative transitions of the vector charmonium(-like) states---including the $\psi(4040)$, $\psi(4160)$, $\psi(4415)$ and $Y(4260)$~\cite{Guo:2013zbw} within NREFT according to the powers of the velocities in Eq.~\eqref{eq:guoNREFT_pw}. The Feynman diagrams are presented in Fig.~\ref{fig:4.x_psitoXgamma}. With the naturalness assumption of the vertices, the production of the $Y(4260) \to X(3872)\gamma$ is enhanced as compared to those of the other channels if the $Y(4260)$ is treated as the $D\bar{D}_1$ molecule. A similar approach was adopted to explore the production of the $X_b$---the HQFS partner of the $X(3872)$ from the decays of $\Upsilon(5S,6S)\to X_b \gamma$~\cite{Wu:2016dws}. In Ref.~\cite{Guo:2014taa}, the same NREFT was used to explore the radiative decays of the $X(3872)$, i.e., $X\to J/\psi \gamma$ and $X\to \psi(2S) \gamma$. 
With the naturalness assumption for
the coupling constants, their results indicate that the experimental ratio in Eq.~\eqref{eq:Rxtopsigamma} does not conflict with the molecule-dominant structure of the $X(3872)$.

In Refs.~\cite{Braaten:2004fk,Braaten:2004ai,Braaten:2004jg,Braaten:2007dw,Braaten:2007ft,Stapleton:2009ey,Braaten:2013poa}, the productions of the $X(3872)$ and the related line shapes were investigated with the factorization formula, e.g., see Fig.~\ref{fig:4.x_factation}. In fact, a heated topic about the $X(3872)$ is its substantial prompt production~\cite{CDF:2003cab,CMS:2013fpt,ATLAS:2016kwu,D0:2020nce,LHCb:2020sey,LHCb:2021ten}, which was often used as an evidence against the molecular interpretation of the $X(3872)$~\cite{Bignamini:2009sk}. In the hadron collider, the $X(3872)$ can be produced from either bottom hadron decays or QCD mechanism. If $X$ is produced by the QCD mechanism, its production vertices will be very close to the collision point, which is refereed as the prompt production. On the contrast, the production vertex from bottom hadron will be displace from the collision point (non-prompt production). Intuitively, the prompt production is the high energy process, in which only the almost relatively static two hadrons can form into loosely bound states. Thus, one would expect the suppressed prompt production for hadronic molecules. In general, the inclusive production of the $X(3872)$ can be formulated as follows (\cvII in Table~\ref{tab:two-convention}),
\begin{eqnarray}
	\begin{aligned}d\sigma[X(3872)]= & \frac{1}{\text{flux}}\sum_{y}\int d\Phi_{\left(D^{*}\bar{D}\right)+y}\left|\int\frac{d^{3}k}{(2\pi)^{3}}\psi_{X}(\bm{k})\frac{\mathcal{A}_{D^{\ast 0}\bar{D}^{0}+y}(\boldsymbol{k})+\mathcal{A}_{D^{0}\bar{D}^{*0}+y}(\boldsymbol{k})}{\sqrt{2}}\right|^{2}\frac{1}{2\mu_{D^{*}D}}\\
		= & \frac{1}{\text{flux}}\sum_{y}\int d\Phi_{\left(D^{*}\bar{D}\right)+y}\left|\int\frac{d^{3}k}{(2\pi)^{3}}\psi_{X}(\bm{k})\mathcal{A}_{D^{\ast 0}\bar{D}^{0}+y}(\boldsymbol{k})\right|^{2}\frac{1}{2\mu_{D^{*}D}}
	\end{aligned},	
\end{eqnarray}
where $y$ represents all the possible final states and $\Phi_{\left(D^{*}\bar{D}\right)+y}$ is the phase space for the composite $(D^*\bar{D})$ and $y$. The validity of the second equality arises from the cancelling-out interference effect. In Ref.~\cite{Bignamini:2009sk}, the authors adopted the Schwarz inequality to obtain that
\begin{equation}
\begin{aligned}d\sigma[X(3872)]\leq & \frac{1}{\text{flux}}\sum_{y}\int d\Phi_{\left(D^{*}\bar{D}\right)+y}\int_{|\bm{k}|<k_{max}}\frac{d^{3}k}{(2\pi)^{3}}\left|\psi_{X}(\bm{k})\right|^{2}\times\int_{|\bm{k}|<k_{max}}\frac{d^{3}k}{(2\pi)^{3}}\left|\mathcal{A}_{D^{*0}\bar{D}^{0}+y}(\boldsymbol{k})\right|^{2}\frac{1}{2\mu_{D^{*}D}}\end{aligned}.
\end{equation}
One finally obtains the following inequality due to the normalized wave function $\psi_X(\bm{k})$,
\begin{equation}
	\sigma[X(3872)]<\sigma[D^{*0}\bar{D}^{0}(|\bm{k}|<k_{max})].
\end{equation}
With the Monte Carlo event generators like Herwig and Pythia, the authors obtained $\sigma[D^{*0}\bar{D}^{0}(|\bm{k}|<k_{max})]$ truncated at $k_{max}$ and then estimated the maximum production cross section of the $X(3872)$. The theoretical result is about two orders smaller than the experimental ones. Therefore, the authors concluded that the results refute the molecular nature of the $X(3872)$. However, one subtle issue in the above calculation is the choice of $k_{max}$. In Ref.~\cite{Bignamini:2009sk}, the authors argued that the momentum scale should be set by the binding momentum of the $X(3872)$ and took $k_{max}=35$ MeV. However, the setting is challenged by Albaladejo~\etal~\cite{Albaladejo:2017blx}, Artoisenet~\etal~\cite{Artoisenet:2009wk} and Braaten~\etal~\cite{Braaten:2018eov}.

\begin{figure}
	\centering
	\includegraphics[width = 0.4\textwidth]{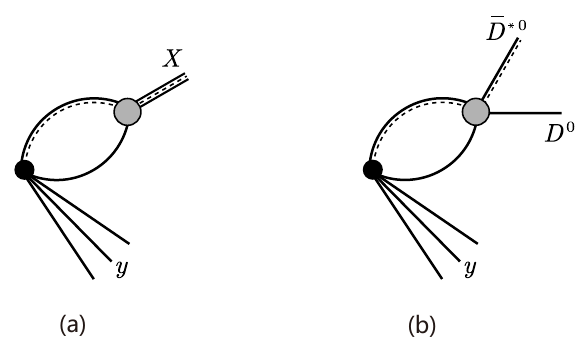}
	\caption{The Feynman diagrams for the inclusive productions of the $X(3872)$ and $\xthn$. }\label{fig:4.x_prompt_prdct}
\end{figure}

In Ref.~\cite{Albaladejo:2017blx}, the authors argued that for the deuteron the $k_{max}\sim 300 ~\text{MeV}\sim 2 m_\pi$ could be a good approximation of the effect of its wave function and the same $k_{max}$ was expected for $X(3872)$. Meanwhile, the authors calculated the inclusive production of the $X(3872)$ by simulating the short-range dynamics with Monte Carlo event generators and combining the long-range interaction in an EFT with a Gaussian regulator cutoff $\Lambda$. The $\Lambda$ can  roughly amount to $2\sqrt{2/\pi}k_{max}$ in the sharp regulator. With a cutoff corresponding to $k_{max}\in [300,600]$ MeV, the cross section is consistent with the experimental results. In Ref.~\cite{Esposito:2017qef}, the authors defended their own conclusions in Ref.~\cite{Bignamini:2009sk} and argued that the $k_{max}$ should be determined independently without surmising the explicit form of the wave function.

In Ref.~\cite{Artoisenet:2009wk}, the authors challenged Ref.~\cite{Bignamini:2009sk} and pointed out that the upper bound should be at the scale of $m_\pi$ due to the rescattering effect of $D^*\bar{D}$, which leads to the consistent result with the experimental measurement. In Ref.~\cite{Braaten:2018eov}, the authors established the relation of the production of the $X(3872)$ and $D\bar{D}^*$ as shown in Fig.~\ref{fig:4.x_prompt_prdct} with the factorization formula. We show the derivation in Ref.~\cite{Braaten:2018eov} with the notation in Sec.~\ref{sec:1chanel_pionless}.  For the long-range dynamics, one has
\begin{eqnarray}
{\cal M}(X)&\sim&\int d^{3}q\psi_{X}(q)\sim\int d^{3}q\frac{g}{E_{X}-\frac{q^{2}}{2\mu}+i\epsilon},\nonumber\\
{\cal M}\left[\{D^0\bar{D}^{*0}(E)\}_{+}\right]&\sim&\int d^{3}q\frac{1}{E-\frac{q^{2}}{2\mu}+i\epsilon}T(E),
\end{eqnarray}
where $g\sim \sqrt{\gamma_X}$ and $T\sim (i\sqrt{2\mu E}+\gamma_X)^{-1}$ are given in Eqs.~\eqref{eq:t_matrix_x_pole} and~\eqref{eq:4.1_tMtx_biding}, respectively. The integrals are linear divergent as shown in Eq.~\eqref{eq:4.1F}, thus one can eliminate the cutoff $\Lambda$ by calculating the ratio of $\mathcal{M}(X)$ and ${\cal M}[\{D\bar{D}^{0*}(E)\}_{+}]$. One can set up the following relation,
\begin{equation}
	{\cal M}\left[\{D^0\bar{D}^{*0}(E)\}_{+}\right]\sim\frac{{\cal M}(X)}{g}T(E)\sim\frac{{\cal M}(X)/\sqrt{\gamma_{X}}}{i\sqrt{2\mu E}+\gamma_{X}}.
\end{equation}
The explicit result for the cross section was given in Ref.~\cite{Braaten:2018eov}, which reads
\begin{equation}
d \sigma\left[D^{0} \bar{D}^{\ast0}\right]=d \sigma[X(3872)] \frac{\pi / \gamma_{X}}{\gamma_{X}^{2}+k^{2}} \frac{d^{3} k}{(2 \pi)^{3}}.~\label{eq:4.x_cross_section}
\end{equation}
In the derivation of Eq.~\eqref{eq:4.x_cross_section}, an implicit constraint is the cutoff $\Lambda$ to regulate the linear divergence, which is at the scale of $m_\pi$. From Eq.~\eqref{eq:4.x_cross_section}, the authors obtained that $\sigma[X(3872)]=\sigma[D^{0}\bar{D}^{\ast0}(k<7.73\gamma_X)]$. Thus, the upper bound is $k_{max}=7.73\gamma_X$, in which the theoretical cross section for the prompt production of $X$ is consistent with the experimental measurements.

\begin{figure}
	\centering
	\includegraphics[width = 0.45\textwidth]{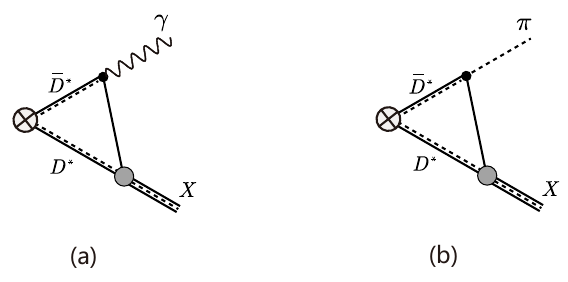}
	\caption{Two production processes for the $X(3872)$ with the triangle singularities. }\label{fig:4.x_TS}
\end{figure}

Guo pointed out that the short-distance creation of the $S$-wave $D^{0}\bar{D}^{\ast0}$ pair [the Feynman diagram in Fig.~\ref{fig:4.x_TS}(a)] can produce a narrow peak in the $X(3872) \gamma$ invariant mass spectrum due to the triangle singularity~\cite{Guo:2019qcn}. Another similar narrow peak in the $X(3872)\pi$ from triangle singularity was pointed out by Braaten \etal~\cite{Braaten:2019yua,Braaten:2019sxh} due to the short-range creation of the $D^*\bar{D}^*$ pair as shown in Fig.~\ref{fig:4.x_TS}(b). Recently, the similar triangle singularity was investigated for the production of the $T_{cc}^+$~\cite{Braaten:2022elw}. We refer to the review~\cite{Guo:2019twa} for details of triangle singularity. In order to investigate the triangle singularity, one has to adopt the non-factorization formula~\cite{Guo:2019qcn,Braaten:2019gfj,Braaten:2019yua,Braaten:2019sxh,Braaten:2019gwc}. In Refs.~\cite{Braaten:2019yua,Braaten:2019sxh}, the productions of the $X(3872)$ accompanied by a soft pion at hadron colliders and  in $B$ meson decays were investigated, where the rescattering of $D^{*}\bar{D}^{*}\to X \pi$ is embedded in XEFT (which is similar to Ref.~\cite{Braaten:2010mg}). The production of the $T_{cc}^+$ in heavy-ion collisions and in hadron colliders were investigated in Refs.~\cite{Abreu:2022lfy,Jin:2021cxj}, respectively.

\subsection{Charged charmoniumlike and bottomoniumlike states without and with the strangeness}\label{sec:sec5.6}

In this section, we review the applications of $\slashed{\pi}$EFT, $\chi$EFT and other related calculations in heavy hadron systems. The $Z_c(3900)$, $Z_c(4020)$, $Z_{cs}(3985)$, $Z_b(10610)$, $Z_b(10650)$ and so on are very good candidates of hadronic molecules due to their proximity to the corresponding thresholds. There are extensive studies and intensive debates on the inner configurations of these states, see Sec.~\ref{sec1.3.2} and some reviews~\cite{Chen:2016qju,Guo:2017jvc,Liu:2019zoy,Lebed:2016hpi,Esposito:2016noz,Brambilla:2019esw,Hosaka:2016pey,Ali:2017jda,Olsen:2014qna,Olsen:2017bmm}. In the following we review the progress on how the EFTs help us understand the nature of these states (see an analogical analysis of nuclear forces and their counterparts in heavy hadron systems~\cite{PavonValderrama:2019ixb}).

\subsubsection{$Z_c$, $Z_b$ and partners}

{In the BW fits which ignored the nearby thresholds}, the $Z_c(3900)$, $Z_c(4020)$, $Z_b(10610)$ and $Z_b(10650)$ (in what follows we will denote them as $Z_c$, $Z_c^\prime$, $Z_b$, and $Z_b^\prime$, respectively) lie few MeVs above the $D\bar{D}^\ast$, $D^\ast\bar{D}^\ast$, $B\bar{B}^\ast$ and $B^\ast\bar{B}^\ast$ thresholds, respectively.  In addition to the similarities of their decay patterns, the mass differences of $(Z_c,Z_c^\prime)$ and $(Z_b,Z_b^\prime)$ almost equal to those of $(D,D^\ast)$ and $(B,B^\ast)$, respectively. Thus, they are suggested to be the molecular twin partners in HQSS and HQFS. It will be very instructive to classify these states into one group and study their behaviors in a uniform framework.

With the EFT formalism, there are two directions toward understanding the nature of these states. The first one is to calculate the mass spectrum with $\slashed{\pi}$EFT, in which the LO LECs are related with HQS and they are fixed with the help of the assumption that 
$X(3872)$ and $Z_b$ are the $D\bar{D}^{\ast}$ and $B\bar{B}^{\ast}$ bound state respectively 
~\cite{Guo:2013sya}. The second one resorts to fitting the invariant mass spectrum from which the particle is observed~\cite{Albaladejo:2015lob,Mehen:2013mva,Hanhart:2015cua,Guo:2016bjq,Aceti:2014uea,He:2015mja,Wang:2018jlv,Wang:2020dko}. In this case, the $D^{\ast}\bar{D}^{(\ast)}/B^\ast\bar{B}^{(\ast)}$ rescatterings are treated perturbatively or nonperturbatively. For the former case, one deems that the $Z_Q^{(\prime)}$ ($Q\equiv c,b$) states are generated from the kinematic effect~\cite{Bugg:2011jr,Chen:2011pv,Chen:2013coa,Swanson:2014tra}. But this is criticized in Ref.~\cite{Guo:2014iya}, in which Guo \etal built a solvable model to fit the elastic ($D\bar{D}^\ast$) and inelastic ($J/\psi\pi$) invariant mass distributions of the $Z_c$. In order to fit the experimental line shapes, the interaction strength of $D\bar{D}^\ast$ needs to be adjusted to large values, which in turn negates the perturbative assumptions. Therefore, the pronounced near-threshold narrow peaks in experiments cannot be ascribed to the pure kinematic effect. The strong interacting $D\bar{D}^\ast$ should be resummed in an infinite series of loops, which generates pole(s) in the $S$-matrix.

There are several lattice QCD studies on the $Z_c$ states. The corresponding results disfavored the $Z_c$ state as a conventional resonance. The first kind is based on the finite volume energy levels. For example, Ref.~\cite{Prelovsek:2013xba} considered the $D\bar{D}^{\ast}$ and $J/\psi\pi$ channels with $m_\pi=266$ MeV, but did not find the signal of the $Z_c$ except the non-interacting two meson energy levels (see also~\cite{Prelovsek:2014swa,Lee:2014uta} for similar results). Ref.~\cite{Chen:2014afa} simulated the single-channel---$(D\bar{D}^\ast)^\pm$ scattering with $m_\pi=485,420,300$ MeV, where the authors found the $D\bar{D}^\ast$ interaction with three different $m_\pi$ is weakly repulsive. In Refs.~\cite{CLQCD:2019npr} and~\cite{Liu:2019gmh}, the coupled-channel L\"uscher formula together with the Ross-Shaw theory is used to study the near-threshold scattering of the $D\bar{D}^{\ast}$. The results indicated that neither the threshold effect interpretation nor the resonance interpretation is favored. HAL QCD Collaboration considered the three channel couplings among $J/\psi\pi$, $D\bar{D}^\ast$, and $\eta_c\rho$  with $m_\pi=410-700$ MeV~\cite{HALQCD:2016ofq,Ikeda:2017mee}. Their simulations revealed that the coupled-channel potentials are dominated by the off-diagonal terms, i.e., the couplings of $J/\psi\pi$-$D\bar{D}^\ast$ as well as $\eta_c\rho$-$D\bar{D}^\ast$, which makes the $Z_c$ look more like a cusp effect but not a conventional resonance state.

Unlike the lattice QCD results, the combinations of EFT and experimental data do support the $Z_Q^{(\prime)}$ to be the virtual states or resonances generated from the nonrelativistic $D^{\ast}\bar{D}^{(\ast)}/B^\ast\bar{B}^{(\ast)}$ interactions. 
From Eq.~\eqref{eq:V1pmepsV2pm}, one obtains that
\begin{eqnarray}\label{eq:hqfsrelation}
V_{1^{+-}}^\alpha(B\bar{B}^\ast)=V_{1^{+-}}^\alpha(B^\ast\bar{B}^\ast)\overset{\mathrm{HQFS}}{=}V_{1^{+-}}^\alpha(D\bar{D}^\ast)=V_{1^{+-}}^\alpha(D^\ast\bar{D}^\ast),
\end{eqnarray}
where the HQSS and HQFS are employed to relate the potentials of $[D^{\ast}\bar{D}^{(\ast)},B^\ast\bar{B}^{(\ast)}]_{1^{+-}}$.  For the single channel case, there is only one LEC if we define $C_{\alpha z}=(\mathcal{C}_0^\alpha+\mathcal{C}_1^\alpha)/2$. Assuming the $Z_b$ as a $[B\bar{B}^\ast]_{1^{+-}}$ bound state, Ref.~\cite{Guo:2013sya} obtained the virtual state solution for the $Z_c^{(\prime)}$. {However, one should be cautious about the HQFS relation in Eq.~\eqref{eq:hqfsrelation}. For example, in Ref.~\cite{AlFiky:2005jd}, a general dimensional analysis shows that the LO contact LECs are scaled as $1/M$, where $M$ is proportional to the heavy meson mass. In Ref.~\cite{Baru:2018qkb}, it is argued that the renormalizibility requires that the LO LECs introduced in Sec.~\ref{sec:HQSinHHM} decreases at least as $1/M$ in the limit $M\to\infty$. Thus, it is even claimed that there cannot be a common EFT for different heavy quark masses. A coercive using of the HQFS to relate the double-charm and -bottom sectors may lead to predictions with uncontrolled uncertainties.}

Albaladejo {\it et al} analyzed the pole distributions of the $Z_c$ via fitting the $D\bar{D}^\ast$ and $J/\psi\pi$ invariant mass spectra~\cite{Albaladejo:2015lob}, in which they modeled a coupled-channel---$J/\psi\pi$ and $D\bar{D}^\ast$ effective potential in the framework of $\slashed{\pi}$EFT,
\begin{eqnarray}\label{eq:NLOcontactFormCoupledchannel}
V_{ij}=\left[\begin{array}{cc}
0& \tilde{\mathcal{C}}\\
\tilde{\mathcal{C}} & \mathcal{C}+\mathcal{C}^\prime \bm p^2
\end{array}\right],
\end{eqnarray}
where the interaction of $J/\psi\pi$ is set to be zero due to their weak coupling~\cite{Detmold:2012pi,Liu:2012dv}. The $D\bar{D}^\ast$-$J/\psi\pi$ coupling is momentum independent assuming it is dominated by the $S$-wave interaction, while the $D\bar{D}^\ast$ interaction is kept to the NLO. In Ref.~\cite{Albaladejo:2015lob}, different explanations for the $Z_c$ are given when the $\mathcal{C}^\prime \bm p^2$ term is switched on and off. When $\mathcal{C}^\prime\neq0$, the $Z_c$ state is a resonance that lies above the $D\bar{D}^\ast$ threshold. If $\mathcal{C}^\prime=0$, the $Z_c$ becomes a virtual state that lies below the $D\bar{D}^\ast$ threshold (with a small width from the $J/\psi\pi$ decay). For the case of the virtual state, the line shape of the inelastic channel---$J/\psi\pi$ peaks exactly at the threshold of the $D\bar{D}^\ast$, which is a generic feature that has been discussed by Frazer {\it et al}~\cite{Frazer:1964zz} (for more general discussions on the pole distributions and classifications in different Riemann sheets, we refer to~\cite{Frazer:1964zz,Eden:1964zz,Badalian:1981xj}). Besides, this approach is also reformulated in a box~\cite{Albaladejo:2016jsg} to compare with the lattice QCD energy levels~\cite{Prelovsek:2014swa}.

\begin{figure}[htbp]
 	\centering
 	\includegraphics[width = 0.8\textwidth]{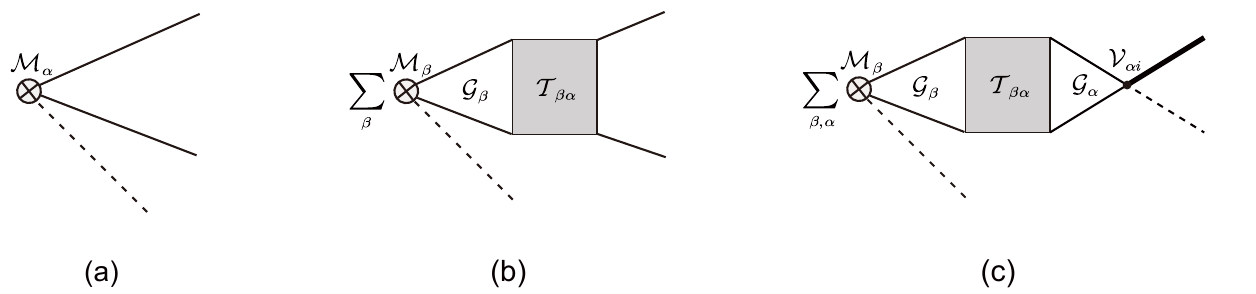}
 	\caption{A diagrammatic representation of Eqs.~\eqref{eq:prodampela} and~\eqref{eq:proampinela}, where the (thin) solid and dashed lines denote the vector (pseudoscalar) $B$ mesons and pion, respectively, while the thick solid line denotes the bottomonium. The circled cross represents the production source. The shaded box stands for the rescattering $T$-matrix of the elastic channels.}\label{fig:ProductionFig}
 \end{figure}

An elaborated approach to parameterizing the line shapes of the near-threshold states was developed in Refs.~\cite{Hanhart:2015cua,Guo:2016bjq} based on the coupled-channel LSEs. The $Z_b^{(\prime)}$ states were observed in the $\Upsilon(5S)\to B^\ast\bar{B}^{(\ast)}\pi$, $\Upsilon(nS)\pi\pi$, and $h_b(mP)\pi\pi$ decays in the $B^\ast\bar{B}^{(\ast)}$,  $\Upsilon(nS)\pi$ and $h_b(mP)\pi$ invariant mass spectra. For the $Z_b^{(\prime)}$ states, the relevant elastic and inelastic channels are $B\bar{B}^\ast$, $B^\ast\bar{B}^\ast$, and $\Upsilon(nS)\pi~(n=1,2,3)$, $h_b(mP)\pi~(m=1,2)$, respectively. The coupled-channel effective potential (the bare pole is not considered here, see discussions in~\cite{Wang:2018jlv}) is given with the $(N_e+N_i)\times(N_e+N_i)$ matrix (with $N_e=2$ and $N_i=5$ the numbers of the elastic and inelastic channels, respectively),
\begin{eqnarray}
V	=	\left[\begin{array}{cc}
\mathcal{V}_{\alpha\beta}(p,p^{\prime}) & \mathcal{V}_{\alpha i}(p,k_{i})\\
\mathcal{V}_{j\beta}(k_{j}^{\prime},p^{\prime}) & \mathcal{V}_{ji}(k_{j}^{\prime},k_{i})
\end{array}\right],
\end{eqnarray}
where the $\alpha,\beta$ denote the elastic channels and $i,j$ denote the inelastic channels, respectively. In order to simplify the calculation, the authors made some  assumptions and approximations such as the $\mathcal{V}_{ji}\approx0$ because the coupling of the heavy quarkonium and pion is rather weak~\cite{Detmold:2012pi,Liu:2012dv}. With this approximation, the inelastic channels can be disentangled from the elastic ones. The inelastic channel contributions are reduced to the additive terms in the contact potentials of $\mathcal{V}_{\alpha\beta}$~\cite{Hanhart:2015cua,Guo:2016bjq,Wang:2018jlv}, which is equivalent to using the description of the optical (complex) potential\footnote{The optical potential has both the real and imaginary parts, which is usually introduced to describe a process where the explicitly considered flux is not conserved. This is general in hadronic physics, such as the $p\bar{p}$ annihilation at low energies~\cite{Dover:1992vj}, where the opened channels---multi light hadrons cannot be explicitly described by the theory, e.g., the $\chi$EFT~\cite{Kang:2013uia}.}, i.e.,
\begin{eqnarray}\label{eq:opticpotential}
\mathcal{V}_{\alpha\beta}^{\mathrm{ct}}(E,p,p^{\prime})=v_{\alpha\beta}-\frac{i}{2\pi E}\sum_{i}m_{H_{i}}m_{h_{i}}g_{i\alpha}g_{i\beta}k_{i}^{2l_{i}+1},
\end{eqnarray}
with $v_{\alpha\beta}$ the contact potential of the elastic-to-elastic scattering. The $m_{H_i}$ ($m_{h_i}$), $k_i$ and $l_i$ are the mass of the heavy (light) meson, the momentum and the angular momentum in the $i$th inelastic channel, respectively. The $E$ represents the invariant mass of the system. The second term in Eq.~\eqref{eq:opticpotential} comes from the transition vertex of the elastic-to-inelastic 
\begin{eqnarray}
\mathcal{V}_{i\alpha}(k_{i},p)=\mathcal{V}_{\alpha i}(p,k_{i})=g_{i\alpha}k_{i}^{l_{i}},
\end{eqnarray}
and the inelastic loop. For more details, see Refs.~\cite{Hanhart:2015cua,Guo:2016bjq}. If one assumes that the direct production vertex is saturated by the $B^\ast\bar{B}^{(\ast)}\pi$ channels, the production amplitudes for the elastic and inelastic channels read, respectively,
\begin{eqnarray}
\mathcal{U}_{\alpha}(E,\bm{p})&=&\mathcal{M}_{\alpha}(E,\bm{p})+\sum_\beta\int\frac{d^{3}\bm{q}}{(2\pi)^{3}}\mathcal{V}_{\alpha\beta}(E,\bm{p},\bm{q})\mathcal{G}_{\beta}(E,\bm{q})\mathcal{U}_{\beta}(E,\bm{q}),\label{eq:prodampela}\\
\mathcal{U}_{i}(E,\bm{k})&=&\sum_\beta\int\frac{d^{3}\bm{q}}{(2\pi)^{3}}\mathcal{V}_{i\beta}(\bm k)\mathcal{G}_{\beta}(E,\bm{q})\mathcal{U}_{\beta}(E,\bm{q}),\label{eq:proampinela}
\end{eqnarray}
where
\begin{eqnarray}
\mathcal{G}_{\beta}=\frac{2\mu_{\beta}}{\bm p_{\beta}^{2}-\bm q^{2}+i\epsilon},\qquad \bm p_{\beta}^{2}\equiv2\mu_{\beta}(E-m_{\mathrm{th}}^{\beta}),
\end{eqnarray}
with $\mu_\beta$ and $m_{\mathrm{th}}^\beta$ the reduced mass and threshold of the $\beta$th elastic channel, respectively. The Eqs.~\eqref{eq:prodampela} and~\eqref{eq:proampinela} were also illustrated via a diagrammatic representation in Fig.~\ref{fig:ProductionFig}.

\begin{figure}[h]
\centering
\begin{tabular}{ll}
\includegraphics[width=0.4\textwidth,angle=-0]{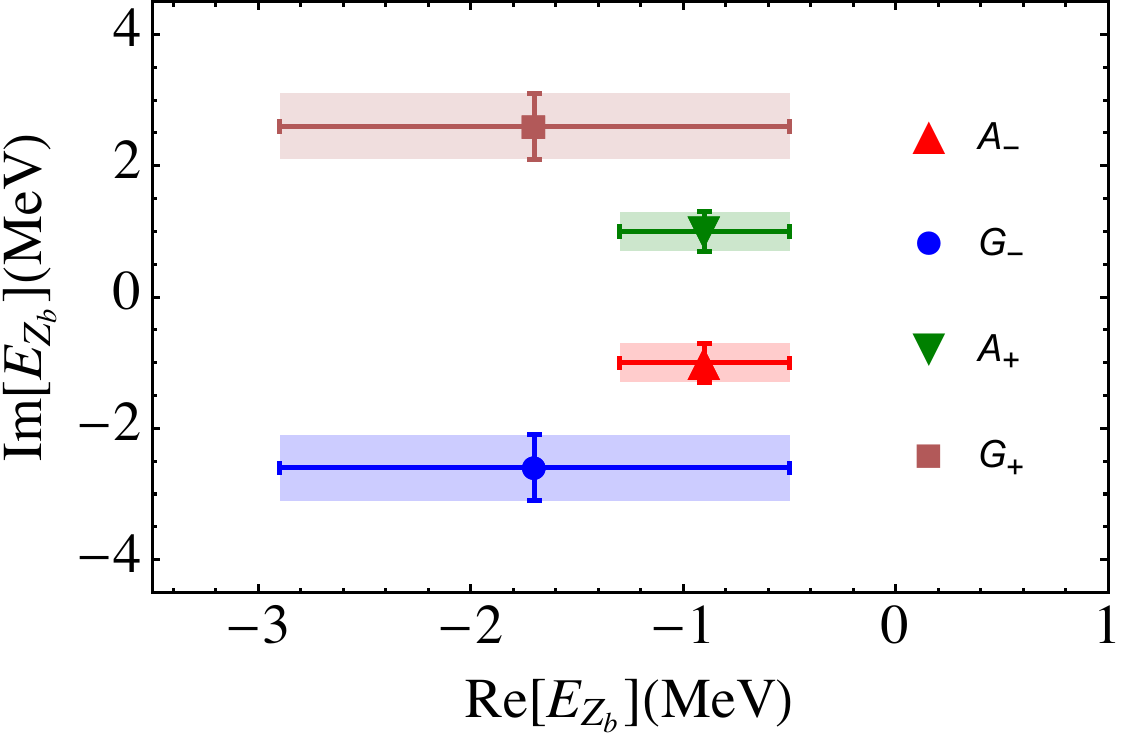}
& \includegraphics[width=0.4\textwidth,angle=-0]{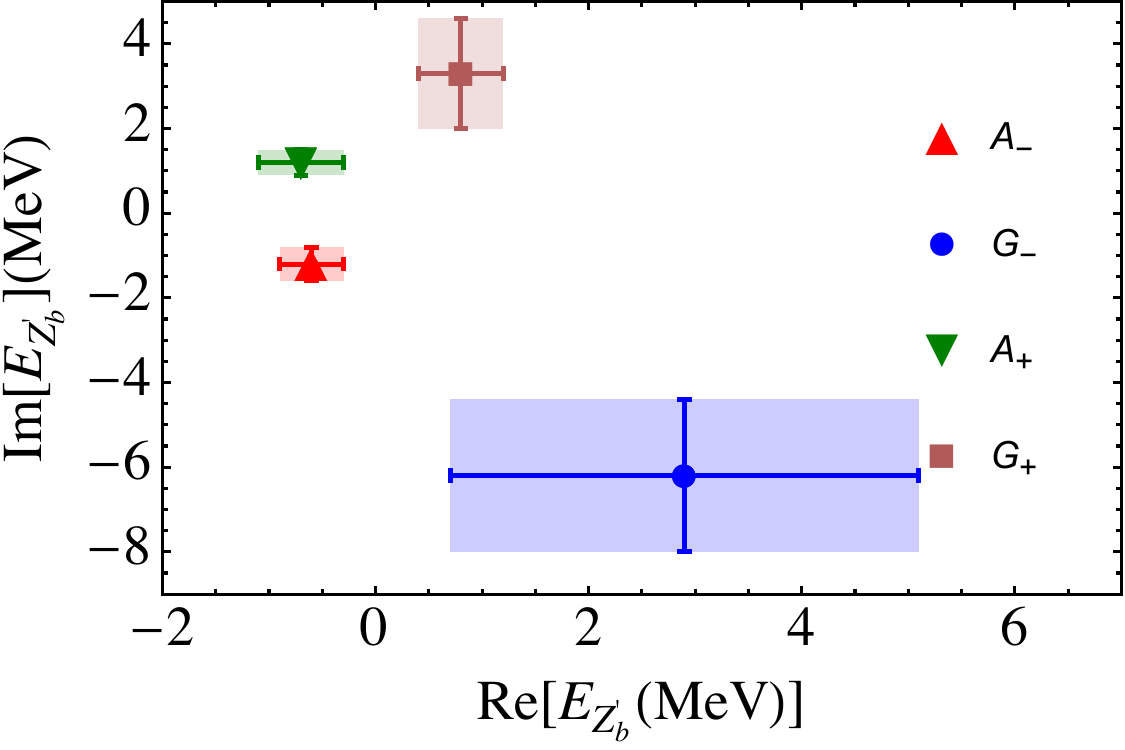}\\
\end{tabular}
\caption{The real and imaginary parts of the poles of the $Z_b$ (left panel) and $Z_b^{\prime}$ (right panel) states obtained in fit Scheme A and G, respectively~\cite{Wang:2018jlv}, where the $x$-axes are defined as $E_{Z_b}=M_{Z_b}^{\mathrm{pole}}-m_B-m_{B^\ast}$, $E_{Z_b^\prime}=M_{Z_b}^{\mathrm{pole}}-2m_{B^\ast}$.\label{fig:fitZbWang}}
\end{figure}

With this approach, Refs.~\cite{Hanhart:2015cua,Guo:2016bjq,Wang:2018jlv} fitted the invariant mass spectrum of the $B^\ast\bar{B}^{(\ast)}$ and $h_b(mP)\pi$, as well as those of the $\Upsilon(nS)\pi$ in Ref.~\cite{Baru:2020ywb} with the inclusion of the $\pi\pi$ final state interaction via the dispersion relation. In Refs.~\cite{Hanhart:2015cua,Guo:2016bjq}, the $B^\ast\bar{B}^{(\ast)}$ interactions are built upon the pionless framework, the LO contact potentials are constrained by the HQS (see Sec.~\ref{sec:HQSinHHM}). A good overall description of the data is consistent with the $Z_b$ as a virtual state and $Z_b^\prime$ as a resonance. Their poles both reside very close to the $B\bar{B}^\ast$ and $B^\ast\bar{B}^\ast$ thresholds, respectively~\cite{Guo:2016bjq}. An improved fitting with the Weinberg scheme is given by Ref.~\cite{Wang:2018jlv} with the potential $\mathcal{V}_{\alpha\beta}$ of the contact terms (LO plus NLO), OPE and one-eta exchange (OEE),
\begin{eqnarray}\label{eq:ValphabetaFull}
\mathcal{V}_{\alpha\beta}=V_{\alpha\beta}^{\mathrm{ct}}+V_{\alpha\beta}^{\mathrm{OPE}}+V_{\alpha\beta}^{\mathrm{OEE}},
\end{eqnarray}
where the three-body dynamics in OPE/OEE is considered via the TOPT formalism (see Sec.~\ref{sec:LOinteractions}). Seven different fitting schemes are considered by including the OPE/OEE, $S$$-$$D$ mixing, HQSS violation and NLO contact potentials by order. They obtained that the line shapes are insensitive to the central part of the $S$-wave OPE potential since it can be absorbed by adjusting the LO LECs, while the effect of the tensor force from OPE (though it is partially balanced by the $S$$-$$D$ transition terms of the NLO contact terms) leads to visible modifications of the line shapes [where the regulator~\eqref{eq:hardcutoff} is used and the (hard) cutoff is around $1.0$ GeV]. The effect of the OEE interaction is rather weak, and the higher order terms for the contact potentials of the inelastic channels seem not necessary. The Fig.~\ref{fig:fitZbWang} shows the real and imaginary parts of the $Z_b$ and $Z_b^\prime$ poles, where the fitting schemes A (only with the LO $V_{\alpha\beta}^{\mathrm{ct}}$) and G [with the full form of Eq.~\eqref{eq:ValphabetaFull} as well as the $S$$-$$D$ mixing] indicate that the $Z_b$ and $Z_b^\prime$ are virtual and resonant states residing below the $B\bar{B}^\ast$ and above the $B^\ast\bar{B}^\ast$ thresholds, respectively.

\begin{figure}[h]
\centering
\begin{minipage}[t]{0.35\linewidth}
\centering
\includegraphics[width=\columnwidth]{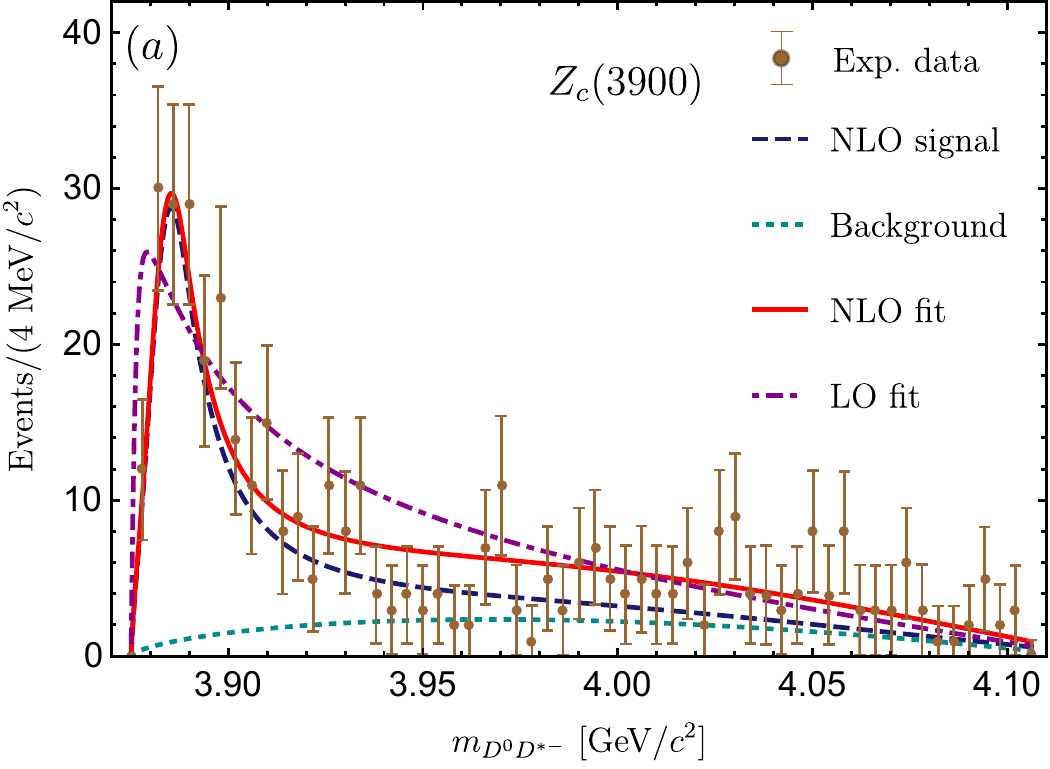}
\end{minipage}%
\hspace{1.0cm}
\begin{minipage}[t]{0.35\linewidth}
\centering
\includegraphics[width=\columnwidth]{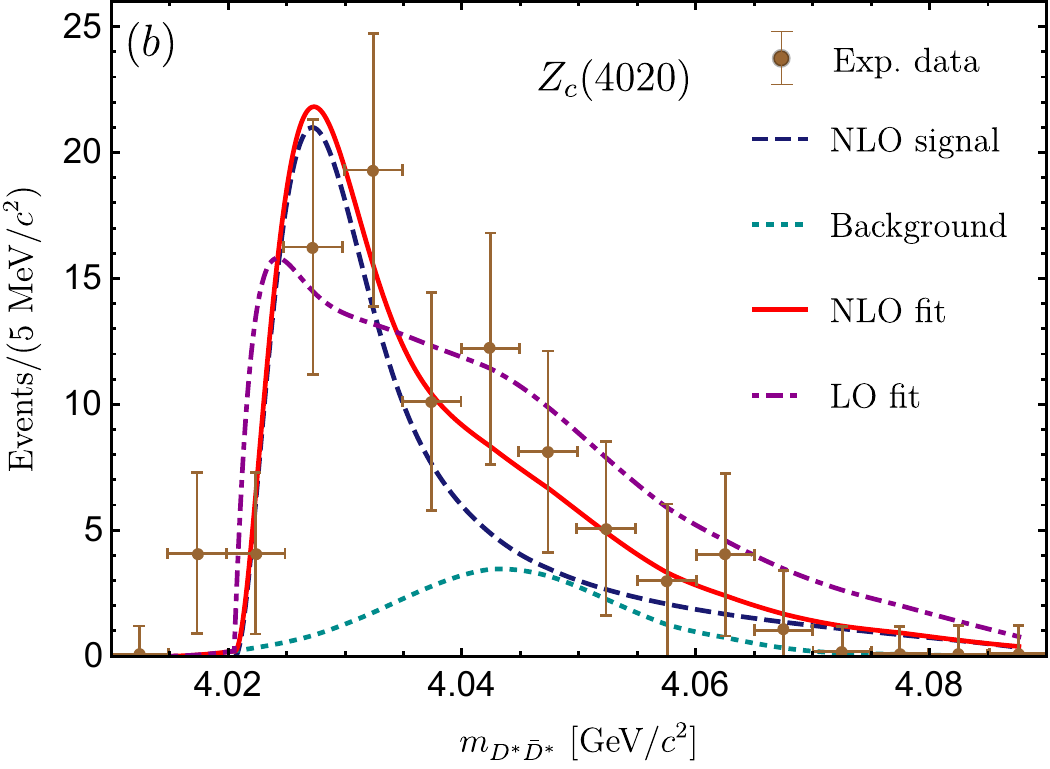}
\end{minipage}
\\
\begin{minipage}[t]{0.376\linewidth}
\centering
\includegraphics[width=\columnwidth]{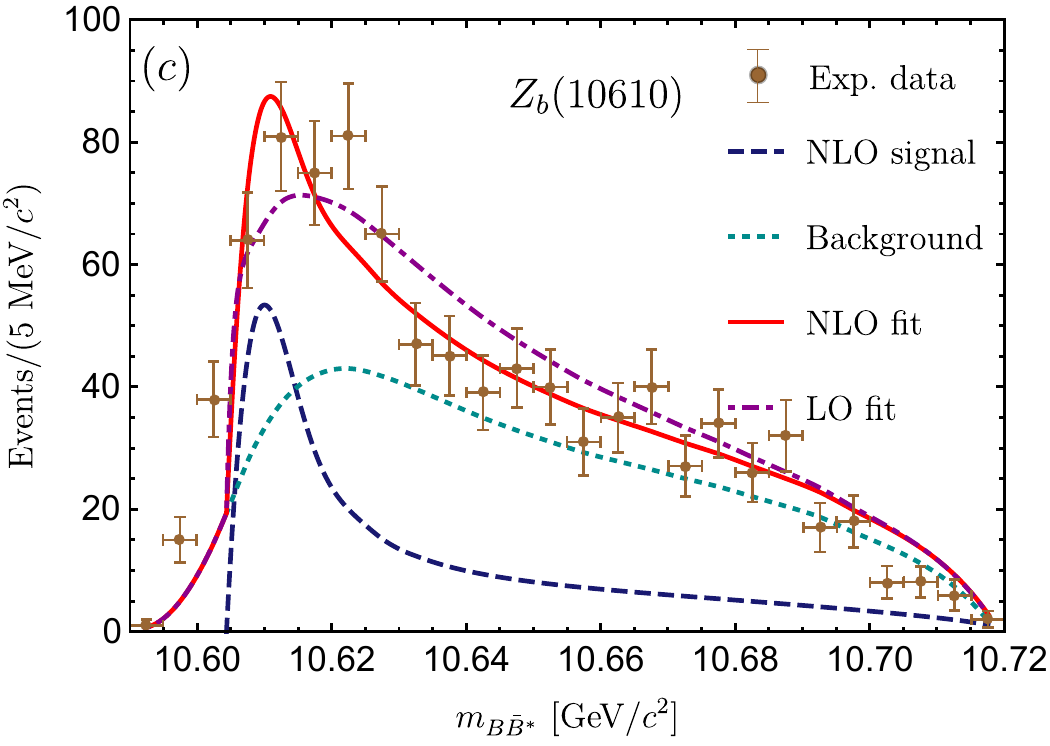}
\end{minipage}
\hspace{0.65cm}
\begin{minipage}[t]{0.366\linewidth}
\centering
\includegraphics[width=\columnwidth]{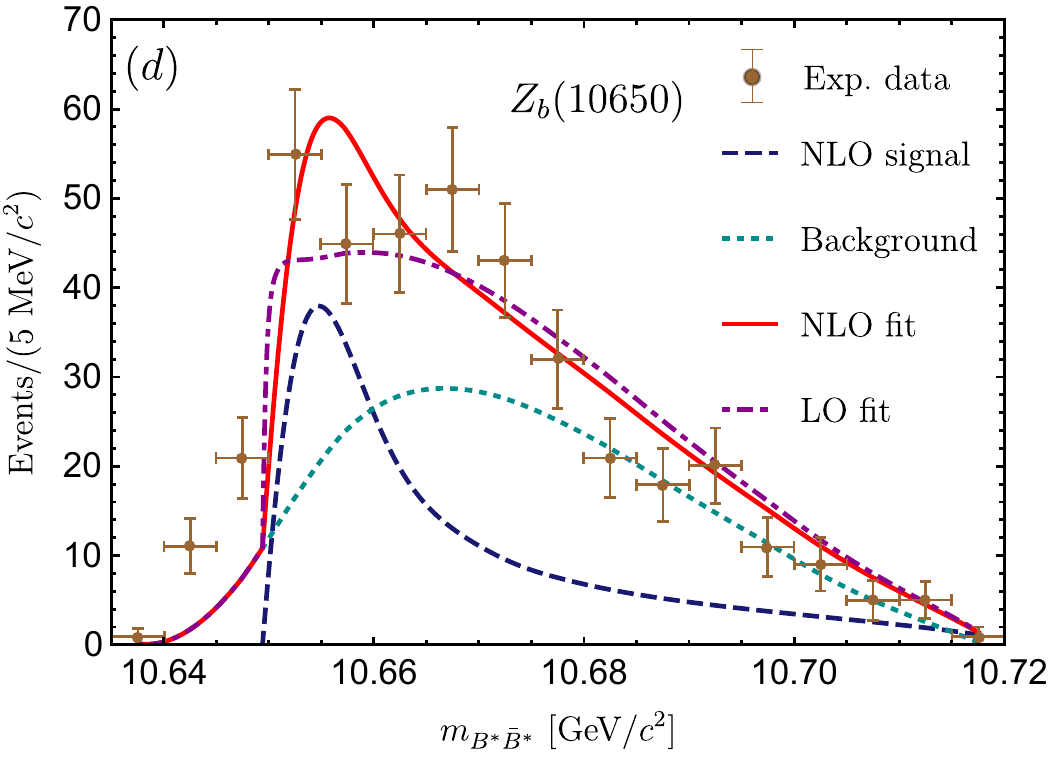}
\end{minipage}
\caption{The fitted invariant mass spectra of the $D^\ast \bar{D}^{(\ast)}$ and $B^\ast\bar{B}^{(\ast)}$ distributions in $e^+e^-\to\pi V(V)P$ (where $V$ and $P$ denote the vector and pseudoscalar $D~(B)$ mesons, respectively) transitions from Ref.~\cite{Wang:2020dko}. The data with error bars in figures (a), (b) and
(c)/(d) are taken from Refs.~\cite{BESIII:2015pqw},~\cite{BESIII:2015tix} and~\cite{Belle:2015upu} at $\sqrt{s}=4.26$, $4.23$, and $10.86$ GeV, respectively. The figures are taken from Ref.~\cite{Wang:2020dko}.
\label{fig:ZcZbWang}}
\end{figure}

An extensive study of the $Z_c^{(\prime)}$ and $Z_b^{(\prime)}$ states was provided by Ref.~\cite{Wang:2020dko} in $\chi$EFT, in which the effective potentials of the $D^\ast\bar{D}^{(\ast)}$ and $B^\ast\bar{B}^{(\ast)}$ were calculated up to NLO (e.g., see Sec.~\ref{sec:NLOinteractions}). Unlike the Refs.~\cite{Hanhart:2015cua,Guo:2016bjq,Wang:2018jlv}, this work is performed in the single-channel framework (ignoring the coupled-channel effect induced by the $S$$-$$D$ mixing), and the inelastic channels are not considered. The coupled-channel effect (e.g., $D\bar{D}^\ast\to D^\ast\bar{D}^\ast$ and $B\bar{B}^\ast\to B^\ast\bar{B}^\ast$) is (partially) included in the TPE diagrams. The corresponding invariant mass distributions of the elastic channels were fitted, see Fig.~\ref{fig:ZcZbWang}, and the extracted poles of the $Z_c^{(\prime)}$ and $Z_b^{(\prime)}$ all reside in the unphysical Riemann sheet and lie above the $D^\ast\bar{D}^{(\ast)}$ and $B^\ast\bar{B}^{(\ast)}$ thresholds, respectively. Therefore, they are explained as the $D^\ast\bar{D}^{(\ast)}$ and $B^\ast\bar{B}^{(\ast)}$ resonances, respectively. One virtue of Ref.~\cite{Wang:2020dko} is that the soft cutoff $\Lambda~(\le0.5\text{ GeV})$ is natural and consistent with the validity of $\chi$EFT (see discussions in Sec.~\ref{sec:NonperturbativeRenor}). In Ref.~\cite{Wang:2018jlv}, the typical momentum scale for the coupled-channel is $p^{\mathrm{ty}}=\sqrt{2\mu_{B^\ast\bar{B}^{(\ast)}}\delta_b}\approx0.5$ GeV. In order to effectively involve the coupled-channels, the relatively hard cutoff ($\sim1.0$ GeV) was used.

In the HQSS limit, there are six $B^{(*)}\bar{B}^{(*)}$ systems in the isovector channel including the $Z_b$ and $Z'_b$, which were proposed by Bondar {\it et al}~\cite{Bondar:2011ev}. The other four isovector states with negative $G$ parities are named as $W_{bJ}^{(\prime)}$~\cite{Voloshin:2011qa}
\begin{eqnarray}
|B\bar{B}\rangle_{1^{-}(0^{++})}&:&W_{b0}\to\eta_{b}\pi,\chi_{b}\pi,\Upsilon\rho,\\
|B^{\ast}\bar{B}^{\ast}\rangle_{1^{-}(0^{++})}&:&W_{b0}^{\prime}\to\eta_{b}\pi,\chi_{b}\pi,\Upsilon\rho,\\
|B\bar{B}^{\ast}\rangle_{1^{-}(1^{++})}&:&W_{b1}\to\chi_{b}\pi,\Upsilon\rho,\\
|B^{\ast}\bar{B}^{\ast}\rangle_{1^{-}(2^{++})}&:&W_{b2}\to\chi_{b}\pi,\Upsilon\rho,
\end{eqnarray}
where their possible decay modes are also presented. Because they have the negative $G$-parity, these states can be detected in the radiative decays of the $\Upsilon(5S)/\Upsilon(6S)$ rather than the of pionic transitions (for a review, see~\cite{Drutskoy:2012gt}). With pionfully nonperturbative treatment of the  $B^\ast\bar{B}^{(\ast)}$ interactions and using the parameters of the $Z_b^{(\prime)}$ as inputs, the $W_{bJ}^{(\prime)}$ states are predicted as the virtual states or resonances that reside near their corresponding thresholds~\cite{Baru:2017gwo,Baru:2019xnh}.

\subsubsection{$Z_{cs}$ and partners}

The above frameworks were also transplanted to study the recently observed $Z_{cs}(3985)$ by the BESIII Collaboration~\cite{BESIII:2020qkh} as well as the $Z_{cs}(4000)$ and $Z_{cs}(4220)$ by the LHCb Collaboration~\cite{LHCb:2021uow}.

In Ref.~\cite{Meng:2020ihj}, the $Z_{cs}(3985)$ was treated as the $U/V$-spin partners of $Z_c$ state in flavor $\SU(3)$ symmetry. The notion of the isospin was extended to the $U$ and $V$-spins in exact $\SU(3)$ symmetry. The corresponding $G_U$ and $G_V$ parities were defined (see Sec.~\ref{sec:Ebocs}). The $Z_{cs}(3985)$ can be connected to the $Z_c$ via
\begin{eqnarray}
Z_{c}^{-}(c\bar{c}d\bar{u})~\underleftrightarrow{\hat{U}}~Z_{cs}^{-}(c\bar{c}s\bar{u}),\qquad\qquad Z_{c}^{+}(c\bar{c}u\bar{d})~\underleftrightarrow{\hat{V}}~\bar{Z}_{cs}^{0}(c\bar{c}s\bar{d}),
\end{eqnarray}
where $\hat{U}~(\hat{V})$ denotes the $U~(V)$-spin rotation. One can consequently define the $G_{U/V}$ parity even and odd basis through the two particle basis $\bar{D}_sD^\ast$ and $\bar{D}_s^\ast D$ (\cvI in Table~\ref{tab:two-convention}) as
\begin{eqnarray}
|G_{U}	=\eta\rangle&=&\frac{1}{\sqrt{2}}\left(|D_{s}^{-}D^{\ast+}\rangle+\eta|D_{s}^{\ast-}D^{+}\rangle\right),\label{eq:GUeta}\\
|G_{V}=\eta\rangle&=&\frac{1}{\sqrt{2}}\left(|D_{s}^{-}D^{\ast0}\rangle+\eta|D_{s}^{\ast-}D^{0}\rangle\right),\label{eq:GVeta}
\end{eqnarray}
where $\eta=+1$ corresponds to the $\bar{Z}_{cs}^0$ and $Z_{cs}^-$, respectively. The potential~\eqref{eq:NLOcontactFormCoupledchannel} was expanded to the three-channel case, $(J/\psi\pi,~D\bar{D}^\ast,~\bar{D}^\ast D^\ast)$ for the $Z_c$ and $(J/\psi K,~\bar{D}_s D^\ast,~\bar{D}_s^\ast D^\ast)$ for the $Z_{cs}(3985)$. The four LECs (see Ref.~\cite{Meng:2020ihj}) were fixed using the mass and width of the $Z_c$ ($Z_b$) and $Z_c^\prime$ ($Z_b^\prime$) as inputs. In addition, the molecular resonances in the $\bar{D}_s^\ast D^\ast$, $B_{s}\bar{B}^{\ast}$ and $B_{s}^{\ast}\bar{B}^{\ast}$ channels were predicted.

\begin{figure}[h]
\centering
\begin{minipage}[t]{0.44\linewidth}
\centering
\includegraphics[width=\columnwidth]{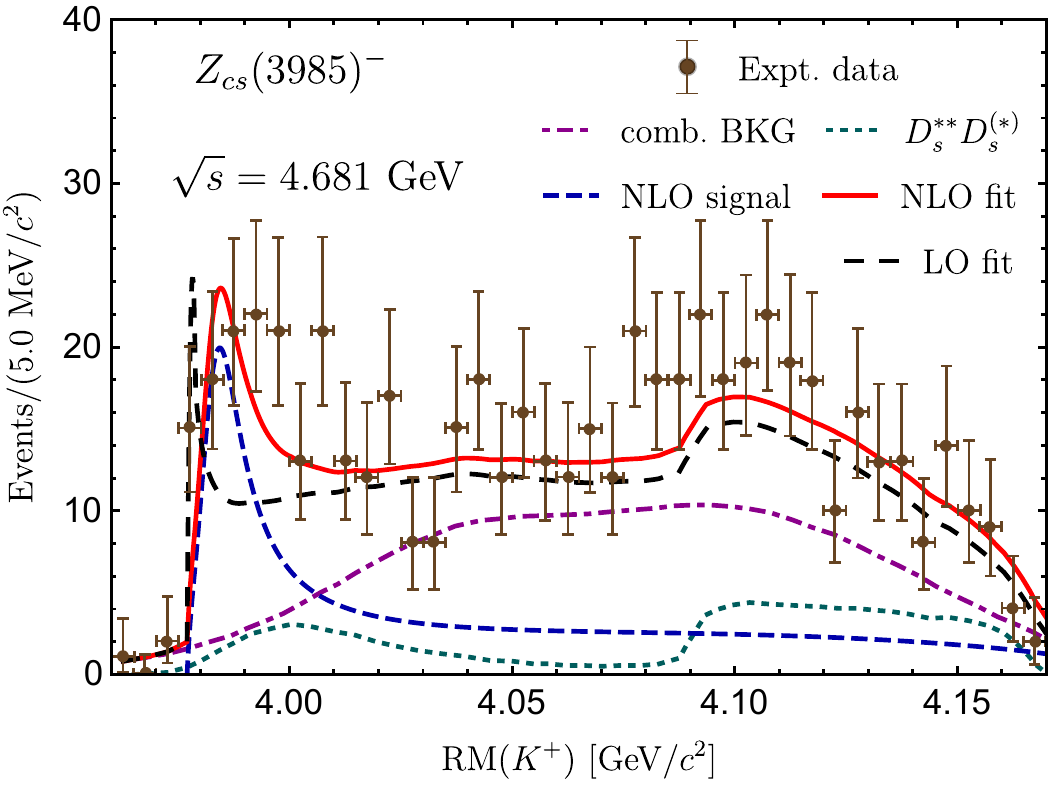}
\end{minipage}%
\hspace{0.5cm}
\begin{minipage}[t]{0.25\linewidth}
\centering
\includegraphics[width=\columnwidth]{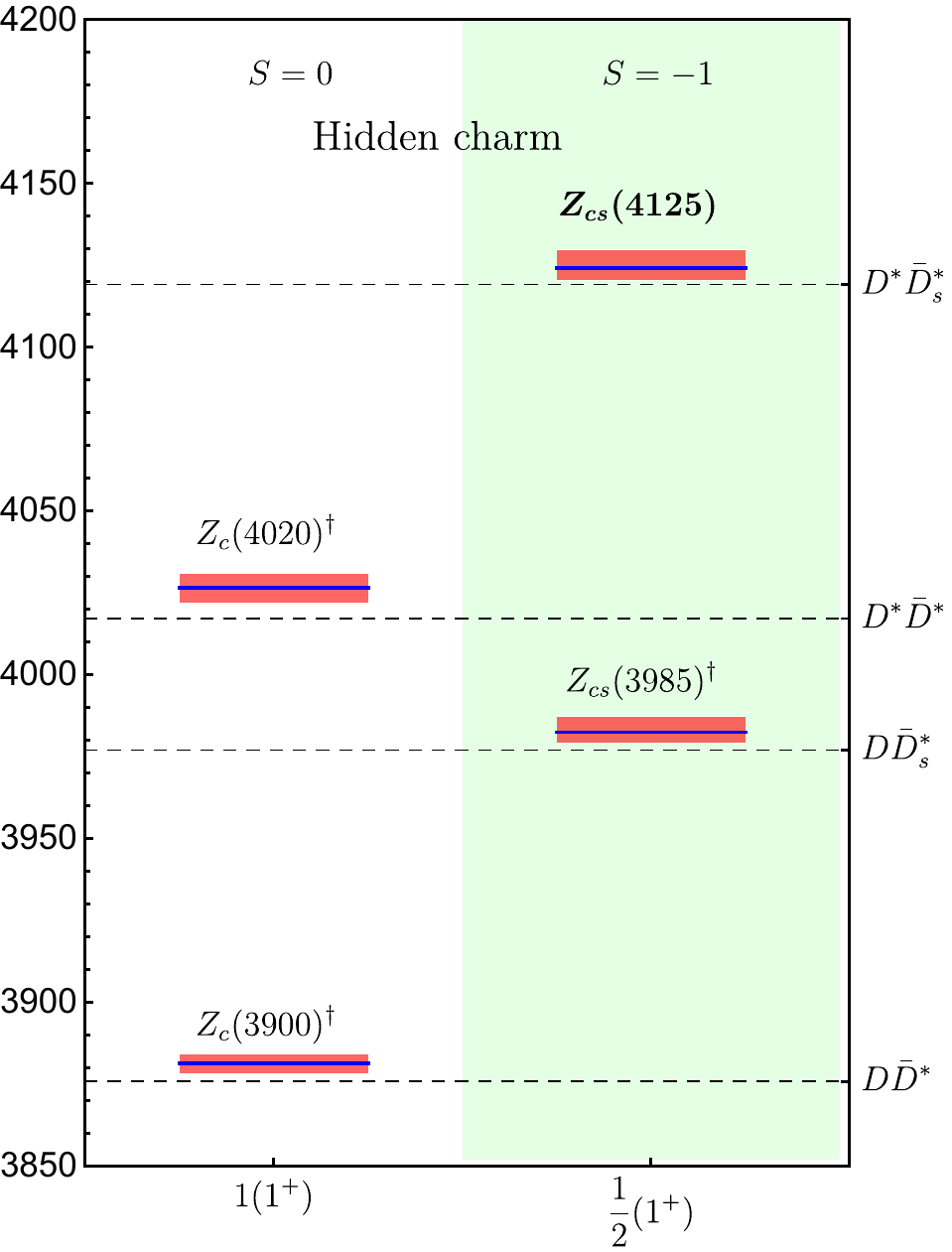}
\end{minipage}
\hspace{0.05cm}
\begin{minipage}[t]{0.255\linewidth}
\centering
\includegraphics[width=\columnwidth]{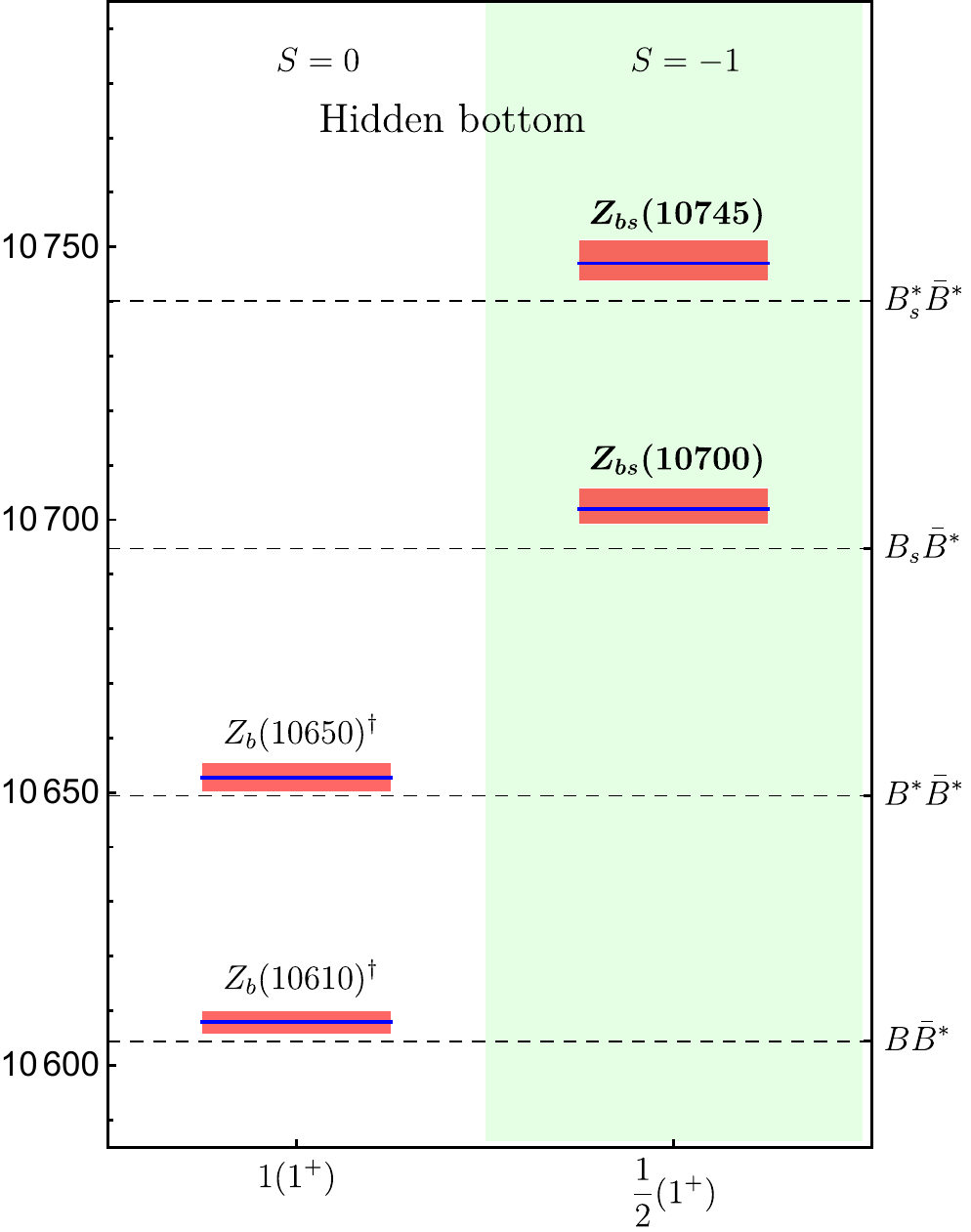}
\end{minipage}
\caption{Left panel: the $K^+$ recoil-mass spectrum distributions obtained with the fitted parameters of the $Z_c(3900)$~\cite{Wang:2020dko} as inputs. The data with error bars are taken from~\cite{BESIII:2020qkh} at $\sqrt{s}=4.681$ GeV. Middle and right panels: the spectrum of the charmoniumlike and bottomoniumlike states with positive $G_{I/U/V}$ parities predicted in Ref.~\cite{Wang:2020htx}. The blue solid line and red band denote the central value and range of errors of the mass. The observed and predicted states are marked with $\dagger$ and boldface, respectively. The figures are taken from Ref.~\cite{Wang:2020htx}.
\label{fig:fitandSpecofZcsZbsWang}}
\end{figure}

The calculation for the $\bar{D}_sD^\ast$ system within $\chi$EFT was performed in Ref.~\cite{Wang:2020htx}, where the effective potential was given up to NLO including the OEE and two-kaon exchange (TKE) interactions. Based on the parameters by fitting the line shapes of the $Z_c^{(\prime)}$ and $Z_b^{(\prime)}$ in Ref.~\cite{Wang:2020dko}, the line shape and resonance parameters of the $Z_{cs}(3985)$ were  reproduced~\cite{Wang:2020htx}. For example, a sharp peak automatically emerges in the $\bar{D}_sD^\ast$ invariant mass spectrum (see Fig.~\ref{fig:fitandSpecofZcsZbsWang}), and the extracted mass and width $(m,\Gamma)=(3982.4_{-3.4}^{+4.8},11.8_{-5.2}^{+5.5})$ MeV are very consistent with the experimental value. The possible resonances in the $\bar{D}_s^\ast D^\ast$, $B_{s}\bar{B}^{\ast}$ and $B_{s}^{\ast}\bar{B}^{\ast}$ systems were also predicted, which established a complete spectrum of the charged heavy quarkoniumlike states with the positive $G_{I/U/V}$ parities (see Fig.~\ref{fig:fitandSpecofZcsZbsWang}).


Other similar studies within the $\slashed{\pi}$EFT were presented in Refs.~\cite{Yang:2020nrt,Du:2020vwb,Baru:2021ddn}. For example, with the framework of Ref.~\cite{Albaladejo:2015lob}, Ref.~\cite{Yang:2020nrt} fitted the invariant mass spectrum of the $\bar{D}_sD^\ast$ with either the LO or NLO effective potential of the $\bar{D}_sD^\ast$ system in the energy range below $4.03$ GeV, where the production is enhanced by the triangle singularity. As a consequence, the $Z_{cs}(3985)$ and its HQSS partner are virtual states or resonances depending on the potential. An extension to the whole energy region in experiments was performed through introducing the coupled-channel $\bar{D}_{s}^{\ast}D^{\ast}$  with the LO contact potential constrained by the HQSS~\cite{Baru:2021ddn}. Ref.~\cite{Du:2020vwb} used the potential~\eqref{eq:V1pmepsV2pm} with the parameters $\mathcal{C}_{0}^{\alpha}$, $\mathcal{C}_{1}^{\alpha}$ fixed by fitting the data of the $Z_c^{(\prime)}$ and $Z_{cs}(3985)$. They predicted the possible existence of the $W_{cJ}^{(\prime)}$, $W_{csJ}^{(\prime)}$ either as the bound states or virtual states.

\begin{figure}[h]
\centering
\begin{minipage}[t]{0.46\linewidth}
\centering
\includegraphics[width=\columnwidth]{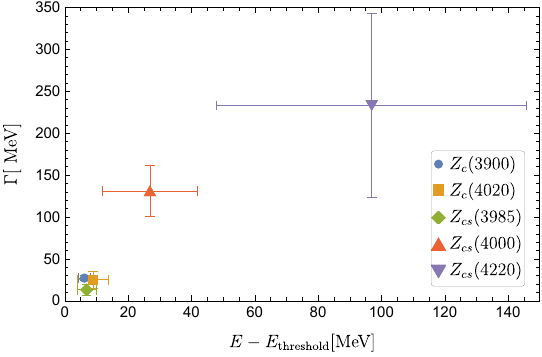}
\end{minipage}%
\hspace{0.5cm}
\begin{minipage}[t]{0.5\linewidth}
\centering
\includegraphics[width=\columnwidth]{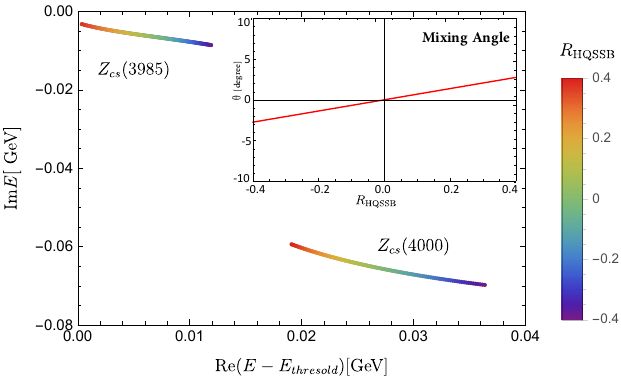}
\end{minipage}
\caption{Left panel: the $x$-axis denotes the masses of the $Z_c(3900)$, $Z_c(4020)$, $Z_{cs}(3985)$, $Z_{cs}(4000)$ and $Z_{cs}(4220)$ with respect to the corresponding thresholds $\bar{D}D^\ast/\bar{D}^\ast D$, $D^\ast\bar{D}^\ast$, $\bar{D}_s D^\ast/\bar{D}_s^\ast D$, $\bar{D}^\ast D_s/\bar{D}D_s^\ast$ and $\bar{D}^\ast D_s^\ast$, respectively, while the $y$-axis represents their widths~\cite{BESIII:2015pqw,BESIII:2013mhi,BESIII:2020qkh,LHCb:2021uow}. Right panel: the pole trajectories and the mixing angle range of $|\bar{D}_sD^*/\bar{D}_s^*D,+\rangle$ and $|\bar{D}_sD^*/\bar{D}_s^*D,-\rangle$ basis for the $Z_{cs}(3985)$ and $Z_{cs}(4000)$ states when varying $R_{\text{HQSSB}}$ from $-0.4$ to $0.4$. The figures are taken from Ref.~\cite{Meng:2021rdg}.\label{fig:MassWidthZcZcsHQSSbreakingEffect}}
\end{figure}

About four months after the observation of $Z_{cs}(3985)$~\cite{BESIII:2020qkh}, the LHCb discovered two states $Z_{cs}(4000)$ and $Z_{cs}(4220)$ in the invariant mass spectrum of $J/\psi K^+$ for the $B^+\to J/\psi K^+\phi$ decay with the significance of $15~\sigma$ and $5.9~\sigma$, respectively~\cite{LHCb:2021uow}. The mass of the $Z_{cs}(4000)$ is also very close to the $\bar{D}_sD^\ast$ threshold but its width is around $130$ MeV, which is about $10$ times larger than that of the $Z_{cs}(3985)$ as shown in Fig.~\ref{fig:MassWidthZcZcsHQSSbreakingEffect}. The $Z_{cs}(4220)$ is also much broader than any theoretical predictions of the HQSS partners of $Z_{cs}(3985)$. Therefore, whether the $Z_{cs}(3985)$ and $Z_{cs}(4000)$ are the same states observed in different processes~\cite{Ortega:2021enc} or they are utterly two different states~\cite{Meng:2021rdg} deserves serious analyses.

\begin{figure}[h]
 	\centering
 	\includegraphics[width = 1.0\textwidth]{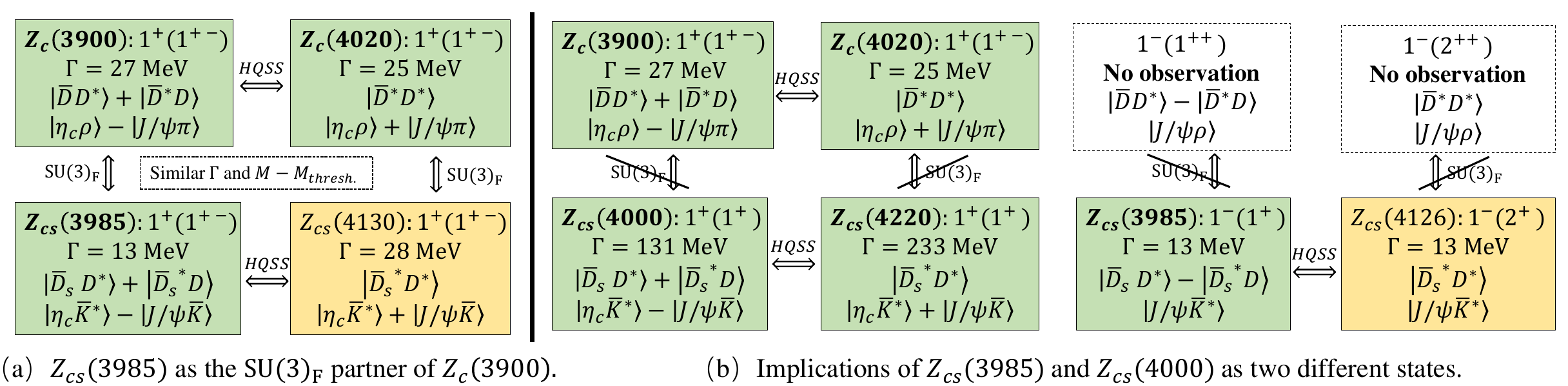}
 	\caption{Two assignments for the $Z_{cs}(3985)$ and $Z_{cs}(4000)$ states in Ref.~\cite{Meng:2021rdg} and the corresponding consequences. Subfigure (a): The $Z_{cs}(3985)$ is treated as the $\SU(3)$ partner of the $Z_c(3900)$ with the positive $G_{I/U/V}$ parities. Subfigure (b): the $\SU(3)$ partner of the $Z_c(3900)$ is $Z_{cs}(4000)$ (with significant symmetry breaking), while the $Z_{cs}(3985)$ is assigned as a negative $G_{U/V}$ parity state. The projected flavor wave function in the open-charm and hidden-charm channels, as well as the corresponding $I^{G_I}/U^{G_U}/V^{G_V}(J^{PC})$ quantum numbers can be traced back to Sec.~\ref{sec:HQSinHHM}. The particles listed in green and yellow cards represent the observed and predicted states, respectively. The double head arrows denote the partner states in HQSS and $\SU(3)$ symmetries (the symmetries hold well in this case), while the ones slashed by the oblique lines represent the breaking of $\SU(3)$ symmetry. The two partner channels given in the white cards are required by the symmetry, but they are not considered in Ref.~\cite{Meng:2021rdg}.}\label{fig:asTwoStates}
 \end{figure}

From Eqs.~\eqref{eq:GUeta} or~\eqref{eq:GVeta} we know that there are two orthogonal basis with $\eta=+1$ and $-1$, respectively. In the above investigations~\cite{Wang:2020htx,Meng:2020ihj,Yang:2020nrt,Du:2020vwb,Baru:2021ddn}, the $Z_{cs}(3985)$ was assigned as the $\eta=+1$ state. Meng {\it et al} derived some important implications by treating the $Z_{cs}(3985)$ and $Z_{cs}(4000)$ as two different states~\cite{Meng:2021rdg} corresponding to the states with $\eta=-1$ and $\eta=+1$, respectively. With this assumption, the $\SU(3)$ partners of the $Z_c$ and $Z_c^\prime$ are assigned as the $Z_{cs}(4000)$ and $Z_{cs}(4220)$, respectively, see details in Fig.~\ref{fig:asTwoStates}. In such a scenario, the $\SU(3)$ breaking effect in the $D^\ast\bar{D}^{(\ast)}$ and $\bar{D}_s^\ast D^{(\ast)}$ systems should be very significant since $\Gamma_{Z_{cs}(4000)}\gg\Gamma_{Z_{c}(3900)}$, $\Gamma_{Z_{cs}(4220)}\gg\Gamma_{Z_{c}(4020)}$, and $m_{Z_{cs}(4220)}-m_{Z_{c}(4020)}\approx2[m_{Z_{cs}(4000)}-m_{Z_c(3900)}]$. With strict HQSS, the $G_{I/U/V}$ parity even and odd states are orthogonal, whereas the mixing occurs when the HQSS is broken to some extent. In Ref.~\cite{Meng:2021rdg} the HQSS breaking effect was investigated via introducing the following coupled-channel potential,
\begin{eqnarray}\label{eq:NLOcontactplusminus}
V_{1^{+}}(\bm p,\bm p^{\prime})_{\{+,-\}} =	\left[\begin{array}{cc}
\frac{c_{a}^{+}+\delta c_{a}}{\Lambda} & \frac{\delta c_{a}}{\Lambda}\\
\frac{\delta c_{a}}{\Lambda} & \frac{c_{a}^{-}+\delta c_{a}}{\Lambda}
\end{array}\right]+\left[\begin{array}{cc}
\frac{c_{b}^{+}(\bm p^{2}+\bm p^{\prime2})}{\Lambda^{3}}\\
 & \frac{c_{b}^{-}(\bm p^{2}+\bm p^{\prime2})}{\Lambda^{3}}
\end{array}\right],
\end{eqnarray}
where the $c_{a}^+~(c_a^-)$ and $c_b^+~(c_b^-)$ denote the LO and NLO LECs for the $\eta=+1~(-1)$ channel, respectively. The $\delta c_a$ designates the strength of the off-diagonal term. A conversion of the potential~\eqref{eq:NLOcontactplusminus} into the $\{\bar{D}_s D^\ast,\bar{D}_s^\ast D\}$ basis defines a measure of the HQSS breaking scale, $R_{\mathrm{HQSSB}}=4\delta c_a/|c_a^++c_a^-|$. Ref.~\cite{Meng:2021rdg} found that the pole trajectories of the $Z_{cs}(3985)$ and $Z_{cs}(4000)$ are not sensitive to $R_{\mathrm{HQSSB}}$ (the corresponding mixing angle $\theta$ changes from $-3^\circ$ to $3^\circ$) when the $R_{\mathrm{HQSSB}}$ is varied from $-0.4$ to $0.4$, see Fig.~\ref{fig:MassWidthZcZcsHQSSbreakingEffect}.

If the $Z_{cs}(3985)$ and $Z_{cs}(4000)$ are different states~\cite{Meng:2021rdg}: (1) HQSS is a good symmetry for the $\bar{D}_s^{(\ast)} D^{(\ast)}$ systems. (2) The $Z_{cs}(4000)$ and $Z_{cs}(4220)$ are the HQSS partners. (3) The $Z_{cs}(3985)$ and $Z_{cs}(4000)$ are pure $G_{U/V}$ parity odd and even states, respectively. (4) There exists a tensor resonance $|\bar{D}_s^\ast D^\ast\rangle_{2^+}$ as the HQSS partner of the $Z_{cs}(3985)$. Its mass and width are predicted to be $(m,\Gamma)=(4126\pm3,13\pm6)$ MeV. (5) For the $Z_{cs}(3985)$ and $Z_{cs}(4000)$, the branching ratio $\mathscr{R}(Z_{cs}\to\bar{D}_s^\ast D/Z_{cs}\to \bar{D}_sD^\ast)\approx0.5$. (6) The decay mode $Z_{cs}(3985)\to J/\psi K$ is suppressed in the HQS limit [see Eq.~\eqref{eq:BB1plusplus}]. (7) The $\SU(3)$ breaking effect is significant for the systems without and with the strangeness. The implications (4) and (6) could serve as a criterion for experiments to test the scenario proposed in Ref.~\cite{Meng:2021rdg} .

\subsection{Hidden-charm molecular pentaquarks without and with the strangeness} \label{sec:Pc}

\subsubsection{$P_c$ and partners  }\label{sec:PcNoStrangeness}

In 2019, the LHCb analyzed both the run-I and run-II data~\cite{LHCb:2019kea}, and found the previously observed $P_c(4450)$ signal~\cite{LHCb:2015yax} actually contains two substructures---the $P_c(4440)$ and $P_c(4457)$. Meanwhile, another new structure $P_c(4312)$ was discovered, while the broad $P_c(4380)$~\cite{LHCb:2015yax} lost its significance in the $J/\psi p$ invariant mass spectrum. {One may still notice a bump around the $4.38$ GeV. However, this bump is too narrow (see Ref.~\cite{Du:2021fmf} for a quantitative fit) to be identified as the same state in the first analysis.} Their isospin $I=1/2$ can be easily inferred from the final state $J/\psi p$. The community instantly noticed that the $P_c(4312)$ and $P_c(4440)$, $P_c(4457)$ reside very close to (and below) the $\Sigma_c\bar{D}$ and $\Sigma_c\bar{D}^\ast$ thresholds, respectively. This triggered a flood of works toward the molecular explanations with various models, e.g., see Sec.~\ref{sec1.3.3}.

In Ref.~\cite{Liu:2019tjn}, the $\slashed{\pi}$EFT and HQSS was combined to parameterize the interactions of the $\Sigma_c\bar{D}$, $\Sigma_c\bar{D}^\ast$, $\Sigma_c^\ast\bar{D}$ and $\Sigma_c^\ast\bar{D}^\ast$ by two LECs, which are equivalent to the $\mathcal{C}_{1/2}^\alpha$ and $\mathcal{C}_{3/2}^\alpha$ of Eqs.~\eqref{eq:C12alpha} and~\eqref{eq:C32alpha} but within the uncoupled-channel framework. The two LECs are fixed using the binding energies of the $P_c(4440)$ and $P_c(4457)$ via treating them as the $S$-wave $\Sigma_c\bar{D}^\ast$ molecules with $J^P=\frac{1}{2}^-$ and $\frac{3}{2}^-$. The $J^P$ quantum numbers of the $P_c(4440)$ and $P_c(4457)$ are not determined yet, so there are two options:
\begin{eqnarray}
\text{canonical spin order: }&P_c(4440)\text{: }J^P=\frac{1}{2}^-,\quad P_c(4457)\text{: }J^P=\frac{3}{2}^-,\label{eq:canonicalorder}\\
\text{non-canonical spin order: }&P_c(4440)\text{: }J^P=\frac{3}{2}^-,\quad P_c(4457)\text{: }J^P=\frac{1}{2}^-,\label{eq:noncanonicalorder}
\end{eqnarray} 
where the spin order in Eq.~\eqref{eq:canonicalorder} is called {\it canonical} because the empirical pattern from the hadron spectra is that the higher spin states usually have larger masses~\cite{ParticleDataGroup:2022pth}. It was shown that the canonical order is more preferable considering the induced binding energy of the $P_c(4312)$ is closer to the experimental data. Another four bound states in the $\Sigma_c^\ast\bar{D}$ and $\Sigma_c^\ast\bar{D}^\ast$ systems with $J^P=\frac{3}{2}^-$ and $\frac{1}{2}^-$, $\frac{3}{2}^-$, $\frac{5}{2}^-$ were predicted with binding energies around $3-27$ MeV. {The fits in Ref.~\cite{Du:2019pij} show that the two arrangements in Eqs.~\eqref{eq:canonicalorder} and~\eqref{eq:noncanonicalorder} work equally well without pions. However, in a recent study based on the machine learning~\cite{Zhang:2023czx}, the authors claimed that the $J^P$ quantum numbers of $P_c(4440)$ and $P_c(4457)$ can be discriminated even within the $\slashed{\pi}$EFT, and the canonical spin order is supported in the neural network-based approach.}

The OPE interaction for the $\Sigma_c\bar{D}^\ast$ system was included in Refs.~\cite{Yamaguchi:2019seo,PavonValderrama:2019nbk} (see a more general discussion on the role of OPE for heavy hadron systems~\cite{Karliner:2015ina}), in which the hard cutoff $\Lambda\gtrsim1.0$ GeV was used to regularize the Schr\"odinger equation. The results showed that the $J^P$ of the $P_c(4440)$ and $P_c(4457)$ turns into the non-canonical order of Eq.~\eqref{eq:noncanonicalorder}. A systematic study of the $\Sigma_c^{(\ast)}\bar{D}^{(\ast)}$ interactions within the $\chi$EFT was performed in Refs.~\cite{Meng:2019ilv,Wang:2019ato}, where the contact interactions, the OPE and TPE were considered. The contact Lagrangian of the $\Sigma_c^{(\ast)}\bar{D}^{(\ast)}$ is given as
\begin{eqnarray}\label{eq:LOcontactHB}
\mathcal{L}&=&\tilde{D}_{a}\langle\bar{\tilde{\mathcal{H}}}\tilde{\mathcal{H}}\rangle\mathrm{Tr}(\bar{\psi}_{Q}^{\mu}\psi_{Q\mu})+i\tilde{D}_{b}\epsilon_{\sigma\mu\nu\rho}v^{\sigma}\langle\bar{\tilde{\mathcal{H}}}\gamma^{\rho}\gamma_{5}\tilde{\mathcal{H}}\rangle\mathrm{Tr}(\bar{\psi}_{Q}^{\mu}\psi_{Q}^{\nu}) \nonumber\\
&&+\tilde{E}_{a}\langle\bar{\tilde{\mathcal{H}}}\tau_{i}\tilde{\mathcal{H}}\rangle\mathrm{Tr}(\bar{\psi}_{Q}^{\mu}\tau_{i}\psi_{Q\mu})+ i\tilde{E}_{b}\epsilon_{\sigma\mu\nu\rho}v^{\sigma}\langle\bar{\tilde{\mathcal{H}}}\gamma^{\rho}\gamma_{5}\tau_{i}\tilde{\mathcal{H}}\rangle\mathrm{Tr}(\bar{\psi}_{Q}^{\mu}\tau_{i}\psi_{Q}^{\nu}),
\end{eqnarray}
where the corresponding LECs are denoted as $\tilde{D}_{a}$, $\tilde{D}_b$, $\tilde{E}_a$ and $\tilde{E}_b$, respectively. The $\psi_Q^\mu$ represents the superfield of the singly heavy baryons in Sec.~\ref{sec:singlyHB}. Expanding Eq.~\eqref{eq:LOcontactHB} one obtains
\begin{eqnarray}
\mathcal{V}_{\Sigma_c\bar{D}}^{\mathrm{ct}}&=&-\tilde{D}_a-2\tilde{E}_a(\bm{I}_1\cdot\bm{I}_2),\label{eq:LO_SigmacDb_Con}\\
\mathcal{V}_{\Sigma_c\bar{D}^\ast}^{\mathrm{ct}}&=&-\tilde{D}_a-2\tilde{E}_a(\bm{I}_1\cdot\bm{I}_2)+\frac{2}{3}\Big[-\tilde{D}_b-2\tilde{E}_b(\bm{I}_1\cdot\bm{I}_2)\Big]\boldsymbol{\sigma}\cdot\bm{T},\label{eq:X21_Potential}\\
\mathcal{V}_{\Sigma_c^\ast\bar{D}}^{\mathrm{ct}}&=&-\tilde{D}_a-2\tilde{E}_a(\bm{I}_1\cdot\bm{I}_2),\\
\mathcal{V}_{\Sigma_c^\ast \bar{D}^\ast}^{\mathrm{ct}}&=&-\tilde{D}_a-2\tilde{E}_a(\bm{I}_1\cdot\bm{I}_2)+\Big[-\tilde{D}_b-2\tilde{E}_b(\bm{I}_1\cdot\bm{I}_2)\Big]\boldsymbol{\sigma}_{rs}\cdot\bm{T},\label{eq:X41_Potential}
\end{eqnarray}
where $\bm{I}_1~(\bm{I}_2)$ denotes the isospin operator of the $\Sigma_c^{(\ast)}~(\bar{D}^{(\ast)})$. The spin operator of the $\Sigma_c$ ($\Sigma_c^\ast$) is related to the $\bm\sigma$ ($\bm\sigma_{rs}$) via $\bm S_1=\bm\sigma/2$ ($\bm S_1^\prime=3\bm\sigma_{rs}/2$), while the spin operator of the $\bar{D}^\ast$ is defined as $\bm S_2=-\bm T=i\bm{\varepsilon}^\dagger\times\bm{\varepsilon}$, e.g., see the Appendix C of Ref.~\cite{Wang:2019ato}. Meanwhile, the effective potentials from the OPE and NLO TPE interactions can be derived from Eqs.~\eqref{eq:app1:lagDbar} and~\eqref{eq:app1:lagBc2}. The OPE only survives in the $\Sigma_c\bar{D}^\ast$ and $\Sigma_c^\ast\bar{D}^\ast$ systems, which read
\begin{eqnarray}
\mathcal{V}_{\Sigma_c\bar{D}^\ast}^{\mathrm{OPE}}&=&-(\bm{I}_1\cdot\bm{I}_2)\frac{gg_1}{2f_\pi^2}\frac{(\bm{q}\cdot\boldsymbol{\sigma})(\bm{q}\cdot\bm{T})}{\bm{q}^2+m_\pi^2},\label{H21_Potential}\\
\mathcal{V}_{\Sigma_c^\ast\bar{D}^\ast}^{\mathrm{OPE}}&=&(\bm{I}_1\cdot\bm{I}_2)\frac{gg_5}{2f_\pi^2}\frac{(\bm{q}\cdot\boldsymbol{\sigma}_{rs})(\bm{q}\cdot\bm{T})}{\bm{q}^2+m_\pi^2}.
\end{eqnarray}  {As in the $NN$ interactions within the $\chi$EFT~\cite{Epelbaum:2008ga,Machleidt:2011zz}, the static OPE potential is introduced for the $\Sigma_c\bar{D}^\ast$ and $\Sigma_c^\ast\bar{D}^\ast$ systems. For the OPE potential with the kinetic energies of the heavy particles in the framework of TOPT, e.g., the expressions like those in Eqs.~\eqref{OPEel} and \eqref{OPEin}, one can consult the supplemental materials of Ref.~\cite{Du:2019pij}}.

The TPE expressions including the mass splittings are lengthy and can be found in Refs.~\cite{Meng:2019ilv,Wang:2019ato}. For the specified total isospin $\bm I=\bm I_1+\bm I_2$, the matrix element $\langle\bm I_1\cdot\bm I_2\rangle=[I(I+1)-I_1(I_1+1)-I_2(I_2+1)]/2$. From Eqs.~\eqref{eq:LO_SigmacDb_Con}--\eqref{eq:X41_Potential}, we can redefine
\begin{eqnarray}\label{eq:LECs_redefine}
\mathbb{D}_1=\tilde{D}_a+2\tilde{E}_a(\bm{I}_1\cdot\bm{I}_2),\qquad\qquad
\mathbb{D}_2=\tilde{D}_b+2\tilde{E}_b(\bm{I}_1\cdot\bm{I}_2),
\end{eqnarray}
where $\mathbb{D}_1$ and $\mathbb{D}_2$ represent the strength of the central potential and spin-spin interaction of the light d.o.f, respectively. One can build a connection between $\mathbb{D}_1$, $\mathbb{D}_2$ and $\mathcal{C}_{1/2}^\alpha$, $\mathcal{C}_{3/2}^\alpha$ through matching the potentials in the corresponding channels.

\begin{figure}
\centering
\begin{minipage}[t]{0.33\linewidth}
\centering
\includegraphics[width=\columnwidth]{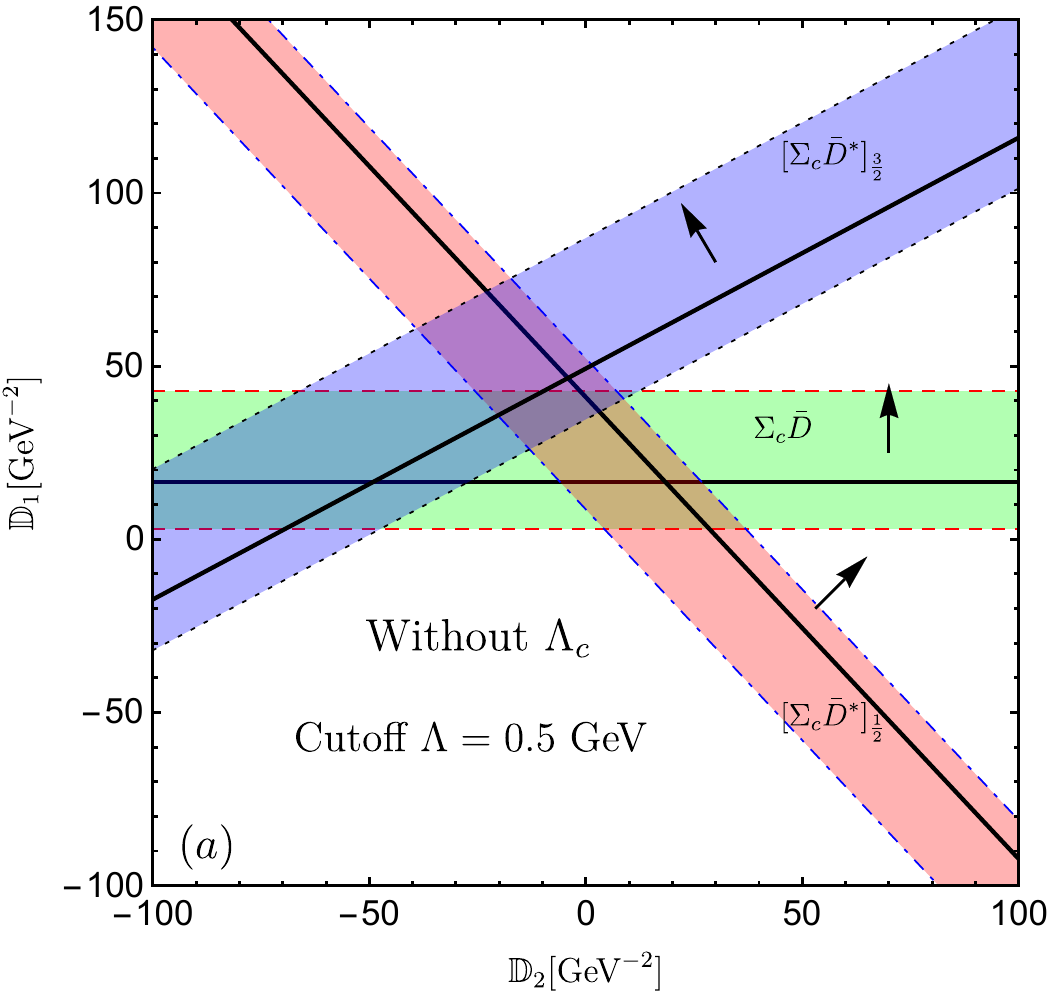}
\end{minipage}%
\begin{minipage}[t]{0.33\linewidth}
\centering
\includegraphics[width=\columnwidth]{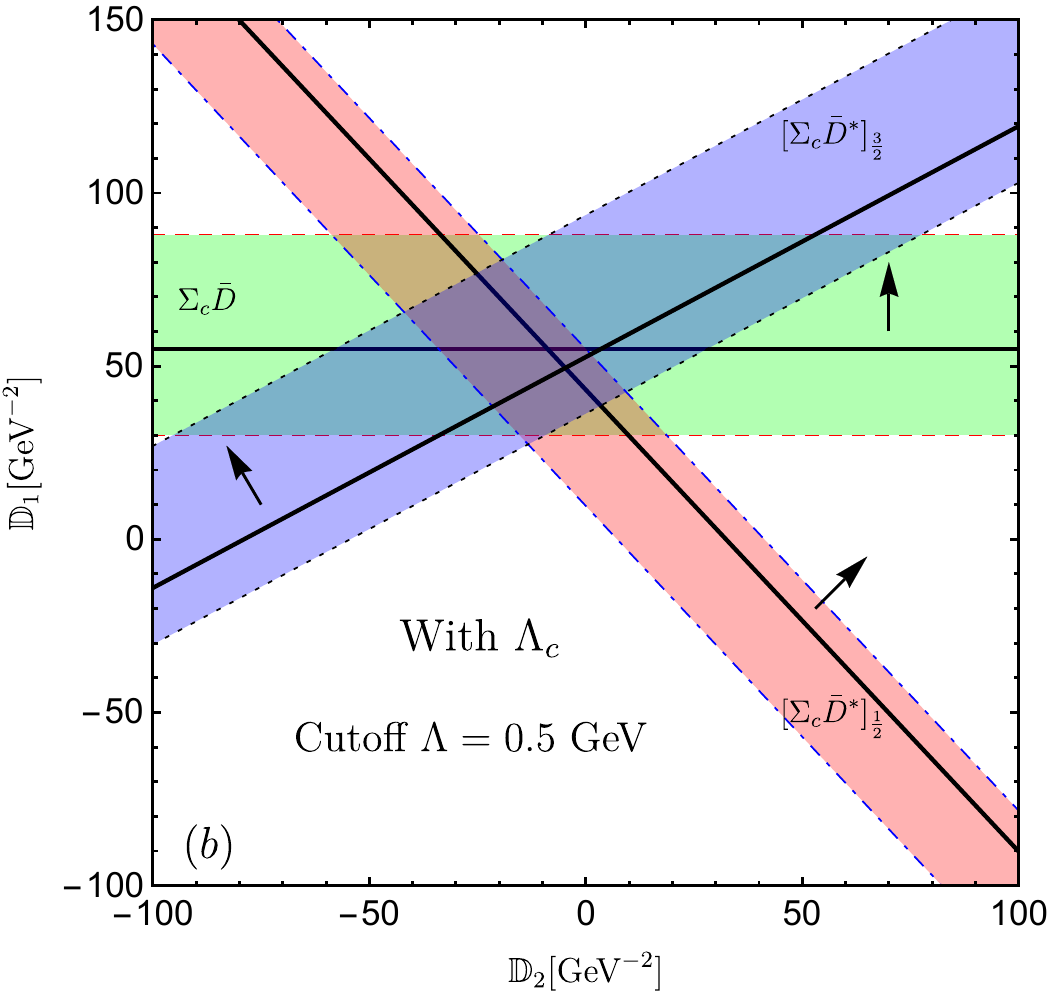}
\end{minipage}
\begin{minipage}[t]{0.33\linewidth}
\centering
\includegraphics[width=\columnwidth]{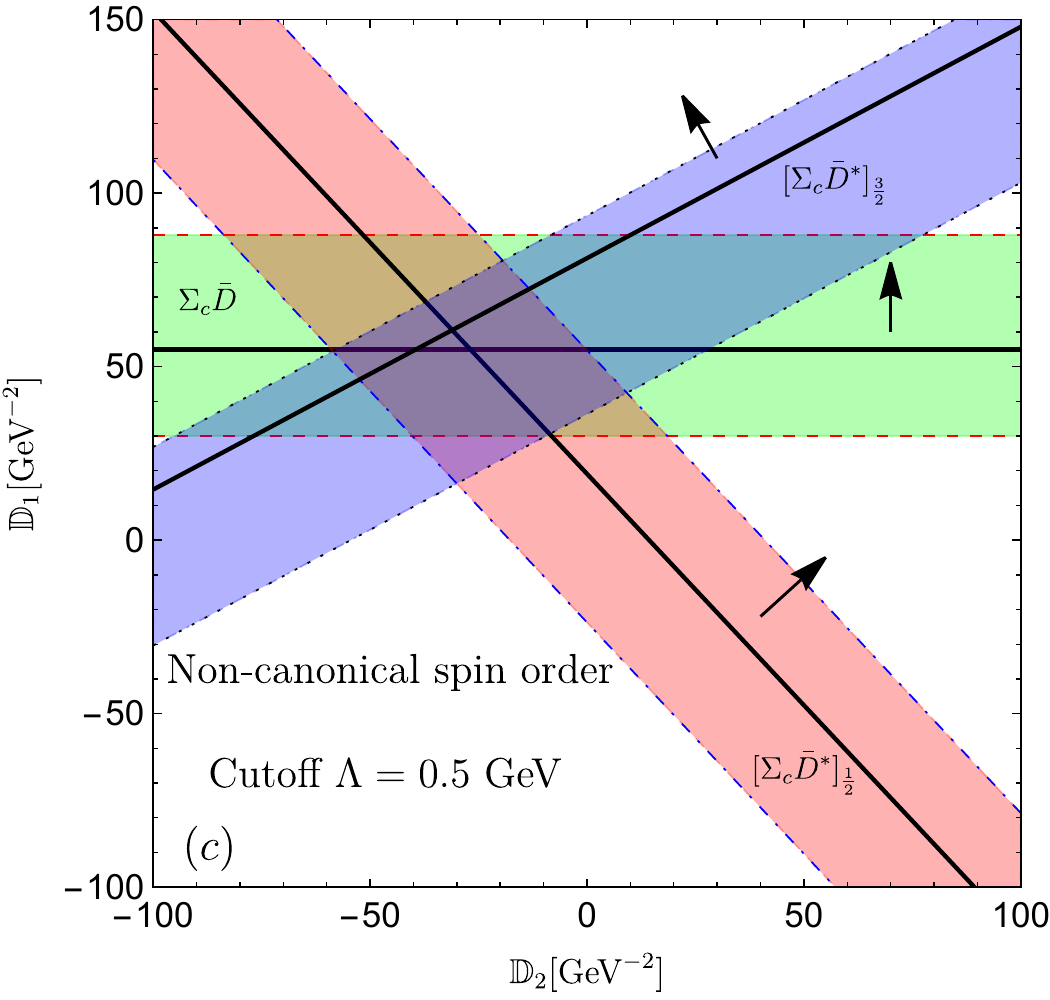}
\end{minipage}
\caption{The parameter regions of three $P_c$ states in the cases of (a)-not considering the contribution of the $\Lambda_c$ in the loop diagrams, (b)-considering the contribution of the $\Lambda_c$ in the loop diagrams, (c)-using the non-canonical spin order for the $P_c(4440)$ and $P_c(4457)$. The green, red and blue bands denote the binding energies of the $[\Sigma_c\bar{D}]_{J=1/2}$, $[\Sigma_c\bar{D}^\ast]_{J=1/2}$ and $[\Sigma_c\bar{D}^\ast]_{J=3/2}$ systems in the region $[-30,0]$ MeV, respectively, while the arrow for each one denotes the direction that the binding becomes deeper. The corresponding three black straight lines represent the central values of the binding energies obtained from the experimental data~\cite{LHCb:2019kea}.\label{fig:Results_WithoutWith_Lambdac}}
\end{figure}

In Refs.~\cite{Meng:2019ilv,Wang:2019ato}, the values of $\mathbb{D}_1$ and $\mathbb{D}_2$ were varied in the ranges $[-100,150]$ GeV$^{-2}$ and $[-100,100]$ GeV$^{-2}$ with the cutoff $\Lambda=0.5$ GeV.  They investigated the results without and with the $\Lambda_c$ as the intermediate state in the TPE loop diagrams in the case of the canonical spin order for the $P_c(4440)$ and $P_c(4457)$, see Figs.~\ref{fig:Results_WithoutWith_Lambdac}(a) and~\ref{fig:Results_WithoutWith_Lambdac}(b). The result with the contributions of the $\Lambda_c$ is better than that of without the $\Lambda_c$. This is a natural consequence of the strong coupling between $\Sigma_c^{(\ast)}$ and $\Lambda_c\pi$ as well as the accidental degeneration of the $\Sigma_c\bar{D}$ and $\Lambda_c\bar{D}^\ast$ (the mass difference $m_{\Sigma_c\bar{D}}-m_{\Lambda_c\bar{D}^\ast}\simeq28\text{ MeV}\ll m_\pi$). With the configuration of the canonical spin order, one gets $|\mathbb{D}_1|\gg|\mathbb{D}_2|$ [see Fig.~\ref{fig:Results_WithoutWith_Lambdac}(b)], which implies that the potential is dominated by the central term and the spin-spin interaction only serves as a `perturbation' to tune the binding energies of the $[\Sigma_c\bar{D}^\ast]_{1/2}$ and $[\Sigma_c\bar{D}^\ast]_{3/2}$ slightly (where the subscript denotes the total angular momentum $J$). {It is worth noting that the role of $\Lambda_c$ is still under debate. For example, it turns out that the inclusion of the $\Lambda_c$ channels does not necessarily lead to improved fits in an updated fit in Ref.~\cite{Du:2021fmf}.}

Wang {\it et al} also explored the possibility of the non-canonical spin order~\cite{Wang:2019ato}, see Fig.~\ref{fig:Results_WithoutWith_Lambdac}(c). It was shown that the experimental data can be equally well reproduced as that in Fig.~\ref{fig:Results_WithoutWith_Lambdac}(b), but one has to largely enhance the spin-spin interaction, such as $|\mathbb{D}_1|\sim|\mathbb{D}_2|$ in this case. From the experience of the $NN$ interaction, the LECs $C_T$ for the spin-spin interaction is much smaller than the $C_S$ for the central interaction, which is a manifestation of the Wigner symmetry in the large $N_c$ limit~\cite{Kaplan:1995yg}. With the help of the quark model as shown in the Appendix of Ref.~\cite{Wang:2019nvm}, one can always relate the contact terms of the $\Sigma_{c}^{(*)}\bar{D}^{(*)}$ to those of the $NN$ systems. Thus, one expects the suppression of the $\mathbb{D}_2$. Therefore, the canonical spin order is favored in Ref.~\cite{Wang:2019ato}.

\begin{figure}
\centering
\begin{minipage}[t]{0.33\linewidth}
\centering
\includegraphics[width=\columnwidth]{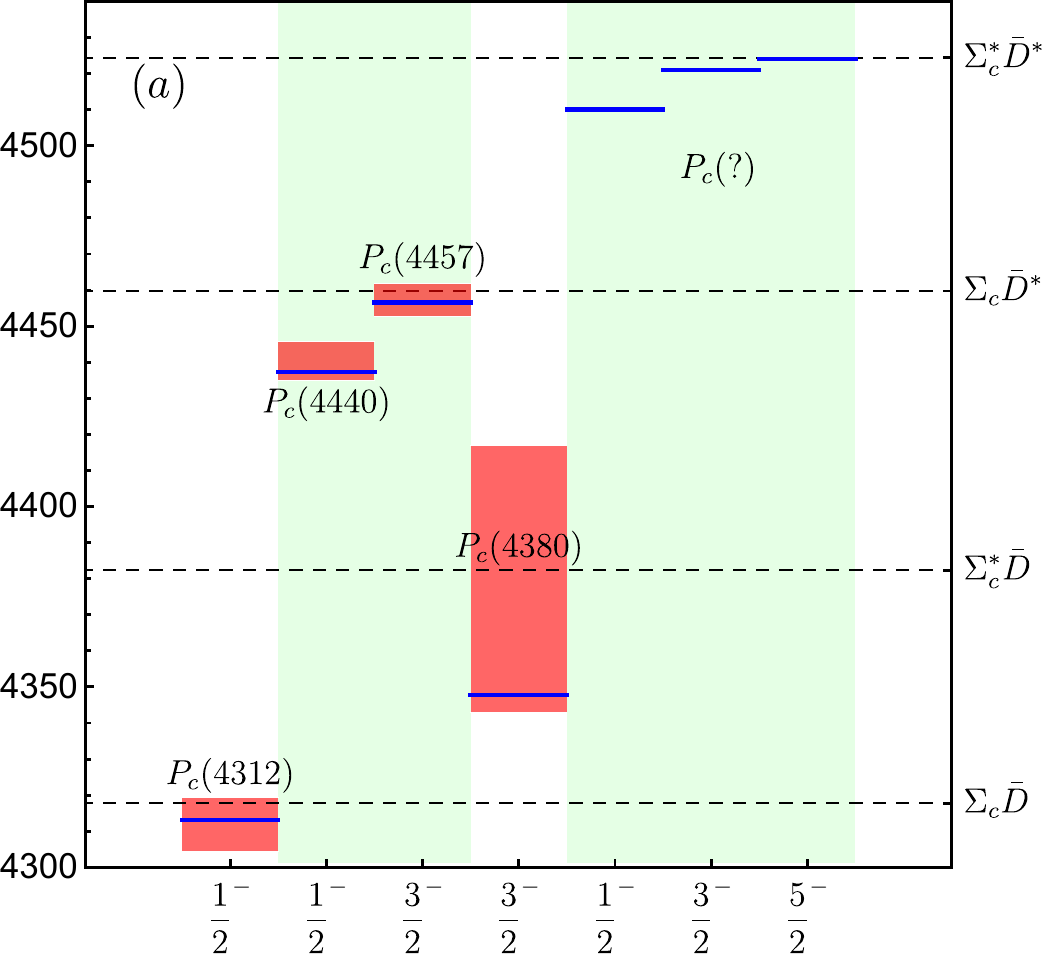}
\end{minipage}%
\hspace{2.5cm}
\begin{minipage}[t]{0.33\linewidth}
\centering
\includegraphics[width=\columnwidth]{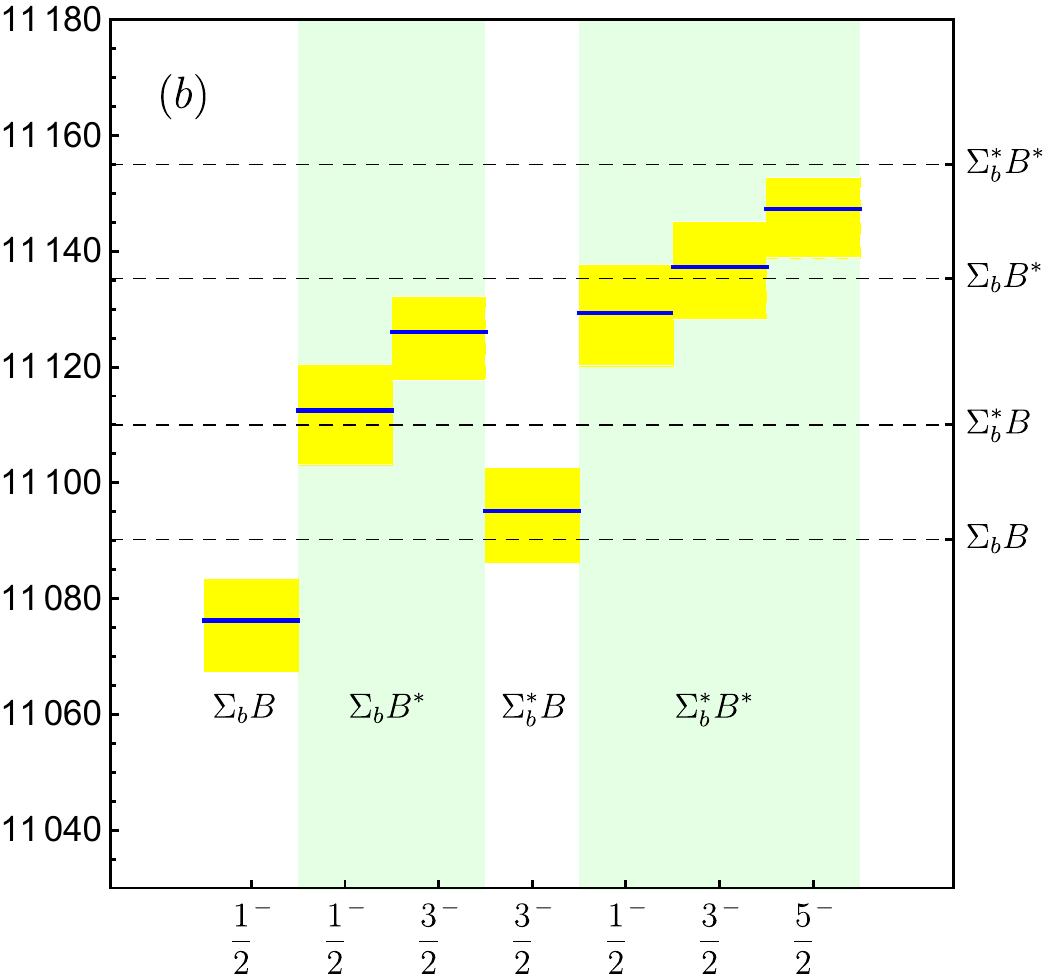}
\end{minipage}
\caption{Left panel: The mass spectrum of the $\Sigma_c^{(\ast)}\bar{D}^{(\ast)}$ molecular pentaquarks. Right panel: The predicted mass spectrum of the $\Sigma_b^{(\ast)}B^{(\ast)}$ molecular pentaquarks. The mass ranges obtained from the experimental
measurements and theoretical estimations are denoted by the red and yellow bands, respectively, while the central values calculated in Ref.~\cite{Wang:2019ato} are represented by the blue solid line. \label{fig:MassSpectraPentaquark}}
\end{figure}

With the fixed $\mathbb{D}_1$ and $\mathbb{D}_2$, the full spectrum of the $\Sigma_c^{(\ast)}\bar{D}^{(\ast)}$ hadronic molecules with $I=1/2$ was established, see Fig.~\ref{fig:MassSpectraPentaquark}(a). The $[\Sigma_c^\ast\bar{D}]_{3/2}$ system is the most deeply bound one with a mass at $4348$ MeV (but this result is sensitive to the cutoff, see Table $3$ of Ref.~\cite{Wang:2019ato}), which might correspond to the previously reported $P_c(4380)$~\cite{LHCb:2015yax} or the recently announced $P_c(4337)$~\cite{LHCb:2021chn} (for alternative explanations of the $P_c(4337)$, see~\cite{Yan:2021nio,Nakamura:2021dix,Wang:2021crr}). A prediction of the possible molecular pentaquarks in the hidden-bottom $\Sigma_b^{(\ast)}B^{(\ast)}$ systems was also given, see Fig.~\ref{fig:MassSpectraPentaquark}(b).

\begin{figure}
 	\centering
 	\includegraphics[width = 0.9\textwidth]{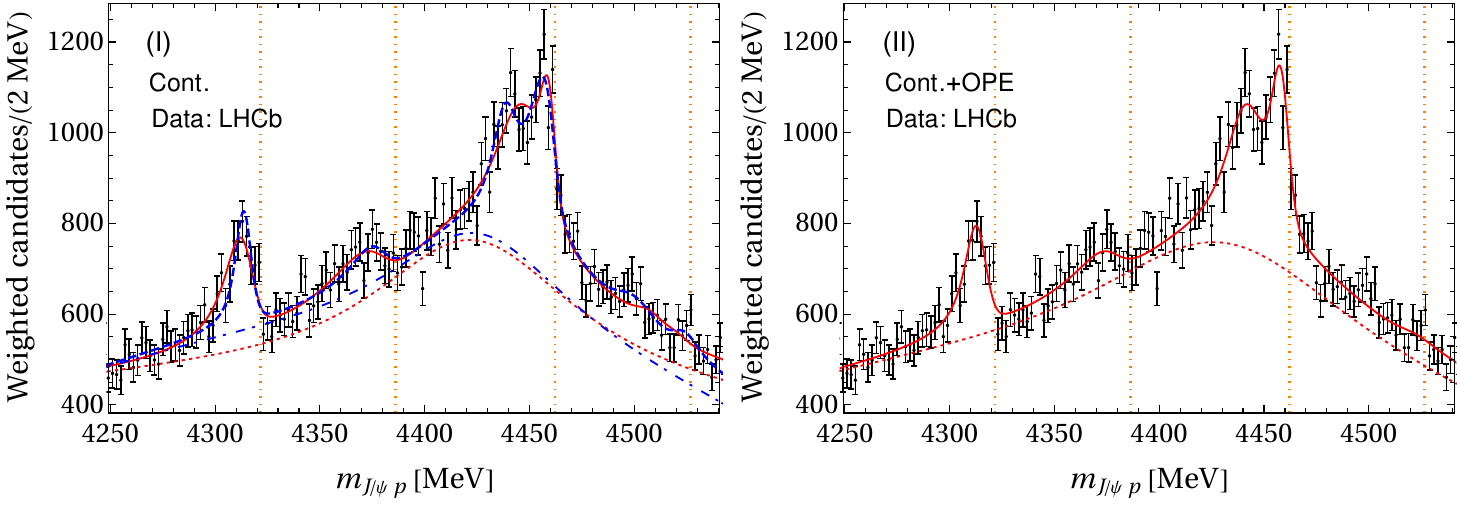}
 	\caption{The fitted invariant mass spectrum of the $J/\psi p$ in Ref.~\cite{Du:2019pij}. The data with error bars are the weighted experimental data from Ref.~\cite{LHCb:2019kea}.  Left panel: Only the contact potentials were considered in the fit. There exist two possible solutions with almost equal $\chi^2/\mathrm{d.o.f}$ and they correspond to canonical (blue dashed curves) and non-canonical (red solid curves) spin orders for the $P_c(4440)$ and $P_c(4457)$. The blue dot-dashed and red dotted curves are the corresponding backgrounds, respectively. Right panel: Apart from the contact terms, the OPE interactions were also included in the fit (shown with the red solid curves).}\label{fig:FitPcDu}
 \end{figure}

The invariant mass spectrum of the $J/\psi p$ was fitted by Du {\it et al}~\cite{Du:2019pij} within the framework of Refs.~\cite{Hanhart:2015cua,Guo:2016bjq,Wang:2018jlv}. The LO contact potentials $V_{\alpha\beta}$ of the elastic channels $\Sigma_c^{(\ast)}\bar{D}^{(\ast)}$ were derived from Eqs.~\eqref{eq:LOcontactJ12}-\eqref{eq:C32alpha}, and the contribution of the inelastic channel $J/\psi p$ was added into the $\mathcal{V}_{\alpha\beta}$ via an imaginary term due to the weak coupling of $J/\psi p$~\cite{Skerbis:2018lew}, which is analogous to Eq.~\eqref{eq:opticpotential}. Besides, the OPE for the $\Sigma_c\bar{D}^\ast$ and $\Sigma_c^\ast\bar{D}^\ast$ systems was considered and treated nonperturbatively. The LSEs~\eqref{eq:prodampela} and~\eqref{eq:proampinela} were regularized with the hard cutoff [see Eq.~\eqref{eq:hardcutoff}], where $\Lambda$ is chosen to be in the range $1.0-1.5$ GeV. The fitted line shape is shown in Fig.~\ref{fig:FitPcDu}. There are two schemes to fit the data: Scheme I---only the LO contact potential was considered; Scheme II---the OPE was incorporated. The results in Scheme I are shown in Fig.~\ref{fig:FitPcDu}(I), where there are two solutions as those of Ref.~\cite{Liu:2019tjn}, i.e., two optional spin orders in Eqs.~\eqref{eq:canonicalorder} and~\eqref{eq:noncanonicalorder}, and the corresponding fitting quality $\chi^2/\mathrm{d.o.f}=1.01$ and $1.03$, respectively. However, including the OPE (Scheme II) gives the unique solution with the slightly improved $\chi^2/\mathrm{d.o.f}=0.98$. In this case, the spin-parity of the $P_c(4440)$ and $P_c(4457)$ abide by the non-canonical spin order. In both schemes, a bound state was produced in the $[\Sigma_c^\ast\bar{D}]_{3/2}$ channel with a narrow width, which qualitatively matches the inconspicuous bump around $4.38$ GeV.
An improved fitting with the additional inelastic $\eta_c p$ and $\Lambda_c\bar{D}^{(\ast)}$ channels was done in Ref.~\cite{Du:2021fmf} (see also Ref.~\cite{Xiao:2020frg}), which supported the conclusion of Ref.~\cite{Du:2019pij}, and predicted the line shapes in the invariant mass spectra of the $\eta_c p$ and $\Lambda_c\bar{D}^{(\ast)}$, respectively.

The results in Refs.~\cite{Wang:2019ato,Yamaguchi:2019seo,Liu:2019tjn,PavonValderrama:2019nbk,Du:2019pij,Du:2021fmf} showed the $P_c(4440)$ and $P_c(4457)$ in the non-canonical spin order imply that the spin order in the $\Sigma^\ast\bar{D}^\ast$ system is also non-canonical, i.e., $m_{[\Sigma^\ast\bar{D}^\ast]_{1/2}}>m_{[\Sigma^\ast\bar{D}^\ast]_{3/2}}>m_{[\Sigma^\ast\bar{D}^\ast]_{5/2}}$. In Refs.~\cite{Yamaguchi:2019seo,PavonValderrama:2019nbk,Du:2019pij,Du:2021fmf}, the OPE was all included with a hard cutoff $\Lambda\gtrsim1.0$ GeV to regularize the iterative equation {with the coupled-channel dynamics}, which is indeed a very hard scale from the experience of the {single-channel} $NN$ interaction~\cite{Epelbaum:2008ga,Machleidt:2011zz}. In the {single-channel} $NN$ case, choosing $\Lambda\sim m_\rho$ or even larger values was already found to result in spurious deeply bound states in the $NN$ systems~\cite{Epelbaum:2014efa}.  Besides, the nuclear lattice simulations of Refs.~\cite{Epelbaum:2009pd,Borasoy:2006qn} also correspond to smaller cutoff values.

The decays of the $P_c$ states into the $J/\psi\Delta$, $J/\psi N$, and $\eta_c N$ were investigated within $\slashed{\pi}$EFT in Refs.~\cite{Guo:2019fdo,Sakai:2019qph}. The $P_c(4457)$ is much closer to the $\Sigma_c\bar{D}^\ast$ threshold than the other $P_c$ states to their corresponding thresholds. The isospin of the $\Sigma_c\bar{D}^\ast$ can either be $1/2$ or $3/2$, which can be expressed with the two particle state,
\begin{eqnarray}
\left[\begin{array}{c}
|\Sigma_{c}\bar{D}^{\ast};I=\frac{1}{2},I_{3}=\frac{1}{2}\rangle\\
|\Sigma_{c}\bar{D}^{\ast};I=\frac{3}{2},I_{3}=\frac{1}{2}\rangle
\end{array}\right]&=&\left[\begin{array}{cc}
\sqrt{\frac{2}{3}} & -\frac{1}{\sqrt{3}}\\
\frac{1}{\sqrt{3}} & \sqrt{\frac{2}{3}}
\end{array}\right]\left[\begin{array}{c}
|\Sigma_{c}^{++}D^{\ast-}\rangle\\
|\Sigma_{c}^{+}\bar{D}^{\ast0}\rangle
\end{array}\right].
\end{eqnarray}
Because the $P_c(4457)$ is very close to the thresholds, the difference of thresholds has a significant effect on the component of the $P_{c}(4457)$. The binding energies of the $P_c(4457)$ with respect to the $\Sigma_{c}^{++}D^{\ast-}$ and $\Sigma_{c}^{+}\bar{D}^{\ast0}$ thresholds read~\cite{Guo:2019fdo},
\begin{eqnarray}
m_{\Sigma_c^{++}D^{\ast-}}-m_{P_c(4457)}=6.9^{+1.8}_{-4.1},\qquad m_{\Sigma_c^{+}\bar{D}^{\ast0}}-m_{P_c(4457)}=2.5^{+1.8}_{-4.2}.
\end{eqnarray}
Since $m_{\Sigma_c^{+}\bar{D}^{\ast0}}-m_{P_c(4457)}\ll m_{\Sigma_c^{++}D^{\ast-}}-m_{P_c(4457)}$, one may expect the significant isospin violating decays for the $P_c(4457)$ as for the $X(3872)$ [for the isospin violating decays of the $X(3872)$, see Sec.~\ref{sec:X-short}]. In Ref.~\cite{Guo:2019fdo}, Guo found that the isospin breaking ratio
\begin{eqnarray}
\mathcal{R}_{\Delta^+/p}=\frac{\mathcal{B}[P_c(4457)^+\to J/\psi\Delta^+]}{\mathcal{B}[P_c(4457)^+\to J/\psi p]},
\end{eqnarray}
is in the range from $\mathcal{O}(10^{-2})$ to $30\%$, where the large uncertainty mainly comes from the mass of the $P_c(4457)$.

With the HQSS inspired potential for the $\Sigma_c^{(\ast)}\bar{D}^{(\ast)}$ systems~\cite{Liu:2019tjn}, Sakai \etal  calculated their $S$-wave decays into the $J/\psi N$ and $\eta_c N$, respectively~\cite{Sakai:2019qph}, i.e.,
\begin{eqnarray}
\left\{[\Sigma_c\bar{D}]_{1/2},[\Sigma_c^\ast\bar{D}]_{3/2},[\Sigma_c\bar{D}^\ast]_{1/2},[\Sigma_c\bar{D}^\ast]_{3/2},[\Sigma_c^\ast\bar{D}^\ast]_{1/2},[\Sigma_c^\ast\bar{D}^\ast]_{3/2}\right\}&\to& J/\psi N,\label{eq:sixPcdecayintoJN}\\
\left\{[\Sigma_c\bar{D}]_{1/2},[\Sigma_c\bar{D}^\ast]_{1/2},[\Sigma_c^\ast\bar{D}^\ast]_{1/2}\right\}&\to& \eta_c N.
\end{eqnarray}
They found that among the six states in Eq.~\eqref{eq:sixPcdecayintoJN}, at least four states (including two $[\Sigma_c^\ast\bar{D}^\ast]_{1/2}$ and $[\Sigma_c^\ast\bar{D}^\ast]_{3/2}$) decay more easily into $J/\psi N$ than the $P_c(4312)$. Meanwhile, the partial width of the $P_c(4312)\to\eta_c N$ is larger than that of the $J/\psi N$. The ratio $\Gamma([\Sigma_c\bar{D}^\ast]_{1/2}\to J/\psi N)/\Gamma([\Sigma_c\bar{D}^\ast]_{1/2}\to \eta_c N)$ differs a lot for the canonical and non-canonical spin orders. See also the related calculations using a quark interchange model~\cite{Wang:2019spc}.

\subsubsection{$P_{cs}$ and partners}

Based on Ref.~\cite{Wang:2019ato}, Wang {\it et al} studied the interactions of the $\Xi_c\bar{D}^{(\ast)}$, $\Xi_c^\prime\bar{D}^{(\ast)}$ and $\Xi_c^\ast\bar{D}^{(\ast)}$ up to NLO~\cite{Wang:2019nvm}, where the flavor symmetry group was enlarged to the $\SU(3)$ [the isospin Pauli matrix $\tau_i$ in Eq.~\eqref{eq:LOcontactHB} should be replaced by $\lambda_i$ in this case]. In order to delineate the short-range interactions between the antitriplet singly heavy baryon $B_{\bar{\bm3}}$ and anticharmed mesons, they constructed the following contact Lagrangian,
\begin{eqnarray}\label{eq:Contact_Lag_B3M}
\mathcal{L}_{\tilde{\mathcal{H}}{B}_{\bar{\bm3}}}&=&\tilde{D}_a^\prime \langle\bar{\tilde{\mathcal{H}}}\tilde{\mathcal{H}}\rangle\mathrm{Tr}(\bar{{B}}_{\bar{\bm3}} {B}_{\bar{\bm3}})+\tilde{D}_b^\prime\langle\bar{\tilde{\mathcal{H}}}\gamma^\rho\gamma_5\tilde{\mathcal{H}}\rangle\mathrm{Tr}(\bar{{B}}_{\bar{\bm3}}\gamma_\rho\gamma_5 {B}_{\bar{\bm3}})\nonumber\\
&&+\tilde{E}_a^\prime\langle\bar{\tilde{\mathcal{H}}}\lambda_i\tilde{\mathcal{H}}\rangle\mathrm{Tr}(\bar{{B}}_{\bar{\bm3}}\lambda_i
{B}_{\bar{\bm3}})+\tilde{E}_b^\prime\langle\bar{\tilde{\mathcal{H}}}\gamma^\rho\gamma_5\lambda_i\tilde{\mathcal{H}}\rangle\mathrm{Tr}(\bar{{B}}_{\bar{\bm3}}\gamma_\rho\gamma_5\lambda_i
{B}_{\bar{\bm3}}),
\end{eqnarray}
from which the contact potentials of the $\Xi_c \bar{D}$ and $\Xi_c\bar{D}^\ast$ are derived as
\begin{eqnarray}
\mathcal{V}_{\Xi_c \bar{D}}&=&2\tilde{D}_a^\prime+4\left(\bm I_1\cdot\bm I_2-1/12\right)\tilde{E}_a^\prime,\\
\mathcal{V}_{\Xi_c\bar{D}^\ast}&=&2\tilde{D}_a^\prime+4\left(\bm I_1\cdot\bm I_2-1/12\right)\tilde{E}_a^\prime+\left[2\tilde{D}_b^\prime+4\left(\bm I_1\cdot\bm I_2-1/12\right)\tilde{E}_b^\prime\right]\boldsymbol{\sigma}\cdot\bm{T}.\label{eq:PpXicDast}
\end{eqnarray}
The potentials of the $\Xi_c^\prime \bar{D}^{(\ast)}$ and $\Xi_c^\ast\bar{D}^{(\ast)}$ are similar to those of the $\Sigma_c\bar{D}^{(\ast)}$ and $\Sigma_c^\ast\bar{D}^{(\ast)}$, respectively [the $\bm I_1\cdot\bm I_2$ in Eqs.~\eqref{eq:LO_SigmacDb_Con}--\eqref{eq:X41_Potential} should be replaced by $(\bm I_1\cdot\bm I_2-1/12)$].
The HQSS cannot be used to relate the $\tilde{D}_a^\prime,\dots,\tilde{E}_a^\prime$ to the $\tilde{D}_a,\dots,\tilde{E}_a$, since the $B_{\bar{\bm3}}$ and $B_{\bm6}^{(\ast)}$ are not HQSS partners. In Refs.~\cite{Meng:2019nzy,Wang:2019nvm}, Meng {\it et al} proposed a `microscopic' approach to build connections between the LECs in different systems containing the light quarks, e.g., Eqs.~\eqref{eq:Contact_Lag_B3M} and~\eqref{eq:LOcontactHB}. Such an approach is analogous to the resonance saturation model~\cite{Epelbaum:2001fm}, but carries the concepts of the quark-hadron duality and quark model. They constructed a quark level Lagrangian,
\begin{eqnarray}\label{eq:QLlAG}
\mathcal{L}_{qq}=\tilde{g}_s\bar{q}\mathcal{S}q+\tilde{g}_a\bar{q}\gamma_\mu\gamma^5\mathcal{A}^\mu q,
\end{eqnarray}
where $q=(u,d,s)^T$. Unlike the OBE model, the exchanged particles are not specified (e.g, $\rho,\omega,f_0,\dots$). The scalar field $\mathcal{S}$ and axial-vector field $\mathcal{A}^\mu$ are two fictitious fields ($\tilde{g}_s$ and $\tilde{g}_a$ are the corresponding coupling constants). They are introduced to produce the central potential and spin-spin interaction between two light quarks, respectively. They are assumed to take the same matrix form as that of Eq.~\eqref{eq:Upionmatrix}, but carry the quantum numbers of the scalar and axial-vector particles, respectively.  {The second terms is analogous to the axial-vector coupling of the light quarks and Goldstone bosons which stem from the chiral symmetry, e.g., see Sec.~\ref{sec:CSSB}. The toy Lagrangian in Eq.~\eqref{eq:QLlAG} is introduced to predict the spectrum of $P_{cs}$ states with the data of the $P_c$ states as input. In principle, one should also include the vector-exchange interactions. However,  if we only focus on the HQSS symmetry, the unit operator and spin-spin operator in the spin space have been included by the two terms in Eq.~\eqref{eq:QLlAG} after nonrelativistic reduction. The effect of the vector meson exchange interaction will be absorbed by the present two terms to some extent. For the predictions of the heavy flavor hadronic molecules using the above toy Lagrangian, see Refs.~\cite{Chen:2021cfl,Chen:2021spf}}.

With the Lagrangian~\eqref{eq:QLlAG}, the short-range interaction of the $\Sigma_c\bar{D}^\ast$ system can be written as
\begin{eqnarray}\label{eq:Pquark}
\mathcal{V}_{\Sigma_c\bar{D}^\ast}=\langle\Sigma_c\bar{D}^\ast|\mathcal{L}_{\bar{q}q}|\Sigma_c\bar{D}^\ast\rangle=-\frac{\tilde{g}_s^2}{6m_{\mathcal{S}}^2}-\frac{\tilde{g}_s^2}{m_{\mathcal{S}}^2}(\bm{I}_1\cdot\bm{I}_2)
+\frac{\tilde{g}_a^2}{9m_{\mathcal{A}}^2}\bm{\sigma}\cdot\bm{T}+\frac{2\tilde{g}_a^2}{3m_{\mathcal{A}}^2}(\bm{\sigma}\cdot\bm{T})(\bm{I}_1\cdot\bm{I}_2),
\end{eqnarray}
where $q^2\ll m_{\mathcal{S},\mathcal{A}}^2$ is assumed and $m_{\mathcal{S}}~(m_{\mathcal{A}})$ denotes the effective mass of the $\mathcal{S}~(\mathcal{A}^\mu)$ field. Therefore, if one determines the square of the ``charge-to-mass ratios" $\tilde{g}_s^2/m_{\mathcal{S}}^2$ and $\tilde{g}_a^2/m_{\mathcal{A}}^2$, one could use them to determine the short-range potentials of the other systems containing light quarks. The $\tilde{g}_s^2/m_{\mathcal{S}}^2$ and $\tilde{g}_a^2/m_{\mathcal{A}}^2$ are fixed by matching Eqs.~\eqref{eq:Pquark} and~\eqref{eq:X21_Potential} [also using the redefinition of Eq.~\eqref{eq:LECs_redefine}],
\begin{eqnarray}
\frac{\tilde{g}_s^2}{m_{\mathcal{S}}^2}=-\frac{6}{5}\mathbb{D}_1,\qquad\qquad \frac{\tilde{g}_a^2}{m_{\mathcal{S}}^2}=\frac{6}{5}\mathbb{D}_2,
\end{eqnarray}
where the $\mathbb{D}_1$ and $\mathbb{D}_2$ are fixed using the data of the $P_c$ states. Similarly, one can calculate the contact potentials of the $\Xi_c^\prime\bar{D}^\ast$ and $\Xi_c\bar{D}^\ast$ with the quark level Lagrangian~\eqref{eq:QLlAG}, and then match with the potentials obtained with the Lagrangians~\eqref{eq:LOcontactHB} and~\eqref{eq:Contact_Lag_B3M} in the $\SU(3)$ case, respectively (for more details, see Refs.~\cite{Meng:2019nzy,Wang:2019nvm}).

Using the approach described above and the $P_c$ data as inputs, Wang \etal predicted ten molecular pentaquarks in the isoscalar  $\Xi_c\bar{D}^{(\ast)}$, $\Xi_c^\prime\bar{D}^{(\ast)}$ and $\Xi_c^\ast\bar{D}^{(\ast)}$ systems~\cite{Wang:2019nvm}, see Table~\ref{tab:Csyslabel2}. The binding solutions in the $\Xi_c\bar{D}^{(\ast)}$ systems were also supported by the following works~\cite{Xiao:2019gjd,Liu:2020hcv,Xiao:2021rgp,Peng:2020hql}. The authors of Ref.~\cite{Wang:2019nvm} found that the contribution of the OEE is rather feeble, but the contribution of the TPE is very significant for the $\Xi_c^\prime\bar{D}$ and $\Xi_c\bar{D}^\ast$ systems due to their approximate accidental degeneration, e.g., $m_{\Xi_c\bar{D}^\ast}-m_{\Xi_c^\prime\bar{D}}\simeq32$ MeV$\ll m_\pi$. They also found that no bound states exist in the isovector channels. They proposed to search for these isoscalar strange hidden-charm molecular pentaquarks in the $J/\psi\Lambda$ invariant mass spectrum from the decays $\Lambda_b(\Xi_b)\to J/\psi\Lambda K(\eta)$~\cite{Lu:2016roh,Feijoo:2015kts,Chen:2015sxa}. There do not exist bound states in the $\Lambda_c\bar{D}_s^{(\ast)}$,  $\Sigma_c^{(\ast)}\bar{D}_s^{(\ast)}$ and $\Omega_c^{(\ast)}\bar{D}_s^{(\ast)}$ systems~\cite{Wang:2019nvm}. The non-existence of the binding solution of the $\Lambda_c\bar{D}^{(\ast)}$ systems were pointed out long time ago in Refs.~\cite{Wang:2011rga,Yang:2011wz}. As for the $\Omega_c^{(\ast)}\bar{D}_s^{(\ast)}$ system, the bound states were obtained with a large cutoff in the OBE model in Ref.~\cite{Wang:2021hql}.

Very recently, the LHCb Collaboration reported a structure in the $J/\psi\Lambda$ invariant mass spectrum in the $\Xi_b^-\to J/\psi\Lambda K^-$ decay with a significance of $3.1~\sigma$~\cite{LHCb:2020jpq}. The mass and width were measured to be $(m,\Gamma)=(4458.8\pm2.9^{+4.7}_{-1.1},17.3\pm6.5^{+8.0}_{-5.7})$ MeV. The mass of this structure is very consistent with that of the predicted state $[\Xi_c\bar{D}^\ast]_{1/2}$ in Ref.~\cite{Wang:2019nvm}, see Table~\ref{tab:Csyslabel2}. There are two states in the $\Xi_c\bar{D}^\ast$ system with the total angular momentum $J=1/2$ and $3/2$, respectively. The LHCb also tested the two-peak hypothesis, and found that in this case the masses and widths of these two states are $(m,\Gamma)=(4454.9\pm2.7,7.5\pm9.7)$ MeV and $(4467.8\pm3.7,5.2\pm5.3)$ MeV, respectively. The data obtained with two peaks also support the predictions in Ref.~\cite{Wang:2019nvm}, but the analysis of the current
data sample cannot confirm or refute the two-peak hypothesis~\cite{LHCb:2020jpq}.
\begin{table*}[htbp]
\centering
\renewcommand{\arraystretch}{1.5}
\caption{The binding energies $\Delta E$ and masses $M$ for the isoscalar $[\Xi_c^\prime\bar{D}^{(\ast)}]_J$, $[\Xi_c^\ast\bar{D}^{(\ast)}]_J$ and $[\Xi_c\bar{D}^{(\ast)}]_J$ (where the subscript ``$J$" denotes the total spin of the system) systems predicted in Ref.~\cite{Wang:2019nvm}. The state denoted by ``$\sharp$" may be nonexistent at the upper limit. The results are given in units of MeV. \label{tab:Csyslabel2}}
\setlength{\tabcolsep}{0.4mm}
{
\begin{tabular}{c|cccccccccc}
\hline
System&$[\Xi_c^\prime\bar{D}]_{\frac{1}{2}}$&$[\Xi_c^\prime\bar{D}^\ast]_{\frac{1}{2}}$&$[\Xi_c^\prime\bar{D}^\ast]_{\frac{3}{2}}$&$[\Xi_c^\ast\bar{D}]_{\frac{3}{2}}$&$[\Xi_c^\ast\bar{D}^\ast]_{\frac{1}{2}}$
&$[\Xi_c^\ast\bar{D}^\ast]_{\frac{3}{2}}$&$[\Xi_c^\ast\bar{D}^\ast]_{\frac{5}{2}}^\sharp$&$[\bm{\Xi_c\bar{D}}]_{\frac{1}{2}}$&$[\bm{\Xi_c\bar{D}^\ast}]_{\frac{1}{2}}$&$[\bm{\Xi_c\bar{D}^\ast}]_{\frac{3}{2}}$\\
\hline
$\Delta E$&$-18.5^{+6.4}_{-6.8}$&$-15.6^{+6.4}_{-7.2}$&$-2.0^{+1.8}_{-3.3}$&$-7.5^{+4.2}_{-5.3}$&$-17.0^{+6.7}_{-7.5}$&$-8.0^{+4.5}_{-5.6}$&$-0.7^{+0.7}_{-2.2}$&$-13.3^{+2.8}_{-3.0}$&$-17.8^{+3.2}_{-3.3}$&$-11.8^{+2.8}_{-3.0}$\\
$M$&$4423.7^{+6.4}_{-6.8}$&$4568.7^{+6.4}_{-7.2}$&$4582.3^{+1.8}_{-3.3}$&$4502.9^{+4.2}_{-5.3}$&$4635.4^{+6.7}_{-7.5}$&$4644.4^{+4.5}_{-5.6}$&$4651.7^{+0.7}_{-2.2}$&$4319.4^{+2.8}_{-3.0}$&$4456.9^{+3.2}_{-3.3}$&$4463.0^{+2.8}_{-3.0}$\\
\hline
\end{tabular}
}
\end{table*}

\begin{figure}
\centering
\begin{tabular}{ll}
\includegraphics[width=0.4\textwidth,angle=-0]{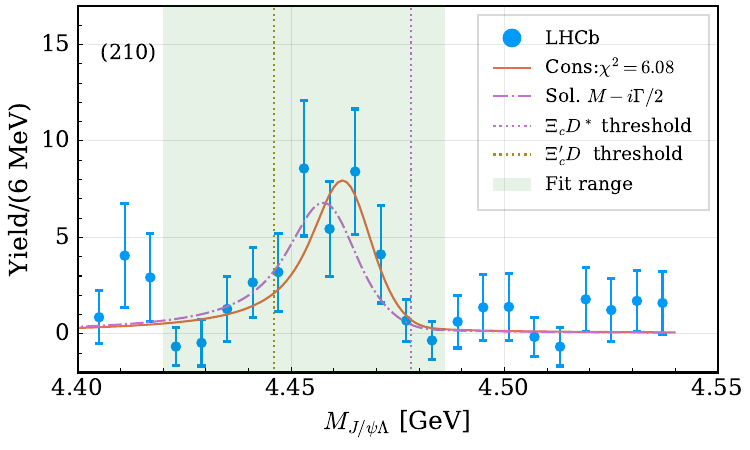}
& \includegraphics[width=0.4\textwidth,angle=-0]{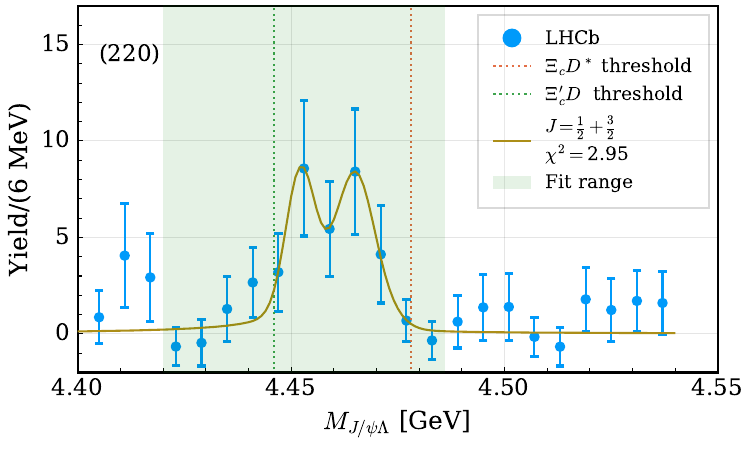}\\
\includegraphics[width=0.4\textwidth,angle=-0]{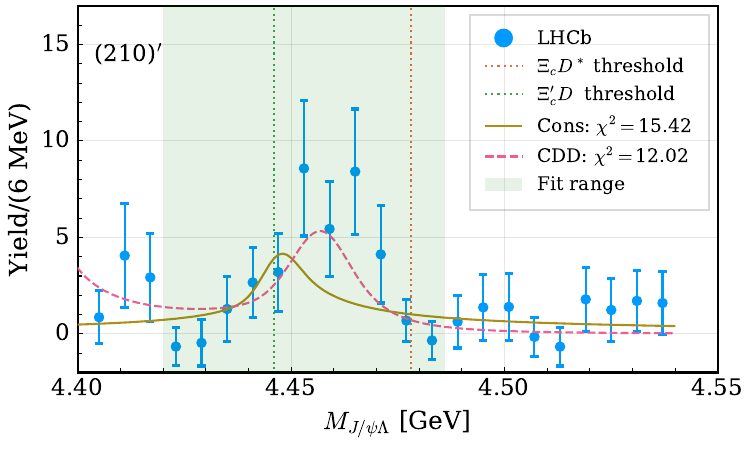}
& \includegraphics[width=0.4\textwidth,angle=-0]{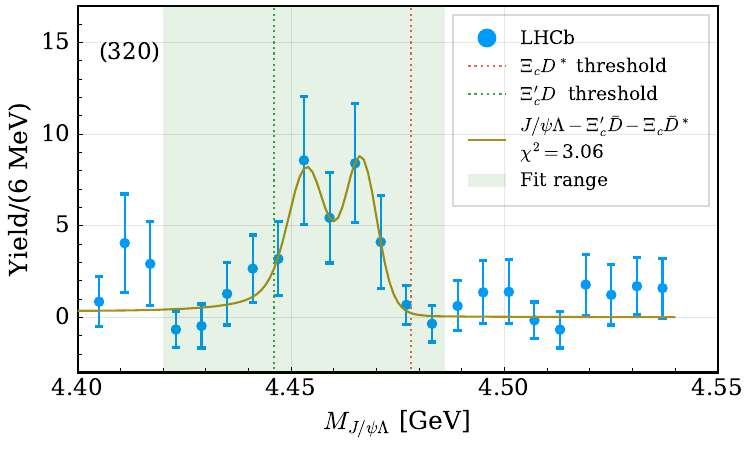}
\\
\end{tabular}
\caption{The fitted invariant mass spectrum of the $J/\psi\Lambda$ in Ref.~\cite{Du:2021bgb}, where the data with error bars are taken from Ref.~\cite{LHCb:2020jpq}. The greenish area corresponds to the energy region used in the fit. The two dotted vertical lines from left to right correspond to the $\Xi_c'\bar{D}$ and $\Xi_c\bar{D}^*$ thresholds, respectively. The indications of the fitted curves are marked in each subfigure, and more explanations are given in the context. \label{fig:FitPcsDu}}
\end{figure}

Following the experiments of LHCb~\cite{LHCb:2020jpq}, the newly observed $P_{cs}(4459)$ was systematically studied in Ref.~\cite{Du:2021bgb} with three closely connected methods: the effective range expansion~(see Sec.~\ref{sec:ere}), the extended Weinberg's compositeness relation~\cite{Guo:2015daa}, and fitting the line shape within $\slashed{\pi}$EFT, see Fig.~\ref{fig:FitPcsDu}. The authors considered the interplay between the three channels, $J/\psi\Lambda$, $\Xi_c^\prime\bar{D}$ and $\Xi_c\bar{D}^\ast$, where their interactions are constrained by the HQSS. The two particle states $\Xi_c^\prime\bar{D}$ and $\Xi_c\bar{D}^\ast$ are expanded with the $|S_h\otimes S_\ell\rangle$ basis as
\begin{eqnarray}
|\Xi_{c}^{\prime}\bar{D}\rangle_{J=\frac{1}{2}}&=&-\frac{1}{2}|0_{h}\otimes\frac{1}{2}_{\ell}\rangle+\frac{\sqrt{3}}{2}|1_{h}\otimes\frac{1}{2}_{\ell}\rangle,\\
|\Xi_{c}\bar{D}^{\ast}\rangle_{J=\frac{1}{2}}&=&\frac{\sqrt{3}}{2}|0_{h}\otimes\frac{1}{2}_{\ell}\rangle-\frac{1}{2}|1_{h}\otimes\frac{1}{2}_{\ell}\rangle,\\
|\Xi_{c}\bar{D}^{\ast}\rangle_{J=\frac{3}{2}}&=&|1_{h}\otimes\frac{1}{2}_{\ell}\rangle.
\end{eqnarray}
According to Sec.~\ref{sec:HQSinHHM}, there is only one LEC in principle, i.e., the $\mathcal{C}_{1/2}^\alpha$. In Ref.~\cite{Du:2021bgb}, the following $S$-wave channel couplings were considered,
\begin{eqnarray}\label{eq:threecasesCouple}
 \mathrm{I}:~J/\psi\Lambda-\Xi_{c}\bar{D}^{\ast},~J=\frac{1}{2},\frac{3}{2}\qquad \mathrm{II}:~J/\psi\Lambda-\Xi_{c}^{\prime}\bar{D},~J=\frac{1}{2}\qquad \mathrm{III}:~J/\psi\Lambda-\Xi_{c}^{\prime}\bar{D}-\Xi_{c}\bar{D}^{\ast},~J=\frac{1}{2},
\end{eqnarray}
where $J$ denotes the total angular momentum of the coupled system. The corresponding effective potentials for the $J=1/2~(3/2)$ in the two-channel and three-channel couplings may be read from Ref.~\cite{Du:2021bgb}. The fitted line shapes of the three cases in Eq.~\eqref{eq:threecasesCouple} are shown in Fig.~\ref{fig:FitPcsDu}, where each subfigure is marked with $(abc)$, and $a,b,c$ refer to
\begin{eqnarray}
&&a-\text{the number of channels}: 2\equiv(J/\psi \Lambda-\Xi_c\bar{D}^\ast\text{ or }J/\psi \Lambda-\Xi_c^\prime\bar{D}),\quad 3\equiv(J/\psi\Lambda-\Xi_{c}^{\prime}\bar{D}-\Xi_{c}\bar{D}^{\ast}),\\
&&b-\text{the number of partial waves}:1\equiv(J=\frac{1}{2}\text{ or }J=\frac{3}{2}),\quad 2\equiv(J=\frac{1}{2}\text{ and }J=\frac{3}{2}),\\
&&c-\text{without or with the energy-dependent term in the potentials}:0\equiv(\text{without}),\quad 1\equiv(\text{with}).
\end{eqnarray}
In Figs.~\ref{fig:FitPcsDu}$(210)$ and~\ref{fig:FitPcsDu}$(220)$, the authors considered the two-channel $J/\psi \Lambda-\Xi_c\bar{D}^\ast$ coupling. The dot-dashed line in Fig.~\ref{fig:FitPcsDu}$(210)$ is obtained by solving the pole of the $T$-matrix with the mass and width of the $P_{cs}(4459)$ as inputs, while the solid line comes from the fit. The authors also introduced an energy-dependent term to mimic the role of the CDD pole in the case of the $\Xi_c\bar{D}^\ast$ single-channel scattering, but the resulting outputs are unacceptable, which implies the dynamically generated nature of the $P_{cs}(4459)$. In Fig.~\ref{fig:FitPcsDu}$(220)$, the fit result of including the $J=3/2$ partial wave is shown. The result is better than that of Fig.~\ref{fig:FitPcsDu}$(210)$. There are two peaks in the line shape, which correspond to two poles of the $T$-matrix. In Fig.~\ref{fig:FitPcsDu}$(210)^\prime$, the authors considered the $J/\psi \Lambda-\Xi_c^\prime\bar{D}$ coupling. They tested permitting constant terms (the solid line) and CDD pole (dashed line) in the potential. Neither of them can well reproduce the data. In Fig.~\ref{fig:FitPcsDu}$(320)$, the result with the three-channel $J/\psi\Lambda-\Xi_{c}^{\prime}\bar{D}-\Xi_{c}\bar{D}^{\ast}$ coupling is illustrated. The fitting result contains two peaks as that of~\ref{fig:FitPcsDu}$(220)$. They found the role of the $\Xi_c^\prime\bar{D}$ channel is perturbative but it leads to the mass splitting between $J=1/2$ and $J=3/2$ states (see also Ref.~\cite{Peng:2020hql}). {From an overview of Fig.~\ref{fig:FitPcsDu}, the two fits in the right column are much better than those in the left one, which indicates both the partial waves $J=1/2$ and $J=3/2$ are important. However, in the signal region, the fits in the right column are both far too good. The lines almost go through all data points, which indicates it is hard to tell the different coupled-channel dynamic mechanisms forming the $P_{cs}(4459)$ under the present statistic significance.} They also found the $J=1/2$ and $J=3/2$ states obey the canonical spin order, i.e., $m_{[\Xi_c\bar{D}^\ast]_{1/2}}<m_{[\Xi_c\bar{D}^\ast]_{3/2}}$, and the corresponding masses are consistent with those from the two-peak hypothesis of LHCb~\cite{LHCb:2020jpq}, but with smaller widths.

\subsection{Other similar systems}
\label{sec:othersystems}

In the above sections, we mainly focused on the experimentally observed molecular candidates. However, there exist theoretical predictions of many other hadronic molecules. Some of these predicted states may potentially be observed in the future experiments. Searching for these states would help us to assemble the jigsaw puzzles of the {\it hadronic molecular physics}. Therefore, we briefly review these efforts in this section.

It is instructive to enlarge the basis of the heavy matter fields. We consider the ground-state heavy matter fields containing light quarks, which include the nucleon\footnote{Sometimes, the $K^\ast$ meson may be regarded as the heavy matter field to some extent ($m_{K^\ast}\sim m_N$). The $X_0(2900)$ and $X_1(2900)$ observed by the LHCb are very close to the $\bar{D}^\ast K^\ast$ threshold~\cite{LHCb:2020bls,LHCb:2020pxc}. They are explained as the $S$- and $P$-wave $\bar{D}^\ast K^\ast$ molecules, respectively~\cite{Wang:2021lwy}.}, singly heavy mesons, singly heavy baryons, doubly heavy baryons and their antiparticles. Their possible combinations may lead to the molecular matrix with the following form,

\begin{eqnarray}\label{eq:HMFantiHMF}
\left[\begin{array}{c}
\mathtt{N}\\
\mathtt{M}_{\mathtt{Q}}\\
\mathtt{B}_{\mathtt{Q}}\\
\mathtt{B}_{\mathtt{QQ}}
\end{array}\right]&\bm{\hat{\otimes}}&\left[\begin{array}{c}
\mathtt{N}\\
\mathtt{M}_{\mathtt{Q}}\\
\mathtt{B}_{\mathtt{Q}}\\
\mathtt{B}_{\mathtt{QQ}}\\
\bar{\mathtt{N}}\\
\bar{\mathtt{M}}_{\mathtt{Q}}\\
\bar{\mathtt{B}}_{\mathtt{Q}}\\
\bar{\mathtt{B}}_{\mathtt{QQ}}
\end{array}\right]^T\Rightarrow\left[\begin{array}{cccccccc}
\mathtt{NN} & \mathtt{NM_{Q}} & \mathtt{NB_{Q}} & \mathtt{NB_{QQ}} & \mathtt{N\bar{N}}\\
 & \mathtt{M_{Q}M_{Q}} & \mathtt{M_{Q}B_{Q}} & \mathtt{M_{Q}B_{QQ}} & \mathtt{M_{Q}\bar{N}} & \mathtt{M_{Q}\bar{M}_{Q}}\\
 &  & \mathtt{B_{Q}B_{Q}} & \mathtt{B_{Q}B_{QQ}} & \mathtt{B_{Q}\bar{N}} & \mathtt{B_{Q}\bar{M}_{Q}} & \mathtt{B_{Q}\bar{B}_{Q}}\\
 &  &  & \mathtt{B_{QQ}B_{QQ}} & \mathtt{B_{QQ}\bar{N}} & \mathtt{B_{QQ}\bar{M}_{Q}} & \mathtt{B_{QQ}\bar{B}_{Q}} & \mathtt{B_{QQ}\bar{B}_{QQ}}
\end{array}\right]\nonumber\\
&\Rightarrow&\left[\begin{array}{cccccccc}
\mathrm{deuteron} & \Lambda_c(2940)?,\dots & \Box & \Box & X(1835)?\\
 & T_{cc},\dots & \Box & \Box & \Box & X(3872),Z_{c(s)}^{(\prime)},Z_{b}^{(\prime)},\dots\\
 &  & \Box & \Box & \Box & P_{c(s)},\dots & \Box\\
 &  &  & \Box & \Box & \Box & \Box & \Box
\end{array}\right],
\end{eqnarray}
where we use the $\mathtt{N}$, $\mathtt{M}_{\mathtt{Q}}$, $\mathtt{B}_{\mathtt{Q}}$, and $\mathtt{B}_{\mathtt{QQ}}$ to denote the sets of the nucleons, singly heavy mesons, singly heavy baryons and doubly heavy baryons in order, while the overhead bar represents their antiparticles. The charge conjugate is implied in the matrix of the first line, e.g., we only show the $\mathtt{M_Q\bar{N}}$ explicitly, while its charge conjugate $\mathtt{N\bar{M}_Q}$ is not given. The states in the matrix of the second line stand for the observed molecular candidates in the corresponding combinations, and the ellipses indicate that there might exist more states. The empty box denotes the observation in the corresponding combinations is still absent in experiments. If the future experiments could further fill up the empty boxes in Eq.~\eqref{eq:HMFantiHMF} or give crucial evidences for the molecular nature of the observed states, the general pattern of the molecular spectrum would provide crucial insight into the underlying dynamics of the hadronic molecules.

\subsubsection{$ND^{(\ast)}$, $N\Sigma_c^{(\ast)}$, $N\Xi_{cc}^{(\ast)}$} \label{sec:5.8.1}
\vspace{0.2cm}
\noindent
{\it $ND^{(\ast)}$ systems}
\vspace{0.2cm}

The interactions of the $ND$ and $ND^\ast$ are closely related to the inner structures of the charmed baryons $\Sigma_c(2800)$ and $\Lambda_c(2940)$, since these two states are very close to the thresholds of the $ND$ and $ND^\ast$ ($m_{ND}=2805$ MeV, $m_{ND^\ast}=2946$ MeV), respectively. The $I(J^P)$ quantum numbers of the $\Sigma_c(2800)$ and $\Lambda_c(2940)$ are  $1(?^?)$ and $0(\frac{3}{2}^-)$ [note that the $J=3/2$ for $\Lambda_c(2940)$ is favored but not certain], respectively~\cite{ParticleDataGroup:2022pth}. Especially for the $\Lambda_c(2940)$, if it is the $2P$ state in the charmed baryon spectroscopy, its mass is about $60-100$ MeV smaller than the quark model calculations~\cite{Capstick:1986ter,Ebert:2011kk,Chen:2014nyo,Lu:2016ctt}, which is very similar to what happened for the $\Lambda(1405)$, $D_{s0}^\ast(2317)$ and $X(3872)$. For more details, see Sec.~\ref{sec1.2} and the extensive discussions in the recent review~\cite{Chen:2022asf}. Moreover, understanding the $ND^{(\ast)}$ interactions is important for investigating the charmed mesic nuclei~\cite{Tsushima:1998ru,Garcia-Recio:2010fiq} and the properties of the charmed mesons in the nuclear matter~\cite{Hosaka:2016ypm,Krein:2017usp}. Haidenbauer {\it et al} have constructed the $N\bar{D}$ and $ND$ interactions based on the meson-exchange model~\cite{Haidenbauer:2007jq,Haidenbauer:2008ff,Haidenbauer:2010ch}.

In Ref.~\cite{Wang:2020dhf}, Wang {\it et al} studied the $ND^{(\ast)}$ interactions in $\chi$EFT up to NLO, where the contribution of the $\Delta(1232)$ resonance was also included in the loops of the TPE. The LO LECs are fixed from the $N\bar{N}$ interactions~\cite{Kang:2013uia} with the help of quark model and SU(3) symmetry as in Eq.~\eqref{eq:QLlAG}. Three singly charmed molecular pentaquarks in the isoscalar $ND^{(\ast)}$ [as well as the $\bar{B}^{(\ast)}N$] systems were predicted. However, there are no bound states in the isovector channels, which supports the interpretation of the $\Sigma_c(2800)$ as the compact charmed baryon. In Ref.~\cite{Sakai:2020psu}, by fitting the invariant mass spectrum of the $pD^0$ in the decay $\Lambda_b\to pD^0\pi^-$\cite{LHCb:2017jym}, Sakai {\it et al} extracted the scattering length of the $ND$ system ($nD^+$ and $pD^0$) in the coupled-channels~\cite{Cohen:2004kf}. They found the scattering length in the isovector channel is very large, and obtained a bound state pole in the isospin symmetry limit, which is assigned as the $\Sigma_c(2800)$.

If the $\Lambda_c(2940)$ is indeed the molecule of the $ND^\ast$~\cite{Wang:2020dhf}, the observed peak of the $\Lambda_c(2940)$ by BaBar~\cite{BaBar:2006itc}, Belle~\cite{Belle:2006xni} and LHCb~\cite{LHCb:2017jym} should contain two subpeaks with $J=1/2$ and $3/2$, respectively (recall that the similar story has happened for the $P_c$ states after increasing the data sample, e.g., see Sec.~\ref{sec:PcNoStrangeness}). In this case, in contrast to the $J=3/2$ one, the $J=1/2$ component can easily decay into the $pD^0$ and $\Sigma_c\pi$ modes via the $S$-wave. 

{In Ref.~\cite{Zhang:2022pxc}, the authors considered the interplay between the compact $udc$ core and $D^\ast N$ channel in an unquenched framework. They interpreted the recently observed $\Lambda_c(2910)$~\cite{Belle:2022hnm} and the $\Lambda_c(2940)$ as the conventional charmed baryons but dressed with the $D^\ast N$ channel. They also showed that the spin-parity of $\Lambda_c(2910)$ prefers $3/2^-$, while the $\Lambda_c(2940)$ is more likely to be $J^P=1/2^-$ state}.


\vspace{0.2cm}
\noindent
{\it $N\Sigma_c^{(\ast)}$ systems}
\vspace{0.2cm}

The $NY_c$ ($Y_c=\Sigma_c,\Lambda_c$) interactions are essential for understanding the charmed hypernuclei~\cite{Bhamathi:1981yu,Bando:1983yt,Starkov:1986ye} and the in-medium properties of the charmed baryons~\cite{Hosaka:2016ypm}. In recent years, the experimental proposals at the J-PARC~\cite{Fujioka:2017gzp} and GIS-FAIR~\cite{Wiedner:2011mf} have stimulated many investigations on the $NY_c$ interactions and charmed hypernuclei~\cite{Garcilazo:2015qha,Vidana:2019amb,Liu:2011xc,Maeda:2015hxa,Maeda:2018xcl,Huang:2013zva,Garcilazo:2019ryw,Miyamoto:2017tjs,Haidenbauer:2017dua}. In Refs.~\cite{Miyamoto:2017tjs,Miyamoto:2017ynx}, the HAL QCD Collaboration calculated the phase shifts of the $N\Lambda_c$ and $N\Sigma_c$ scatterings from lattice QCD at the unphysical pion mass $m_\pi=410-570$ MeV. The extrapolations of lattice QCD results to the physical pion mass were done in Refs.~\cite{Haidenbauer:2017dua,Meng:2019nzy,Haidenbauer:2020uci,Haidenbauer:2020kwo,Song:2020isu,Haidenbauer:2021tlk,Song:2021war} with (covariant) $\chi$EFT. It was shown that the attractive interaction of the $N\Lambda_c$ is too moderate to form a bound state in this system\footnote{However, the phenomenological calculations in Refs.~\cite{Liu:2011xc,Maeda:2015hxa} yielded bound states in the $N\Lambda_c$ system with $J^P=0^+$ and $1^+$ once the coupled-channel effect among $\Lambda_c$, $\Sigma_c$ and $\Sigma_c^\ast$ are taken into account.}, while the bound states in the $4N\Lambda_c$ or $5N\Lambda_c$ might be possible~\cite{Haidenbauer:2017dua}.

\begin{figure*}
\centering
\begin{tabular}{lll}
\includegraphics[width=0.3\textwidth,angle=-0]{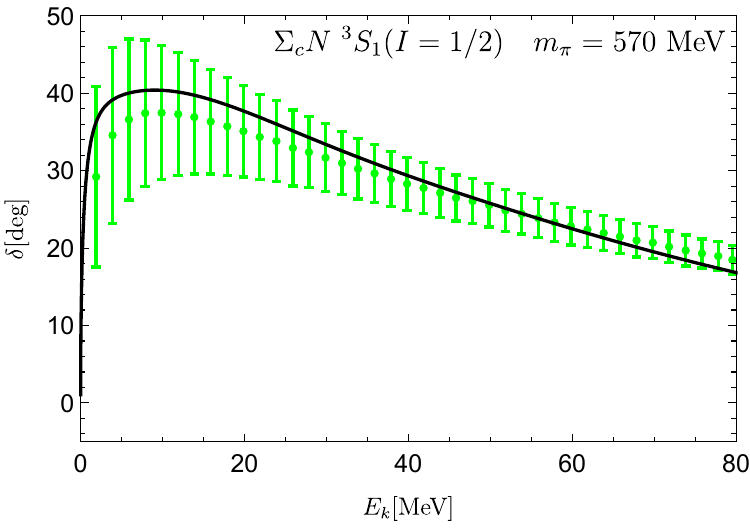}
&\includegraphics[width=0.3\textwidth,angle=-0]{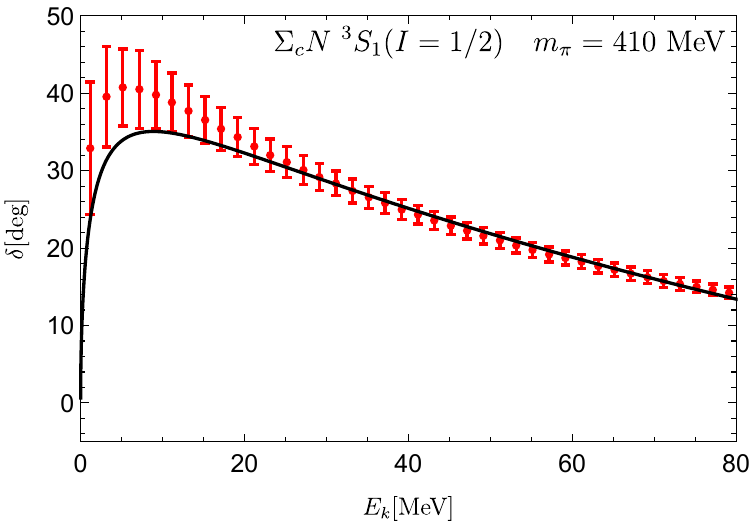}
&\includegraphics[width=0.3\textwidth,angle=-0]{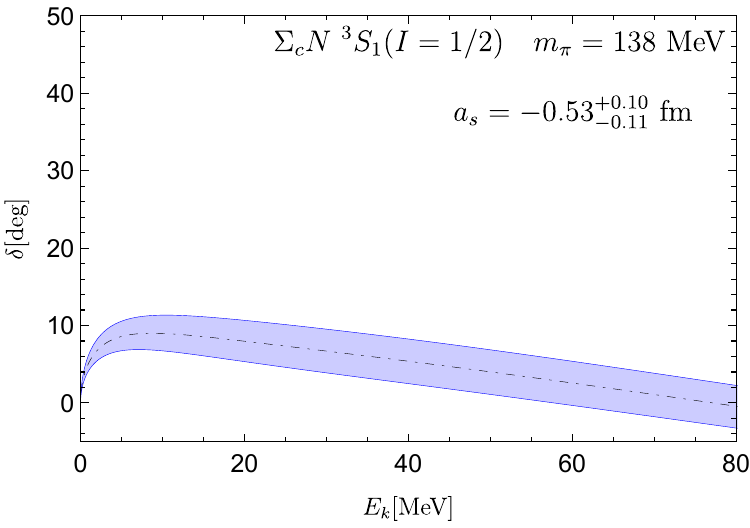}\\
\includegraphics[width=0.3\textwidth,angle=-0]{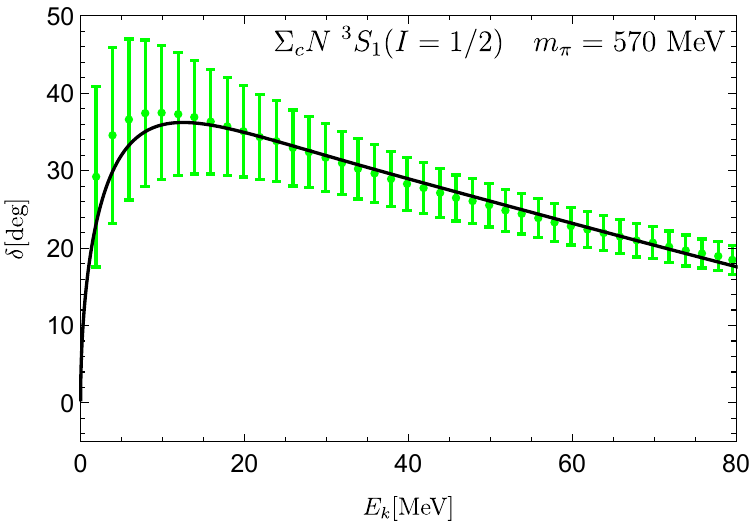}
&\includegraphics[width=0.3\textwidth,angle=-0]{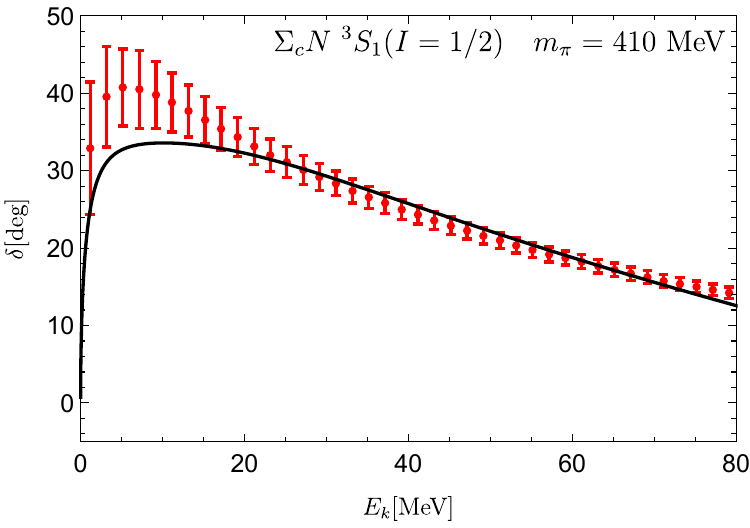}
&\includegraphics[width=0.3\textwidth,angle=-0]{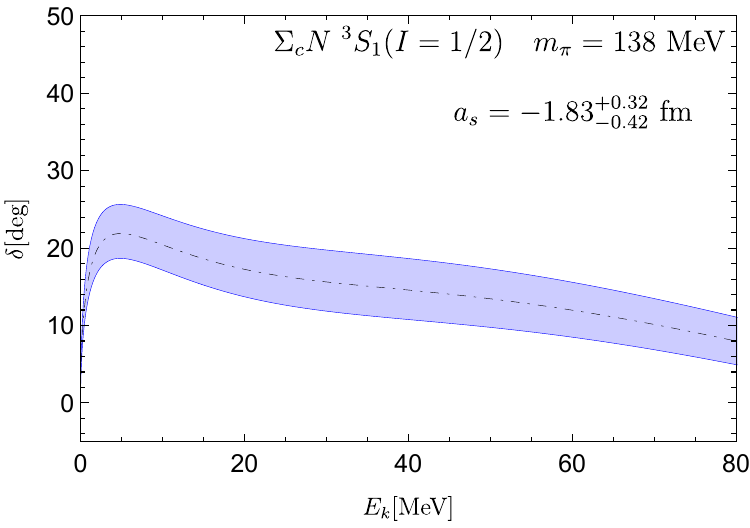}\\
\end{tabular}
\caption{The fitted phase shifts of the $\Sigma_c N$ scattering in the $I=1/2$ and $^3S_1$ channels at unphysical pion mass (black solid curves), and the chiral extrapolation of the phase shifts to the physical pion mass (given in the third column)~\cite{Meng:2019nzy}. The data with error bars are the phase shifts from lattice QCD calculations~\cite{Miyamoto:2017ynx}. The results in the first and second rows are fitted with $n=1,\Lambda=0.8$ GeV and $n=2,\Lambda=1.0$ GeV, respectively. \label{fig:FitSigmacNMeng}}
\end{figure*}

In Ref.~\cite{Meng:2019nzy}, Meng {\it et al} performed the chiral extrapolation of the $N\Sigma_c$ interaction in $\chi$EFT up to NLO, where the one-loop corrections to OPE and LO contact interactions are explicitly considered due to their $m_\pi$-dependence. They introduced a Gauss form factor $\mathcal{F}(\bm q)=\exp(-\bm q^{2n}/\Lambda^{2n})$ to transform the momentum-space potential to the coordinate space. They considered two scenarios to fit the phase shifts of the $N\Sigma_c$ scatterings from lattice QCD~\cite{Miyamoto:2017ynx}. In scenario I: $n=1$ and $\Lambda=0.8$ GeV, while in scenario II: $n=2$ and $\Lambda=1.0$ GeV. The corresponding results from scenario I and II are shown in the first and second row of Fig.~\ref{fig:FitSigmacNMeng}. An extrapolation of the $N\Sigma_c$ interaction to the physical pion mass gives the scattering length in scenario I and II as
\begin{eqnarray}
\text{scenario I}: a_s&=&-0.53^{+0.10}_{-0.11}\text{ fm },\nonumber\\
\text{scenario II}: a_s&=&-1.83^{+0.32}_{-0.42}\text{ fm}.
\end{eqnarray}
The result shows that the interaction in the $^3S_1(I=1/2)$ $N\Sigma_c$ system is weakly attractive, but no bound solution can be found.

However, one should be aware that only the naive extrapolation was done in the above work, e.g., the finite volume effects were not considered. On the other hand, the lattice calculation was performed at a very large pion mass region (about $3m_\pi-4m_\pi$). The chiral extrapolation uncertainties at such a large pion mass are hard to control.

\vspace{0.2cm}
\noindent
{\it $N\Xi_{cc}^{(\ast)}$ systems}
\vspace{0.2cm}

The first calculation of the $N\Xi_{cc}$ interaction was given in Refs.~\cite{Julia-Diaz:2004ict,Froemel:2004ea}, in which $N\Xi_{cc}$ was linked to the $NN$ system with the quark model, and the binding energy was predicted to be from several to several hundred MeVs depending on the concrete potential models. A recent calculation from the OBE model also predicted the bound states in the $N\Xi_{cc}$ and $\bar{N}\Xi_{cc}$ systems~\cite{Meng:2017udf}.

\subsubsection{$\Sigma_{c}^{(\ast)}D^{(\ast)}$, $\Sigma_{c}^{(\ast)}\bar{\Sigma}_{c}^{(\ast)}$, $\Sigma_{c}^{(\ast)}\Sigma_{c}^{(\ast)}$}

\vspace{0.2cm}
\noindent
{\it $\Sigma_{c}^{(\ast)}D^{(\ast)}$ systems}
\vspace{0.2cm}

\begin{figure*}[htbp]
\centering
\includegraphics[width=0.45\linewidth]{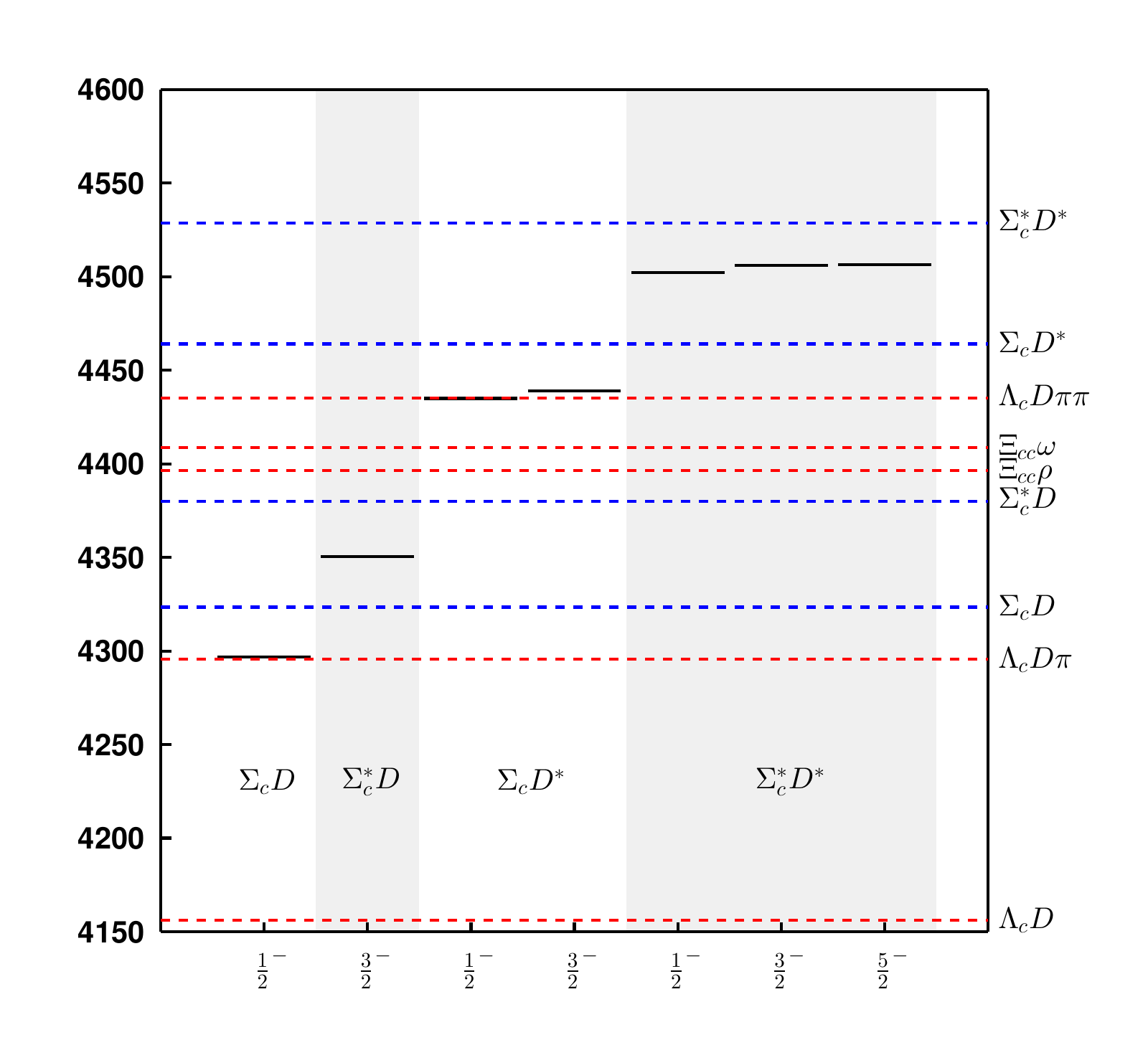}\\
\caption{The mass spectrum of the $P_{cc}$ molecular pentaquarks in the $I=1/2$ $\Sigma_c^{(\ast)}D^{(\ast)}$ systems predicted in Ref.~\cite{Chen:2021htr}. The thresholds of the $\Sigma^{(*)}_cD^{(*)}$ and possible decay channels are denoted with the blue and red dashed lines, respectively. \label{fig:SpectrumPcc}}
\end{figure*}

Based on Refs.~\cite{Meng:2019ilv,Wang:2019ato}, the doubly charmed molecular pentaquarks $P_{cc}$ composed of the $\Sigma_{c}^{(\ast)}D^{(\ast)}$ systems were explored by Chen {\it et al}~\cite{Chen:2021htr}. They predicted seven analogous molecules as those of $\Sigma_c^{(\ast)}\bar{D}^{(\ast)}$ in the isospin $I=1/2$ channels, see Fig.~\ref{fig:SpectrumPcc}, while no bound states were found in the $I=3/2$ channels. Since the quark components of the $P_{cc}$ is $[cqq][c\bar{q}]$, the decay patterns are very different from those of the $P_c$ states. There exist two types of decay modes for the $P_{cc}$ states, i.e., the
$[cqq]+[c\bar{q}]$ and $[ccq]+[q\bar{q}]$ modes. The $\Lambda_c\pi$ and $D\pi$ are the dominant decay modes of the $\Sigma_c^{(\ast)}$ and $D^\ast$, respectively. Thus the $P_{cc}$ states with $J=1/2$ can easily decay into $\Lambda_c D$ through $S$-wave via the OPE interaction. The lowest decay channel is $\Xi_{cc}\pi$ with the threshold at $3760$ MeV, which is much lower than the $P_{cc}$ states and it is not shown in Fig.~\ref{fig:SpectrumPcc}. Besides, the $P_{cc}$ states with the components $\Sigma_c^{(\ast)}D^\ast$ can also decay into $\Xi_{cc}\rho$ and $\Xi_{cc}\omega$. For more thresholds of the possible decay channels, such as the $\Lambda_c D\pi$ and $\Lambda_c D\pi\pi$, see Fig.~\ref{fig:SpectrumPcc}. Ref.~\cite{Chen:2021htr} also predicted possible molecules in the charmed-bottom $\Sigma_c^{(*)}\bar{B}^{(*)}$, $\Sigma_b^{(*)}D^{(*)}$, and doubly bottom $\Sigma_b^{(*)}\bar{B}^{(*)}$ systems.

The doubly charmed pentaquarks were also investigated in the color-magnetic interaction model~\cite{Zhou:2018bkn,Park:2018oib}, the chiral quark model~\cite{Yang:2020twg}, the QCD sum rule~\cite{Wang:2018lhz}, and the meson-exchange model~\cite{Shimizu:2017xrg,Dias:2018qhp,Chen:2021kad,Dong:2021bvy}.

\vspace{0.2cm}
\noindent
{\it $\Sigma_{c}^{(\ast)}\bar{\Sigma}_c^{(\ast)}$ and $\Sigma_{c}^{(\ast)}\Sigma_{c}^{(\ast)}$ systems}
\vspace{0.2cm}

The existence of the hidden-charm pentaquarks $P_{c}$ stimulates the theorist to investigate whether the (double) hidden-charm hexaquarks exist likewise. In Ref.~\cite{Lu:2017dvm}, the authors found that the isoscalar $\Lambda_c\bar{\Lambda}_c$, $\Sigma_c^{(\ast)}\bar{\Sigma}_c^{(\ast)}$ and isovector $\Lambda_c\bar{\Sigma}_c^{(\ast)}$ as well as their doubly charmed and doubly bottom counterparts are good candidates of the molecular hexaquarks. The calculations in various models~\cite{Lee:2011rka,Meguro:2011nr,Huang:2013rla,Carames:2015sya,Garcilazo:2020acl,Dong:2021bvy,Dong:2021juy,Chen:2021cfl,Chen:2021spf,Ling:2021asz,Liu:2021gva} do indicate that the interactions in the isoscalar and isovector $\Sigma_c^{(\ast)}\bar{\Sigma}_c^{(\ast)}$ and $\Sigma_c^{(\ast)}\Sigma_c^{(\ast)}$ systems are strong enough to form bound states.

\subsubsection{$\Xi_{cc}^{(\ast)}{D}^{(\ast)}$, $\Xi_{cc}^{(\ast)}\Sigma_{c}^{(\ast)}$, $\Xi_{cc}^{(\ast)}\bar{\Xi}_{cc}^{(\ast)}$}

The $\Xi_{cc}^{(\ast)}~[\bar{\Xi}_{cc}^{(\ast)}]$ can be related to the $\bar{D}^{(\ast)}~[D^{(\ast)}]$ with the heavy diquark-antiquark symmetry,
\begin{eqnarray}
\Xi_{cc}^{(\ast)}~~\underleftrightarrow{\scriptsize\text{HDAS}}~~\bar{D}^{(\ast)},\qquad\qquad\bar{\Xi}_{cc}^{(\ast)}~~\underleftrightarrow{\scriptsize\text{HDAS}}~~D^{(\ast)}.
\end{eqnarray}
Therefore, the systems containing the doubly charmed baryons can be regarded as an extension of the systems with the charmed mesons.
As a consequence of the HDAS, the $\Xi_{cc}^{(\ast)}{D}^{(\ast)}$, $\Xi_{cc}^{(\ast)}\Sigma_{c}^{(\ast)}$ and $\Xi_{cc}^{(\ast)}\bar{\Xi}_{cc}^{(\ast)}$ systems can be related to the $\bar{D}^{(\ast)}{D}^{(\ast)}$, $\bar{D}^{(\ast)}\Sigma_{c}^{(\ast)}$ and $\bar{D}^{(\ast)}{D}^{(\ast)}$, respectively. Thus, the existence of the molecular states in the $\bar{D}^{(\ast)}{D}^{(\ast)}$ and $\bar{D}^{(\ast)}\Sigma_{c}^{(\ast)}$ systems should also imply the existence of the molecular states in the $\Xi_{cc}^{(\ast)}{D}^{(\ast)}$, $\Xi_{cc}^{(\ast)}\Sigma_{c}^{(\ast)}$ and $\Xi_{cc}^{(\ast)}\bar{\Xi}_{cc}^{(\ast)}$ systems.

In Ref.~\cite{Guo:2013xga}, the authors predicted the triply heavy pentaquarks with $I(J^P)=0(\frac{3}{2}^-),0(\frac{5}{2}^-)$ with the $X(3872)$ as input, as well as the $1(\frac{1}{2}^-)$ and $1(\frac{3}{2}^-)$ ones with the $Z_b(10650)$ as input (see also the calculations of Chen {\it et al} with the OBE model~\cite{Chen:2017jjn}).

In Ref.~\cite{Pan:2019skd}, the authors proposed an alternative way to determine the spins of the $P_c(4440)$ and $P_c(4457)$ from the spectrum of the $\Xi_{cc}^{(\ast)}\Sigma_{c}^{(\ast)}$ systems with the help of lattice QCD~\cite{Junnarkar:2019equ}. They predicted ten bound states in the $\Xi_{cc}^{(\ast)}\Sigma_{c}^{(\ast)}$ systems with the $P_c$ states as inputs. Different spin orders of the $P_c(4440)$ and $P_c(4457)$ are also reflected in the $\Xi_{cc}^{(\ast)}\Sigma_{c}^{(\ast)}$ spectrum. For the possible molecular states in $\Xi_{cc}^{(\ast)}\Sigma_{c}^{(\ast)}$ systems, see also Ref.~\cite{Chen:2018pzd}.

In Ref.~\cite{Yang:2019rgw}, Yang {\it et al} investigated the possible bound states in the $\Xi_{cc}^{(\ast)}\Xi_{cc}^{(\ast)}/\bar{\Xi}_{cc}^{(\ast)}$ systems, and predicted the molecular candidates in the isoscalar and isovector channels.

\section{Summary and outlook}

The chiral dynamics is very important not only for the hadrons composed of light quarks (e.g. pion and nucleon) but also for the heavy flavor hadrons with one or two light quarks (e.g. heavy-light meson and singly heavy baryons). Combining the heavy quark symmetry, the chiral perturbation theory was successfully extended to the singly heavy systems  to calculate the chiral correction to their masses, axial vector current, electromagnetic form factors and so on. Inspired by the recently observed $\Xi_{cc}$ state, the $\chi$PT was also employed to investigate the chiral dynamics of the doubly heavy systems, especially the doubly charmed baryons. The heavy diquark-antiquark symmetry enhances the prediction powers of the $\chi$PT in the doubly heavy sector.

Since 2003, many exotic hadrons with heavy flavor were observed in experiments, some of which lie very close to the two-hadron threshold and are good candidates of the loosely bound hadronic molecules. 
The chiral effective field theories helped us understand the $\pi\pi$ scattering, $\pi N$ scattering, $NN$ scattering and related resonances in the past decades. The same chiral dynamics which plays a pivotal role in the formation of the $f_0(500)$, $\Lambda(1405)$ and deuteron also manifests themselves in the heavy flavor sector.
The chiral effective field theories were also extended to explore these heavy flavor hadronic molecules. 

Combining the lattice QCD simulation, the $\varphi D^{(*)}_{(s)}$ scattering were investigated with the unitarized $\chi$PT. The low mass puzzle of the $\DsI$ and $\DsII$ could be resolved either through the dynamical generation of the resonance in the molecular picture or through the channel couplings between the $c\bar s$ core and $\varphi D^{(*)}_{(s)}$ scattering state. Especially, the molecular picture leads to the two-pole structure of the $D_{0}^*(2300)$ and $D_1(2430)$, which awaits the confirmation of the experimental measurements and lattice QCD simulations in the future. 

In the two matter field sector, there have been impressive progresses in the consistent description of the near-threshold states such as $X(3872)$, $T_{cc}^+$, $Z_b$, $Z_c$, $Z_{cs}$, $P_{c}$ within the frameworks of the chiral effective field theory. Their masses, decays, lineshapes, production rates, etc. were reproduced well. The $\chi$EFT gives very useful predictions including their spin and flavor partner states, unobserved decay modes, branch ratios, production rates etc, which provide important clues for the future experiments.

However, the natures of the above exotic candidates have not been pinned down unambiguously. From the theoretical perspective, the nonperturbative  QCD cannot be solved analytically. {Except the \textit{ab initio} lattice QCD formalism}, it is still hard to set up the bottom-up theoretical frameworks  {with usability and predictive power}.  Meanwhile, the heavy flavor mesons and baryons are not available as the scattering target. The heavy flavor exotica were mainly observed in the complicated reactions or decay processes, which limit the experimental precision. Thus, in order to reveal their underlying structures, the joint efforts from the effective field theories, phenomenological models, lattice QCD simulations and more experimental measurements are necessary. In the following, we will summarize the wish list to clarify the main puzzles and challenges, focusing on the EFT perspective combining the lattice QCD and experiments.

\begin{itemize}
    \item $X(3872)$
    \begin{itemize}
        \item The masses of the $X(3872)$ coincides exactly with the $\xthn$ threshold. Such a fine tuning is very {interesting}, which becomes acute when the similar fine tuning appears again in the $T_{cc}$ system. 
        \item The refined measurement of the $X(3872)$ mass with respect to the $\xthn$ threshold {and the precise measurement of the lineshape of $X(3872)\to \xthn$} will be very helpful.  
        {Now, the limiting factor for the mass measurement is the experimental resolution. Thus, the measurement needed seems to be possible only at PANDA, once it will be in operation}.

        \item The experimental search of the partner states of the $X(3872)$ in the heavy quark symmetry and SU(3) flavor symmetry will help to explore its inner structure.

        \item It should be figured out  what clues the production of the $X(3872)$ in heavy ion collisions can provide regarding its underlying structure. 
    \end{itemize}
    \item $T_{cc}$
    \begin{itemize}
        \item {The combined investigation and comparison of prompt productions of the doubly charm family, i.e.  $T_{cc}^+$, $\Xi_{cc}^{++}$ and $\Xi_{cc}^{+}$ (absent in the present observations) in $pp$ collision may provide the clue to their structures.  }
        \item The similarity of the $T_{cc}$ and $X(3872)$, especially their similar fine tuning, needs to be { understood in an unified framework}.

        \item The spin and flavor partner states of the $T_{cc}^+$.
    \end{itemize}
    \item $\DsI$ and $\DsII$
    \begin{itemize}
        \item More independent lattice QCD simulations about the $\DsI/\DsII$ or $D^{*}K$ scattering.
        \item There is only one scalar charm meson in quark model around 2.3 GeV, while the chiral unitary approaches generate two poles. These two models can be distinguished from the identification of the higher pole of the $D_0^*$ in experiments and lattice QCD simulations.
    \end{itemize}
    \item $Z_c$, $Z_b$ and $Z_{cs}$
    \begin{itemize}
        \item The precise measurements of the lineshapes of the heavy quarkonia plus the pion can help to resolve the virtual state and resonance controversies of the $Z_{Q(s)}^{(\prime)}$ ($Q=c,b$).
        \item The experimental search of the HQSS partners of the $Z_{Q(s)}^{(\prime)}$, such as the $W_{cJ}^{(\prime)}$, $W_{bJ}^{(\prime)}$, $Z_{cs}^\prime$ and $Z_{bs}^{(\prime)}$.
        \item Lattice QCD simulations of the  $D\bar{D}^\ast/B\bar{B}^\ast$ scattering near the physical pion mass will be very helpful.
        \item Whether the $Z_{cs}(3985)$ from BESIII and $Z_{cs}(4000)$ from LHCb are the same   or different states should be checked carefully with the criteria in Ref. ~\cite{Meng:2021rdg}.
    \end{itemize}
\item $P_c$ and $P_{cs}$
    \begin{itemize}
        \item The measurement of the $J^P$ quantum numbers of the $P_c$ states can resolve the spin-order problem of the $P_c(4440)$ and $P_c(4457)$.
        \item The experimental search of the predicted $\Sigma_c^\ast\bar{D}$ [$P_c(4380)$?] and $\Sigma_c^\ast\bar{D}^\ast$ molecular states.
        \item The experimental clarification of the relationship between the $P_c(4337)$ and $P_c(4312)/P_c(4380)$.
        \item The confirmation of the $P_{cs}(4459)$ as well as the search for the other  $\Xi_c\bar{D}^{(\ast)}$, $\Xi_c^\prime\bar{D}^{(\ast)}$ and $\Xi_c^\ast\bar{D}^{(\ast)}$ molecular candidates.
    \end{itemize}
    \item General aspects of EFTs
    \begin{itemize}
\item The EFTs with the resonances should be further developed (see~\cite{Habashi:2020ofb} for an example).
\item The criteria to discern the ``elementary" and ``composite" particles in spirit of EFT, {namely the decoupling of the high energy and low energy physics,} need to be further developed. { Distinguishing the ``elementary" and ``composite" particles should depend on the resolving scale. In the Weinberg's compositeness criterion, the low-energy scattering information (scattering length and effective range) is used.   In principle, the short-range structure of a loosely bound state cannot be probed by the low-energy scattering. A quantity measuring compositeness should be defined in a resolving-scale-dependent way, which has been addressed in Refs.~\cite{Li:2021cue,Bruns:2019xgo,Song:2022yvz} to some extent.} 
\item   EFT based frameworks to extract the interactions and  structures from the lattice QCD data~(see~\cite{Doring:2011vk,Doring:2012eu,Meng:2021uhz,Liu:2015ktc} for examples) should be further developed, especially for the coupled-channel systems.
\item EFT based amplitude analysis methods~(see~\cite{LHCb:2020xds,LHCb:2021auc} for two examples and the related review~\cite{JPAC:2021rxu}) should be further developed.
\item Higher order calculations from $\chi$EFT are indispensable. The role of the large mass intermediate states should be examined carefully.
    \end{itemize}
\end{itemize}

The discovery of the $X(3872)$ in 2003 launched the new era of hadron spectroscopy. The striking experimental progresses, like the discovery of many $X,Y,Z$ states, the $P_c$ states and $T_{cc}$ state, remind us once again that this is a golden age of ambition and wisdom. We are essentially working on the clustering phenomena of the quark and gluon degrees of freedom of QCD and following the footsteps of the pioneers uncovering the structures at other levels such as the molecules, atoms, and nucleus. The few body physics of the quarks and gluons is very different from either the electron-electron interaction within the atom or the nucleon-nucleon interaction within the nucleus because the color confinement renders the basic d.o.fs nontransparent. Meanwhile, the underlying interactions (QCD) of hadrons are highly nonperturbative at the low energy scale. Although the endeavors in the past decades have brought us huge amounts of fresh perspectives, there is still a long long way to fully understand their internal structures. However, in the coming future, the accumulated data from the BaBar, Belle, BESIII, CDF, D0, LHCb, ATLAS and CMS Collaborations will continuously contribute to the discovery of the exotica with heavy flavors. At the same times, the ongoing, upgrading and upcoming experiments, including the GlueX~\cite{GlueX:2019xxx}, Belle II~\cite{Belle-II:2018jsg}, BESIII~\cite{BESIII:2020nme}, LHCb~\cite{LHCb:2018roe} and PANDA~\cite{PANDA:2009yku,PANDA:2021ozp} shall bring us more surprises. The amazing promotion of the computational capabilities will make more precise lattice QCD simulations available. The developments of theoretical frameworks, especially the EFTs, shall refresh our understanding about the exotic hadrons and add the new knowledge of the nonperturbative QCD dynamics.

\vspace{0.2cm}

\section*{Acknowledgments}

We would like to thank all the collaborators who contributed to the present investigations, in particular to Dian-Yong Chen, Hua-Xing Chen, Kan Chen, Rui Chen, Wei Chen, Xiao-Lin Chen, Cheng-Rong Deng, Wei-Zhen Deng, Meng-Lin Du, Jun He, Bo-Lin Huang, Peng-Zhi Huang, Nan Jiang, Hao-Song Li, Ning Li, Zi-Yang Lin, Xiang Liu, Xiao-Hai Liu, Yan-Rui Liu, Zhan-Wei Liu, Zhi-Gang Luo, Li Ma, Zhi-Feng Sun, Xin-Zhen Weng, Li-Ye Xiao, Hao Xu, Zhong-Cheng Yang, Lu Zhao. L.M. is grateful to the helpful discussions with Evgeny Epelbaum and Jambul Gegelia. G.J.W. thanks the useful discussions with Makoto Oka. S.L.Z. dedicates this review to his beloved mother Bao-Feng Huang. This project is supported by
the National Natural Science Foundation of China under Grants No.~11975033 and No.~12070131001 and No.~12105072, the Youth Funds of Hebei Province (No. A2021201027) and the Start-up Funds for Young Talents of Hebei University (No. 521100221021). This project is also funded by the Deutsche Forschungsgemeinschaft (DFG, German Research Foundation, Project ID 196253076-TRR 110) and JSPS KAKENHI under Grant No. 20F20026.

\vspace{0.2cm}

\begin{appendix}

\section{Building blocks and the superfields}\label{app:1}

\subsection{Building blocks}\label{Sect.1.1}

In order to fulfill the chiral symmetry, $C$, $P$, $T$ symmetries and Hermitian, the chiral Lagrangians are often constructed using several `building blocks' with known transformation behaviors under these symmetries. To avoid confusions, we use
$\text{Tr}(X)$ to represent the trace in the flavor space and $\langle X\rangle$ to represent the trace in the spinor space. The traceless building block $\hat{X}$ in the SU(3) case are defined as
\begin{equation}
\hat{X}=X-\frac{\mathbf{I}}{3}\mathrm{Tr}(X).
\end{equation}
The corresponding results in $\SU(2)$ flavor symmetry can be deduced easily.

The building blocks to accommodate the Goldstone bosons were constructed as follows,
\begin{eqnarray}
	U=\xi^{2}=\exp(i\varphi/f_{\varphi}), \quad \varphi=\left(\begin{array}{ccc}
		\pi^{0}+\frac{1}{\sqrt{3}}\eta & \sqrt{2}\pi^{+} & \sqrt{2}K^{+}\\
		\sqrt{2}\pi^{-} & -\pi^{0}+\frac{1}{\sqrt{3}}\eta & \sqrt{2}K^{0}\\
		\sqrt{2}K^{-} & \sqrt{2}\bar{K}^{0} & -\frac{2}{\sqrt{3}}\eta
	\end{array}\right),
\end{eqnarray}
where $f_\varphi$ is the decay constant in the chiral limit. With the Goldstone bosons,  left-handed external fields $l_\mu$, right-handed external fields  $r_\mu$, scalar external fields $s$ and pseudoscalar external fields $p$, one can construct the following building blocks.
\begin{eqnarray}
	\Gamma_{\mu}&=&\frac{1}{2}\left[\xi^{\dagger}(\partial_{\mu}-ir_{\mu})\xi+\xi(\partial_{\mu}-il_{\mu})\xi^{\dagger}\right],\label{eq:connection}\\
    u_{\mu}&=& {i \over 2}\left[\xi^{\dagger}(\partial_{\mu}-ir_{\mu})\xi-\xi(\partial_{\mu}-il_{\mu})\xi^{\dagger}\right], \label{eq:psef}\\
	\chi&=&2B_{0}(s+ip),\\
\chi_{\pm}&=&\xi^{\dagger}\chi \xi^{\dagger}\pm \xi\chi^{\dagger}\xi,\\
f_{\mu\nu}^{R}&=&\partial_{\mu}r_{\nu}-\partial_{\nu}r_{\mu}-i[r_{\mu},r_{\nu}],\\
f_{\mu\nu}^{L}&=&\partial_{\mu}l_{\nu}-\partial_{\nu}l_{\mu}-i[l_{\mu},l_{\nu}],\\
f_{\mu\nu}^{\pm}&=&\xi^{\dagger}f_{\mu\nu}^{R}\xi\pm \xi f_{\mu\nu}^{L}\xi^{\dagger},\label{eq:fmnpm}\\
\hat{f}_{\mu\nu}^{\pm}&=&f_{\mu\nu}^{\pm}-\frac{1}{3}\mathrm{Tr}(f_{\mu\nu}^{\pm})\label{eq:fmnt}.
\end{eqnarray}
{The scalar external field $s=\mathrm{diag}(m_u,m_d,m_s)$ is used to introduce the quark mass effect. If the pseudoscalar external field $p=0$, the expansion of the $\chi_+$  yields
\begin{eqnarray}\label{eq:chipt}
\chi_+&=&2B_0\mathrm{diag}(m_u,m_d,m_s)=\mathrm{diag}(2m_\pi^2,2m_\pi^2,4m_K^2-2m_\pi^2),
\end{eqnarray}
where we have kept the leading order terms only. $B_0$ is related to the quark condensate, see Eq.~\eqref{eq:explicit_breaking} and its context for details.}

If the external source is the electromagnetic field ($r_\mu=l_\mu=-e\mathscr{Q} A_\mu$),  the right- and left- field strength tensors $f_{\mu\nu}^{R}$ and $f_{\mu\nu}^{L}$ are given as
\begin{eqnarray}
f_{\mu\nu}^R=f_{\mu\nu}^L=-e\mathscr{Q}\left(\partial_\mu A_\nu-\partial_\nu A_\mu\right).\label{eq:fmn}
\end{eqnarray}
For the charmed and bottom mesons, $\mathscr{Q}$ denotes the electric charge matrix of their SU(3) multiplets, {which reads $\mathscr{Q}=\mathrm{diag}(0,-1,-1)$ for the $(\bar{D}^{(\ast)0}, D^{(\ast)-},D_s^{(\ast)-})$, and $\mathscr{Q}=\mathrm{diag}(1,0,0)$ for the $(B^{(\ast)+},B^{(\ast)0}, B_s^{(\ast)0})$, respectively}.  Expanding the terms in Eqs.~\eqref{eq:fmnpm} and~\eqref{eq:fmnt}, respectively, one notes that $f_{\mu\nu}^+$ contains the $\mathscr{Q}$, while $\hat{f}_{\mu\nu}^+$ is proportional to the electric charge matrix $\mathscr{Q}_l$ of the light quark current $\mathscr{J}^\ell_\mu=\frac{2}{3}\bar{u}\gamma_\mu u-\frac{1}{3}\bar{d}\gamma_\mu d-\frac{1}{3}\bar{s}\gamma_\mu s$, i.e., $\mathscr{Q}_l=\rm{diag}(2/3,-1/3,-1/3)$.

In order to investigate the isospin violation from the electromagnetic interaction, the left and right charges were often introduced~\cite{Meissner:1997fa,Meissner:1997ii}. Their chiral transformations read
\begin{equation}
Q_{L/R}\to g_{L/R}Q_{L/R}g_{L/R}^{\dagger}
\end{equation}
One can define the building block $Q_{\pm}$,
\begin{eqnarray}
	Q_{\pm}=\xi^{\dagger}Q_{R}\xi\pm \xi Q_{L}\xi^{\dagger}.
\end{eqnarray}

The chiral orders, chiral transformation, space inversion, charge conjugation and Hermitian conjugation of the above building blocks are given in Table~\ref{tab:building}. The time reversal symmetry is usually guaranteed by the CPT theorem. In the practical calculation, there are some techniques to reduce the independent terms of the Lagrangians (see Refs.~\cite{Fettes:1998ud,Fettes:2000gb,Jiang:2019hgs,Qiu:2020omj} for examples).

\begin{table}[!htbp]
	\centering
	\renewcommand{\arraystretch}{1.2}
	\caption{The chiral order (D) and transformation properties of the building blocks under the chiral transformation (CH), space inversion (P), charge conjugation (C) and Hermitian conjugation (H.c.). We take the doublet nucleon $\psi$ as an example, in which the covariant derivatives are defined as $\mathcal{D}_{\mu}=\partial_{\mu}+\Gamma_{\mu}$ and $\mathcal{D}_{\mu}^{\prime}=\partial_{\mu}-\Gamma_{\mu}$. The covariant derivatives for the other mater fields are listed in the main text.}\label{tab:building}
	\setlength{\tabcolsep}{6.7mm}
{
	\begin{tabular}{cccccc}
		\hline
		& $U$ & $\xi$ & $\chi$ & $f_{\mu\nu}^{R}$ & $f_{\mu\nu}^{L}$\tabularnewline
		\midrule
		D & 0 & 0 & 2 & 2 & 2\tabularnewline
		CH & $g_{R}Ug_{L}^{\dagger}$ & $g_{R}\xi K^{\dagger}\text{or }K \xi g_{L}^{\dagger}$ & $g_{R}\chi g_{L}^{\dagger}$ & $g_{R}f_{\mu\nu}^{R}g_{R}^{\dagger}$ & $g_{L}f_{\mu\nu}^{L}g_{L}^{\dagger}$\tabularnewline
		P & $U^{\dagger}$ & $\xi^{\dagger}$ & $\chi^{\dagger}$ & $f^{L\mu\nu}$ & $f^{R\mu\nu}$\tabularnewline
		C & $U^{T}$ & $\xi^{T}$ & $\chi^{T}$ & $-(f_{\mu\nu}^{L})^{T}$ & $-(f_{\mu\nu}^{R})^{T}$\tabularnewline
		H.c. & $U^{\dagger}$ & $\xi^{\dagger}$ & $\chi^{\dagger}$ & $f_{\mu\nu}^{R}$ & $f_{\mu\nu}^{L}$\tabularnewline
		\midrule
		& $\chi_{\pm}$ & $f_{\mu\nu}^{\pm}$ & $u_{\mu}$ & $Q_{\pm}$ & $h_{\mu\nu}$\tabularnewline
		\midrule
		D & 2 & 2 & 1 & 1 & 2 \tabularnewline
		CH & $K\chi_{\pm}K^{\dagger}$ & $Kf_{\mu\nu}^{\pm}K^{\dagger}$ & $K u_{\mu}K^{\dagger}$ & $KQ_{\pm}K^{\dagger}$ & $Kh_{\mu\nu}K^{\dagger}$\tabularnewline
		P & $\pm\chi_{\pm}$ & $\pm f^{\pm\mu\nu}$ & $-u^{\mu}$ & $\pm Q_{\pm}$ & $-h^{\mu\nu}$ \tabularnewline
		C & $\chi_{\pm}^{T}$ & $\mp(f_{\mu\nu}^{\pm})^T$ & $(u_{\mu})^{T}$ & $Q_{\pm}^{T}$ & $(h_{\mu\nu})^{T}$ \tabularnewline
		H.c. & $\pm\chi_{\pm}$ & $f_{\mu\nu}^{\pm}$ & $u_{\mu}$ & $Q_{\pm}$ & $h_{\mu\nu}$ \tabularnewline
		\midrule
		& $\Gamma_{\mu}$ & $\psi$ & $\bar{\psi}$ & $\mathcal{D}_{\mu}\psi$ & $g_{\mu\nu}$\tabularnewline
		\midrule
		D & 1 & 0 & 0 & 0 & 0\tabularnewline
		CH & $K\Gamma^{\mu}K^{\dagger}-\partial^{\mu}KK^{\dagger}$ & $K\psi$ & $\bar{\psi}K^{\dagger}$ & $K\mathcal{D}_{\mu}\psi$ & $g_{\mu\nu}$\tabularnewline
		P & $\Gamma^{\mu}$ & $\gamma^{0}\psi$ & $\bar{\psi}\gamma^{0}$ & $\gamma^{0}\mathcal{D}^{\mu}\psi$ & $g^{\mu\nu}$\tabularnewline
		C & $-(\Gamma_{\mu})^{T}$ & $C\bar{\psi}^{T}$ & $\psi^{T}C$ & $C\mathcal{D}_{\mu}^{\prime T}\bar{\psi}^{T}$ & $g_{\mu\nu}$\tabularnewline
		H.c. & $-\Gamma_{\mu}$ & $\bar{\psi}\gamma_{0}$ & $\gamma^{0}\psi$ & $\psi^{\dagger}\overleftarrow{\mathcal{D}}_{\mu}^{\prime}$ & $g_{\mu\nu}$\tabularnewline
		\midrule
		& $\bar{\psi}\gamma^{\mu}\psi$ & $\bar{\psi}\gamma^{\mu}\gamma_{5}\psi$ & $\bar{\psi}\gamma_{5}\psi$ & $\bar{\psi}\sigma_{\mu\nu}\psi$ & $\epsilon_{\mu\nu\rho\sigma}$\tabularnewline
		\midrule
		D & 1 & 0 & 1 & 0 & 0\tabularnewline
		CH & $\bar{\psi}\gamma^{\mu}\psi$ & $\bar{\psi}\gamma^{\mu}\gamma_{5}\psi$ & $\bar{\psi}\gamma_{5}\psi$ & $\bar{\psi}\sigma_{\mu\nu}\psi$ & $\epsilon_{\mu\nu\rho\sigma}$\tabularnewline
		P & $\bar{\psi}\gamma_{\mu}\psi$ & $-\bar{\psi}\gamma_{\mu}\gamma_{5}\psi$ & $-\bar{\psi}\gamma_{5}\psi$ & $\bar{\psi}\sigma^{\mu\nu}\psi$ & $-\epsilon^{\mu\nu\rho\sigma}$\tabularnewline
		C & $\bar{-\psi}\gamma^{\mu}\psi$ & $\bar{\psi}\gamma^{\mu}\gamma_{5}\psi$ & $\bar{\psi}\gamma_{5}\psi$ & $-\bar{\psi}\sigma_{\mu\nu}\psi$ & $\epsilon_{\mu\nu\rho\sigma}$\tabularnewline
		H.c. & $\bar{\psi}\gamma^{\mu}\psi$ & $\bar{\psi}\gamma^{\mu}\gamma_{5}\psi$ & $-\bar{\psi}\gamma_{5}\psi$ & $\bar{\psi}\sigma_{\mu\nu}\psi$ & $\epsilon_{\mu\nu\rho\sigma}$\tabularnewline
		\hline
	\end{tabular}
	}
\end{table}

The representation theory of SU(3) group can guide the construction of the Lagrangians to avoid omitting some terms. For example, for the heavy flavor system, the common and nontrivial reductions of the representation are as follows, 
\begin{eqnarray}
\bm3\otimes\bm6&=&\bm8\oplus\bm{10},\\
\bar{\bm6}\otimes\bm6&=&\bm1\oplus\bm8\oplus\bm{27},\\
\bm8\otimes\bm8&=&\bm1\oplus\bm8_{1}\oplus\bm8_{2}\oplus\bm{10}\oplus\overline{\bm{10}}\oplus\bm{27},\\
\bar{\bm 3} \otimes {\bm 8}&=&\overline{\bm{15}} \oplus {\bm 6} \oplus \bar {\bm 3}.
\end{eqnarray}
{The above $\bm8_{1}$ and $\bm8_2$ are actually the same irreducible representation (irrep.) of the SU(3) group, namely, $\{p,q\}=\{1,1\}$ irrep., where the first row has one more box than the second row and the second row has one more box than the third row in Young diagram. However, in the tensor method to perform reduction (see~\cite{Lee:1981} for details), the irrep. with eight dimensions corresponds to a tensor with rank $(1,1)$. The two $(1,1)$ tensors $A^i_j$ and $B^i_j$ have two ways to form the new $(1,1)$ tensors,
\begin{equation}
    F^i_a=A^i_jB^j_a-B^i_jA^j_a,\quad D^i_a=A^i_jB^j_a+B^i_jA^j_a-{2\over 3}\delta ^i_aS \label{eq:88reduction}
\end{equation}
where $S=A^i_aB^a_i$ is the tensor of rank $(0,0)$. Apparently, the $F$ and $D$ are the antisymmetric and symmetric reductions, respectively. Therefore, we use $\bm8_1$ and $\bm8_2$ to label the two different reductions. } In the Table \ref{tab:app1:g_irp}, we illustrate the reductions of two nonets $X$ and $Y$, $\bar{B}_{\bm6}$ and $B_{\bm6}$, $\bar{B}_{\bar{\bm3}}$ and $B_{\bm6}$, respectively.

\begin{table}[!htbp]
	\centering
	\renewcommand{\arraystretch}{1.5}
	\caption{Reductions of two nonets $X$ and $Y$, $\bar{B}_{\bm6}$ and $B_{\bm6}$, $\bar{B}_{\bar{\bm3}}$ and $B_{\bm6}$. The notations of the (anti)symmetrization are  $(XY)_{\{a,b\}}=X_{a}Y_{b}+X_{b}Y_{a}$ and $(XY)_{[a,b]}=X_{a}Y_{b}-X_{b}Y_{a}$.}\label{tab:app1:g_irp}
	\setlength{\tabcolsep}{3.72mm}
{
	\begin{tabular}{c|ccccc}
		\hline
		Group representation & $\bm1\otimes\bm1\to\bm1$ & $\bm1\otimes\bm8\to\bm8$ & $\bm8\otimes\bm8\to\bm1$ & $\bm8\otimes\bm8\to\bm8_{1}$ & \tabularnewline
		Flavor structure & $\text{Tr}(X)\text{Tr}(Y)$ & $\text{Tr}(X)\hat{Y}$ or $\text{Tr}(Y)\hat{X}$ & $\text{Tr}(\hat{X}\hat{Y})$ & $[\hat{X},\hat{Y}]$ & \tabularnewline
		\hline
		Group representation & $\bm8\otimes\bm8\to\bm8_{2}$ & $\bm8\otimes\bm8\to\bm{10}$ & $\bm8\otimes\bm8\to\overline{\bm{10}}$ & $\bm8\otimes\bm8\to\bm{27}$ & \tabularnewline
		Flavor structure & $\{\hat{X},\hat{Y}\}$ & $(\hat X\hat Y)_{[a,b]}^{\{i,j\}}$ & $(\hat X\hat Y)_{\{a,b\}}^{[i,j]}$ & $(\hat X\hat Y)_{\{a,b\}}^{\{i,j\}}$ & \tabularnewline
		\hline
		Group representation & $\bar{\bm6}\otimes\bm6\to\bm1$ & $\bar{\bm6}\otimes\bm6\to\bm8$ & $\bar{\bm6}\otimes\bm6\to\bm{27}$ & $\bm3\otimes\bm6\to\bm{8}$ & $\bm3\otimes\bm6\to\bm{10}$\tabularnewline
		Flavor structure & $\text{Tr}(\bar{B}_{\bm6}B_{\bm6})$ & $\bar{B}_{\bm6ab}B_{\bm6}^{ca}$ & $\bar{B}_{\bm6ab}B_{\bm6}^{ij}$ & $\bar{B}_{\bar{\bm3}ab}B_{\bm6}^{ca}$ & $\bar{B}_{\bar{\bm3}ij}B_{\bm6}^{ab}$\tabularnewline
		\hline
	\end{tabular}
	}
\end{table}

In literature, there are two conventions of the charge conjugation transformation for $D^*$ in constructing the di-meson states with fixed $C$-parity. The two conventions were spelled out in Ref.~\cite{Liu:2008fh} and Refs.~\cite{Thomas:2008ja,Baru:2011rs}, respectively. We summarize the them in Table~\ref{tab:two-convention}. In the main body, we will label the specific convention when the related Lagrangians or wave functions appear.

\begin{table}[hbtp]
    \centering
    \renewcommand{\arraystretch}{1.5}
    \caption{Two conventions for the charge conjugation of $D^*$. The convention-I and -II are spelled out in Ref.~\cite{Liu:2008fh} and Refs.~\cite{Thomas:2008ja,Baru:2011rs}, respectively.}
    \label{tab:two-convention}
    \setlength{\tabcolsep}{10.72mm}
{
\begin{tabular}{c|c|c}
\hline
 & Interpolating current & Charge conjugation\tabularnewline
\hline 
convention-I & \multirow{2}{*}{$D=\bar{q}\gamma_{5}c,\;\bar{D}=\bar{c}\gamma_{5}u$} & \multirow{2}{*}{$\hat{C}D\hat{C}^{-1}=\bar{D},\;|\bar{D}\rangle=\hat{C}|D\rangle$}\tabularnewline
\cline{1-1} 
convention-II &  & \tabularnewline
\hline 
convention-I & $D^{*}=\bar{q}\gamma_{\mu}c,\;\bar{D}^{*}=\bar{c}\gamma_{\mu}q$ & $\hat{C}D^{*}\hat{C}^{-1}=-\bar{D}^{*},\;|\bar{D}^{*}\rangle=-\hat{C}|D^{*}\rangle$\tabularnewline
\hline 
convention-II & $D^{*}=\bar{q}\gamma_{\mu}c,\;\bar{D}^{*}=-\bar{c}\gamma_{\mu}q,$ & $\hat{C}D^{*}\hat{C}^{-1}=\bar{D}^{*},\;|\bar{D}^{*}\rangle=\hat{C}|D^{*}\rangle$\tabularnewline
\hline
\end{tabular}
}
\end{table}

\subsection{Superfields}

The superfield is a technique to embed the heavy quark symmetry into the Lagrangians. In Ref.~\cite{Falk:1991nq}, Falk illustrated the constructions of the superfields with arbitrary spins. In this section, we present some commonly used superfields. In principle, it is free to choose the relative phase of different components in the superfield. Therefore, one may see different definitions of superfields up to a phase in literature. We first introduce two projection operators, $\Lambda_{+}=(1+\slashed v)/2$ and $\Lambda_{-}=(1-\slashed v)/2$.

The superfields for the ground state heavy mesons can be understood as the (tensor) product of the two spinors of the quark and antiquark~\cite{Falk:1991nq},
\begin{equation}
{\cal H}\sim u_{h}\bar{v}_{l},\quad {\cal \tilde{H}}\sim u_{l}\bar{v}_{h},\quad	\bar{{\cal H}}\sim v_{l}\bar{u}_{h},\quad \bar{\tilde{\mathcal{H}}}\sim v_{h}\bar{u}_{l},
\end{equation}
where $l$ and $h$ represent the light and heavy quarks, respectively. At the hadronic level, the superfields read
\begin{eqnarray}
	{\cal H}=\Lambda_{+}(P_{\mu}^\ast\gamma^{\mu}+iP\gamma_{5}),\qquad \tilde{{\cal H}}=(\tilde{P}_{\mu}^\ast\gamma^{\mu}+i\tilde{P}\gamma_{5})\Lambda_{-}.
\end{eqnarray}
The $\tilde{{\cal H}}$ and $\cal{H}$ are related to each other through
\begin{equation}
	 \tilde{\cal H}=C[{\cal C}{\cal H}{\cal C}^{-1}]^{T}C^{-1},
\end{equation}
where $\cal{C}$ and $C$ are the charge conjugation acting on the field operator and Dirac matrix with
\begin{equation}
	{\cal C}P_{\mu}{\cal C}^{-1}=-\tilde{P_{\mu}},\quad{\cal C}P{\cal C}^{-1}=\tilde{P},\quad\ensuremath{C=i\gamma^{2}\gamma^{0}}.
\end{equation}
The conjugation of the fields are defined by $	\bar{{\cal H}}=\gamma_{0}{\cal H}^\dagger\gamma_{0}$ and $\bar{\tilde{\mathcal{H}}}=\gamma_{0}\tilde{\cal H}^\dagger\gamma_{0}$.
The general Lagrangians using the superfield were [e.g., see Eqs.~\eqref{eq:app1:lagD} and~\eqref{eq:app1:lagDbar}]
\begin{eqnarray}
{\cal L}&\sim&\langle\bar{u}_{l}\Gamma_{l}u_{l}\bar{v}_{h}\Gamma_{h}v_{h}\rangle=\langle v_{h}\bar{u}_{l}\Gamma_{l}u_{l}\bar{v}_{h}\Gamma_{h}\rangle\sim\langle\bar{\tilde{{\cal H}}}\Gamma_{l}\tilde{{\cal H}}\Gamma_{h}\rangle,\label{eq:app1:hlHHt}\\
{\cal L}&\sim&\langle\bar{u}_{h}\Gamma_{h}u_{h}\bar{v}_{l}\Gamma_{l}v_{l}\rangle=\langle u_{h}\bar{v}_{l}\Gamma_{l}v_{l}\bar{u}_{h}\Gamma_{h}\rangle\sim\langle{\cal H}\Gamma_{l}\bar{{\cal H}}\Gamma_{h}\rangle,\label{eq:app1:hlHH}
\end{eqnarray}
where $\Gamma_{h}$ and $\Gamma_{l}$ are the Dirac matrices in the bilinear forms of the heavy and light (anti)quarks, respectively . With Eqs.~\eqref{eq:app1:hlHHt} and~\eqref{eq:app1:hlHH}, one can easily construct the Lagrangians to keep or break the heavy quark spin symmetry by choosing the pertinent $\Gamma_h$.

For the excited $P$-wave heavy mesons, the superfield of the $(0^+,1^+)$ doublet reads
\begin{eqnarray}
	{\cal S}=\Lambda_{+}\left[R^{*\mu}\gamma_{\mu}\gamma_{5}-R\right],\quad\text{  with  }
	{\cal S}\sim u_{h}\bar{v}_{l}\gamma_{5},\quad\bar{{\cal S}}\sim-\gamma_{5}v_{l}\bar{u}_{h},
\end{eqnarray}
where the $v_l$ is the effective spinor corresponding to combination of the light antiquark and the $l=1$ angular momentum. The general form of the Lagrangians read [e.g., see Eq.~\eqref{eq:app1:lagS}]
\begin{eqnarray}
	{\cal L}\sim\langle\bar{v}_{l}\Gamma_{l}v_{l}\bar{u}_{h}\Gamma_{h}u_{h}\rangle=\langle u_{h}\bar{v}_{l}\gamma_{5}\gamma_{5}\Gamma_{l}\gamma_{5}\gamma_{5}v_{l}\bar{u}_{h}\Gamma_{h}\rangle=-\langle{\cal S}\gamma_{5}\Gamma_{l}\gamma_{5}\bar{{\cal S}}\Gamma_{h}\rangle=-\langle\bar{{\cal S}}\Gamma_{h}{\cal S}\gamma_{5}\Gamma_{l}\gamma_{5}\rangle.
\end{eqnarray}

Meanwhile, the superfield of the $(1^+,2^+)$ doublet reads
\begin{equation}
	{\cal T^{\mu}}	=\Lambda_{+}\left\{ Y^{\mu\nu}\gamma_{\nu}-\sqrt{\frac{3}{2}}Y_{\nu}\gamma_{5}\left[g^{\mu\nu}-\frac{1}{3}(\gamma^{\mu}-v^{\mu})\gamma^{\nu}\right]\right\},\quad
\text{ with }
{\cal T}\sim u_{h}\bar{v}_{l}^{\mu},\quad\bar{{\cal T}}\sim v_{l}^{\mu}\bar{u}_{h},
\end{equation}
where the $v_l^\mu$ denotes the vector-spinor of the light d.o.f with $j_l=\frac{3}{2}$. Similarly, the most general Lagrangians should have the following form [e.g., see Eq.~\eqref{eq:app1:lagTT}],
\begin{equation}
\mathcal{L}\sim\langle\bar{u}_{h}\Gamma_{h}u_{h}\bar{v}_{l}^{\mu}\Gamma_{l}^{\mu\nu}v_{l}^{\nu}\rangle	=\langle v_{l}^{\nu}\bar{u}_{h}\Gamma_{h}u_{h}\bar{v}_{l}^{\mu}\Gamma_{l}^{\mu\nu}\rangle=\langle\bar{{\cal T}}^{\nu}\Gamma_{h}{\cal T}^{\mu}\Gamma_{l}^{u\nu}\rangle=\langle{\cal T}^{\mu}\Gamma_{l}^{u\nu}\bar{{\cal T}}^{\nu}\Gamma_{h}\rangle.
\end{equation}

For the singly heavy baryon with the symmetric light diquark (flavor sextet), the spin-$1\over 2$ and -$3\over 2$ states form the doublet in the heavy quark spin symmetry. The superfield is constructed as~\cite{Falk:1991nq}
\begin{equation}\label{eq:sfofshb}
	\psi_{Q}^{\mu}={\mathcal{B}^{*}_{\bm6}}^{\mu}+\sqrt{\frac{1}{3}}(\gamma^{\mu}+v^{\mu})\gamma^{5}{\mathcal{B}_{\bm6}}.
\end{equation}
The superfield can be understood as the
\begin{equation}
	\psi_Q^{\mu} \sim u_{h}A_{l}^{\mu},
\end{equation}
where $u_h$ and $A_l^\mu$ represent the spinor of the heavy quark and vector light diquark, respectively. A general interaction Lagrangian containing the $\psi_Q^{\mu}$ only has the following form [e.g., see Eq.~\eqref{eq:app1:lagBc2}],
\begin{equation}
	\mathcal{L}\sim\Tr[(\bar{u}_h\Gamma_h u_h)(A_l^{\ast\mu}\Gamma_{l\mu\nu}A_l^\nu)]\sim\Tr[\bar{u}_h A_l^{\ast\mu}\Gamma_h\Gamma_{l\mu\nu}u_h A_l^\nu)]\sim \Tr(\bar{\psi}_Q^\mu \Gamma_h\Gamma_{l\mu\nu}\psi_Q^\nu).
\end{equation}
in which the Lorentz index of $\Gamma_{l\mu\nu}$ contracts with that of the light d.o.f. If one wants to keep the heavy quark spin symmetry, the $\Gamma_h$ should be the unit matrix.

For the ground state doubly heavy baryons, two identical heavy quarks form a vector diquark. The spin-$1\over 2$ and spin-$3\over 2$ states form the spin doublet in the heavy quark spin symmetry. The superfield is constructed as~\cite{Meng:2018zbl}
\begin{equation}\label{eq:sfofdhb}
\psi_{QQ}^{\mu}={\mathcal{B}_{QQ}^{*}}^{\mu}+\sqrt{\frac{1}{3}}(\gamma^{\mu}+v^{\mu})\gamma^{5}{\mathcal{B}_{QQ}}.
\end{equation}
The superfield can be understood as 
\begin{equation}
	\psi_{QQ}^{\mu} \sim u_{l}A_{h}^{\mu},
\end{equation}
where the $u_l$ and $A_{h}^{\mu}$ denote the spinor of the light quark and vector heavy diquark, respectively.
The general interaction may have the following form [e.g., see Eq.~\eqref{eq:app1:lagBcc2}],
\begin{equation}
	\mathcal{L}\sim(\bar{u}_l\Gamma_l u_l)(A_h^{\ast\mu}\Gamma_{h\mu\nu}A_h^\nu)\sim(\bar{u}_l A_h^{\ast\mu}\Gamma_l \Gamma_{h\mu\nu} u_l A_h^\nu)\sim \bar{\psi}_{QQ}^\mu\Gamma_l \Gamma_{h\mu\nu}\psi_{QQ}^\nu.
\end{equation}
If one wants to keep the heavy diquark spin symmetry, the Lorentz indices should contract with each other directly, i.e., $\Gamma_{h\mu\nu}=\alpha_h g_{\mu\nu}$ (with $\alpha_h$ a constant).

Although the superfields of the singly heavy and doubly heavy baryons have very similar forms, there are essential difference considering the light and heavy d.o.fs. The rules to construct the Lagrangians in the heavy (di)quark spin symmetry are totally different.

\section{Heavy field expansion}\label{app:HFE}

We take the spin-$1\over 2$ fermion field $\Psi$ as an example to illustrate the heavy field expansion. The momentum $p$ of the field can be decomposed into the mass term and the residual momentum term,
\begin{equation}
		p_{\mu}=Mv_{\mu}+q_{\mu},
\end{equation}
where $M$ is the mass, $v_\mu$ satisfies $v^2=1$, and $q$ is the residual momentum. The field $\Psi$ is divided into the light field $H$ and the heavy field $h$,
\begin{equation}
	H=e^{iMv\cdot x}\Lambda_+\Psi, \quad h=e^{iMv\cdot x}\Lambda_-\Psi,
\end{equation}
where the projectors are defined as $\Lambda_\pm=\frac{1}{2}(1\pm \slashed v)$. With this decomposition, the most general Lagrangian becomes
\begin{equation}
	\mathcal{L}=\bar{H}{\cal A}H+\bar{h}{\cal B}H+\bar{H}\gamma_{0}{\cal B}^{\dagger}\gamma_{0}h-\bar{h}{\cal C}h.~\label{eq:app2:generalL}
\end{equation}
Using the free Lagrangian $\ensuremath{\mathcal{L}=\bar{\Psi}(i\slashed{\mathcal{D}}-M)\Psi}$ as an illustration, the corresponding $\mathcal{A}$, $\mathcal{B}$ and $\mathcal{C}$ are
\begin{equation}
	{\cal A}=iv\cdot \mathcal{D},\quad{\cal B}=i\slashed{\mathcal{D}}_{\perp},\quad{\cal C}=i(v\cdot\mathcal{D})+2M,~\label{eq:heavyFE}
\end{equation}
where $X_\perp^\mu \equiv X^\mu-(v\cdot X)v^\mu$. One can see that the $H$ is massless, while the mass of the $h$ is $2M$.

The meanings of the $H$ and $h$ is clear with the Dirac representation of the spinor and gamma matrices. The  positive energy plane wave solution to the free Dirac equation reads
\begin{eqnarray}
	\psi_{\bm p}^s(x,t)=\sqrt{E(\bm{p})+M}\left[\begin{array}{c}
		\chi^{s}\\
		\frac{\bm{\sigma}\cdot\bm{p}}{E(\bm{p})+M}\chi^{s}
	\end{array}\right]e^{-ip\cdot x},
\end{eqnarray}
with $\chi^s$ the spin wave functions of two components. With a convenient choice $v^\mu=(1,\bm 0)$, one gets
\begin{eqnarray}
H(x,t)&=&\sqrt{E(\bm{p})+M}\left[\begin{array}{c}
	\chi^{s}\\
	0
\end{array}\right]e^{-i[E(\bm{p})-M]t+i\bm{p}\cdot\bm{x}},\nonumber\\h(x,t)&=&\sqrt{E(\bm{p})+M}\left[\begin{array}{c}
	0\\
	\frac{\bm{\sigma}\cdot\bm{p}}{E(\bm{p})+M}\chi^{s}
\end{array}\right]e^{-i[E(\bm{p})-M]t+i\bm{p}\cdot\bm{x}}.
\end{eqnarray}
Therefore, the light field $H$ and heavy field $h$ correspond to the large and small components of the wave function, respectively, in which $h$ is $1/M$ suppressed in comparison with $H$. 

 In order to obtain the effective field theory with the light field $H$ only, one can either substitute the $h$ with the assistance of the equation of motion~\cite{Neubert:1993mb} (in the classical sense) or integrate out the heavy field $h$ in the path integral approach~\cite{Bernard:1992qa,Neubert:1993mb} (in the quantum sense). We here present the latter derivation explicitly. The generating functional of the Lagrangians is
 \begin{equation}
 	Z[\eta,\bar{\eta},v,a,s,p]=\int[d\Psi][d\bar{\Psi}][du]e^{ i \left[S+\int d^{4}x(\bar{\eta}\Psi+\bar{\Psi}\eta)\right]},
 \end{equation}
where the $\eta $ and $\bar{\eta}$ are the external fields. The $v$, $a$, $s$ and $p$ are possible external fields in the action $S$. $u$ is the field (e.g., the Goldstone field) other than the fermion field $\Psi$. The external field $\eta$ can be decomposed into the light part and heavy part,
	\begin{equation}
		R=\frac{1}{2}(1+\slashed{v})e^{iMv\cdot x}\eta,\quad\rho=\frac{1}{2}(1-\slashed{v})e^{iMv\cdot x}\eta.
	\end{equation}
 The action corresponding to Lagrangian~\eqref{eq:app2:generalL} can be rewritten as
 \begin{equation}
	S+\int d^{4}x(\bar{\eta}\Psi+\bar{\Psi}\eta)=\int d^{4}x[\bar{H}{\cal A}H+\bar{H}\gamma^{0}{\cal B}^{\dagger}\gamma_{0}{\cal C}^{-1}{\cal B}H-\bar{h}'{\cal C}h'+\bar{R}H+\bar{H}R],
\end{equation}
with the substitution $h'=h-{\cal C}^{-1}({\cal B}H+\rho)$. Integrating out $h'$, one obtains
 \begin{equation}
 	Z[\eta,\bar{\eta},v,a,s,p]=\int[dH][d\bar{H}][du]\Delta_{h}e^{ i\int d^{4}x[\bar{H}{\cal A}H+\bar{H}\gamma^{0}{\cal B}^{\dagger}\gamma_{0}{\cal C}^{-1}{\cal B}H+\bar{R}H+\bar{H}R]},
 \end{equation}
where $\Delta_{h}$ is the resulting constant by integrating out $h'$. Therefore, we obtain the effective Lagrangian,
\begin{equation}
	{\cal L}_{\mathrm{eff}}=\bar{H}({\cal A}+\gamma^{0}{\cal B}^{\dagger}\gamma_{0}{\cal C}^{-1}{\cal B})H,~\label{eq:sec1.5:lightandheavy}
\end{equation}
where the $\bar{H}\mathcal{A}H$ is the LO in the heavy field expansion. The second term is the recoiling effect suppressed by the powers of $1/M$. In the example in Eq.~\eqref{eq:heavyFE}, the expansion of $\mathcal{C}^{-1}$ read,
\begin{eqnarray}
	{\cal C}^{-1}=\frac{1}{2M}-\frac{i(v\cdot\mathcal{D})}{(2M)^{2}}+....
\end{eqnarray}

In the definition of $H$ and $h$, we introduce a specific $v$ with $v^2=1$. In the strict notation, the light field and heavy filed should be $H_v$ and $h_v$ depending on the choice of $v$. The introduction of $v$ will break the Lorentz invariance. However, one can constrain the theory independent of the choice of $v$ to restore the Lorentz symmetry partially, which is called the reparameterization invariance~\cite{Luke:1992cs,Boyd:1994pa}. For example, $i\bar{H}_{v}v\cdot{\cal D}H_{v}$, the Lagrangian in the heavy field limit of Eq.~\eqref{eq:heavyFE} cannot fulfill the reprarameterization invariance. One can introduce the flollowing tranformation,
\begin{equation}
    i\bar{H}_{v}v\cdot DH_{v}\to i\bar{H}_{v}v\cdot DH_{v}+\bar{H}_{v}\frac{(iD)^{2}}{M}H_{v}+\mathcal{O}\left(\frac{1}{M^{2}}\right),
\end{equation}
where the recoiling effect is introduced to make the term the reparameterization invariant up to the discrepancy at $\mathcal{O}(1/M^2)$.

\begin{table}
	\centering
	\renewcommand{\arraystretch}{1.5}
	\caption{The reductions of the Dirac matrices in heavy baryon formalism (H.B.F) via $\Lambda_{+}\Gamma\Lambda_+$, where $\Lambda_{+}={1\over 2}(1+\slashed{v})$.}\label{fig:dirac}
	\setlength{\tabcolsep}{6.7mm}
{
	\begin{tabular}{c|cccccc}
		\hline
		$\Gamma$ & $1_{4\times4}$ & $\gamma_{5}$ & $\gamma^{\mu}$ & $\gamma^{\mu}\gamma_{5}$ & $\sigma^{\mu\nu}$ & $\sigma^{\mu\nu}\gamma_{5}$\tabularnewline
		\hline
		H.B.F & $1_{4\times4}$ & 0 & $v^{\mu}$ & $2S^{\mu}$ & $-2\varepsilon^{\mu\nu\rho\sigma}v_{\rho}S_{\sigma}$ & $2i(v^{\mu}S^{\nu}-v^{\nu}S^{\mu})$\tabularnewline
		\hline
	\end{tabular}
	}
\end{table}

In the heavy field expansion, it is very convenient to introduce the Pauli-Lubanski vector
\begin{equation}
	S^{\mu}=\frac{i}{2}\gamma_{5}\sigma^{\mu\nu}v_{\nu}=-\frac{1}{2}\gamma_{5}(\gamma^{\mu}\slashed{v}-v^{\mu}).
	\end{equation}
The Dirac matrices can be reexpressed with the relations in Table~\ref{fig:dirac}. In the spacetime with $d=4$,
\begin{equation}
v\cdot S=0,\quad
\{S^\mu, S^\nu\}=\frac{1}{2}(v^\mu v^\nu-g^{\mu\nu}),\quad
[S^\mu_v,S^\nu_v]=-i\varepsilon^{\mu\nu\rho\sigma}
v_\rho S_\sigma^v.	
\end{equation}
where the Levi-Civita symbol is defined as~\footnote{The Levi-Civita symbol is different from that in Refs.~\cite{Scherer:2002tk,Scherer:2012xha} by a sign.}:
\begin{equation}
	\varepsilon^{\mu\nu\alpha\beta}=\begin{cases}
		+1 & \text{even permutation}\\
		-1 & \text{odd permutation}\\
		0 & \text{others}
	\end{cases}.
\end{equation}

Under the heavy field reduction, the propagators of the spin-$1\over 2$ and spin-$3\over 2$ particles are
\begin{equation}
	\mathcal{P}_{1/2}=\frac{i}{v\cdot q+i\varepsilon},\qquad \mathcal{P}_{3/2}^{\mu\nu}=\frac{-iP^{\mu\nu}}{v\cdot q+i\varepsilon},
\end{equation}
where $P^{\mu\nu}=g^{\mu\nu}-v^{\mu}v^{\nu}+\frac{4}{d-1}S^{\mu}S^{\nu}$ is the projection operator which singles out the spin-$3\over 2$ component from the Rarita-Schwinger field.

For the scalar and vector fields, the free Lagrangian is
\begin{eqnarray}{\cal L}=\mathcal{D}_{\mu}\Phi \mathcal{D}^{\mu}\Phi^{\dagger}-M^{2}\Phi\Phi^{\dagger}-\frac{1}{2}F^{\mu\nu}F_{\mu\nu}^{\dagger}+M^{2}\Phi^{\ast\mu}\Phi_{\mu}^{\ast\dagger},
\end{eqnarray}
where $F_{\mu\nu}=\partial_{\mu}\Phi_{\nu}^\ast-\partial_{\nu}\Phi_{\mu}^\ast$. The heavy field expansions for the particle $P^{(\ast\mu)}$ and antiparticle $\tilde{P}^{(\ast\mu)}$ are
\begin{eqnarray}
 	\sqrt{M}\Phi^{(\ast\mu)}&=&e^{-iMv\cdot x}P^{(\ast\mu)},~~\Rightarrow {\cal L}=-2Piv\cdot \mathcal{D}P^{\dagger}+2P^{\ast\mu}iv\cdot \mathcal{D}P_{\mu}^{\ast\dagger},\\
  	\sqrt{M}\Phi^{(\ast\mu)}&=&e^{iMv\cdot x}\tilde{P}^{(\ast\mu)\dagger},~~\Rightarrow
 {\cal L}=2\tilde{P}{}^{\dagger}iv\cdot \mathcal{D}\tilde{P}-2\tilde{P}^{\ast\dagger\mu}iv\cdot \mathcal{D}\tilde{P}_{\mu}^\ast.
\end{eqnarray}
With the above reductions, the propagators of the scalar and vector heavy particles are
\begin{equation}
	\mathcal{P}_{0}=\frac{i}{2v\cdot q+i\varepsilon},\qquad \mathcal{P}_{1}=-\frac{i(g_{\mu\nu}-v_\mu v_\nu)}{2v\cdot q+i\varepsilon}.
\end{equation}

\section{Electromagnetic form factors}~\label{app:EM_FF}
\subsection{Vector mesons of spin-$1$}
The magnetic moment of vector state is extracted from the matrix element of electromagnetic current as defined in Eq.~\eqref{eq:emcurrent}. The  explicit expression parameterized in a Lorentz covariant form reads~\cite{Arnold:1979cg},
\begin{eqnarray}\label{eq:FormPV}
\mathscr{G}^\mu(q^2)&=&\langle V(p^\prime,\varepsilon^{\prime\ast})|\mathscr{J}_{\text{em}}^\mu(q^2)|V(p,\varepsilon)\rangle \nonumber\\
&=&-\mathscr{G}_1(q^2)(\varepsilon\cdot\varepsilon^{\prime\ast})(p+p^\prime)^\mu 
+\mathscr{G}_2(q^2)\left[(\varepsilon\cdot q)\varepsilon^{\prime\ast\mu}-(\varepsilon^{\prime\ast}\cdot q)\varepsilon^{\mu}\right] 
+\mathscr{G}_3(q^2)\frac{(\varepsilon\cdot q)(\varepsilon^{\prime\ast}\cdot q)}{2m_V^2}(p+p^\prime)^\mu,
\end{eqnarray}
where $p(p^\prime)$ and $\varepsilon(\varepsilon^\prime)$ are the momentum and polarization vector of initial (final) state. $m_V$ is the mass of vector meson $V$. $\mathscr{G}_i(q^2)$ are called electromagnetic form factors that can be related to the charge, quadrupole and magnetic dipole form factors in the Breit frame. The kinetics in Breit frame are given as
\begin{eqnarray}\label{eq:Breitf}
&q^\mu=(p-p^\prime)^\mu=(0,\boldsymbol{Q}),\qquad\boldsymbol{Q}=Q\hat{z},\qquad p^\mu=(p^0, \frac{1}{2}\boldsymbol{Q}),&\nonumber\\
&p^{\prime\mu}=(p^0, -\frac{1}{2}\boldsymbol{Q}),\qquad
-q^2=Q^2\ge0,\qquad p^0=\sqrt{m_V^2+\frac{1}{4}Q^2}.
\end{eqnarray}
The time and space components of Lorentz vector $\mathscr{G}^\mu(q^2)$ in Breit frame are derived as
\begin{eqnarray}
\mathscr{G}^0(Q^2)&=&2p^0\Bigg\{\mathscr{G}_C(Q^2)(\boldsymbol{\varepsilon}\cdot\boldsymbol{\varepsilon}^{\prime\ast})+\frac{\mathscr{G}_Q(Q^2)}{2m_V^2}\Big[(\boldsymbol{\varepsilon}\cdot\boldsymbol{Q})(\boldsymbol{\varepsilon}^{\prime\ast}\cdot\boldsymbol{Q})
-\frac{1}{3}(\boldsymbol{\varepsilon}\cdot\boldsymbol{\varepsilon}^{\prime\ast})Q^2\Big]\Bigg\},\label{eq:GTime}\\
\boldsymbol{\mathscr{G}}(Q^2)&=&\mathscr{G}_2(Q^2)\left[(\boldsymbol{\varepsilon}^{\prime\ast}\cdot\boldsymbol{Q})\boldsymbol{\varepsilon}-(\boldsymbol{\varepsilon}\cdot\boldsymbol{Q})\boldsymbol{\varepsilon}^{\prime\ast}\right]
=2p^0\frac{\mathscr{G}_M(Q^2)}{2m_V}\left[(\boldsymbol{\varepsilon}^{\prime\ast}\cdot\boldsymbol{Q})\boldsymbol{\varepsilon}-(\boldsymbol{\varepsilon}\cdot\boldsymbol{Q})\boldsymbol{\varepsilon}^{\prime\ast}\right],\label{eq:GSpace}
\end{eqnarray}
where $\mathscr{G}_C$, $\mathscr{G}_Q$ and $\mathscr{G}_M$ are the so-called charge, electric quadrupole and magnetic dipole form factors. They are bridged to $\mathscr{G}_i$ as defined in Eq.~\eqref{eq:FormPV} via
\begin{eqnarray}
\mathscr{G}_C&=&\mathscr{G}_1+\frac{2}{3}\eta\mathscr{G}_Q,\qquad
\mathscr{G}_Q=\mathscr{G}_3+\mathscr{G}_2(1+\eta)^{-1}+\frac{1}{2}\mathscr{G}_1(1+\eta)^{-1},\qquad
\mathscr{G}_M=\mathscr{G}_2,
\end{eqnarray}
where $\eta={Q^2}/{(4m_V^2)}$. Note that in deriving Eqs.~\eqref{eq:GTime} and ~\eqref{eq:GSpace}, the transverse conditions $p\cdot\varepsilon=0$, and $p^\prime\cdot\varepsilon^{\prime\ast}=0$ have been used. Then the magnetic moment of a vector meson is defined as
\begin{eqnarray}
\mu_V=\mathscr{G}_M(Q^2)\big|_{Q^2\to 0}.
\end{eqnarray}

The radiative decay width can be expressed in terms of the transition magnetic moment $\mu^\prime(q^2)$, and which is extracted from the following transition matrix element,
\begin{equation}\label{eq:FormPVNon}
\langle P(p^\prime)|\mathscr{J}_{\text{em}}^\mu(q^2)|V(p,\varepsilon)\rangle=e\sqrt{m_Vm_P} \mu^\prime(q^2) \epsilon^{\mu\nu\alpha\beta} v_\nu q_\alpha\varepsilon_{\beta}.
\end{equation}
Performing the standard procedure for calculating the decay width of $1\to 2$ process, the radiative decay width of $V\to P\gamma$ is given as
\begin{equation}
\Gamma\left[V\to P\gamma\right]=\frac{\alpha}{3}\frac{m_P}{m_V}\left|\mu^\prime(0)\right|^2|\boldsymbol{q}|^3,
\end{equation}
with $\alpha=1/137$ the fine structure constant. The transition magnetic moment $\mu_{V\to P\gamma}$ is defined as
\begin{equation}
\mu_{V\to P\gamma}=\frac{e}{2}\mu^\prime(0).
\end{equation}

\subsection{Baryons of spin-$\frac{1}{2}$ and spin-$\frac{3}{2}$}

The parameterizations of the electromagnetic currents for the spin-$\frac{1}{2}$~\cite{Wang:2018gpl} and spin-$\frac{3}{2}$~\cite{Meng:2018gan} baryons (or the transition processes~\cite{Wang:2018cre})  have been given with the relativistic and nonrelativistic forms (or see~\cite{Jones:1972ky,Nozawa:1990gt,Faessler:2006ky,Ledwig:2011cx}). The (transition) magnetic moments can be extracted from the corresponding electromagnetic form factors. We summarize the nonrelativistic froms in the following,
\begin{itemize}

  \item[(1)] {\it Spin-$\frac{1}{2}$ baryons}:
  \begin{eqnarray}
\langle B_{\bar{\bm3}}\left(p^{\prime}\right)\left|\mathscr{J}^{\mu}_{\text{em}}\right| B_{\bar{\bm3}}(p)\rangle=\bar{u}\left(p^{\prime}\right)\left[v_{\mu} \mathscr{G}_{E}(q^{2})+\frac{[S_{\mu}, S \cdot q]}{m_{B}} \mathscr{G}_{M}(q^{2})\right] u(p),
\end{eqnarray}
where $\mathscr{G}_E$ and $\mathscr{G}_M$ are the electric and magnetic form factors, respectively. $S_\mu=\frac{i}{2}\gamma^5\sigma_{\mu\nu}v^\nu$ is the covariant spin operator. The magnetic moment is given by $\mu_B=\frac{e}{2m_B}\mathscr{G}_M(0)$. Although it is given with the $\bar{\bm 3}_f$ baryons, the $\bm 6_f$ ones take the same form.

  \item[(2)] {\it Spin-$\frac{3}{2}$ baryons}:
  \begin{equation}
\langle B_{\bm6}^\ast\left(p^{\prime}\right)\left|\mathscr{J}^{\mu}_{\text{em}}\right| B_{\bm6}^\ast\rangle=\bar{u}^{\rho}\left(p^{\prime}\right) \mathscr{O}_{\rho \mu \sigma}\left(p^{\prime}, p\right) u^{\sigma}(p),
\end{equation}
with
\begin{equation}
\mathscr{O}_{\rho \mu \sigma}\left(p^{\prime}, p\right)=-g_{\rho \sigma}\left[v_{\mu}\left(\mathscr{F}_{1}-\tau \mathscr{F}_{2}\right)+\frac{[S_{\mu}, S_{\alpha}]}{m_{B}} q^{\alpha}\left(\mathscr{F}_{1}+\mathscr{F}_{2}\right)\right]
-\frac{q^{\rho} q^{\sigma}}{4 m_{B}^{2}}\left[v_{\mu}\left(\mathscr{F}_{3}-\tau \mathscr{F}_{4}\right)+\frac{[S_{\mu}, S_{\alpha}]}{m_{B}} q^{\alpha}\left(\mathscr{F}_{3}+\mathscr{F}_{4}\right)\right],
\end{equation}
from which the charge, electric quadrupole, magnetic dipole, and magnetic octupole form factors are defined as
\begin{eqnarray}
\mathscr{G}_{C}(q^{2})&=&\mathscr{F}_{1}-\tau \mathscr{F}_{2}+\frac{2}{3} \tau \mathscr{G}_{E 2},\qquad
\mathscr{G}_{Q}(q^{2})=\mathscr{F}_{1}-\tau \mathscr{F}_{2}-\frac{1}{2}(1+\tau)\left(\mathscr{F}_{3}-\tau \mathscr{F}_{4}\right),\nonumber \\
\mathscr{G}_{M}(q^{2})&=&\mathscr{F}_{1}+\mathscr{F}_{2}+\frac{4}{5} \tau \mathscr{G}_{M 3},\qquad~
\mathscr{G}_{M 3}(q^{2})=\mathscr{F}_{1}+\mathscr{F}_{2}-\frac{1}{2}(1+\tau)\left(\mathscr{F}_{3}+\mathscr{F}_{4}\right).
\end{eqnarray}
where $\tau=-q^2/(4m_B^2)$. The magnetic moment is then given as $\mu_B=\frac{e}{2m_B}\mathscr{G}_{M}(0)$.
 \item[(3)] {\it Spin-$\frac{1}{2}$$\to$spin-$\frac{1}{2}+\gamma$ transitions}:
 \begin{equation}
\langle\psi\left(p^{\prime}\right)\left|\mathscr{J}^{\mu}_{\text{em}}\right| \psi(p)\rangle=e \bar{u}\left(p^{\prime}\right)\left[\left(v_{\mu}-\frac{\delta}{q^{2}} q_{\mu}\right) \mathscr{G}_{E}(q^{2})\right.\left.+\frac{2\left[S^{\mu}, S^{\nu}\right] q_{\nu}}{m+m^{\prime}} \mathscr{G}_{M}(q^{2})\right] u(p).
\end{equation}
The decay width is expressed by the magnetic form factor $\mathscr{G}_M(0)$ as
\begin{equation}
\Gamma=\frac{4 \alpha |\bm{p}_{\gamma}|^{3}}{\left(m+m^{\prime}\right)^{2}}\left|\mathscr{G}_{M}(0)\right|^{2}.
\end{equation}
\item[(4)] {\it Spin-$\frac{3}{2}$$\to$spin-$\frac{1}{2}+\gamma$ transitions}:
\begin{equation}
\langle\psi_{6^{*}}\left|\mathscr{J}_{\text{em}\mu}\right| \psi\rangle=e \bar{u}^{\rho}\left(p^{\prime}\right) \left[2 \mathscr{G}_{1}(q^{2})\left(q_{\rho} S_{\mu}-q \cdot S g_{\rho \mu}\right)+\mathscr{G}_{2}(q^{2}) \frac{2 m^{\prime}}{m+m^{\prime}}\left(q_{\rho} v_{\mu}-q \cdot v g_{\rho \mu}\right) q \cdot S\right] u(p),
\end{equation}
which is suitable for the on-shell transitions (for complete form, see Ref.~\cite{Wang:2018cre}). This form factor corresponding to leading  M1 and E2 transitions are given by
\begin{equation}
\begin{aligned}
\mathscr{G}_{M}(q^{2})=& \frac{1}{4}\left[\mathscr{G}_{1} \frac{m_{+}\left(3 m^{\prime}+m\right)-q^{2}}{m^{\prime}}+\mathscr{G}_{2}\left(m_{+} m_{-}-q^{2}\right)\right.
\left.+2\left(\mathscr{G}_{3}+\mathscr{G}_{2}\right) q^{2}\right], \\
\mathscr{G}_{Q}(q^{2})=& \frac{1}{4}\left[\mathscr{G}_{1} \frac{m_{+} m_{-}+q^{2}}{m^{\prime}}+\mathscr{G}_{2}\left(m_{+} m_{-}-q^{2}\right)\right.
\left.+2\left(\mathscr{G}_{2}+\mathscr{G}_{3}\right) q^{2}\right],
\end{aligned}
\end{equation}
with $m_{\pm}=m^\prime\pm m$. Then one can find the ratio
\begin{equation}
\frac{\mathscr{G}_{Q}(0)}{\mathscr{G}_{M}(0)}=\frac{m_-}{m}\left(\frac{1}{4}+\frac{\mathscr{G}_2}{4\mathscr{G}_1}\right)
\end{equation}
shall be very small due to a suppressed factor $m_-/m$, which can qualitatively interpret the ratio obtained by Savage~~\cite{Savage:1994wa}. The decay widths are given with the helicity amplitudes as
\begin{equation}
\Gamma=\frac{m m^{\prime}}{8 \pi}\left(1-\frac{m^{2}}{m^{\prime 2}}\right)^{2}\left[A_{3 / 2}^{2}(0)+A_{1 / 2}^{2}(0)\right],
\end{equation}
with the helicity amplitudes
\begin{equation}
A_{3 / 2}\left(q^{2}\right)=-\sqrt{\frac{\pi \alpha \omega}{2 m^{2}}}\left[\mathscr{G}_{M}\left(q^{2}\right)+\mathscr{G}_{Q}\left(q^{2}\right)\right],\qquad
A_{1 / 2}\left(q^{2}\right)=-\sqrt{\frac{\pi \alpha \omega}{6 m^{2}}}\left[\mathscr{G}_{M}\left(q^{2}\right)-3 \mathscr{G}_{Q}\left(q^{2}\right)\right],
\end{equation}
where $\omega=(m^{\prime2}-m^2+q^2)/2m^\prime$.
\end{itemize} 
\end{appendix}

\bibliographystyle{elsarticle-num}
\bibliography{ref}


\end{document}